%% file: Oxford_Thesis.tex
\documentclass[a4paper,oneside]{ociamthesis}

\correctionstrue

\usepackage[numbers,sort&compress]{natbib}  

	\newcommand{\rd}{\textsuperscript{rd}}
	\newcommand{\JHEP}{\textit{JHEP }}
	\newcommand{\PRD}{\textit{Phys. Rev. D }}
	\newcommand{\EH}{\text{EH}}
	
\DeclarePairedDelimiter{\ceil}{\lceil}{\rceil}
\newcommand{\ssc}{\scriptscriptstyle}
\newcommand{\diff}{\mathrm{d}}

\newcommand{\diffmone}{{\diff^{d-1}}}
\newcommand{\req}[1]{(\ref{#1})}
\newcommand{\ie}{{i.e.,}\ }
\newcommand{\e}[1]{\text{e}^{#1}}
\newcommand{\eg}{{e.g.,}\ }
\newtheorem{theorem}{Theorem}
\newtheorem{corollary}{Corollary}

\newtheorem{defi}{Definition}

\newtheorem{conjecture}{Conjecture}
\newcommand{\X}{\pazocal{X}}
\newcommand{\I}{\pazocal{I}}
\newcommand{\K}{\pazocal{K}}
\newcommand{\W}{\pazocal{W}}
\newcommand{\R}{\pazocal{R}}
\newcommand{\Z}{\pazocal{Z}}
\newcommand{\Y}{\pazocal{Y}}

\newcommand{\B}{\pazocal{B}}
\newcommand{\C}{\pazocal{C}}
\newcommand{\F}{\pazocal{F}}
\newcommand{\D}{\pazocal{D}}
\newcommand{\E}{\pazocal{E}}
\newcommand{\EE}{\text{EE}}

\newcommand{\M}{\pazocal{M}}

\newcommand{\rt}{\text{RT}}
\newcommand{\ren}{\text{ren}}
\newcommand{\vol}{\text{Vol}}
\newcommand{\Area}{\text{Area}}
\newcommand{\GN}{G_\text{N}}

\newcommand{\lnr}{^{\mathrm{L}}}

\newcommand{\iu}{\text{i}}

\newcommand{\deth}{\sqrt{h}}
\newcommand{\SEE}{S_{\text{EE}}}
\newcommand{\see}{S_{\text{EE}}}
\newcommand{\suniv}{s^{\text{univ}}}
\newcommand{\SEMI}{S_{\text{EE}}^{\rm \ssc EMI}}

\newcommand{\IEren}{I^{\text{ren}}_{\text{E}}}

\newcommand{\RR}{\pazocal{R}_2}
\newcommand{\RRR}{\pazocal{R}_3}
\newcommand{\Ls}{L_\star}
\newcommand{\bF}[1]{\bar{\pazocal{F}}_{#1}}

\newcommand{\rhor}{r_{\mathrm{h}}}
\newcommand{\yhor}{y_{\mathrm{h}}}
\newcommand{\xhor}{x_{\mathrm{h}}}
\newcommand{\IE}{I_\text{E}}

\makeatletter
\newcommand{\dal}{\mathop{\mathpalette\dal@\relax}}
\newcommand{\dal@}[2]{%
  \begingroup
  \sbox\z@{$\m@th#1\square$}%
  \dimen0=\fontdimen8
    \ifx#1\displaystyle\textfont\else
    \ifx#1\textstyle\textfont\else
    \ifx#1\scriptstyle\scriptfont\else
    \scriptscriptfont\fi\fi\fi3
  \makebox[\wd\z@]{%
    \hbox to \ht\z@{%
      \vrule width \dimen0
      \kern-\dimen0
      \vbox to \ht\z@{
        \hrule height \dimen0 width \ht\z@
        \vss
        \hrule height 2\dimen0
      }%
      \kern-2.5\dimen0
      \vrule width 2.5\dimen0
    }%
  }%
  \endgroup
}
\makeatother

\newcommand{\SKt}{S_{\mathrm{KT}}}

\newcommand{\detsigtil}{\sqrt{|\tilde{\sigma}|}}

\usepackage{calrsfs}
\DeclareMathAlphabet{\pazocal}{OMS}{zplm}{m}{n}
\begin{document}

\setlength{\textbaselineskip}{15pt plus2pt minus1pt}

\setlength{\frontmatterbaselineskip}{15pt plus1pt minus1pt}

\setlength{\baselineskip}{\textbaselineskip}


\setcounter{secnumdepth}{2}
\setcounter{tocdepth}{2}


\begin{romanpages}

\input{text/titlepage}


\newpage
\thispagestyle{empty}
\null
\vfill
\begin{center}
    { \sffamily\large Director de Programa de Doctorado en Ciencias Físicas\\[1 ex] 
        \textbf{Dr. Nelson Videla} (PUCV)\\[3 ex]}
    {\sffamily\large Comisión de Evaluación de la Tesis\\[1 ex] 
        \textbf{Dr. Ignacio J. Araya} (UNAP)\\[1 ex]
        \textbf{Dr. Dumitru Astefanesei} (PUCV)\\[1 ex]
        \textbf{Dr. Pablo Bueno} (CERN)\\[1 ex]} 
\end{center}

\begin{acknowledgements}
 	\input{text/acknowledgements}
\end{acknowledgements}

\begin{abstract}
	\input{text/abstract}
\end{abstract}

\begin{resumen}
	\input{text/resumen}
\end{resumen}


\begin{listofp}
The following articles, listed in inverse chronological order, were published by the candidate and their collaborators during the realization of the thesis:
\begin{enumerate}[label={[\arabic*]}]

\item Pablo Bueno, Pablo A. Cano, R. A. Hennigar, Mengqi Lu, Javier Moreno, \textit{Generalized quasi-topological gravities: the whole shebang}, \href{https://arxiv.org/abs/2203.05589}{2203.05589}. 

\item Pablo Bueno, Pablo A. Cano, Quim Llorens, Javier Moreno, Guido van der Welde, \textit{Aspects of three-dimensional higher curvatures gravities}, \textit{Class. Quant. Grav.} \textbf{39} (2022) 125002, \href{https://arxiv.org/abs/2201.07266}{[2201.07266]}. 

\item Pablo Bueno, Horacio Casini, Oscar Lasso Andino, Javier Moreno,  \textit{Disks globally maximize the entanglement entropy in 2+1 dimensions}, \JHEP \textbf{10} (2021) 179 [\href{https://arxiv.org/abs/2107.12394}{2107.12394}]

\item Pablo Bueno, Pablo A. Cano, Javier Moreno, Guido van der Velde, \textit{Regular black holes in three dimensions}, \PRD \textbf{104} (2021) L021501 [\href{https://arxiv.org/abs/2104.10172}{2104.10172}]

\item Giorgos Anastasiou, Ignacio J. Araya, Javier Moreno, Rodrigo Olea, David Rivera-Betancour, \textit{Renormalized holographic entanglement entropy for Quadratic Curvature Gravity}, \PRD \textbf{104} (2021) 086003 [\href{https://arxiv.org/abs/2102.11242}{2102.11242}]

\item Giorgos Anastasiou, Javier Moreno, Rodrigo Olea, David Rivera-Betancour, \textit{Shape dependence of renormalized holographic entanglement entropy}, \JHEP \textbf{09} (2020) 173 [\href{https://arxiv.org/abs/2002.06111}{2002.06111}]

\item Pablo Bueno, Pablo A. Cano, Javier Moreno, \'Angel Murcia, \textit{All higher-curvature gravities as Generalized quasi-topological gravities}, \JHEP \textbf{11} (2019) 062 [\href{https://arxiv.org/abs/1906.00987}{1906.00987}]
 \end{enumerate}
\end{listofp}

\begin{conventions}
Throughout the thesis we use the following conventions. We work in natural units $c=\hbar=1$ but the Newton constant $\GN$ is shown explicitly. Lorentzian manifolds are equipped with metrics with mostly plus signs, assigning the minus sign to the timelike coordinate, \ie $(-,+,+,\ldots)$. The number of dimensions is denoted by $D=d+1$. The left hand side is mostly used when discussing gravity theories in generality and the right hand side in the context of the gauge/gravity duality, as $d$ stands for the number of dimensions of the dual conformal field theory. We also make explicit distinction when referring to tensors and indices in both frameworks. In the gravity side, tensors are written in uppercase and labeled by Roman indices $a,b,c,d,e,f,m,n$. In the conformal field theory side defined on the boundary, tensors are written calligraphically, with Greek indices $\mu,\nu,\rho,\sigma,\iota,\kappa$. Indices $i,j,k,l$ and $\alpha,\beta,\gamma,\delta$ are used for codimension-two objects with respect to the bulk- and the boundary theory respectively. Other conventions on different tensors include:
\begin{itemize}
  \item[$\cdot$] Symmetric and antisymmetric parts of a tensor $T_{a_1\ldots a_p a_{p+1}\ldots a_q}$ can be written between parenthesis and squared brackets as
  \begin{align}
      T_{(a_1\ldots a_p)a_{p+1}\ldots a_{q}}&\equiv\frac{1}{p!}\sum_{\sigma}T_{a_{\sigma(1)}\ldots a_{\sigma(p)}a_{p+1}\ldots a_{q}},\\ T_{[a_1\ldots a_p]a_{p+1}\ldots a_{q}}&\equiv\frac{1}{p!}\sum_{\sigma}\mathrm{sgn}(\sigma)T_{a_{\sigma(1)}\ldots a_{\sigma(p)}a_{p+1}\ldots a_{q}},
  \end{align}
  where $\sigma$ ranges over the permutations of the numbers 1 to $p$ and sgn$(\sigma)$ denotes the signature of the permutation. If the (anti-)symmetrized indices are not contiguous, the excluded indices lie between bars, \eg $T_{(a_1| a_2|\ldots a_p)a_{p+1}\ldots a_{q}}$.
  
 \item[$\cdot$] The covariant derivative of a certain tensor with mixed indices $T\indices{_a^b}$ reads
 \begin{equation}
     \nabla_c T\indices{_a^b}\equiv \partial_c T\indices{_a^b}-\Gamma_{c a}^dT\indices{_d^b}+\Gamma_{c d}^bT\indices{_a^d}, 
 \end{equation}
 where $\Gamma_{ab}^c\equiv\frac{1}{2}g^{cd}\left(\partial_a g_{bd}+\partial_b g_{da}-\partial_dg_{ab}\right)$ is the Levi-Civita connection (or Christoffel symbols) and $g_{ab}$ the metric.
 
\item[$\cdot$] Given a vector $V^c$, the commutator of the covariant derivative defines the Riemann tensor of the spacetime as
    \begin{equation}
        \left[\nabla_a,\nabla_b\right]V^c=R\indices{^c_d_a_b}V^d,
    \end{equation}
where we identified $R\indices{^c_d_a_b}=2\partial_{[a}\Gamma_{b]d}^c+\Gamma_{e [a}^c\Gamma_{b] d}^e$.

\item[$\cdot$] The extrinsic curvature of a $d$-dimensional surface $\Sigma$ with metric $\gamma_{\mu\nu}$ embedded in a spacetime one dimension higher is given by
\begin{equation}
    K_{\mu\nu}=\nabla_\mu n_\nu,
\end{equation}
where $n_\nu$ is the unit normal to $\Sigma$.
\end{itemize}

\end{conventions}


\flushbottom

\tableofcontents




\end{romanpages}

\flushbottom
\include{text/ch1-intro}

\begin{part}{Higher-curvature gravity}

\include{text/ch2-gqtorder}

\include{text/ch3-fieldred}

\include{text/ch4-hcg3d}

\include{text/ch5-bh3d}

\end{part}

\begin{part}{Entanglement entropy}

\include{text/ch5-hentang}

\include{text/ch6-hentangqcg}

\include{text/ch7-entang}

\include{text/ch8-concl}

\end{part}


\startappendices
\include{text/appendices}


%
%

\bibliographystyle{JHEP}
\bibliography{references}

\end{document}

%% file: text/titlepage.tex
\begin{titlepage}

    \sffamily
    \begin{center}
        \vfil
 {\Huge
            \rule[1 ex]{\textwidth}{2.5 pt}
            \onehalfspacing\textbf{Higher-Curvature Gravity and Entanglement Entropy}\\
            \rule[-1 ex]{\textwidth}{2.5 pt}
        }
        \vspace{1cm}
        \vfil
        {\large Tesis entregada a la Pontifica Universidad Católica de Valparaíso\\[1.25 ex]
        \large en cumplimiento parcial de los requisitos para optar al grado de\\[1.25 ex]
        \large Doctor en Ciencias Físicas\\[1.25 ex]
        \large por\\[1.25 ex]
        \textbf{Francisco Javier Moreno Gonz\'alez}\\[1.25 ex]
        \large Julio 2022
       }
        \vfil
        {\large Directores de tesis\\[1 ex] 
        \textbf{Dra. Olivera Mi\v skovi\'c} (PUCV)\\[1 ex]
        \textbf{Dr. Rodrigo Olea} (UNAB)\\[1.25 ex] }
        \vspace{1cm}
        \vfil
 \includegraphics[height = 4 cm]{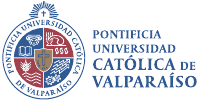}\\
        \vfil
    \end{center}

\end{titlepage}

%% file: text/acknowledgements.tex
Cruzar el atlántico para realizar los estudios de doctorado no es algo muy común entre mis conocidos, menos aún si el país de destino está en América del Sur. A uno siempre le puede inquietar si va a encontrar comodidad o si va a perder lazos con la gente que conformaba su vida anteriormente. Por suerte, en mi caso puedo decir que 
que no solo he conseguido mantener a mis círculo previo sino que además he encontrado nuevos amigos y amigas por el camino que deseo mantener. Esta sección es para agradecer su participación, apoyo y hospitalidad en algunas de las fases de la presente tesis. Por ello, voy a tratar de nombrarlos por orden de aparición.

Por supuesto, nada de esto habría sido posible de no ser por mis supervisores, Rodrigo Olea y Olivera Mi\v skovi\'c, a quienes debo agradecer que me acogieron como uno más desde el primer día y cuyo apoyo e implicación en mi formación como físico ha sido determinante para que haya podido culminar la presente tesis. En los inicios, también debo agradecer la amistad y las interacciones con Georgios, que me ayudó mucho a tomar mis primeros pasos en investigación. Posteriormente, en este hemisferio también tuve la suerte de encontrarme con Pablo Bueno, que también ha sido uno de los pilares de esta tesis, ofreciéndome apoyo constantemente tanto académico como personal. Tampoco puedo dejar de agradecer la frecuente ayuda de Pablo Cano, quien me propuso un proyecto en mi primer año de doctorado y puso la primera piedra de una de las líneas de investigación que conforman esta tesis. Finalmente, también me gustaría agradecer la atención de Ignacio Araya, que también tuvo que soportar durante un tiempo mi asedio de preguntas. Todos estos son los actores que más han contribuido al desarrollo de la tesis, a mi formación como físico y como inspiración para la labor de un buen docente, por eso los tendré presente en toda mi trayectoria profesional. Me gustaría también agradecer los largos días de trabajo en nuestros proyectos a otros colaboradores ---en la cercanía y la distancia impuesta por la pandemia--- como son Horacio Casini, Robie Hennigar, Mengqi Lu, Óscar Lasso Andino, Quim Llorens, Ángel Murcia, David Rivera-Betancour y Guido van der Velde. 

Asimismo, he encontrado gente maravillosa tanto en Chile como en cada uno de los lugares que he realizado estancias. Respecto a Chile, me gustaría agradecerle a muchos de mis compañeros, compañeras y profesores tales como  Apratim, Arindam, Elram, Fabián, Felipe Díaz, Felipe Olivares, Gabriel, Jeremy, Jorge, Juan Calles, Juan Marín, Lina, Pau, Paula, Maritza, Menka, Clément, Omar, Radouane, Romina y Yolbeiker. Algunos me enseñaron física, otros me aportaron una inestimable hospitalidad, y en numerosas ocasiones, ambas.

Acabando mi primer año realicé mi primera estancia en Madrid gracias a Tomás Ortín, que nunca ha dudado en ayudarme desde que me decanté por estudiar física teórica. Allí me acogieron y me hicieron sentir como en casa, Alejandro, Ángel, David y el susodicho Pablo Cano. Posteriormente conocí a nuevos miembros de la familia como Matteo y Zach, quienes pienso que han tenido la enorme suerte de estudiar junto a Tomás y los demás miembros de su grupo. 

Al año siguiente, ya tras haber coincidido con Pablo Bueno, me brindó la fantástica oportunidad de visitarle en Bariloche e involucrarme en su grupo. Conservo un gran recuerdo de este período, en el que tuve la suerte de encontrarme con Andrea, Felipe, Gonzalo, Horacio Casini, Marina, Mario, Martín y Valentín. 
En este entorno fue cuando me encontré con un compañero y colaborador habitual Guido van der Velde, que ya mencioné anteriormente.

También me gustaría agradecer la invitación para visitar Davis a Mukund y Veronika. Por desgracia estuvo acompañada de una constante incertidumbre por causa de la pandemia pero pude volver a un anhelado ambiente de trabajo agradable. Esto fue posible gracias a Andrew, César, Chris, Kento, Sean, Shruti, Temple y en especial Julio, que tuvo que soportar mis innumerables preguntas y lo hizo siempre con la mayor dedicación. 

La última de mis visitas, a Leuven, fue posible gracias a la invitación de Pablo Cano y Thomas Hertog. En ella, tanto ellos como Kwinten, Robert y Sergio, entre otros, me acogieron cálidamente y me involucraron en las actividades cotidianas del departamento. En este lugar fue donde la presente tesis ha sido en gran parte redactada y por eso se me quedarán asociadas para la posteridad. En este tiempo, también tuve la gran suerte de coincidir con una de mis personas favoritas, Marta Acedo, que me apoyó enormemente durante toda mi estancia.

Por otro lado, también quiero agradecer el apoyo incondicional a mis amigos y amigas, brindado a través de Discord siempre que lo necesitaba y con quienes, a pesar de la distancia, no he perdido ni un ápice de confianza. Son muchas las personas que habría que nombrar aquí. Seguro que muchos se sienten interpelados con el término englobador  de ``Kultura'', pero me voy a permitir mencionar explícitamente a Abraham, Ale Pliego, Alejandra, Carlos, Carmen Belloso, Carmen Gómez, Chaquetas, Chema, Dani, Duntless, Elena Arjona, Elena López, Elva, Ester, Íñigo, Jesús, José Alberto, Juanpe, Juan Pablo, Laura, Maite, María José, Marta Ternero, Mercerces, Migue, Núñez y Pedro, entre muchos otros.

Para concluir los agradecimientos personales, quiero expresar mi agradecimiento a mi familia, incluyendo a mis abuelos, tíos, primos ---incluyendo a los nuevos miembros y a los que quedan en la memoria--- con especial mención a mis padres, hermanos y mi tía. Ellos se esforzaron en inculcarme la inquietud para crecer como persona y por extensión en el ámbito académico. El mismo objetivo lo ha compartido Mari, mi mujer, a quien he reservado nombrarla para el final, para que el lector recuerde especialmente su sacrificio, dedicación y apoyo fundamental que me ha aportado para escribir esta tesis. Gracias.

Esta tesis también ha sido posible gracias al apoyo financiero aportado por la Pontificia Universidad Católica de Valparaíso y por la Agencia Nacional de Investigación de Desarrollo (ANID), a través del programa de becas de doctorado nacional, folio No. 21190234. También agradezco el apoyo de Fondo Nacional de Desarrollo Científico y Tecnológico (FONDECYT) Regular No. 1190533 \emph{Black holes and asymptotic symmetries} (Investigadora Responsable (IR) Olivera Mi\v skovi\'c), FONDECYT Regular No. 1170765 \emph{Boundary dynamics in anti-de Sitter gravity and gauge/gravity duality} (IR Rodrigo Olea) y el proyecto ANID-SCIA-ANILLO ACT210100 \emph{Holography and its applications to high energy physics, quantum gravity and condensed matter systems}.

%% file: text/abstract.tex
Higher-curvature extensions of Einstein gravity have been considered in different contexts in order to address unsolved questions in high energy physics. In this thesis we focus on their applications as toy models to probe phenomena in strongly correlated systems using the gauge/gravity duality, as they represent inequivalent theories to Einstein gravity that allows to gain insights into universal properties of the conformal field theory (CFT) under consideration. In this context, we are particularly interested in the family of theories known as generalized quasi-topological gravities, whose defining property is that their equations of motion for statically spherically symmetric solutions are of second order, at most and lack of ghost modes in their linearized spectrum. Here, we characterize the number of theories of these types existing at a given curvature order and dimensions. Moreover, we claim that any effective higher-curvature theory is connected, via field redefinitions to some generalized quasi-topological gravity. This situation is special in the case of three spacetime dimensions, as theories of this type have trivial equations of motion. However, when matter fields are added into the picture, the equations of motion become non-trivial, describing, among other solutions, black holes that represent a multiparametric generalization of the Bañados-Teitelboim-Zanelli black hole. 

From the CFT side, entanglement entropy arises as a prominent quantity that encodes important information about the field theory, such as the type A and type B trace anomalies in even dimensions and the sphere free energy of the theory in odd dimensions when considering spherical entangling regions. As entanglement entropy also includes divergences, we employ the Kounterterms scheme to extract the physically relevant quantities. In the case of three-dimensional CFTs dual to Einstein gravity, we show that the finite part is isolated and can be written in terms of the Willmore energy. Exploiting its properties, we manage to give an upper bound on holographic entanglement entropy. We extend this remarkable result to arbitrary CFTs under consideration. Besides, we show the validity of the Kounterterms scheme for general quadratic curvature gravity, extracting the type A, type B anomalies of the theory in even dimensions and the sphere free energy in odd ones.

%% file: text/resumen.tex
Las extensiones de mayor curvature a la gravedad de Einstein han sido consideradas en diferentes contextos con el fin de abordar problemas abiertos en física de altas energías. En esta tesis nos enfocamos en su uso como modelos de juguete para explorar fenómenos en sistemas fuertemente acoplados usando la dualidad gravedad/gauge. Dado que proporcionan teorías diferentes a gravedad de Einstein, podemos adquirir una intuición acerca de propiedades universales de la teoría conforme de campos de interés. En este contexto, estamos particularmente interesados en una familia de teorías conocidas como gravedades cuasitopológicas generalizadas, cuya propiedad fundamental es el hecho de que sus ecuaciones de movimiento son de segundo orden en la curvatura, como mucho, y carecen de modos fantasma en su espectro linealizado. En este trabajo, caracterizamos el número de teorías de este tipo existentes a un cierto orden de curvatura y dimensión. Asimismo, mostramos que cualquier teoría efectiva de mayor curvatura está conectada, a través de redifiniciones del campo a alguna gravedad cuasitopológica generalizada. Mención aparte merece el caso de espaciotiempos de tres dimensiones, ya que este tipo de familias tienen ecuaciones de movimiento trivial. No obstante, cuando se añade materia, las ecuaciones de movimiento se vuelven no triviales, describiendo, entre otras soluciones, agujeros negros que representan generalizaciones multiparamétricas del agujero negro Bañados-Teitelboim-Zanelli.

En el lado de las CFTs, entropía de entrelazamiento aparece como una cantidad interesante que incluye información importante acerca de la teoría de campos, tales como sus anomalías de traza tipo A y B en dimensiones pares y la energía libre de la esfera en dimensiones impares cuando la superficie de entrelazamiento es una región esférica. Dado que entropía de entrelazamiento es una medida contaminada por la existencia de divergencias, empleamos el esquema de los Kontratérminos para extraer las magnitudes físicas relevanets. En el caso de teorías de tres dimensiones duales a gravedad de Einstein, mostramos que la parte finita es aislada correctamente y que puede escribirse en términos de la energía de Willmore. Aprovechando sus propiedades, proporcionamos una cota superior para la entropía de entrelazamiento holográfica. Este interesante resultado es posteriormente extendido a cualquier CFT considerada. Además, mostramos que la validez del esquema de Kontratérminos para gravedad de curvatura cuadrática, extrayengo las anomalías de tipo A y B en dimensiones pares y la energía libre de la esfera en dimensiones impares.

%% file: text/ch1-intro.tex
%

\chapter{Introduction}\label{ch:intro}

General relativity (GR), also referred as Einstein gravity, is the most successful theory of gravity to date. In this theory, gravitational interaction between bodies is described by spacetime curvature through the Einstein field equations. These equations of motion are determined applying the variational principle to the \emph{Einstein-Hilbert action}, given by
\begin{equation}\label{eq:EH}
    I_\EH=\frac{1}{16\pi \GN}\int_{\B}\diff^D x\sqrt{-g}R,
\end{equation}
where $\GN$ is the gravitational constant, $\B$ is a $D$-dimensional, torsion-free, Lorentzian manifold equipped with the non-degenerate, smooth, symmetric metric tensor $g_{ab}$ and $R$ its Ricci scalar. The Einstein-Hilbert action \eqref{eq:EH} can be supplemented with matter fields and a cosmological constant $\Lambda$ to account for the vacuum energy $\pazocal{L}=\pazocal{L}_{\text{EH}}-\frac{\Lambda}{8\pi \GN}+\pazocal{L}_{\text{matter}}$. With this in mind, the Einstein field equations of $g_{ab}$ read
\begin{equation}\label{eq:EFE}
\pazocal{E}_{ab}\equiv\frac{1}{\sqrt{-g}}\frac{\updelta I}{\updelta g^{ab}}=G_{ab}+\Lambda g_{ab}=8\pi\GN T_{ab}^\text{matter},
\end{equation}
where $G_{ab}=R_{ab}-\frac{1}{2}g_{ab}$ is the Einstein tensor, $R_{ab}$ the Ricci tensor, whose contraction is the Ricci scalar and $T_{ab}^\text{matter}\equiv-\frac{2}{\sqrt{-g}}\frac{\updelta \pazocal{L}_{\text{matter}}}{\updelta g^{ab}}$ is the stress-energy tensor. As a result of diffeormorphism invariance of the theory, the equations of motion obey the Bianchi identity $\nabla^a \pazocal{E}_{ab}=0$. 

After varying the action and finding the equations of motion \eqref{eq:EFE}, an additional boundary term is required. Using Dirichlet boundary conditions
\begin{equation}
\updelta g_{ab}\big|_{\partial\pazocal{M}}=0,
\end{equation}
and leaving the normal derivative free to fluctuate leaves us with a not well posed variational problem. This inconsistency can be solved by introducing a suitable surface term, which, in the case of Einstein gravity,
corresponds to the Gibbons-Hawking-York (GHY) term
\begin{equation}\label{eq:GHY}
I_{\text{GHY}}=\frac{1}{8\pi \GN}\int_{\partial\B}\diff^{D-1}x\sqrt{-\gamma} K,
\end{equation}
where $K$ is the trace of the extrinsic curvature of the induced metric $\gamma_{\mu\nu}$ on the hypersurface defined at $\partial\B$. Interestingly, the GHY is not the only boundary term that can be added without spoiling the Dirichlet boundary conditions. We will discuss this later on.

For now, let us focus on vacuum theories, \ie $T_{ab}=0$. There are a number of spacetimes that solve the equations of motion \eqref{eq:EFE}. An useful tool to count the number of symmetries of the different solutions is given by the Killing vectors $\xi_a$, which preserve the metric and satisfy the Killing equation
\begin{equation}
\nabla_{(a}\xi_{b)}=0.
\end{equation}
Generally speaking, a metric is said to be stationary if it admits a timelike Killing vector $\xi_t$ and static if there exists a codimension-one surface  that is orthogonal to $\xi_t$ everywhere.

Solutions with the maximum number of Killing vectors $D(D+1)/2$ are called maximally symmetric backgrounds. These are, Minkowski $\mathbb{R}^{1,D-1}$, de Sitter (dS) and Anti-de Sitter (AdS) with isometry groups SO$(1,D-1)$, SO$(1,D)$ and SO$(2,D-1)$, respectively. As a consequence, the geometry of (A)dS can be embedded in $(D+1)$-dimensional flat spacetime ($\mathbb{R}^{1,D}$ or $\mathbb{R}^{2,D-1}$) in a similar way a sphere or a hyperboloid can be embedded in an Euclidean space, respectively. The surfaces follow a constraint characterized by the (A)dS radius $L$, related to the cosmological constant as
\begin{equation}\label{eq:cosmoconst}
\Lambda=\pm\frac{(D-1)(D-2)}{2L^2},
\end{equation} 
where the positive sign corresponds to dS and the negative to AdS.

Another set of important spacetimes are \emph{static and spherically symmetric} (SSS) \textit{ans\"atze}. In general, these can be written in terms of two undetermined functions of the radial coordinate $N=N(r)$ and $f=f(r)$,
\begin{equation}
\label{eq:SSS}
\diff s^2_{N,f}=-N^2 f\diff t^2+\frac{\diff r^2}{f}+r^2\diff \Omega^2_{D-2},
\end{equation}
where $\diff\Omega_{D-2}^2$ is the metric of the unit $(D-2)$-dimensional sphere $\mathbb{S}^{D-2}$. However, these metrics solve the Einstein field equations \eqref{eq:EFE} when $N=1$, which are often referred as ``single-function'' SSS solutions. The spherical symmetry admits $D-1$ Killing vectors, satisfying the $ \mathfrak{so}(D-1)$ algebra. These spacetimes can include event horizons, that disconnect causally its interior with the exterior. A notable example of single-function SSS solution is the Schwarzschild-Tangherlini black hole \cite{Schwarzschild:1916uq,Tangherlini:1963bw}
\begin{equation}\label{eq:SchwTang}
f=1-\frac{16\pi\GN M}{(D-2)\Omega_{D-2}r^{D-3}}-\frac{2\Lambda r^2}{(D-1)(D-2)},
\end{equation}
where $M$ is the total mass of the spacetime, that can be computed using the ADM prescription \cite{Arnowitt:1960es,Arnowitt:1960zzc,Arnowitt:1961zz}. 

If we are interested in studying the spacetimes including electric charge $Q$, the appropriate metric is provided by the Reissner-Nordstr\"om black hole \cite{reissner1916eigengravitation,nordstrom1918energy}, which solves the Einstein-Maxwell equations. These are obtained after applying the variational principle to the Einstein-Hilbert action supplemented with the Maxwell action functional $\int_\B\diff^Dx\sqrt{-g}F^2$, where $F^2\equiv F_{ab}F^{ab}$ is the squared field strength tensor --- see more in sec. \ref{sec:EGQTg}.\footnote{If we relax the static condition, we find other examples of stationary, axisymmetric black holes such as Kerr \cite{Kerr:1963ud} (including angular momentum $J$) and Kerr-Newman solutions \cite{Newman:1965my} (including charge and angular momentum)} 

If we restrict ourselves to flat spacetimes, \ie $\Lambda=0$, Schwarzschild and Reissner-Nordstr\"om black holes are the unique SSS solutions for vacuum Einstein-Maxwell field equations, according to the Birkhoff's theorem \cite{jebsen2005general,birkhoff1923relativity,pappas1984proof}.
 
Since its publication, predictions of Einstein gravity have been verified in numerous experimental tests in our $D=4$ world. The large number of them allows us to differentiate between two types. On the one hand we have the classical ones, proposed by Einstein in his seminal paper such as the perihelion precession of Mercury or the deflection of light by the Sun and gravitational redshift. On the other hand, a longer list comprises the so-called modern ones. Among them, two tests that were of utter importance in recent years further confirmed predictions of GR. The first one is the detection of gravitational waves coming from merging of several black hole and neutron star binaries by LIGO and Virgo collaboration as soon as 2016 \cite{LIGOScientific:2016aoc, LIGOScientific:2016sjg,LIGOScientific:2017bnn,LIGOScientific:2017vwq,LIGOScientific:2018mvr}. The other one is the first image obtained  of M87${}^*$ ---and more recently another one of Sagitarius A${}^*$--- by the Event Horizon Telescope using radiotelescopes. These two supermassive black holes are located at the center of M87 and Milky Way galaxies respectively \cite{EventHorizonTelescope:2019dse,EventHorizonTelescope:2019uob,EventHorizonTelescope:2019jan,EventHorizonTelescope:2019ths,EventHorizonTelescope:2019pgp,EventHorizonTelescope:2019ggy,EventHorizonTelescope:2022xnr,EventHorizonTelescope:2022vjs,EventHorizonTelescope:2022wok,EventHorizonTelescope:2022exc,EventHorizonTelescope:2022urf,EventHorizonTelescope:2022xqj}.

Despite its great success, GR leaves many unresolved phenomena and leads to inconsistencies both at cosmological and quantum level. For the former, astrophysical observations measuring galaxies velocities and its rotation curves \cite{zwicky1933rotverschiebung,rubin1970rotation,roberts1973comparison,rubin1980rotational}, gravitational lenses \cite{walsh19790957,bartelmann2001weak} and Cosmic Microwave Background (CMB) \cite{Planck:2013oqw, Planck:2018vyg}, among others, do not fit GR predictions. Some directions to solve this discrepancy involve modifying Einstein gravity or introducing the so-called \emph{dark matter} whose distribution restores Einstein gravity results. In this context, the most prominent cosmological model, referred as $\Lambda$CDM, predicts about five times more dark matter distribution than baryonic matter in our Universe. $\Lambda$CDM model also introduces a large quantity ---about fourteen times baryonic matter density distribution--- of unknown energy distribution in our Universe. This is referred as \emph{dark energy} in an attempt to fit the accelerated expansion measured experimentally from supernovae, as they are standard candles across cosmological distances \cite{SupernovaSearchTeam:1998fmf}, CMB \cite{Planck:2018vyg} and large scale structure \cite{drinkwater2010wigglez}, among others. However, $\Lambda$CDM model does not provide any hint regarding the origin of dark matter and dark energy. Extensive work has been carried in this direction but no satisfactory results have been found yet.

At the quantum level, the gravitational interaction becomes more relevant. In this regime, fundamental problems appear, raising doubts on the validity of GR. Among the most prominent ones, we draw attention to three of them. First, the appearance of singularities either in the black hole interior and cosmological solutions. Second, the cosmological constant problem which describes the lack of contribution from the zero point energies of the Standard Model fields to $\Lambda$ even though they are different in more than twenty orders of magnitude \cite{Weinberg:2000yb}. The last one is the non-renormalizability by power-counting of EH action after applying standard quantization procedures \cite{tHooft:1974toh,Deser:1974cz}. 
This observation can be reconciled if we view GR as some effective field theory that is only valid at low energy regime. As we explore higher energies scales ---such as the Planck scale, for example--- we expect higher-curvature terms\footnote{By higher-curvature terms we refer to terms in our Lagrangian that can be constructed from the metric and verify the principle of general covariance. At order one, the simplest curvature invariant of a Riemann manifold is the Ricci scalar} to become relevant. This is the case in supergravity theories, the low energy effective action of supersymmetric String theory (Superstring, for short) \cite{Schwarz:1982jn,Polchinski:1998rq,Polchinski:1998rr,Becker:2006dvp}. 

From a different perspective, particular modified gravities presenting special properties have often been considered in the holographic context through the concrete realization of the anti-de Sitter/conformal field theory (AdS/CFT) correspondence \cite{Maldacena:1997re,Gubser:1998bc,Witten:1998qj}. They define useful toy models of strongly coupled CFTs inequivalent to Einstein gravity ---see \eg \cite{Hofman:2008ar,Camanho:2009vw,deBoer:2009pn,Buchel:2009sk,deBoer:2009gx,Grozdanov:2014kva,Andrade:2016rln,Konoplya:2017zwo,Myers:2010ru,Myers:2010jv,Hung:2011nu,Hung:2014npa,deBoer:2011wk,Bianchi:2016xvf,Bueno:2018xqc} and references therein--- and they have been crucial in the discovery of certain universal results valid for general CFTs \cite{Myers:2010tj,Myers:2010xs,Mezei:2014zla,Bueno:2015rda,Bueno:2015xda,CanoMolina-Ninirola:2019uzm,Murcia:2022XXX} ---or to raise doubts on the possible universality of others \cite{Buchel:2004di,Kats:2007mq,Brigante:2007nu,Myers:2008yi,Cai:2008ph,Ge:2008ni}.

Taking these motivations into account, different modifications of Einstein gravity attempted to solve particular problems either cosmological, fundamental or in the holographic context. In next section, we review some generalities of a certain subset, the so-called higher-curvature extensions, with a main focus on purely geometric modifications. We also provide notable examples that will be mentioned throughout this thesis, such as quadratic curvature, Lovelock and generalized quasi-topological and  gravities, with some of their most relevant properties.



\section{Higher-curvature extensions to Einstein gravity}

Let us consider the most general metric-covariant, parity preserving, theory of gravity consisting on the Riemann tensors, its contractions using the metric $g_{ab}$ and an undetermined number of covariant derivatives. We denote such a theory by its Lagrangian as
\begin{equation}\label{eq:generalhdg}
\pazocal{L}=\pazocal{L}\left(R_{abcd}, \nabla_{e}R_{abcd},  \nabla_{e}\nabla_{f}R_{abcd},\ldots\right).
\end{equation}
We can classify different higher-order Lagrangian densities belonging to this class of theories based on the number of derivatives of the metric. For example, the four- and six-derivative order Lagrangians are given by
\begin{align}\label{s4}
\pazocal{L}^{(2)}=&\alpha_1^{(2)}R^2+\alpha_2^{(2)} R_{ab}R^{ab}+\alpha_3^{(2)} R_{abcd}R^{abcd}\, ,\\ \label{s6}
\pazocal{L}^{(3)}=&\alpha_1^{(3)}\tensor{R}{_{a}^{c}_{b}^{d}}\tensor{R}{_{c}^{e}_{d}^{f}}\tensor{R}{_{e}^{a}_{f}^{b}}+\alpha_2^{(3)} \tensor{R}{_{ab}^{cd}}\tensor{R}{_{cd}^{ef}}\tensor{R}{_{ef}^{ab}}+\alpha_3^{(3)} \tensor{R}{_{a bcd}}\tensor{R}{^{a bc}_{e}}R^{d e}\\ \notag
&+\alpha_4^{(3)}\tensor{R}{_{abcd}}\tensor{R}{^{abcd}}R+\alpha_5^{(3)}\tensor{R}{_{abcd}}\tensor{R}{^{ac}}\tensor{R}{^{bd}}+\alpha_6^{(3)}R_{a}^{\ b}R_{b}^{\ c}R_{c}^{\ a}+\alpha_7^{(3)}R_{ab }R^{ab }R\\ \notag
&+\beta_8R^3+\alpha_9^{(3)} \nabla_{d}R_{ab} \nabla^{d}R^{ab}+\alpha_{10}^{(3)}\nabla_{a}R\nabla^{a}R,
\end{align}
where $\alpha_{i_{n}}^{(n)}$ are some coupling constants. If we are interested in the subset of Lagrangians $\pazocal{L}(R_{abcd})$, then we can also distinguish them by curvature order $n$ instead of the number of derivatives in the metric. In this language, EH Lagrangian \eqref{eq:EH}, $\pazocal{L}^{(2)}$ and $\pazocal{L}^{(3)}$ (modulo the covariant derivative terms) are linear, second and third order in curvature, respectively. In what follows this denomination is more frequently used. Moreover, whenever we refer to the most general curvature order $n$ Lagrangian, we also include all densities up to order $n$.

For gravities of the type $\pazocal{L}(R_{abcd})$, we find that the variational problem is not well posed again. On the one hand, the equations of motion in the bulk, given by\footnote{In the case of theories of the type \eqref{eq:generalhdg}, the equations of motion can be found in ref.~\cite{Edelstein:2022lco}}
\begin{equation}\label{eq:eomgen} 
\pazocal{E}_{ab} = P_{a}{}^{cde}R_{bcde} - \frac{1}{2} g_{ab} \pazocal{L} - 2 \nabla^c \nabla^d P_{a c d b} = 0\, , \quad \text{with} \quad
P^{abcd} \equiv \frac{\partial \pazocal{L}}{\partial R_{abcd}},
\end{equation}
include fourth-order derivative terms coming from $\nabla^c \nabla^d P_{a c d b} $ \cite{Padmanabhan:2011ex}. In general, we find that the Hamiltonian associated with Lagrangian via a Legendre transform is unbounded from below, as the Ostrogradsky theorem states \cite{Ostrogradsky:1850fid,Woodard:2015zca}. This instability gives rise to ghost fields propagating in our theory. However, certain theories satisfy $\nabla^c \nabla^d P_{a c d b}=0$, avoiding this situation. This is case of Lovelock gravity  \cite{lovelock1970divergence,Lovelock:1971yv} and generalized quasitopological gravity \cite{Bueno:2016xff,Hennigar:2016gkm,Bueno:2016lrh,Hennigar:2017ego,Bueno:2017sui,Ahmed:2017jod,Bueno:2017qce}, the latter when evaluated on a SSS solution \eqref{eq:SSS} ---see also \cite{Teimouri:2016ulk}. We discuss this in detail below. In the case of truncated higher-order gravities including covariant derivatives \eqref{eq:generalhdg} the equations of motion are far more involved than \eqref{eq:eomgen} and always involve ghost fields \cite{Edelstein:2022lco}.

Regarding the boundary, finding the required surface term that generalizes the GHY term for arbitrary theories in order to find a well posed variational problem is generally quite complicated. There are particular examples that we mention below, namely, quadratic curvature, Lovelock and generalized quasitopological gravities. However, a general treatment of the problem was considered in ref. \cite{Deruelle:2009zk,Teimouri:2016ulk} in the context of Hamiltionian formulation.

In order to visualize easily this discussion, let us present some famous theories that have been subject of thoroughly study. We also address some of its more relevant properties and contextualize its relevance.

\subsection{Quadratic curvature gravity}

The most immediate example of higher-curvature gravity is quadratic curvature (QC) gravity, given by Lagrangian \eqref{s4}. Interestingly, when supplemented to the EH term \eqref{eq:EH}, the theory becomes is power-counting renormalizable \cite{Stelle:1976gc,Julve:1978xn}. Besides, QC terms appear in the ten-dimensional effective action of Heterotic superstring theory \cite{Bergshoeff:1989de,Cano:2019ore}. However, in the absence of other higher-order terms, arbitrary values of the coupling constants $\alpha_1^{(2)}$, $\alpha_2^{(2)}$ and $\alpha_3^{(2)}$ imply the existence of massive ghost field in the theory. Let us discuss further on this. Without loss of generality, we can rewrite the QC Lagrangian in terms of a particular combination of second order densities, given by
\begin{equation}\label{eq:GB} 
\pazocal{X}_4=R^2-4R_{ab}R^{ab}+R_{abcd}R^{abcd}.
\end{equation}
In this convenient writing, the most general QC Lagrangian $\pazocal{L}(R_{abcd})$ can be written as
\begin{equation}\label{eq:QCGgen}
\pazocal{L}_{\text{\text{QC}}}=\frac{1}{16\pi \GN}\left[R-2\Lambda+L^{2}\left(\tilde{\alpha}_1^{(2)}R^2+\tilde{\alpha}_2^{(2)} R_{ab}R^{ab}+\tilde{\alpha}_3^{(2)}\pazocal{X}_4\right)\right],
\end{equation}
where $\tilde{\alpha}_{i_2}^{(2)}$ are some algebraically redefined dimensionless coupling constant with respect to $\alpha_{i_2}^{(2)}$.\footnote{For simplicity, from now on we drop the tilde in the redefined coupling constants.}

The interest of expressing the QC gravity Lagrangian as in eqs. \eqref{eq:GB} becomes evident after having a look on the equations of motion, given by
\begin{align}
\pazocal{E}_{ab}=&+\alpha_1^{(2)}\left[2R\left(R_{ab}-\frac{1}{4}g_{ab}R\right)+2\left(g_{ab}\dal R-\nabla_a\nabla_b R\right)\right]\\
&+\alpha_2^{(2)}\left[2\left(R_{ac}R\indices{_b^c}-\frac{1}{2}g_{ab}R_{cd}R^{cd}\right)+\left(\dal R_{ab}+\frac{1}{2}g_{ab}\dal R-2\nabla_{c}\nabla_{(a}R\indices{_{b)}^c}\right)\right],\notag\\
&+\alpha_3^{(2)}\left[2\left(RR_{ab}-2R_{ac}R\indices{_b^c}-2R^{cd}R_{acbd}+R\indices{_a^{cde}}R_{bcde}\right)-\frac{1}{2}\X_{4}\right].\notag
\end{align}
If we choose $\alpha_1^{(2)}=\alpha_2^{(2)}=0$ and $\alpha_3^{(2)}\neq0$, we observe that the equations of motion are second order in the metric. This is because the Gauss-Bonnet term is the four dimensional Euler density and it is related to the Euler characteristic of the manifold. As such, in the critical dimension $D=4$, its contribution to the equations of motion is trivial, being dynamical in higher dimensions and identically vanishing in lower dimension. 

This discussion can be extended to the $n$-th order generalization of Euler densities, which are the terms corresponding to Lovelock gravity. Because of this, let us contextualize the general properties of Gauss-Bonnet theory  in the framework provided by Lovelock.

\subsection{Lovelock gravity}

As we said, we can generalize to an arbitrary order the Lagrangian that yields second order equations of motion, based on a series of dimensionally extended Euler densities. This series is known as \emph{Lovelock gravity} (or L\'anczos-Lovelock), whose Lagrangian is given by the expression
\begin{equation}\label{eq:LovelockLag}
\pazocal{L}_{\mathrm{Lovelock}}=\sum_{n=0}^{\lfloor\frac{D-1}{2}\rfloor}\alpha_n L^{2k-2}\X_{2n},
\end{equation}
for any value of the dimensionless coupling constants $\alpha_n$ and where the Euler densities are defined by \cite{lovelock1970divergence,Lovelock:1971yv}
\begin{equation}\label{eq:Eulerdens}
\X_{2n}\equiv\frac{1}{2n}\delta_{c_1d_1\ldots c_nd_n}^{a_1b_1\ldots a_nb_n}R\indices{_{a_1}_{b_1}^{c_1}^{d_1}}\cdots R\indices{_{a_n}_{b_n}^{c_n}^{d_k}},
\end{equation}
where $\delta_{c_1d_1\ldots c_nd_n}^{a_1b_1\ldots a_nb_n}\equiv (2n)!\delta^{a_1}_{[c_1}\delta^{b_1}_{d_1}\cdots\delta^{a_n}_{c_k}\delta^{b_n}_{d_n]}$ is the generalized Kronecker symbol and $\lfloor x\rfloor$ is the floor function of $x$. The powerful Lovelock theorem states that the series \eqref{eq:LovelockLag} is the most general Lagrangian of the type $\pazocal{L}(R_{abcd})$ that admits second order differential equations of motion \cite{lovelock1970divergence,Lovelock:1971yv}. 

Some low order example of Lovelock densities are $\X_2$ and $\X_4$, which of course correspond to the EH \eqref{eq:EH} and the Gauss-Bonnet term \eqref{eq:GB}, respectively. Another example is the cubic order Lovelock term, given by
\begin{align}
\X_6=&-8\tensor{R}{_{a}^{c}_{b}^{d}}\tensor{R}{_{c}^{e}_{d}^{f}}\tensor{R}{_{e}^{a}_{f}^{b}}+4\tensor{R}{_{ab}^{cd}}\tensor{R}{_{cd}^{ef}}\tensor{R}{_{ef}^{ab}} -24\tensor{R}{_{abcd}}\tensor{R}{^{a bc}_{e}}R^{d e}\label{eq:Lovthird}\\
&+3\tensor{R}{_{abcd}}\tensor{R}{^{abcd}}R+24\tensor{R}{_{abcd}}\tensor{R}{^{ac}}\tensor{R}{^{bd}}
+16 R^{b}_{a} R_{b}^{c} R_{c}^{a}
-12R_{ab}R^{ab} R+R^3. \notag
\end{align}

As anticipated when discussing the case of the Gauss-Bonnet term, the behavior of each term in the Lovelock series depends on the dimensionality of the theory. The Euler density $\X_{2n}$ is identically trivial if $D<2n$ and dynamical in $D>2n$, as it contributes to the equations of motion. In the critical dimension $D=2n$, the term becomes a topological invariant, as it can see from its relation to the Euler characteristic of the manifold $\chi(\B)$ through the Euler-Chern theorem \cite{Chern:1945curvatura}
\begin{equation}\label{eq:Cherntheo}
\int_{\B}\diff^{2n}x\sqrt{-g}\X_{2n}=(4\pi)^n\Gamma(n+1)\chi(\B)+\int_{\partial\B}\diff^{D-1}x\sqrt{-\gamma} B_{2n-1},
\end{equation}
where $B_{2n-1}$ corresponds to the Chern form at the boundary whose explicit expression is shown in \eqref{eq:Bdeven}. The equations of motion in the bulk are given by the compact expression \cite{Padmanabhan:2013xyr}
\begin{equation}
\E\indices{^a_c}=\sum_{k=0}^{\lfloor\frac{D-1}{2}\rfloor}\frac{\alpha_nL^{2n-2}}{2^{n+1}}
\delta_{c c_1d_1\ldots c_nd_n}^{a a_1b_1\ldots a_kb_k}R\indices{_{a_1}_{b_1}^{c_1}^{d_1}}\cdots R\indices{_{a_n}_{b_k}^{c_n}^{d_n}}=0,
\end{equation}
showing no higher order dependence in the metric. As the propagating modes in Lovelock gravity are the same as in GR, Lagrangian \eqref{eq:LovelockLag} is often referred as the most natural generalization of Einstein-Hilbert theory. Regarding the surface term, the generalization of the GHY for each of the Lovelock terms was constructed in refs. \cite{Myers:1987yn,Teitelboim:1987zz}, namely
\begin{align}
Q_n=-2n\int\limits_0^1\diff s\,&\delta^{\mu_1\cdots \mu_{2n-1}}_{\nu_1\cdots \nu_{2n-1}}K^{\nu_1}_{\mu_1}\left(\frac{1}{2}\pazocal{R}\indices{^{\nu_2}^{\nu_3}_{\mu_2}_{\mu_3}}-\epsilon s^2K^{\nu_2}_{\mu_2}K^{\nu_3}_{\mu_3}\right)\times\cdots\nonumber\\
&\cdots\times\left(\frac{1}{2}\pazocal{R}\indices{^{\nu_{2n-2}}^{\nu_{2n-1}}_{\mu_{2n-2}}_{\mu_{2n-1}}}-\epsilon s^2K^{\nu_{2n-2}}_{\mu_{{2n-2}}}K^{\nu_{2n-1}}_{\mu_{2n-1}}\right),\label{eq:GGHYterm}
\end{align}
where $\pazocal{R}\indices{^\mu^\nu_\rho_\sigma}$ is the is the intrinsic Riemann curvature tensor of mtric $\gamma_{\mu\nu}$, defined at the boundary, and $\epsilon=n_an^a=\pm 1$ is the norm of the normal vector to the boundary, whose sign is positive in the case of timelike hypersurface and negative in the case of a spacelike one ---in the case $\epsilon=1$, the expression of the generalized GHY term \eqref{eq:GGHYterm} coincides with the Chern form $B_{2n-1}$ in eq. \eqref{eq:Cherntheo}. Once we have the generalized GHY term, the action
\begin{equation}
I=\frac{1}{16\pi G}\sum_{n=0}^{\lfloor\frac{D-1}{2}\rfloor}\left(\alpha_n L^{2k-2}\int_\B\diff^{D}x\sqrt{-g}\,\X_{2n}+\int_{\partial\B}\diff^{D-1}x\sqrt{-\gamma}\,Q_n\right),
\end{equation}
yields the well posed Lovelock action.

Since Gauss-Bonnet and Lovelock gravities are free of ghost fields, evading any problem with unitarity \cite{Zwiebach:1985uq}, they have been studied extensively and many solutions have been found \cite{Wheeler:1985nh,Wheeler:1985qd,Myers:1988ze,Cai:2001dz,Zegers:2005vx,Camanho:2011rj}, along with pure Lovelock gravity \cite{Cai:2006pq,Gannouji:2013eka,Dadhich:2015ivt,Gannouji:2019gnb}---which consists on the addition only of the $n$-th order Euler density to the EH action. Moreover, as Lovelock terms appear in String Theory, they have also attracted the interest in the context of the AdS/CFT correspondence \cite{deBoer:2009gx,Camanho:2009vw,Camanho:2009hu,Camanho:2010ru,Camanho:2013pda}.

\subsection{Generalized quasi-topological gravities}\label{sec:introGQT}

Let us now discuss an interesting family of higher-curvature theories of gravity identified in recent years \cite{Bueno:2016xff,Hennigar:2016gkm,Bueno:2016lrh,Hennigar:2017ego,Bueno:2017sui,Ahmed:2017jod,Bueno:2017qce}. The Lagrangian of the so-called \emph{generalized quasi-topological} (GQT) \textit{gravities}  \cite{Hennigar:2017ego} can be written schematically as
\begin{equation}\label{eq:priA}
\pazocal{L}_{\text{GQT}}=\frac{1}{16\pi \GN}\left(-2\Lambda+R+\sum_{n=2} \sum_{i_n} L^{2(n-1)}\alpha_{i_n}^{(n)}\mathfrak{R}_{i_n}^{(n)}\right),
\end{equation}
where $\mathfrak{R}^{(n)}_{i_n}$ are densities constructed from $n$ Riemann tensors and its contraction, the $\alpha^{(n)}_{i_n}$ are dimensionless couplings, and $L$ is some length scale. The subindex $i_n$ refers to the number of independent GQT invariants at each order $n$. 

The technical requirement which makes a generic Lagrangian belong to the GQT class is the following. Let $L_{N,f}\equiv L\big|_{N,f}$ be the effective Lagrangian which results from evaluating $\sqrt{-g}\pazocal{L}$ in SSS ansatz\footnote{The discussion below also applies to solutions in which the horizon is planar or hyperbolic instead of spherical. The different cases are parametrized by a constant $k$ taking values $k=1,0,-1$ for spherical, planar and hyperbolic horizons respectively.} \eqref{eq:SSS}, namely
\begin{equation}\label{ansS}
L_{N,f}(r,f,N,f',N',\dots)= N r^{D-2} \pazocal{L}|_{N,f},
\end{equation}
up to an angular contribution that is irrelevant for the definition. Also, let $L_f\equiv L_{1,f}$, \ie the expression resulting from imposing $N=1$ in $L_{N,f}$.

\begin{defi}\label{Def0}
We say that $\pazocal{L}(R_{abcd})$ belongs to the GQT family if the Euler-Lagrange equation of $L_f$ vanishes identically, \ie if
\begin{equation}\label{eq:GQTcond}
\frac{\delta L_f}{\delta f} = 
\frac{\partial L_f}{\partial f}-\frac{\diff}{\diff r} \frac{\partial L_f}{\partial f'}+\frac{\diff^2}{\diff r^2}  \frac{\partial L_f}{\partial f''}-\ldots=0\, , \quad \forall \, \, f\, .
\end{equation}
\end{defi}
GQTs have by now been the subject of quite intensive investigation and many of the interesting properties of these theories are now well-understood. Here we summarize some particularly relevant ones:

\begin{enumerate}
\item When linearized around any maximally symmetric background, the equations of motion of GQT theories become second-order \ie they only propagate the usual massless and traceless graviton characteristic of Einstein gravity on such backgrounds \cite{Bueno:2016xff,Hennigar:2016gkm,Bueno:2016lrh,Hennigar:2017ego,Bueno:2017sui,Ahmed:2017jod,Bueno:2017qce}.\footnote{Note that higher-curvature gravities satisfying property ``1.'' ---and not necessarily the rest of properties appearing in the list, nor condition \req{eq:GQTcond}--- have been studied in several other papers, \eg \cite{Bueno:2016ypa,Tekin:2016vli,Sisman:2011gz,Karasu:2016ifk,Bueno:2016dol,Ghodsi:2017iee,Li:2017ncu,Li:2017txk,Li:2018drw,Li:2019auk,Lu:2019urr}.}

\item  They have a continuous and well-defined Einstein gravity limit, which corresponds to  setting $\alpha^{(n)}_{i_n}\rightarrow 0$ for all $n$ and $i_n$.

\item They admit generalizations of the (asymptotically flat, de Sitter or Anti-de Sitter) Schwarzschild-Tangherlini black hole \eqref{eq:SchwTang}  ---\ie solutions which reduce to it in the Einstein gravity limit--- characterized by a single function $f$ \cite{Hennigar:2016gkm,Bueno:2016lrh,Hennigar:2017ego,Bueno:2017sui,Ahmed:2017jod,Bueno:2017qce}. For them, $N=1$ (or some other constant) in eq. \req{eq:SSS} and $g_{tt}g_{rr}=-1$.

\item The metric function $f$ is determined from a differential equation of order $\leq 2$. They can be obtained from the Euler-Lagrange equation of $N$ associated to the effective Lagrangian $L_{N,f}$ defined in eq. \req{ansS}, this is
\begin{equation}\label{eq:eqN}
\left.\frac{\updelta L_{N,f}}{\updelta N}\right|_{N=1}=0,
\end{equation}
as long as the action does not include covariant derivatives of the Riemann tensor\footnote{When it does, the differential equation would be of order $\leq 2m+2$, where $m$ is the number of covariant derivatives of the term with the greatest number of them.}. Schematically, $\pazocal{E}[ r,f,f',f'';\alpha_{i_n}^{(n)}]=0$.
In that case, there are typically three situations:
\begin{itemize}
\item  The corresponding density does not contribute at all to the equation and then we call it ``trivial''.
\item The density contributes to the equation with an algebraic dependence on $f$ ---namely, with terms involving powers of $f$. This is the case of quasi-topological (QT) \cite{Oliva:2010eb,Myers:2010ru,Dehghani:2011vu,Ahmed:2017jod,Cisterna:2017umf} and Lovelock \cite{lovelock1970divergence,Lovelock:1971yv} terms. This kind of contributions only exist for $D\geq 5$.
\item The density contributes to the equation with terms containing up to two derivatives of $f$. This is the case \eg of Einsteinian cubic gravity in $D=4$ \cite{Bueno:2016xff,Hennigar:2016gkm,Bueno:2016lrh}.
\end{itemize}


\item The thermodynamic properties of black holes can be computed analytically by solving a system of algebraic equations without free parameters. At least in $D=4$, black holes typically become stable below certain mass, which substantially modifies their evaporation process \cite{Bueno:2017qce}.


\item A (generally) different subset of four-dimensional GQT gravities also gives rise to second-order equations for the scale factor when evaluated on a Friedmann-Lema\^itre-Robertson-Walker ansatz, giving rise to a well-posed cosmological evolution \cite{Arciniega:2018fxj,Cisterna:2018tgx,Arciniega:2018tnn,Arciniega:2019oxa,Edelstein:2020nhg,Edelstein:2020lgv}. Remarkably, an inflationary period smoothly connected with late-time standard $\Lambda$CDM evolution is naturally generated by the higher-curvature terms.

\item Densities belonging to the subclass QT have been identified for $D\geq5$, whereas GQT densities exist as well in $D=4$ \cite{Bueno:2016xff}. Moreover, it can be show, using recurring relations that the can be systematically constructed at all orders $n$ in curvature \cite{Bueno:2019ycr}. In ch.~\ref{ch:ateach} we will argue that in $D\geq5$ there is only one GQT density at each order $n$ whereas we can find $n-2$ ``genuine'' GQT densities ---by genuine we mean that they do not belong to the QT subclass.

\item Extensions away from pure metric theories, including scalars or vector fields, while preserving the main properties are possible \cite{Cano:2020qhy, Cano:2022ord}. In ch.~\ref{ch:hc3d} address this case in three dimensions.


\end{enumerate}
In addition to these more or less structural properties, GQT gravities have been considered in various  contexts, and many interesting additional properties and applications explored ---see \eg \cite{Dey:2016pei,Feng:2017tev,Hennigar:2017umz,Hennigar:2018hza,Bueno:2018xqc,Bueno:2018yzo,Poshteh:2018wqy,Mir:2019ecg,Mir:2019rik,Mehdizadeh:2019qvc,Erices:2019mkd,Emond:2019crr,BeltranJimenez:2020lee,Fierro:2020wps,Bakhtiarizadeh:2021vdo,Gray:2021roq,Edelstein:2022xlb}. We will explicitly mention some of them in the context of black hole thermodynamics and its utility as toy models through the AdS/CFT correspondence in due time.

Before concluding, let us present a concrete example of these theories. At cubic order in curvature, the most general (nontrivial) GQT theory can be written as
\begin{equation}\label{cubiGQT}
\pazocal{L}^{(3)}=\frac{1}{16\pi \GN}\left[-2\Lambda+R+L^2 \alpha^{(2)}_1 \pazocal{X}_4+ L^4 \left(\alpha^{(3)}_1 \pazocal{X}_{6}+\alpha^{(3)}_2\pazocal{Z}_{D}+\alpha^{(3)}_3\pazocal{S}_{D}\right)\right],
\end{equation}
where we used the notation of eq. \req{eq:priA} to denote the couplings. Here, 
$\pazocal{Z}_D$ is the so-called QT gravity density \cite{Oliva:2010eb,Myers:2010ru}
\begin{align}
\pazocal{Z}_D=&+{{{R_a}^b}_c{}^d} {{{R_b}^e}_d}^f {{{R_e}^a}_f}^c
               + \frac{1}{(2D - 3)(D - 4)} \left[ \frac{3(3D - 8)}{8} R_{a b c d} R^{a b c d} R \nonumber \right. \\ \label{ZD}
              & \left. - \frac{3(3D-4)}{2} {R_a}^c {R_c}^a R  - 3(D-2) R_{a c b d} {R^{a c b}}_e R^{d e} + 3D R_{a c b d} R^{a b} R^{c d} \right. \\ \nonumber
              & \left.
                + 6(D-2) {R_a}^c {R_c}^b {R_b}^a  + \frac{3D}{8} R^3 \right]\, .
\end{align}
When the terms $\X_4$, $\X_6$ and $\Z_D$ are included in addition to the usual Einstein-Hilbert action, the equation satisfied by the metric function $f$ is algebraic ---which partially explains why they were identified before the last term, $\pazocal{S}_D$. We also stress that for $D\geq 6$, $\pazocal{Z}_D$ affects the equation of $f$ in the same way as $\pazocal{X}_6$ does. For $D=5$, $\pazocal{X}_6$ is trivial, and the effect of $\pazocal{Z}_5$ is nontrivial ---from this perspective, we could have just omitted $\pazocal{X}_6$ from \req{cubiGQT}. 

When the density $\pazocal{S}_D$ is included, the equation becomes differential of second order. The explicitly form of this density can be chosen to be \cite{Hennigar:2017ego}
\begin{equation}
\begin{aligned}
\pazocal{S}_D=&14\tensor{R}{_{a}^{c}_{b}^{d}}\tensor{R}{_{c}^{e}_{d}^{f}}\tensor{R}{_{e}^{a}_{f}^{b}} +2\tensor{R}{_{a bcd}}\tensor{R}{^{a bc}_{e}}R^{d e}-\frac{(38-29 D+4 D^2)}{4(D-2)(2D-1)}\tensor{R}{_{abcd}}\tensor{R}{^{abcd}}R\\
&-\frac{2(-30+9D+4D^2)}{(D-2)(2D-1)}\tensor{R}{_{abcd}}\tensor{R}{^{ac}}\tensor{R}{^{bd}}
-\frac{4(66-35D+2D^2))}{3(D-2)(2D-1)}R^{b}_{a} R_{b}^{c} R_{c}^{a}\\
&+\frac{(34-21D+4D^2)}{(D-2)(2D-1)}R_{ab}R^{ab} R-\frac{(30-13D+4D^2)}{12(D-2)(2D-1)} R^3\, .
\end{aligned}
\label{SD}
\end{equation}
The equation of $f$ corresponding to eq. \req{cubiGQT} can be found \eg in ref. \cite{Hennigar:2017ego}.
In $D=4$, $\pazocal{S}_4$ is usually rewritten in terms of the so-called Einsteinian cubic gravity density,\footnote{The original construction of Einsteinian cubic gravity in ref. \cite{Bueno:2016xff} was based on the fact that it satisfies properties ``1.'' and  ``2.'' for general dimensions, and it does so in a way such that the relative coefficients appearing in its definition in \req{ECG} are the same for general $D$ ---just like for Lovelock theories. It was later realized that the four-dimensional version of the theory satisfies the rest of properties listed.} defined as \cite{Bueno:2016xff}
\begin{equation}\label{ECG}
\pazocal{P}=12 R_{a\ b}^{\ c \ d}R_{c\ d}^{\ e \ f}R_{e\ f}^{\ a \ b}+R_{ab}^{cd}R_{cd}^{ef}R_{ef}^{ab}-12R_{abcd}R^{ac}R^{bd}+8R_{a}^{b}R_{b}^{c}R_{c}^{a}\, ,
\end{equation}
which was in fact the first GQT gravity identified beyond the Lovelock and QT ones \cite{Hennigar:2016gkm,Bueno:2016lrh}. Both densities are connected through \cite{Hennigar:2017ego}
\begin{equation}\label{s4ECG}
\pazocal{S}_4-\frac{1}{4}\pazocal{X}_6+4\pazocal{C}=\pazocal{P}\, ,
\end{equation}
where $\pazocal{C}$ is an example of a trivial GQT, in the sense that it has no effect on the equation of $f$, as its contribution to it vanishes identically. It is given by
\begin{equation}\label{s4C}
\pazocal{C}=\frac{1}{2}R_{a}^b R_b^a R-2R^{ac}R^{bd}R_{abcd}-\frac{1}{4}R R_{abcd}R^{abcd}+R^{de}R_{abcd}R^{abc}\,_e\, .
\end{equation}

We will address the appropriate surface term for this type of theories in sec.~\ref{sec:holoren}.

\subsection{Electromagnetic generalized quasi-topological gravities}\label{sec:EGQTg}

Recently, an extension of GQT gravities, non-minimally coupled to Maxwell theory was constructed \cite{Cano:2020qhy,Cano:2021tfs,Cano:2021hje,Cano:2022ord}. These theories share many of the properties stated above with the previous theories without electromagnetism. Here, we succinctly review its construction, as it will be relevant for us in sec.~\ref{sec:3DBH}. Let us begin by expanding the symmetries of Lagrangian \eqref{eq:generalhdg} to include a U$(1)$ gauge field $A_a$, admiting an expansion of the form
\begin{equation}
\pazocal{L}(R_{abcd},F_{ab})=R-F^2+\ldots,
\end{equation}
where the ellipsis denotes higher-derivative terms formed from monomials of the Riemann tensor and the Field strength $F=\diff A$. On general grounds, the equation of motion of the theory are given by
\begin{align}
\E_{ab}&=P\indices{_{(a}^c^d^e}R_{b)cde}-\frac{1}{2}g_{ab}\pazocal{L}+2\nabla_c\nabla_dP\indices{_{(a}^c_{b)}^d}-2M\indices{_{(a}^c}F_{b)c},\quad \text{with }  M^{ab}\equiv\frac{\partial\pazocal{L}}{\partial F_{ab}},\\
\E_a&=\nabla^bM_{ab},
\end{align}
and $P_{abcd}$ is given in eq.~\eqref{eq:eomgen}. However, we are again interested in theories that are at most second order equations of motion when evaluated in a charged SSS solution. Before continuing, let us mention that we can work equivalently with the dual $(D-2)$-form $H=\diff B$ of $F$, with $B$ being an auxiliary $(D-3)$-form whose equations of motion are given by the Bianchi identity $\diff F=0$. The motivation for the dual frame is that a magnetic ansatz allows for a much simpler expression than its electric counterpart ---we will see an example of this in sec.~\ref{sec:3DBH}. Therefore, we consider SSS solutions given by eq.~\eqref{eq:SSS} and a magnetic ansatz of the form
\begin{equation}\label{eq:magansatz}
H=p\,\omega_{D-2},
\end{equation}
where $p$ is a constant related to the magnetic charge and $\omega_{D-2}$ is the volume form of $\mathbb{S}^{D-2}$. Following the discussion of previous section, we can define an effective Lagrangian $L_{f,H}\equiv L\big|_{N=1,f,H}$ as a result of evaluating $\sqrt{-g}\pazocal{L}$ in a magnetically charged SSS solution, given by eqs. \eqref{eq:SSS} and \eqref{eq:magansatz}. Now, we are in conditions of making the formal definition of these theories.
\begin{defi}
A theory of the type $\pazocal{L}(R_{abcd},F_{ab})$ is said to be electromagnetic generalized quasi-topological (EGQT) gravity if the Euler-Lagrange equation of the effective Lagrangian evaluated in a magnetically charged SSS solution vanishes identically, \ie 
\begin{equation}
\frac{\updelta L_{f,H}}{\updelta f}=0.
\end{equation}
\end{defi}
As in the case of GQT gravities, the equations of motion for EGQT gravities can be found from the Euler-Lagrange equation of $N$ associated to the effective Lagrangian $ L_{N,f,H}$.

We conclude the general discussion of these theories by saying some of its applications. On the one hand, solutions of these theories have been found to be free of singularities, describing a globally regular geometry \cite{Cano:2020qhy,Cano:2020ezi,Cano:2021hje}. On the other hand they represent useful setups in the holographic context, providing new insights on phenomena that its uncharged counterpart is agnostic about \cite{Cano:2022ord,Bueno:2022jbl}.

\section{Black hole thermodynamics}

In Einstein-Maxwell theory, black holes are completely characterized by their mass $M$, charge $Q$ and angular momentum $J$ \cite{Israel1967event,Israel1968event,Carter1971axisymmetric}. This is often referred as the \textit{no-hair theorem}\footnote{Funnily, the proposition has not a rigorous proof and the hair is a metaphor, referring to the resemblance of the black hole to a bald head.} \cite{Misner1973}. The conserved quantities (or conserved charges) satisfy the so-called laws of black hole mechanics, which read \cite{Bardeen:1973gs}
\begin{flushleft}
\textbf{Zeroth law}\ \ \textit{For stationary black holes, the event horizon has constant surface gravity $\kappa$}.
\end{flushleft}
\begin{flushleft}
\textbf{First law}\ \ \textit{If a stationary black hole is perturbed, the mass of the system changes as}
\begin{equation}\label{eq:firstlaw}
\updelta M=\frac{\kappa\updelta A}{8\pi \GN}+\Omega\updelta J+\Phi\updelta Q,
\end{equation}
\textit{where $A$ is the area of the black hole horizon, $\Omega$ its angular velocity and $\Phi$ its electric potential}\footnote{If additional variables are required to characterized the black hole, the first law is modified. This is the case \eg for hairy black holes \cite{Anabalon:2015ija,Astefanesei:2018vga,Astefanesei:2021vcp}}.
\end{flushleft}
\begin{flushleft}
\textbf{Second law}\ \ \textit{As time evolves, the area $A$ of a black hole does not decrease
\begin{equation}\label{eq:secondlaw}
\Delta A\geq0
\end{equation}
}
\end{flushleft}
\begin{flushleft}
\textbf{Third law}\ \ \textit{It is not possible to form a black hole with vanishing surface gravity.}
\end{flushleft}
The flagrant parallelism between the laws of black hole mechanics and the laws of thermodynamics allowed to identify the area of the horizon with the black hole entropy, this is what we call today the Bekenstein-Hawking formula (or simply area law) \cite{Bekenstein:1973ur}
\begin{equation}\label{eq:BHform}
S_{\text{BH}}=\frac{A}{4\GN}.
\end{equation}
Following this interpretation, the second law implied that the area of the event horizon never decreases $\Delta S_{\text{BH}}\geq0$ . On the other hand, Hawking identified the surface gravity with the temperature in one of the earliest realizations of quantum gravity. In his derivation, he showed that black holes emit thermal radiation arising from the conversion of quantum vacuum fluctuations into pairs of particles. 
Whereas one of them is trapped inside the black hole horizon, the other escapes to infinity, reaching an observer that measures is temperature as \cite{Hawking:1974sw}
\begin{equation}\label{eq:Hawktemp}
T=\frac{\kappa}{2\pi}.
\end{equation}
There are many different approaches to derive the temperature of the Hawking radiation \cite{Hartle:1976tp,Brout:1995rd,Visser:2001kq}. Here, we review the Euclidean action approach \cite{Gibbons:1976ue,Gibbons:1977mu}, as we will use it afterwards. 

\subsection{Euclidean derivation of black hole temperature}
Starting from the general SSS ansatz \eqref{eq:SSS} and its horizon radius $\rhor$, we can perform a Wick rotation on the time coordinate\footnote{Conversely, when we work in setups in which the time coordinate did not undergo a Wick rotation, we can refer them as \emph{real-time} approach.} $t\rightarrow-\iu t_\text{E}$ and study the near-horizon limit $r\to\rhor$. By doing so, the time and radial components of the metric read
\begin{equation}
\diff s^2_\text{E}=N(\rhor)f'(\rhor)(r-\rhor)\diff t^2_\text{E}+\frac{\diff r^2}{f'(\rhor)(r-\rhor)}+r^2\diff\Omega_{D-2}^2.
\end{equation} 
If we introduce in our metric the change of coordinates $\rho\equiv2\sqrt{\frac{(r-\rhor)}{f'(\rhor)}}$, with $\rho\ll \rhor$, we arrive to
\begin{equation}
\diff s^2_\text{E}\approx\diff \rho^2+\rho^2\left(\frac{N(\rhor)f'(\rhor)}{2}\right)^2\diff t^2_\text{E}+\rhor^2\diff\Omega_{D-2}^2,
\end{equation}
which corresponds to the metric of $\mathbb{R}^2\times\mathbb{S}^{D-2}$. The sector $\mathbb{R}^2$ is written in polar coordinates and in order to avoid a conical singularity in the near-horizon geometry we must identify $t_\text{E}\sim t_\text{E}+\frac{4\pi}{N(\rhor)f'(\rhor)}$. In consequence, the time coordinate in Lorentzian signature is periodic in the imaginary direction $t\sim t+\frac{4\pi\iu}{N(\rhor)f'(\rhor)}$. With this in mind, we recall that in quantum field theory, a state in thermal equilibrium satisfies the Kubo-Martin-Schwinger condition \cite{Kubo:1957mj,Martin:1959jp,Haag:1967sg}, which imposes periodicity in the imaginary time for the thermodynamic Green's function 
\begin{equation}
G_\beta(t_1,\mathbf{x}_1;t_2,\mathbf{x}_2)=G_\beta(t_1+\iu \beta,\mathbf{x}_1;t_2,\mathbf{x}_2),
\end{equation}
where $\beta=T^{-1}$ is the inverse of the temperature. Quantum fields placed in a periodic spacetime are also subject to the same condition \cite{Gibbons:1976pt} and hence, the SSS black hole \eqref{eq:SSS} has a temperature, given by
\begin{equation}
T=\frac{N(\rhor)f'(\rhor)}{4\pi}=\frac{\kappa}{2\pi}.
\end{equation}	

Once the temperature is found and the entropy fixed following the Bekenstein-Hawking formula \eqref{eq:BHform}, we can define the rest of thermodynamic quantities using the gravitational Euclidean path integral
\begin{equation}
Z=\int\pazocal{D}[g_{ab}]\e{-I_{\text{E}}[g_{ab}]}.
\end{equation}
We can approximate the partition function by working around a classical saddle point, \ie a solution of the classical equations of motion $I_{\text{E}}[g_{ab}]=I_{\text{E}}[g_{ab}]\Big|_{\text{on-shell}}+\ldots$, where the leading contribution is evaluated on-shell. In consequence, the partition function reads
\begin{equation}\label{eq:saddle}
Z\approx \e{-I_{\text{E}}[g_{ab}]\Big|_{\text{on-shell}}}
\end{equation}
in the semiclassical regime. In this expression, we have to include the GHY term \eqref{eq:GHY} so that the variational problem is well posed.  However, the story is not yet complete, as after evaluating the Euclidean on-shell action we will find a divergent quantity. Because of this, we also need to add counterterms to renormalize it. We will address this problem in next sec.~\ref{sec:holoren} as it of relevance in the context of the AdS/CFT correspondence.

Based on these considerations, in the case of GR, the Euclidean action is given by\footnote{For now we will keep denoting $I_\text{E}$ as some Euclidean action that includes its appropriate counterterm. However, later we will be interested in making explicit distinction whether it is renormalized. In that case we denote it by $I_\text{E}^{\text{ren}}$.}
\begin{equation}\label{eq:Euaction}
I_{\text{E}}=-\frac{1}{16\pi\GN}\int_{\B}\diff^Dx\sqrt{g}\left(R-2\Lambda\right)-\frac{1}{8\pi \GN}\int_{\partial\B}\diff^{D-1}x\sqrt{\gamma}K+I_{\text{ct}},
\end{equation}
where $I_{\text{ct}}$ indicates the presence of a counterterm to renormalize the on-shell action. 

Once we have the partition function, from statistical mechanics, we know that the Helmholtz free energy (or free energy for short) of the black hole is given by
\begin{equation}
F=-\frac{1}{\beta}\log Z\approx \frac{1}{\beta}I_{\text{E}}[g_{ab}]\Big|_{\text{on-shell}}.
\end{equation}
With all these ingredients, we are able to find closed expressions for the thermodynamic quantities
\begin{equation}
M=-\frac{\partial}{\partial\beta}\log Z,\quad S=\beta M+\log Z,
\end{equation}
which satisfy the relation
\begin{equation}
F=M-TS,
\end{equation}
in the absence of chemical potential like the electric charge or angular momentum.

After giving some general notions of black hole thermodynamics in GR, we move to solutions in higher-curvature gravity. However, let us mention before that there exists an extended thermodynamics approach in which the the mass of the black hole is identified with the enthalpy of spacetime and the cosmological constant $\Lambda$ is regarded as the pressure of the black hole, with the volume as its conjugate thermodynamic value \cite{Kubiznak:2014zwa,Kubiznak:2016qmn,Visser:2021eqk,Cong:2021jgb}. This interpretation holds in the case of hairy black holes \cite{Astefanesei:2018vga,Astefanesei:2019ehu,Astefanesei:2021vcp}.

\subsection{Wald entropy}

For higher-curvature gravities, the area law \eqref{eq:BHform} is no longer true and corrections arise. For a general diffeomorphism invariant theory with arbitrary higher-curvature terms $\pazocal{L}$, the black hole entropy is computed using the Wald's formula \cite{Wald:1993nt,Jacobson:1993vj,Iyer:1994ys}, given by
\begin{equation}\label{eq:WaldS}
S_\text{Wald} = -2\pi \int_{\pazocal{H}} \diff^{D-2}  x  \sqrt{{}_{\pazocal{H}}g} \, \tilde{P}_{ab}{}^{cd} \varepsilon^{ab}\varepsilon_{cd} \, ,
\end{equation}
where integration is taken over the binormal surface of the horizon $\pazocal{H}$, ${}_{\pazocal{H}}g$ is determinant of the induced metric and the tensor $\tilde{P}_{ab}{}^{cd}$ results from the variation of the Lagrangian considering the Riemann tensor as an independent variable, this is
\begin{equation}\label{eq:Ptensor}
\tilde{P}_{ab}{}^{cd}\equiv\frac{\updelta\pazocal{L}}{\updelta R^{ab}{}_{cd}}=\frac{\partial \pazocal{L}}{\partial R^{ab}{}_{cd}}-\nabla_e\left(\frac{\partial \pazocal{L}}{\partial \nabla_e R^{ab}{}_{cd}}\right)+\ldots
\end{equation}
Besides, $\varepsilon_{ab}$ is the binormal vector to the horizon $\pazocal{H}$, normalized such that  $\varepsilon_{ab}\varepsilon^{ab}=-2$. For stationary black holes, $\pazocal{H}$ is a bifurcation surface, which is a fixed point locus of the time translation symmetry.

Using the Wald entropy, it can be shown that the first law holds for theories of the type $\pazocal{L}(R_{abcd})$, using $S_{\text{W}}$ instead of $S_{\text{BH}}$ in eq. \eqref{eq:BHform}. As long as we restrict ourselves to stationary black holes, the second law \eqref{eq:secondlaw}
will also hold. If this is not the case, the suitable entropy for the generalized second law \cite{Wall:2011hj,Sarkar:2013swa,Wall:2015raa} coincides with the Camps-Dong formula for entanglement entropy \cite{Dong:2013qoa,Camps:2013zua}. We discuss further on this in sec.~\ref{sec:heehcg}.

%

\section{The AdS/CFT correspondence}

%

One of the most famous predictions of String Theory is the holographic principle, which conjectures the duality between a the propagating fields of quantum gravity living in a $(D=d+1)$-dimensional bulk and the observables of a theory defined in its boundary \cite{Thorn:1991fv,tHooft:1993dmi,Susskind:1994vu}, \ie in one dimension lower. A concrete realization of this principle is provided by the AdS/CFT correspondence \cite{Maldacena:1997re,Gubser:1998bc,Witten:1998qj}, which has been subject of extensive research since its inception. Among the many forms of the correspondence, the most famous one is given by the duality between type IIB superstring theory on AdS$_{5}\times \mathbb{S}^5$, with a strongly coupled $\pazocal{N}=4$ supersymmetric Yang-Mills theory with gauge group SU$(N)$ in four dimensions, where $\pazocal{N}$ being the number of supercharges of the theory. The realization of the correspondence becomes clear when the theory is compactified on the sphere $\mathbb{S}^5$ and the effective five-dimensional gravity theory in AdS${}_5$ is considered, in whose boundary lives the four-dimensional gauge theory.

The fundamental identity is the equality of the partition functions of both theories is the Gubser-Klebanov-Polchinski-Witten (GKP-W) relation, namely \cite{Gubser:1998bc,Witten:1998qj}
\begin{equation}\label{eq:GKPW}
Z_\text{CFT}=Z_\text{ST}\approx\e{-I_{\text{E}}[g_{ab}]\Big|_{\text{on-shell}}},
\end{equation}
where in the last equality, the saddle-point approximation \eqref{eq:saddle} ---contrary to the expression before, $I_{\text{E}}[g_{ab}]\Big|_{\text{on-shell}}$ stands for some unspecified Euclidean gravity on-shell action not necessarily evaluated at the SSS solution \eqref{eq:SSS}. From this expression, we are able to compute the quantities in any of the theories and the set of relations between them is known as \textit{holographic dictionary}.

\subsection{Anti-de Sitter spacetime}

Before discussing anything regarding the holographic dictionary, we make a brief review of AdS and CFT so that we acquire some vocabulary first. We have already discussed that an ambient $(d+2)$ flat spacetime with two timelike coordinates $\mathbb{R}^{2,d}$, defined by the metric
\begin{equation}\label{eq:flatdp2}
\diff s^2=-\diff y_{-1}^2-\diff y_0^2+\delta_{\mu\nu}\diff y^{\mu}\diff y^{\nu},
\end{equation}
allows AdS$_{d+1}$ spacetime with radius $L$ as an embedding hypersurface, which satisfies the constraint
\begin{equation}
-y_{-1}^2-x^2_0+y_1^2+\sum_{i=1}^{d}\diff y^2_{i}=-L^2.
\end{equation} 
From this definition it is manifest that SO$(2,d)$ is the isometry group of AdS$_{d+1}$, as stated above. We are particularly interested in two sets of coordinates that describe this spacetime, namely, global coordinates and Poincar\'e coordinates. For the former, we introduce the coordinate transformations
\begin{align}
y_{-1}&\equiv L\cosh \rho\sin t,\notag\\
y_{0}&\equiv L\cosh \rho\cos t,\\
y_{i}&\equiv L y_i \cosh \rho\cos t,\notag
\end{align}
where $y_i$ define a $d$-dimensional sphere, provided that they satisfy the relation $\sum_{i=1}^d y_i=1$. This already defines global coordinates but if we further perform the coordinate transformation $r\equiv\sinh \rho$ we arrive to its popular expression
\begin{equation}\label{eq:globalcoord}
\diff s^2=-\left(\frac{r^2}{L^2}+1\right)\diff t^2+\frac{\diff r^2}{r^2+1}+r^2\diff\Omega_{d-1}^2.
\end{equation}
This coordinate patch cover the AdS$_{d+1}$ spacetime entirely, whose boundary, given by $r\rightarrow\infty$,  corresponds to $\mathbb{R}\times \mathbb{S}^{d-1}$ spacetime. On the other hand, if starting from metric \eqref{eq:flatdp2} we change of coordinates
\begin{align}
y_{-1}&\equiv\frac{z}{2}+\frac{1}{2z}\left(L^2-t^2+\sum_{i=1}^{d-1}x^2_i\right),\notag\\
y_{0}&\equiv\frac{L t}{z},\\
y_{i}&\equiv  \frac{Lx_i}{z},\quad y_d\equiv\frac{1}{2r}+\frac{r}{2}\left(-L^2-t^2+\sum_{i=1}^{d-1}x^2_i\right)\notag,
\end{align}
we arrive to the Poincar\'e coordinates, given by
\begin{equation}\label{eq:Poincare}
\diff s^2=\frac{L^2}{z^2}\left(\diff z^2+\eta_{\mu\nu}\diff x^{\mu}\diff x^{\nu}\right),
\end{equation}
covering half of AdS$_{d+1}$ spacetime and whose boundary at $z=0$ is the Minkowski space $\mathbb{R}^{1,d-1}$, denoted by $\eta_{\mu\nu}$.

Since AdS is a maximally symmetric negative constant curvature spacetime, we can write the Riemann tensor as a symmetrized combination of metric tensors as
\begin{equation}\label{eq:RAdS}
R\indices{_a_b^c^d}=-\frac{1}{L^2}\delta_{ab}^{cd},
\end{equation}
and its contraction, the Ricci tensor and the Ricci scalar are given by $R\indices{_a^b}=-\frac{d}{L^2}\delta_a^b$ and $R=-\frac{d(d+1)}{L^2}$, respectively.

In the Poincar\'e patch, the boundary of AdS$_{d+1}$ is a flat spacetime with a second order pole. However, the boundary metric can have a different geometry and still solve the Einstein field equations \eqref{eq:EFE}. These spacetimes are known as asymptotically locally AdS (AlAdS) and its metric can be written in terms of the Fefferman-Graham coordinates
\begin{equation}\label{eq:FG}
\diff s^2=\frac{L^2}{z^2}\left(\diff z^2+\gamma_{\mu\nu}(z,x^\mu)\diff x^\mu \diff x^\nu\right),
\end{equation}
where $\gamma_{\mu\nu}$ is regular in the spacetimes under consideration. By construction, $\gamma_{\mu\nu}$ should admit an expansion in non-negative powers of the holographic coordinate $z$, as
\begin{equation}
\gamma_{\mu\nu}(z,x^\mu)=\tilde{\gamma}_{\mu\nu}^{(0)}+z\tilde{\gamma}_{\mu\nu}^{(1)}+z^2\tilde{\gamma}_{\mu\nu}^{(2)}\ldots,
\end{equation}
which is known as the \emph{Fefferman-Graham expansion}. The terms of the expansion can be computed order by order by plugging them into the Einstein field equations \eqref{eq:EFE}. Recursion relations are available for odd and even terms, but the latter are more involved as logarithmic coefficients appear at order $\pazocal{O}(z^d)$ in the expansion. We will come back to these construction when discussing holographic renormalization in the future.

Once we have discussed coordinates of AdS$_{d+1}$ of our interest, we turn our attention to CFTs. 

\subsection{Conformal field theory}\label{sec:CFT}

Let us consider a general $d$-dimensional quantum field theory (QFT$_d$). If the QFT is invariant under the conformal transformation of the metric
\begin{equation}
\gamma_{\mu\nu}(x)\rightarrow\Omega\left(x\right)\gamma_{\mu\nu}(x),
\end{equation}
with $x=x^\mu$ being the spacetime coordinates, then we say it is a CFT$_d$. The whole group of symmetries, known as conformal group, is an extension of the Poincar\'e group, including dilatations $x^\mu\rightarrow \lambda x^\mu$, for $\lambda\in\mathbb{R}$ a certain constant, and special conformal transformations $x^\mu=(x^\mu+b^\mu x^2)/(1+2b_\mu x^\mu+b^2x^2)$ for a certain vector $b^\mu$ \cite{francesco2012conformal}. Among other practicalities of CFT, we can mention that they are relevant in the study of systems undergoing continuous phase transitions in condensed matter physics, as they display scale invariance. Besides, we will discuss its applications in the context of renormalization group flows in sec.~\ref{sec:RGflows}. 

In connection with our previous discussion, we mention that the boundary of Poincar\'e patch of AdS is connected through a conformal transformation of the metric to Minkowski spacetime. This argument can be extended to AlAdS spacetimes, as seen in the Fefferman-Graham coordinates \eqref{eq:FG}.

For now let us continue exploring features of CFTs. If we consider a general Euclidean action $I_\text{E}$, the variation of the action under the infinitesimal conformal transformation $\updelta \gamma_{\mu\nu}=2\Omega\, \gamma_{\mu\nu}$ is given by
\begin{equation}\label{eq:traceanomaly}
\updelta I_{\text{E}}=\int\diff^dx\sqrt{g}\,T\indices{_\mu^\mu}\updelta\Omega,
\end{equation}
where $T\indices{_\mu^\mu}$ is the trace of the stress-energy tensor, defined in eq. \eqref{eq:EFE}. If the theory is conformally inviant, the variation of the action must vanish, which imposes $T\indices{_\mu^\mu}=0$. The traceless property at classical level is violated at quantum level in the case of even-dimensional CFTs, giving rise to the trace anomalies. Its expression is given by \cite{Bonora:1985cq,Deser:1993yx,Duff:1993wm}
\begin{equation}\label{eq:traceanomaly2}
\langle T\indices{_\mu^\mu}\rangle=-2(-1)^{d/2}A\X_{d}+\sum_iB_i\pazocal{I}_i+\nabla_\mu J^\mu,
\end{equation}
where $\X_{d}$ is the dimensionally extended Euler density as defined in eq. \eqref{eq:Eulerdens}, $\pazocal{I}_d$ are a set of independent Weyl invariants, built from the Weyl tensor and related to the Riemann tensor $\R_{\mu\nu\rho\sigma}$ of metric $\gamma_{\mu\nu}$ as
\begin{equation}\label{eq:Weyl}
   \W_{\mu\nu\rho\sigma}\equiv\R_{\mu\nu\rho\sigma}-\frac{2}{(d-1)}\left(\gamma_{\mu[\rho}\R_{\sigma]\nu}-\gamma_{\nu[\rho}\R_{\sigma]\mu}\right)+\frac{2}{d(d-1)}\R \gamma_{\mu[\rho}\gamma_{\sigma]\nu},
\end{equation}
and $\nabla_aJ^a$ are total derivatives terms that depend on the renormalization scheme. The coefficients $A$ and $B_i$ are referred as the type $A$ and $B_i$ central charges of the conformal field theory. In sec.~\ref{sec:RGflows} we will see the relevance of these coefficients in the context of renormalization group flows. 

In the $d=2$ case, the Euler density is given by $\X_2=\R$ while there are no Weyl invariants. Thus, the trace anomaly reads
\begin{equation}
\langle T\indices{_\mu^\mu}\rangle=\frac{c}{24\pi}\X_2,
\end{equation}
where the type-A central charge is a rescaled as $A=\frac{c}{48\pi}$, following the standard convention in CFT${}_2$. This quantity has a remarkable property in renormalization group flows, as we will see below in sec.~\ref{sec:RGflows}. On the other hand, in the next case, corresponding to $d=4$, the trace anomaly reads
\begin{equation}
\langle T\indices{_\mu^\mu}\rangle=-\frac{a}{16\pi^2}\X_4+\frac{c}{16\pi^2}\pazocal{I}_4,
\end{equation}
where $a$ refers to the type A central charge, $c$ to the type B and $\pazocal{I}_4=\W_{abcd}\W^{abcd}$ is the Weyl-squared invariant. In general, both the value of the coefficients can be constrained imposing physical requirements for the theories considered. One example is, given a initial state, imposing positivity on the total energy flux measured at infinity leads to the so-called conformal bounds \cite{Hofman:2008ar}
\begin{equation}\label{eq:confbound}
3\geq\frac{c}{a}\geq\frac{18}{31},
\end{equation}
which hold for any unitary four-dimensional CFT \cite{Hofman:2016awc}.

Contrary to type A central charge, the number of type B charges grows as we consider higher dimensions. For instance in six dimensions there are three Weyl invariants \cite{Bonora:1985cq}. This is represented in eq.~\eqref{eq:traceanomaly2} by the sum term.

Another quantity of interest to the holographic dictionary is the stress-tensor two-point correlation function. Remarkably, its form is heavily constrained by conformal symmetry as~\cite{Osborn:1993cr}
\begin{equation}\label{eq:ctdef}
\langle T_{\mu\nu}(x)T_{\rho\sigma}(0)\rangle=\frac{C_{\ssc T}}{x^{2d}}\left(I_{\mu(\rho}I_{\sigma)\nu}-\frac{\eta_{\mu\nu}\eta_{\rho\sigma}}{d}\right)
\end{equation}
where $I_{\mu\nu}\equiv\eta_{\mu\nu}-2\frac{x_\mu x_\nu}{x^2}$ and  $C_{\ssc T}$, referred as the coefficient of the two-point stress-tensor function, is the only theory dependent quantity. In the gravity picture, the metric perturbations around Minkowski $\eta_{\mu\nu}\rightarrow\eta_{\mu\nu}+h_{\mu\nu}(x)$ with $h_{\mu\nu}\ll \eta_{\mu\nu}$ appearing in the boundary of AdS in the Poincar\'e patch \eqref{eq:Poincare}, allows to compute the two-point correlation function, from the expression
\begin{equation}
\langle T_{\mu\nu}(x)T_{\rho\sigma}(x')\rangle=-\left.\frac{\updelta I_{\text{gravity}}}{\updelta h^{\mu\nu}(x)\updelta h^{\rho\sigma}(x')}\right|_{h_{\mu\nu}=0}.
\end{equation}
Thus, we can read the coefficient $C_{\ssc T}$ for CFTs that are dual the gravity theory $I_{\text{gravity}}$.

\subsection{Higher-curvature gravities in holography}

Motivated by the success of the AdS/CFT correspondence in the original form, a natural question is whether this framework allows to explore phenomena in more general strongly correlated systems. In other words, if the AdS/CFT correspondence is a concrete realization of a more general \emph{gauge/gravity duality}. With this in mind, we can interpret that Einstein-AdS gravity is the result of a truncation of a larger bulk theory. The enlarged one may include higher-curvature corrections in the bulk, additional matter fields or other additional features. In this perspective, theories including these additional degrees of freedom allow us to define toy models of strongly coupled CFTs inequivalent to Einstein-AdS gravity. Thus, higher-derivative theories provide a playground to study general results, find universal features or counterexamples to previous conjectures in the literature. 

To illustrate this we will mention some notable examples, but first let us discuss that when higher-curvature densities are considered, the theory posses several maximally symmetric solutions in general. The AdS vacua, whose bare radius $L$ is modified in terms of an effective one $L_{\star}=L/\sqrt{f_\infty}$, of any theory of the type \eqref{eq:priA} can be obtained by solving the equation \cite{Bueno:2016ypa,Bueno:2018yzo}
\begin{equation}\label{eq:embdf}
h(f_\infty)\equiv\frac{16\pi \GN L^2}{d(d-1)}\left[\pazocal{L}(f_\infty)-\frac{2f_\infty}{d+1}\pazocal{L}'(f_\infty)\right]=0,
\end{equation}
where $\pazocal{L}'(f_\infty)\equiv\frac{\diff \pazocal{L}(f_{\infty})}{\diff f_\infty}$ and $\pazocal{L}(f_\infty)$ is the on-shell Lagrangian evaluated on pure AdS${}_{d+1}$ with the effective AdS radius. This amounts to evaluating all Riemann tensors appearing in the Lagrangian \eqref{eq:priA} as eq. \eqref{eq:RAdS}, with $L/\sqrt{f_\infty}$ instead of $L$. By doing so, the ``characteristic polynomial'' $h(f_\infty)$ or ``embedding function'', reads \cite{Bueno:2016ypa,Bueno:2018yzo}
\begin{equation}\label{eq:polych}
h(f_\infty)=1-f_\infty+\sum_{n=2}\sum_{i_n}\alpha_{i_n}^{(n)}\mathfrak{R}_{i_n}^{(n)},
\end{equation}
where the coefficients $\alpha_{i_n}^{(n)}$ are conveniently rescaled. It is clear that in the case of Einstein gravity $f_\infty=1$ and in consequence $L=L_\star$. A less obvious example is the case of quadratic gravity \eqref{eq:QCGgen}, whose characteristic polynomial reads
\begin{equation}\label{eq:pchQC}
h(f_\infty)=1-f_\infty+(d-3)f_\infty^2\left[\frac{d(d+1)}{d-1}\alpha_1^{(2)}+\frac{d}{d-1}\alpha_2^{(2)}+(d-2)\alpha_3^{(2)}\right].
\end{equation}
An explicit example of rescaling the coupling constant to make the polynomial look like eq. \eqref{eq:polych} becomes evident when restricting ourselves to the Gauss-Bonnet term. In this case $\alpha_1^{(2)}=\alpha_2^{(2)}=0$ and thus transforming $\alpha_3^{(2)}\to\alpha_3^{(2)}/[(d-2)(d-3)]$ we obtain the well known relation \begin{equation}
f_\infty=\frac{1\pm\sqrt{1-4\alpha_{3}^{(2)}}}{2\alpha_{3}^{(2)}}.
\end{equation}
The auxiliary function $h(f_\infty)$ has other applications in the context of black hole thermodynamics, as we will see in chapter \ref{ch:ateach}.

Taking the existence of an effective AdS radius into account, the first example of inequivalent CFT we discuss pertains the computation of type A and type B central charges in $d=4$ CFTs. In the case of field theories dual to five-dimensional Einstein-AdS gravity the coefficients coincide \cite{Henningson:1998ey,Henningson:1998gx,Schwimmer:2008yh}
\begin{equation}
a=c=\frac{\pi L^3}{8 \GN},
\end{equation}
 corresponding to a single value in the conformal bound \eqref{eq:confbound}. However, when a higher-curvature correction is included in the action, the central charges no longer coincide. Following the discussion before, in the case of theories dual to Gauss-Bonnet, the charges read \cite{Myers:2010jv}
\begin{equation}
a=\frac{\pi L^3_\star}{8\GN}\left(1-6\alpha^{(2)}_3f_\infty\right),\quad c=\frac{\pi L^3_\star}{8\GN}\left(1-2\alpha^{(2)}_3f_\infty\right),
\end{equation}
allowing to consider a wide range of dual CFTs, depending on the coupling constants. From them, only certain values correspond to well-defined theories, following the conformal bounds \eqref{eq:confbound}. Central charges have been computed for a number of theories, including Lovelock, GQT gravities given its interest in renormalization group flows as we will see below \cite{Myers:2010jv,Myers:2010tj,Myers:2010xs,Bueno:2020uxs}. In this context, we will study the general quadratic curvature case in ch.~\ref{chap:heeqcg}.

Another example of the relevance of higher-curvature gravities is provided in the context of the study of the coefficient of higher-$n$ point stress-tensor correlation function. As we have seen before, the case $n=2$ is controlled by the coefficient $C_{\ssc T}$, which, for theories dual to Einstein-AdS gravity is given by
\begin{equation}
C_{\ssc T}^{\text{E}}=\frac{\Gamma(d+2)L^{d-1}}{8\pi^{\frac{d+2}{2}}(d-1)\Gamma(d/2)\GN}.
\end{equation}
In the case of the $n=3$ correlation function, two new characteristic coefficients, $t_2$ and $t_4$ for $d\geq4$ ---in the case of $d=3$, only $t_4$ is present. As they vanish in the case of CFT dual to Einstein-AdS gravity, we conclude that this theory provides a particular form of the three-point function. The situations differs for higher-curvature gravity, and we can gain access to $t_2$ in the case of Lovelock gravities \cite{Camanho:2009vw,Buchel:2009sk,Buchel:2009tt,Camanho:2009hu,Camanho:2010ru,Camanho:2013pda} and $t_4$ when considering theories dual to GQT gravities \cite{Myers:2010jv,Bueno:2018yzo}. It should be noted that not all coupling constants define well-behaved CFTs. As we argued above \eqref{eq:confbound}, positivity of the CFT imposes bounds on the values of the central charges $a$ and $c$. Similarly, causality and unitarity constraint the value of the coefficients $t_2$ and $t_4$. In turn, this restricts the range of the coupling constants of the higher-curvature Lagrangian under consideration. In the case of Lovelock and GQT gravities, this situation has been studied in refs.~\cite{Camanho:2009vw,Camanho:2009hu,Camanho:2013pda}.

Lastly, we mention an example on how higher-curvature gravities can provide counterexamples to previous conjectures. A famous case involves the Kovtun-Son-Starinets bound for the shear viscosity over entropy density ratio \cite{Kovtun:2004de}. Even though the value predicted by Einstein-AdS gravity is in accordance with the bound, when higher-curvature terms are included in the picture, the bound is violated in some cases \cite{Buchel:2004di,Kats:2007mq,Myers:2008yi,Brigante:2008gz,Ge:2008ni}.

These succinct list of examples is intended to  motivate the study of higher-curvature terms in the context of the gauge/gravity duality. Throughout the rest of the thesis we will mention some other particular cases that are relevant for our purposes

\subsection{Infrared renormalization of gravity}\label{sec:holoren}

As we pointed out before, the Einstein-AdS on-shell action is a divergent quantity ---which can be inferred from the conformal factor $1/z^2$ in the Poincar\'e patch of AdS \eqref{eq:Poincare}. In the early days of the AdS/CFT correspondence, the renormalization  of AdS gravity was achieved by the addition of intrinsic counterterms, such that the Dirichlet boundary condition was not spoiled \cite{Emparan:1999pm,Kraus:1999di,deHaro:2000vlm,Balasubramanian:1999re,Henningson:1998gx,Papadimitriou:2004ap,Papadimitriou:2005ii}. This prescription, valid in Einstein-AdS gravity in its original formulation and generalized afterwards for GQT gravities when evaluated in AdS spacetimes. It consist in the term \cite{Myers:2010tj,Myers:2010xs,Schwimmer:2008yh,Bueno:2018xqc}
\begin{equation}
I_\text{ct}=\frac{2a^*}{\Omega_{d-1}L^{d-1}_\star}\int_{\partial \B}\diff^{d}x\sqrt{\gamma}\left(\frac{d-1}{L_\star}+\frac{L_\star}{2(d-2)}\pazocal{R}+\ldots\right),
\end{equation}
where $a^*$, related to the entanglement entropy as we will discuss below, is given by the evaluation of the graviational Lagrangian in AdS
\begin{equation}
a^*=-\frac{\pi^{d/2}L^{d+1}_\star}{d\, \Gamma(d/2)}\pazocal{L}\big|_{\text{AdS}}.
\end{equation}

However, a Dirichlet condition on the boundary metric $\gamma_{\mu\nu}$ does not make sense in AlAdS spacetimes, as it was later pointed out in ref.  \cite{Papadimitriou:2005ii}. As a consequence, the only way to have a well-posed variational principle in AdS gravity is fixing (instead) the metric $\gamma_{(0)\mu\nu}$ at the conformal boundary. This argument implies that the addition of counterterms is required not only for canceling divergent terms in the variation of the action, but also for the consistency of the variational problem on $\gamma_{(0)\mu\nu}$.

A remarkable property of AdS is that the leading order in the asymptotic expansion of the extrinsic curvature is (up to a numerical factor) the same as the one in $\gamma_{\mu\nu}$. Indeed, in the Fefferman-Graham frame, $K_{\mu\nu}=\frac{1}{L}\frac{\tilde{\gamma}_{(0)\mu\nu}}{z^2}+...$, what has been recently emphasized in ref.\cite{Witten:2018lgb}. This simple observation means that one can express also variations of $K_{\mu\nu}$ in terms of variations of $\tilde{\gamma}_{(0)\mu\nu}$. In turn, this implies that we can consider also surface terms which depend on the
extrinsic curvature and act as counterterms, in the sense that they cancel divergent contributions in the AdS gravity action. So, even though they are plain incompatible with a Dirichlet condition for the full boundary metric $\gamma_{\mu\nu}$, they can still
reproduce the correct holographic stress tensor varying with respect to $\gamma_{(0)\mu\nu}$.

The above reasoning, which opens the possibility to look for an alternative sort of counterterms, is justified by the lack of a closed expression  for the series in arbitrary dimensions.
More than twenty years ago, evidence was provided on the fact that topological terms were able to regulate the
variation of the AdS gravity action in even dimensions \cite{Aros:1999id,Aros:1999kt}, though based on the study
of particular solutions.
As for the Euclidean action, the addition of the Euler term at the boundary of $d=2n$ dimensions, with $n\in\mathbb{N}$, renders it finite in AAdS solutions
if the coupling is adequately chosen \cite{Olea:2005gb}. However, it was not clear what this prescription to
renormalize AdS gravity had to do with Holographic Renormalization and the addition of standard counterterms.

A first step towards the understanding of this issue was given in ref. \cite{Olea:2005gb}, where topological terms in the bulk are equivalently written as the corresponding Chern form $B_{2n-1}$ at the boundary,
\begin{align} \label{Bdodd}
B_{2n-1}=&-2n\int\limits_0^1\diff s\,\delta^{\mu_1\cdots \mu_{2n-1}}_{\nu_1\cdots \nu_{2n-1}}K\indices{^{\nu_1}_{\mu_1}}\left(\frac{1}{2}\pazocal{R}\indices{^{\nu_2}^{\nu_3}_{\mu_2}_{\mu_3}}- s^2K\indices{^{\nu_2}_{\mu_2}}K\indices{^{\nu_3}_{\mu_3}}\right)\times\cdots\nonumber\\
&\cdots\times\left(\frac{1}{2}\pazocal{R}\indices{^{\nu_{2n-2}}^{\nu_{2n-1}}_{\mu_{2n-2}}_{\mu_{2n-1}}}-s^2K\indices{^{\nu_{2n-2}}_{\mu_{{2n-2}}}}K\indices{^{\nu_{2n-1}}_{\mu_{2n-1}}}\right),
\end{align}
For the first time, counterterms which depend on $K_{\mu\nu}$ were proposed to deal with the renormalization of AdS gravity. 

A similar structure at the boundary of $d+1=2n+1$ dimensions was far more involved to obtain. In particular, due to the fact that there is no equivalent form in the bulk for such boundary term. The extensive use
of field-theory tools in the context of anomalies (Chern-Simons and transgression forms, homotopy operator, etc.)
allows to make a concrete proposal for that case. The resulting term, in essence, shares common properties with the Chern form, as it is a given polynomial of the extrinsic and intrinsic curvatures \cite{Olea:2006vd},
but it did not exist in the mathematical literature before. Its specific form is given by the following expression
\begin{align}\label{eq:Bdeven}
B_{2n}=&\int_0^1\diff s_1\int_0^{s_1}\diff s_2\delta^{\mu_1\cdots \mu_{2n-1}}_{\nu_1\cdots \nu_{2n-1}}K\indices{^{\nu_1}_{\mu_1}}\left(\frac{1}{2}\pazocal{R}\indices{^{\nu_2}^{\nu_3}_{\mu_2}_{\mu_3}}- s^2_1K\indices{^{\nu_2}_{\mu_2}}K\indices{^{\nu_3}_{\mu_3}}+\frac{s^2_2}{L^2_\star}\delta_{\mu_2}^{\nu_2}\delta_{\mu_3}^{\nu_3}\right)\times\cdots\nonumber\\
&\cdots\times\left(\frac{1}{2}\pazocal{R}\indices{^{\nu_{2n-2}}^{\nu_{2n-1}}_{\mu_{2n-2}}_{\mu_{2n-1}}}-s^2_1K\indices{^{\nu_{2n-2}}_{\mu_{{2n-2}}}}K\indices{^{\nu_{2n-1}}_{\mu_{2n-1}}}+\frac{s_2^2}{L^2_\star}\delta_{\mu_{2n-2}}^{\nu_{2n-2}}\delta_{\mu_{2n-1}}^{\nu_{2n-1}}\right),
\end{align}
where $\Ls$ is the effective AdS radius of the theory. We might think of this renormalization scheme, dubbed \emph{Kounterterms}, may lead to a variational principle which is at odds with the holographic description of AdS gravity in terms of the boundary source $\gamma_{(0)ab}$, as it seems to require a different boundary condition on the extrinsic curvature. But the analysis portrayed above gives a firmer ground to the addition of Kounterterms to the gravitational action: the total action is consistent with a holographic description, as its variation is both finite and given in terms of $\updelta \tilde{\gamma}_{(0)\mu\nu}$.

This simple reasoning suggests the resummation of the counterterm series as an expression in terms of the extrinsic curvature. As a matter of fact, an asymptotic expansion of the term $B_{d}$ reproduces the counterterms, once the GHY term is correctly isolated \cite{Miskovic:2009bm,Anastasiou:2020zwc}\footnote{In ref. \cite{Anastasiou:2020zwc}, it  was shown that, in Einstein-AdS gravity, the Kounterterms are the resummation of the counterterms for asymptotically conformally flat manifolds in arbitrary dimensions. 
For a generic AlAdS space, there is a mismatch between counterterms and Kounterterms, consisting on terms which are the dimensional continuation of conformal invariants at the boundary. At the lowest order, this difference is a Weyl-squared term, which is identically vanishing for a conformally flat boundary. However, it may be the case that this condition is relaxed by taking Weyl$^{2}=0$ instead, what would be the analogue of demanding local flatness ($\text{Riem}=0)$ \emph{vs} a vanishing Kretschmann scalar for a given spacetime.
}. In addition, earlier works in the mathematical literature \cite{2000math.....11051A,2005math......4161A}, indicate that the Chern form is fundamental in defining the renormalized volume of an Einstein space.

The Kounterterm method has been used to deal with the construction of conserved quantities and the thermodynamic description
of black holes in Einstein-Gauss-Bonnet AdS and, in general, Lovelock AdS gravity \cite{Kofinas:2006hr,Kofinas:2007ns}. Furthermore, it has also linked
the concept of Conformal Mass to the addition of boundary terms in Einstein-Hilbert \cite{Jatkar:2014npa} and higher-curvature gravity \cite{Arenas-Henriquez:2017xnr,Arenas-Henriquez:2019rph,Arenas-Henriquez:2019syw}. 
Evidence has been given that Kounterterms can provide finite conserved charges in QCG, as well \cite{Giribet:2018hck,Giribet:2020aks,Miskovic:2022mqv}. 

In chapters \ref{chap:sdhee} and \ref{chap:heeqcg} of this thesis, we will extensively use its properties to deal with the problem of renormalization of holographic entanglement entropy dual to Einstein gravity and QC gravity. This quantity corresponds to another entry of the holographic dictionary so in next section we introduce it and contextualize its notable relevance in recent times. Preliminary results on renormalized holographic entanglement entropy using Kounterterms can be found in refs. \cite{Anastasiou:2017xjr,Anastasiou:2018rla,Anastasiou:2018mfk,Anastasiou:2019ldc,Anastasiou:2021jcv}.

\section{Quantum entanglement}

Entanglement is a fundamental phenomenon present in quantum physics. It refers to the nonclassical correlations that can appear between two spacelike-separated quantum systems. This counterintuitive behavior led Einstein, Podolsky, Rosen to argue that the quantum description of reality was ``incomplete'' \cite{einstein1935can}, famously describing entanglement as a \emph{spooky action at a distance}. They argued for the existence of ``elements of reality'' that were not included in the quantum theory, speculating the possibility of constructing theories containing them. These are the so-called \emph{hidden-variable theories}.
After years of extensive research and with the great contribution of John Bell, whose results are known as Bell's theorem \cite{Bell:1964fg,Bell:1964kc,Aspect:1981zz}, it was elucidated that quantum correlations are incompatible with local hidden variables\footnote{However, Bell's inequalities do not rule out the possibility of nonlocal hidden-variables. A prominent example of this is the De Broglie-Bohm theory \cite{Bohm:1951xw,Bohm:1951xx}.}. This implies that indeed the quantum theory involves nonlocal interactions in some sense. Since this seminal discovery there has been extensive discussion by physicists and philosophers regarding its implications. The goals here are more humble and we will study its relevance  as a tool to understand discrete and continuum systems, in the context of gauge/gravity duality and in renormalization group flows.

\begin{figure}
\begin{minipage}{.4\linewidth}
\begin{center}
\includegraphics[width=.25\linewidth,valign=c]{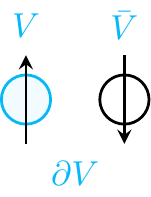} 
\end{center}
\hfill\vline\hfill
\end{minipage}
\begin{minipage}{.54\linewidth}
\centering
\includegraphics[width=.8\linewidth]{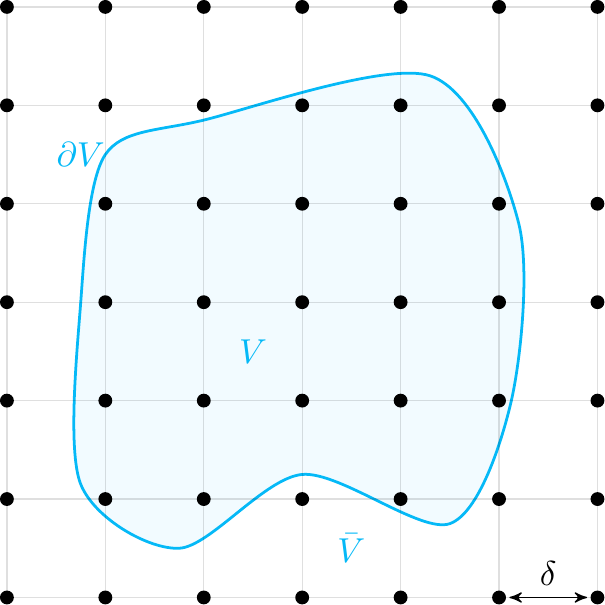}
\end{minipage}
\caption{\textsf{Two bipartite systems, divided by $\partial V$ in the region $V$ and its complementary $\bar{V}$. Left: a system consisting in two qubits. Right: a two-dimensional square lattice with spacing $\delta$.}}\label{fig:bip}
\end{figure}

Let us first illustrate the phenomenon of entanglement considering a bipartite quantum system ---this is a system that is divided into a subset $V$ and its complementary $\bar{V}$. In what follows, we will denote the region $V$ as the \emph{entangling region} and its boundary $\partial V$ as the \emph{entangling surface}. In figure \ref{fig:bip} we show two examples of discrete bipartite systems. They are, respectively, two quantum bits (or qubits, for short) and a two-dimensional square lattice with spacing $\delta$. 

The Hilbert space of these systems is given by the tensor product of each one of the qubits individually
\begin{equation}\label{eq:fact}
\pazocal{H}_\text{tot}=\pazocal{H}_V\otimes\pazocal{H}_{\bar{V}}.
\end{equation}
A general pure state\footnote{We require that the state are normalized, namely $\langle\Psi|\Psi\rangle=1$.} $\ket{\Psi}\in\pazocal{H}_\text{tot}$ can be written in terms of the bases of each of the Hilbert spaces  $\{\ket{i}_V\}\in\pazocal{H}_{V}$ and $\{\ket{j}_{\bar{V}}\}\in\pazocal{H}_{\bar{V}}$ as $\ket{\Psi}=\sum_{i,j}c_{i,j}\ket{i}_{V}\otimes\ket{j}_{\bar{V}}$, where the coefficients $c_i$ must satisfy $\sum_{i,j}\abs{c_{i,j}}^2=1$.

If the state $\ket{\Psi}$ can be written as a tensor product of the states of the subsystems, \ie $\ket{\Psi}=\ket{\Phi}_V\otimes\ket{\Phi}_{\bar{V}}$ with $\ket{\Phi}_V\in\pazocal{H}_V$ and $\ket{\Phi}_{\bar{V}}\in\pazocal{H}_{\bar{V}}$ the state is said to be \emph{separable}. Otherwise, the state is known as \emph{non-separable}, or \emph{entangled}. In the case of the qubits system examples of these states are
\begin{align}
& \ket{\Psi_1}\equiv\frac{1}{\sqrt{2}}\left(\ket{00}+\ket{01}\right)=\frac{1}{\sqrt{2}}\ket{0}_V\otimes\left(\ket{0}_{\bar{V}}+\ket{1}_{\bar{V}}\right) &\quad \text{separable},\\
& \ket{\Psi_2}\equiv\frac{1}{\sqrt{2}}\left(\ket{10}+\ket{01}\right) &\quad \text{entangled},\label{eq:entangqubit}
\end{align}
where we used the notation $\ket{ij}=\ket{i}_V+\ket{j}_{\bar{V}}$. Needless to say, any state of the system evolves temporally in the Schr\"odinger picture according to the Schr\"odinger equation $\iu \frac{\diff}{\diff t}\ket{\Psi_i(t)}=H \ket{\Psi_i(t)}$, where $H$ is the Hamiltonian of the system. 

\subsection{Entanglement entropy}

In order to give a measure of the degree of the entanglement of the system, we need to define first the density matrix operator of a pure state, given by
\begin{equation}
\varrho_\text{tot}=\ket{\Psi}\bra{\Psi},
\end{equation}
which can be used to compute the expectation values of observables $\pazocal{O}$, through the relation $\tr(\varrho \pazocal{O})=\bra{\Psi}\pazocal{O}\ket{\Psi}$.

In the bipartite system, we can also define a \emph{reduced density matrix} $\varrho_V$ of the region $V$ by partially tracing over the complementary region $\bar{V}$, this is
\begin{equation}
\varrho_V\equiv\tr_{\bar{V}}(\varrho_\text{tot})=\sum_i {}_{\bar{V}}\bra{i}\varrho_{\text{tot}}\ket{i}_{\bar{V}}.
\end{equation}
In the case of a separable state ---such as $\ket{\Psi_1}$ above---, the tensor product of the reduced density matrices of the two subsystems recovers the total density matrix $\varrho_{\text{tot}}=\varrho_V\otimes\varrho_{\bar{V}}$. However, when the state is entangled, this is no longer true.


Based on these considerations, we can introduce the first measure of the degree of non-separability in a system. The \emph{entanglement entropy} of the subsystem $V$ by the \emph{von Neumann entropy} of the reduced density matrix $\varrho_V$
\begin{equation}\label{eq:EE}
\SEE(V)=-\tr_V(\rho_V\log \rho_V).
\end{equation}
Notice that the entanglement entropy of the total system vanishes for a pure ground state, this is $\SEE(V\cup\bar{V})=0$. On the other hand, following the example of the qubits system \eqref{eq:entangqubit}, we find that partially tracing $\ket{\Psi_2}$ with respect to $\bar{V}$ yields the maximal entanglement entropy $\SEE(V)=\log 2$ ---by maximal we mean that the entanglement entropy of any other state of the system must verify $\SEE(V)\leq \log 2$. Conversely, if we partial trace $V$, we find the same result, \ie
\begin{equation}\label{eq:purestate}
\SEE(V)=\SEE(\bar{V}).
\end{equation}
This result is a fundamental property of pure ground states of every quantum system and we will exploit it in the future (see section \ref{ch:gen}). Entanglement entropy has many other useful properties ---see \cite{nielsen2002quantum} and references therein for more details ---, but we will only mention an additional one that will be relevant for our purposes. Let us consider now a tripartite system, \ie divided in three disjoint regions $V_1$, $V_2$ and $V_3$. The following relations \cite{Araki:1970ba,Lieb:1973cp}
\begin{align}
&\SEE(V_1\cup V_2\cup V_3)+\SEE(V_2)\leq \SEE(V_1\cup V_2)+\SEE(V_2\cup V_3), \label{eq:SSEE}\\
&\SEE(V_1)+\SEE(V_3)\leq\SEE(V_1\cup V_2)+\SEE(V_2\cup V_3),
\end{align}
are known as the strong subadditivity property of entanglement entropy. This inequality can be used to derive other interesting properties such as the subadditivity property for bipartite systems $\SEE(V \cup W)\leq \SEE(V)+\SEE(W)$ (denoting $V\equiv V_1$ and $W\equiv V_3$). This is achieved by choosing $V_2$ as the empty set $\varnothing$. On the other hand, if $V_2=V_1\cap V_3$ we obtain the relation for intersecting regions
\begin{equation}\label{eq:SSEEi}
\SEE(V)+\SEE(W)\geq\SEE(V\cap W)+\SEE(V\cup W),
\end{equation}
which will be used in ch.~\eqref{ch:gen}.

The strong subadditivity property of entanglement entropy is essential in the context of renormalization group flows in QFT, as we will see below. Until now, we have discussed only discrete systems but we will describe the situation in the continuum limit. Before that, let us mention other entanglement measures that are relevant for us.

\subsection{Other entanglement measures}

Different measures of entanglement are useful depending on the area of interest. Here we will only review two of them, namely, R\'enyi entropy, modular entropy and mutual information, as we will be using afterwards. For reviews covering more entanglement measures in a holographic context see ref.~\cite{Rangamani:2016dms,Nishioka:2018khk}.

A straightforward generalization of entanglement entropy \eqref{eq:EE} is given by considering the moments of the reduced density matrix. This quantity is known as \emph{R\'enyi entropy} and it is defined as \cite{renyi1961}
\begin{equation}\label{eq:Renyi}
S_m(V)\equiv\frac{1}{1-m}\log\tr_V(\varrho^m_V),
\end{equation}
where $n\in\mathbb{Z}_+$ is known as the replica parameter. Even though $m$ is integer, we may analytically continue to reals, \ie $n\in\mathbb{R}_+$ and retrieve entanglement entropy in the limit $m\rightarrow1$ with the normalization $\tr_V(\varrho_V)=1$,
\begin{equation}\label{eq:EEfromRenyi}
\SEE(V)=\lim_{m\to 1}S_m(V)=-\lim_{m\to 1}\partial_m \log \tr_V(\varrho^m_V).
\end{equation}
This is known as the \emph{replica trick}. In sec.~\ref{sec:LMp} we will review its relevance in the holographic derivation of entanglement entropy. Among other applications we can mention its relation to thermodynamic quantities. This is achieved through the definition of the modular Hamiltonian 
\begin{equation}\label{eq:modH}
H_V\equiv-\frac{1}{2\pi}\log \varrho_V,
\end{equation}
and considering the partition function $Z(\beta)\equiv\tr_V(\varrho_V^n),$ yielding the inverse temperature $\beta=2\pi m$. This allows to define a \emph{modular entropy}
\begin{equation}
\tilde{S}_m(V)=m^2\partial_m\left[\frac{m-1}{n}S_m(V)\right],
\end{equation}
together with other quantities such as the modular energy, modular capacity, modular free energy, etc, that verify the usual thermodynamic relations \cite{zyczkowski2003renyi,Hung:2011nu,Nakaguchi:2016zqi}.

Another measure of our interest is known as \emph{mutual information}, which quantifies the amount of correlation between the subsystems. Its definition in terms of the entanglement entropy is given by\footnote{Notice that mutual information can be defined without any reference to entanglement entropy. Another definition involves relative entropy \cite{Araki:1970ba,Narnhofer:2011zz}, but we do not cover this measure here.}
\begin{equation}
I^\text{MI}(V,W)\equiv \SEE(V)+\SEE(W)-\SEE(V\cup W).
\end{equation}
This quantity will be useful when studying entanglement entropy in QFT as it is free from ultraviolet divergences ---shortly we will see this is the case of entanglement entropy. Moreover, from the subadditivity property of entanglement entropy we can check that mutual information is positive definite $I^\text{MI}(V,W)\geq0$.

In ch.~\ref{ch:gen}, we will be interested in yet another measure, defined for tripartite systems, given by
\begin{equation}
I_3\left(V_1,V_2,V_3\right)\equiv I^\text{MI}\left(V_1,V_2\right)+I^\text{MI}\left(V_1,V_3\right)-I^\text{MI}\left(V_1,V_2\cup V_3\right),
\end{equation}
More concretely, we will work within a model, known as ``Extensive Mutual Information'' (EMI) model, in which the tripartite information vanishes
 \begin{equation}\label{eq:ExMI}
I_3\left(V_1,V_2,V_3\right)=0.
\end{equation}

\subsection{Entanglement in continuum systems}

So far we have discussed the phenomenon of entanglement in arbitrary systems and we have provided explicit examples in a discrete one, such as the two qubits. However, we are ultimately interested in studying entanglement in continuum systems, \ie systems in which the lattice spacing becomes arbitrarily small $\delta\to0$. In such theories, the appropriate language is the one provided by QFT.

Extensive work has been carried out to provide a rigorous formulation of QFT. To this end, the \emph{Wightman axioms} are an attempt to formalize the notion of a quantum field theory on Minkowski spacetime $\mathbb{R}^{d-1,1}$ as a fixed background \cite{Wick:1952nb,wightman1965fields}. The number of axioms is not standardized and some of them appear combined in different references. Here we present them as follows.
\begin{flushleft}
\textbf{Axiom 1}\ \  \textit{There is a Hilbert space $\pazocal{H}$ in which acts the Poincar\'e spinor group.}
\end{flushleft}

\begin{flushleft}
\textbf{Axiom 2}\ \ \textit{There is a unique state $\ket{\Omega}\in\pazocal{H}$ of minimal energy denoted ``vacuum state''.}
\end{flushleft}

\begin{flushleft}
\textbf{Axiom 3}\ \  \textit{ There is a collection of quantum field operators $\phi(x)$, which transform under a unitary representation of the Poincar\'e spinor group.}
\end{flushleft}
However, formulation of QFT in terms of $\phi(x)$ can lead to singularities in certain situations and thus, it is sometimes more convenient to work with the smeared field $\phi(f_i)\equiv\int\diff^{d}x\, f_i(x)\phi(x)$ for smooth functions with compact support, like Gaussian functions\footnote{During this thesis, we use mostly $\phi(x)$ instead of $\phi(f)$ as it is customary. However, we must take into account that this is abuse of notation.}. 

\begin{flushleft}
\textbf{Axiom 4} \ \  \textit{Any two fields $\phi(f_1)$ and $\phi(f_2)$ (anti-)commute under spacelike-separated supports of $f_1(x)$ and $f_2(x)$. This is $\left[\phi(f_1),\phi(f_2)\right]_{\pm}=0$.}
\end{flushleft}

\begin{flushleft}
\textbf{Axiom 5} \ \  \textit{There is a dynamical law that allows to compute fields at an arbitrary time in terms of the fields in a Cauchy slice $S_{\mathrm{Cauchy}}=\{x|\epsilon>|x^0-t|\}$.}
\end{flushleft}


This axioms allows us to properly define a QFT. Moreover, an important result is the Wightman's theorem, which states that we are able to reconstruct the quantum fields and the Hilbert space from the vacuum expectation value of product of fields, \ie the correlation functions,
\begin{equation}
    \{\phi(x),\pazocal{H}_{\text{tot}}\}\Longleftrightarrow \bra{\Omega}\phi(x_1)\cdots\phi(x_n)\ket{\Omega}.
\end{equation}
This means that we can write any state $\ket{\Psi}\in\pazocal{H}_{\text{tot}}$ as a linear combination $\ket{\Psi}=$ L.C.$\left[\phi(x_1)\cdots\phi(x_n)\ket{\Omega}\right]$.

\begin{figure}
\centering
\includegraphics[width=.9\linewidth]{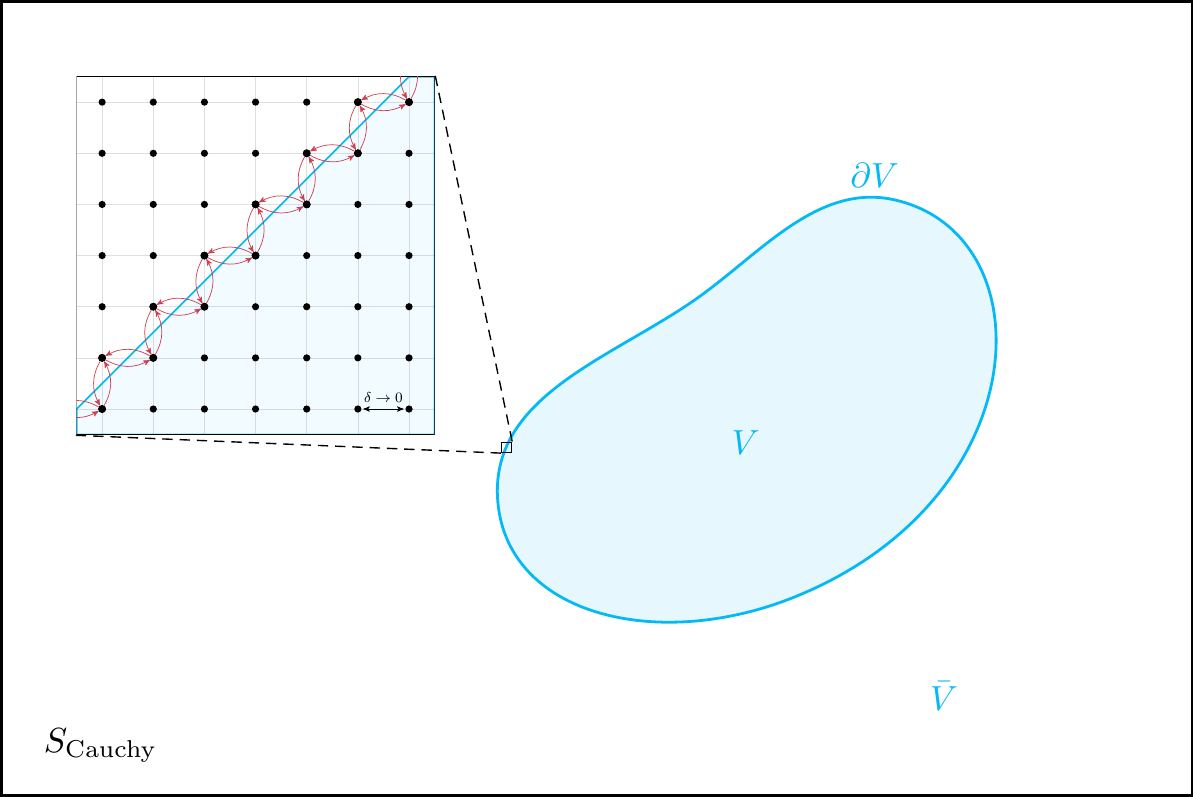}
\caption{\textsf{The bipartite system $V$ and $\bar{V}$ in a Cauchy slice $S_{\text{Cauchy}}$ of a continuum QFT. In the inset, the infinite correlations between neighbors across the entangling surface $\partial V$ are represented by red arrows.}}\label{fig:infcorrel}
\end{figure}

Based on these considerations, let us now consider some 
state satisfying the Schr\"odinger equation is the eigenvector of the field operator $\ket{\phi(x)}=\ket{\phi(t,\mathbf{x})}$. Besides, the wavefunction of a state $\ket{\Psi}$ corresponds to the wave functional $\Psi\left[\phi(x)\right]$. These are the ingredients needed to give a path integral description of the theory.
In the Cauchy slice , we can either have a pure state represented by the wavefunctional $\Psi\left[\phi(\mathbf{x})\right]$ or, more generally a total density matrix $\varrho$. In this setup, we are able to define a codimension-one region $V$, whose complementary $\bar{V}$ covers the entirety of the rest of $S_{\text{Cauchy}}$ and with $\partial V$ as its boundary. This construction is represented in  fig.~\ref{fig:infcorrel}.

A first difference between discrete and continuum systems is that the total Hilbert space is no longer factorized in a product tensor of the subsystems as in eq. \eqref{eq:fact}. It is often argued in the literature that it is possible as long as the field theory does not include gauge symmetries\footnote{The same situation applies in lattice systems, as the gauge symetries obstruct factorization of Hilbert spaces of the subregions \cite{Casini:2013rba,Donnelly:2011hn}. To achieve it, the Hilbert space must be extended to include the states of the electric strings crossing $\partial V$.} \cite{Donnelly:2014gva,Harlow:2015lma,Rangamani:2016dms}. The argument is that elementary excitations in gauge theories are associated with closed loops belonging to $V$ and $\partial{V}$ \cite{Buividovich:2008gq}. However, factorization is not even possible even for free fields. The reason is that observables are generally believed to be self-adjoint operators on a Hilbert space. However, rigorously they are elements of a $C^*$-algebra. Instead of associating Hilbert spaces to regions, we should be assigning algebras\footnote{An axiomatic approach of QFT in terms of algebras of local operators, known as Haag–Kastler formulation, is found in ref.~\cite{Haag:1963dh,Haag:1992hx}.}. Algebras associated to spatial regions in QFT have an ill-defined notion of trace ---these are known as \emph{type-III von Neumann algebras}. As a consequence, Hilbert spaces cannot be factorized in any QFT, regardless its symmetries. The interested reader in this subject should check refs. \cite{Haag:1992hx,Witten:2018zxz,Casini:2022rlv}

Physically, we can interpret the divergence appearing in entanglement entropy for a spatial subregion of a QFT as the manifestation of infinite correlations between the degrees of freedom living in both sides of the entangling region $\partial V$. This is shown in the inset of fig.~\ref{fig:infcorrel}.

In order to characterize the divergence appearing in the entanglement entropy we can start by considering a lattice with a finite $\delta$, as the one appearing in fig.~\ref{fig:bip} and compute the quantity explicitly. In this approach, $\delta$ acquires the role of a UV cutoff and the result for any QFT takes the form \cite{Srednicki:1993im}
\begin{equation}\label{eq:arealaw}
S_{\EE} \left (V\right) =\frac{\text{Area}\left (\partial V\right )}{\delta^{d-2}}+\ldots,
\end{equation}
where the ellipsis stand for subleading contributions. This universal result, is called the \emph{area-law} of entanglement entropy as it is remarkable that scales with the area of $\partial V$ and not with the volume of $V$. This should remind us the situation with the thermodynamic entropy that we associated to the black holes in gravity. As we will see in next subsection, entanglement entropy somewhat generalizes the concept of gravitational entropy.

\subsubsection{Replica trick in quantum field theory}\label{sec:ReplicaQFT}

A popular approach to explicitly compute entanglement entropy in QFTs is exploiting the replica trick \eqref{eq:EEfromRenyi} from R\'enyi entropy
\cite{Holzhey:1994we,Calabrese:2004eu}. Here, we sketch this prescription and focus on the particularities that will be specially relevant in the context of the holographic derivation of entanglement entropy.

Let us start by considering the wavefunctional of the ground state $\Psi\left[\phi_0(\mathbf{x})\right]\equiv\braket{\phi(t=0,\mathbf{x})|\Psi}$ in the Euclidean path integral representation, given by
\begin{equation}
\Psi\left[\phi_0,\mathbf{x}\right]=\int_{t\to-\infty}^{t=0,\, \phi_0}\D[\phi]\e{-\IE[\phi]}=\vspace{1cm}\includegraphics[valign=c]{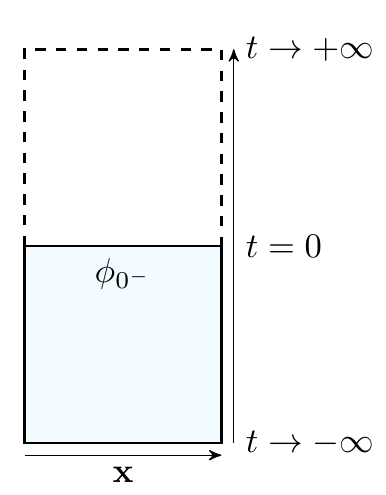},
\end{equation}
where the last equality represents pictorially the path integral, which is performed over the colored region.  On the same footing, the conjugate of the wavefunctional reads
\begin{equation}
\Psi^*[\phi_{0}(\mathbf{x})]=\int^{t\to+\infty}_{t=0,\,\phi_{0}}\D[\phi]\e{-\IE[\phi]}=\includegraphics[valign=c]{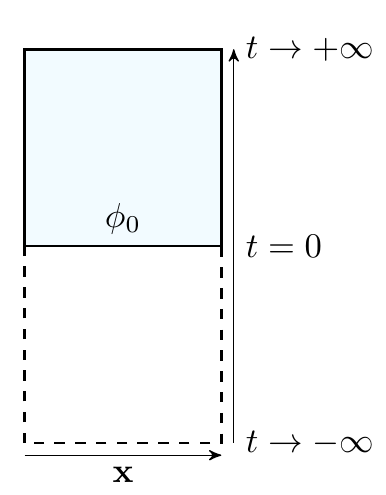}.
\end{equation}
This allows to define the Euclidean partition function at a certain time, which is computed over the whole Euclidean space as
\begin{equation}
Z=\int\D [\phi_0]\braket{\phi_0|\Psi}\braket{\Psi|\phi_0}.
\end{equation}

In the bipartite system, partially tracing over $\bar{V}$ amounts for integrating the total density matrix over the fields with support only on $\mathbf{x}\in \bar{V}$, which we denote as $\phi_0^{\bar{V}}\equiv \phi(t=0,\mathbf{x}\in\bar{V})$. Pictorially, this is equivalent to glue the two sheets along the boundary $\partial V$, and leaving a cut in the region $V$. Then, the reduced density matrix is given by $(\varrho_V)_{ab}\equiv\bra{\phi^V_{a}}\varrho_V\ket{\phi^V_b}$, where the indices $a,b$ specify the boundary conditions on $V$ at $t=0^-$ and $t=0^-$, respectively, this is
\begin{equation}
(\varrho_V)_{ab}=\frac{1}{Z}\int\D[\phi]\delta^V_{ab}\e{-\IE}=\includegraphics[valign=c]{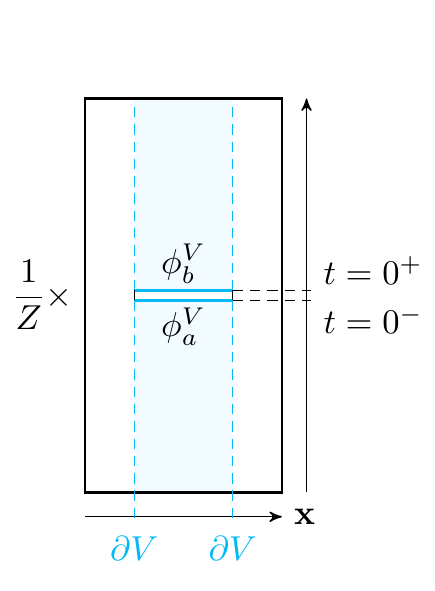}.
\end{equation}
Here, the quantity $\delta^V_{ab}$ corresponds to a Dirac delta introduced to isolate the elements of the  matrix, whose explicit expression reads
\begin{equation}
    \delta^V_{ab}\equiv\prod_{\mathbf{x}\in V}\delta\left[\phi_{0^-}(\mathbf{x})-\phi_{a}^V(\mathbf{x})\right]\times\delta\left[\phi_{0^+}(\mathbf{x})-\phi_{b}^V(\mathbf{x})\right]
\end{equation}

Now that we have a reduced density matrix, its $m$-th power is obtained by gluing $m$ copies of the manifold $\M$, resulting in the construction of a new cover, $\M_m$, with a cut along the region $V$. 

During this process, in each copy we identified the fields $\phi_{a}^V$ an $\phi_b^V$ with the ones belonging to the previous copy and the next one respectively. This includes identifying field $\phi_1^V$, which appears in both the first and the last copy. By doing so, we obtain a partition function, that we denote $Z_m$, of all the fields living on the background $\M_m$. The resulting cover presents a conical singularity  with a deficit angle $2\pi(1-m)$ along the codimension-two entangling surface $\Sigma$. This is
\begin{equation}\label{eq:mcopies}
\tr_V(\varrho_V^m)=\includegraphics[scale=1.1,valign=c]{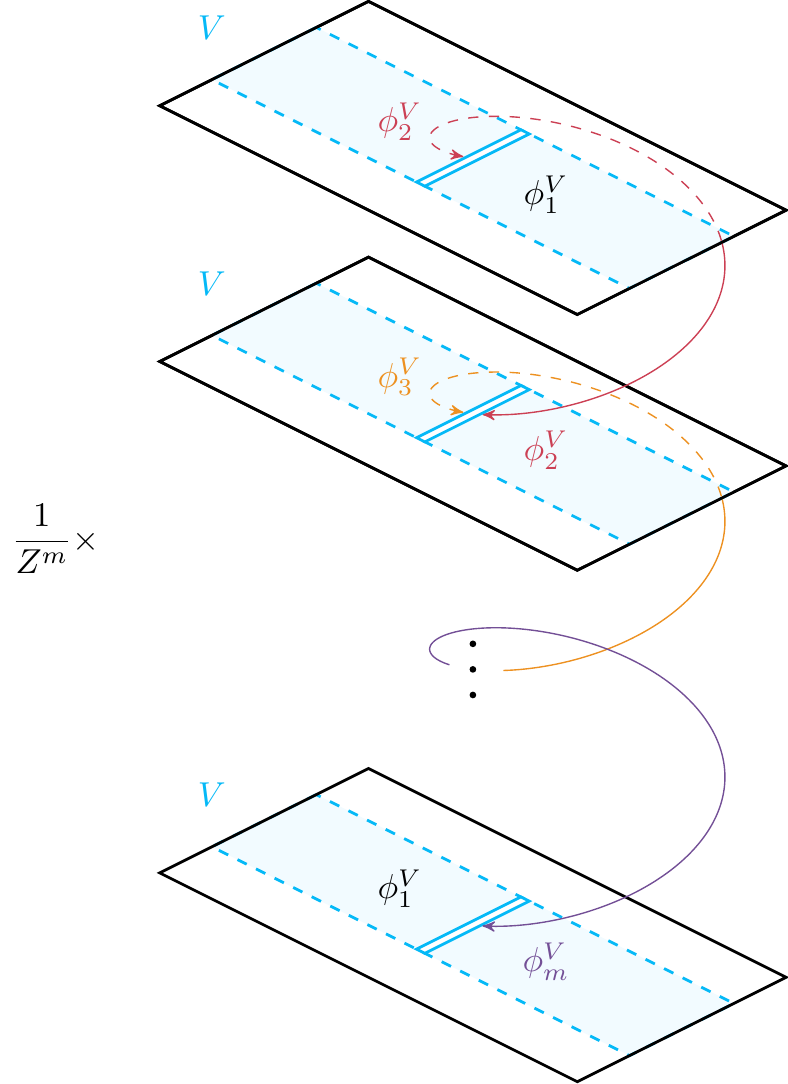}=\frac{Z_m}{Z^m}.
\end{equation}
Besides, the conical singularity, $\pazocal{M}_m$, presents a $\mathbb{Z}_m$ symmetry ---this is, we observe that we can define an angular coordinate, say $\tau$, revolving around the entangling surface. As we glue the $m$ copies, its range is $0\leq\tau\leq2\pi m$ with the symmetry $\tau=\tau+2\pi$. We can combine eqs. \eqref{eq:mcopies} with \eqref{eq:Renyi} to obtain the R\'enyi entropy associated to the region $V$, given by
\begin{equation}\label{eq:Renyim}
S_m(V)=\frac{1}{1-m}\log\frac{Z_m}{Z^m}.
\end{equation}
From this expression, we can analytically continue the replica parameter $m$ and retrieve entanglement entropy \eqref{eq:EEfromRenyi}, finding
\begin{equation}\label{eq:ReplicaQFT}
\SEE(V)=-\lim_{m\to 1}\partial_m\left(\log Z_m-m\log Z\right).
\end{equation}
This expression will be relevant for us when deriving entanglement entropy holographically in sec. \ref{sec:LMp}

Notice that this construction is also possible in the real-time approach and it is the appropriate one for non-trivial time dependent states. In this case, one must make use of the Schwinger-Keldysh formalism~\cite{schwinger1961brownian,keldysh1965diagram} to characterize the time-dependence of the density matrix $\varrho(t)$. See refs.~\cite{Haehl:2016pec,Haehl:2016uah} for modern reviews and \cite{Rangamani:2016dms} for discussion on the replica trick in time-dependent systems.

Let us discuss further on the geometry of $\Sigma$ after this construction of $\pazocal{M}_m$.

\subsubsection{Geometry of the entangling region}\label{sec:geoentang}

We mentioned above that the replica manifold $\pazocal{M}_m$ has a discrete $\mathbb{Z}_m$ symmetry and a conical singularity $\pazocal{C}_m$ along the entangling surface $\Sigma\equiv\partial V$. The conical singularity may present a U$(1)$ symmetry or not. In the former case, the geometry of $\pazocal{M}_m$ can be written as
\begin{equation}
\diff s^2_{\pazocal{M}_m}=\e{\sigma}\left[\diff r^2+r^2\diff \tau^2+{}_{\Sigma}\gamma_{\alpha\beta}(r,x)\diff x^\alpha\diff x^\beta\right],
\end{equation}
where $r>0$ describe the radius and $\tau$ the angle of the cone with singularity at $r=0$. As stated above, $\tau$ is an angular coordinate and should not be confused with the Euclidean time. Instead, it is related to the modular time appearing in the modular Hamiltonian \eqref{eq:modH}. The coordinates $x^\alpha$, with $(\alpha=1,\ldots d-2)$ represent the coordinates of the entangling surface located at the singularity. The conformal factor $\sigma=\sigma(r)$ allows the expansion $\sigma=\sigma_1 r^2+\sigma_2 r^4+\ldots$ in the vicinity of $r=0$.

As we approach the singularity, we can introduce the regulator $\delta$ that deforms the tip of the geometry, defining a \emph{squashed conical manifold} or a regularized space $\tilde{\pazocal{C}}_m$. 
By doing so, the metric of the regularized singular part reads
\begin{equation}
\diff s^2_{\tilde{\pazocal{C}}_m}=\e{\sigma}(f_{\delta}\diff r^2+r^2\diff\tau^2),
\end{equation}
where $f_\delta=f_\delta(r)$ is a smooth function that satisfies $f_\delta(r\to0)=m^2$ and $f_\delta(r>\delta)=1$.

In two-dimensional spacetimes, the Riemann tensor has only one independent component so it is equivalent to the Ricci scalar up to an overall constant, 
which reads
\begin{equation}\label{eq:Riccicone}
{}_{\tilde{\C}_m}R=\e{-\sigma}\left(\frac{\partial_r f_\delta}{rf_\delta^2}-\frac{\partial_r^2\sigma}{f_\delta}\right).
\end{equation}
According to the behavior of the function $f_\delta$ when $\delta\to0$, one would naively expect that the contribution coming from the first term in eq.~\eqref{eq:Riccicone} vanishes. Nonetheless, it does contribute as a surface term
\begin{equation}
\lim_{\delta\to0}\int_{\tilde{\C}_m}\diff^2x\sqrt{\tilde{\varsigma}}\, {}_{\tilde{\C}_m}R=4\pi (1-m)+\int_{\C}\diff x^2\sqrt{\varsigma}\, {}_{\C}R,
\end{equation}
where we denoted $\tilde{\varsigma}$ and $\varsigma$ as the metrics of $\C_m$ and $\C$ respectively. The resulting integral removes the origin $\C\equiv\C_m\setminus\Sigma$ in the limit $\delta\to 0$. The final result does not depend whatsoever on the regulator $\delta$, so it is an universal contribution.

This computation can be generalized to an arbitrary higher-dimensional replica manifold $\M_m$, finding \cite{Fursaev:1995ef}
\begin{align}
{}_{\M_m}\R&=\R+4\pi(1-m)\delta_\Sigma, \\
{}_{\M_m}\R_{\mu\nu}&=\R_{\mu\nu}+2\pi(1-m)N_{\mu\nu}\delta_\Sigma, \\
{}_{\M_m}\R\indices{_\mu_\nu^\rho^\sigma}&= \R\indices{^\mu^\nu_\rho_\sigma}+ 2\pi \left(1-m\right) N^{\mu \nu}_{\rho\sigma} \delta_{\Sigma} \label{eq:Riemannm}
\end{align}
where $N^{\mu \nu}_{\rho\sigma}=2n\indices{^{(\iota)}^\mu}n\indices{^{\left(\iota\right)}_{[\rho}} n\indices{^{(\kappa)}^{\nu}}n\indices{^{\left(j\right)}_{\sigma]}}$ and $N^{\mu \nu}=n_\mu^{(\iota)}n_\nu^{(\iota)}$ are linear combinations of the $i$-th normal vector to the surface $\Sigma$, $n^{\left(\iota\right)\mu}$. The quantities $R$, $R_{\mu\nu}$ and $R\indices{_\mu_\nu^\rho^\sigma}$ are evaluated at the singled-valued geometry $\M\equiv\M_m\setminus\Sigma$, which is far away from the singularity and $\delta_\Sigma$ as the $(d-1)$-dimensional Dirac delta localized at the conical singularity, satisfying
\begin{equation}
\int_{\M_m}\diff^{d}x\sqrt{\gamma}\,\delta_\Sigma=\int_{\Sigma}\diff^{d-2}x\sqrt{{}_\Sigma\gamma}.
\end{equation}
Interestingly, combining expression \eqref{eq:Riemannm},  \eqref{eq:Eulerdens} and \eqref{eq:Cherntheo} modulo the boundary term, we can check that the Euler density $\X_{d}$ self-replicates in codimension-two, that is \cite{Fursaev:1995ef}
\begin{equation}\label{eq:Eulerself}
{}_{\M_m}\X_{d}=\X_{d}+(1-m){}_\Sigma\X_{d-2}+\pazocal{O}\left[(1-m)^2\right],
\end{equation}
at leading order in the replica parameter.

So far we have discussed squashed conical manifolds with U$(1)$ symmetry around the entangling surface $\Sigma$. The metric of such object can be written as
\begin{equation}
\diff s^2_{\tilde{\M}_m}=f_\delta \diff r^2+r^2\diff\tau^2+\left[{}_{\Sigma}\gamma_{\alpha\beta}(x)+2r^nc^{1-n}\left(K_{\alpha\beta}^{\tau}\cos\tau+K_{\alpha\beta}^{r}\sin\tau\right)\right],
\end{equation}
where $K_{\mu\nu}^{(\iota)}\equiv {}_{\Sigma}\gamma_{\mu\rho}{}_{\Sigma}\gamma_{\nu\sigma}\nabla^\rho n^{\sigma(\iota)}$ is the extrinsic curvature normal the codimension-two surface $\Sigma$, whose induced metric is given by ${}_\Sigma \gamma_{\mu\nu}=\gamma_{\mu\nu}-n_\mu^{(\iota)}n_\nu^{(\iota)}$ and $c$ is some length scale. In this expression we see that the geometry no longer exhibits U$(1)$ symmetry because of the angular dependence. Following the same procedure as before and after a lengthy computation that is detailed in ref.~\cite{Fursaev:2013fta} one finds the following relations
\begin{align}
\int_{\M_m}&\diff^dx\sqrt{\gamma}\, \R^2=\int_{\M}\diff^{d} x\sqrt{\gamma}\ \R^2
\label{eq:RicciS2Cone}
\\
&+8\pi(1-m)\int_\Sigma\diff^{d-2} x\sqrt{{}_{\Sigma}\gamma}\left({}_{\Sigma}\R
+2\R^{(\iota)(\iota)}-\R^{(\iota)(\kappa)(\iota)(\kappa)}-2{K^{(\iota)}}\indices{^\mu_{[\mu}}\, {K^{(\iota)}}\indices{^\nu_{\nu]}}\right)+\ldots
\nonumber\\
\int_{\M_m}&\diff^dx\sqrt{\gamma}\, \R_{\mu\nu}\R^{\mu\nu}=\int_{\M}\diff^{d} x\sqrt{\gamma}\, \R_{\mu\nu}\R^{\mu\nu}
\\
&+4\pi(1-m)\int_\Sigma \diff^{d-2}x\sqrt{{}_\Sigma\gamma}\left(\R^{(\iota)(\iota)}-\frac{1}{2}{K^{(i)}}^2\right)+\ldots
\nonumber,\\
\int_{\M_m}&\diff^d x\sqrt{\gamma}\, \R_{\mu\nu\rho\sigma}\R^{\mu\nu\rho\sigma}=\int_{\M}\diff^{d} x\sqrt{\gamma}\, \R_{\mu\nu\rho\sigma}\R^{\mu\nu\rho\sigma} \\
&+8\pi(1-m)\int_\Sigma \diff^{d-2} x\sqrt{{}_\Sigma\gamma}\left(\R^{(\iota)(\kappa)(\iota)(\kappa)}-{K^{(\iota)}}^2\right)+\ldots\nonumber,
\label{eq:Riem2Cone}
\end{align}
where $\R^{(\iota)(\kappa)(\kappa)(\kappa)}=n_\mu^{(\iota)}n_\nu^{(\iota)}n_\rho^{(\kappa)}n_\sigma^{(\iota)}\R^{\mu\nu\rho\sigma}$ and $\R^{(\iota)(\iota)}=n_\mu^{(\iota)}n_\nu^{(\iota)}\R^{\mu\nu}$ are the induced Riemann tensor and Ricci tensor, respectively, and the ellipsis denote subleading terms depending on $(1-m)^2$.  If desired, the final expression of the Ricci scalar squared \eqref{eq:RicciS2Cone} can be rewritten as
\begin{equation}
\int_{\M_m}\diff^dx\sqrt{\gamma}\, \R^2=\int_{\M}\diff^{d} x\sqrt{\gamma}\ \R^2+8\pi(1-m)\int_\Sigma\diff^{d-2} x\sqrt{{}_{\Sigma}\gamma}\R+\ldots,
\end{equation}
using the Gauss-Codazzi equations
\begin{equation}
\R={}_{\Sigma}\R
+2\R^{(i)(i)}-\R^{(i)(j)(i)(j)}-2{K^{(i)}}\indices{^\mu_{[\mu}}{K^{(i)}}\indices{^\nu_{\nu]}}.
\end{equation}
These results allow to decompose the Euler characteristic of the manifold $\M_m$ as
\begin{equation}
\chi\left[\M_m\right]=\chi\left[\M\right]+(1-m)\chi\left[\Sigma\right]+\ldots
\end{equation}
We will discuss this relations again when addressing the generalized gravitational entropy. Before that, we review the structure of entanglement entropy when considering CFTs.

\subsection{Structure of entanglement entropy in CFTs}\label{sec:structent}

Let us consider a $d$-dimensional CFT. The entanglement entropy associated to a smooth\footnote{When the entangling regions are not smooth and singularities are present, the structure of divergences is modified. An example of this situation is the presence of a corner in three-dimensional CFT. In this case a new logarithmic divergence appears, whose coefficient depends on the opening of the angle \cite{Hirata:2006jx,Casini:2006hu,Casini:2008as,Casini:2009sr}. The very sharp and almost-smooth limits are controlled by universal coefficients, such as the corresponding to the slab region $k$ \cite{Myers:2012vs,Bueno:2015xda} and the stress-energy tensor two-point function coefficient $C_{\ssc T}$ \cite{Bueno:2015rda,Faulkner:2015csl}, respectively. The situation is different for conical singularities in $d=4$. In this case, there appears a squared logarithmic divergence whose coefficient depends on the central charge $c$ \cite{Myers:2012vs,Klebanov:2012yf}. This structure is preserved at higher-dimensional singular regions \cite{Bueno:2019mex}} region $V$ is given by  ---see \eg \cite{Grover:2011fa,Liu:2012eea}
\begin{equation}\label{eq:EEexp}
\SEE(V)=\ldots+\begin{cases*}
                    b_{(1)}\frac{H}{\delta}+(-1)^{\frac{d-1}{2}}s^{\text{univ}}(V) & if odd $d$,  \\
                    b_{(2)}\frac{H^2}{\delta^2}+(-1)^{\frac{d-2}{2}}s^{\text{univ}}(V)\log\frac{H}{\delta}+b_{(0)} & if even $d$,
                 \end{cases*}
\end{equation}
where now the ellipsis include the area-law term \eqref{eq:arealaw} as well as other terms that are regularized by different powers of the regulator $\delta$, $H$ is a characteristic length appearing in the entangling region $V$, $b_{(0)}$ is a constant coefficients and $s^{\text{univ}}$ is some universal contribution appearing in any CFT. The nature of each of the terms appearing in the expansion is the following. Those that appear along with regulators, \ie $b_{(1)}$, $b_{(2)}$ and other $b_{(d)}$'s included in the ellipsis, are scheme-dependent and thus physically irrelevant. This can be seen from the fact that we can always reparametrize the regulator $\delta$ and modify the corresponding term. Moreover, they have a local nature as they depend on the shape of the entangling surface $\Sigma$. Regarding $b_{(0)}$, its nature is nonlocal, but it can be polluted with local contributions. On the other hand, the finite quantity in odd dimensions and the coefficient of the logarithmic divergence in even ones ---both denoted $\suniv$--- are universal, carrying important information of the CFT under consideration. For the latter, it encodes information of the trace anomaly charges of the theory, as
\begin{equation}
\suniv(V)=\left.\sum_iB_i\left(\partial_m\int_{\M_m}\I_i\right)\right|_{m\to1}-2(-1)^{\frac{d}{2}}A\chi(\Sigma) \quad \text{if even }d.
\end{equation}
Following the examples in sec.~\ref{sec:CFT}, in a two-dimensional CFT, the entanglement entropy associated to an interval $V$ of length $H$ is given by \cite{Calabrese:2004eu,Calabrese:2009qy}
\begin{equation}\label{eq:EE2d}
\SEE^{(2)}(V)=\frac{c}{3}\log \frac{H}{\delta}+b_0,
\end{equation}
where $c$ is the Virasoro central charge of the two-dimensional CFT. In the case of four-dimensional CFTs we find \cite{Solodukhin:2008dh,Fursaev:2012mp}
\begin{align}
\SEE^{(4)}(V)=&-\left[\frac{a}{2\pi}\int_{\Sigma}\diff^2 x\sqrt{{}_\Sigma\gamma} {}_{\Sigma}\R+\frac{c}{2\pi}\int_\Sigma\diff^2x\sqrt{{}_\Sigma\gamma}\left({K^{(\iota)}}_{\mu\nu}{K^{(\iota)}}^{\mu\nu}-\frac{1}{2}{K^{(\iota)}}^2\right)\right]\log\frac{H}{\delta},\notag\\
&+b_2\frac{H_2}{\delta^2}+b_0,
\end{align}
where the last part is found by recalling that $\I_4=\W_{\mu\nu\rho\sigma}\W^{\mu\nu\rho\sigma}$, expressing it in terms of the Riemann tensor and its contractions through \eqref{eq:Weyl} and employing relations \eqref{eq:RicciS2Cone} to \eqref{eq:RicciS2Cone}.

Remarkably, if in a certain $d$-dimensional CFT the entangling surface is a sphere $\partial V=\mathbb{S}^{d-2}$ ---which defines a $(d-1)$-ball as entangling region $V=\mathbb{B}^{d-1}$---, the type A anomaly of the CFT is the only one that contributes to $\suniv$. In contrast, if the entangling region is a cylinder $V=\mathbb{R}\times\mathbb{B}^{d-2}$, the type A part vanishes the contribution is only given by the type B anomaly. Thus, in $d=4$, by computing entanglement entropy across a sphere of radius $R$ and a cylinder of radius $R$ and length $L$, we can isolate charges $a$ and $c$ respectively as
\begin{equation}
\suniv(\mathbb{B}^3)=-4a\log \frac{R}{\delta},\quad \suniv(\mathbb{R}\times\mathbb{B}^2)=-\frac{c}{2} \frac{L}{R}\log\frac{R}{\delta}.
\end{equation}

The situation in odd-dimensional CFTs differs as there are no anomaly terms. In this case, the finite part represents the universal contribution $\suniv$, with a non-local character. The factor of $(-1)^{\frac{d-1}{2}}$ is introduced to impose positivity of the quantity regardless the dimensionality of the CFT \cite{Klebanov:2011gs}.

\begin{figure}[t] \centering
	\includegraphics[scale=0.35]{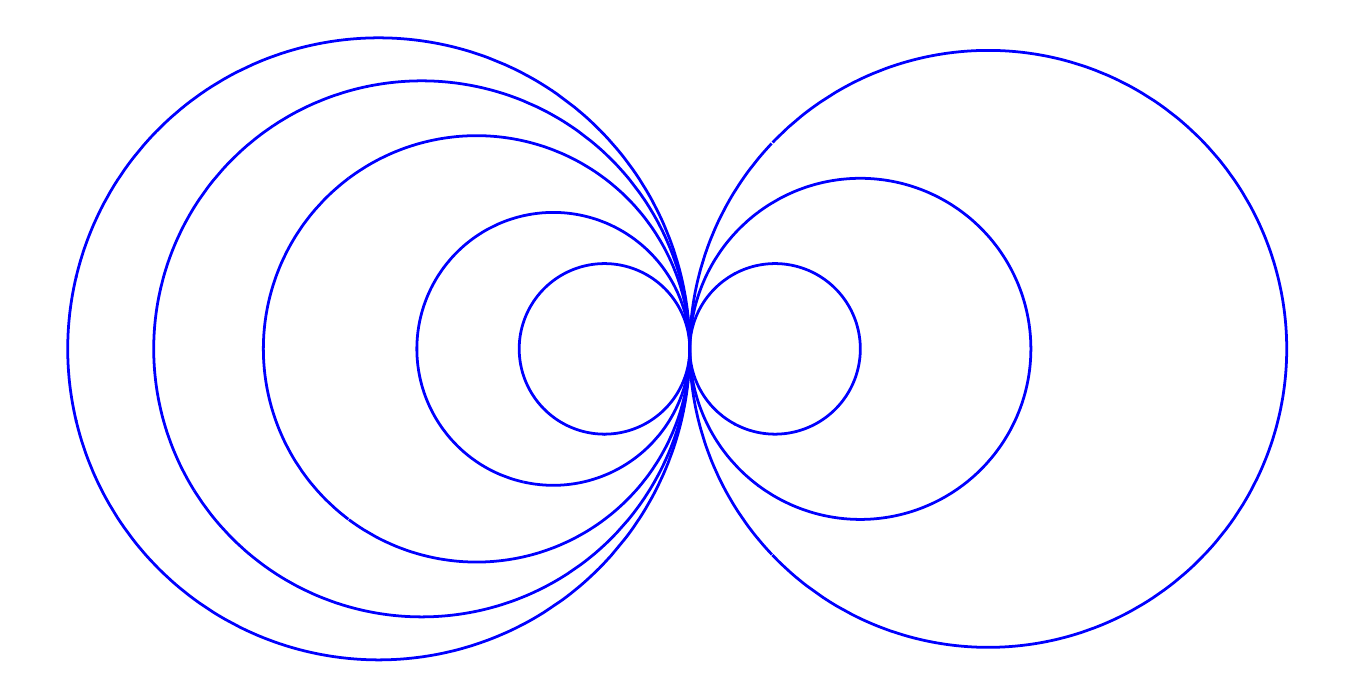}\hspace{0.2cm}	
	\includegraphics[scale=0.4]{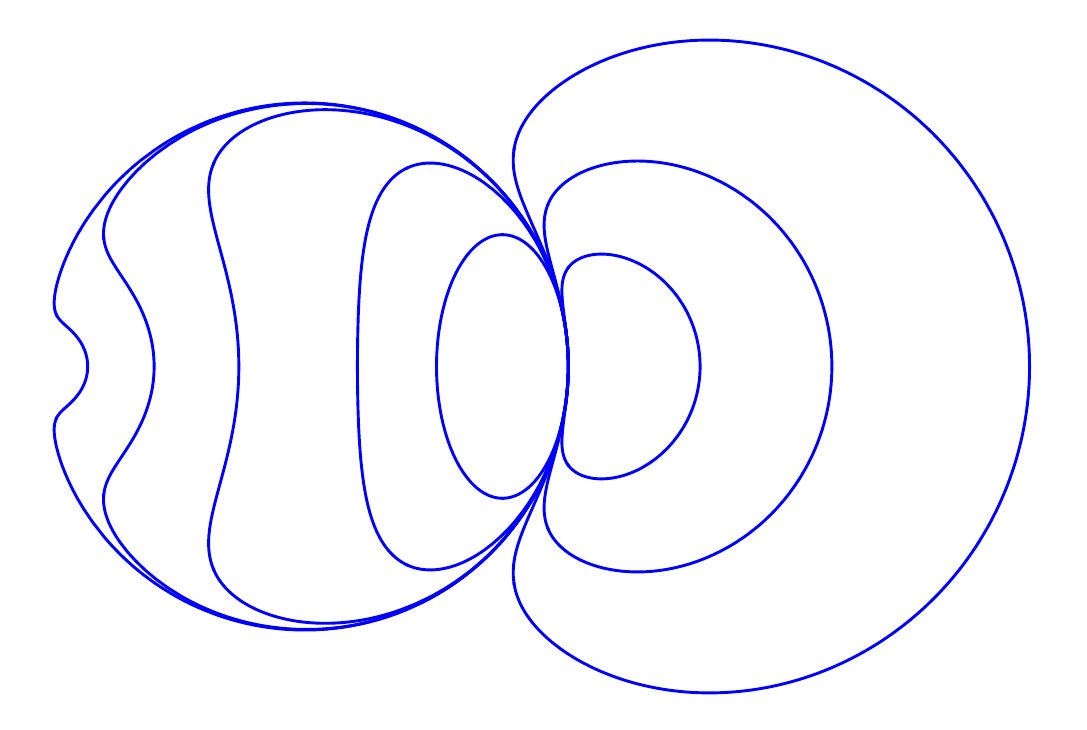}\hspace{0.2cm}
	\includegraphics[scale=0.37]{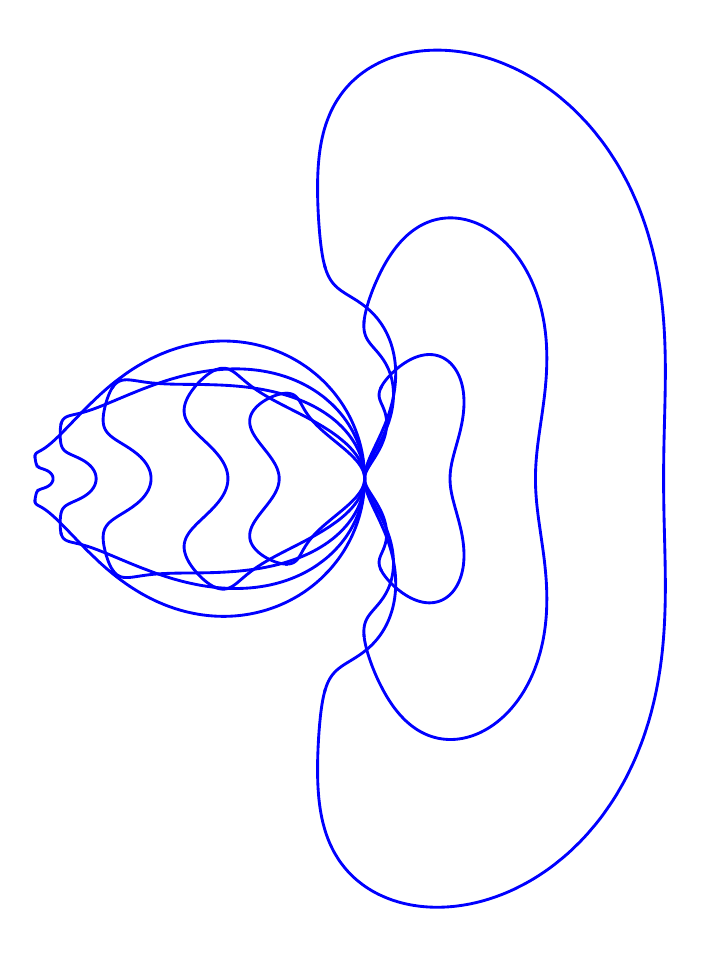}
	\caption{ \textsf{We show three sets of entangling regions (left, middle and right plots, respectively). In each set, all the different regions are related to each other through conformal transformations (the same transformations are applied in each set), and therefore share the same $F$ for a given CFT. Disks are naturally unchanged, but notice for instance how an ellipse (innermost figure in middle plot) can be deformed, without modifying $F$, into a disk with a small depression.} }
	\label{refiss376}
\end{figure}

In the particular example of three-dimensional CFTs, the entanglement entropy of a smooth region takes the form
\begin{equation}\label{entro}
\SEE(V) = b_1\frac{{\rm Area}(\partial V)}{\delta} - F(V),
\end{equation}
where we denoted $F(V)\equiv\suniv(V)$. Importantly, $F(V)$ is a conformal invariant quantity ---see fig.~\ref{refiss376}. As discussed above, the $F$ depends on the theory under consideration. When one of the dimensions of the entangling region becomes very thin compared to the other, the universal contribution approaches the one corresponding to a strip ---whose boundary is defined by two parallel straight lines of length $L$ and separated a distance $r\ll L$. In that situation, $F$ tends to diverge as
\begin{equation}\label{strip}
F(\text{strip}) \simeq k^{(3)} \frac{L}{r} +\dots
\end{equation}
where $k^{(3)}$ is a positive coefficient characteristic of the CFT under consideration.

For arbitrarily high odd dimensions, it turns out that when the entangling region is a sphere, the finite part of entanglement entropy is identified with the free energy of the system. This remarkable result is known as the Casini-Huerta-Myers map \cite{Casini:2011kv}
and consists in the following. First, start with the free energy $F_m\equiv-\log Z_m$ of the replica manifold $\M_m$, which is related to the R\'enyi entropy as
\begin{equation}
S_m(V)=\frac{m F_1-F_m}{1-m},
\end{equation}
using eq.~\eqref{eq:Renyim}. Now, consider a spherical entangling surface $\mathbb{S}^{d-2}$ at $\{t,r\}=\{0,R\}$ in flat space
\begin{equation}\label{eq:CHMmap1}
\diff s^2_{\mathbb{R}^d}=\diff t^2+\diff r^2+r^2\diff\Omega_{d-2}^2.
\end{equation}
The conformal symmetry allows to connect this geometry with convenient ones, such as hyperbolic $\mathbb{S}\times \mathbb{H}^{d-1}$ and spherical $\mathbb{S}^d$ spacetimes. For the former, the change of coordinates $t=R\frac{\sin\tau}{\cosh u+\cosh \tau}$ and $r=R\frac{\sinh u}{\cosh u+\cos \tau}$, with ranges $0\leq\tau\leq2\pi$ and $0\leq u\leq \infty$, maps metric \eqref{eq:CHMmap1} to
\begin{equation}\label{eq:CHMmap2}
\diff s^2_{\mathbb{R}^d}=\frac{R}{\cosh u+\cos \tau}\left(\diff\tau^2+\diff u^2+\sinh^2 u\, \diff\Omega_{d-2}^2\right),
\end{equation}
which is a hyperbolic spacetime $\mathbb{S}\times \mathbb{H}^{d-1}$ up to a conformal factor. In the case of the replica manifold $\M_m$, this associates a temperature $T_m=1/(2\pi m)$ as extends the range of the angular coordinate $\tau$ that we previously mentioned acts as modular time. Then, we can relate the free energy $F_m$ and the thermal free energy $F(T)$ at temperature $T_m$ on $\mathbb{H}^{d-1}$ as $F_m=F(T_m)/T_m$. Exploiting that the thermal free energy is related to thermal entropy through the Maxwell relation $S(T)=\partial F/\partial T$, we can relate the free energy to R\'enyi entropy in eq. \eqref{eq:Renyim} as \cite{Hung:2011nu}
\begin{equation}
S_m=\frac{m}{m-1}\frac{1}{T}\int_{T_m}^T S(T')\diff T'.
\end{equation}
Notice that in the limit $m\to1$, this relation implies that for spherical entangling regions of odd-dimensional CFTs at finite temperature, the entanglement entropy is identified with the thermal entropy
\begin{equation}\label{eq:EEequalsS}
\SEE=S(T).
\end{equation}

is proportional to the trace anomaly, following the general expression
\begin{equation}
E=-\int\diff^{d-1}x\sqrt{\gamma}T_{\mu\nu}
\end{equation}

On the other hand, starting from metric \eqref{eq:CHMmap2}, the change of coordinates $\sinh u=\cot\theta$ with $0\leq\theta\leq\pi/2$ allows to visualize that flat spacetime and spherical one are equivalent up to a conformal factor as
\begin{equation}
\diff s^2_{\mathbb{R}^d}=\frac{R}{1+\sin\theta\cos\tau}\left(\diff\theta^2+\sin^2\theta\diff\tau^2+\cos^2\theta\diff\Omega_{d-2}^2\right),
\end{equation}
indicating the equivalence of the replica manifold $\M_m$ on $\mathbb{R}^d$ and on the $d$-sphere $\mathbb{S}^d_m$ with the extended angular coordinate $0\leq\tau\leq2\pi m$. 

Now, we recall our recent equality between the thermal entropy and entanglement entropy \eqref{eq:EEequalsS}. The former is given in terms of the total energy of the system and the free energy as
\begin{equation}
S(T)=\frac{E(T)-F(T)}{T}.
\end{equation}
In this expression, the total energy $E$ is given by the expectation value of the operator which generates the $\tau$ translations, which correspond to a Killing symmetry of the spherical patch. In consequence, the energy is given by
\begin{equation}
E=\int_V\diff^{d-1}x\sqrt{\gamma}\langle T_{\mu\nu}\rangle\xi^{\mu}n^{\nu}=\int_V\diff^{d-1}x\sqrt{\gamma}\langle \, T\indices{_\tau_\tau}\rangle,
\end{equation} 
where we used $\xi^{\nu}n^{\nu}=\sqrt{g^{\tau\tau}}\partial_\tau$. As we discussed above, the trace anomaly for spherical entangling regions is proportional to the type A trace anomaly so this term vanishes in odd dimensions, implying $E=0$ after renormalizing all divergences. In consequence, we find that in odd-dimensional CFTs the finite piece of entanglement entropy is given by the free energy when placing the theory on a spherical background, this is\cite{Casini:2011kv}
\begin{equation}\label{eq:CHMmap}
\suniv(\mathbb{B}^{d-1})=(-1)^{\frac{d-1}{2}}F_0,
\end{equation}
where we denoted $F_0\equiv\log Z[\mathbb{S}^d]$ for simplicity. This result is of extreme relevance in the context of renormalization group flows. 

\subsection{Renormalization group flows and entanglement entropy}\label{sec:RGflows}

CFTs appear naturally in the context of renormalization group (RG) flows and we can exploit the previous results on entanglement entropy to find interesting relations in this field. Before addressing them, let us mention the most fundamental concept of renormalization transformation. An interested reader should check ref.~\cite{Nishioka:2018khk} for a more comprehensive review with focus on the connections with entanglement entropy.

Consider two QFT theories $Q_1$ and $Q_2$ that can be represented as points in a theory space $\mathfrak{T}$ and whose coordinates are given by the coupling constants $\{g_i\}$ of each of the theory. Following the Wilsonian perspective of QFT \cite{Wilson:1973jj}, we can assume that $Q_1$ and $Q_2$ have a different energy scale, \ie the degrees of freedom that have been integrated out in a coarse-graining process are different. A renormalization transformation $R_t$ maps $Q_1$ to $Q_2$ defining a trajectory in $\mathfrak{T}$ parametrized by change in energy scale $\mu_{\text{s}}$. We refer to such trajectory as a RG flow. For convenience, we can parametrize the RG flow with the RG time $t_{\text{RG}}=-\log\frac{\mu_{\text{E}}}{\delta}$, which is proportional to the number of times we coarse-grained a QFT.

When certain coordinates of our theory space are scale-independent, we refer to them as fixed points and they correspond to CFTs. If we know what CFTs are defined in the IR and UV limit, we can learn universal properties about the QFTs that flow to these fixed points in both limits. Starting from the UV fixed point, as the RG time progresses, we integrate out the high energy degrees of freedom until we arrive to the IR limit. In order to characterize such process, it would be interesting to find a quantity $\C$ that depends on the coupling constants $g_i$ and the RG time $t_{\text{RG}}$ and whose values in the fixed points defined in the limits satisfy \cite{Barnes:2004jj}
\begin{equation}\label{eq:weakc}
\C_{\text{UV}}\geq \C_{\text{IR}}.
\end{equation}
In that case, $\C$ is referred as a \emph{$\C$-function} in the weak version. In the stronger version, $\C$ is monotonically decreasing along the RG flow, this is 
\begin{equation}\label{eq:monotone}
\frac{\diff \C}{\diff t_{\text{RG}}}\leq0.
\end{equation}
Finally, we can define a strongest version if $\C$ defines a potential satisfying the relation
\begin{equation}
\beta_i(g)=G_{ij}(g)\frac{\partial \C}{\partial g_j},
\end{equation}
where $\beta_{i}(g)\equiv \frac{\diff g_i}{\diff t_{\text{RG}}}$ is the beta function of Callan-Symanzik \cite{Callan:1970yg}.

The first example was found in two-dimensions field theories by Zamolodchikov and constitutes what we call today the \emph{c-theorem} \cite{Zamolodchikov:1986gt}. Remarkably, the c-function $c(g_i,\mu_{\text{s}})$, besides satisfying the monotonicity condition \eqref{eq:monotone}, takes the constant value of the Virasoro central charge at the fixed points of the RG flow
\begin{equation}
c(g_i,\mu_{\text{s}})\Big|_{\text{fixed point}}=c.
\end{equation}
From this observation, we already see the relevance of entanglement entropy in CFT as in two dimensions the logarithmic divergence features the central charge \eqref{eq:EE2d}. In fact, one can alternatively show an \emph{entropic c-theorem} constructing a $\C$-function from entanglement entropy of an interval $S(R)$, which is known as ``entropic c-function'' \cite{Casini:2004bw}. The idea to prove monotonicity of the entropic c-function is to exploit  the strong subadditivity property of this quantity \eqref{eq:SSEE}.

The situation in three dimensions is different as there are no conformal anomalies. Instead, the role of $C$-function is played by a quantity $\F=\F(g_i,\mu_{\text{s}})$ that coincides with the sphere free energy of the CFT at the corresponding fixed point
\begin{equation}
\F(g_i,\mu_{\text{s}})\Big|_{\text{fixed point}}=-F_0\Big|_{d=3}.
\end{equation}
This result, known as \emph{F-theorem} was proven perturbatively \cite{Klebanov:2011gs,Yonekura:2012kb} in the weak version \eqref{eq:weakc} and shortly after non-perturbatively in the strong version \eqref{eq:monotone} by Casini and Huerta \cite{Casini:2012ei} using the CHM map \cite{Casini:2011kv} and again relying on strong subadditity property \eqref{eq:SSEE}. It is interesting to notice that this result was anticipated from holographic results considering CFTs dual to Einstein gravity, Gauss-Bonnet and quasi-topological gravity \cite{Myers:2010xs,Myers:2010tj}.

Shortly after the discovery of the $c$-theorem, it was conjectured that the type A anomaly coefficient played the role of $\C$-function in arbitrary high even dimension \cite{Cardy:1988cwa}. In the case of four dimensions, it was soon proven perturbatively \cite{Osborn:1989td,Jack:1990eb}. The strong version of this \emph{$a$-theorem} was recently proven exploring the connection with the scattering amplitudes of four dilaton fields that couple to the stress-energy tensor \cite{Komargodski:2011vj}. An alternative proof can be found from entanglement entropy \cite{Casini:2017vbe}. For higher even-dimensional QFTs, there are no definitive proofs, with supporting evidence coming from holography \cite{Myers:2010xs,Myers:2010tj}. The same situation applies for higher odd-dimensional field theories which was suggested in ref.~\cite{Myers:2010xs,Myers:2010tj} and remains without a general proof beyond the holographic case.

\subsubsection{Holographic renormalization group flows}\label{sec:hrgf}

The conjecture provided by Myers and Sinha \cite{Myers:2010xs,Myers:2010tj} is another example of how holographic techniques can provide us new conjectures phenomena involving in strongly correlated systems. Holographic RG flows is one of the first applications found in the AdS/CFT correspondence \cite{Freedman:1999gp}. The idea to study RG flows from the gravity side is the following. First, suppose a solution of some $(d+1)$-dimensional gravity action, described by the metric
\begin{equation}\label{eq:RGmetric}
\diff s^2=\diff r^2+A(r)^2(-\diff t^2+\delta_{ij}\diff x^i\diff x^j).
\end{equation}
Supplementing the action with an appropriate stress-tensor, the metric can be made to interpolate between two asymptotic AdS$_3$ regions \cite{Freedman:1999gp,Girardello:1998pd} which, from the CFT point of view would represent IR and UV fixed points. Intermediate values of the holographic coordinate are then interpreted as representing the RG flow between both CFTs.

Following the previous discussion, the idea of a holographic c-theorem involves constructing a function $\C_{\text{holo}}=\C_{\text{holo}}(r)$ which decreases monotonously along the holographic coordinate, whose  In the present holographic context, the fixed points can be chosen to be  $r_{\rm UV}\to\infty$  and   $r_{\rm IR}\to-\infty$, so a function satisfying
\begin{equation}
    \frac{\diff\C_\text{holo}}{\diff r} \geq 0,
\end{equation}
does the job. Now, the usual holographic c-theorem construction involves considering a function $\C_\text{holo}$ such  that its derivative is  proportional to the combination of stress-tensor components $T\indices{_t
^t}-T\indices{_{r}^{r}}$. Then, imposing that the stress-tensor satisfies the \emph{null energy condition}, such combination has a sign, namely, \cite{Hawking:1973uf}
\begin{equation}
    T\indices{_t^t}-T\indices{_{r}^{r}}  \overset{\rm  \ssc NEC}{\leq} 0\, .
\end{equation}
Therefore, any $\C_\text{holo}$ such that  $\frac{\diff\C_\text{holo}}{\diff r}\propto -(T\indices{_t^t}-T\indices{_{r}^{r}})$, up to an overall positive-definite constant, satisfies the requirement.

From a different perspective, given the relation between $\C$-functions and entanglement entropy from the entropic c-theorems, one can approach the study of renormalization group flows from entanglement entropy.

\section{Holographic entanglement entropy}\label{sec:LMp}

The formula of entanglement entropy in a time-independent setup from holography was first proposed in the celebrated Ryu-Takayanagi (RT) conjecture\footnote{The prescription was extended to consider time-dependent states in ref.~\cite{Hubeny:2007xt}} \cite{Ryu:2006bv,Ryu:2006ef,Nishioka:2009un}. The proposal was to consider a spatial slice of an AdS spacetime dual to the CFT under consideration and relate the entanglement entropy of a region $V$ with the minimal area of a surface $\Gamma_V$ in the bulk that was cobordant to the entangling surface $\Sigma$ and homologous to $V$ (see eq.~\eqref{eq:RT} below). The setup is shown in fig.~\ref{fig:RTgeo}. In order to show the RT formula, Lewkowycz and Maldacena in ref.~\cite{Lewkowycz:2013nqa} provided a derivation in which the GKP-W relation \eqref{eq:GKPW} and the replica trick \eqref{eq:EEfromRenyi} are used. 

\begin{figure}[t] \centering
	\includegraphics[width=0.5\textwidth]{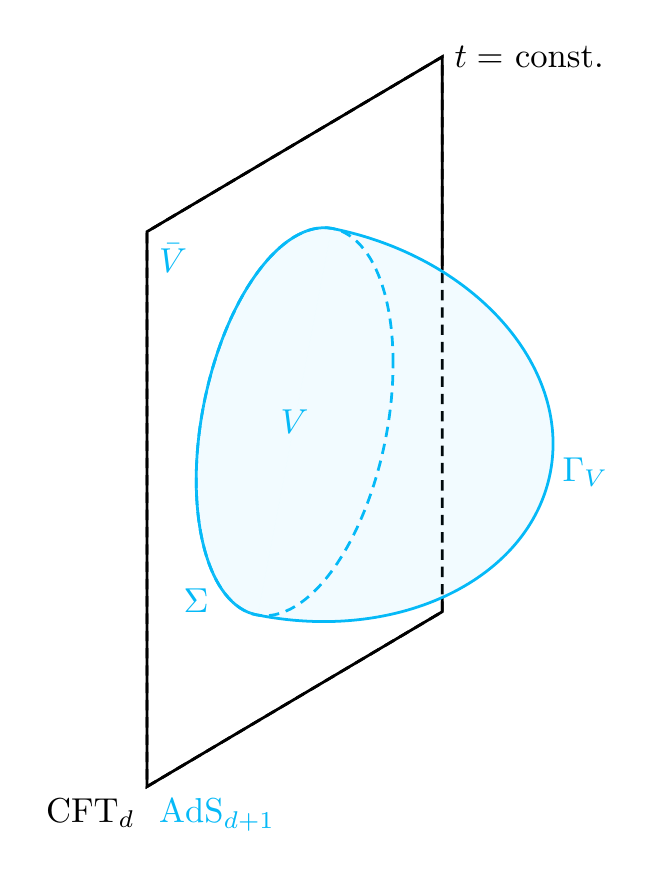}
	\caption{ \textsf{RT surface $\Gamma_V$ in the bulk theory AdS$_{d+1}$ for an entangling region $V$ in a $t=$ const. slice of a CFT$_{d}$. Both the entangling region and the RT surface are cobordant $\partial V=\partial \Gamma_V=\Sigma$.}}
	\label{fig:RTgeo}
\end{figure}

\subsection{Generalized gravitational entropy}

Consider a Cauchy slice of a $d$-dimensional CFT defined on $\M$ and a codimension-two entangling surface $\Sigma$ separating the system into the spacial subregions $V$ and $\partial {V}$. As we saw in sec.~\ref{sec:ReplicaQFT}, we can compute entanglement entropy from R\'enyi entropy using the replica trick \eqref{eq:ReplicaQFT} defining an $m$-fold cover $\M_m$. In the gravitational picture, we  consider a dual gravitational action on an $(d+1)$-dimensional asymptotically AdS spacetime $\B$ described by Einstein gravity, whose boundary is $\partial\B=\M$. The geometrical setup is depicted in fig.~\ref{fig:RTgeo}. We further assume that there is bulk solution $\B_m$ such that its boundary is the replica manifold of the CFT, $\partial\B_m=\M_m$. Following the GKP-W relation \eqref{eq:GKPW}, in the saddle-point approximation, we find that holographic entanglement entropy is given in terms of the Euclidean gravity action as \cite{Lewkowycz:2013nqa}
\begin{equation}\label{eq:LMp}
\SEE(V)=\lim_{m\to1}\partial_m\left(I_\text{E}[\B_m]-m I_{\text{E}}[\B]\right).
\end{equation}
As $\M_m$ is symmetric under $\mathbb{Z}_m$, we further assume that $\B_m$ is as well. The fixed locus of $\mathbb{Z}_m$ on $\M_m$ is the entangling surface $\Sigma$. A such, it extends to a codimension-two surface $\Gamma^{(m)}_V$ of $\B_m$. By rescaling the modular time $\vartheta\equiv \tau/m$, we make $\vartheta$ periodic in the range $0\leq\vartheta\leq2\pi$. Now, we have to take into account that the bulk geometry $\B_m$ is a smooth solution to Einstein gravity and, in consequence, there is no deficit angle in the surface $\Gamma^{(m)}_V$. Because of this, we define a replica orbifold
\begin{equation}
\tilde{\B}_m\equiv\B_m\setminus\mathbb{Z}_m,
\end{equation}
which has a deficit angle $2\pi(1-\frac{1}{m})=2\pi(1-\vartheta)$. It will be convenient to define the quantity $\tilde{I}_\text{E}[\tilde{\B}_\vartheta]\equiv \vartheta \tilde{I}_\text{E}[\tilde{\B}_\vartheta]$, which can be interpreted as a bulk density per replica unit. This allows to rewrite eq.~\eqref{eq:LMp} as
\begin{equation}\label{eq:finalLM}
\SEE(V)=-\partial_\vartheta \tilde{I}_{\text{E}}[\tilde{\B}_\vartheta]\Big|_{\vartheta=1}.
\end{equation}
Notice that, following the discussion in sec.~\ref{sec:geoentang} the conical singularity of $\tilde{\B}_\vartheta$ provides an additional contribution. In the case of Einstein gravity, the contribution comes from the Ricci scalar and amounts for
\begin{equation}\label{eq:Riccidec}
\int_{\tilde{\B}_\vartheta}\diff^{d+1}x\sqrt{g}\, R=\int_{\B}\diff^{d+1}x\sqrt{g}R+(4\pi-\vartheta)\int_{\Gamma_V}\diff^{d-2}x\sqrt{{}_{\Gamma_V} g}
\end{equation}
where we adapted our results in the case of squashed cone and took into account that the deficit angle of the orbifold $\tilde{\B}_\vartheta$ is $2\pi(1-1/m)=2\pi(1-\vartheta)$. The second term in eq.~\eqref{eq:Riccidec}, is the area of the codimension-2 surface $\Gamma_V$, which is often denoted as \emph{RT surface}. Taking this into account, we can decompose the orbifold as \cite{Dong:2016fnf}
\begin{equation}
\tilde{I}_{\text{E}}[\B_\vartheta]=I_{\text{E}}[\B_\vartheta]+T_\vartheta \text{Area}\left(\Gamma_V\right),
\end{equation}
where the coefficient $T_\vartheta=\frac{1-\vartheta}{8\GN}$ represents a tension as the second term can be interpreted as the action of the Einstein gravity coupled to a codimension-two cosmic brane. Finally, inserting this result in eq.~\eqref{eq:finalLM} we obtain
\begin{equation}\label{eq:RT}
\SEE(V)=\frac{\text{Area}(\Gamma_V)}{4\GN}.
\end{equation}
In this expression, it can be shown that in order to avoid singularities in the equations of motion of the orbifold, one must impose the constraint \cite{Bhattacharyya:2013sia,Bhattacharyya:2014yga}
\begin{equation}\label{eq:minK}
\K^{(\alpha)}=0
\end{equation}
on the traces of the extrinsic curvatures along the transverse directions to $\Gamma_V$. Expression \eqref{eq:RT} together with the minimality condition is known as \emph{RT formula} \cite{Ryu:2006bv,Ryu:2006ef}.

Based on the equality of entanglement entropy and thermal entropy \eqref{eq:EEequalsS}, for solutions $\B$ with U$(1)$ symmetry, the RT formula reduces to the Bekenstein-Hawking one \eqref{eq:BHform}. In this sense, entanglement entropy can be regarded as a generalized gravitational entropy \cite{Lewkowycz:2013nqa}.


\subsection{Renormalization of holographic entanglement entropy}\label{sec:holorenS}

The presence of divergences in entanglement entropy in the CFT, described in sec.~\ref{sec:structent}, is manifest in the holographic picture by the fact that AAdS spacetimes have infinite volume. In the Kounterterms scheme\footnote{Alternatively, one can renormalize holographic entanglement entropy using the counterterms scheme, see refs.~\cite{Taylor:2016aoi,Taylor:2020uwf}}, renormalized holographic entanglement entropy for CFTs dual to Einstein-AdS gravity in arbitrary dimensions is obtained in the following way. First, one adds the appropriate boundary term to the bare Euclidean bulk action, namely \cite{Anastasiou:2017xjr,Anastasiou:2018mfk,Anastasiou:2018rla,Anastasiou:2019ldc}
\begin{equation}\label{eq:Irenwkt}
\IE^{\text{ren}}=\IE+\frac{c_d}{16\pi\GN}\int_{\M}B_d,
\end{equation}
where $B_d$ is given in eqs.~\eqref{Bdodd} and \eqref{eq:Bdeven} depending whether $d$ is odd or even, respectively, and $c_d$ is given by
\begin{equation}\label{eq:cd}
c_{d}=\begin{dcases*}\frac{\left(  -1\right)  ^{\frac{d+1}{2}}L^{d-1}}
{\left(  \frac{d+1}{2}\right)  \Gamma(d)}& \text{if $d$ odd,} \\
\frac{\left(  -1\right)  ^{\frac{d}{2}}L^{d-2}}
{2^{d-3}d\, \Gamma(d/2)^2}&  \text{if $d$ even.}\end{dcases*} 
\end{equation}
Once we have the renormalized form of the Euclidean action $\IE^{\text{ren}}$, we evaluate it on the orbifold $\tilde{\B}$ in order to use eq.~\eqref{eq:finalLM}. A remarkable property of the Kounterterm $B_d$ is that ---when evaluated in the orbifold--- the singular part self-replicates in codimension-two \cite{Anastasiou:2018rla,Anastasiou:2018mfk}
\begin{equation} \label{eq:Kountereroddorbifold}
\int_{\M_\vartheta}\diff^{d}x\sqrt{\gamma} B_{d}=\int_{\M}\diff^{d}x\sqrt{\gamma} B_{d}+4\pi\left\lfloor\frac{(d+1)}{2}\right\rfloor (1-\vartheta)\int_{\Sigma}\diff^{d}x\sqrt{{}_\Sigma\gamma} B_{d-2},
\end{equation}
This can be seen in the case of odd $d$ by relating the Chern form to the Euler-density through the Chern-theorem \eqref{eq:Cherntheo} and recalling that the Euler density self-replicates in codimension-two \eqref{eq:Eulerself}. Unlike the Chern form, Kounterterms for odd bulk dimensions are not associated to the Euler density. Therefore, the previous analysis for orbifolds cannot be repeated verbatim. However, even though $B_{d}$ for $d$ even and odd are different geometrical objects, it can be shown that eq.~\eqref{eq:Kountereroddorbifold} holds in presence of squashed conical singularities regardless the dimensionality \cite{Anastasiou:2019ldc}. Their explicit expression is given by
\begin{align}
 B_{d-2}=&
-(d-1)\int\limits_0^1\diff s\sqrt{{}_{\Sigma}\gamma}\delta^{\alpha_1\cdots \alpha_{d-2}}_{\beta_1\cdots \beta_{d-2}}\K\indices{_{\alpha_1}^{\beta_1}}\left(\frac{1}{2}{}_{\Sigma}\R\indices{_{\alpha_2}_{\alpha_3}^{\beta_2}^{\beta_3}}-s^2\K\indices{_{\alpha_2}^{\beta_2}}\K\indices{_{\alpha_3}^{\beta_3}}\right)\times\cdots\notag\\
    & \cdots\times\left(\frac{1}{2}{}_{\Sigma}\R\indices{_{\alpha_{d-3}}_{\alpha_{d-2}}^{\beta_{d-3}}^{\beta_{d-2}}}-s^2\K\indices{_{\alpha_{d-3}}^{\beta_{d-2}}}\K\indices{_{\alpha_{d-3}}^{\beta_{d-2}}}\right),\label{eq:Bdminus2odd}
\end{align}
for $d$ odd and 
\begin{small}
\begin{align}
 B_{d-2}=&
-(d-2)\int\limits_0^1\diff s \int\limits_0^s \diff t  \sqrt{{}_{\Sigma}\gamma}\delta_{\beta_1\cdots \beta_{d-3}}^{\alpha_1\cdots \alpha_{d-3}}\K\indices{_{\alpha_1}^{\beta_1}}\left(\frac{1}{2}{}_{\Sigma}\R\indices{_{\alpha_2}_{\alpha_3}^{\beta_2}^{\beta_3}}-s^2\K\indices{_{\alpha_2}^{\beta_2}}\K\indices{_{\alpha_3}^{\beta_3}}+\frac{t^2}{L^2}\delta^{\beta_2}_{\alpha_2}\updelta^{\beta_3}_{\alpha_3}\right)\times\cdots\notag\\
    &\cdots\times\left(\frac{1}{2}{}_{\Sigma}\R\indices{_{\alpha_{d-4}}_{\alpha_{d-3}}^{\beta_{d-4}}^{\beta_{d-3}}}-s^2\K\indices{_{\alpha_{d-4}}^{\beta_{d-3}}}\K\indices{_{\alpha_{d-4}}^{\beta_{d-3}}}+\frac{t^2}{L^2}\updelta^{\beta_{d-4}}_{\alpha_{d-4}}\updelta^{\beta_{d-3}}_{\alpha_{d-3}}\right)\label{eq:Bdminus2even}
\end{align}
\end{small}
for $d$ even, where $\K\indices{_\alpha^\beta}$ denotes the codimension-one extrinsic curvature of $\Sigma$.

Finally, introducing the resulting expressions in eq.~\eqref{eq:finalLM}, we obtain the renormalized holographic entanglement entropy \cite{Anastasiou:2019ldc}
\begin{equation}\label{eq:SEEren}
\SEE^{\text{ren}}(V)=\SEE(V)+S_{\text{kt}}(V)=\begin{cases*}
\suniv(V) & \text{if odd }d,\\
\suniv(V)\log\frac{H}{\delta} & \text{if even }d,\\
\end{cases*}
\end{equation}
where the Kounterterm
\begin{equation}\label{eq:ktS}
S_{\text{kt}}(V)=\frac{c_d}{4\GN}\left\lfloor\frac{(d+1)}{2}\right\rfloor\int_\Sigma B_{d-2},
\end{equation}
cancels the divergences coming from the set of points anchoring the surface $\Sigma$ to the conformal boundary ---leaving the logarithmic divergence untouched in even-dimensional CFTs. From here, It has been shown that they correctly isolate the universal terms of the entanglement and R\'enyi entropies for CFTs which are dual to Einstein gravity, one can read off the universal contributions in each case \cite{Anastasiou:2017xjr,Anastasiou:2018mfk,Anastasiou:2018rla,Anastasiou:2019ldc}. In that situation, the relation between entanglement entropy and the notion of renormalized area was made manifest. 

\subsection{Holographic entanglement entropy dual to higher-curvature gravity}\label{sec:heehcg}

During the derivation of the RT formula \eqref{eq:RT}, we have been completely general regarding the dual gravity theory until eq.~\eqref{eq:finalLM}. If we are interested in studying CFTs dual to higher-order gravity, then we need to modify the subsequent discussion. Similarly to the case of the Bekenstein-Hawking entropy \cite{Bekenstein:1973ur,Hawking:1974sw}, when the Wald entropy formula gives the correct entropy functional when considering higher-curvature gravity \cite{Wald:1993nt,Iyer:1994ys}, the holographic entanglement entropy functional acquires corrections as well. One might expect that the Wald functional would suffice to render the correct result \cite{Hung:2011xb}. However, this is not the case and additional contributions depending on extrinsic curvatures are arise. In the case of theories of the type $\pazocal{L}(R_{abcd})$\footnote{The correct holographic entanglement entropy functional in the case of theories including covariant derives of the Riemann tensor is given in ref.~\cite{Miao:2014nxa}.}, the holographic entanglement entropy functional, known as Camps-Dong formula is given by \cite{Dong:2013qoa,Camps:2013zua}
\begin{equation}\label{eq:CDEE}
S_{\text{CD}}(V)=S_{\text{Wald}}(V)+S_{\text{anomaly}}(V),
\end{equation} 
where the first term, corresponds to the usual Wald entropy \eqref{eq:WaldS}. The second term, known as anomaly term, is given by
\begin{align}
S_{\text{anomaly}}=&\int_{\Gamma_V}\diff^{d-1}x\sqrt{{}_{\Gamma_V}g}\sum_m\frac{2}{q_m+1}\left(\frac{\updelta^2 \pazocal{L}}{\updelta R_{a_1b_1c_1d_1}\updelta R_{a_2b_2c_2d_2}}\right)K_{e_1b_1d_1}K_{e_2b_2d_2}\notag\\
&\times\left[\left(n_{a_1a_2}n_{c_1c_2}-\epsilon_{a_1a_2}\epsilon_{c_1c_2}\right)n^{e_1e_2}+\left(n_{a_1a_2}\epsilon_{c_1c_2}+\epsilon_{a_1a_2}n_{c_1c_2}\right)\epsilon^{e_1e_2}\right],
\end{align}
where $K_{abc}=n_a^{(\iota)}K_{bc}^{(\iota)}$, $\epsilon_{ab}=n_a^{(\iota)}n_a^{(\kappa)}\epsilon_{\iota\kappa}$ is the unit binormal to $\Gamma_V$. Notice that in the case of Killing horizons ---or equivalently, when there is a bifurcation surface---, the extrinsic curvature vanishes, and we recover Wald's formula.

Similarly to the case of Einstein gravity, the generalized RT surface $\Gamma_V$ must satisfy the extremality condition \cite{Dong:2017xht}. However, contributions involving Riemann tensors coming from the anomaly term must be split into a sum of pieces with different weights $q_m$ according to some prescription. It turns out that this procedure depends on the regulation scheme employed to deal with the conical singularities \cite{Miao:2014nxa,Camps:2014voa,Miao:2015iba,Camps:2016gfs}. This uncomfortable situations is referred as the \emph{splitting problem}. While it does not show up in the for $f(R)$, quadratic and Lovelock\footnote{In the case of Lovelock gravity, the Dong-Camps functional becomes the Jacobson-Myers functional \cite{Jacobson:1993xs}, which was derived using a Hamiltonian approach} gravities, it makes unclear how to compute the extremal surface in a general higher-curvature gravity beyond quadratic order. A different approach to avoid the splitting problem in the anomaly term is to consider the RT surface as the extremal one. By doing so, one obtains the holographic entanglement entropy of the higher-curvature gravity that is valid at perturbative order in the couplings. Following this prescription, in ref.~\cite{Bueno:2020uxs} results of universal terms for CFTs dual to $f(R)$, $f(R_{\mu\nu})$, cubic, quartic gravities are obtained.

\section{Summary}

Until now, we have merely discussed previous results which are available in the literature. The rest of the thesis is based on works published as refs.~\cite{Bueno:2019ltp,Anastasiou:2020smm,Anastasiou:2021swo,Bueno:2021fxb,Bueno:2021krl,Bueno:2022lhf,Bueno:2022res}, which provide new insights on open questions. Some parts are transcribed from the published articles, while others are written originally for this thesis. In both cases, the notations and conventions have been adapted to provide an uniform and fluid presentation, as well as the figures.

The results have been segmented into two distinguishable parts. Part 1 entails original work regarding several aspects of high-curvature gravity, both in three- and arbitrarily high dimensions, with focus on GQT gravity theories. On the other hand, part 2 focus on entanglement entropy for CFTs dual to Einstein-AdS gravity, higher-curvature gravity and general CFTs. Throughout the introduction, we have stressed their interplay and connections in the context of the gauge/gravity duality.

We now present summary of each of the chapters and mention the main original results included in them.

\subsection*{Chapter 2 (Based on ref.~\cite{Bueno:2022res})}

In chapter \ref{ch:ateach}, we demonstrate that, for dimensions $D\geq5$, there are exactly $n-1$ different Lagrangians of curvature $n$ that can be identified as distinct ---or inequivalent---  classes of GQT gravities. Furthermore, we show the uniqueness of a QT theory at each order. These results do not apply for low dimensions. For instance, in the case of $D=4$, there seems to be a single unique GQT density at each order and no QT densities whatsoever.

Based on these results, we compute the thermodynamic charges of the most general $D$-dimensional order-$n$ GQTG. We verify that they satisfy the first law \eqref{eq:firstlaw} and provide evidence that they can be entirely written in terms of the embedding function \eqref{eq:polych}.

\subsection*{Chapter 3 (Based on ref.~\cite{Bueno:2019ltp})}

In chapter \ref{sec:fredef} we show that any gravitational effective action involving higher-curvature corrections of the type $\pazocal{L}(R_{abcd})$ is equivalent, via metric redefinitions, to some GQT gravity \eqref{eq:priA}. We show that this is also the case for effective actions that include up to eight derivatives of the metric as well as for terms involving arbitrary contractions of two covariant derivatives of the Riemann tensor and any number of Riemann tensors. Based on this, we conjecture that the equivalence also applies to theories involving an arbitrary number of covariant derivatives.

This result can be exploited to study the physics generic higher-curvature gravity black hole, which is captured by its GQT gravity counterpart, dramatically easier to characterize and universal.

\subsection*{Chapter 4 (Based on refs.~\cite{Bueno:2022lhf,Bueno:2021fxb})}

In chapter \ref{ch:hc3d}, we turn our attention to three-dimensional gravities and present new results involving general higher-curvature gravity theories. First, we provide a general formula for the exact number of independent order-$n$ densities, $\#(n)$, which satisfies the identity $\#(n-6)=\#(n)-n$. Then, we show that, linearized around a general Einstein solution, a generic order-$n\geq 2$ density can be written as a linear combination of three type of densities which, individually, does not propagate either massive, scalar, or any mode at all. Next, we find a recursive formula as well as a general closed expression for order-$n$ densities which non-trivially satisfy an holographic c-theorem, clarify their relation with Born-Infeld gravities and prove that the scalar mode is always absent from their spectrum. We show that, at each order $n \geq 6$, there exist $\#(n-6)$ densities which satisfy the holographic c-theorem in a trivial way and that all of them are proportional to a single sextic density $\Y_{(6)}$ ---see definition in eq.~\eqref{Omeg6} below. Remarkably, $\Y_{(6)}$ satisfies as well the GQT condition \eqref{eq:GQTcond} with trivial equations of motion. We also mention that adding matter fields into the picture enables the existence of non-trivial EGQT theories, which are constructed from positive powers $(\partial\phi)^2$ and certain linear combinations of $R^{ab}\partial_a {\phi}\partial_b{\phi}$ and $(\partial\phi)^2R$. Finally, comment on the meaning of $\Y_{(6)}$ and its relation to the Segre classification of three-dimensional metrics.

\subsection*{Chapter 5 (Based on refs.~\cite{Bueno:2022lhf,Bueno:2021fxb})}

In chapter \ref{ch:bhs3d} we explore black holes in three dimensions. Based on the results of chapter \ref{ch:hc3d} we study the thermodynamics, quasinormal modes and frequencies of the BTZ black hole as a function of the masses of the graviton and scalar modes for a general theory. 

When matter is included in the picture, we find a plethora of new analytic black holes that are continuous generalization of the BTZ one, as well as globally regular horizonless spacetimes. The solutions involve a single real scalar field $\phi$ which always admits a magnetic-like expression proportional to the angular coordinate. The new metrics solve EQT gravities. 

Some of the solutions obtained describe black holes with one or several horizons. A set of them possesses curvature singularities, while others have conical or BTZ-like ones. Interestingly, in some cases the black holes have no singularity at all, being completely regular. Some of the latter are achieved without any kind of fine tuning or constraint between the action parameters and/or the physical charges of the solution. An additional class of solutions describes globally regular and horizonless geometries.

\subsection*{Chapter 6 (Based on ref.~\cite{Anastasiou:2020smm})}

In chapter \ref{chap:sdhee}, we study the shape-dependence ---in the absence of singular regions--- of holographic entanglement entropy of deformed entangling regions in three-dimensional CFTs dual to Einstein-AdS gravity, using the Kounterterms renormalization scheme, discussed in secs.~\ref{sec:holoren} and \ref{sec:holorenS}. In this prescription, the contribution to the renormalized entanglement entropy coming from the deformation of the entangling surface is encoded purely in the curvature contribution. In turn, as the topological part is given by the Euler characteristic of the RT surface, it remains shape-independent, manifesting its non-local nature. Exploiting the covariant character of the Kounterterms, we apply the renormalization scheme for the case of deformed entangling regions in AdS${}_4$/CFT${}_3$, recovering the results found in the literature \cite{Fonda:2014cca}. We observe when the entangling region is a perfect disk, the geometrical part vanishes and entanglement entropy is maximized.

Finally, we provide a derivation of the relation between renormalized entanglement entropy and Willmore energy ---see definition below in eq.~\eqref{eq:Willmore}. Exploiting its well known properties, we manage to give a relation between the degree of the deformation of the Ryu-Takayanagi surface and the strong subadditivity property \eqref{eq:SSEE}.

\subsection*{Chapter 7 (Based on ref.~\cite{Anastasiou:2021swo})}

In chapter \ref{chap:heeqcg} we turn our attention to CFTs dual to QC gravity \eqref{eq:QCGgen} and derive a covariant expression for the renormalized holographic entanglement entropy in arbitrary dimensions using the Kounterterms scheme \ref{sec:holorenS}. This expression is written as the sum of the bare entanglement entropy functional ---obtained using standard conical defect techniques discussed in sec.~\ref{sec:geoentang}--- plus an appropriate Kounterterm. 

As mentioned before, this renormalization method isolates the universal terms of the holographic entanglement entropy functional. This allows us to compute the standard $\C$-function candidate for CFTs of arbitrary dimension, and the type-B anomaly coefficient $c$ in four-dimensional CFTs.

\subsection*{Chapter 8 (Based on ref.~\cite{Bueno:2021krl})}

Finally, continuing with the studies of holographic shape-dependence in chapter \ref{chap:sdhee}, we study the situation in general three-dimensional CFTs. In this context, we show the finite part of entanglement entropy $F(V)$ of entanglement entropy is minimized when the entangling region is a disk $F_0\equiv F(\mathbb{B}^2)$ regardless the CFT under consideration. The proof makes use of the strong subadditivity property \eqref{eq:SSEE} and the geometric fact that one can always place an osculating circle within a given smooth entangling region. For topologically non-trivial entangling regions with $n_B$ boundaries, the general bound can be improved to $F(V) \geq n_BF_0$. Moreover, we provide accurate approximations to $F(V)$ valid for general CFTs, when $V$ is an elliptic region for arbitrary values of the eccentricity which we check against lattice calculations for free fields. We also evaluate $F$ numerically for more general shapes in the so-called ``Extensive Mutual Information model'' ---see sec.~\ref{secemi} below---, verifying the general bound.

%% file: text/ch2-gqtorder.tex
\chapter{(Generalized) quasi-topological gravities at each order}\label{ch:ateach}

As stated in sec.~\ref{sec:introGQT}, we know there GQT densities exist at all orders in curvature for $D\geq4$ \cite{Bueno:2019ycr}. In this chapter we study the number of different possible GQT densities that give different equations of motion for $f$. To do so, we introduce the concept of ``inequivalent'' GQT in order to distinguish the equations of two given Lagrangians and prove that there exist exactly $(n-2)$ inequivalent genuine GQT densities and only one inequivalent QT density at a given curvature order $n$ in $D\geq 5$. In $D=4$ there are no QT and we argue that our proof for the existence of $(n-2)$ genuine GQT densities fails in that case, illustrating the fact that a single genuine GQT density exists in $D=4$ for $n\geq 3$.

Furthermore, we study the thermodynamics of the most general $D$-dimensional order-$n$ GQT gravity, verify that they satisfy the first law and provide evidence that they can be written in terms of the embedding function which determines the maximally symmetric vacua of the theory

\section{(In)equivalent GQT gravities}

Consider a certain curvature invariant of order $n$, $\mathfrak{R}^{(n)}$, following the notation presented in sec.~\ref{sec:introGQT}. We can define an effective on-shell action $I_{N,f}$ with results from integrating the effective Lagrangian $L_{N,f}$, this is
\begin{equation}
I_{N,f}\equiv\Omega_{D-2}\int\diff t\int\diff r L_{N,f}, \quad \text{where } \Omega_{D-2}=\frac{2\pi^{\frac{D-1}{2}}}{\Gamma\left(\frac{D-1}{2}\right)},
\end{equation}
with $I_{f}\equiv I_{1,f}$. The fact that a certain density $\mathfrak{R}^{(n)}$ satisfies the GQT condition \eqref{eq:GQTcond} is tantamount to asking $L_f$ to be a total derivative 
\begin{equation}\label{eq:tdT0}
L_f=T_0',
\end{equation}
for some function $T_0(r,f,f')$. As stated in the previous chapter, the equation satisfied by $f$ for a given GQT density can be obtained from the variation of $L_{N,f}$ with respect to $N$. As long as eq. \eqref{eq:tdT0} holds, the effective Lagrangian  $L_{N,f}$ takes the form \cite{Bueno:2017sui}
\begin{equation}\label{fofwo}
L_{N,f}=N T_0' +  N' T_1 +  N'' T_2 +\pazocal{O}(N'^2/N)\, ,
\end{equation}
where $T_{1}$, $T_2$  are functions of $f(r)$ and its derivatives, and $\pazocal{O}(N'^2/N)$ is a sum of terms all of which are at least quadratic in derivatives of $N$. Integrating by parts, it follows that
\begin{equation}
I_{N,f} = \Omega_{(D-2)}  \int \diff t \int \diff r \left[N\left(T_0-T_1+T_2' \right)' +\pazocal{O}(N'^2/N) \right]\, .
\end{equation}
So it is possible to write all terms  involving one power of $N$ or its derivatives as a product of $N$ and a total derivative which depends on $f$ alone. Now, it follows straightforwardly that condition \eqref{eq:GQTcond} equates that total derivative to zero. If we integrate it once, we are left with \cite{Bueno:2017sui}
\begin{equation} \label{eq:FTC}
\pazocal{F}_{\mathfrak{R}^{(n)}}  \equiv T_0-T_1+T_2'=\EuScript{C} ,
\end{equation}
where $\EuScript{C}$ is an integration constant related to the ADM mass of the solution \cite{Arnowitt:1960es,Arnowitt:1960zzc,Arnowitt:1961zz,Deser:2002jk}. In particular, for spherical horizons, the precise relation reads 
\begin{equation}
\EuScript{C}= \frac{M}{\Omega_{(D-2)}}.
\end{equation}
Hence, given some linear combination of GQT densities, obtaining the equation satisfied by the metric function $f$ amounts to evaluating $L_{N,f}$ as defined in eq. \eqref{ansS} and then identifying the functions $T_{i=0,1,2}$ from eq. \eqref{fofwo}. The equation is then given by expression \eqref{eq:FTC}).\footnote{Sometimes we will refer to this equation as the ``integrated equation'' of $f$ to emphasize the fact that it follows from integrating once (on $r$) the only non-vanishing component of the actual equations of motion of the theory evaluated on the single-function SSS ansatz.}

Once we have discussed how to obtain the equations of motion of a GQT gravity, a natural question arises. Given a fixed spacetime dimension $D$ and a curvature order $n$, are the integrated equations corresponding to different genuine GQT densities $\left\{ \mathfrak{R}_{\text{I}}^{(n)},\ \mathfrak{R}_{\text{II}}^{(n)},\dots  , \ \mathfrak{R}_{i_n}^{(n)}\right\}$ proportional to each other ---\ie are the functional dependence on $r$, $f$, $f'$ and $f''$ of the equations  identical--- for the various densities? If not, how many inequivalent contributions to the equation of $f$ are there at a given order in curvature? Analogous questions can be asked fixing $D$ and $n$ for theories belonging to the quasi-topological class. These questions motivate the definition of inequivalent GQT densities.

\begin{defi}
Given two GQT densities of order $n$, we say they are inequivalent if the quotient of their integrated equations of motion when evaluated on SSS \eqref{eq:SSS} is not a constant,
\begin{equation}
\mathfrak{R}_{\mathrm{I}}^{(n)} \quad \text{inequivalent from} \quad \mathfrak{R}_{\mathrm{II}}^{(n)} \quad \Leftrightarrow \quad  \frac{\pazocal{F}_{\mathfrak{R}_I^{(n)}}(r,f,f',f'') }{\pazocal{F}_{\mathfrak{R}_{\mathrm{II}}^{(n)}} \left(r,f,f',f''\right)} \neq \text{constant}.
\end{equation}
If the quotient is constant, then two GQT densities are said to be equivalent.
\end{defi}
We could define as well inequivalent QT densities of order $n$ performing an analogous definition. However, we will show in sec.~ \ref{sec:onlyoneQT} that, in fact, all QT densities of a given order are equivalent. That will not be the case for genuine GQT densities, in whose case we will prove that there exist $(n-2)$ inequivalent densities for $D\geq 5$.\footnote{The existence of multiple types of GQT densities was first pointed out in ref.~\cite{Bueno:2020odt}, where two inequivalent quintic densities were explicitly constructed in $D=6$.}

\section{Number of inequivalent theories}\label{sec:numbertheo}

\subsection{At most $(n+1)$ order-$n$ densities}

Let us start our study by giving an upper bound on the possible number of inequivalent GQT densities existing at a given curvature order $n$. As argued in  ref. \cite{Deser:2005pc}, when these GQT densities are evaluated on a metric of the form \eqref{eq:SSS} with $N=1$, the Riemann tensor can be written as 
\begin{equation}\label{rieef}
\left.\tensor{R}{^{ab}_{cd}}\right|_f=2\left(-\EuScript{A} \tau^{[a}_{[c}\tau^{b]}_{d]}+2\EuScript{B} \tau^{[a}_{[c}\sigma^{b]}_{d]}+\psi \sigma^{[a}_{[c}\sigma^{b]}_{d]}\right)\, ,
\end{equation}
where $\sigma_a^b$ and $\tau_a^b$ are projectors on the angular and ($t$, $r$) directions, respectively. These projectors satisfy the algebra relations
\begin{equation}\label{eq:proj}
\tau_a^b \tau_b^c=\tau_a^c,\quad \sigma_a^b \sigma_b^c=\sigma_a^c,\quad \sigma_a^b \tau_b^c=0,\quad \tau_{a}^a=2,\quad \sigma_a^a=(D-2),\quad \delta^{a}_{b}=T^{a}_{b}+\sigma^{a}_{b}.
\end{equation}
On the other hand, the dependence on the radial   coordinate appears exclusively through the three functions  $\EuScript{A}$, $\EuScript{B}$ and $\psi$, which read
\begin{equation}\label{eq:threefun}
\EuScript{A}\equiv \frac{f''}{2}\, ,   \quad   \EuScript{B}\equiv -\frac{f'}{2r}\, , \quad \psi\equiv \frac{k-f}{r^2}\, ,
\end{equation}
where  $k={1,0, -1}$ for spherical, planar and  hyperbolic horizons respectively.

Now, GQT densities are built from contractions of the metric and the Riemann tensor, so any order-$n$ density of that type will become some polynomial of these objects when evaluated on the single function metric \eqref{eq:SSS} , namely,
\begin{equation}\label{eq:Sf}
\pazocal{S}\big|_f=\sum_{l=0}^n\sum_{k=0}^l c_{k,l}\EuScript{A}^{n-l}\EuScript{B}^l\psi^{l-k}\, ,
\end{equation}
for some constants $c_{k,l}$. The idea is now to determine the most general constants  $c_{k,l}$ consistent with the GQT requirement, which demands $r^{D-2} \pazocal{S}|_f $ to be a total derivative, \ie
\begin{equation}\label{pazo}
r^{D-2} \pazocal{S}|_f = T_0'(r)\, .
\end{equation}
Note that imposing this condition on the denstity \eqref{eq:Sf} and finding the compatible values of $c_{k,l}$ does not guarantee that the corresponding GQT theory actually exist, as this does not provide an explicit construction of covariant curvature densities. Doing this does impose, nonetheless, a necessary condition which all actual densities must satisfy. Given a GQT density, $\pazocal{S}$, it is useful to define the object $\EuScript{T}$ through the relation
\begin{equation}\label{ttau}
\EuScript{T} \equiv \frac{T_0}{r^{D-1}} \, , \quad \text{so that} \quad  \pazocal{S}\big|_f= \frac{1}{r^{D-2}} \frac{\diff }{\diff r} \left[ r^{D-1} \EuScript{T}\right] \, .
\end{equation}
In a sense, $\EuScript{T}$ is the fundamental building block as long as on-shell GQT densities are concerned. Observe that since
\begin{equation}
 \sum_i \alpha_i  \pazocal{S}_i\big|_f = \frac{1}{r^{D-2}} \frac{\diff }{\diff r} \left( r^{D-1} \sum_i \alpha_i \EuScript{T}_{(i)}\right) \, ,
\end{equation}
linear combinations of the $\EuScript{T}_{(i)}$ give rise to linear combinations of GQT densities in an obvious way.

Now, imposing relation \eqref{pazo} on densities of the form \eqref{eq:Sf}, we find that there are $(n+1)$ independent possible densities at a given order $n$. In terms of the $\EuScript{T}$, the possibilities turn out to be simply given by $\EuScript{T}=\EuScript{T}_{(n,j)}$, where we defined
\begin{equation}
\EuScript{T}_{(n,j)}\equiv \psi^{n-j} \EuScript{B}^j \, , \quad \text{where} \quad j=0,1,\dots,n\, .
\end{equation}
The corresponding putative on-shell densities read\footnote{Note that for the objects $\pazocal{S}_{(n,j)}$ we omit the $|_f$. By this we mean that we literally define $\pazocal{S}_{(n,j)}$ to be the expression that appears in the right-hand side. Actual densities evaluated on the single-function SSS ansatz will reduce to linear combinations of the $\pazocal{S}_{(n,j)}$.  }
\begin{equation}\label{rj}
\pazocal{S}_{(n,j)} \equiv \frac{1}{r^{D-2}} \frac{\diff}{\diff r} \left( r^{D-1} \EuScript{T}\right) \, , \quad j=0,1,\dots,n\, .
\end{equation}
Observe that $\pazocal{S}_{(n,j)}$ is either linear in $\EuScript{A}$ or without dependence on it at all, which is like restricting the sum in $l$ appearing in eq.~\eqref{eq:Sf} to $l=\{n-1,n\}$.
It follows that any GQT density in any number of dimensions and at any order in curvature must necessarily be expressible as a linear combination of the above densities when evaluated on the single-function SSS ansatz, namely
\begin{equation}\label{genS}
\pazocal{S}|_f=  \frac{1}{r^{D-2}} \frac{\diff}{\diff r} \left(r^{D-1} \sum_{j=0}^n \alpha_{(n,j)} \EuScript{T}_{(n,j)}\right) \, ,
\end{equation}
for certain constants $\alpha_{(n,j)}$. 

Using the methods developed in ref.~\cite{Bueno:2019ycr} ---cf. section 5 of that work--- it is possible to compute the field equations for the putative theory~\eqref{genS} despite the fact that a covariant form of the action is not known. The integrated equation for the metric function $f$ corresponding to a putative density $\pazocal{S}_{(n,j)}$ is given, in the notation of \eqref{eq:FTC}, by\footnote{So, for a linear combination of densities, the equation would read $\sum_j \alpha_{(n,j)} \pazocal{F}_{(j)}=\EuScript{C} $ where $\EuScript{C}$ is an integration constant related to the mass of the solution.}
\begin{align} \label{fnj}
&\pazocal{F}_{(n,j)}=  \frac{(-1)^{j+1}}{2^{j+1}}r^{D-2+j-2n} (k-f)^{n-j-1}(f')^{j-2} \times \\ &\left[ f' \Big[j(D-1+j-2n)(k-f)f-(j-1)r (k+(n-j-1)f) f' \right] +j(j-1)  r (k-f) f f''\Big] \, . \notag
\end{align}
Observe that this simplifies considerably both for $j=0$ and $j=1$. In those cases the dependence on $f'$ and $f''$ disappears and one finds algebraic equations for $f$,
\begin{align}
\pazocal{F}_{(n,0)}&=  -\frac{r^{D-1-2n}}{2} (k-f)^{n-1} [k+(n-1)f],\\ \pazocal{F}_{(n,1)}&=\frac{(D-2n)r^{D-1-2n}}{4}(k-f)^{n-1}f \, .
\end{align}
An obvious question at this point is: which of these possible densities actually corresponds to the Einstein-Hilbert one, if any. In that case we have $n=1$, and the two possible densities and their integrated equations of motion read, respectively,
\begin{align}
\pazocal{S}_{(1,0)}&=-\frac{1}{r^2} \left[(D-3) (f-k)+r f' \right]\, , \quad \pazocal{F}_{(1,0)}=-\frac{r^{D-3} k }{2} \, , \\
\pazocal{S}_{(1,1)}&=-\frac{1}{2r^2} \left[(D-2)r f'+r^2 f'' \right]\, , \quad \pazocal{F}_{(1,1)}=\frac{(D-2)r^{D-3}f }{4} \, .
\end{align}
Now, the corresponding expressions for the Einstein-Hilbert action (\ie for a density given by the Ricci scalar $\pazocal{S}_{\rm \ssc EH}\equiv R$) read
\begin{align}
\pazocal{S}_{\rm \ssc EH}|_f&=-\frac{1}{r^2} \left[(D-2)(D-3)(f-k)+ 2(D-2)r f'+r^2 f''\right]\, , \\
 \pazocal{F}_{\rm \ssc EH}&=-(D-2)(f-k) r^{D-3} \, .
\end{align}
Hence, none of the putative densities coincides with the Einstein-Hilbert one. Rather, it is a linear combination of the two which does, namely,
\begin{equation}
\pazocal{S}_{\rm \ssc EH}|_f= (D-2) \pazocal{S}_{(1,0)} + 2 \pazocal{S}_{(1,1)}\, .
\end{equation}
Even though our approach has selected two possible independent densities susceptible of giving rise to GQT densities at linear order in curvature, there (obviously) exists a unique possibility corresponding to an actual density, given by the Ricci scalar, which therefore is given by a linear combination of the two. While the $n=1$ case is somewhat special, this already illustrates the fact that our upper bound of $(n+1)$ densities at order $n$ is not tight and can be improved. For higher $n$, the only known examples of densities which give rise to algebraic integrated equations for $f$ are Lovelock and Quasi-topological gravities. From our perspective, at a given order $n$ in $D$ dimensions, all available Lovelock and Quasi-topological gravities for such $n$ and $D$ are ``equivalent'' as far as the equation of $f$ is concerned, which means that they should correspond to a fixed linear combination of $\pazocal{S}_{(n,0)}$ and $\pazocal{S}_{(n,1)}$. In the next subsections we argue that, indeed, the bound of $(n+1)$ densities can be lowered to at most $(n-1)$  GQT densities of order $n\geq 2$. While amongst the $(n+1)$ candidates identified here there are two which produce algebraic equations, we will see that only a linear combination of the two survives, precisely corresponding to the known Lovelock and Quasi-topological case. The additional putative $(n-2)$ densities would give rise to distinct second-order differential equations for $f$.



\subsection{At most $(n-1)$ order-$n$ densities}\label{atmostnm1}
In order to lower our upper bound on the number of available GQT densities existing at a given order, we can impose some further conditions on our candidate on-shell densities $\pazocal{S}_{(n,j)}$. The first condition comes from imposing that the equations of motion associated to them admit maximally symmetric solutions. When evaluated on such backgrounds, the equations of motion of actual higher-curvature densities reduce to an algebraic equation which involves the cosmological constant, the curvature scale of the background (\eg the AdS radius) as well as the higher-curvature couplings. More precisely, consider a gravitational Lagrangian consisting of a linear combination of generic higher-curvature densities of the form given in eq.~\eqref{eq:priA}.
The result for the equations of motion when evaluated for
\begin{equation}
f=\frac{r^2}{L_{\star}^2}+k\, ,
\end{equation}
with $k$ parametrizing the sphere, planar or hyperbolic horizon, corresponding to pure AdS$_{D}$ with effective radius $L_{\star}$, given by
\begin{equation}
\frac{r^{D-1}}{16\pi \GN} \left[\frac{(D-2)}{L^2}-\frac{(D-2)}{L_{\star}^2}+ \sum_{n=2} \sum_{i_n} \frac{L^{2(n-1)}}{L_{\star}^{2n}}  \alpha_{i_n}^{(n)} a_{i_n}^{(n)}\right]=0\, , 
\end{equation}
for certain constants $a_{i_n}^{(n)}$. Interestingly, as we will see below, this same equation which determines the vacua, also appears to play a key role in the thermodynamics of black holes in the theory. Naturally, the solution for Einstein gravity is simply $L^2=L_{\star}^2$, which relates the action scale to the AdS radius in the usual way.

Now, what happens when we consider the integrated equations of a linear combination of candidate on-shell GQT densities, each contributing as in \req{fnj}, on such a background? It turns out that the result $\sum_j \alpha_{(n,j)} \pazocal{F}_{(n,j)}$ contains two different kinds of terms, one which goes with a power of $r^{D-1}$, and one which goes with a power of $r^{D-3}$. As we have seen, actual densities contribute with a single power of the type  $r^{D-1}$, so we must impose that the second kind of term is absent for our putative densities. Removing such a piece amounts to imposing the condition
\begin{equation}\label{cond1}
\sum_{j=0}^{n} \alpha_{(n,j)} (2n-Dj)=0\, .
\end{equation}
Hence, we learn that not all the candidate densities can be independent and we reduce the number from $(n+1)$ to $n$. 

There is another condition we can easily impose on our candidate densities. As explained in the first section, GQT densities have second-order linearized equations around general maximally symmetric backgrounds. This is in contradistinction to most higher-curvature gravities, whose linearized equations involve up to four derivatives of the metric ---see \eg \cite{Bueno:2016ypa} for general formulas. Suppose then that we consider a small radial perturbation on AdS space such that the metric function becomes
\begin{equation}\label{perv}
f=\frac{r^2}{L_{\star}^2}+k+\epsilon w\, ,
\end{equation}
where $\epsilon \ll 1$ and $w=w(r)$ is some function of the radial coordinate. Now, observe that in our general discussion, the integrated equation of motion for a GQT density, $\pazocal{F}_{\pazocal{S}_n}$, has been integrated once (on $r$) with respect to the actual equations of motion of the corresponding density. Hence, the fact that the actual (linearized) equations of motion for GQT densities are second order for any perturbation on a maximally symmetric background implies that the integrated equations cannot contain terms involving $w''$ (or more derivatives) at leading order in $\epsilon$. If they did, the actual linearized equations would involve terms of the form $\sim \epsilon w'''$, in contradiction with the linearized second-order behavior. With this in mind, our strategy now is to insert \req{perv} in a linear combination of integrated equations for our candidate on-shell densities (\ref{fnj}) and impose that no terms involving $w''$ appear at leading order in $\epsilon$. By doing so, we find an additional (remarkably simple) condition, which reads
\begin{equation}\label{cond2}
\sum_{j=0}^{n} \alpha_{(n,j)}j (j-1)=0\, .
\end{equation}
Imposing it further reduces the number of independent densities from $n$ to $(n-1)$. Hence, we conclude that in $D$ dimensions there exist at most $(n-1)$ inequivalent GQT theories of order $n$. Later in sec.~\ref{proofexa} we will prove that, in fact, there exist exactly $(n-1)$ inequivalent densities for $D\geq 5$. There are many possible ways to choose a basis of on-shell densities so that eqs. \eqref{cond1} and \eqref{cond2} are implemented. For instance, we may choose for the $\EuScript{T}$ functions defined in eq. \eqref{ttau}
\begin{align}\label{qt}
\EuScript{T}^{\rm QT}_{\{n\}}\equiv & (2n-D) \EuScript{T}_{(n,0)}-2n \EuScript{T}_{(n,1)}\, ,\\ 
\EuScript{T}^{\rm GQT}_{\{n,j\}}\equiv& (j+1)(D j -4 n) \EuScript{T}_{(n,j+1)} \\ \notag &+ \left[2D(1-j^2)-4n(1-2j) \right]\EuScript{T}_{(n,j)} + j [D(j+1)-4n]\EuScript{T}_{(n,j-1)}\, ,
\end{align}
with $j=2,\dots,n-1$, where we isolated the QT class combination in the first line ---see next subsection.

Naturally, constructing actual covariant densities of each of the classes is a non-trivial problem on its own. Explicit formulas for order-$n$ GQT densities in arbitrary dimensions $D\geq 4$ as well as for order-$n$ QT densities in $D\geq 5$ were presented in ref.~\cite{Bueno:2019ycr}. However, these cases only exhausted two of the $(n-1)$ classes which we show to exist for $D\geq 5$ in the present paper (one of the GQT types and the Quasi-topological one). In app.~\ref{densiii} we present explicit formulas for the $(n-2)$ different types of GQT densities for $n=4,5,6$ in $D=5$ and $D=6$.

\subsection{Uniqueness of QT density at each order}\label{sec:onlyoneQT}
As mentioned above, QT densities are a subclass of GQT ones characterized by having an algebraic (as opposed to second-order differential) integrated equation of motion for the metric function $f$ \cite{Oliva:2010eb, Myers:2010ru, Dehghani:2011vu,Ahmed:2017jod,Cisterna:2017umf}. Theories of that kind are required to satisfy an additional condition besides  
\eqref{eq:GQTcond}, namely \cite{Bueno:2019ycr} 
\begin{equation}
\left[ \frac{D-2}{r} \frac{\partial}{\partial f''}+\frac{\diff }{\diff  r} \frac{\partial}{\partial f''}+\frac{(D-3)}{2} \frac{\partial}{\partial f'}+\frac{r}{2} \frac{\diff }{\diff  r} \frac{\partial}{\partial f'}-r \frac{\partial}{\partial f}\right] \pazocal{Z}|_f=0\, ,
\end{equation}
which is equivalent to enforcing that the term $\nabla^d P_{acdb}$ in from the equations of motion \eqref{eq:eomgen} vanishes on a SSS metric ansatz.
Imposing this condition on a general linear combination of our candidate densities  \eqref{genS} severely constrains the values of the $\alpha_j$, and we find that $\EuScript{T}^{\rm QT}_{\{n\}}$ as defined in expression \eqref{qt}
is in fact the only possibility. Hence, we learn that the only combination of putative densities compatible with the QT condition is given by
\begin{equation}\label{zf}
\pazocal{Z}_{(n)}|_f= \frac{1}{r^{D-2}} \frac{\diff}{\diff r} \left\{ r^{D-1} \left[ (2n-D) \EuScript{T}_{(n,0)}-2n \EuScript{T}_{(n,1)}\right] \right\}\, .
\end{equation}
As anticipated in sec.~\ref{sec:introGQT}, QT gravities with precisely this structure were shown to exist in ref.~\cite{Bueno:2019ycr} at all orders in $n$ and for all $D\geq 5$. Therefore, we conclude that the only possible on-shell structure of a QT density is given by \eqref{zf}. There are no additional inequivalent QT densities besides the known ones. In consequence, if a given higher-curvature density possesses second-order linearized equations around maximally symmetric backgrounds and admits black hole solutions satisfying $g_{tt}g_{rr}=-1$, such that, the equation for $f$ is algebraic, then, the equation which determines such a function is uniquely determined to be
\begin{equation}
\pazocal{F}_{\pazocal{Z}_n}=\frac{(D-2n)}{2} r^{D-2n-1} (k-f)^n\, .
\end{equation}
This naturally includes the subcases of Einstein and Lovelock gravities. 

\subsection{Exactly $(n-1)$ order-$n$ densities}\label{proofexa}
Let us finally proceed to prove that there exist exactly $(n-1)$ inequivalent GQT densities of order $n$ in dimensions higher than four.

Consider the following combination of ``on-shell densities''
\begin{equation}
\pazocal{S}^{(k)}_{p}=\sum_{i=0}^{p}\alpha^{(k)}_{p,i}  \pazocal{S}_{(p,i)}\, ,\quad k=1,\ldots,k _p\equiv \max\,(1,p-1)\, ,
\end{equation}
where the $\pazocal{S}_{(p,i)}$ are defined in \req{rj} and where we assume the constants $\alpha^{(k)}_{p,i}$ to satisfy the constraints found in sec.~\ref{atmostnm1}, namely,
\begin{equation}\label{alphaconstrv2}
\sum_{j=0}^{p} \alpha^{(k)}_{p,j} (2p - D j )=0\, ,\quad \sum_{j=0}^{p} \alpha^{(k)}_{p,j} j(j-1)=0\, .
\end{equation}
At each curvature order $p$, there are $k_{p}$ linearly independent solutions and the index $k$ labels each of them.  

Now, let us assume that for $p=1,2,\ldots ,n$ we have proven that all of these on-shell densities correspond to the evaluation of actual higher-curvature densities on the single-function SSS ansatz.  Namely, there exists a set of Lagrangians $\mathfrak{R}_{p}^{(k)}$ such that
\begin{equation}
\mathfrak{R}_{p}^{(k)}\Big|_{f}=\pazocal{S}^{(k)}_{p}\, ,\quad p=1,\ldots n\, , \quad k=1,\ldots, k_{p}\, .
\end{equation}
With this in mind, let us now consider an order-$(n+1)$ density built from a general linear combination of products of all these lower-order densities, \ie
\begin{align}
\tilde{\mathfrak{R}}_{n+1}=\sum_{m=1}^{n}\sum_{k=1}^{k_m} \sum_{k'=1}^{k_{n+1-m}}C_{m,k,k'}\pazocal{R}^{(k)}_{m}\pazocal{R}^{(k')}_{n+1-m}\, ,
\end{align}
where we introduced the constants $C_{m,k,k'}$. 

We can ask now: is it possible to generate $n$ inequivalent GQT densities of order $(n+1)$ in this way? In order to answer this question, let us evaluate $\tilde{\pazocal{R}}_{n+1}$ on the single-function SSS ansatz and try to obtain all the possible on-shell GQT structures. The evaluation yields
\begin{align}
\tilde{\mathfrak{R}}_{n+1}\Big|_{f}&=\sum_{m=1}^{n}\sum_{k=1}^{k_m} \sum_{k'=1}^{k_{n+1-m}}\sum_{i=0}^{m} \sum_{j=0}^{n+1-m}\alpha^{(k)}_{m,i} \alpha^{(k')}_{n+1-m,j}C_{m,k,k'} \pazocal{S}_{(m,i)}\pazocal{S}_{(n+1-m, j)}\\ \label{ramong}
&=\sum_{m=1}^{n}\sum_{i=0}^{m} \sum_{j=0}^{n+1-m}\tilde{C}_{m,i,j} \pazocal{S}_{(m,i)}\pazocal{S}_{(n+1-m, j)}\, ,
\end{align}
where we defined
\begin{equation}
\tilde{C}_{m,i,j}\equiv \sum_{k=1}^{k_m} \sum_{k'=1}^{k_{n+1-m}}\alpha^{(k)}_{m,i} \alpha^{(k')}_{n+1-m,j}C_{m,k,k'}\, .
\end{equation}
Now, since we are summing over all the $\alpha^{(k)}_{n,j}$ satisfying \eqref{alphaconstrv2} and $C_{m,k,k'}$ is an arbitrary tensor, note that this equality is equivalent to demanding that $\tilde{C}_{m,i,j}$ is an arbitrary tensor satisfying the following constraints
\begin{align}\label{Cconstr1}
\sum_{j=0}^{n+1-m}\tilde{C}_{m,i,j} [2(n+1-m) - D j ]&=0\, ,\quad \sum_{j=0}^{n+1-m} \tilde{C}_{m,i,j} j(j-1)=0\, ,\\
\label{Cconstr2}
\sum_{i=0}^{m}\tilde{C}_{m,i,j} (2m - Di)&=0\,,\quad \sum_{i=0}^{m} \tilde{C}_{m,i,j} i(i-1)=0\, .
\end{align}
In this way, we do not need to make reference to the  $\alpha^{(k)}_{n,j}$ anymore.  

Next, it is convenient to rearrange the sum in the following form, in terms of the index $l\equiv i+j$,
\begin{align}
\tilde{\mathfrak{R}}_{n+1}\Big|_{f}&=\sum_{m=1}^{n}\sum_{l=0}^{n+1} \sum_{j=\max(l-m,0)}^{\min(l,n+1-m)}\tilde{C}_{m,l-j,j} \pazocal{S}_{(m,l-j)}\pazocal{S}_{(n+1-m, j)}\\
&=\sum_{l=0}^{n+1}\sum_{m=1}^{n} \sum_{j=0}^{n+1-m}\theta(l-j)\theta(j+m-l)\tilde{C}_{m,l-j,j} \pazocal{S}_{(m,l-j)}\pazocal{S}_{(n+1-m, j)}\, ,
\end{align}
where $\theta(x)$ is the Heaviside step function, verifying $\theta(x)= 1$ if $x\ge0$ and $\theta(x)= 0$ if $x<0$. Observe that the effect of this function is to enforce that $i \geq 0$ and $i\leq m $, respectively, which in eq. \eqref{ramong} is explicit from the $i$ sum.  Expanding the product $\pazocal{S}_{(m,l-j)}\pazocal{S}_{(n+1-m, j)}$ we get the following expression,
\begin{align}
\tilde{\mathfrak{R}}_{n+1}\Big|_{f}=&\sum_{l=0}^{n+1}\sum_{m=1}^{n} \sum_{j=0}^{n+1-m}\theta(l-j)\theta(j+m-l)\tilde{C}_{m,l-j,j} \\\notag
&  \times \bigg[\alpha_{l,m,j} B^{2+l}\psi ^{n-1-l}  +\beta_{l,m,j} B^{1+l} \psi ^{n-l}+\gamma_{l,m,j}B^l\psi ^{1-l+n}  
 \\
&\quad \, \, +\sigma_{l,m,j}r B' B^l \psi^{n-l} +\zeta_{l,m,j}rB'B^{l-1} \psi^{1-l+n} +\omega_{l,m,j}r^2\left(B'\right)^2B^{l-2}\psi ^{1-l+n} \bigg]\, ,\notag
\end{align}
where
\begin{align}
\alpha_{l,m,j} \equiv &-4 (j-l+m) (-1+j+m-n)\, ,\\\notag
\beta_{l,m,j}  \equiv& -2 \big[1-4 j^2-5 l+4 m+D (-1+l-n)+4(l-m) (m-n)+n\\
&+4 j (1+l-2 m+n)\big]\, ,\\
\gamma_{l,m,j}  \equiv&+(-1+D-2 j+2 l-2 m) (-3+D+2 j+2m-2 n)\, ,\\
\sigma_{l,m,j}  \equiv&+2\left[2 j^2-j (1+2 l-2 m+n)+l (1-m+n)\right]\, ,\\
\zeta_{l,m,j}  \equiv&-\left[4 j^2-l (-3+D+2 m-2 n)-2 j (1+2 l-2 m+n)\right]\, ,\\
\omega_{l,m,j}  \equiv&- j (j-l)\, .
\end{align}

Finally, this can be recast as follows,
\begin{align}
\tilde{\mathfrak{R}}_{n+1}\Big|_{f}=\sum_{l=0}^{n+1} \left[\Gamma_{l}B^l\psi ^{1-l+n}+\Upsilon_{l}rB' B^{l-1}\psi ^{1-l+n}+\Omega_{l}r^2\left(B'\right)^2B^{l-2}\psi ^{1-l+n}\right]\, ,
\end{align}
where
\begin{align}
\Gamma_{l}&\equiv \sum_{m=1}^{n}\sum_{j=0}^{n+1-m}\Bigg[\theta(l-2-j)\theta(j+m-l+2)\tilde{C}_{m,l-2-j,j} \alpha_{l-2,m,j}\\
&\quad +\theta(l-1-j)\theta(j+m-l+1)\tilde{C}_{m,l-1-j,j} \beta_{l-1,m,j} +\theta(l-j)\theta(j+m-l)\tilde{C}_{m,l-j,j} \gamma_{l,m,j} \Bigg]\, ,\\
\notag
\Upsilon_{l}&\equiv \sum_{m=1}^{n}\sum_{j=0}^{n+1-m}\Bigg[\theta(l-1-j)\theta(j+m-l+1)\tilde{C}_{m,l-1-j,j} \sigma_{l-1,m,j}\\
 & \quad +\theta(l-j)\theta(j+m-l)\tilde{C}_{m,l-j,j} \zeta_{l,m,j} \Bigg]\, ,\\ 
\Omega_{l}&\equiv \sum_{m=1}^{n}\sum_{j=0}^{n+1-m}\theta(l-j)\theta(j+m-l)\tilde{C}_{m,l-j,j} \omega_{l,m,j} \, .\notag
\end{align}
Now, in order for this to be a GQT gravity we must have 
\begin{align}
\tilde{\mathfrak{R}}_{n+1}\Big|_{f}&=\pazocal{S}^{(k)}_{n+1}=\sum_{l=0}^{n+1}\alpha^{(k)}_{n+1,l}  \pazocal{S}_{(n+1,l)}\\
&=\sum_{l=0}^{n+1}\alpha^{(k)}_{n+1,l} B^{l-1} \psi^{n-l} \left(l r \psi B'+B \psi (D+2 l-2 n-3)-2
   B^2 (l-n-1)\right)\\
   &=\sum_{l=0}^{n+1} \bigg[ B^{l} \psi^{n-l+1}\left(\alpha^{(k)}_{n+1,l}(D+2 l-2 n-3)-\alpha^{(k)}_{n+1,l-1}2(l-n-2)\right)\\
   &\quad \quad \quad +\alpha^{(k)}_{n+1,l}l r B'B^{l-1} \psi^{n-l+1} \bigg]\, ,
\end{align}
for some coefficients $\alpha^{(k)}_{n+1,l}$. Therefore, we have the equations
\begin{align}\label{rec1}
\Gamma_{l}=&\alpha^{(k)}_{n+1,l}(D+2 l-2 n-3)-\alpha^{(k)}_{n+1,l-1}2(l-n-2)\, , \\
\label{rec2}
\Upsilon_{l}=&l\alpha^{(k)}_{n+1,l}\, ,\\
\label{rec3}
\Omega_{l}=&0\, ,
\end{align}
for $l=0,\ldots , n+1$. In addition, the coefficients $\alpha^{(k)}_{n+1,l}$ should satisfy the constraints 
\begin{equation}\label{alphaconstrv3}
\sum_{l=0}^{n+1} \alpha^{(k)}_{n+1,l} (2n+2 - Dl)=0\, ,\quad \quad \sum_{l=0}^{n+1} \alpha^{(k)}_{n+1,l} l(l-1)=0\, ,
\end{equation}
but note that these must arise as consistency conditions in order for the system of equations to have solutions.
Then, the question is whether the system of equations for the tensor $\tilde C_{m,i,j}$ given by eqs.~\eqref{Cconstr1},  \eqref{Cconstr2}, \eqref{rec1}, \eqref{rec2}, \eqref{rec3} has solutions for \emph{any} value of the $\alpha^{(k)}_{n+1,l}$ satisfying the constraints \eqref{alphaconstrv3}. If that is the case, then we have proven the existence of $n$ different GQT densities at order $n+1$ which, as we saw earlier, is the maximum possible number of GQT densities at that order. 

The number of equations to be solved for fixed $n$ ---namely, the number of equations required for establishing the existence of $n$ densities of order $(n+1)$--- and the number of unknowns ($\tilde C_{m,i,j}$)  read, respectively
\begin{equation}
\# \text{ equations} = \frac{12+n(11+3n)}{2}\, , \quad \# \text{ unknowns} =\frac{n(n+2)(n+7)}{6}\, .
\end{equation}
The former is greater than the latter as long as $n<5.10421$ and smaller for greater values of $n$. Observe that while the number of equations grows as $\sim n^2$, the number of unknowns grows as $\sim n^3$. Here, the number of unknowns is the number of constants available to be fixed in order for the GQT conditions to be satisfied, and so having more unknowns than equations means that we have more than enough freedom to impose all the conditions. Hence, as long as we are able to show that the $(n-1)$ different classes of GQT exist for $n\leq 6$ using other methods, this result shows that they will generally exist for $n > 6$.  In practice, solving this system of equations explicitly for any $n\geq 6$ and $D$ is challenging, 
Nevertheless, the resolution for explicit values of $n$ and $D$ is straightforward with the help of a computer algebra system. Doing this, we have checked that there is a solution for any consistent value of the $\alpha^{(k)}_{n+1,l}$ in any $D$ as long as $n\ge 6$.\footnote{In practice, we have checked this explicitly for $n=6$ and general $D$ and for $n=7,\dots, 20$ in $D=5,6,7$. } 

In sum, our results here imply that, if $(n-1)$ inequivalent GQT gravities exist for $n=1,\ldots, 6$, then, $(n-1)$ inequivalent densities will exist for every order $n\geq 6$. In app.~\ref{densiii} we have provided explicit examples of all the inequivalent classes of GQT gravities up to $n=6$ for $D=5,6$, so  this proves that there are $(n-1)$ inequivalent GQT gravities at every order $n\ge 2$ in those cases. The construction of explicit $n\leq 6$ densities of all the different classes for other values of $D$ can be analogously performed (although it requires some non-trivial computational effort in each case) so we are highly confident that our results apply for general $D\geq 7$ as well.

On the other hand, note that our argument here does not work in $D=4$. Indeed, we have found no evidence for the existence of additional inequivalent GQT densities (besides the one known prior to this paper \cite{Hennigar:2016gkm,Bueno:2016lrh,Bueno:2017qce,Bueno:2019ycr}) up to order $6$ in that case. This strongly suggests that in $D=4$ there is a single type of GQT gravity at every curvature order although a rigorous proof of this fact would require some additional work.

\section{Generalities of black holes thermodynamics}\label{GQTGthermo}
In this section we study thermodynamic aspects of GQT gravities in an as general as possible fashion. First we show that the first law of black hole mechanics is satisfied by the black hole solutions of general GQT gravities. Then, we will show that thermodynamic magnitudes of at least one class of genuine GQT gravities can be, similarly to the Lovelock and QT cases, expressed in terms of the characteristic polynomial which embeds maximally symmetric backgrounds in the theory and the on-shell Lagrangian. 

\subsection{The first law for GQT gravities}

Here we wish to understand the first law of thermodynamics for all possible GQT densities. We will begin by working directly with eq.~\eqref{fnj}, without imposing the constraints on the couplings given in eqs.~\eqref{cond1} and \eqref{cond2} at this time.
The integrated field equations of the putative theory can be written in the form
\begin{equation} 
\sum_{n=0}^{n_{\rm max}} \sum_{j=0}^{n} \alpha_{n,j} \pazocal{F}_{(n,j)} = - \frac{8 \pi  \GN M}{\Omega_{D-2}} \, ,
\end{equation}
where the parameter $M$ is the black hole mass \cite{Arnowitt:1960es,Arnowitt:1960zzc,Arnowitt:1961zz,Deser:2002jk}. At a black hole horizon, where $f(\rhor) = 0$, the above equation can be expanded to yield the following constraints:
\begin{align}
M &= \frac{\Omega_{D-2}}{16 \pi \GN} \sum_{n=0}^{n_{\rm max}} \sum_{j=0}^{n} \alpha_{n,j} (j-1) k^{n-j} \rhor^{D-2n - 1} (-2 \pi \rhor T)^j \, ,
\\
0 &= \sum_{n=0}^{n_{\rm max}} \sum_{j=0}^{n} \alpha_{n,j} (D-2n +j-1) k^{n-j} (-2 \pi \rhor T)^j \rhor^{D-2n-2} \, .
\end{align}
where the temperature satisfies $T=f'(\rhor)/(4\pi)$.
The first equation gives the black hole mass in terms of the temperature $T$ and the horizon radius $\rhor$, while the second provides a relationship between $T$ and $\rhor$. 

The other ingredient we need is the black hole entropy, computed with the Wald's formula given in eq.~\eqref{eq:WaldS}. Using the technology introduced in ref.~\cite{Bueno:2019ycr}, this can be computed without knowledge of the covariant form of the Lagrangian. The key insight is that the tensor $P_{ab}{}^{cd}$ in eq.~\eqref{eq:Ptensor} can be computed from the on-shell Lagrangian and must take the form
\begin{equation}\label{Pabcd}
\left.\tensor{P}{_{cd}^{ab}}\right|_{f} = P_1 T^{[a}_{[c} T^{b]}_{d]} +P_2   T^{[a}_{[c} \sigma^{b]}_{d]} +P_3 \sigma^{[a}_{[c} \sigma^{b]}_{d]} \, ,
\end{equation}
where
\begin{equation}
P_1\equiv  -  \frac{\partial {\mathfrak{R}}_{(n)}|_f}{\partial f''} \, ,\quad P_2 \equiv - \frac{r}{D-2} \frac{\partial {\mathfrak{R}}_{(n)}|_f}{\partial f'}  \, , \quad P_3\equiv - \frac{ r^2 }{(D-2)(D-3)} \frac{\partial {\mathfrak{R}}_{(n)}|_f}{\partial f }\, .
\end{equation}
For the case of the static and spherically symmetric black holes considered here, the horizon binormal is given by $\varepsilon_{ab} = 
 2 r_{[a} t_{b]}$ with $r^a$ and $t^b$ the unit spacelike and timelike normal vectors. A calculation then gives
\begin{equation}
S = -4 \pi \Omega_{D-2} \rhor^{D-2} \left.\frac{\partial \pazocal{L}}{\partial f''} \right|_{r = \rhor} = \frac{\Omega_{D-2}}{8 \GN} \sum_{n=0}^{n_{\rm max}} \sum_{j=0}^{n} \alpha_{n,j} j k^{n-j} (-2 \pi \rhor T)^{j-1} \rhor^{D-2n} \, .
\end{equation}
It is then straight-forward to show that the first law of thermodynamics
\begin{equation}
\diff M = T \diff S,
\end{equation}
holds independent of any conditions placed on the couplings $\alpha_{n,j}$. This fact is somewhat surprising because, as discussed earlier, it is only when certain constraints are obeyed by the couplings that a genuine, covariant construction for the Lagrangian can be built based on curvature invariants. However, these same constraints are unnecessary to obtain a valid first law. 

Despite the fact that the coupling constraints are not necessary to obtain a valid first law, it is still possible to understand them from a thermodynamic perspective. For this, the natural starting point is the free energy, which reads
\begin{equation}
F = \frac{\Omega_{D-2}}{16 \pi \GN} \sum_{n, j} \alpha_{n, j} k^{n-j}(-2 \pi T)^j \rhor^{D-1 -2n+j} \, .
\end{equation}
From the free energy, the equation that relates the temperature and horizon radius can be obtained according to
\begin{equation}
\frac{\partial F}{\partial \rhor} = 0 \, ,
\end{equation}
while the mass and entropy can then be verified to follow in the usual way. The constraints on the couplings enforce the following conditions on the free energy:
\begin{eqnarray}
	F-T\frac{\partial F}{\partial T}-\frac{\rhor}{D-1}\frac{\partial F}{\partial \rhor}\Bigg|_{2\pi T \rhor=-k}&=&0\, ,\\
	\frac{\partial^2 F}{\partial T^2}\Bigg|_{2\pi T \rhor =-k}&=&0\, ,
\end{eqnarray}
where it is to be noted that the derivatives here are to be computed without assuming any relationship between $\rhor$ and $T$.

These expressions above, phrasing the coupling constraints in as properties of the free energy, can be reinterpreted as statements about massless hyperbolic black holes. The static black hole with metric function
\begin{equation}
f=-1 + \frac{r^2}{L_{\star}^2} 
\end{equation}
is pure AdS space in a particular slicing. In terms of the parameters we have been using, this corresponds to $k=-1$, $\rhor = L_\star$ and $T = 1/(2 \pi L_\star)$, therefore satisfying the condition $2 \pi T \rhor = - k$. In this language, as we will see explicitly below, the first of the two constraints on the free energy actually ensures that the mass of this black hole vanishes. The second constraint on the free energy does not have as direct of an interpretation in terms of the thermodynamic properties of this black hole, but one could imagine it is a statement about fluctuations.

\subsection{A unified picture of the thermodynamics?}

Lovelock and quasi-topological gravities are, by comparison to alternatives, rather simple extensions of general relativity, especially in the context of static, spherically symmetric black holes. Within our parameterization, the coupling constants $\alpha_{n,j}$ to achieve the on-shell Lagrangian for Lovelock and quasi-topological theories amounts to the choice~\eqref{qt}. For these theories, as has long been known in the case of Lovelock \cite{Boulware:1985wk,Wheeler:1985nh,Wheeler:1985qd}, the field equations for a SSS black hole take the form
\begin{equation}
M = \frac{(D-2)\Omega_{D-2} r^{D-1}}{16 \pi G L^2} h(y) \, , \quad y \equiv \frac{(f -k) L^2}{r^2} \, .
	\end{equation}
The function $h(y)$ appearing here is the same function that determines the vacua of the theory \eqref{eq:embdf}, \ie the field equations for the maximally symmetric solutions of the theory. The fact that the field equations can be written in terms of the embedding function naturally leads to some simple and universal  expressions for black hole thermodynamics:
\begin{align}
M &= \frac{(D-2)\Omega_{D-2} \rhor^{D-1}}{16 \pi G L^2} h(\yhor) \, , \quad \yhor \equiv -\frac{kL^2}{\rhor^2} \, , 
\nonumber\\
S &= - \frac{4 \pi \Omega_{D-2} L^2 \rhor^{D-2}}{D(D-1) } \pazocal{L}'(\yhor) \, .
\end{align}
These relationships are expressed here in their simplest possible forms, but of course can be massaged using the identity~\eqref{eq:embdf} and its derivatives, along with the constraint 
\begin{equation}
0 = (D-1)k h(\yhor) - 2 \yhor (2 \pi \rhor T + k) h'(\yhor) \, ,
\end{equation}
which can be used to isolate for the temperature, if desired. 

It is natural to wonder whether similar relationships hold for the more complicated generalized quasi-topological theories, or whether this result for Lovelock and quasi-topological theories was an artefact of their simplicity. Here we will provide evidence that this is indeed possible, though the situation is more involved than the Lovelock and quasi-topological cases.

Consider the family of theories identified according to the following choices of couplings:
\begin{align}
\alpha_{n, n-j} = \frac{(D-4)^{j-1} n! \left[\left(n - j - 2 \right) D - 4(n-2) \right]}{2^{2j} \,  j! \, (n-j)! \, (n-2)} \alpha_{n,n} \, .
\end{align}
In general dimensions, this corresponds to the family of theories for which an explicit covariant formulation was identified in~\cite{Bueno:2019ycr}. These couplings satisfy the necessary constraints~\eqref{cond1} and~\eqref{cond2}, and in addition define a family of GQT theories for which the free energy can be written as,
\begin{equation}
F = - \frac{(D-2) \Omega_{D-2} \rhor^{D-1}}{16 \pi \GN L^2} h(\xhor) - \frac{4 L^2 \rhor^{D-3}}{D^2(D-1)} \left[(D-2)k + (D-4) \pi \rhor T \right] \pazocal{L}'(\xhor)
\end{equation}
where
\begin{equation}
\xhor  \equiv \frac{8 \pi T L^2}{\rhor D} - \frac{(D-4) k L^2}{\rhor^2 D} \, .
\end{equation}
From this form of the free energy, the full thermodynamic properties for this class of theories can be derived. We obtain for the mass and relationship between the temperature and horizon radius the following two results:
\begin{align}
M =& \, \frac{(D-2) \Omega_{D-2} \rhor^{D-1}}{16 \pi \GN L^2} h(\xhor) - \frac{(D-2) \Omega_{D-2} \rhor^{D-3}}{4 \pi \GN D} (2 \pi \rhor T + k) h'(xhor) 
\nonumber\\
&+ \frac{(D-4) \Omega_{D-2} L^4 \rhor^{D-5}}{D^3(D-1)} (2 \pi \rhor T + k)^2 \pazocal{L}''(\xhor) \, ,
\\
0 =&\, (D-1)(D-2) h(\xhor) - \frac{2 (D-2)^2 L^2}{\rhor^2 D} (2 \pi \rhor T + k) h'(\xhor) 
\nonumber\\
&- \frac{8 (D-4) L^6}{D^3 (D-1) \rhor^4} (2\pi \rhor T + k)^2 \left[ 16 \pi G \pazocal{L}''(\xhor)\right]  \, ,
\end{align}
while the entropy can be simply obtained from the above according to $S = (M-F)/T$.

It is a bit interesting that the thermodynamic properties of black holes can be encoded in terms of the embedding function $h(x)$ and the Lagrangian of the theory $\pazocal{L}(x)$ evaluated on an auxiliary maximally symmetric vacuum spacetime with curvature given by $\xhor/L^2$. There is one case where this result is somewhat natural, and this is the case of massless hyperbolic black holes where $f = -1 + r^2/L_{\star}^2$. Of course, this choice of metric function amounts to a pure AdS space in a particular slicing. One has $k=-1$, $T = 1/(2 \pi L_{\star})$, and $\xhor = L^2/L_{\star}^2$. In this case, the only non-trivial field equation demands that $h(\xhor) = 0$, which in turn demands that $M = 0$.

Next, note that considerable simplification occurs in $D = 4$. In this case, the situation reduces to that first studied in~\cite{Bueno:2017qce}. In that case, the couplings are given by
\begin{equation}
\alpha_{n, n-1} = -\frac{n }{n-2} \alpha_{n, n} \, , \quad \alpha_{2, j} = 0 \quad \forall j \, , \quad \text{and} \quad \alpha_{n, j} = 0 \quad \forall j \neq n, n-1 \, , \forall n \ge 3 \, .
\end{equation} 
The thermodynamic relations in this case simplify to
\begin{align}
M &= \frac{\Omega_{D-2} \rhor^3}{8 \pi \GN L^2} h(\xhor) - \frac{\Omega_{D-2} \rhor}{8 \pi \GN} \left(2 \pi \rhor T + k \right) h'(\xhor) \, , 
\\
S &= \frac{\Omega_{D-2} k \rhor L^2}{6 T} \pazocal{L}'(\xhor) - \frac{\Omega_{D-2} \rhor}{8 \pi \GN T} \left(2 \pi \rhor T + k \right) h'(\xhor) \, ,
\end{align}
and the constraint that determines the temperature in terms of the horizon radius reads
\begin{equation}
0 = \frac{-3 \rhor^2}{L^2} h(\xhor) + \left[2 \pi \rhor T + k \right] h'(\xhor) \, .
\end{equation}

It seems likely that the thermodynamics of each family of GQT gravity can be obtained in this way, though we will leave that full analysis for future work. Nonetheless, we can make a few general remarks, based on the connection with massless hyperbolic black holes. For any given family of GQT theories, the mass must have a term proportional to $h(x)$ followed by a series of terms with powers that vanish for the massless hyperbolic black hole. For example, the simplest possibility would be $(2 \pi \rhor T + k)$ raised to various powers, multiplying derivatives of $h$ and $\pazocal{L}$. Similarly, the entropy must have a term proportional to $\pazocal{L}'(x)$, followed by a series of terms that vanish for the massless hyperbolic black hole, just as above. Lastly, the argument $x$ must be a function of $\rhor$, $T$ and $k$ that limits to $L^2/L_{\star}^2$ for the massless hyperbolic black hole. For example, allowing for a linear dependence on the parameters, the most general option is the one-parameter family
\begin{equation}
\xhor = \frac{2 \pi T L^2 \beta}{\rhor} + \frac{(\upsilon-1) k L^2}{\rhor^2} \, .
\end{equation}
This linear relationship recovers the result for Lovelock/quasi-topological gravity (with $\upsilon = 0$) and the GQT family we have presented above (with $\upsilon = 4/D$). Preliminary calculations have suggested that other GQT families may require a more complicated dependence than this. 

\section{Discussion}

In this chapter, we carried out a structural analysis of GQT gravities, proving that at order $n$ in curvature there exist $n-1$ distinct GQT gravities provided $D > 4$. In the case of $D = 4$, our results strongly suggest that there is a single  (unique up to addition of trivial densities) GQT family corresponding to that identified in~\cite{Bueno:2017qce}. To achieve this, we first derived an upper bound, based on the fact that an on-shell GQT density must be a polynomial in the three independent terms appearing in the Riemann curvature for a SSS background. This upper bound, which holds independent of any knowledge of the covariant form of the densities, was then refined by demanding of the putative theories additional properties that must hold for a true covariant density. Finally, we proved the refined estimate to be exact using arguments based on recurrence formulas, like those introduced in~\cite{Bueno:2019ycr}. In order for our argument to hold, it is required that $n-1$ densities exist for $n=2,3,4,5,6$, which then implies existence for all $n>6$. Such $n-1$ densities for the lowest curvature orders can be constructed explicitly for $D\geq 5$ but not for $D=4$, in which case we have verified that there is always a unique density for every $n=2,\dots,6$. The argument for higher $n$ then fails for $D=4$. While it could in principle be possible that additional inequivalent densities exist in $D=4$ for higher orders ---and our construction involving products of lower-order densities was not general enough to capture them--- we find this possibility highly unlikely.

In addition, we have provided a basic analysis of the thermodynamic properties of black holes in all possible theories, confirming that the first law is satisfied. Perhaps the most interesting result in this direction is the strong evidence that the thermodynamics of black holes in any GQT  gravity may be expressible in terms of the same function that determines the vacua of the theory, just like in Lovelock and QT gravities. Why the thermodynamics of black holes in these theories is encoded in the curvature of some axillary maximally symmetric space remains mysterious to us, and may be worth further investigation. More pragmatically, such closed-form and universal expressions provide a simple means by which the thermodynamics could be studied when an infinite number of higher-curvature corrections are simultaneously included. 

As a by-product, our work has identified $(n-2)$ hitherto unknown families of GQT gravities in $D > 4$. Going forward, it would be interesting to understand how the properties of black hole solutions differ between these different families, or whether there exist universal features, such as occurs in $D = 4$~\cite{Bueno:2017qce}. Moreover, the methods we have used to upper bound the number of distinct theories may generalize to allow for a similar analysis to be carried out when there is non-minimal coupling between gravity and matter fields.

%% file: text/ch3-fieldred.tex
\chapter{Field redefinitions and effective field theory}\label{sec:fredef}

In this chapter we explore some of the effects resulting from redefining the metric tensor on higher-curvature gravities. In  sec.~\ref{osin}, we make some technical comments regarding metric redefinitions involving derivatives of the metric  itself and explain how on-shell actions evaluated on solutions related by metric redefinitions agree with each other. Then, in sec.~\ref{reduci}, we explain how higher-curvature densities involving Ricci curvatures ---or, more generally, densities which become a total derivative when evaluated on Ricc-flat metrics--- can be removed from the gravitational effective action by convenient metric redefinitions.

\section{Field redefinitions in higher-curvature gravity}\label{riccis}

\subsection{On-shell action invariance}\label{osin}
We are interested in determining how the general higher-curvature action \req{eq:generalhdg} transforms under a redefinition of the metric tensor $g_{ab}$ of the form
\begin{equation}\label{eq:metricredef}
g_{ab}=\tilde g_{ab}+\tilde Q_{ab}\, ,
\end{equation}
where $\tilde Q_{ab}$ is a symmetric tensor constructed from the new metric $\tilde g_{ab}$. Ideally, we would like the field redefinition to be algebraic, so that the relation between $g_{ab}$ and $\tilde g_{ab}$ is functional. However, the most general tensor we can build using the metric without introducing higher derivatives is proportional to the metric itself. Hence, $\tilde Q_{ab}$ generically involves curvature tensors, and eq.~\eqref{eq:metricredef} is a differential relation. 
The action $\tilde I$ for the new metric $\tilde g_{ab}$ is simply obtained by substituting eq.~\eqref{eq:metricredef} in the original action, namely
\begin{equation}
\tilde I[\tilde g_{ab}]=I[\tilde g_{ab}+\tilde Q_{ab}]\, .
\end{equation}
Observe that since eq.~\eqref{eq:metricredef} involves derivatives of the metric, extremizing the action with respect to $\tilde g_{ab}$ is, in general, inequivalent from extremizing it with respect to $g_{ab}$.
Whenever $g_{ab}^{\rm sol}$ is a solution of the original theory, the relation \req{eq:metricredef} always produces a solution $\tilde g_{ab}^{\rm sol}$ of the transformed theory when we invert it. However, the converse is not true: there exist solutions of the equations of motion obtained from the variation with respect to $\tilde g_{ab}$ which do not produce a solution of the original theory when we apply the map \req{eq:metricredef}. The reason behind this is the presence of extra derivatives in the field redefinition. This increases the number of derivatives in the equations of motion derived from $\tilde I$, which introduces spurious solutions that need be discarded. This issue is further discussed in app.~\ref{App:2}. Provided it is taken into account, both theories, $I$  and $\tilde I$, are equivalent.

Note that when we keep only the meaningful solutions ---\ie those which are related by eq.~\eqref{eq:metricredef}--- the corresponding on-shell actions match,
\begin{equation}\label{tiss}
\tilde{I}\left[\tilde{g}_{ab}^{\rm sol}\right]=I\left[g_{ab}^{\rm sol}\right]\, .
\end{equation}
Since, \eg black hole thermodynamics can be determined ---in the Euclidean path-integral approach \cite{Gibbons:1976ue}--- by evaluating the on-shell action, this simple observation proves that black hole thermodynamics can be equivalently computed in both frames. 
The same conclusion can be reached \cite{Jacobson:1993vj} using Wald's formula \cite{Wald:1993nt} ---see  refs. \cite{Exirifard:2006wa,Sen:2007qy,Liu:2008kt,Castro:2013pqa,Mozaffar:2016hmg} for additional discussions regarding this issue.\footnote{In order to prove this statement rigorously, it is necessary to assume some mild conditions on $\tilde Q_{ab}$, namely, its fall-off at infinity should be fast enough. All redefinitions  we will consider are well-behaved in this sense.}

We are particularly interested in situations in which both $g_{ab}^{\rm sol}$ and $\tilde g_{ab}^{\rm sol}$ represent static and spherically symmetric black holes. As argued in ref. \cite{Jacobson:1993vj}, field redefinitions given by eq. \req{eq:metricredef} preserve both the asymptotic and horizon structures of $g_{ab}^{\rm sol}$, so they map black holes into black holes. Particularizing even more, from the following section on, we will consider higher-derivative theories controlled by small parameters and perturbative field redefinitions weighted by them.  Ultimately, one of the reasons for considering redefinitions mapping generic higher-derivative theories to GQT terms is the fact that the equations of motion of the latter on static and spherically symmetric configurations become particularly simple and universal. In this particular setup, eq. \req{tiss} will relate the on-shell action corresponding to a certain generalization of the Schwarzschild-(A)dS black hole (continuously connected to it) for a given higher-derivative theory at leading order in the corresponding coupling to the on-shell action of the black hole solution corresponding to the transformed GQT. An explicit example of this match between on-shell actions in ref. \cite{Bueno:2019ltp}.

\subsection{Ricci curvatures and reducible densities}\label{reduci}
Let us now determine how the redefinition \req{eq:metricredef} changes the action \req{eq:generalhdg}. For that, we assume the redefinition to be perturbative, \ie we treat $\tilde Q_{ab}$ as a perturbation and we work at linear order. This is enough for our purposes, since, following the EFT approach, we will also expand the action in a perturbative series of higher-derivative terms. Observe that in this case the relation \req{eq:metricredef} can be inverted as
\begin{equation}\label{eq:metricredefinv}
\tilde g_{ab}=g_{ab}-Q_{ab}+\pazocal{O}(Q^2)\, ,
\end{equation}
where $Q_{ab}$ has the same expression as $\tilde Q_{ab}$ but replacing $\tilde g_{ab}\rightarrow g_{ab}$. 
Let us introduce the equations of motion of the original theory as
\begin{equation}
\E_{ab}=\frac{1}{\sqrt{-g}}\frac{\delta I}{\delta g^{ab}}\, .
\end{equation}
Then, at linear order in $\tilde Q_{ab}$, the transformed action reads
\begin{equation}\label{eq:tildeaction}
\tilde I=\int \diff^Dx\sqrt{-\tilde g}\left[\tilde{\pazocal{L}}-\tilde{\E}_{ab}\tilde Q^{ab}+\pazocal{O}(Q^2)\right]\, .
\end{equation}
where the tildes denote evaluation on $\tilde g_{ab}$. Thus, the redefinition introduces a term in the action proportional to the equations of motion of the original theory 
. Let us be more explicit about the form of the Lagrangian by expanding it as a sum over all possible higher-derivative terms
\begin{equation}\label{eq:faction}
I=\frac{1}{16\pi \GN}\int \diff^Dx\sqrt{-g}\left[ R+\sum_{n=2} L^{2(n-1)}\pazocal{L}^{(n)}\right]\, ,
\end{equation}
where $\pazocal{L}^{(n)}$ represents the most general Lagrangian of order $n$, \ie involving $2n$ derivatives of the metric. The explicit form of the invariants at orders $n=2$ and $n=3$ can be found below in eqs. \req{s4} and \req{s6} respectively. 
The number of terms grows very rapidly, and the $n=4$ Lagrangian already contains 92 terms \cite{Fulling:1992vm}.\footnote{Ref.~\cite{Fulling:1992vm} provides the number of \emph{linearly independent} invariants, but many of them differ by total derivative terms, which are irrelevant for the action. The number of relevant terms is, in general, much smaller ---yet quite large. For instance, besides the 3 quadratic densities and the 10 cubic densities which we include in eqs. \req{s4} and \req{s6}, \cite{Fulling:1992vm} adds $\nabla_a \nabla^a R$ to the former list, and 7 more terms of the form: $\nabla_a\nabla^a\nabla_b\nabla^b R$, $R\nabla_a \nabla^a R$, $\nabla^{a}\nabla^b R R_{ab}$, $R^{ab}\nabla_c \nabla^c R_{ab}$, $\nabla^a\nabla^b R^{cd}R_{cadb}$, $\nabla^a R^{bc}\nabla_{c}R_{ba}$, $\nabla^aR^{bcde}\nabla_a R_{bcde}$ to the latter. All these terms are either total derivatives or can be written in terms of the others plus total derivatives, so they can be discarded ---see \eg \cite{Oliva:2010zd,Myers:2010ru}. }

Let $\tilde Q_{ab}^{(k)}$ be a symmetric tensor containing $2k$ derivatives of the metric. Performing the field redefinition
\begin{equation}\label{eq:metricredefk}
g_{ab}=\tilde g_{ab}+L ^{2k}\tilde Q_{ab}^{(k)}\, , 
\end{equation}
the transformed action \req{eq:tildeaction} reads
\begin{align}\label{eq:faction2}
\tilde I=&\frac{1}{{16\pi \GN}}\int\diff^Dx\sqrt{-\tilde g} \left[\tilde R+\sum_{n=2}^{k} L^{2(n-1)}\tilde{\pazocal{L}}^{(n)}+ L^{2k}\left(\tilde{\pazocal{L}}^{(k+1)}-\tilde{R}^{ab}\hat Q_{ab}^{(k)}\right)\right.\notag\\
&\left.+\sum_{n=k+2}  L^{2(n-1)}\tilde{\pazocal{L}}'^{(n)}\right],
\end{align}
where all quantities are evaluated on $\tilde g_{ab}$, and\footnote{We have $\E^{ab}\tilde Q_{ab}^{(k)}=(R^{ab}-\frac{1}{2}g^{ab}R)\tilde Q_{ab}^{(k)}=R^{ab}\hat Q_{ab}^{(k)}$. 
}
\begin{equation}\label{qk}
\hat Q_{ab}^{(k)}=\tilde Q_{ab}^{(k)}-\frac{1}{2}\tilde g_{ab}\tilde Q^{(k)}\, ,\quad  \tilde Q^{(k)}=\tilde g^{ab}\tilde Q_{ab}^{(k)}\, .
\end{equation}
Hence, all terms containing up to $2k$ derivatives of the metric remain unaffected, while those with $2(k+1)$ derivatives receive a correction of the form $-\tilde{R}^{ab}\hat Q_{ab}^{(k)}$. The higher-order terms also get corrections which depend in a more complicated way on $\tilde Q_{ab}^{(k)}$. If the starting action already contained all possible terms, the net effect of these corrections is just to change the couplings in the Lagrangian. We denote these modified terms as $\tilde{\pazocal{L}}'^{(n)}$.

From this, it is clear that performing this type of field redefinitions order by order, starting at $k=1$, we can remove all terms in the action which involve contractions of the Ricci tensor ---except, of course, the Einstein-Hilbert term. At each order, it suffices to choose $\tilde Q_{ab}^{(k)}$ in eq. \req{eq:metricredefk} such that $\hat Q_{ab}^{(k)}$  equals  the tensorial structure which appears contracted with $R^{ab}$ in the corresponding density.
In other words, any term containing Ricci curvatures is meaningless from the EFT point of view,  and we are free to add or remove terms of that type.  From a different perspective, it has been argued ---\eg in ref. \cite{Endlich:2017tqa}--- that if some higher-curvature correction controlled by $L^{2k}$ involves operators which vanish on the equations of motion produced by the lower-order action, the relevant physics is not affected at $\pazocal{O}(L^{2k})$, and we can just ignore it. For the gravitational effective action, this is equivalent to the possibility of removing all terms involving Ricci curvatures.

Observe that in eq.~\req{eq:faction} we did not include a cosmological constant. When we add it, the effect of the redefinition \req{eq:metricredefk} is 
\begin{equation}\label{eq:factioncosmo}
\begin{aligned}
\tilde I=&\int \frac{\diff^Dx\sqrt{-\tilde g}}{16\pi \GN} \left[-2\Lambda+\tilde R+\sum_{n=2}^{k-1} L^{2(n-1)}\tilde{\pazocal{L}}^{(n)}+ L^{2(k-1)}\left(\tilde{\pazocal{L}}^{(k)}+\frac{2(\Lambda L^2)}{(D-2)}\hat Q^{(k)}\right)\right.\\
&\left.+L^{2k}\left(\tilde{\pazocal{L}}^{(k+1)}-\tilde{R}^{ab}\hat Q_{ab}^{(k)}\right)+\sum_{n=k+2} L^{2n}\tilde{\pazocal{L}}'^{(n)}\right]\, .
\end{aligned}
\end{equation}
Namely, not only the terms involving $2(k+1)$ derivatives of the metric get modified, those involving $2k$ derivatives also receive a correction.
This is a complication with respect to the case without cosmological constant. 
 If we remove terms involving Ricci curvatures at a given order, the field redefinition of the following order will introduce a correction of the form $\frac{2(\Lambda L^2)}{(D-2)}\hat Q^{(k)}$ which will generically include again terms involving Ricci curvatures. Hence, the process cannot be carried out order-by-order because all steps are coupled. If one wants to remove all the terms with Ricci curvature up to order $2k$, it is necessary to consider the most general field redefinition up to that order, \ie including all the terms $\tilde Q_{ab}^{(m)}$ of order $m\le k$ at the same time. Nevertheless, we stress that this is just a technical complication: finding the precise field redefinition that removes the corresponding Ricci curvature terms is more involved, but it can certainly be done. 

Motivated by the above analysis, let us close this section with a definition which will be useful in the remainder of the paper. 
\begin{defi}\label{Def1}
A curvature invariant is said to be ``reducible'' if it is a total derivative when evaluated on any Ricci-flat metric. The rest of them are said to be ``irreducible''.
\end{defi}
Note that this trivially contains  the case in which the invariant vanishes on Ricci-flat metrics. Intuitively, the irreducible terms correspond to those formed purely from contractions of the Riemann tensor, without explicit factors of Ricci curvature.   As we have explained, all reducible terms can be removed or introduced by using field redefinitions, whereas the irreducible ones cannot.
Therefore, the most general higher-derivative gravitational effective action is obtained by including all possible irreducible terms. Then, we are free to add as many reducible terms as we wish: these would simply correspond to different frame choices.

\section{$\pazocal{L}(R_{abcd})$ theories as GQT gravities}\label{alGT}
In the absence of cosmological constant, the gravitational effective action can be written as a series of operators with an increasing number of derivatives of the metric, namely,
\begin{equation}\label{eq:effectiveaction1}
I=\int\frac{\diff^Dx \sqrt{-g}}{16\pi \GN} \left( R+ \sum_{n=2}^{\infty}L^{2n-2}\pazocal{L}^{(n)}\right),
\end{equation}
where $\pazocal{L}^{(n)}$ is the most general Lagrangian involving curvature invariants of order $n$. Let us first focus on the four- and six-derivative actions \eqref{s4} and \eqref{s6}, respectively.
In the case of the four-derivative action, the Riemann-squared term can be traded by the Gauss-Bonnet density \eqref{eq:GB}
so that the most general action reads\footnote{The coefficients $\alpha_i$ are not the same as in eq.~\eqref{s4}, but we prefer not to introduce additional unnecessary notation whenever possible.}
\begin{equation}
\pazocal{L}^{(2)}=\alpha_1R^2+\alpha_2 R_{ab}R^{ab}+\alpha_3 \pazocal{X}_4\, .
\end{equation}
Similarly to the quadratic case, we can trade two of the cubic invariants involving contractions of the Riemann tensor alone by  the cubic Lovelock density $\pazocal{X}_6$, defined in eq.~\eqref{eq:Lovthird}, and one of the cubic Generalized QT densities, $\pazocal{S}_D$, defined in eq.~\eqref{SD}.
Therefore, $\pazocal{L}^{(3)}$ can be alternatively written as
\begin{align}\notag 
\pazocal{L}^{(3)}=&\beta_1\pazocal{X}_{6}+\beta_2 \pazocal{S}_{D}+\beta_3 \tensor{R}{_{a bcd}}\tensor{R}{^{a bc}_{e}}R^{d e}+\beta_4\tensor{R}{_{abcd}}\tensor{R}{^{abcd}}R+\beta_5\tensor{R}{_{abcd}}\tensor{R}{^{ac}}\tensor{R}{^{bd}}\\ \label{sixs}
&+\beta_6R_{a}^{\ b}R_{b}^{\ c}R_{c}^{\ a}+\beta_7R_{ab }R^{ab }R+\beta_8R^3+\beta_9 \nabla_{d}R_{ab} \nabla^{d}R^{ab}+\beta_{10}\nabla_{a}R\nabla^{a}R\, .
\end{align}
Note that in $D\ge 5$, we can alternatively replace either $\pazocal{S}_{D}$ or $\pazocal{X}_6$ by the cubic quasi-topological term $\pazocal{Z}_{D}$ defined in eq.~\eqref{ZD}. Also, in $D=4$ we can replace  $\pazocal{S}_{4}$ by the Einsteinian cubic gravity density \req{ECG} using eq.~\eqref{s4ECG}.
Regardless of these choices, we observe that in addition to the first two terms, belonging to the GQT family, we are left  with a series of reducible terms which, as we have argued in the previous section, can be removed by convenient field redefinitions of the metric.  

The explicit redefinition which removes all terms but $\pazocal{X}_{4}$, $\pazocal{X}_{6}$  and $\pazocal{S}_{D}$ goes as follows.
First, in order to remove the $R^2$ and $R_{ab}R^{ab}$ terms, we perform 
\begin{equation}
g_{ab}=\tilde g_{ab}+\alpha_2L^2 \tilde R_{ab}-\frac{L^2 \tilde R}{D-2}\tilde g_{ab}(2\alpha_1+\alpha_2)\, .
\end{equation}
Then, the Lagrangian of transforms to
\begin{equation}
\pazocal{L}^{(2)}\rightarrow \tilde{\pazocal{L}}^{(2)}=\alpha_3 \tilde{\pazocal{X}}_4\, .
\end{equation}
Now, this redefinition also affects the higher-order terms, but since we are starting from the most general theory, the only effect is to change the coefficients of these terms. In particular, for the six-derivative ones: $\beta_i\rightarrow\tilde\beta_{i}$. Then, the following redefinition of the metric
\begin{align}
&\tilde g_{ab}=\tilde{\tilde{g}}_{ab}+L^4\left[\tilde\beta_3\tensor{\tilde{\tilde R}}{_{a e cd}}\tensor{\tilde{\tilde R}}{_{b}^{e cd}}+\tilde\beta_5 \tilde{\tilde R}^{ef}\tilde{\tilde R}_{ae bf}+\tilde\beta_6 \tensor{\tilde{\tilde R}}{_{a}^{e}}\tilde{\tilde R}_{be}+\tilde\beta_{7}\tilde{\tilde R}\tilde{\tilde R}_{ab}-\tilde\beta_{9}\tilde{\tilde{\nabla}}^2\tilde{\tilde R}_{ab}\right.\\ \notag 
&\left. -\frac{\tilde{\tilde{g}}_{ab}}{D-2}\left(\tensor{\tilde{\tilde R}}{_{ef cd}}\tensor{\tilde{\tilde R}}{^{ef cd}}(\tilde\beta_3+2\tilde\beta_4)+\tilde{\tilde R}_{ef}\tilde{\tilde R}^{ef}(\tilde\beta_5+\tilde\beta_6)+\tilde{\tilde R}^2(\tilde\beta_7+2\tilde\beta_8)-\tilde{\tilde{\nabla}}^2 \tilde{\tilde R}(\tilde\beta_9-2\tilde\beta_{10})\right)
\right]\, ,
\end{align}
leaves the four-derivative terms unaffected, while canceling all six-derivative terms that contain Ricci curvatures,
\begin{equation}
\pazocal{L}^{(3)}\rightarrow \tilde{\tilde{\pazocal{L}}}^{(3)}=\tilde{\beta_1}\tilde{\tilde{\pazocal{X}}}_{6}+\tilde\beta_2\tilde{\tilde{\pazocal{S}}}_{D}.
\end{equation}
Hence, the most general action can be written, after all, as
\begin{equation}\label{eq:effectiveaction11}
\tilde{\tilde I}=\frac{1}{16\pi \GN}\int \diff^Dx \sqrt{-\tilde{\tilde g}}\left[\tilde{\tilde{R}}+L^2 \alpha_3 \tilde{\tilde{\pazocal{X}}}_4+ L^4 \left( \tilde{\beta_1}\tilde{\tilde{\pazocal{X}}}_{6}+\tilde\beta_2\tilde{\tilde{\pazocal{S}}}_{D}\right)+\pazocal{O}(L^6)\right]\, ,
\end{equation}
which only contains GQT terms, as anticipated ---compare with eq.~\eqref{cubiGQT}. 
In $D=4$, the cubic Lovelock density vanishes identically and the Gauss-Bonnet term is topological, which leaves us with 
\begin{equation}
I=\frac{1}{16\pi \GN} \int \diff^4x \sqrt{-g} \left[ R + \beta  L^4 \pazocal{P} + \pazocal{O}(L^6)\right]\, ,
\end{equation}
where we traded $\pazocal{S}_4$ by the ECG density $\pazocal{P}$ using expression \req{s4ECG} and we renamed the gravitational coupling. Hence, Einsteinian cubic gravity  \cite{Bueno:2016xff} is (up to field redefinitions) the most general four-dimensional gravitational effective action we can write including up to six derivatives of the metric.\footnote{This is consistent with the result in ref.~\cite{Endlich:2017tqa}, where $\pazocal{P}$ appears traded by the density $\sim R_{ab}^{cd}R^{ab}_{ef}R^{ef}_{cd}$. That is also the kind of term which appears in the two-loop effective action of perturbative quantum gravity \cite{Goroff:1985th,vandeVen:1991gw}.}

Let us now move on to a more general case, namely, general higher-curvature gravities constructed from arbitrary contractions of the metric and the Riemann tensor. In addition to the notion of ``reducible'' densities introduced in Section \ref{riccis}, it is convenient to  define here another concept:

\begin{defi}\label{Def2} 
We say that a curvature invariant $\pazocal{L}$ is  completable to a Generalized quasi-topological density'' (or just ``completable'' for short), if there exists a GQT density $\pazocal{Q}$ such that $\pazocal{L}-\pazocal{Q}$ is reducible.
\end{defi}
In other words, $\pazocal{L}$ is completable if by adding reducible terms to it, we are able to obtain a GQT term. Note that reducible terms are trivially completable to $0$.  Then, the question whether any higher-derivative gravity can be expressed as a sum of GQT terms is equivalent to the following question:
\emph{Are all irreducible densities completable to a GQT?}
 We have just found that the answer is positive at least up to six-derivative terms. The reason is that there exist more independent GQT densities than irreducible terms, which allowed us to ``complete'' all of them. In the case of the four-derivative terms, the only irreducible density is the Riemann-squared term, and this can be completed to the Gauss-Bonnet density. For the six-derivative terms, we saw that all terms containing derivatives of the Riemann tensor are reducible, and that the only irreducible terms are the two Riemann-cube contributions respectively controlled by $\beta_1$ and $\beta_2$ in \req{s6}. In general dimensions $D$ there are three GQT involving different combinations of these cubic terms, so they can always be completed. 

Observe that the problem of completing irreducible invariants depends on the number of spacetime dimensions. In lower dimensions, many of the densities  are not linearly independent, so the number of irreducible densities is significantly smaller, and this simplifies the problem of completing them to GQTs. As a consequence, on general grounds we expect that if all irreducible invariants are completable for high enough $D$, they will also be completable for smaller $D$. For instance, going back to the six-derivative example, we find that the two cubic densities are independent when $D\ge 6$. In $D=4, 5$ only one of them is linearly independent, and in $D<4$ there is only Ricci curvature so all theories are reducible to Einstein gravity. On the other hand, the number of independent GQTs in $D=4$ is four, whereas in $D>4$ there are only three of them. Therefore, in lower dimensions there are less irreducible terms and more ways to complete them to a GQT theory. The lower the dimension, the easier the task. 

As we will see in a moment, the problem of completing all invariants constructed from an arbitrary contraction of metrics and $n$ Riemann tensors ---a number which grows very rapidly with $n$--- can be drastically simplified. In order to formulate this result, we will need the following somewhat surprising result:

\begin{theorem}[Deser, Ryzhov, 2005 \cite{Deser:2005pc}]\label{th1} 
When evaluated on a general static and spherically symmetric ansatz \req{eq:SSS}, all possible contractions of $n$ Weyl tensors \eqref{eq:Weyl}
are proportional to each other. More precisely, let $(W^n)_i$ be one of the possible independent ways of contracting $n$ Weyl tensors, then for all $i$
\begin{equation}\label{Dser}
(W^n)_i|_{\rm SSS} = F(r)^n c_i  \, ,
\end{equation}
where $c_i$ is some constant which depends on the particular contraction, and $F(r)$ is an $i$-independent  function of $r$ given in terms of the functions appearing in the SSS ansatz \req{eq:SSS}.  In other words, the ratio $[(W^n)_1/(W^n)_2]|_{\rm SSS}$ for any pair of contractions of $n$ Weyl tensors is a constant which does not depend on the radial coordinate $r$.
\end{theorem}

\noindent
\textit{Proof.} 
The Weyl tensor, when evaluated on the SSS ansatz \eqref{eq:SSS}, reads
\begin{equation}\label{fack}
\left. \tensor{ W}{^{ab}_{cd}} \right|_{\rm SSS}=-2\chi(r)\frac{(D-3)}{(D-1)} \tensor{ w}{^{ab}_{cd}}\, ,
\end{equation}
where 
\begin{equation}
\chi=\chi(r)\equiv\frac{(-2+2f-2r f'+r^2f'')}{2r^2}+\frac{N'}{2r N}(-2f +3r f')+\frac{f N''}{N}
\end{equation}
is a function which contains the full dependence on the radial coordinate. On the other hand, $\tensor{ w}{^{ab}_{cd}}$ is a $r$-independent tensorial structure which can be written as \cite{Deser:2005pc}
\begin{equation}\label{ww}
 \tensor{ w}{^{ab}_{cd}}=2\tau^{[a}_{[c} \tau^{b]}_{d]}-\frac{2}{(D-2)} \tau^{[a}_{[c} \sigma^{b]}_{d]}+\frac{2}{(D-2)(D-3)} \sigma^{[a}_{[c} \sigma^{b]}_{d]}\, ,
\end{equation}
where $\tau$ and $\sigma$ are orthogonal projectors defined in section \ref{sec:numbertheo}, satisfying relations \eqref{eq:proj}. Any possible invariant $(W^n)_i$ constructed from the contraction of $n$ Weyl tensors will be given by
\begin{equation}\label{scs}
(W^n)_i|_{\rm SSS} =\left(-2 \chi\frac{(D-3)}{(D-1)} \right)^n (w^n)_i\, ,
\end{equation}
where $ (w^n)_i$ stands for the constant resulting from the  contraction induced on the $w$ tensors, which we can identify with $c_i$ in  \req{Dser}. Therefore, $(W^n)_i|_{\rm SSS}$ takes the form \req{Dser} with $F(r)$ given by the function between brackets. $\, \square$



Now, we are ready to formulate one of the main results of the paper:
\begin{theorem}\label{th2} Let us consider the set of all irreducible curvature invariants of a given order which do not involve covariant derivatives of the curvature. If one of these invariants is completable to a GQT and it does not vanish when evaluated on a static and spherically symmetric ansatz \req{eq:SSS}, then all the invariants are completable.
\end{theorem}

\noindent
\textit{Proof.} 
Let the order of these invariants be $2n$ in derivatives of the metric, \ie $n$ in curvature. Since they are irreducible and they do not contain derivatives of the curvature, they are formed from contractions of a product of $n$ Riemann tensors. We can write schematically $\pazocal{L}_{i}=\left(\text{Riem}^n\right)_i$, where the subscript $i$ denotes again a specific way of contracting the indices. We can consider an alternative basis by replacing the Riemann tensor by the Weyl tensor in the expressions of these densities. Both ways of expressing these invariants are equivalent since they differ by terms containing Ricci curvatures, which are reducible. We denote the densities resulting from replacing $R_{abcd}\rightarrow W_{abcd}$ everywhere in the $\pazocal{L}_i$ by $\tilde{\pazocal{L}}_i=\left(W^n\right)_i$. Next, let us use the hypothesis of Theorem \ref{th2}, which consists in assuming that one of the densities, which we denote $\tilde{\pazocal{L}}_{i_0}$, is completable to a GQT.  As we explained in Section \ref{GQTss}, the condition that determines if a given density belongs to the GQT class exclusively depends on the evaluation of the density on the general (SSS) metric ansatz \req{eq:SSS}, \ie on the way the corresponding density depends on the radial coordinate $r$. But from Theorem \ref{th1} we know that all order-$n$ invariants constructed from contractions of the Weyl tensor are proportional to each other when evaluated on \req{eq:SSS}, in the sense that the dependence on the radial coordinate is identical for all $i$, and given by a fixed function ---which we called $F(r)^n$ in \req{Dser}. Then, since by assumption $\tilde{\pazocal{L}}_{i_0}\big|_{\rm SSS}\neq 0$, all invariants $\tilde{\pazocal{L}}_i$ are proportional to $\tilde{\pazocal{L}}_{i_0}$ when evaluated on SSS metrics. As a consequence, the fact that $\tilde{\pazocal{L}}_{i_0}\big|_{\rm SSS}$ is completable implies that all the rest of densities of order $n$ are, which concludes the proof. $\, \square$

The result can be reformulated as follows:
\begin{corollary} \label{coro1}
Any higher-derivative gravity Lagrangian of the type $\pazocal{L}(R_{abcd})$ can be mapped, order by order, to a sum of GQT terms through a metric redefinition $g_{ab}=\tilde{g}_{ab}+\tilde{Q}_{ab}$ is a symmetric tensor constructed from $\tilde{g}_{ab}$ and its derivatives.
\end{corollary}
Recall that ``irreducible'' means that the density does not vanish on Ricci-flat metrics up to total derivatives whereas ``non-trivial'' means that it does not vanish for SSS metrics. As shown in ref. \cite{Bueno:2019ycr}, GQT gravities exist at every order $n$ in curvature and for all $D$. Before the recipe for the systematic constructions of $n$-th order QT and GQT densities, examples were constructed in a case-to-case basis, contained for some order or dimensionality. For instance, in $D=4$ examples of GQT have been constructed explicitly up to $n=10$ in ref.~\cite{Bueno:2017qce,Arciniega:2018tnn}, where the general form of the equation satisfied by the metric function $f$ in the SSS ansatz \req{eq:SSS} was shown to have a simple dependence on the curvature order $n$. Besides, in that case, the $n>3$ terms were constructed from products of a few $n=2$ and $n=3$ densities, and they already sufficed to produce examples of GQT densities. Many more GQTs could have been constructed had we not restricted the analysis to those building blocks (or even with different combinations of the same densities).
Additional examples of GQTs in $D>4$ and various curvature orders have also appeared in various papers \cite{Dehghani:2011vu,Hennigar:2017ego,Ahmed:2017jod,Cisterna:2017umf} ---\eg cubic and quartic densities up to $D=19$ have been explicitly verified to exist in refs.~\cite{Hennigar:2017ego} and \cite{Ahmed:2017jod}.

Let us also note that the result above shows the existence of a field redefinition that takes the Lagrangian $\pazocal{L}(g^{ab},R_{abcd})$ to a sum of GQTs, but it does not guarantee unicity. Indeed, if at a given order one has several types of non-trivial GQTs ---namely, QT and proper GQT--- it is possible to map the Lagrangian to a sum of terms whose equations for SSS metrics match the ones of the chosen theory (again, QT or proper GQT). More generally, the Lagrangian can be mapped to a combination of those terms. Note that this implies that QT and GQT gravities are related by field redefinitions.\footnote{Imagine, for instance, that we start with a QT density $\pazocal{Z}$ and a GQT density $\pazocal{S}$ of certain order. Replacing all Riemann tensors by Weyl tensors gives rise to new densities $\tilde{\pazocal{Z}}=\pazocal{Z}+\text{RC}_{\pazocal{Z}}$ and $\tilde{\pazocal{S}}=\pazocal{S}+\text{RC}_{\pazocal{S}}$ where RC$_{\pazocal{S},\pazocal{Z}}$ are certain reducible densities involving Ricci curvatures. Now, from Theorem \ref{th1} we know that $\tilde{\pazocal{Z}}|_{\rm SSS}=c \tilde{\pazocal{S}}|_{\rm SSS}$, for some constant $c$. Then, it follows that $\pazocal{Z}= c(\pazocal{S}+\text{RC}_{\pazocal{S}})-\text{RC}_{\pazocal{Z}}+\pazocal{T}$, where $\pazocal{T}$ is a trivial GQT density, \ie one such that $\pazocal{T}|_{\rm SSS}=0$. Naturally, $\pazocal{S}'\equiv c \pazocal{S}+\pazocal{T}$ is another GQT density.  It follows that QT densities can be mapped to GQT densities of the same order ---and viceversa--- via field redefinitions, the mapping generically involving trivial GQT densities (which play no role as far as the equations of SSS metrics are concerned).  }

Before closing this section, let us mention that our conclusions also hold if one includes parity-breaking terms in the effective action, \ie those that involve the Levi-Civita symbol $\epsilon_{a_1\ldots a_D}$. In fact, all such terms vanish for spherically symmetric configurations, hence all of them trivially belong to the GQT family.


\section{Lagrangians including covariant derivatives of the Riemann tensor}\label{covdiv}
In the previous section, we showed that all $\pazocal{L}(R_{abcd})$ gravities can be either removed from the action or written as GQT gravities using field redefinitions. Let us now see what happens with higher-curvature terms involving covariant derivatives of the Riemann tensor. As they have not been used to systematically construct GQT gravities so far, its role is less clear. On the other hand, as we saw in Section \ref{alGT}, up to six-order in derivatives all these terms are actually reducible. However,this is no longer the case at quartic order in curvature. In order to gain some insight about the general behavior of this kind of terms, let us study this order in detail. There exist twenty six independent quartic invariants which do not involve covariant derivatives of the Riemann tensor, namely ---see \eg \cite{Fulling:1992vm,Bueno:2016ypa,Ahmed:2017jod},
\begin{align}
\label{s8}
\pazocal{L}^{(8)}=&\gamma_1 \tensor{R}{^{abcd}}\tensor{R}{_a^e_c^f}\tensor{R}{_e^g_b^h}\tensor{R}{_{fgdh}}+\gamma_2R^{abcd}\tensor{R}{_a^e_c^f} \tensor{R}{_{e}^g_f^h} \tensor{R}{_{bgch}} \\ \notag &+\gamma_3 R^{abcd} \tensor{R}{_{ab}^{ef}}\tensor{R}{_{c}^{g}_{e}^{h}} \tensor{R}{_{dgfh}}  +\gamma_4  R^{abcd} \tensor{R}{_{ab}^{ef}}  \tensor{R}{_{ce}^{gh}} \tensor{R}{_{dfgh}}
+\gamma_5  R^{abcd} \tensor{R}{_{ab}^{ef}}\tensor{R}{_{ef}^{gh}}R_{cdgh}
\\ \notag & +\gamma_{6}R^{abcd} \tensor{R}{_{abc}^{e}}\tensor{R}{_{fghd}}\tensor{R}{^{fgh}_e}  +\gamma_7 (R_{abcd}R^{abcd})^2+\gamma_8 R^{ab}\tensor{R}{^{cdef}}\tensor{R}{_c^g_{ea}}\tensor{R}{_{dgfb}}
\\ \notag & +\gamma_9R^{ab}\tensor{R}{^{cdef}}\tensor{R}{_{cd}^g_a}\tensor{R}{_{efgb}}   +\gamma_{10}R^{ab} \tensor{R}{_a^c_b^d} \tensor{R}{_{efgc}} \tensor{R}{^{efg}_{d}} +\gamma_{11} R \tensor{R}{_a^c_b^d} \tensor{R}{_c^e_d^f}\tensor{R}{_e^a_f^b}
\\ \notag &
+\gamma_{12}R R_{ab}^{cd} R_{cd}^{ef}R_{ef}^{ab}+\gamma_{13} R^{ab}R^{cd}\tensor{R}{^{e}_a^f_c}\tensor{R}{_{ebfd}}  +  \gamma_{14} R^{ab}R^{cd}\tensor{R}{^{e}_a^f_b}\tensor{R}{_{ecfd}}
\\ \notag &+\gamma_{15}R^{ab}R^{cd}\tensor{R}{^{ef}_{ac}}\tensor{R}{_{efbd}}+\gamma_{16}R^{ab}R_b^c \tensor{R}{^{def}_a}\tensor{R}{_{defc}}+\gamma_{17} R_{ef}R^{ef} R_{abcd}R^{abcd}
\\ \notag &+\gamma_{18}R R_{abcd} \tensor{R}{^{abc}_{e}}R^{de}+\gamma_{19}R^2 R_{abcd}R^{abcd}  +\gamma_{20} R^{ab}R_{acbd}R^{ec} R_{e}^d
\\ \notag &+\gamma_{21} R R_{abcd}R^{ac}R^{bd}+ \gamma_{22} R_a^b R_b^cR_c^dR_d^a+ \gamma_{23} \left(R_{ab}R^{ab} \right)^2 +\gamma_{24} R R_a^b R_b^cR_c^a \\ \notag & +\gamma_{25} R^2 R_{ab}R^{ab}+ \gamma_{26} R^4\, .
\end{align}
Of these, at most the first seven are irreducible ---this happens for $D>7$. Now, in ref.~\cite{Ahmed:2017jod} several non-trivial and irreducible GQT theories were constructed using those invariants. Since by virtue of corollary \ref{coro1} we only need one, this immediately implies that the twenty six invariants can always be written as a sum of GQTs using field redefinitions. Hence, just like in the quadratic and cubic cases, all quartic gravities of the form $\pazocal{L}(R_{abcd})$ can be written as GQT densities.

Let us now focus on the terms including covariant derivatves. According to ref.~\cite{Fulling:1992vm}, we find five apparently irreducible terms of that kind, namely 
\begin{eqnarray}
\pazocal{L}_{1}&=&\tensor{R}{^{abcd}}\nabla_{b}\tensor{R}{^{efg}_{a}}\nabla_{d}\tensor{R}{_{efg c}}\, ,\\
\pazocal{L}_{2}&=&\tensor{R}{^{abcd}}\nabla_{c}\tensor{R}{^{efg}_{a}}\nabla_{d}\tensor{R}{_{efgb}}\, ,\\
\pazocal{L}_{3}&=&\tensor{R}{^{abcd}}\nabla^{g}\tensor{R}{^{e}_{a}^{f}_{c}}\nabla_{g}\tensor{R}{_{ebfd}}\, ,\\
\pazocal{L}_{4}&=&\tensor{R}{^{abcd}}\tensor{R}{_{a}^{efg}}\nabla_{d}\nabla_{g}\tensor{R}{_{becf}}\, ,\\
\pazocal{L}_{5}&=&\nabla_{e}\nabla_{f}\tensor{R}{_{abcd}}\nabla^{e}\nabla^{f}\tensor{R}{^{abcd}}\, .
\end{eqnarray}
However, a careful analysis ---using commutation of covariant derivatives, the symmetries of the Riemann tensor and the Bianchi identities\footnote{Recall that these read: $
R_{abcd}+R_{acdb}+R_{adbc}=0$ and  $\nabla_{e} R_{abcd}+\nabla_cR_{abde}+\nabla_d R_{abec}=0$ respectively.}--- reveals that all of them can be decomposed as a sum of total derivative terms plus quartic curvature terms (without covariant derivatives) plus terms with Ricci curvature (hence reducible). This is, they can be expressed as
\begin{equation}\label{eq:Lred}
\pazocal{L}_{i}=\nabla_{a}\pazocal{J}_{(i)}^{a}+\pazocal{Q}_{(i)}+R_{ab}\pazocal{T}_{(i)}^{ab}\, ,
\end{equation}
for certain tensors $\pazocal{J}_{(i)}^{a}$ and $\pazocal{T}_{(i)}^{ab}$ and some quartic density $\pazocal{Q}_{(i)}$.
In order to illustrate this, let us show how $\pazocal{L}_1$ is reduced to an expression of the form \req{eq:Lred}. First, we have
\begin{align}
\pazocal{L}_{1}&=\tensor{R}{^{abcd}}\nabla_{b}\tensor{R}{^{efg}_{a}}\nabla_{d}\tensor{R}{_{efgc}}=\frac{1}{4}\tensor{R}{^{abcd}}\nabla^{g}\tensor{R}{^{ef}_{ab}}\nabla_{g}\tensor{R}{_{efcd}}\\ \notag
&=\frac{1}{4}\nabla_{g}\left(\tensor{R}{^{abcd}}\nabla^{g}\tensor{R}{^{ef}_{ab}}\tensor{R}{_{efcd}}\right)-\frac{1}{4}\tensor{R}{^{abcd}}\nabla^2\tensor{R}{^{ef}_{ab}}\tensor{R}{_{efcd}}-\frac{1}{4}\nabla_{g}\tensor{R}{^{abcd}}\nabla^{g}\tensor{R}{^{ef}_{ab}}\tensor{R}{_{efcd}}\, ,
\end{align}
where in the first equality we applied the differential Bianchi identity twice, and in the second we integrated by parts.  Now we note that the last term in the second line is actually $-\pazocal{L}_{1}$, so we get
\begin{align}
\pazocal{L}_{1}&=\frac{1}{8}\nabla_{g}\left(\tensor{R}{^{abcd}}\nabla^{g}\tensor{R}{^{ef}_{ab}}\tensor{R}{_{efcd}}\right)-\frac{1}{8}\tensor{R}{^{abcd}}\nabla^g\nabla_g\tensor{R}{^{ef}_{ab}}\tensor{R}{_{efcd}}\, .
\end{align}
Then we are done, because the Laplacian of the Riemann tensor decomposes as \cite{andrews2010ricci}
\begin{align}
\nabla^e\nabla_eR_{abcd}=&+2\nabla_{[a|}\nabla_{c}R_{|b] d}+2\nabla_{[b|}\nabla_{d}R_{|a] c}-4\left(\tensor{R}{^p_a^q_b} R_{p[c| q |d]}+ \tensor{R}{^p_a^q_{[c|}} R_{pbq|d]}\right)\\ \notag &+g^{pq}\left(R_{qbcd}R_{pa}+R_{aqcd}R_{pb}\right),
\end{align}
so we can indeed express $\pazocal{L}_{1}$ as in eq.~\eqref{eq:Lred}.
Proceeding similarly with the rest of terms we arrive at the same conclusion. 

Since total derivatives are irrelevant for the action, and since we can remove all terms containing Ricci curvatures by means of field redefinitions, the terms with covariant derivatives of the Riemann tensor only change the coefficients of the quartic terms, which are already present in the action. Hence, from the point of view of effective field theory, these densities are meaningless and can be removed. In addition, we conclude that all eight-derivative terms can be recast as a sum of GQT densities by implementing field redefinitions. 

Let us now extend our argument to a general case. Any higher-derivative gravity can be written as the span of all monomials formed from contractions of $\nabla_a$, $W_{abcd}$ and $R_{ab}$. Such a set can be written schematically as $\mathcal{A}=\cup_{q,n,r\in\mathbb{N}}\mathcal{A}_{q,n,r}$ where $\mathcal{A}_{q,n,r}=\{\nabla^{q}\times W^{n}\times {\rm Ric}^{r}\}$. Out of these subsets, the only ones susceptible of containing irreducible terms are $\mathcal{A}_{q,n,0}$, so the ultimate goal would be to prove that all elements in
\begin{equation}
\mathcal{J}_q=\bigcup_{n\in\mathbb{N}}\mathcal{A}_{q,n,0}
\end{equation}
are completable to a GQT density. First, let us note that these sets can be split according to the partitions of the number of covariant derivatives, $q$,
\begin{equation}
\mathcal{J}_q=\bigcup_{k=1}^{p(q)}\mathcal{J}^{P_k(q)}_q\, ,
\end{equation}
where $p(q)$ is the the number of partitions of $q$ and $P_k(q)$ denotes the $k$-th partition of $q$ (we assume  partitions to be ordered in some way). 
For instance, the first few cases are: $\mathcal{J}_0$, which is the set of monomials formed from general contractions of Weyl tensors; $\mathcal{J}_2$, which is the set of monomials formed from Weyl tensors and two covariant derivatives ---this can be in turn split as the union of $\mathcal{J}_2^{\{1,1\}}$ and $\mathcal{J}_2^{\{2\}}$: in the former set the two covariant derivatives act on two different Weyl tensors, while in the second the two derivatives act on the same Weyl; $\mathcal{J}_4$, which contains terms with four covariant derivatives and an arbitrary number of Weyl tensors ---this can be decomposed as $\mathcal{J}_4=\mathcal{J}_2^{\{1,1,1,1\}}\cup \mathcal{J}_2^{\{2,1,1\}}\cup  \mathcal{J}_2^{\{2,2\}}\cup  \mathcal{J}_2^{\{3,1\}}\cup  \mathcal{J}_2^{\{4\}}$.
Observe that not all subsets are independent. For example, we see that any term belonging to $\mathcal{J}_2^{\{2\}}$ can be written as a sum of terms in $\mathcal{J}_2^{\{1,1\}}$ upon integration by parts. For the same reason, for $q=4$ it is enough to keep the subsets $\mathcal{J}_2^{\{1,1,1,1\}}$, $\mathcal{J}_2^{\{2,1,1\}}$ and $\mathcal{J}_2^{\{2,2\}}$.

We know that all terms in $\mathcal{J}_0$ can be completed to GQTs, and the purpose of the remainder of this section is to show explicitly that all terms in $\mathcal{J}_2$ satisfy the same property.
We expect the trend to go on for all sets $\mathcal{J}_q$ but a general proof seems quite challenging ---not so much a case-by-case partial proof for the following $\mathcal{J}_{q\geq 4}$. 

As we have said, the only subset of $\mathcal{J}_{2}$ which needs to be considered is $\mathcal{J}_2^{\{1,1\}}$. Any term belonging to this subset can be written schematically as
\begin{equation}\label{eq:Inv2}
\mathcal{J}_2^{\{1,1\}}\ni\mathcal{R}_2^{\{1,1\}}=W^n \nabla W\nabla W\, ,
\end{equation}
for some value of $n$. We saw in eq.~\eqref{fack} that, when evaluated on a SSS metric the Weyl tensor has a very simple structure so that any scalar formed from it is proportional to the same quantity. In app.~\ref{apricot} we show that any term of the form \req{eq:Inv2} can be written in turn as
\begin{equation}\label{eq:Inv3}
\left.\mathcal{R}_2^{\{1,1\}} \right|_{\rm SSS}=f\chi^n\left(c_1(\chi')^2+c_2\frac{\chi\chi'}{r}+c_3\frac{\chi^2}{r^2}\right)\, ,
\end{equation}
where $\chi' =\diff \chi/\diff  r$ and $c_{1,2,3}$ are constants. Thus, there are at most three linearly independent terms in $\mathcal{J}_2^{\{1,1\}}$ when one considers SSS metrics. Hence, if we are able to find three independent terms in $\mathcal{J}_2^{\{1,1\}}$ which are completable to a GQT, that will imply that all densities in $\mathcal{J}_2^{\{1,1\}}$ are completable.
Three possible terms of that type are 
\begin{align}
\mathcal{W}^{\{1,1\}}_1=&\sum_{k=0}^n\nabla_{b}\tensor{W}{_{a_1a_2}^{a_3a_4}}\tensor{(W^{n-k})}{_{a_3 a_4}^{a_5 a_6}}\nabla^{b}\tensor{W}{_{a_5 a_6}^{a_7 a_8}} \tensor{(W^{n})}{_{a_7 a_8}^{a_1 a_2}}\, ,\\
\mathcal{W}^{\{1,1\}}_2=&\nabla_{b}\tensor{W}{^{bcd}_{a_1}}\nabla_{c}\tensor{W}{_{da_2}^{a_3a_4}}\tensor{W}{_{a_3a_4}^{a_5a_6}}\ldots \tensor{W}{_{a_{2n+1}a_{2n+2}}^{a_1a_2}}\, ,\\
\mathcal{W}^{\{1,1\}}_3=&\nabla_{b}\tensor{W}{^{bcde}}\nabla_{f}\tensor{W}{^{f}_{cde}}\tensor{W}{_{a_1a_2}^{a_3a_4}}\ldots \tensor{W}{_{a_{2n-1}a_{2n}}^{a_1a_2}} \, ,
\end{align}
where $\tensor{(W^{n})}{_{b c}^{d f}}$ denotes a $n$-Weyl product of the form $\tensor{W}{_{b c}^{a_1 a_2}} \tensor{W}{_{a_1 a_2}^{a_3 a_4}} \ldots \tensor{W}{_{a_{2n} a_{2n+1}}^{d f}}$. 
We can check that when evaluated on a SSS metric the previous terms are linearly independent. For instance, in $D=4$ we obtain the expressions
\begin{align}
\mathcal{W}^{\{1,1\}}_1=&\frac{3^{-n-2} 4 f\left[(-1)^n+2^{n+1}\right] (-\chi )^n \left( (n+1) r^2 \left(\chi '\right)^2+6 \chi ^2\right)}{r^2} \, ,\\
\mathcal{W}^{\{1,1\}}_2=&3^{-n-1} f\left(2^n-(-1)^n\right)   (-\chi )^n \left[\frac{\chi '\chi}{r} +3 \frac{\chi^2}{r^2} \right]\, ,\\
\mathcal{W}^{\{1,1\}}_3=&f \left( \chi '+3 \frac{\chi}{r} \right)^2 (-\chi)^n \frac{(2-(-1)^{n-1}2^{2-n})}{3} \, ,
\end{align}
which are linearly independent for any integer value of $n$. Hence, any term of the form \req{eq:Inv3} can be expressed a sum of these three combinations (the same conclusion holds for arbitrary $D$). Therefore all invariants in $\mathcal{J}_2^{\{1,1\}}$ can be expressed as a linear combination of these terms when evaluated on SSS metrics. This can be alternatively written as 
\begin{equation}\label{eq:Inv2b}
\mathcal{R}_2^{\{1,1\}}=C_1 \mathcal{W}^{\{1,1\}}_1+C_2\mathcal{W}^{\{1,1\}}_2+C_3\mathcal{W}^{\{1,1\}}_3+\ldots \, ,
\end{equation}
where the ellipsis denote terms that vanish on SSS metrics ---which are trivially completable to a GQT. Now, it is easy to check that, by means of field redefinitions, the densities $\mathcal{W}^{\{1,1\}}_{1,2,3}$ are completable. Actually, both $\mathcal{W}^{\{1,1\}}_2$ and $\mathcal{W}^{\{1,1\}}_3$ are reducible because they are proportional to the divergence of Weyl tensor, which depends only on Ricci curvatures.
On the other hand, $\mathcal{W}^{\{1,1\}}_1$ can be written as
\begin{equation}
\begin{aligned}
\mathcal{W}^{\{1,1\}}_1&=\nabla_{b}\left(\nabla^{b}\tensor{W}{_{a_1a_2}^{a_3a_4}}\tensor{W}{_{a_3a_4}^{a_5a_6}}\tensor{W}{_{a_5a_6}^{a_7a_8}}\ldots\tensor{W}{_{a_{2n+3}a_{2n+4}}^{a_1a_2}}\right)\\
&-\nabla^2\tensor{W}{_{a_1a_2}^{a_3a_4}}\tensor{W}{_{a_3a_4}^{a_5a_6}}\tensor{W}{_{a_5a_6}^{a_7a_8}}\ldots\tensor{W}{_{a_{2n+3}a_{2n+4}}^{a_1a_2}}\, .
\end{aligned}
\end{equation}
Since the Laplacian of the Weyl tensor can be expressed as $\nabla^2\text{Weyl}=\nabla\nabla \text{Ricci}+\text{Riem}^2$, we conclude that, by means of field redefinitions, $\mathcal{W}^{\{1,1\}}_1$ can be reduced to a sum of terms without covariant derivatives. We know that those terms are completable, so the densities $\mathcal{W}^{\{1,1\}}_{1,2,3}$ and any other $\mathcal{R}_2^{\{1,1\}}$ are also completable. The result is actually stronger than that: since the densities $\mathcal{W}^{\{1,1\}}_{1,2,3}$ can be completed to a GQT without covariant derivatives of the Riemann tensor, this implies that any other $\mathcal{R}_2^{\{1,1\}}$ can be completed to a GQT density which, when evaluated on a SSS metric, is equivalent to a GQT density without covariant derivatives.

In sum, we have shown that, at least for densities including eight (or less) derivatives of the metric as well as for densities constructed from an arbitrary number of Riemann tensors and two covariant derivatives, all densities can be mapped to GQT gravities. In all cases, those GQT densities become equivalent to GQT densities which do not involve covariant derivatives when evaluated on SSS metrics.  

\section{Discussion}

Our main result is Theorem \ref{th2} and corollary \ref{coro1}, which essentially tells us that densities of the type $\pazocal{L}(R_{abcd})$are completable to a GQT theory.

On the other hand we have seen that, interestingly, densities containing explicit covariant derivatives of the Riemann tensor do not seem to play any role. In fact, we have checked that, up to eighth order, all terms involving derivatives of the Riemann tensor are irrelevant ---they can always be mapped to other terms which already appear in the action. More generally, we have been able to prove that any term with two covariant derivatives can be completed to a GQT density which is equivalent to a GQT gravity of the form $\pazocal{L}(R_{abcd})$ when evaluated on a SSS metric. Note that the last claim is slightly different from stating that the original term can be completed to a GQT theory of the form $\pazocal{L}(R_{abcd})$. It means that the GQT density to which the original density is completed may, in principle, contain covariant derivatives of the curvature, but it is guaranteed that those terms vanish for a SSS metric. We argued that corollary \ref{coro1} may, very likely, extend to densities with an arbitrary number of covariant derivatives, which suggests a stronger conjecture:

\begin{conjecture}\label{cj2} Any higher-derivative gravity Lagrangian can be mapped, order by order, to a sum of GQT terms which, when evaluated on a SSS metric, are equivalent to GQT gravities of the $\pazocal{L}(R_{abcd})$ type.
\end{conjecture}

If true, the second statement in this conjecture implies that we can study SSS black holes of the most general higher-derivative gravity effective action by analyzing only the solutions of the GQT gravity of the form $\pazocal{L}(R_{abcd})$. A concrete example of this analysis was performed in the gravitational sector of the Type II-B String Theory effective action on AdS$_5\times\mathbb{S}^5$ truncated at subleading order in the string tension $\alpha'$ \cite{Gross:1986iv,Grisaru:1986px}. We compared the black hole solutions at leading order in the higher-curvature coupling in both frames and show that their thermodynamic properties match \cite{Bueno:2019ltp}. While, in general, the profile of the solutions will be different in every frame, recall that black hole thermodynamics is invariant under the change of frame. 

The conclusion is that theories of the GQT class are not just toy models with interesting properties. According to our results, they capture, at the very least, a very large part of all possible effective theories of gravity, and very likely ---if conjecture \ref{cj2} is true--- they capture all of them. From this point of view, we could think of  GQT gravities  as the most general EFT expressed in a frame in which the study of SSS black holes is particularly simple and universal.

%% file: text/ch4-hcg3d.tex
\chapter{Higher-curvature gravity in three dimensions}\label{ch:hc3d}

So far we have studied gravity in either four or arbitrary higher dimensions. However, the situation becomes special when we move to three dimensions. Firstly, the Weyl tensor vanishes identically, implying that all curvatures are Ricci curvatures. This means that all solutions of three-dimensional Einstein gravity are locally equivalent to maximally symmetric backgrounds and that no gravitational waves propagate. In spite of this, global differences between spacetimes  do appear and prevent the theory from being ``trivial'', even at the classical level. In particular, in the presence of a cosmological constant, the theory admits black hole solutions \cite{Banados:1992wn,Banados:1992gq}, known as Ba\~nados-Teitelboim-Zanelli (BTZ) black hole. Despite important differences with their higher-dimensional counterparts , they share many of their properties ---including the existence of event and Cauchy horizons, thermodynamic properties, holographic interpretation, etc. See refs.~\cite{Barrow1986three,Carlip:1995qv,Mann:1995eu} for reviews on the subject.

The local equivalence of all classical solutions allows for a characterization of the phase space of the theory \cite{Witten:1988hc}.  In addition ---up to non-negligible details--- three-dimensional Einstein gravity is classically equivalent to a Chern-Simons gauge theory \cite{Achucarro:1986uwr}. From an holographic point of view \cite{Maldacena:1997re,Witten:1998qj,Gubser:1998bc}, these qualitative changes with respect to higher dimensions are manifest in the distinct nature of conformal field theories in two dimensions. In fact, while the observation that the symmetry algebra  of AdS$_3$ spaces is generated  by two copies of the conformal algebra in two dimensions  \cite{Brown:1986nw}   is often considered to be  a precursor of AdS/CFT, the nature of the putative holographic theory ---or ensemble of theories--- dual to pure Einstein gravity is still subject of debate \cite{Maloney:2007ud,Keller:2014xba,Benjamin:2019stq,Alday:2019vdr,Cotler:2020ugk,Maxfield:2020ale}.

The above simplifications also affect higher-curvature modifications of Einstein gravity. In particular, all theories can be constructed exclusively from contractions of the Ricci tensor, which reduces the number of independent densities drastically. Similarly, the usual arguments for considering higher-curvature corrections ---which involve their appearance in the form of infinite towers of terms coming from stringy corrections--- do not make much sense in three-dimensions. This is because all non-Riemann curvatures can be removed via field redefinitions, and hence one is left again with Einstein gravity ---plus cosmological constant and a possible gravitational Chern-Simons term \cite{Gupta:2007th}. However, there is a different reason to consider higher-curvature gravities with non-perturbative couplings in three dimensions. This is the fact that, as opposed to Einstein gravity, they can give rise to non-trivial local dynamics. This  appears in the form of a massive graviton and/or a scalar mode ---see \eg \cite{Gullu:2010sd}.


 By far, the best known higher-curvature modification of Einstein gravity in three dimensions is the so-called ``New Massive Gravity'' (NMG) \cite{Bergshoeff:2009hq}.\footnote{Interestingly, the NMG quadratic density constructed in \cite{Bergshoeff:2009hq} had been identified in the mathematical literature \cite{Gursky} years before the seminal paper appeared. We thank Bayram Tekin for pointing this out to us. } At the linearized level, the theory describes a massive graviton  with the same dynamics of a Fierz-Pauli theory. In addition, the theory is distinguished by possessing second-order traced equations \cite{Oliva:2010zd}, by admitting an holographic c-theorem \cite{Sinha:2010ai} and by admitting a Chern-Simons description \cite{Hohm:2012vh}. Unfortunately, demanding unitarity of the bulk theory spoils the unitarity of the boundary theory and viceversa \cite{Bergshoeff:2009aq}, a problem which has been argued to be unavoidable for general higher-curvature theories sharing the spectrum of NMG \cite{Gullu:2010vw}.
 
Moving from quadratic to higher orders, one can use some of the above criteria to select special theories. One possibility is to demand that the corresponding theories admit an holographic c-theorem \cite{Sinha:2010ai,Paulos:2010ke}. Alternatively, one can look for additional theories which admit a Chern-Simons description \cite{Afshar:2014ffa,Bergshoeff:2014bia,Bergshoeff:2021tbz}. A different route involves considering special $D\rightarrow 3$ limits of higher-dimensional theories with special properties \cite{Alkac:2020zhg}. Often, the densities resulting from these different approaches coincide with each other. Alternative routes are described in refs.~\cite{Banados:2009it,Paulos:2012xe,Bergshoeff:2013xma,Bergshoeff:2014pca,Alkac:2017vgg,Alkac:2018eck,Ozkan:2018cxj}. 
 
While higher-curvature modifications of three-dimensional Einstein gravity have been studied extensively by now, most of the results are only valid for the lowest curvature orders or for particular theories ---see \eg \cite{Paulos:2010ke,Gurses:2011fv,Gurses:2015zia} for exceptions. In this chapter, we present a collection of new results for general-order higher-curvature theories. 

First, we obtain a formula for the exact number of independent order-$n$ densities, $\#(n)$, satisfying the interesting recursive relation $\#(n-6)=\#(n)-n$, which says that the number of order-$n$ densities minus $n$ equals the number of densities of six orders less. Then, we present the equations of motion for a general higher-curvature gravity and the algebraic equations these reduce to when evaluated for Einstein metrics. Furthermore, we obtain the linearized equations of a general higher-curvature gravity around an Einstein spacetime as a function of the effective Newton constantand the masses of the new spin-2 and spin-0 modes generically propagated. After that, we study which theories satisfy an holographic c-theorem and its relation to Born-Infeld gravity Lagrangian of \cite{Gullu:2010pc}. Finally, we discuss the existence of GQT gravities in three dimensions. We show that there exist $\#(n)-n$ theories of that kind, and that all of them are ``trivial'' ---in the sense of making no contribution to the equation of the black hole metric function--- and again proportional to the a sextic density that appears in the study of the theories that admit holographic c-theorem as well. We also provide some comments of the sextic density and the Segre classification of three-dimensional spacetimes.

\section{Counting higher-curvature densities}\label{counting}

In this section, we compute the exact number of independent densities of order $n$ constructed from arbitrary contractions of the Riemann tensor and the metric.  The vanishing of the Weyl tensor in three dimensions reduces the analysis to theories constructed from contractions of the Ricci tensor and the metric,  $\pazocal{L}\left(g_{ab},R_{ab}\right)$. Let us denote the possible contraction of $n$ Ricci tensors as
\begin{equation}
\left(\mathcal{R}_n\right)_a^b\equiv R^b_{i_1}\cdots R_a^{i_n-1},\quad \text{and } \left(\mathcal{R}_n\right)\equiv \left(\mathcal{R}_n\right)_a^a.
\end{equation}
The ``Schouten identities'' drastically reduce the number of independent densities of a given order. These identities take the form \cite{Paulos:2010ke}
\begin{equation}\label{schou}
\delta_{b_1 \dots b_n}^{a_1\dots a_n} R_{a_1}^{b_1}R_{a_2}^{b_2}\cdots R_{a_n}^{b_n}=0\, ,\quad \text{for} \quad n>3\, ,
\end{equation}
and rely on the fact that totally antisymmetric tensors with ranks higher than $3$ vanish identically in $D=3$. From eq. \req{schou}, it follows that the cyclic contraction of $n>3$ Riccis can be written in terms of lower-order densities, and hence the generality of eq. \req{action3}. One finds, for instance
\begin{align}
R_a^bR_b^cR_c^dR_d^a &=  \frac{1}{6} R^4+ \frac{4}{3} R \mathcal{R}_3 + \frac{1}{2} \mathcal{R}_2^2- \mathcal{R}_2 R^2 \, , \\  R_a^bR_b^cR_c^dR_d^e R_e^a&= \frac{1}{6}R^5+\frac{5}{6}\left(\mathcal{R}_3\mathcal{R}_2+\mathcal{R}_3 R^2-\mathcal{R}_2 R^3\right)\, , \\
R_a^bR_b^cR_c^dR_d^e R_e^fR_f^a&= \frac{1}{12}R^6+\mathcal{R}_3\mathcal{R}_2 R+\frac{1}{3}\mathcal{R}_3 R^3-\frac{1}{4}\mathcal{R}_2 R^4 -\frac{3}{4}\mathcal{R}_2^2 R^2+\frac{1}{4}\mathcal{R}_2^3+ \frac{1}{3} \mathcal{R}_3^2\, . 
\end{align}
The existence of these Schouten identities, namely implies that the most general higher-curvature action can be written as \cite{Paulos:2010ke,Gurses:2011fv}
\begin{equation}\label{action3}
I_{(\mathcal{R})}=\frac{1}{16\pi\GN} \int \diff ^3x \sqrt{-g} \pazocal{L}_{(\mathcal{R})}\, , \quad \pazocal{L}_{(\mathcal{R})}\equiv \frac{2}{L^2}+R+\pazocal{F}\left(R,\mathcal{R}_2,\mathcal{R}_3 \right) \, ,
\end{equation}
where we chose a negative cosmological constant. Often we will assume $\pazocal{F}\left(R,\mathcal{R}_2,\mathcal{R}_3 \right) $ to be either an analytic function of its arguments, or a series of the form
\begin{equation}\label{Fseries}
\pazocal{F}\left(R,\mathcal{R}_2,\mathcal{R}_3 \right) =\sum_{i,j,k}L^{2(i+2j+3k-1)}\,  \alpha_{ijk}  R^i \mathcal{R}_2^j \mathcal{R}_3^k\, ,
\end{equation}
for some dimensionless coefficients $\alpha_{ijk}$.

It is often convenient to use a basis of invariants involving the traceless part of the Ricci tensor,
\begin{equation}\label{tracelessR}
 R_{\langle ab \rangle}\equiv R_{ab}-\frac{1}{3}g_{ab} R\, .
\end{equation}
Then, we can define 
\begin{equation}
\mathcal{S}_2\equiv R\indices{_{\langle a}^{b\rangle } }R\indices{_{\langle b}^{a\rangle}}=\mathcal{R}_2-\frac{1}{3}R^2\, , \quad  \mathcal{S}_3\equiv R\indices{_{\langle a}^{b\rangle} }R\indices{_{\langle b}^{c\rangle}} R\indices{_{\langle c}^{a\rangle}}=\mathcal{R}_3-R \mathcal{R}_2+\frac{2}{9}R^3\, ,
\end{equation}
and alternatively write the most general theory replacing $\pazocal{F}\left(R,\mathcal{R}_2,\mathcal{R}_3 \right) $ by $\pazocal{G}\left(R,\mathcal{S}_2,\mathcal{S}_3 \right)$ in eq.~\eqref{action3} \cite{Gurses:2011fv}, namely
\begin{equation}\label{action4}
I_{(\mathcal{S})}=\frac{1}{16\pi\GN} \int \diff ^3x \sqrt{-g} \pazocal{L}_{(\mathcal{S})}\, , \quad \pazocal{L}_{(\mathcal{S})}\equiv \frac{2}{L^2}+R+\pazocal{G}\left(R,\mathcal{S}_2,\mathcal{S}_3 \right) \, .
\end{equation}
We will write the polynomial version of $\pazocal{G}$ as
\begin{equation}\label{Gseries}
\pazocal{G}(R,\mathcal{S}_2,\mathcal{S}_3)=\sum_{i,j,k}L^{2(i+2j+3k-1)}\,  \beta_{ijk}  R^i \mathcal{S}_2^j \mathcal{S}_3^k\, .
\end{equation}
While eq. \eqref{action3}  feels like a  more natural choice from a higher-dimensional perspective, it turns out that many formulas simplify considerably when expressed in terms of $\tilde R_{ab}$ instead. We will try to present most of our results in both bases.

Let us consider the case in which the theory is a power series of the building blocks $R,\mathcal{R}_2,\mathcal{R}_3$ (or, alternatively, $R,\mathcal{S}_2,\mathcal{S}_3$). The order $n$ of a certain combination of scalar invariants is related to the powers of the individual components through $n=i+2j+3k$. One finds the following possible invariants at the first orders,
\begin{align}
R\, , \quad \text{for}\quad &n=1\, ,\\
R^2\, , \quad \mathcal{R}_2\, , \quad \text{for}\quad &n=2\, ,\\
R^3\, , \quad R \mathcal{R}_2\, , \quad \mathcal{R}_3\, ,\quad \text{for}\quad &n=3\, ,\\
R^4\, , \quad R^2 \mathcal{R}_2\, , \quad R \mathcal{R}_3\, , \quad \mathcal{R}_2^2 \, , \quad \text{for}\quad &n=4\, ,\\
R^5\, , \quad R^3 \mathcal{R}_2\, , \quad R^2 \mathcal{R}_3\, , \quad R \mathcal{R}_2^2 \, , \quad \mathcal{R}_2 \mathcal{R}_3\, , \quad \text{for}\quad &n=5\, ,\\
R^6\, , \quad R^4 \mathcal{R}_2\, , \quad R^3 \mathcal{R}_3\, , \quad R^2 \mathcal{R}_2^2 \, , \quad R \mathcal{R}_2 \mathcal{R}_3\, , \quad \mathcal{R}_2^3\, ,\quad \mathcal{R}_3^2\, , \quad \text{for}\quad &n=6\, ,
\end{align}
and so on. Then, the function $\#(n)$ counting the number of invariants of order $n$ takes the values $\#(1)=1$, $\#(2)=2$, $\#(3)=3$, $\#(4)=4$, $\#(5)=5$, $\#(6)=7$. 

In order to find the explicit form of $\#(n)$ as a function of $n$, we can proceed as follows. If we understand the number of elements constructed from powers of $R$ alone up to order $n$ as the coefficients of a power series, we can define the generating function $f^{(R)}(x)$ as 
\begin{equation}
f^{(R)}(x)\equiv \frac{1}{1-x}\sim 1+x+x^2+x^3+\dots\, ,
\end{equation}
\ie such that the right hand side, which is the Maclaurin series of the left hand side, has coefficient $1$ for all powers. This is because at every order $n$ there is a single density we can construct with $R$ alone, namely, $R^n$. Now, if we want to do the same for $\mathcal{R}_2$, we need to take into account that the corresponding coefficients should be $1$ when $n$ is even, and $0$ otherwise. We define then
\begin{equation}
f^{(\mathcal{R}_2)}(x)\equiv  \frac{1}{1-x^2}\sim 1+x^2+x^4+x^6+\dots\, 
\end{equation}
Following the same reasoning for $\mathcal{R}_3$, we define
\begin{equation}\label{eq:GenFR}
f^{(\mathcal{R}_3)}(x)\equiv  \frac{1}{1-x^3}\sim 1+x^3+x^6+x^9+\dots\, 
\end{equation}  
 Now, we can obtain $\#(n)$ as the coefficient of the Maclaurin series corresponding to the generating function which results from the product of the three generating functions previously defined, namely
 \begin{equation}
 f^{(R)}(x) f^{(\mathcal{R}_2)}(x) f^{(\mathcal{R}_3)}(x)=\frac{1}{(1-x)(1-x^2)(1-x^3)}\sim \sum_n \#(n) x^n \, .
 \end{equation}
The result can be written explicitly as
\begin{equation}\label{num1}
\#(n)=\frac{1}{72}\left[47+(-1)^n9+6n(6+n)+16\cos\left(\frac{2n\pi}{3}\right)\right]\, .
\end{equation}
This gives the exact number of independent  three-dimensional order $n$ densities. It is easy to verify that this yields the same values obtained above for the first $n$'s. Note that $\#(n)$ is not an analytic function, but it is still easy to see that it goes as
 \begin{equation}\label{apro}
 \#(n)\sim \frac{n}{2} \left(\frac{n}{6}+1 \right)\, ,
  \end{equation}
  for $n\gg 1$. The fact that $ \#(n)$ scales with $\sim n^2$ for large $n$ had been previously observed in ref. \cite{Paulos:2010ke}. It can be shown that $\#(n)$ can be alternatively written exactly (for integer $n$, which is the relevant case) as
  \begin{equation}\label{num2}
  \#(n) = \ceil[\Big]{\frac{n}{2} \left(\frac{n}{6}+1 \right)+\epsilon}\, ,
  \end{equation}
  where $\ceil[]{x}\equiv {\rm min} \left\{k \in \mathbb{Z}\, |\, k\geq x \right\}$ is the usual ceiling function and $\epsilon$ is any positive number such that $\epsilon \ll 1$. For instance, at order $n=1729$, one has $\#(1729)=249985$ independent densities, as one can easily verify both from eqs.~\req{num1} or \req{num2}. 
  
 The function $\#(n)$ satisfies several relations which connect its values at different orders. A particularly suggestive one is the recursive relation
 \begin{align}\label{properttt}
   \#(n-6) &=  \#(n)-n\, ,
 \end{align}
 which connects the number of densities of a given order with the number of densities  of six orders less. This follows straightforwardly from the general expression of $\#(n)$ in eqref.~\eqref{num2}. We will use this relation in secs. \ref{ctheorem} and \ref{GQTss} to prove a couple of results concerning the general form of densities which trivially satisfy an holographic c-theorem and of densities which belong to the GQT class.



  

\section{Equations of motion and Einstein solutions}\label{eomEs}
The equations of motion of a general higher-curvature theory are simpler than in eq. \eqref{eq:eomgen}. In this case we only have to consider arbitrary contractions of the Ricci tensor and the metric. In consequence, they can be written as \cite{Padmanabhan:2011ex}
\begin{equation}\label{eomsss}
\pazocal{E}_{ab}\equiv P_{a}^{c}R_{b c} -\frac{1}{2}g_{a b}\pazocal{L}-\nabla_{(a}\nabla_c P_{b)}^c+\frac{1}{2}\dal P_{a b}+\frac{1}{2}g_{a b} \nabla_c\nabla_d P^{c d}=0,
\end{equation}
with $ P^{a b}\equiv \left.\frac{\partial \pazocal{L}}{\partial R_{a b}}\right\vert_{g^{c d}}$. In the case of Lagragians of the type~\req{action3}, the explicit form of these equations reads
\begin{equation}\label{eq:EOM2}
\begin{aligned}
\pazocal{E}_{ab}^{(\mathcal{R})}\equiv &+R_{a b}(1+\pazocal{F}_R)-\frac{1}{2}g_{a b}\left(R+\frac{2}{L^2}+\pazocal{F}\right)+\left(g_{a b}\dal - \nabla_{a}\nabla_{b}\right) \pazocal{F}_R\\ &+ 2 \pazocal{F}_{\mathcal{R}_2}  R_{a}^{c}R_{c b}+ 3 \pazocal{F}_{\mathcal{R}_3}R_{a}^{c}R_{c d}R_{b}^{d}+g_{a b}\nabla_{c}\nabla_{d}\left(\pazocal{F}_{\mathcal{R}_2}R^{c d}+\frac{3}{2}\pazocal{F}_{\mathcal{R}_3}R^{c f}R_{f}^{d}\right)\\
&+\dal \left( \pazocal{F}_{\mathcal{R}_2}R_{a b}+\frac{3}{2}\pazocal{F}_{\mathcal{R}_3}R_{a}^{c}R_{c b}\right)-2\nabla_{c}\nabla_{(a}\left(R_{b)}^{c}\pazocal{F}_{\mathcal{R}_2}+\frac{3}{2}R_{b)}^{d}R_{d}^{c}\pazocal{F}_{\mathcal{R}_3}\right)=0 \, ,
\end{aligned}
\end{equation}
In the $R,\mathcal{S}_2,\mathcal{S}_3$ basis, the equations of motion read instead \cite{Gurses:2011fv}
\begin{equation}\label{eq:EOM3}
\begin{aligned}
\pazocal{E}_{ab}^{(\mathcal{S})}\equiv &+\left(\tilde{R}_{a b}+\frac{1}{3}g_{a b}R\right)-\frac{1}{2}g_{a b}\left(R+\frac{2}{L^2}+\pazocal{G}\right)+2 \pazocal{G}_{\mathcal{S}_2}  \tilde{R}_{a}^{c}\tilde{R}_{c b}+ 3 \pazocal{G}_{\mathcal{S}_3}\tilde{R}_{a}^{c}\tilde{R}_{c d}\tilde{R}_{b}^{d}\\ &+\left(g_{a b}\dal - \nabla_{a}\nabla_{b}+\tilde{R}_{a b}+\frac{1}{3}g_{a b}R\right) \left(\pazocal{G}_R-\pazocal{G}_{\mathcal{S}_3}\mathcal{S}_2\right)+g_{a b}\nabla_{c}\nabla_{d}\left(\pazocal{G}_{\mathcal{S}_2}S^{c d}+\frac{3}{2}\pazocal{G}_{\mathcal{S}_3}S^{c f}\tilde{R}_{f}^{d}\right)\\
&+\left(\dal+\frac{2}{3}R\right) \left( \pazocal{G}_{\mathcal{S}_2}\tilde{R}_{a b}+\frac{3}{2}\pazocal{G}_{\mathcal{S}_3}\tilde{R}_{a}^{c}\tilde{R}_{c b}\right)-2\nabla_{c}\nabla_{(a}\left(\tilde{R}_{b)}^{c}\pazocal{G}_{\mathcal{S}_2}+\frac{3}{2}\tilde{R}_{b)}^{d}\tilde{R}_{d}^{c}\pazocal{G}_{\mathcal{S}_3}\right)=0 \, .
\end{aligned}
\end{equation}

Solutions of Einstein gravity plus cosmological constant can be easily embedded in the general higher-curvature theory 
\req{action3} or \req{action4}. These include, for instance, pure AdS$_3$ and the BTZ black hole. Indeed, consider Einstein metrics of the form 
\begin{equation}\label{constantr}
\bar R_{ab}=-\frac{2}{L_{\star}^2} \bar g_{ab}\, .
\end{equation}
In that case, one has
\begin{equation}\label{RR2R3S2S3}
\bar R=-\frac{6}{L_{\star}^2}\, , \quad \bar{\mathcal{R}}_2=\frac{12}{L_{\star}^ 4}\, , \quad   \bar{\mathcal{R}}_3=-\frac{24}{L_{\star}^ 6}\, , \quad \bar{\mathcal{S}}_2=0\, , \quad \bar{\mathcal{S}}_3=0\, .
\end{equation}
Hence, eq. \req{constantr} satisfies the equations of motion \req{eq:EOM2} provided
\begin{equation} \label{sks}
\frac{6}{ L^2}-\frac{6}{L_{\star}^2}\left(1-2\bar{\pazocal{F}}_R+\frac{8}{L_{\star}^2 } \bar{\pazocal{F}}_{\mathcal{R}_2}-\frac{2 4}{L_{\star}^4 } \bar{\pazocal{F}}_{\mathcal{R}_3} \right) +3\bar{\pazocal{F}}=0
\, ,
\end{equation}
is satisfied. In the alternative formulation in terms of traceless Ricci tensors, the analogous equation is considerably simpler and reads \cite{Gurses:2011fv}\begin{equation}\label{vacuS}
\frac{6}{ L^2}-\frac{6}{L_{\star}^2}\left[1- 2 \bar{\pazocal{G}}_R \right] +3\bar{\pazocal{G}}=0\,.
\end{equation}
For Einstein gravity, this simply reduces to $L^2=L_{\star}^2$, which just says that the AdS$_3$ radius coincides with the cosmological constant scale. In general, expressions \req{sks} and \req{vacuS} are equations for $f_\infty$. 
If the series form \eqref{Fseries} is assumed, eq. \eqref{sks} takes the form 
\begin{equation}
1-f_\infty+\sum_{n} a_n  f_\infty^n=0\, ,\quad \text{where}\quad a_n\equiv (-1)^n 6^{n-1} (3-2n) \sum_{j,k} \frac{\alpha_{n-2j-3k,j,k} }{3^{j+2k}} \, .
\end{equation}
Similarly, eq.~\req{vacuS} takes the form
\begin{equation}
1-f_\infty+\sum_{n} b_n  f_\infty^n=0\, ,\quad \text{where}\quad b_n\equiv (-1)^n 6^{n-1} (3-2n) \beta_{n00} \,,
\end{equation}
where observe that terms involving $\mathcal{S}_2$ and $\mathcal{S}_3$ make no contribution to the equation.


On general grounds, the above polynomial equations will possibly have several positive solutions for $\chi_{0}$, so the corresponding theories will possess several AdS$_3$ vacua.  Finding higher-curvature theories with a single vacuum in three and higher dimensions has been subject of study of numerous papers ---see \eg \cite{Crisostomo:2000bb,Gullu:2015cha,Karasu:2016ifk} and references therein. In the present case, a complete analysis of the conditions which lead to a single vacuum can be easily performed in a case-by-case basis, but not so much for a completely general theory, so we will not pursue it here.  Let us nonetheless make a couple of comments. First, observe that all extensions of Einstein gravity with terms involving either $\mathcal{S}_2$ and/or $\mathcal{S}_3$ will have a single vacuum, since for those the Einstein gravity solution $f_\infty=1$ will be the only one.  A different possibility for single-vacuum theories would correspond to an order-$n$ degeneration of the solutions of the above polynomial equations, \ie to the cases in which these become
\begin{equation}
\left(1-\frac{f_\infty}{n}\right)^n=0\, .
\end{equation} 
Observe that this  involves $n-1$ conditions for a theory containing densities of order $n$ and lower and these will necessarily mix couplings of different orders. In particular, for a theory written in the $\{R,\mathcal{S}_2,\mathcal{S}_3 \}$ basis involving densities of order up to $n$, these read
\begin{equation}
\beta_{i00}=\binom{n}{i}\frac{1}{n^i 6^{i-1}(3-2i)}\, , \quad i=2,\dots, n\,.
\end{equation}
Hence, a Lagrangian of the form
\begin{small}
\begin{align}
\pazocal{L}_{(n)}^{\text{s.v.}}&=\frac{2}{L^2}+R+\sum_{i=2}^n \binom{n}{i}\frac{L^{2(i-1)}}{n^i 6^{i-1}(3-2i)}R^i + \mathcal{S}_2 h_2(R,\mathcal{S}_2,\mathcal{S}_3)+ \mathcal{S}_3 h_3(R,\mathcal{S}_2,\mathcal{S}_3)\, ,
\end{align}
\end{small}
where $h_{2,3}$ are any analytic functions of their arguments will have a single AdS$_3$ vacuum.


\section{Linearized equations}\label{seclineq}
The linearized equations of motion around maximally symmetric backgrounds of higher-curvature gravities involving general contractions of the Riemann tensor and the metric  were obtained in refs.~\cite{Bueno:2017qce,Bueno:2016ypa} ---see also refs.~\cite{Tekin:2016vli,Sisman:2011gz}. The resulting expression was expressed in terms of four parameters, $a$, $b$, $c$ and $e$, and a simple method for computing such coefficients for a given theory was also provided, along with the connection between them and the relevant physical parameters ---namely, the effective Newton constant and the masses of the additional modes. In this section we apply this method to a general higher-curvature theory in three dimensions and classify theories according to the content of their linearized spectrum. 

Let $g_{ab}=\bar g_{ab}+h_{ab}$ where the background metric is an Einstein spacetime satisfying eq. \req{constantr} and $h_{ab}\ll 1$, $\forall a,b=0,1,2$. Then, restricted to a general three-dimensional higher-curvature gravity of the form \req{action3}, the equations of motion of the theory read, at leading order in the perturbation
 \cite{Bueno:2016ypa}  
 \begin{align}\label{line}
\frac{1}{32\pi\GN}\pazocal{E}\lnr_{ab}\equiv&\left[e+c\left(\bar{\dal}+\frac{2}{L_{\star}^2}\right)\right]G\lnr_{ab}+(2b+c)\left(\bar{g}_{ab}\bar{\dal}-\bar{\nabla}_a\bar{\nabla}_b\right)R\lnr \notag\\
&-\frac{1}{L_{\star}^2}\left(4b+c\right)\bar{g}_{ab}R\lnr=\frac{1}{4}T_{ab}\lnr\, ,
\end{align}
where we included a putative matter stress-tensor for clarity purposes and where the linearized Einstein and Ricci tensors, and Ricci scalar read
 \begin{IEEEeqnarray}{ll}
G\lnr_{ab}&=R\lnr_{ab}-\frac{1}{2}\bar{g}_{ab}R\lnr+\frac{2}{L_{\star}^2} h_{ab}\, ,\\
R\lnr_{ab}&=\bar{\nabla}_{\left(a\right|}\bar{\nabla}_{c}h\indices{^c_{\left|b\right)}}-\frac{1}{2}\bar{\dal}h_{ab}-\frac{1}{2}\bar{\nabla}_a\bar{\nabla}_b h-\frac{3}{L_{\star}^2} h_{ab}+\frac{1}{L_{\star}^2} h \bar{g}_{ab}\, ,\\
R\lnr &=\bar{\nabla}^a\bar{\nabla}^b h_{ab}-\bar{\dal} h+\frac{2}{L_{\star}^2} h\, .
\end{IEEEeqnarray}
 In higher dimensions there is an additional parameter ---denoted ``$a$'' in ref.~\cite{Bueno:2016ypa}--- appearing in the linearized equations. However, this turns out to be nonzero only for densities which involve Riemann curvatures, and so we have $a=0$ for all three-dimensional theories. For a generic higher-curvature theory in that case, eq.~\eqref{line} describes three propagating degrees of freedom corresponding to a massive ghost-like spin-2 mode plus a spin-0 mode.  The parameters $e$, $c$ and $b$ above can be related to the effective Newton constant $\GN^{\rm eff}$ and the masses (squared) of such modes, which we denote $m_g^2$ and $m_s^2$, as
\begin{equation}\label{phypara}
\GN^{\rm eff}=\frac{1}{32\pi e}\, , \quad m_g^2=-\frac{e}{c}\, , \quad m_s^2=\frac{e+\frac{8}{L_{\star}^2}(3b+c)}{3c+8b}\, .
\end{equation}
In subsection \ref{classifi} we explain how to compute these parameters for a general higher-curvature theory and do this explicitly in our three-dimensional context.

In terms of the physical quantities, the linearized equations read
\begin{equation}
\begin{aligned}\label{linEQ}
\frac{\GN^{\rm eff}}{\GN} m_g^2\cdot \pazocal{E}\lnr_{ab}\equiv &+ \left(m_g^2-\frac{2}{L_{\star}^2}-\bar{\dal}\right)G\lnr_{ab}- \frac{1}{L_{\star}^2}\left(\frac{m_g^2+m_s^2-\frac{2}{L_{\star}^2}}{2(m_s^2-\frac{3}{L_{\star}^2})} \right) \bar g_{ab} R\lnr \\  &+ \left(\frac{m_g^2-m_s^2+\frac{4}{L_{\star}^2}}{4(m_s^2-\frac{3}{L_{\star}^2})} \right) (\bar g_{ab} \bar \dal- \bar\nabla_a\bar\nabla_b)R\lnr=\ell_{\ssc \rm P}^{\rm eff}m_g^2\cdot T\lnr_{ab}\, .
\end{aligned}
\end{equation}

\subsection{Physical modes}\label{physmod}
From what we have said so far, it is not obvious that eq.~\eqref{linEQ} describes the aforementioned modes of masses $m_s$, $m_g$. In order to see this, it is convenient to decompose the metric perturbation as 
\begin{equation}
h_{ab}=\hat{h}_{ab}+\frac{\bar\nabla_{\langle a}\bar\nabla_{b\rangle} h}{\left(m_s^2-\frac{3}{L_{\star}^2}\right)}+\frac{1}{3}\bar{g}_{ab}h\, ,
\end{equation}
where $\langle ab\rangle$ denotes the traceless part, and $\hat{h}_{ab}$ satisfies 
\begin{equation}
\bar{g}^{ab}\hat{h}_{ab}=0\, ,\quad \bar\nabla^{a}\hat h_{ab}=0\, ,
\end{equation}
where the second condition is imposed using gauge freedom. Let us note that this decomposition fails in the special case $m_s^2=\frac{3}{L_{\star}^2}$. In that situation, it is not possible to decouple the trace and traceless parts of $h_{ab}$. However, from eq.~\eqref{phypara}, it follows that in this case $m_g^2=-1/L_\star^2$, so the spin-2 mode is a tachyon. Hence, we will assume that $m_s^2\neq\frac{3}{L_{\star}^2}$ to avoid this problematic situation. 

Then, the trace and the traceless part of the linearized equations become, respectively, \cite{Bueno:2016ypa}
\begin{align}\label{hmode}
\frac{1}{L_{\star}^2} \frac{\left(1+\frac{1}{m_g^2L_{\star}^2}\right)}{\left(m_s^2-\frac{3}{L_{\star}^2}\right)} (\bar \dal - m_s^2)h &= 4\pi\GN^{\rm eff}T^{\rm L}\, , \\ \label{hmode2}
\frac{1}{2m_g^2} \left(\bar \dal +\frac{2}{L_{\star}^2} \right)\left(\bar \dal +\frac{2}{L_{\star}^2}-m_g^2\right) \hat{h}_{ab}&= 8\pi\GN^{\rm eff} T^{\rm{L,eff}}_{\langle ab\rangle}\, ,
\end{align}
where $T^L\equiv \bar g^{ab}T_{ab}^L$ and 
\begin{equation}
T^{\rm{L,eff}}_{\langle ab\rangle}\equiv T^{\rm{L}}_{\langle ab\rangle}-\frac{L_{\star}^2}{2}\frac{\left(\bar \dal +\frac{1}{L_{\star}^2}-m_g^2 \right)}{\left(m_g^2+\frac{1}{L_{\star}^2} \right)}\bar\nabla_{\langle a}\bar\nabla_{b\rangle} T^{\rm L}\, .
\end{equation}
Eq.~\eqref{hmode} describes a spin-0 mode corresponding to the trace of the perturbation. On the other hand, eq.~\eqref{hmode2} can be further rewritten by defining $\hat h_{ab}\equiv \hat h_{ab}^{(m)} +\hat h_{ab}^{(M)}$, where
\begin{equation}
\hat h_{ab}^{(m)}\equiv -\frac{1}{m_g^2}\left[ \bar \dal +\frac{2}{L_{\star}^2}-m_g^2\right] \hat h_{ab}\, , \quad \hat h_{ab}^{(M)}\equiv \frac{1}{m_g^2}\left[ \bar \dal +\frac{2}{L_{\star}^2}\right] \hat h_{ab}\, , 
\end{equation}
as
\begin{align}
-\left(\bar \dal +\frac{2}{L_{\star}^2}\right)\hat h_{ab}^{(m)} &= 8\pi\GN^{\rm eff} T^{\rm{L,eff}}_{\langle ab\rangle} \, , \\ \label{spin2m}
\left(\bar \dal +\frac{2}{L_{\star}^2}-m_g^2\right)\hat h_{ab}^{(M)} &=  8\pi\GN^{\rm eff} T^{\rm{L,eff}}_{\langle ab\rangle}\, .
\end{align}
These describe two traceless spin-2 modes which couple to matter with opposite signs. However, as opposed to higher dimensions, only the massive one is propagating in $D=3$. The would-be massless spin-2 mode is pure gauge (whenever $T^{L}_{ab}=0$) in this number of dimensions  ---see \eg \cite{Deser:1983tn,Nakasone:2009bn,Myung:2011bn,Moynihan:2020ejh}.\footnote{Massless and massive gravitons in $D$ dimensions propagate  $\frac{D(D-3)}{2}$ and $\frac{(D+1)(D-2)}{2}$ degrees of freedom, respectively, which means $0$ and $2$ degrees of freedom respectively for $D=3$.}  Hence, the relevant equations are \req{hmode} and \req{spin2m} which describe a maximum of three degrees of freedom ---one from the scalar mode and two from the spin-2 one--- propagated around Einstein solutions by higher-curvature gravities in the most general case. 

When $\GN^{\rm eff}>0$, the massive graviton is a ghost and the scalar mode has positive energy, but since there is no massless graviton, one could also consider $\GN^{\rm eff}<0$, so that the massive graviton has positive energy and the scalar is a ghost. As we will see below, there are theories that only propagate either the scalar mode or the massive spin-2 mode, and these can be made unitary by taking $\GN^{\rm eff}>0$ or $\GN^{\rm eff}<0$, respectively. An example of the latter is NMG as introduced in \cite{Bergshoeff:2009hq}, in which the Ricci scalar appears with the ``wrong'' sign, hence implying $\GN^{\rm eff}<0$.

\subsection{Identification of physical parameters}
Given a higher-curvature theory, one can linearize its equations and deduce the values of the parameters $b,c,e$ (and consequently $\GN^{\rm eff},m_g^2,m_s^2$) by comparing them with the above general expressions. A much faster way of performing this  identification was proposed in ref.~\cite{Bueno:2016ypa}, which we adapt here to our three-dimensional setup. One starts by replacing all Ricci tensors in the Lagrangian by
\begin{equation}
R^{\rm aux}_{ab}=-\frac{2}{L_{\star}^2}g_{ab}+\alpha (x -1)k_{ab} \, ,
\end{equation}
where $x$ is an arbitrary integer constant and the symmetric tensor $k_{ab}$ is defined such that $k_a^a\equiv x$ and $k_a^b k_b^c=k_a^c$. Then, the parameters can be unambiguously extracted from the general formulas \cite{Bueno:2016ypa}
\begin{equation}
\left. \frac{\partial \pazocal L(R^{\rm aux}_{ab})}{\partial \alpha}\right|_{\alpha=0}=2e\, x(x-1)\, , \quad \left. \frac{\partial^2 \pazocal L(R^{\rm aux}_{ab})}{\partial \alpha^2}\right|_{\alpha=0}=4x(x-1)^2 (c+ b x)\, .
\end{equation}
It is straightforward to do this for our general three-dimensional actions. When the theory is expressed in terms of the traceless Ricci tensor as in eq.~\eqref{action4}, the resulting parameters take a particularly simple form
\begin{equation}\label{GGL}
e=\frac{1}{32\pi\GN} (1+\bar{\pazocal{G}}_R)\, , \quad b=\frac{1}{32\pi\GN} \left(\frac{1}{2}\bar{\pazocal{G}}_{R,R}-\frac{1}{3} \bar{\pazocal{G}}_{\mathcal{S}_2} \right)\, , \quad c=\frac{1}{32\pi\GN} \bar{\pazocal{G}}_{\mathcal{S}_2}\, ,
\end{equation}
where $\pazocal{G}_X\equiv \partial \pazocal{G}/\partial X$, $\pazocal{G}_{X,X}\equiv \partial^2 \pazocal{G}/\partial X^2$ and the bar means that we are evaluating the resulting expressions on the background geometry, which is implemented through expression \req{RR2R3S2S3}. 
If we assume that the Lagrangian allows for a polynomial expansion, it is useful to decompose $\pazocal{G}$ in the following way,
\begin{equation}\label{GS2}
\pazocal{G}(R,\mathcal{S}_2,\mathcal{S}_3)=f(R)+\mathcal{S}_2 g(R)+\pazocal{G}_{\rm triv}\, ,
\end{equation}
where $\pazocal{G}_{\rm triv}\equiv \mathcal{S}_2^2 h(R,\mathcal{S}_2)+\mathcal{S}_3 l(R,\mathcal{S}_2,\mathcal{S}_3)$ includes all terms which do not contribute to the linearized equations around any constant curvature solution. That is the case of any density involving any power of $\mathcal{S}_2$ greater or equal than two and any power of $\mathcal{S}_3$ (different from zero).
With the Lagrangian expressed in this way, the background equation \req{vacuS}, reduces to
\begin{equation}
\frac{6}{L^2}-\frac{6}{L_{\star}^2}(1-2 \bar f_R)+3\bar f=0\, ,
\end{equation}
and using eq. \eqref{phypara} we find the physical quantities of the linearized spectrum, 
\begin{align}\label{ppog}
\GN^{\rm eff}=\frac{\GN}{1+\bar f_R}\, ,\quad
m_g^2=-\frac{1+\bar f_R}{ \bar g}\, ,\quad
m_s^2=\frac{1+\bar f_R+\frac{12}{L_{\star}^2} \bar f_{R,R}}{4\bar f_{R,R}+\frac{1}{3}\bar g}\, .
\end{align}
When expressed explicitly in terms of the gravitational couplings in an expansion of the form (\ref{Gseries}) these read 
\begin{equation}
\begin{aligned}\label{ppog2}
\GN^{\rm eff}=&\frac{\GN}{1+\sum_i\beta_{i00} i(-6 \chi_{0})^{i-1}}\, ,\quad
m_g^2=-\frac{ 1+\sum_i\beta_{i00} i(-6\chi_{0})^{i-1} }{\Ls^2\sum_i\beta_{i10}(-6)^i}\\
m_s^2=&\frac{1-\sum_i\beta_{i00} i (2 i-3)(-6\chi_{0})^{n-1} }{4\Ls^2\sum_i(-6 \chi_{0})^{i-2}  [(i-1) i \beta_{i00}+3\beta_{i10}]}\, .
\end{aligned}
\end{equation}

\begin{normalsize}
Using expressions above, it is easy to classify the different theories according to the presence or absence of the massive graviton and scalar modes in their spectrum. Before doing so, let us present in passing the expressions analogous to eq.~\eqref{ppog} when the analysis is performed for a theory expressed in the 
$\{R,\mathcal{R}_2,\mathcal{R}_3\}$ basis instead. In that case, the equations become more involved and a decomposition of the form \eqref{GS2} is not available. We have
\end{normalsize}
\begin{small}
\begin{align}\label{555}
\GN^{\rm eff}=&\frac{\GN\Ls^4}{\Ls^4(1+\bF{R})-4\Ls^2\bF{\RR}+12\bF{\RRR}}\, , \quad
m_g^2=\frac{\Ls^4(1+\bF{R})-4\Ls^2\bF{\RR}+12\bF{\RRR}}{6\Ls^2\bF{\RRR}-\Ls^4\bF{\RR}}\, , \\
m_s^2=&\frac{3}{\Ls^2}+\left[\Ls^8\left(1+\bF{R}\right)-5\Ls^6\bF{\RR}-18\Ls^4\bF{\RRR}\right]/\left(3\Ls^8\bF{\RR}-18\Ls^6\bF{\RRR}+4\Ls^8\bF{R,R}\right.\notag\\
&\left.-32\Ls^6\bF{R,\RR}+96\Ls^4\bF{R,\RRR}-384\Ls^2\bF{\RR,\RRR}+4\Ls^8\bF{R,R}+64\Ls^4\bF{\RR,\RR}+576\bF{\RRR,\RRR}\right)\, .\notag
\end{align}
\end{small}
The polynomial form is straightforward to obtain from these expressions (and as ugly as one may anticipate).

\subsection{Classification of theories}\label{classifi}
Expressions \eqref{GS2} and \eqref{ppog} allow for a simple classification of all theories depending on the mode content of their linearized spectrum. The three sets of theories we consider here are: theories which are equivalent to Einstein gravity at the linearized level, theories which do not propagate the massive graviton, and theories which do propagate the scalar mode.

\subsubsection{Einstein-like theories}
A first group of densities are those for which $m_g^2,m_s^2\rightarrow \infty$, namely, densities in whose spectrum both the massive graviton and the scalar mode are absent. These are theories which, at the level of the linearized equations, are identical to Einstein gravity ---up to, at most, a change in the effective Newton constant.  As we mentioned earlier, a large set of densities do not contribute whatsoever to the linearized equations. These are given by 
\begin{equation}\label{gtriv}
\pazocal{G}|_{\text{trivial linearized equations}}= \mathcal{S}_2^2 h(R,\mathcal{S}_2)+\mathcal{S}_3 l(R,\mathcal{S}_2,\mathcal{S}_3)\, .
\end{equation}
It is not difficult to see that there are $\#(n)-2$ densities of this kind at order $n$. Namely, all order-$n$ densities but those of the forms $R^n$ and $\mathcal{S}_2 R^{n-2}$ contribute trivially to the linearized equations. While there are no ``trivial'' densities for $n=1,2$, they start to proliferate for $n\geq 3$, becoming the vast majority for higher orders. As it turns out, these ``trivial densities'' are the only Einstein-like theories which exist beyond Einstein gravity itself. The reason is that removing both the massive graviton and the scalar from the spectrum amounts at imposing $c=b=0$, which implies $\bar g=\bar f_{R,R}=0$. These are on-shell conditions, but if we want to avoid relations between densities of different orders, we must force them to hold for any value of $\bar R$. Hence, the conditions become $g(R)=f_{R,R}(R)\equiv 0$, whose only non-trivial solution besides \req{gtriv} is Einstein gravity plus a cosmological constant.  Hence, most higher-curvature densities have in fact trivial linearized equations.

It is a remarkable ---and exclusively three-dimensional--- fact that Einstein gravity is unique in this sense. Observe that starting in four dimensions and for higher $D$ there are generally several Einstein-like densities with non-trivial linearized equations at each curvature order. Examples are Lovelock \cite{lovelock1970divergence,Lovelock:1971yv} and some $f(\text{Lovelock})$ densities \cite{Bueno:2016dol}, Einsteinian cubic gravity \cite{Bueno:2016xff}, QT \cite{Oliva:2010eb,Myers:2010ru,Dehghani:2011vu,Cisterna:2017umf} and GQT gravities \cite{Hennigar:2017ego,Bueno:2017sui,Bueno:2017qce,Bueno:2019ycr}, among others \cite{Li:2017ncu,Karasu:2016ifk,Li:2017txk}.

\subsubsection{Theories without massive graviton}
Theories for which  $m^2_g\rightarrow\infty$ do not propagate the massive graviton. In terms of our parameters $e$, $b$ and $c$, this condition is given by $c=0$.

From eq.~\eqref{ppog} it is clear that this set of theories are those with $\bar g\equiv g(\bar R)=0$. Again, in order to impose this condition at each curvature order we must demand $g(R)\equiv 0$. Hence, the most general (polynomial) density which makes a non-trivial contribution to the linearized equations and which does not propagate the massive graviton in three-dimensions is $f(R)$ gravity
\begin{equation}
\pazocal{G}|_{\text{no massive graviton}}=f(R)\, ,
\end{equation}
Obviously, at  order $n$ there is $1$ such density, corresponding to $R^n$. Of course, one can obtain more complicated densities satisfying the $m^2_g\rightarrow\infty$ condition by combining some of the trivial Einstein-like densities with the $f(R)$ ones. Hence, there are actually $\#(n)-1$ independent densities which do not propagate the massive graviton at order $n$.

For comparison, observe that in $D\geq 4$ there is a large set of higher-curvature theories which do not have the massive graviton in their spectrum. This is the case, in particular, of all $f(\text{Lovelock})$ theories \cite{Bueno:2016dol} ---the set also includes all the Einstein-like theories mentioned in the last paragraph of the previous subsubsection. 

\subsubsection{Theories without scalar mode}
The condition for the scalar mode to be absent from the spectrum, $m_s^2\rightarrow \infty$,  reads instead $3c+8b=0$, which is satisfied by theories for which $12\bar f_{R,R}+\bar g=0$. From this we learn that the most general class theories of this kind contributing non-trivially to the linearized equations reads
\begin{equation}
\pazocal{G}|_{\text{no scalar mode}}=f(R)-12f_{R,R}(R)\mathcal{S}_2\, ,\qquad (f_{R,R}(R)\neq 0)
\end{equation}
Again, there is a single order-$n$ density of this kind, corresponding to
\begin{align}
\pazocal{G}^{(n)}|_{\text{no scalar mode}} &=R^n-12 n (n-1) R^{n-2}\mathcal{S}_2 \, , \\  &= [1+4n(n-1)]R^n-12 n (n-1) R^{n-2}\mathcal{R}_2\, .
\end{align}
For $n=2$, the above density is nothing but the New Massive Gravity one \cite{Bergshoeff:2009hq}.  Once again, we can combine the above order-$n$ densities with the $\#(n)-2$ ``trivial'' densities to obtain additional densities which do not propagate the scalar mode. There are then $\#(n)-1$ densities which do not propagate the scalar mode at each order.

In higher dimensions, a prototypical example of a theory which satisfies this condition  is conformal gravity \cite{Hassan:2013pca,Bueno:2016ypa}, which can be thought of as a natural $D$-dimensional extension of NMG. 

In sum, in $D=3$, at any order  $n\geq 2$ we can always decompose the most general linear combination of higher-curvature densities as a sum of a term which by itself would not propagate the massive graviton, plus a term which by itself would not propagate the scalar mode, plus  $\#(n)-2$ densities which do not contribute to the masses of any of them. 

\section{Theories satisfying an holographic c-theorem}\label{ctheorem}
Interesting extensions of Einstein and NMG to higher orders can be obtained by demanding that the corresponding densities satisfy a simple holographic c-theorem \cite{Sinha:2010ai,Paulos:2010ke}. This set of theories is defined by the property that they yield second-order equations when evaluated on the ansatz \eqref{eq:RGmetric}, together with the considerations described in sec.~\ref{sec:hrgf}.

For theories of the type considered above, it is straightforward to construct an appropriate c-function such  that \cite{Freedman:1999gp,Myers:2010xs,Myers:2010tj} 
\begin{equation}
    c'(r)=-\frac{A^2}{8\GN A'^{2}}\,(T_t^t-T_{r}^{r}),
\end{equation}
where $A'=\frac{\diff A}{\diff r}$. This can be obtained from the Wald-like \cite{Wald:1993nt} formula \cite{Sinha:2010ai,Myers:2010tj}
\begin{equation}
c(r)\equiv \frac{\pi A}{2 A'}\frac{\partial \pazocal{L}}{\partial R^{t r}\,_{tr}}\, ,
\end{equation}
where the Lagrangian derivative components are evaluated on \req{eq:RGmetric}. By construction, $c(r)$ coincides with the Virasoro central charges of the fixed-point theories.

As argued in ref.~\cite{Paulos:2010ke}, demanding second-order equations for the ansatz \req{eq:RGmetric} for a set of order-$n$ densities amounts at imposing $n-1$ conditions. The idea is to consider the on-shell evaluation of the corresponding Lagrangian densities and impose that neither terms involving derivatives of $a(\rho)$ higher than two, nor powers of  $a''(\rho)$ higher than one appear in the resulting expression. This enforces the corresponding equations of motion to be second-order and that a simple c-function can be defined from the above formulas.

As we have shown, there are $\#(n)$ independent densities at order $n$, which means that there are $\#(n)-(n-1)$ independent order-$n$ densities which satisfy a simple holographic c-theorem. Hence, for $n=1,\dots,5$, there is a single such density at each order, but degeneracies start to appear at order six. As observed in ref.~\cite{Paulos:2010ke}, it is always possible to write the corresponding linear combination of order-$n$ densities satisfying a simple holographic c-theorem as a single density which has a non-trivial on-shell action when evaluated on metric \req{eq:RGmetric}, plus densities which simply vanish when evaluated on such ansatz. Hence, we learn that there are $\#(n)-n$ independent  order-$n$ densities which are trivial on the ansatz \req{eq:RGmetric}. Remarkably, as we show below, all such densities of arbitrary orders turn out to be proportional to a single sextic density which identically vanishes on the metric \eqref{eq:RGmetric}. As for the densities which contribute non-trivially to the holographic c-function we find a new recursive formula which allows for the construction of the corresponding order-$n$ density from the order-$(n-1)$, the order-$(n-2)$, the Einstein gravity and the NMG densities. The recurrence can be solved explicitly, and so we are able to provide an explicit formula for a general order density which non-trivially satisfies the holographic c-theorem. Finally, we explore the relation between such general order density and the one resulting from the expansion of previously proposed Born-Infeld gravities which also satisfy the holographic c-theorem. Naturally, the relation always involves densities trivially satisfying the holographic c-theorem.  



\subsection{Recursive formula}



As we have mentioned, at each order there is a single possible functional dependence on $a(\rho)$ of the on-shell action of theories satisfying the holographic c-theorem. Then, up to terms which do not contribute when evaluated on ansatz \req{eq:RGmetric}, there is a unique  such density at each curvature order. The on-shell expressions for $R,\mathcal{S}_2,\mathcal{S}_3$ read
\begin{align}
   \left. R\right|_a  
    = -\frac{2(a'^2 +2 a a'') }{ a^2} \, , \quad
  \left.  \mathcal{S}_2\right|_a  
    = \frac{2(a'^2 - a a'')^2}{3 a^4} \, , \quad
  \left.  \mathcal{S}_3\right|_a  
    = \frac{2(a'^2 - a a'')^3}{9 a^6}\, .
\end{align}
As observed in ref.~\cite{Paulos:2010ke}, the on-shell Lagrangian of densities satisfying the holographic c-theorem in a non-trivial fashion follows the simple pattern 
\begin{equation}\label{eq:Lnonshell}
\left. \pazocal{C}_{(n)}\right|_{a}=\left(\frac{a'}{a}\right)^{2(n-1)}\left[\frac{a''}{a}+\frac{3-2n}{2n}\left(\frac{a'}{a}\right)^{2}\right]\, .
\end{equation}
With this choice of normalization, the first three densities read
\begin{align}
\pazocal{C}_{(1)}&=-\frac{1}{4}R\, ,\\
\pazocal{C}_{(2)}&=+\frac{3 R^2}{16}-\frac{\mathcal{R}_2}{2}\\ &=+\frac{R^2}{48}-\frac{\mathcal{S}_2}{2}\, ,\\
\pazocal{C}_{(3)}&=-\frac{17 R^3}{48}+\frac{3 R \mathcal{R}_2}{2}-\frac{4 \mathcal{R}_3}{3}\\ &=-\frac{R^3}{432}+\frac{R \mathcal{S}_2}{6}-\frac{4 \mathcal{S}_3}{3}\, .
\end{align}
Now, an easy way to prove that instances of non-trivial densities actually exist at arbitrarily high orders is by finding a recursive relation.
Since, essentially, these densities are defined by the form of their on-shell Lagrangian on the RG-flow metric \eqref{eq:RGmetric}, we can try to derive such recursive relations by using eq.~ \req{eq:Lnonshell}. We find the particularly simple relation,
\begin{equation}\label{recuu}
\pazocal{C}_{(n)}=\frac{4(n-1)(n-2)}{3n(n-3)}\left(\pazocal{C}_{(n-1)}\pazocal{C}_{(1)}-\pazocal{C}_{(n-2)}\pazocal{C}_{(2)}\right)\, .
\end{equation} 
This expression allows us to generate holographic $c$-theorem satisfying densities of arbitrary orders once we know $\pazocal{C}_{(1)}$, $\pazocal{C}_{(2)}$ and $\pazocal{C}_{(3)}$, which are given above. Since $\pazocal{C}_{(4)}$ and $\pazocal{C}_{(5)}$ are unique, this formula should give precisely those densities. This is indeed the case, and one finds
\begin{align}
    \pazocal{C}_{(4)} & =+ \frac{41R^4}{384} - \frac{3R^2\mathcal{R}_2}{8} + \frac{2R\mathcal{R}_3}{3} - \frac{\mathcal{R}_2^2}{2}
    \\ &=+ \frac{R^4}{3456} - \frac{R^2\mathcal{S}_2}{24} + \frac{2R\mathcal{S}_3}{3} - \frac{\mathcal{S}_2^2}{2},
\end{align}
and
\begin{align}
    \pazocal{C}_{(5)} & =+ \frac{61R^5}{960} - \frac{7R^3\mathcal{R}_2}{12} + \frac{2R^2\mathcal{R}_3}{15} + \frac{7R\mathcal{R}_2^2}{5} - \frac{16\mathcal{R}_2\mathcal{R}_3}{15}
    \\ & = -\frac{R^5}{25920} + \frac{R^3\mathcal{S}_2}{108} - \frac{2R^2\mathcal{S}_3}{9} + \frac{R\mathcal{S}_2^2}{3} - \frac{16\mathcal{S}_2\mathcal{S}_3}{15},
\end{align}
which agree with the results previously reported in refs.~\cite{Sinha:2010ai,Paulos:2010ke}. On the other hand, for $n\ge 6$ the recursion produces a single representative non-trivial density. For example, for $n=6$ ---which is the order at which degeneracies start to appear due to the existence of densities trivially satisfying the holographic c-theorem--- we find from the recursive formula
\begin{align}
    \pazocal{C}_{(6)} & = -\frac{1103 R^6}{20736} + \frac{115R^4\mathcal{R}_2}{288} - \frac{19R^3\mathcal{R}_3}{81} - \frac{71R^2\mathcal{R}_2^2}{108} + \frac{8R\mathcal{R}_2\mathcal{R}_3}{9} - \frac{10 \mathcal{R}_2^3}{27}
   \\ & = +\frac{R^6}{186624} - \frac{5R^4\mathcal{S}_2}{2592} + \frac{5R^3\mathcal{S}_3}{81} - \frac{5R^2\mathcal{S}_2^2}{36} + \frac{8R\mathcal{S}_2\mathcal{S}_3}{9} - \frac{10 \mathcal{S}_2^3}{27}\, .
\end{align}

\subsection{General formula for order-$n$ densities}
Interestingly, it is possible to solve the two-term recurrence relation (\ref{recuu}) analytically and obtain an explicit expression for the order-$n$ density non-trivially satisfying the holographic c-theorem. The result  which, once again, takes a simpler form in terms of the $\{R,\mathcal{S}_{2},\mathcal{S}_3\}$ set, reads,
\begin{equation}\label{eq:cdensityn}
\begin{aligned}
\pazocal{C}_{(n)}=\frac{3 (-1)^{n}}{4\cdot 6^n n}\Bigg\{&\left(R+\sqrt{24\mathcal{S}_2}\right)^{n-1}\left(R-(n-1)\sqrt{24\mathcal{S}_2}\right)\left(1-\sqrt{6}\frac{\mathcal{S}_3}{\mathcal{S}_2^{3/2}}\right)\\
+&\left(R-\sqrt{24\mathcal{S}_2}\right)^{n-1}\left(R+(n-1)\sqrt{24\mathcal{S}_2}\right)\left(1+\sqrt{6}\frac{\mathcal{S}_3}{\mathcal{S}_2^{3/2}}\right)
\Bigg\}\, .
\end{aligned}
\end{equation}
Even though this expression may look odd because it depends in a non-polynomial way on the densities, it does reduce to a polynomial expression when we evaluate it for any integer $n\ge 1$. One can check this by expanding the $\left(R\pm \sqrt{24\mathcal{S}_2}\right)^{n-1}$ terms using the binomial coefficients.  In particular, note that this formula is even under the exchange $\mathcal{S}_2^{1/2}\rightarrow- \mathcal{S}_2^{1/2}$, and therefore $\mathcal{S}_2^{1/2}$ always appears with even powers ---\ie, there are no square roots---. Explicitly, the result of this expansion reads
\begin{equation}\label{CnFormula}
\begin{aligned}
\pazocal{C}_{(n)}=\frac{3 (-1)^{n}}{2\cdot 6^n n}\Bigg\{&\sum_{k=0}^{\lfloor \frac{n}{2} \rfloor}(24 \mathcal{S}_2)^k R^{n-2k}\left[\binom{n-1}{2k}-(n-1)\binom{n-1}{2k-1}\right]\\
&-288\mathcal{S}_3\sum_{k=0}^{\lfloor\frac{n-3}{2}\rfloor}(24 \mathcal{S}_2)^k R^{n-3-2k}\left[\binom{n-1}{2k+3}-(n-1)\binom{n-1}{2k+2}\right]\Bigg\}\, ,
\end{aligned}
\end{equation}
which is valid whenever $n\in \mathbb{N}$. 

Interestingly, the density \eqref{eq:cdensityn} can also be applied for non-integer $n$, since it always yields the result \eqref{eq:Lnonshell} when evaluated on the metric \eqref{eq:RGmetric}, and therefore it yields second-order equations for the RG-flow metric.  Hence, these Lagrangians provide a generalization of the  holographic $c$-theorem-satisfying densities for arbitrary real values of $n$. 

\subsection{All densities with a trivial c-function emanate from a single sextic density }
For the first five curvature orders, there exists a single density which satisfies the holographic c-theorem condition. Now, for $n=6$, there exists an additional density,
\begin{align}\label{Omeg6}
\pazocal{Y}_{(6)} &\equiv 6 \mathcal{S}_3 ^2-\mathcal{S}_2 ^3\\ &=\frac{1}{3} \left[R^6-9R^4\mathcal{R}_2+8R^3\mathcal{R}_3+21R^2\mathcal{R}_2^2-36R\mathcal{R}_2\mathcal{R}_3-3\mathcal{R}_2^3+18\mathcal{R}_3^2\right]\, ,
\end{align}
with the property of being identically vanishing when evaluated on the c-theorem ansatz (\ref{eq:RGmetric}) and which therefore does not
contribute to the equations of motion for that ansatz. 

An immediate consequence is that any product of $\pazocal{Y}_{(6)} $ with any other density also satisfies trivially the holographic c-theorem.
 Therefore, for $n\ge 6$ we have, at least, the following set of densities which satisfy the holographic c-theorem
\begin{equation}\label{cThLs}
\pazocal{L}_{(n)}^{\rm c-theorem}=\alpha_n \pazocal{C}_{(n)}+\pazocal{Y}_{(6)}\cdot \pazocal{L}^{\rm general}_{(n-6)}\, ,
\end{equation}
where $\pazocal{L}^{\rm general}_{(n-6)}$ is the general Lagrangian of order $n-6$ in the curvature. Remarkably, these are all the densities of this type that exist. 

This can be proven as follows. First, observe that there exist $\#(n-6)$ densities of order $n-6$. Hence, there exists the same number of order-$n$ densities in the set $\pazocal{Y}_{(6)} \cdot \pazocal{L}^{\rm general}_{(n-6)}$. Now, as observed earlier, there exist $\#(n)-(n-1) $ independent order-$n$ densities which satisfy the holographic c-theorem, one of which does so in a non-trivial fashion. The latter can be chosen to be $\pazocal{C}_{(n)}$ and we are left with $\#(n)-n$ independent densities which trivially satisfy the holographic c-theorem. Now, invoking the result in eq.~\eqref{properttt}, we observe that this number exactly matches the number of densities in the set $\pazocal{Y}_{(6)} \cdot \pazocal{L}^{\rm general}_{(n-6)}$.

In sum, $\pazocal{L}_{(n)}^{\rm c-theorem}$ as defined above is the most general higher-curvature order-$n$ density satisfying the holographic c-theorem and all densities satisfying it in a trivial fashion emanate from the sextic density $\pazocal{Y}_{(6)}$. This is a rather intriguing result which suggests that there may be something more fundamentally special about this density. As a matter of fact, this will not be the last time we encounter it.


\subsection{Absence of scalar mode in the spectrum}
An immediate consequence of decomposition \eqref{cThLs} is that none of the densities trivially satisfying the holographic c-theorem contributes to the linearized equations around an Einstein metric. This is because all densities involved take the form $\pazocal{Y}_{(6)} \cdot \pazocal{L}^{\rm general}_{(n-6)}$  and therefore belong to the set $\pazocal{G}|_{\text{trivial linearized equations}}$ as defined in eq.~\eqref{gtriv}. On the other hand, we can use our previous results to prove that densities which satisfy the holographic c-theorem in a non-trivial fashion do not incorporate the scalar mode in their spectrum. This latter property seems to have been observed in certain particular cases \cite{Afshar:2014ffa} but we have found no general proof in the literature.

We saw in section \ref{seclineq} that the condition for the absence of the scalar mode in the linearized spectrum, $m_s^2 \to \infty$, was satisfied by theories of the form
\begin{equation}
\pazocal{G}(R,\mathcal{S}_2,\mathcal{S}_3) = f(R) + \mathcal{S}_2 g(R) + \pazocal{G}_{\rm triv}
\end{equation}
for which
\begin{equation}\label{12frr}
    12 \bar{f}_{R,R} + \bar{g} = 0.
\end{equation}

For theories where $\pazocal{G}(R,\mathcal{S}_2,\mathcal{S}_3)$ is a polynomial, as the ones we are considering, this cancellation must occur order by order. At any given order $n$ the only possible forms of $f$ and $g$ are $f_{(n)}(R) = \lambda_{(n)} R^n$ and $g_{(n)}(R) = \mu_{(n)}R^{n-2}$ for some constants $ \lambda_{(n)}$ and $ \mu_{(n)}$, and so $12 \bar{f}_{R,R} + \bar{g} = \left[12n(n-1)\lambda_{(n)} + \mu_{(n)} \right] \bar{R}^{n-2}$, and so condition (\ref{12frr}) becomes
\begin{equation}\label{cTnoS}
    12n(n-1)\lambda_{(n)} + \mu_{(n)} = 0,
\end{equation}
where $\lambda_{(n)} = \beta_{n00}$ is the coefficient in front of the $R^n$ term and $\mu_{(n)} = \beta_{(n-2)10}$ is the coefficient in front of the $\mathcal{S}_2R^{n-2}$ term.


Now, expanding eq. \eqref{CnFormula} and keeping only the terms with $k = 0,1$ in the first sum, we see
\begin{equation}
    \pazocal{C}_{(n)} = \frac{3(-1)^n}{2\cdot6^nn} \Big\{ R^n - 12n(n-1) \mathcal{S}_2 R^{n-2} + \cdots \Big\},
\end{equation}
and so 
\begin{equation}
    \lambda_{(n)} = \frac{3(-1)^n}{2\cdot6^nn}, \quad \quad \mu_{(n)} = - 12n(n-1)\frac{3(-1)^n}{2\cdot6^nn},
\end{equation}
which clearly fulfill condition \eqref{cTnoS}. This proves that all theories satisfying the holographic c-theorem have a linearized spectrum which does not include the scalar mode.

\subsection{Born-Infeld gravity }

It was proposed in \cite{Gullu:2010pc} that NMG could also be extended through a Born-Infeld gravity theory with Lagrangian density
\begin{equation}\label{BINMG}    \pazocal{L}_{\text{BI-NMG}} = \sqrt{\det \left( \delta_a^b + \frac{\sigma}{m^2} G_a^b \right) }  - \left( 1 - \frac{\Lambda}{2m^2} \right),
\end{equation}
where $G_{ab} = R_{ab} - \frac{1}{2}g_{ab}R$ is the Einstein tensor and $\sigma = \pm 1$. This theory reproduces NMG when expanded to quadratic order in the curvature. Then, after ref.~\cite{Sinha:2010ai} proved that both NMG and the cubic order term of eq.~\eqref{BINMG} admitted an holographic c-function, it was soon proven in ref.~\cite{Gullu:2010st} that the full theory also satisfied a simple holographic c-theorem of the same kind as the one described in the previous subsections. The cancellations on the on-shell evaluation of these theories required by the c-theorem construction occur order by order, and so the theory defined by eq. \eqref{BINMG} generates an infinite number of higher derivative densities which non-trivially fulfil an holographic c-theorem at any truncated order \cite{Alkac:2018whk}.

Now, in view of our results, it would be interesting to know whether the terms generated by the expansion of Lagrangian \eqref{BINMG} order by order, which we shall call $\pazocal{B}_{(n)}$, are the same ones as the non-trivial densities $\pazocal{C}_{(n)}$ generated by the recursive formula \eqref{recuu}.
Following what we have just learned in the previous section, that should indeed be the case for $n = 1,\ldots,5$. For $n\geq6$ we expect both sets of densities to coincide up to ``trivial'' densities, and we find that to be the case.

Let us expand the density \eqref{BINMG}. We set $\sigma = 1$ and $m^2 = 1$ for simplicity, as they can be easily restored by dimensional analysis. In three dimensions the determinant of any matrix $X$ can be computed as 
\begin{equation}
    \det (X) = \frac{1}{6} \left[\tr (X)^3 - 3 \tr (X) \tr (X^2) + 2 \tr(X^3) \right].
\end{equation}
In our case, we have $A = \mathbf{1} + g^{-1}G$, which gives
\begin{equation}
    \det (\mathbf{1} + g^{-1}G ) = 1 + \frac{-1}{2}R + \frac{1}{4} \mathcal{T}_2 + \frac{1}{24}\mathcal{T}_3
\end{equation}
where we have defined
\begin{align}
    \mathcal{T}_2 & \equiv R^2 - 2 \mathcal{R}_2 = \frac{1}{3} R^2 - 2 \mathcal{S}_2,
    \\
    \mathcal{T}_3 & \equiv  R^3 - 6 R \mathcal{R}_2 + 8 \mathcal{R}_3 = \frac{-1}{9}R^3 + 2 R \mathcal{S}_2 + 8 \mathcal{S}_3.
\end{align}
We can now simply Taylor expand the square root, $\sqrt{1+x} = \sum_{m=0}^{\infty} \binom{1/2}{m} x^m$, with $x = \det (\mathbf{1} + g^{-1}G )-1$ and then collect the relevant terms at each order $n$ to build $\pazocal{B}_{(n)}$. The result is the following,
\begin{equation}\label{Bn}
    \pazocal{B}_{(n)} = \sum_{i+2j+3k=n} \binom{1/2}{i+j+k} \frac{(i+j+k)!}{i!j!k!}\left( \frac{-1}{2}R \right)^i\left( \frac{1}{4}\mathcal{T}_2 \right)^j \left( \frac{1}{24} \mathcal{T}_3 \right)^k.
\end{equation}
The lowest order densities given by the this formula are
\begin{small}
\begin{equation}
    \pazocal{B}_{(1)}  = \pazocal{C}_{(1)}\, ,\quad  
    \pazocal{B}_{(2)}   = \frac{1}{2}\pazocal{C}_{(2)} \, , \quad 
    \pazocal{B}_{(3)}  =  -\frac{1}{8}\pazocal{C}_{(3)}\, , \quad 
    \pazocal{B}_{(4)}  = \frac{1}{16}\pazocal{C}_{(4)}\, , \quad 
    \pazocal{B}_{(5)}  = -\frac{5}{128}\pazocal{C}_{(5)}\, , 
\end{equation}
\end{small}
which are indeed proportional to the densities $\pazocal{C}_{(n)}$ found previously through the recursion relation \eqref{recuu}, as expected.
At the next orders, however, eq. \eqref{Bn} gives a different non-trivial density than the one given by the recursion relation \eqref{recuu}. Following eq. \eqref{cThLs}, we see that the relationship between the densities $\pazocal{B}_{(n)}$ and $\pazocal{C}_{(n)}$ at orders $n \geq 6$ is given by
\begin{equation}
    \pazocal{B}_{(n)} = (-1)^n \frac{(2n-5)!!}{(2(n-1))!!} \pazocal{C}_{(n)} + \pazocal{Y}_{(6)} \cdot \pazocal{L}_{(n-6)},
\end{equation}
for some particular densities $\pazocal{L}_{(n-6)}$.
For example,
\begin{align}
    \pazocal{B}_{(6)} & = \frac{7}{256} \pazocal{C}_{(6)} - \frac{1}{432} \pazocal{Y}_{(6)}, \\
     \pazocal{B}_{(7)} & = -\frac{21}{1024} \pazocal{C}_{(7)} - \frac{R}{576} \pazocal{Y}_{(6)}, \\
     \pazocal{B}_{(8)} & = \frac{33}{2048} \pazocal{C}_{(8)} - \frac{11R^2 + 24\mathcal{S}_2}{13824} \pazocal{Y}_{(6)}.
\end{align}
Hence, both $\pazocal{C}_{(n)} $ and $\pazocal{B}_{(n)} $ provide sets of order-$n$ densities which non-trivially satisfy the holographic c-theorem. While the $\pazocal{C}_{(n)} $ are distinguished by the property of satisfying the simple recurrence relation (\ref{recuu}), the $\pazocal{B}_{(n)} $ have the property of corresponding to the general term in the expansion of the Born-Infeld theory (\ref{BINMG}). Both sets are equal up to terms which identically vanish in the holographic c-theorem ansatz which, as we have seen, are all proportional to the density $\pazocal{Y}_{(6)}$.

Another Born-Infeld theory has been proposed as a non-minimal extension of NMG (nM-BI) \cite{Alkac:2018whk}, with Lagrangian density
\begin{equation}\label{nMBI}
    \pazocal{L}_{\text{nM-BI}} = \sqrt{\det \left( \delta_a^b - \frac{2}{m^2} P_a^b + \frac{1}{m^4} P_a^c P_c^b \right) }  - \left( 1 - \frac{\Lambda}{2m^2} \right),
\end{equation}
where $P_a^b = R_a^b - \frac{1}{4}\delta_a^b R$ is the Schouten tensor. The full theory also allows for an holographic c-function. However, when expanded order by order using a similar method as the one described above, we see that it does not produce an infinite number of higher derivative densities which non-trivially fulfil an holographic c-theorem. At order $n=2$ and $n=3$ we obtain terms proportional to $\pazocal{C}_{(2)}$ and $\pazocal{C}_{(3)}$, as expected, but the terms with $n \geq 4$ all trivialize due to the Schouten identities described in section 2. 
Therefore, the density \eqref{nMBI} is equivalent to the much simpler density 
\begin{equation}\label{nMBIv2}
    \pazocal{L} = R - 2\Lambda + \frac{2}{m^2} \left( \mathcal{R}_2 - \frac{3}{8}R^2 \right) + \frac{1}{m^4} \left( \frac{17}{48} R^3 - \frac{3}{2}R \mathcal{R}_2 + \frac{4}{3} \mathcal{R}_3 \right).
\end{equation}


\section{GQT gravities in three dimensions?}\label{GQTss}
Here we are interested in exploring the possible existence of GQT gravities in three dimensions. In order to do that, we need to determine the set of densities for which eq.~\eqref{eq:GQTcond} holds, if any. As a first step, we need to evaluate our fundamental building-block densities on the single-function SSS ansatz \eqref{eq:SSS} with $\diff\Omega_1^2=\diff\theta^2$. Following the analysis of section \ref{sec:numbertheo}, we can study the dependence on the radial coordinate of the quantities defined in our general Lagrangian \eqref{Gseries} and \eqref{action3}, this is
 \begin{align}
 R\big|_f= -2\EuScript{A}+4\EuScript{B}\, , \quad  \mathcal{S}_2\big|_f =\frac{2}{3}(\EuScript{A}+\EuScript{B})^2 \, , \quad \mathcal{S}_3\big|_f =\frac{2}{9}(\EuScript{A}+\EuScript{B})^3 \, , 
 \end{align}
 and 
\begin{equation}
\mathcal{R}_2\big|_f=2(\EuScript{A}^2-2\EuScript{A} \EuScript{B}+3\EuScript{B}^2)\, , \quad \mathcal{R}_3\big|_f =-2(\EuScript{A}^3-3\EuScript{A}^2 \EuScript{B}+3\EuScript{A} \EuScript{B}^2+5\EuScript{B}^3) \, .
\end{equation}
We immediately observe the absence of $\psi$ in these expressions, which means that three-dimensional on-shell Lagrangians do not depend on the function $f$ explicitly, but only on its first and second derivatives.
Hence, in this case the GQT condition \eqref{eq:GQTcond} becomes simpler, namely,
\begin{equation}\label{integ}
\frac{ \partial L_f}{\partial f'}=\frac{\diff}{\diff r}\frac{\partial L_f}{\partial f''}+c\, ,
\end{equation}
where $c$ is an integration constant.

Evaluating on-shell a general order-$n$ density in the $\{R,\mathcal{S}_2,\mathcal{S}_3\}$ basis, we find
\begin{align}
L_{(n),f} &=r L^{2(n-1)} \sum_{j,k} \beta_{n-2j-3k,j,k} \frac{(-1)^{n-2j-3k}}{6^j 36^k} \left(f''+\frac{2f'}{r}\right)^{n-2j-3k} \left(f''-\frac{f'}{r}\right)^{2j+3k}\, , \\
& =r L^{2(n-1)} \sum_{j,k,l,m} c_{jklm} {f''}^{(l+m)} \left(\frac{f'}{r}\right)^{n-m-l}
\end{align}
where $L_{(n),f}\equiv \sqrt{g}\pazocal{L}_{(n)}|_{f}$ and where we used the binomial expansion twice in the second line and defined the constants
\begin{equation}
c_{jklm}\equiv \frac{ \beta_{n-2j-3k,j,k}}{6^{j+2k}}  (-1)^{n-m} 2^{n-2j-3k-l}\binom{n-2j-3k}{l}\binom{2j+3k}{m} \, .
\end{equation}
The combination $l+m$ takes integer values from $0$ to $n$, and hence $L_{(n),f}$ can be written as a linear combination of terms with different powers of  $f''$ taking such values. Now, in order for  $L_{(n),f}$ to be a total derivative, we need to impose that all terms involving powers of $f''$ higher than one vanish. This implies imposing $n-1$ conditions on the coefficients $\beta_{n-2j-3k,j,k}$. Once this is done, we are left with
\begin{align}
L_{(n),f} = g_1(r,f') + g_2(r,f') f''\, , 
\end{align}
where
\begin{equation}
g_1\equiv r L^{2(n-1)} \sum_{j,k}c_{jk00} \left(\frac{f'}{r}\right)^{n}\, , \quad g_2\equiv r L^{2(n-1)} \sum_{j,k}(c_{jk10} +c_{jk01})  \left(\frac{f'}{r}\right)^{n-1}\, .
\end{equation}
However, the fact that $L_{(n),f}$ is linear in $f''$ does not guarantee that $L_{(n),f}$ is a total derivative. In order for this to be the case, we need to impose the additional condition  given by eq.~\req{integ} which, in terms of $g_1$ and $g_2$ becomes 
\begin{equation}
\frac{\partial g_1}{\partial f'}=\frac{\partial g_2}{\partial r}\, .
\end{equation}
Explicitly, this condition becomes
\begin{equation}\label{condime}
n \sum_{j,k}c_{jk00} = (n-2) \sum_{j,k} (c_{jk10}+c_{jk01})\, ,
\end{equation}
which in terms of the original $\beta_{ijk}$ coefficients reads,
\begin{equation}
\sum_{j,k} \frac{\beta_{n-2j-3k,j,k}}{6^{j+2k}} 2^{n-2j-3k-1}(2-n+6j+9k)=0\, .
\end{equation}
Adding this to the $n-1$ conditions imposed earlier, we find a total of $n$ conditions to be imposed to $L_{(n),f} $ in order for it to be a GQT density. Hence, we have $\#(n)-n=\#(n-6)$ GQT densities at order $n$.   


\subsection{All GQT densities emanate from the same sextic density }
 Interestingly, the number of order-$n$ GQT densities exactly coincides with the number of densities trivially satisfying the holographic c-theorem. More remarkably, the two sets of densities are in fact identical. Indeed, the special sextic density $\pazocal{Y}_{(6)}$ defined in eq.~\req{Omeg6} as the source of all densities trivially satisfying the holographic c-theorem turns out to be also the source of all GQT densities. Indeed, it is not difficult to see that 
 \begin{equation}
\left.  \pazocal{Y}_{(6)}\right|_f=0\, ,
 \end{equation}
 which means that all densities involving $\pazocal{Y}_{(6)}$ identically vanish and are therefore ``trivial'' GQT densities ---in the sense that they make no contribution to the equation of $f(r)$. Since there are $\#(n-6)$ of such densities, we learn that in fact all GQT densities in three dimensions are ``trivial'' and proportional to $\pazocal{Y}_{(6)}$, 
  \begin{equation}
\pazocal{L}_{(n)}^{\rm GQT}=\pazocal{Y}_{(6)}\cdot \pazocal{L}^{\rm general}_{(n-6)}\, ,
\end{equation}
where $\pazocal{L}^{\rm general}_{(n-6)}$ is the most general order-$(n-6)$ density.

In sum, we learn that in three dimensions there  exist no non-trivial GQT densities, opposed to higher dimensions. As we have seen in chapter \ref{ch:ateach}, in $D=4$ there exists one independent non-trivial GQT density for every $n\geq 3$ whereas for $D\geq 5$ there actually exist $n-1$ independent inequivalent  GQT densities for every $n$ ---namely, there exist $n-1$ densities of order $n$ each of which makes a functionally different contribution to the equation of $f$ \cite{Bueno:2022res}. As a matter of fact, the triviality of the three-dimensional case unveiled here is not so surprising given that all higher-curvature theories admit the BTZ solution ---as opposed to non-trivial GQT gravities in higher dimensions, which admit modifications of Schwarzschild as solutions, but not Schwarzschild itself.

\subsection{EMQT gravities in three dimensions}\label{sec:EQT3D}

The triviality of GQT in three dimensions can be circumvented by adding matter fields to our theory. Inspired by their higher-dimensional counterparts \cite{Cano:2020ezi,Cano:2020qhy,Cano:2022ord} presented in sec.~\ref{sec:EGQTg},  we can define EMQT gravities in three dimensions by the condition that a general Lagrangian $\sqrt{-g}\pazocal{L}(R_{ab},\partial_a \phi)$  becomes a total derivative when evaluated on the single-function SSS with a magnetic ansatz for the scalar field,
\begin{equation}\label{eq:EMQTSSSmetric}
\diff s^2=-N^2f\diff t^2+\frac{\diff r^2}{f}+r^2\diff \theta^2\, ,\quad \phi=p \theta\, ,
\end{equation}
where $p$ is an arbitrary dimensionless constant and setting $N=1$. These theories are characterized by possessing an integrated equation depending only on $f$. The following family of densities belongs to the three-dimensional satisfy this requirement\footnote{As a matter of fact, we could consider an even more general density of the form 
\begin{align}
 \tilde{\pazocal{Q}}  = K(X)+X F(X) R  -[3F(X)+2X F'(X)]R^{ab}\partial_a\phi\partial_b\phi \, ,
\end{align}
where $X\equiv (\partial\phi)^2$ and $K$ and $F$ are two arbitrary differentiable functions.  }
\begin{equation}\label{eq:EQTG3D}
I_{\rm \ssc EMQT}=\frac{1}{16\pi \GN}\int \diff ^3x\sqrt{-g}\left(R+\frac{2}{L^2}- \pazocal{Q} \right)\, ,
\end{equation}
where
\begin{small}
  \begin{equation}
\pazocal{Q}  \equiv \sum_{n=1} \alpha_n L^{2(n-1)} (\partial \phi)^{2n}-\sum_{m=0} \beta_m L^{2(m+1)}(\partial \phi )^{2m}\left[ (3+2m) R^{bc} \partial_b \phi \partial_c \phi- (\partial \phi )^2 R \right],
\end{equation} 
\end{small}  
and  where we used the notation $(\partial \phi)^2 \equiv (g^{ab}\partial_a \phi \partial_b \phi)$.
In this expression, the $\alpha_n$, $\beta_m$ are arbitrary dimensionless constants. Observe that $\pazocal{Q}$ contains terms which are at most linear in Ricci curvatures. Strictly speaking, a Lagrangian with arbitrary $\alpha_n$ and $\beta_m$ is a higher-derivative theory instead of higher-curvature one. When studying theories including higher powers in the Ricci tensor, densities that satisfy the EMQT condition are not found.

Let us compute the full equations of motion of this theory and check explicitly the equivalence between the on-shell approach in which we vary with respect to $N$, $f$ and $\phi$ in eq.~\eqref{eq:EQTG3D}. For the variation with respect to the inverse metric tensor $g^{ab}$ we find

\begin{align}
&\frac{1}{\sqrt{-g}}\frac{\updelta I_{\ssc \rm EMQT}}{\updelta g^{ab}}=G_{ab}-\frac{1}{L^2}g_{ab}-\sum_{n}\alpha_n (\partial\phi)^{2(n-1)}L^{2(n-1)}\left[n\partial_a\phi\partial_b\phi-\frac{1}{2}g_{ab}(\partial\phi)^2\right]\\\notag
&\quad +\sum_{m}\beta_{m}L^{2(m+1)}(\partial\phi)^{2(m-1)}\left[2(3+2m)R_{c(a}\partial_{b)}\phi\partial^{c}\phi (\partial\phi)^2-R_{ab}(\partial\phi)^4\right. \\
&\quad \left.  +m(3+2m)\partial_a\phi\partial_b\phi R^{cd}\partial_{c}\phi\partial_{d}\phi -(m+1)\partial_{a}\phi\partial_{b}\phi(\partial\phi)^2 R\right]-\frac{1}{2}g_{ab}R^{cd}Q_{cd},\notag\\
&\quad -\nabla^{c}\nabla_{(a}Q_{b)c}+\frac{1}{2}\nabla^2Q_{ab}+\frac{1}{2}g_{ab}\nabla_{cd}Q^{cd} \notag
\end{align}
where $Q_{ab}\equiv \sum_{m}\beta_{m}L^{2(m+1)}(\partial\phi)^{2m}\Big[(3+2m)\partial_{a}\phi\partial_{b}\phi-g_{ab}(\partial\phi)^2\Big].$
The variation with respect to the scalar field $\phi$ reads in turn
\begin{small}
\begin{align}
&\frac{1}{\sqrt{-g}}\frac{\delta I_{\ssc \rm EMQT}}{\delta \phi}=2\nabla_a \Bigg\{ \sum_{n=1}n\alpha_n L^{2n-1}(\partial\phi)^{2(n-1)}\partial^a\phi-\sum_{m=0}\beta_mL^{2(m+1)}(\partial\phi)^{2(m-1)}\times \\ \notag
&\times\left[m(3+2m)\partial^a\phi R^{bc}\partial_b\phi\partial_c\phi+(3+2m)(\partial\phi)^2R^{ab}\partial_b\phi-(m+1)R(\partial\phi)^2\partial^a\phi\right]\Bigg\}.
\end{align}
\end{small}

An alternative route that yields the same solution to the equations of motion involves considering the on-shell effective Lagrangian  $L_{f,N,\phi}=\sqrt{|g|}\pazocal{L} |_{\eqref{eq:EMQTSSSmetric}}$ and taking variations with respect to the undetermined functions, namely 
\begin{equation}
\pazocal{E}_N\equiv\frac{\updelta L_{N,f,\phi}}{\updelta N}\, ,\quad \pazocal{E}_f\equiv\frac{\updelta L_{N,f,\phi}}{\updelta f}\, , \quad \pazocal{E}_\phi\equiv\frac{\updelta L_{N,f,\phi}}{\updelta \phi}\, .
\end{equation}

For each of them we find
\begin{small}
\begin{align}
\pazocal{E}_N =&\frac{2r}{L^2}-\sum_{n=1}\frac{\alpha_n (\dot \phi)^{2n}L^{2(n-1)}}{r^{2(n-1)}}-f'\notag\\
&+\sum_{m=0}\frac{\beta_m(\dot \phi)^{2(m+1)}(2m+1)L^{2(m+1)}}{r^{2m+3}}\left[2(m+1)f-rf'\right]=0 ,\label{eq:EOMEMQT1}\\
\pazocal{E}_f= &N' \left[1+\sum_{m=0}\frac{\beta_m(\dot \phi)^{2(m+1)}(2m+1)L^{2(m+1)}}{r^{2(m+1)}}\right]=0 \label{eq:EOMEMQT2} ,\\
\pazocal{E}_\phi =&\ddot \phi \Bigg[ N\sum_{n=1}\frac{2n (2 n-1)(\dot \phi)^{2(n-1)}\alpha_n  L^{2 (n-1)} }{r^{2(n-1)}} + \sum_{m=0} \frac{2(m+1) (2 m+1)\beta_mL^{2 (m+1)}}{r^{2 (m+1)}}( \dot \phi)^{2m}   \notag\\ 
&\left(N' \left[-3 r f'+(2 m
   +1)f\right]+N \left[(2 m+1) f'-r f''\right]-2 r f N''\right) \Bigg]=0 \label{eq:EOMEMQT3},
\end{align}
\end{small}
where we used the notation $g'\equiv \diff g/ \diff r$ and $\dot \phi \equiv \diff \phi/ \diff \theta $. Now, it is evident that a linear dependence on the angle for $\phi$ and a constant one for $N$ automatically solve the last two equations. On the other hand, the first equation can be integrated once to obtain the algebraic equation for $f$ ---its precise expression is written in next chapter, in eq.~\eqref{eq:fsolEMQT}. Note that there is no dependence at all on $N$ in such equation, which means that the solutions with $N=$ constant, implied by the equation of $f$, are the only possible ones. This is a general property of QT gravities (Electromagnetic or not). 

General three-dimensional theories of the form $\pazocal{L}(R_{ab},\partial_a\phi)$ are dual to theories with an electromagnetic field, $\pazocal{L}_{\rm dual}(R_{ab},F_{ab})$, where $F_{ab}\equiv 2\partial_{[a}A_{b]}$. In particular, the dual field strength is defined by
\begin{equation}
F_{ab}=-\frac{1}{2}\epsilon_{abc}\frac{\partial \pazocal{L}}{\partial\left(\partial_{c}\phi\right)},
\end{equation}
and note that, in this way, the Bianchi identity of $F$ is equivalent to the equation of motion of $\phi$.

 In order to obtain the dual Lagrangian, $\pazocal{L}_{\text{dual}}\equiv \pazocal{L}-F_{ab}\partial_{c}\phi\epsilon^{abc}$, one needs to invert the above relation to get $\partial\phi(F)$. Unfortunately, this cannot be done explicitly for the theories we are considering here. Nevertheless, we can perform a perturbative expansion in powers of $L$, in whose case the dual action reads (for $\alpha_1\neq0$)
\begin{equation}
\pazocal{L}_{\text{dual}} =R+\frac{2}{L^2}-\frac{1}{2\alpha_1} F^2 +L^2\left[-\frac{\alpha_2}{4\alpha_1^4}(F^2)^2+3\frac{\beta_0}{\alpha_1^2}\tensor{F}{_{a}^{c}}F^{ab}R_{\langle cb\rangle}\right]+\pazocal{O}(L^4)\, ,
\end{equation}
where we denoted $F^2\equiv F_{ab}F^{ab}$. We stress that the dual Lagrangian has an infinite number of terms even when the original action only has a finite number of them (except when the only non-vanishing coupling is $\alpha_1$).

Observe that the leading term is the usual Maxwell piece, which explains the match with the charged BTZ metric when only  $\alpha_1$ is active. Indeed, solutions of $\pazocal{L}_{\rm \ssc EMQT}$ are also solutions of $\pazocal{L}_{\text{dual}}$. In the dual frame, the original ``magnetically charged'' solutions become ``electrically charged'', with a field strength $F_{tr}=-\partial A_t(r)/\partial r$, where $A_t(r)$ is the electrostatic potential. Remarkably, this quantity can be obtained exactly and it reads
\begin{equation}\label{at}
A_t(r)=-\alpha_1 p \log\frac{r}{L}+\sum_{i=2}\frac{i \alpha_i p}{2(i-1)}\left(\frac{L p}{r}\right)^{2(i-1)}+f' L\sum_{j=0}\beta_j (j+1)\left(\frac{L p}{r}\right)^{2j+1}.
\end{equation}
Hence, we can think of our new solutions as ``magnetic'' or ``electric'' depending on which frame we consider. For the former, the action is simple, taking the form  and the extra field is a scalar whose equation is solved by $\phi=p \theta$. For the latter, the action can only be accessed perturbatively and the auxiliary field is a standard $1$-form whose equation is solved by $A_t(r)$ as given in  eq.~\req{at}. Given the trivial behavior of $\phi$ in the magnetic frame, from now on we will exclusively analyze the metric behavior ---which is the same in both frames--- leaving a more exhaustive analysis of the behavior of  $A_t(r)$ and the electric frame for future work.

In next chapter we will study the solutions to the equations of motion and present a plethora of black holes and regular spacetimes depending on the choice of the coupling constants $\alpha_i$ and $\beta_j$.

\section{Segre classification of three-dimensional spacetimes}\label{Ommm6}
We have seen that all GQT densities as well as all densities trivially satisfying the holographic c-theorem  emanate from a single sextic density, $\pazocal{Y}_{(6)}$, defined in eq.~\eqref{Omeg6}. The reason for such occurrence can be understood as follows. As we saw earlier, when evaluated on-shell on metrics \req{eq:RGmetric} and single-function \eqref{eq:SSS} respectively, the densities $R,\mathcal{S}_2,\mathcal{S}_3$ read\footnote{As a matter of fact, these two ans\"atze have been previously considered simultaneously before in the four-dimensional case \cite{Arciniega:2018fxj,Arciniega:2018tnn,Cisterna:2018tgx} in a cosmological context. The reason is that the condition for demanding a simple holographic c-theorem can be alternatively understood as the condition that the equations of motion for the scale factor in a standard Friedmann-Lema\^itre-Robertson-Walker ansatz are second order \cite{Sinha:2010pm}.}
\begin{small}
\begin{align}
  \left. R\right|_a      &=   -\frac{2(a'^2 +2 a a'') }{ a^2} \, , \quad
  \left.  \mathcal{S}_2\right|_a  
    = \frac{2(a'^2 - a a'')^2}{3 a^4} \, , \quad
  \left.  \mathcal{S}_3\right|_a  
    = \frac{2(a'^2 - a a'')^3}{9 a^6}\, .\\
\left. R\right|_f &=-\frac{1}{r}\left(r f''+2f'\right) \, , \quad  \left. \mathcal{S}_2\right|_f=\frac{1}{6r^2}\left(r f''- f'\right)^2 \, , \quad  \left.  \mathcal{S}_3\right|_f=\frac{1}{36r^3}\left(r f''- f'\right)^3 \, .
\end{align}
\end{small}
Observe that $\mathcal{S}_2$ and $\mathcal{S}_3$ have in both cases the same functional dependence on $f(r)$ and $a(\rho)$, respectively, up to a power, whereas $R$ has a different dependence from the other two densities in both cases. Now, for $n\leq 5$, there is no way to construct linear combinations of the various order-$n$ densities such that the resulting expression identically vanishes. This is not the case for $n=6$. In that case, $\mathcal{S}_2^3$ and $\mathcal{S}_3^2$ have exactly the same functional dependence on $f(r)$ and $a(\rho)$ respectively, and a particular linear combination of them can be found such that it identically vanishes. This combination is precisely $\pazocal{Y}_{(6)}$ in both cases,
\begin{equation}
\pazocal{Y}_{(6)} = 6 \mathcal{S}_3 ^2-\mathcal{S}_2 ^3\, , \quad \left. \pazocal{Y}_{(6)} \right|_{a}=\left. \pazocal{Y}_{(6)} \right|_{f}=0\, .
\end{equation}
It is obvious that any density multiplied by $\pazocal{Y}_{(6)}$ will similarly vanish for these two ans\"atze. One could wonder what happens for other values of $n$ such as $n=12,18,\dots$, for which there is again a match in the functional dependence of the seed densities to the corresponding powers. It is however easy to see that in those cases the combinations which vanish are precisely the ones given by powers of $\pazocal{Y}_{(6)}$.


It is natural to wonder whether $\pazocal{Y}_{(6)}$ may actually vanish identically for general metrics. This is however not the case. For instance, for a general SSS ansatz \eqref{eq:SSS} in three dimensions, $\pazocal{Y}_{(6)}$ yields a complicated function of $f$ and $N$. 

As it turns out, the particular linear combination $6\mathcal{S}_3 ^2-\mathcal{S}_2^3$ appearing in $\pazocal{Y}_{(6)}$ is in fact connected to the Segre classification of three-dimensional spacetimes \cite{Hall,Sousa:2007ax}.  This consist in characterizing the different types of metrics according to the eigenvalues of the traceless Ricci tensor $R_{\langle ab\rangle}$. There exist three large sets of metrics which are precisely characterized by the relative values of  $6\mathcal{S}_3 ^2$ and $\mathcal{S}_2^3$, namely \cite{Chow:2009km,Gurses:2011fv,Chow:2009vt}: 
\begin{align}
\rm{Group\,\, 1:} \quad &6\mathcal{S}_3 ^2=\mathcal{S}_2^3=0\, , \quad &[\text{Type-O, Type-N, Type-III}] \\ \label{group2}
\rm{Group\,\, 2:} \quad &6\mathcal{S}_3 ^2=\mathcal{S}_2^3 \neq 0\, , \quad &[\text{Type-D}_s \text{, Type-D}_t \text{, Type-II}] \\
\rm{Group\,\,  3:} \quad &6\mathcal{S}_3 ^2 \neq \mathcal{S}_2^3\, , \quad &[\text{Type-I}_{\mathbb{R}}, \text{Type-I}_{\mathbb{C}}]
\end{align}
The first group, which in the ---perhaps more familiar--- Petrov notation includes Type-O, Type-N and Type-III spacetimes corresponds to spacetimes such that both $\mathcal{S}_3$ and $\mathcal{S}_2$ vanish. The second group, which includes Type-D and Type-II spacetimes is the one corresponding to metrics which have non-vanishing $6\mathcal{S}_3^2$ and $\mathcal{S}_2^3$ but such that they are equal to each other, \ie such that $\pazocal{Y}_{(6)}=0$. Finally, spacetimes of Type-I  have a non-vanishing $\pazocal{Y}_{(6)}$. 

Metrics of the Group 1 have traceless Ricci tensors which can be written as
\begin{align}
&R_{\langle ab\rangle}=0 \quad &[\text{Type-O}]\, , \\ 
&R_{\langle ab\rangle}= s \lambda_a \lambda_b\, , \quad &[\text{Type-N}]\, ,\\
&R_{\langle ab\rangle}= 2 s \xi_{(a}\lambda_{b)}\, , \quad & [\text{Type-III}]\, ,
 \end{align}
where $\lambda_a\lambda^a=0$, $\xi_a\xi^a=1$, $\lambda_a\xi^a=0$, and $s=\pm 1$. On the other hand, for metrics of the Group 2 we have 
\begin{align}\label{typeD}
&R_{\langle ab\rangle}=p(x^a) \left(g_{ab}-\frac{3}{\sigma} \xi_a \xi_b\right) \quad &[\text{Type-D}_{s,t}]\, , \\ 
&R_{\langle ab\rangle}=p(x^a) \left(g_{ab}-\frac{3}{\sigma} \xi_a \xi_b\right) + s \lambda_a\lambda_b \quad &[\text{Type-II}]\, , 
 \end{align}
where $p(x^a)$ are scalar functions and $\lambda_a\lambda^a=0$, $\xi_a\xi^a=\sigma=\pm 1$, $\lambda_a\xi^a=0$, and $s=\pm 1$. Finally, metrics of the Group 3 satisfy
\begin{align}
&R_{\langle ab\rangle}=p(x^a) \left(g_{ab}-3 \xi_a \xi_b\right) -q(x^a) (\lambda_a\lambda_b+\nu_a\nu_b)\quad &[\text{Type-I}_{\mathbb{R}}]\, , \\ 
&R_{\langle ab\rangle}=p(x^a) \left(g_{ab}-3 \xi_a \xi_b\right) -q(x^a) (\lambda_a\lambda_b-\nu_a\nu_b)\quad &[\text{Type-I}_{\mathbb{C}}]\, ,
\end{align}
where $p(x^a)$ and $q(x^a)$ are scalar functions (such that $q\neq \pm 3 p$ for Type-I$_{\mathbb{R}}$) and where $\lambda_a\lambda^a=0$, $\xi_a\xi^a=1$, $\nu_a\nu^a=0$, $\lambda_a\xi^a=\lambda_a\nu^a=0$, and $\lambda_a\nu^a=-1$.

For the single-function SSS metric and the holographic c-theorem metric, one finds that the traceless Ricci tensor satisfies eq.~\req{typeD} with
\begin{align}
p(r)=\frac{f'-r f''}{6r}\, , \quad \xi_a =r \delta_{a\theta} \, ,\quad \text{and} \quad
p(\rho)=\frac{a'' a-{a'}^2}{3a^2}\, , \quad \xi_a = \delta_{ar} \, ,
\end{align}
respectively. Hence, both spacetimes are of Type-D$_s$ and from eq.~\eqref{group2} it follows that $\pazocal{Y}_{(6)}=0$ in both cases.

From this we learn that the appearance of $\pazocal{Y}_{(6)}$ as a distinguished density was to be expected for the classes of metrics considered here, and that a similar phenomenon is likely to occur for all metrics of the Types D and II.\footnote{For various papers classifying and obtaining explicit solutions of the Groups 1 and 2 for three-dimensional higher-curvature gravities, see \cite{Chow:2009km,Gurses:2011fv,Chow:2009vt,Ahmedov:2010uk,Ahmedov:2011yd,Ahmedov:2012di,Alkac:2016xlr,Alkac:2017rxr}.} Still, the fact that all densities satisfying the holographic c-theorem in a trivial fashion and that all GQT densities are proportional to this density was far from obvious in advance.

We believe it would be interesting to further study the properties of $\pazocal{Y}_{(6)}$, understood as a higher-curvature density. 
Its equations of motion can be easily computed using expression \eqref{eq:EOM3}, and read
\begin{align}
\frac{\pazocal{E}_{ab}^{\pazocal{Y}_{(6)}}}{3}=&-\frac{1}{6}g_{ab}\left(6\mathcal{S}_3^2-\mathcal{S}_2^3\right)-2\mathcal{S}_2^2\tilde{R}_a^c\tilde{R}_{bc}+12\mathcal{S}_3\tilde{R}_a^c\tilde{R}_{cd}S^d_b\\
&-4\left(g_{ab}\dal-\nabla_a\nabla_b+\tilde{R}_{ab}+\frac{1}{3}g_{ab}R\right)\mathcal{S}_2\mathcal{S}_3-g_{ab}\nabla_c\nabla_d\left(\mathcal{S}_2^2\tilde{R}^{cd}-6\mathcal{S}_3\tilde{R}^{cf}\tilde{R}_f^d\right)\notag\\&-\left(\dal+\frac{2}{3}R\right)\left(\mathcal{S}_2^2\tilde{R}_{ab}-6\mathcal{S}_3\tilde{R}_a^c\tilde{R}_{bc}\right)
+2\nabla_c\nabla_{(a}\left(\tilde{R}_{b)}^c\mathcal{S}_2^2-6\tilde{R}_{b)}^d\tilde{R}_d^c\mathcal{S}_3\right)\, .\notag
\end{align}
These are identically satisfied by all metrics belonging to the Groups 1 and 2, but not for Type-I metrics. It would be then within such set that non-trivial solutions would arise. 

\section{Discussion}

In this chapter we have presented several new results involving general higher-curvature gravities in three dimensions. First, we provide a general formula for the exact number of possible independent order-$n$ densities $\#(n)$. Interestingly, this number satisfies the recursive relation $\#(n-6)=\#(n)-n$. In addition, we presented their corresponding equations of motion and the algebraic equations these reduce to when evaluated for Einstein metrics.

We also provided the linearized equations of a general higher-curvature gravity around an Einstein spacetime as a function of the effective Newton's constant and the masses of the new spin-2 and spin-0 modes generically propagated. Using these results, we showed that the most general order-$n$ density can be written as the sum of three densities. A first one that does not propagate the massive spin-2 mode but which does propagate the scalar one (for $n\geq2$). A second one that by itself does not propagate the scalar mode but which does propagate the spin-2 mode. On the other hand the last one does not contribute at all to the linearized equations.

Afterwards, we turn our attention to higher-curvature theories which satisfy an holographic c-theorem $\C_{(n)}$. First, we provide a recursive formula for densities which satisfy it in a non-trivial fashion ---\ie they contribute non-trivially to the c-function. The recurrence can be explicitly solved, providing a general explicit formula. We argue that there are $\#(n) -(n -1)$ densities of order $n$ which satisfy the holographic c-theorem. Of those, $\#(n) -n$ are trivial in the sense of making no contribution to the c-function. We show that all such trivial densities are proportional to a trivial density $\Y_{(6)}$. In addition, we show that if a theory satisfies the holographic c-theorem, then it does not include the scalar mode in its spectrum. Finally, we study the relation of between $\C_{(n)}$ with the general term obtained from expanding the Born-Infeld gravity Lagrangian.

Then, we showed that there exist $\#(n)-n$ theories of the type GQT gravity, and that all of them are trivial ---in the sense of making no contribution to the equation of the black hole metric function--- and again proportional to the same sextic density $\Y_{(6)}$. This situation changes when matter fields are included in the game, giving rise to EMQT gravities. These theories are constructed from positive powers $(\partial\phi)^2$ and certain linear combinations of $R^{ab}\partial_a {\phi}\partial_b{\phi}$ and $(\partial\phi)^2R$. In next chapter we will discuss the solutions appearing in these new theories and its properties.

To conclude, we made some comments on the relation between $\Y_{(6)}$ and the Segre classification of three-dimensional spacetimes. We explain why the prominent role played by this density in the identification of trivial densities of the types studied in the previous two sections could have been expected ---at least to some extent.

%% file: text/ch5-bh3d.tex
\chapter{Three-dimensional black holes in higher-curvature gravity}\label{ch:bhs3d}

Since the seminal discovery of the BTZ solution \cite{Banados:1992wn,Banados:1992gq, Carlip:1995qv}, the catalog of three-dimensional black holes has grown in different directions. On the one hand, higher-curvature modifications such as NMG \cite{Bergshoeff:2009hq} and its extensions \cite{Gullu:2010pc,Sinha:2010ai,Paulos:2010ke} allow for new solutions which differ from the BTZ one \eg in being locally inequivalent from AdS$_3$, in possessing asymptotically flat, dS$_3$ or Lifshitz  asymptotes, or in including curvature (rather than conical or BTZ-like) singularities \cite{Bergshoeff:2009aq,Oliva:2009ip,Clement:2009gq,Alkac:2016xlr,Barnich:2015dvt,AyonBeato:2009nh,Gabadadze:2012xv,Ayon-Beato:2014wla,Fareghbal:2014kfa,Nam:2010dd,Gurses:2019wpb,Gabadadze:2012xv,Bravo-Gaete:2020ftn,Anastasiou:2013mwa}. Including extra fields also allows for progress, and additional solutions are known for Einstein-Maxwell \cite{Clement:1993kc,Kamata:1995zu,Hirschmann:1995he,Martinez:1999qi,Cataldo:1996yr,Dias:2002ps,Cataldo:2002fh,Cataldo:2004uw,Panah:2022btb,Podolsky:2021gsa} as well as for Einstein-Maxwell-dilaton \cite{Chan:1994qa,Chen:1998sa,Koikawa:1997am,Fernando:1999bh} and Maxwell-Brans-Dicke type \cite{Sa:1995vs,Dias:2001xt,Dias:2002cg}  theories. These typically include logarithmic profiles for some of the fields and curvature singularities. Black hole solutions for minimally and non-minimally coupled scalars have also been constructed \cite{Martinez:1996gn,Henneaux:2002wm,Correa:2011dt,Zhao:2013isa,Tang:2019jkn,Karakasis:2021lnq,Baake:2020tgk,Gonzalez:2021vwp}, including some which exploit well-defined limits of Lovelock theories to three dimensions \cite{Hennigar:2020drx,Hennigar:2020fkv,Ma:2020ufk,Konoplya:2020ibi}. These typically contain curvature singularities and sometimes globally regular scalars.
A final class involves coupling Einstein gravity to non-linear electrodynamics \cite{Cataldo:1999wr,Myung:2008kd,Mazharimousavi:2011nd,Mazharimousavi:2014vza,Hendi:2017mgb,EslamPanah:2021xaf,Panah:2022cay}. For these, examples of regular black holes have been presented for special choices of the modified Maxwell Lagrangian  \cite{Cataldo:2000ns,He:2017ujy,HabibMazharimousavi:2011gh}. See ref.~\cite{Garcia-Diaz:2017cpv} for a comprehensive study of different solutions in three-dimensions.

In this section we employ the results of sec.~\ref{seclineq} to study the quasinormal modes and frequencies of the BTZ hole in a general higher curvature gravity. Moreover, we extend the above catalog with a new family of three-dimensional modifications of Einstein gravity involving a non-minimally coupled scalar which admit a large family of analytic generalizations of the BTZ metric describing black holes and globally regular horizonless spacetimes. The new black holes display one or several horizons and include curvature, conical, BTZ-like or no singularity at all depending on the case.

\section{The BTZ black hole in higher-order gravity}\label{BTZsec}
Since the BTZ black hole \cite{Banados:1992wn,Banados:1992gq} is an Einstein spacetime ---and hence it is locally AdS$_3$--- it is an exact solution of all higher-derivative theories of gravity. The parameters of the solution have different values, though.  To begin with, the AdS scale is not given by the cosmological constant scale $L$, but by $L_{\star}=L/\sqrt{\chi_{0}}$, as we discussed earlier. Thus, the general, rotating BTZ metric can be written as
\begin{equation}\label{eq:rotatingBTZ}
\diff s^2=-f\diff t^2+\frac{\diff r^2}{f}+r^2\left(\diff \theta-\frac{r_{+}r_{-}}{L_{\star} r^2}\diff t\right)^2,
\end{equation}
where $f\equiv \frac{(r^2-r_{+}^2)(r^2-r_{-}^2)}{L_{\star}^2r^2}$ and we assume that $r_{+}^2>r_{-}^2$, without loss of general. In addition, the coordinate $\theta$ has periodicity $2\pi$. 

In order to study the quasinormal modes of the BTZ solutions for general higher-curvature theories, we first need to identify the mass and angular momentum of the solution in this general context, and for that we can resort to thermodynamics. 


\subsection{Thermodynamics}
The horizon of the BTZ black hole is located at $r=r_{+}$, and this leads to a Hawking temperature
\begin{equation}
T=\frac{f'(r_{+})}{4\pi}=\frac{r_{+}^2-r_{-}^2}{2\pi L_{\star}^2r_{+}}\, .
\end{equation}
On the other hand, it will also be important to take note of the angular velocity of the horizon,
\begin{equation}
\Omega=\frac{r_{-}}{L_{\star}r_{+}}\, .
\end{equation}
We can find the free energy by evaluating the Euclidean on-shell action, whose bulk part is 
\begin{equation}
I_{E}^{\rm bulk}=-\frac{1}{16\pi \GN} \int_{\pazocal{M}} \diff ^3x \sqrt{g} \pazocal{L}\, .
\end{equation}
On the other hand, for asymptotically AdS solutions (as in this case), one can use the prescription given in \cite{Bueno:2018xqc} ---see also \cite{Araya:2021atx}--- for the boundary terms and counterterms,
\begin{equation}
I_{E}^{\rm bdry}=-\frac{a^*}{\pi L_{\star}}\int_{\partial \pazocal{M}}\diff^2x\sqrt{h}\left(K-\frac{1}{L_{\star}}\right)\, ,
\quad \text{where} \quad
a^{*}=-\frac{ L_{\star}^3}{32\GN}\bar{\pazocal{L}}\, ,
\end{equation}
and $K$ is the usual Gibbons-Hawking-York term \cite{York:1972sj,Gibbons:1976ue} corresponding to the trace of the extrinsic curvature of $\partial \pazocal{M}$. In the holographic context, the constant $a^{*}$ represents the universal contribution to the entanglement entropy across a spherical entangling region \cite{Myers:2010tj,Myers:2010xs,Casini:2011kv} in general dimensions which, for general higher-curvature gravities, is proportional to the on-shell Lagrangian \cite{Imbimbo:1999bj,Schwimmer:2008yh,Myers:2010tj,Myers:2010xs,Bueno:2018xqc}. In the present case, $a^{*}$ is proportional to the central charge of the two-dimensional dual CFT, $c=12 a^*$. For the Lagrangian expressed as in eq.~\eqref{action4}, we have
\begin{equation}
a^{*}=\frac{ L_{\star}^3}{32\GN}\left(\frac{6}{L_{\star}^2}-\frac{2}{L^2}-\bar{\pazocal{G}}\right)\, ,
\end{equation}
and after making use of the relation between $L$ and $L_{\star}$ \eqref{vacuS}, we can write it as
\begin{equation}
a^{*}=\frac{ L_{\star}}{8\GN}\left(1+\bar{\pazocal{G}}_{R}\right)=\frac{ L_{\star}}{8\GN^{\rm eff}}\, .
\end{equation}

Therefore, the Euclidean action on the BTZ black hole is simply computed by
\begin{equation}
I_{E}=-\frac{\bar{\pazocal{L}}}{16\pi\GN} \left[\int_{\pazocal{M}} \diff ^3x \sqrt{g}-\frac{L_{\star}^2}{2}\int_{\partial M}\diff^2x\sqrt{h}\left(K-\frac{1}{L_{\star}}\right)\right]\, ,
\end{equation}
which gives the following value for the free energy $F\equiv I_{E}/\beta$,
\begin{equation}
F=-\frac{r_{+}^2-r_{-}^2}{8\GN^{\rm eff}L_{\star}^2}\, .
\end{equation}
Now, the entropy can be computed from the Wald's formula \cite{Wald:1993nt,Iyer:1994ys}, which gives
\begin{equation}
S=-\frac{1}{8\GN}\int d\sigma \frac{\partial \pazocal{L}}{\partial R^{ab}}\epsilon_{ac}\tensor{\epsilon}{_{b}^{c}}=\frac{\pi r_{+}}{2\GN^{\rm eff}}.
\end{equation}

Combining the entropy and the free energy, we can obtain the mass, 
\begin{equation}
M=F+TS=\frac{r_{+}^2+r_{-}^2}{8\GN^{\rm eff}L_{\star}^2}\, .
\end{equation}
Finally, from the first law
\begin{equation}
\diff M-T\diff S=\frac{r_{-}}{4\GN^{\rm eff}L_{\star}^2 r_{+}}\left(\diff r_{-}r_{+}+\diff r_{+}r_{-}\right)=\Omega \diff\left(\frac{r_{+}r_{-}}{4\GN^{\rm eff}L_{\star}}\right)\, ,
\end{equation}
we identify the angular momentum,
\begin{equation}
J=\frac{r_{+}r_{-}}{4\GN^{\rm eff}L_{\star}}\, .
\end{equation}
Thus, everything turns out to work out just like for Einstein gravity, with the identifications $\GN\rightarrow \GN^{\rm eff}$ and $L\rightarrow L_{\star}$.

\subsection{Perturbations and quasinormal modes}
While the BTZ black hole is a solution of all higher-order gravities, its perturbations behave differently depending on the theory. In  fact, for Einstein gravity it makes no sense to consider the gravitational perturbations around a BTZ black hole, since all of them are trivial (equivalent to infinitesimal gauge transformations), due to the fact that all solutions of three-dimensional Einstein gravity are locally AdS$_3$. 
However, as we saw, higher-derivative gravities do introduce additional degrees of freedom: in general, a scalar mode and a massive graviton.  Interestingly, since the BTZ is locally AdS$_3$, we can apply our results from sec.~\ref{seclineq} and perform a general analysis valid for all higher-curvature gravities. 

Following the results obtained in sec.~\ref{physmod}, the quasinormal modes are solutions to the equations 
\begin{align}
\left(\bar{\dal}-m_s^2\right)h=0\, ,\quad
\left(\bar{\dal}+\frac{2}{L_{\star}^2}-m_g^2\right)\hat{h}_{ab}=0\, ,
\label{eq:massiveeq}
\end{align}
which satisfy  an outgoing-wave boundary condition at the black hole horizon, and that behave as normalizable modes at infinity. The quasinormal modes of a scalar field in the background of the BTZ black hole are well-known~\cite{Cardoso:2001hn,Birmingham:2001pj,Berti:2009kk}, while those of the massive graviton were computed in ref.~\cite{Myung:2011bn}  in the case of NMG. However, in a general metric theory we have both types of perturbations, with masses that depend on the different parameters of the Lagrangian. Let us then compute the quasinormal modes of these perturbations and their associated frequencies.

\subsection{The unit radius BTZ}
Without loss of generality, we can consider the BTZ black hole with $r_{+}=L_{\star}$, $r_{-}=0$. The general case can be obtained from this one after appropriate coordinate transformations. The metric of this black hole reads
\begin{equation}\label{eq:unitBTZ}
\diff s^2=-\left(\frac{r^2}{L_{\star}^2}-1\right)\diff t^2+\frac{\diff r^2}{\left(\frac{r^2}{L_{\star}^2}-1\right)}+r^2\diff \theta^2,
\end{equation}
and introducing $r=L_{\star}\cosh\rho$, it can be written as
\begin{equation}\label{eq:unitBTZ2}
\diff s^2=-\sinh^2\rho \,\diff t^2+L_{\star}^2\cosh^2\rho \, \diff \theta^2+L_{\star}^2 \diff \rho^2.
\end{equation}
Finally, one can also write this metric in terms of the null coordinates $u=t+L_{\star}\theta$, $v=t-L_{\star}\theta$, so that we have
\begin{equation}
\diff s^2=\frac{1}{4}\left(\diff u^2+\diff v^2\right)-\frac{1}{2}\cosh2\rho\, \diff u \diff v+L_{\star}^2\diff \rho^2\, .
\end{equation}
The interesting property about these coordinates is that the two sets of generators of the $\mathrm{SL}(2,\mathbb{R})$ algebra, $\{L_{k}\}$ and $\{\bar L_{k}\}$, with $k=-1,0,1$, take a symmetric form, namely,
\begin{align}
L_0&=-\partial_{u}\, ,\quad L_{\pm 1}=e^{\pm u}\left(-\frac{\cosh2\rho}{\sinh2\rho}\partial_{u}-\frac{1}{\sinh2\rho}\partial_{v}\pm\frac{1}{2 L_{\star}}\partial_{\rho}\right)\, ,\\
\bar L_0&=-\partial_{v}\, ,\quad \bar L_{\pm 1}=e^{\pm v}\left(-\frac{\cosh2\rho}{\sinh2\rho}\partial_{v}-\frac{1}{\sinh2\rho}\partial_{u}\pm\frac{1}{2 L_{\star}}\partial_{\rho}\right)\, .
\end{align}
Then, the idea of reference \cite{Sachs:2008gt} is that the quasinormal modes of the BTZ black hole can be found as the descendants of ``chiral highest weight" modes, which are those that are either annihilated by $L_{1}$ or by $\bar L_{1}$. These are also the fundamental quasinormal modes, \ie, those with the lowest (in magnitude) imaginary part. 

In the case of a scalar field (the trace of the metric $h$ in our case), we separate the dependence of the field on the $u$ and $v$ variables and consider a perturbation of the form
\begin{equation}
h=e^{-\iu up_{+}-\iu v p_{-}} F(\rho)\, .
\end{equation}
Then we first search for the chiral highest weight modes, $h_{(0)}^{L}$ and $h_{(0)}^{R}$, which satisfy 
\begin{equation}
L_{1}h_{(0)}^{L}=0\, ,\qquad \bar{L}_{1}h_{(0)}^{R}=0\, ,
\end{equation}
corresponding to left-moving and right-moving modes, respectively. These are first order equations for the function $F$, and we find in each case
\begin{equation}
\begin{aligned}\label{eq:Frhosol}
F^{L}(\rho)=&\left(\sinh2\rho\right)^{-\iu L_{\star}p_{+}} \left(\tanh\rho\right)^{-\iu L_{\star}p_{-}}\, ,\\
F^{R}(\rho)=&\left(\sinh2\rho\right)^{-\iu L_{\star}p_{-}} \left(\tanh\rho\right)^{-\iu L_{\star}p_{+}}\, .
\end{aligned}
\end{equation}
Then, it turns out that when one inserts these expressions into the equation of motion \eqref{eq:massiveeq}, one simply finds an algebraic equation either for $p_{-}$ or $p_{+}$, 
\begin{equation}
\begin{aligned}\label{eq:p1p2eq}
(\bar{\dal}-m_s^2)h_{(0)}^{L}=0\, &\Rightarrow\, 4 L_{\star} p_{+}^2+4 \iu p_{+}+L_{\star}m_s^2=0\, ,\\
(\bar{\dal}-m_s^2)h_{(0)}^{R}=0\, &\Rightarrow\, 4 L_{\star} p_{-}^2+4 \iu p_{-}+L_{\star}m_s^2=0\, .
\end{aligned}
\end{equation}
Each of these equations has the following solutions
\begin{align}\label{eq:p1p2sol}
h_{(0)}^{L}\Rightarrow\, p_{+}=-\frac{\iu}{L_{\star}}\mathfrak{h}^{s}_{\pm}(m_s^2)\, ,\quad h_{(0)}^{R}\Rightarrow\,p_{-}=-\frac{\iu}{L_{\star}}\mathfrak{h}^{s}_{\pm}(m_s^2)\, ,
\end{align}
where
\begin{equation}\label{eq:hs}
2\mathfrak{h}^{s}_{\pm}(m_s^2)\equiv 1\pm \sqrt{1+L_{\star}^2m_s^2}\, ,
\end{equation}
are, from an holographic perspective, the conformal dimensions of each of the modes of the scalar field. Now, with this result the asymptotic behavior of $h_{(0)}^{L,R}$ is 
\begin{equation}
h_{(0)}^{L,R}\sim e^{-2\rho\, \mathfrak{h}^{s}_{\pm}(m_s^2)}\, ,
\end{equation}
and, since we are only interested in solutions in which only the normalizable mode is active, we choose the ones with $\mathfrak{h}_{\pm}(m_s^2)>0$. When $m_s^2>0$ this means that we only keep the solution with $\mathfrak{h}^{s}_{+}(m_s^2)$. However, for $-1<L_{\star}^2m_s^2<0$, the mode $\mathfrak{h}^{s}_{-}(m_s^2)$ is also normalizable. 

Now, let us introduce the frequency and the momentum of these modes,
\begin{equation}\label{eq:omegapp}
\omega=p_{+}+p_{-}\, ,\quad m=L_{\star}(p_{-}-p_{+})\, ,
\end{equation}
where $m\in \mathbb{Z}$ is an integer, since we take $\theta$ to have a periodicity of $2\pi$. We also introduce the tortoise coordinate 
\begin{equation}
\rho_{*}=L_{\star}\int\frac{\diff \rho}{\sinh(\rho)}=L_{\star}\log\left(\tanh\rho/2\right)\, ,
\end{equation}
and in terms of this, we see that near the horizon $\rho_{*}\rightarrow -\infty$, these solutions behave as
\begin{equation}
h_{(0)}^{L,R}\sim e^{-\iu\omega(t+\rho_{*})+\iu m\phi}\, .
\end{equation}
This corresponds to waves moving towards the horizon, and hence we have shown that these solutions are quasinormal modes.  Their associated quasinormal mode frequencies (QNFs) are obtained from eqs.~\eqref{eq:omegapp} and \eqref{eq:p1p2sol}, and they read
\begin{equation}
\omega_{s}^{(0)}L_{\star}=\pm m-2\iu\, \mathfrak{h}^{s}_{+}(m_s^2)\, ,
\end{equation}
where $+m$ is for the left-moving modes and $-m$ for the right-moving ones. 

Finally, as shown in ref.~\cite{Sachs:2008gt},  one can generate the infinite set of overtones from these fundamental quasinormal modes by applying the combination $L_{-1}\bar{L}_{-1}$, namely
\begin{equation}
h_{(n)}^{L,R}=(L_{-1}\bar{L}_{-1})^{n}h_{(0)}^{L,R}\, .
\end{equation}
These are also quasinormal modes, because they satisfy the appropriate boundary conditions, and the effect of the operation $L_{-1}\bar{L}_{-1}$ is to shift $\omega\rightarrow \omega-2i/L_{\star}$. Therefore, the complete set of quasinormal frequencies for the scalar field is
\begin{equation}
\omega_{s}^{(n)}L_{\star}=\pm m-2\iu\left(n+ \mathfrak{h}^{s}_{+}(m_s^2)\right)\, .
\end{equation}

Let us now move to the case of the massive graviton. The quasinormal modes of this field were computed in ref.~\cite{Myung:2011bn} in the context of NMG, by using the fact that the equation \eqref{eq:massiveeq} can be written as the square of the equation for a chiral massive graviton, and hence one can apply the results of \cite{Sachs:2008gt} for Topologically Massive Gravity \cite{Deser:1981wh,Deser:1982vy}. Here we follow a direct analysis of eq.~\req{eq:massiveeq}. 

We need to find a symmetric, transverse and traceless field $\hat{h}_{ab}$ that satisfies eq.~\req{eq:massiveeq} with the quasinormal boundary conditions. Again, we can apply the technique of ref.~\cite{Sachs:2008gt} and find the chiral highest weight modes, which by definition satisfy
\begin{equation}\label{eq:L1massive}
L_{1}\hat{h}_{(0)ab}^{L}=0\, ,\qquad \bar{L}_{1}\hat{h}_{(0)ab}^{R}=0\, .
\end{equation}
It turns out that an appropriate ansatz for these modes is the following,
\begin{align}
\hat{h}_{(0)ab}^{L}\diff x^{a}\diff x^{b}&=F^{L}(\rho)e^{-\iu u p_{+}-\iu v p_{-}}\left(\diff v+\frac{2L_{\star}\diff \rho}{\sinh2\rho}\right)^2\, ,\\
\hat{h}_{(0)ab}^{R}\diff x^{a}\diff x^{b}&=F^{R}(\rho)e^{-\iu u p_{+}-\iu v p_{-}}\left(\diff u+\frac{2L_{\star}\diff \rho}{\sinh2\rho}\right)^2\, .
\end{align}
These are traceless and, interestingly, the conditions in \req{eq:L1massive} are equivalent to imposing them to be transverse, $\bar\nabla^{a}\hat{h}_{ab}=0$. Thus, we find the following first-order equations for the functions $F^{L,R}(\rho)$, 
\begin{equation}
\begin{aligned}
\left(\frac{2\iu L_{\star}(p_{-}+p_{+}\cosh2\rho)}{\sinh2\rho}+\partial_{\rho}\right)F^{L}&=0\, ,\\
\left(\frac{2\iu L_{\star}(p_{+}+p_{-}\cosh2\rho)}{\sinh2\rho}+\partial_{\rho}\right)F^{R}&=0\, .
\end{aligned}
\end{equation}
These are the same equations as those for the radial functions in the scalar case, and therefore the solutions are given by \req{eq:Frhosol}.
Then, we have to solve the equations of motion \eqref{eq:massiveeq}, and, again, when we insert the previous result, we see that all the components of the equations are reduced to an algebraic equation either for $p_{+}$ or $p_{-}$, 
\begin{equation}
4 L_{\star} p_{\pm}^2-4 i p_{\pm}+L_{\star}m_g^2=0\, .
\end{equation}
The solutions in this case are:
\begin{align}\label{eq:p1p2solg}
\hat{h}_{(0)ab}^{L}\Rightarrow\, p_{+}=-\frac{\iu}{L_{\star}}\mathfrak{h}^{g}_{\pm}(m_g^2)\, ,\quad \hat{h}_{(0)ab}^{R}\Rightarrow\,p_{-}=-\frac{\iu}{L_{\star}}\mathfrak{h}^{g}_{\pm}(m_g^2)\, ,
\end{align}
where
\begin{equation}
\mathfrak{h}^{g}_{\pm}(m_g^2)=\frac{-1\pm \sqrt{1+L_{\star}^2m_g^2}}{2}\, .
\end{equation}
Note the $-1$ rather than the $+1$ with respect to $\mathfrak{h}^{s}_{\pm}(m_g^2)$ in \req{eq:hs}. 
As in the case of the scalar field, it is immediate to verify that all these modes behave as outgoing waves at the horizon (they fall towards the horizon), but at infinity only the modes associated to $\mathfrak{h}^{g}_{+}(m_g^2)$ are normalizable, and hence only these are quasinormal modes. By using again eq.~\eqref{eq:omegapp}, we obtain the frequencies of these fundamental modes
\begin{equation}
\omega_{g}^{(0)}L_{\star}=\pm m-2\iu\mathfrak{h}^{g}_{+}(m_g^2)\, ,
\end{equation}
where the $+m$ case is for left movers and $-m$ for the right movers. 
We note that, in this case, if $m_g^2<0$, the imaginary part becomes positive and the modes become unstable, unlike in the case of scalar perturbations. 
Finally, the overtones can be obtained by applying the operator $L_{-1}\bar{L}_{-1}$, which shifts the imaginary part of the frequency in $-2/{L_{\star}}$, as happened for the scalar. Therefore, the complete set of quasinormal mode frequencies for the massive graviton is
\begin{equation}
\omega_{g}^{(n)}L_{\star}=\pm m-2\iu\left(n+ \mathfrak{h}^{g}_{+}(m_g^2)\right)\, .
\end{equation}

\subsection{General BTZ}
The results from the previous subsubsection can be easily generalized to the BTZ black hole with arbitrary mass and angular momentum by noticing that the metric \eqref{eq:rotatingBTZ} can be mapped to eq. \eqref{eq:unitBTZ}. Starting with the rotating BTZ black hole \eqref{eq:rotatingBTZ}, we can perform the change of variables $
t=\frac{r_{+}L_{\star}}{r_{+}^2-r_{-}^2}\left(\tilde{t}+\frac{r_{-}L_{\star}}{r_{+}}\tilde{\theta}\right)$, $
\theta=\frac{r_{+}L_{\star}}{r_{+}^2-r_{-}^2}\left(\tilde{\theta}+\frac{r_{-}}{r_{+}L_{\star}}\tilde{t}\right)$, $r=\sqrt{r_{-}^2+(r_{+}^2-r_{-}^2)\cosh^2(\rho)}$, leading to the metric 
\begin{equation}\label{eq:unitBTZ3}
\diff s^2=-\sinh^2(\rho) \diff \tilde{t}^2+L_{\star}^2\cosh^2(\rho) \diff \tilde{\theta}^2+L_{\star}^2\diff \rho^2\, ,
\end{equation}
which is locally the same as eq. \eqref{eq:unitBTZ2}. However, the geometry is different, because the ranges of the coordinates are different. In particular, $\tilde\theta$ is not to be considered an angular coordinate. 
Expressed in this way, we can study the perturbations as we did in the previous subsection by introducing the coordinates $\tilde{u}=\tilde{t}+L_{\star}\tilde\theta$, $\tilde v=\tilde t-L_{\star}\tilde\theta$. Then, we search for solutions with a dependence $\sim e^{-\iu\tilde u p_{+}-\iu\tilde v p_{-}}$, and the result of this analysis is the one we have just seen: either $p_{+}$ (for left-moving modes) or $p_{-}$ (for right-moving ones) must have a certain value for quasinormal modes. This is given by eq.~\req{eq:p1p2sol} for the scalar mode and by eq.~\req{eq:p1p2solg} for the massive graviton. 
However, now the frequency and the momentum are  identified by $-\iu\tilde u p_{+}-\iu\tilde v p_{-}=-\iu\omega t+i m\phi$, where $m$ is an integer. This yields the relations
\begin{align}
\omega L_{\star}&=p_{+}(r_{+}-r_{-})+p_{-}(r_{+}+r_{-})\, ,\\
m&=p_{-}(r_{+}+r_{-})-p_{+}(r_{+}-r_{-})\, .
\end{align}
Inserting the appropriate value of $p_{+}$ or $p_{-}$ in these equations, we get the QNFs as a function of the momentum. In full detail, we obtain the following set of frequencies
\begin{align}
\omega_{s}^{L,R}=&\pm \frac{m}{L_{\star}}-\iu\frac{r_{+}\mp r_{-}}{L_{\star}^2}\left(2n+1+ \sqrt{1+L_{\star}^2m_s^2}\right)\, ,\\
\omega_{g}^{L,R}=&\pm \frac{m}{L_{\star}}-\iu\frac{r_{+}\mp r_{-}}{L_{\star}^2}\left(2n-1+ \sqrt{1+L_{\star}^2m_g^2}\right)\, ,
\end{align}
with $n=0,1,\dots$ Additionally, recall that if $-1\le L_{\star}^2m_{s}^2\le 0$, there is a second family of scalar modes, with $-\sqrt{1+L_{\star}^2m_s^2}$.
From the point of view of holography, these results would be interpreted as the poles of the retarded Green functions of a thermal two-dimensional CFT placed on a circle of radius $L_{\star}$.  This can be generalized to a circle of arbitrary radius $R$ by performing a rescaling of the boundary metric, identifying $\omega'=\omega L_{\star} /R$ and $T'=T L_{\star}/R$. Now, the left- and right-moving Virasoro algebras of the $1+1$ dimensional CFT split the theory in two sectors, that in the background of the BTZ black hole have temperatures \cite{Birmingham:2001pj}
\begin{equation}
T^{L}=\frac{r_{+}-r_{-}}{2\pi L_{\star}^2}\, ,\quad T^{R}=\frac{r_{+}+r_{-}}{2\pi L_{\star}^2}\, .
\end{equation}
We can then write the QNFs as 
\begin{align}
\omega_{s}^{L,R}=&\pm \frac{m}{R}-2i\pi T^{L,R}\left(2n+1+ \sqrt{1+L_{\star}^2m_s^2}\right)\, ,\\
\omega_{g}^{L,R}=&\pm \frac{m}{R}-2i\pi T^{L,R}\left(2n-1+ \sqrt{1+L_{\star}^2m_g^2}\right)\, .
\label{eq:massiveomega}
\end{align}
As noted in ref.~\cite{Birmingham:2001pj}, the scalar QNFs precisely match the field theory computation of the poles of retarded thermal correlators, which is considered to be a remarkable test of the AdS/CFT correspondence. Holography tells us that eq.~\req{eq:massiveomega} should compute the poles of the thermal correlators for a CFT dual to a bulk theory with a massive graviton.

\section{A myriad of black holes in electromagnetic quasi-topological gravity}\label{sec:3DBH}

Let us now consider EMQT gravities given by action \eqref{eq:EQTG3D}. It is possible to verify that a solution of the form \eqref{eq:EMQTSSSmetric} with $\phi = p \theta$, $N=1$ and $f$ given by 
\begin{equation}\label{eq:fsolEMQT}
f= \frac{\displaystyle \frac{r^2}{L^2}-\mu-\alpha_1 p^2 \log \frac{r}{L}+ \sum_{i=2} \frac{\alpha_i  p^{2n} L^{2(i-1)}}{2(i-1)r^{2(i-1)}}}{\displaystyle 1+ \sum_{j=0} \frac{ \beta_j p^{2(j+1)} (2j+1) L^{2(j+1)}}{r^{2(j+1)}}} ,
\end{equation}
solves all the components of the equations of motion of the theory \eqref{eq:EOMEMQT1}, \eqref{eq:EOMEMQT2}, and \eqref{eq:EOMEMQT3}. Obviously, if we set all the $\alpha_i$ and the $\beta_j$ to zero, we are left with the usual static BTZ metric with mass $\mu$. Similarly, when only $\alpha_1$ is active, the metric takes the same form as for the charged BTZ black hole \cite{Clement:1993kc,Martinez:1999qi}, a fact which follows from the electromagnetic-dual description of our EMQT action, on which we comment next. Since the couplings $\alpha_i$ and $\beta_j$ can be tuned at will, our solutions represent continuous multiparametric generalizations of the BTZ metric.

Before starting studying the different possible solutions, let usstart with some general comments. Firstly, the number of horizons depends on the number of positive roots of the equation 
\begin{equation}
\frac{r^2}{L^2}-\mu-\alpha_1 p^2 \log \left(\frac{r}{L}\right)+ \sum_{n=2} \frac{\alpha_n p^{2n}L^{2(n-1)}}{2(n-1)r^{2(n-1)}}=0\, ,
\end{equation}
which in turn depends on the values and signs of the $\alpha_n$. 
Turning off all the $\alpha_n$, the solution describes a black hole with a single horizon of radius $r=L \sqrt{\mu}$ whenever $\mu >0$, analogously to the neutral BTZ case. 

 In the next-to-simplest case, corresponding to $\alpha_2 \neq 0$ and $\alpha_1=\alpha_{i\geq 3}=0$, at least one horizon exists as long as $2 p^4 \alpha_2 /\mu^2 \leq 1$ and $\mu >0$. In addition, if $ \alpha_2> 0$, the solution possesses two horizons, except for the case $\alpha_2=   \mu^2/(2p^4)$, which corresponds to an extremal black hole. When they exist, the outer and inner horizons correspond, respectively, to
\begin{equation}\label{rpm}
r_{\pm} = \frac{L \sqrt{\mu}}{\sqrt{2}} \sqrt{1 \pm \sqrt{1- \frac{2 p^4 \alpha_2 }{\mu^2} } }\, .
\end{equation}

If we turn on $\alpha_1$ and turn off the rest of $\alpha_i$, the numerator in the metric function is identical to the whole $f$ corresponding to the charged BTZ metric with the identification $Q^2\equiv 2\alpha_1 p^2$. In that case, the horizons structure was studied in detail in ref.~\cite{Martinez:1999qi}. Whenever $\mu>1$ the metric describes a black hole with two horizons. For $0<\mu<1$, we have black holes for $2\alpha_1 p^2 \leq Q_1^2$ and for $2\alpha_1 p^2 \geq Q_2^2$ where $Q_{1,2}$ are the roots of the equation $\mu=(Q/2)^2[1-\log (Q /2)^2]$. Whenever, the inequalities are saturated, the black holes are extremal. Interestingly, even for negative values of $\mu$ black holes exist as long as $2\alpha_1 p^2 \geq Q_2^2$ is sufficiently large, in particular, as long as $-|\mu|\geq (Q/2)^2[1-\log (Q/2)^2]$ holds. As more $\alpha_i$ are turned on, the number of horizons can increase and the analysis becomes more involved ---see \eg fig.~\ref{fig:bh1} for a couple of examples with three horizons.

As $r\rightarrow 0$, the spacetime described by eq.~\eqref{eq:fsolEMQT} can look very different, depending on the value of the combination $i_{\rm max}+2-j_{\rm max}$, where we define $i_{\rm max}$ and $j_{\rm max}$ as the largest values of $i$ and $j$ corresponding to non-vanishing $\alpha_i$'s and $\beta_j$'s. In particular, the metric function goes as $f\overset{r\to\infty}{\sim} r^{2(j_{\rm max}+ 2 -i_{\rm max})}$ ---where, for convenience, if all the $\beta_m$ are zero, we define $m_{\rm max}\equiv -1$ and $\beta_{-1}\equiv -1$. 
The case $i_{\rm max}=1$ is slightly different and reads instead $f\overset{r\to\infty}{\sim }r^{2(j_{\rm max}+1)} \log r$. We study the different situations in the following subsections.

\subsection{Black holes with curvature singularities}
\begin{figure}
\begin{minipage}[c]{0.45\textwidth}
\vspace{-1cm}
	\includegraphics[width=1\textwidth]{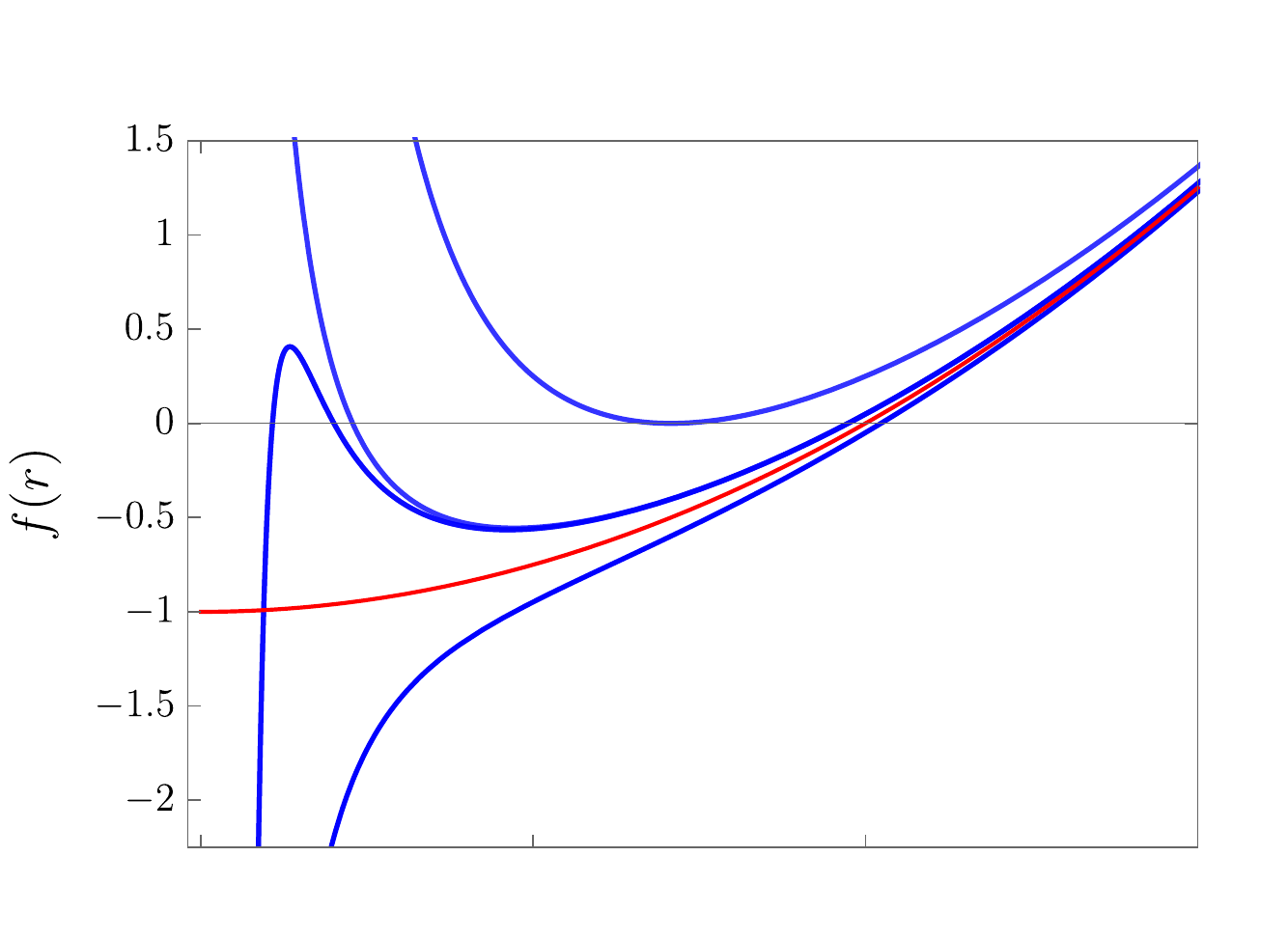}\vspace{-1cm}
	\includegraphics[width=0.975\textwidth]{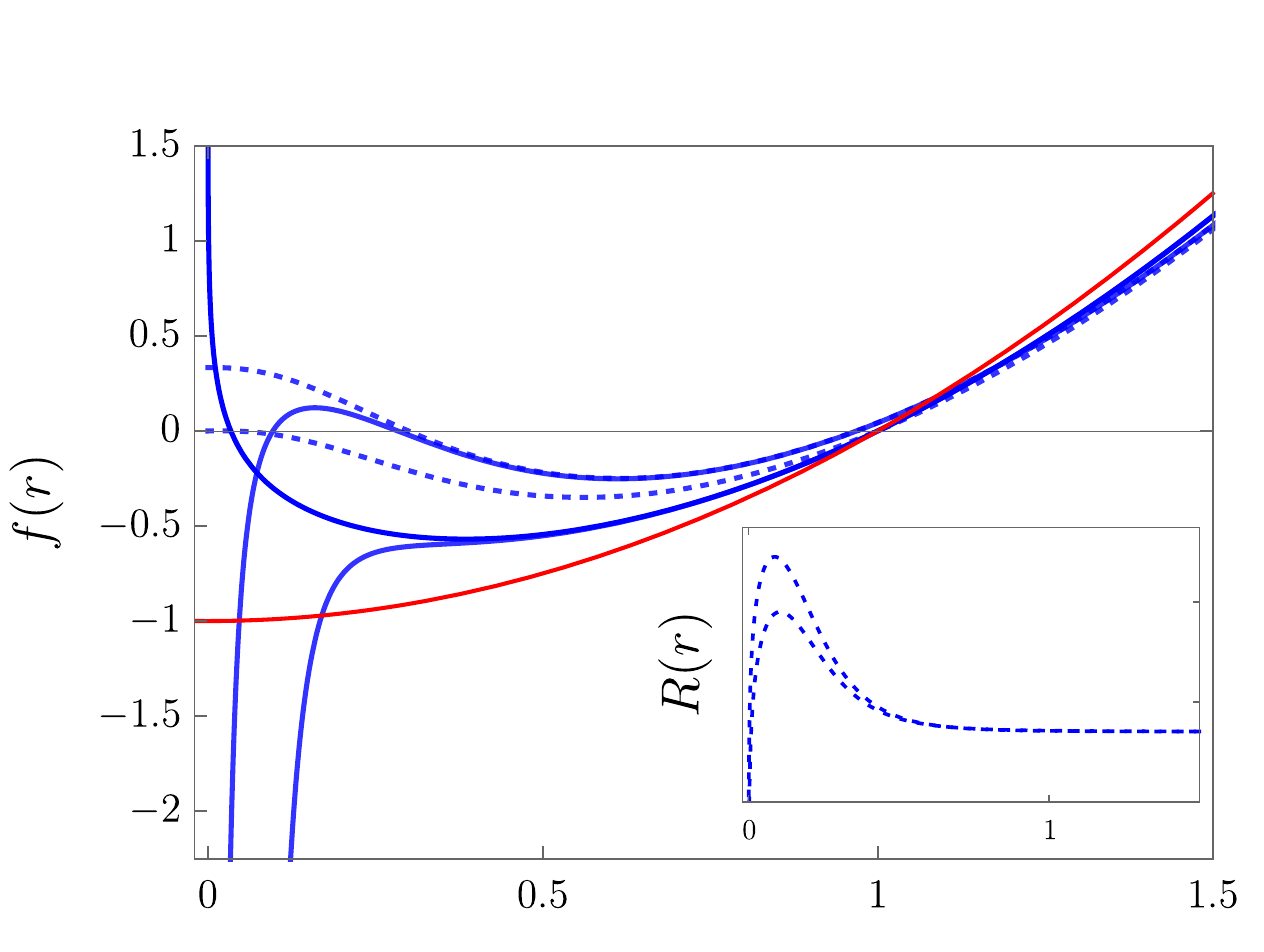}
	\end{minipage}\hfill
	\begin{minipage}[c]{0.545\textwidth}
	\caption{(Top) We plot $f(r)$ for four black hole solutions with curvature singularities at the origin possessing one, three, two and one (degenerate) horizons respectively (blue curves from bottom to top). The curves are obtained for $L=1$, $\mu=1$, $p= 2/3$ in the cases: $\alpha_2=-1/2$; $\alpha_2=1/2$, $\alpha_3=-1/50$; $\alpha_2=1/2$; and $\alpha_2=81/32$ respectively (unspecified coupling values equal zero).  
	(Bottom) We plot $f(r)$ for three black hole solutions possessing two, three and one horizon (thick blue curves moving from upper-left corner towards lower-right corner). The curves are obtained for the same values of $L,\mu,p$ in the cases: $\alpha_1=2/3$; $\alpha_1=2/3$, $\alpha_2=1/2$, $\alpha_3=-1/50$, $\beta_0=1/2$; and $\alpha_1=2/3$, $\alpha_3=-1/50$.  The dotted lines correspond to solutions for which the Ricci scalar diverges logarithmically at the origin, as shown in the inset plot. These correspond to $\alpha_1=2/3$, $\beta_0=1/2$; and $\alpha_1=2/3$, $\alpha_2=1/2$, $\beta_0=1/2$.  The red line corresponds to the usual BTZ black hole in both plots.}
	\label{fig:bh1}
	\end{minipage}
\end{figure}

An important set of solutions corresponds to black holes possessing a curvature singularity at $r=0$, hidden behind one or several horizons. This situation occurs whenever $f$ contains at least a real zero, and either $i_{\rm max} > j_{\rm max} +2$ or $i_{\rm max}=1$, $j_{\rm max}=-1$ hold. We plot examples of configurations of these kinds in fig.~\ref{fig:bh1}. Curvature invariants diverge at the origin in these cases. For instance, the Ricci scalar behaves as
\begin{equation}
R \overset{r\rightarrow 0}{= } - \frac{c_{i_{\rm max},j_{\rm max}} \alpha_{i_{\rm max}} L^{2(i_{\rm max}-j_{\rm max}-2)}}{\beta_{j_{\rm max}} r^{2(i_{\rm max} - j_{\rm max} -1)}} \, ,
\end{equation}
where $c_{i_{\rm max},j_{\rm max}}\equiv (2i_{\rm max}- 2j_{\rm max}- 5)( i_{\rm max}- j_{\rm max}-2)p^{(2i_{\rm max}-2j_{\rm max}-2)}/[(1+2j_{\rm max}) (i_{\rm max}-1)]$  is a positive constant for all $i_{\rm max}$ and $j_{\rm max}$.  

Slightly special are the cases corresponding to $i_{\rm max}=1$, $j_{\rm max}=0$ and $i_{\rm max}=2$, $j_{\rm max}=0$ (with a non-vanishing $\alpha_1$). For those, the Ricci scalar diverges logarithmically as $r\rightarrow 0$ even though $f$ tends there, respectively, to zero and to a constant. The  dotted lines in fig.~\ref{fig:bh1} correspond to these two cases.

\subsection{Black holes with BTZ-like and conical singularities}
The usual static BTZ black hole is locally equivalent to pure AdS$_3$ \cite{Banados:1992wn,Banados:1992gq}. All curvature invariants are constant and the spacetime is therefore very different at the origin from the one of the black holes considered in the previous subsection. For the BTZ, the spacetime contains a ``sort of''  conical singularity for general values of $\mu$ different from $-1$ (which precisely corresponds to pure AdS$_3$), hidden behind a horizon whenever $\mu > 0$ and naked whenever $\mu < 0$ (the $\mu=0$ case describes the so-called ``black hole vacuum'' \cite{Banados:1992gq}).  Indeed, when the mass parameter is negative and different from minus one, the $(r,\theta)$ components of the metric have the same signature, $\diff r^2/|\mu| +r^2 \diff \theta^2 $, which corresponds to a standard conical singularity with deficit angle $2\pi (1- \sqrt{|\mu|})$. On the other hand, when the mass parameter is positive, we have instead $-\diff r^2/|\mu| +r^2 \diff \theta^2 $, which has a singularity in the causal structure at $r=0$ which resembles the one of a Taub-NUT space \cite{Banados:1992gq} ---in particular, the spacetime is no longer Hausdorff. 


Both kinds of singularities ---conical and BTZ-like--- appear for some of the new black holes considered here. The situation described takes place for $i_{\rm max} = j_{\rm max} +2$ (with $i_{\rm max}\geq 3$ if $\alpha_1\neq 0$). In that case, the metric function and the curvature invariants tend to constant values at $r=0$. For instance, if the only active couplings are $\alpha_l$ and $\beta_{l-2}$, the metric function tends to the constant value
\begin{equation}
f\overset{r\rightarrow 0}{=}\frac{\alpha_l p^2 }{2 (l-1)(2l-3) \beta_{l-2}} \, ,
\end{equation}
whereas the Ricci scalar vanishes as  $\sim r^2$. Then, whenever the quotient $\alpha_l/\beta_{l-2}$ is positive, we have a conical singularity, and whenever it is negative, we have a BTZ-like one. The analysis is similar as more couplings are turned on. We plot examples of both kinds of solutions in the upper plot of fig.~\ref{fig:bh2}.


\subsection{Singularity-free black holes}
Whenever the metric function tends to $1$ at $r=0$,
\begin{equation}\label{reg11}
f(r) \overset{r\rightarrow 0}{ =}1\, ,
\end{equation}
 the angular defect present at $r=0$ disappears and the metric becomes regular there ---as mentioned earlier, this is precisely what happens with the BTZ metric for the special value $\mu=-1$, for which it reduces to pure AdS$_3$.

\begin{figure}[t!] \centering \vspace{-0.6cm}
\begin{minipage}[c]{0.45\textwidth}
\includegraphics[width=1\textwidth]{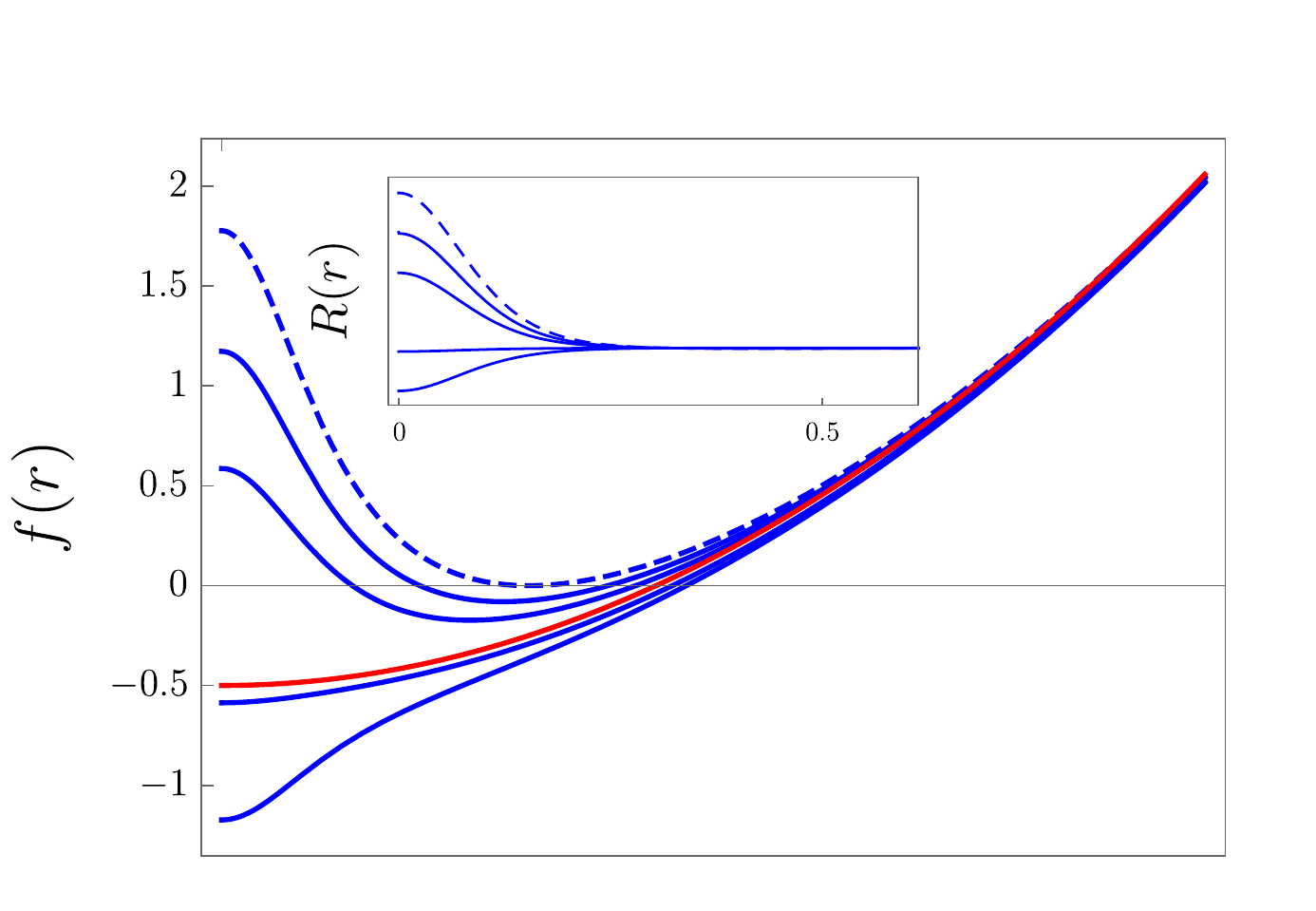}\vspace{-1cm}
\includegraphics[width=1\textwidth]{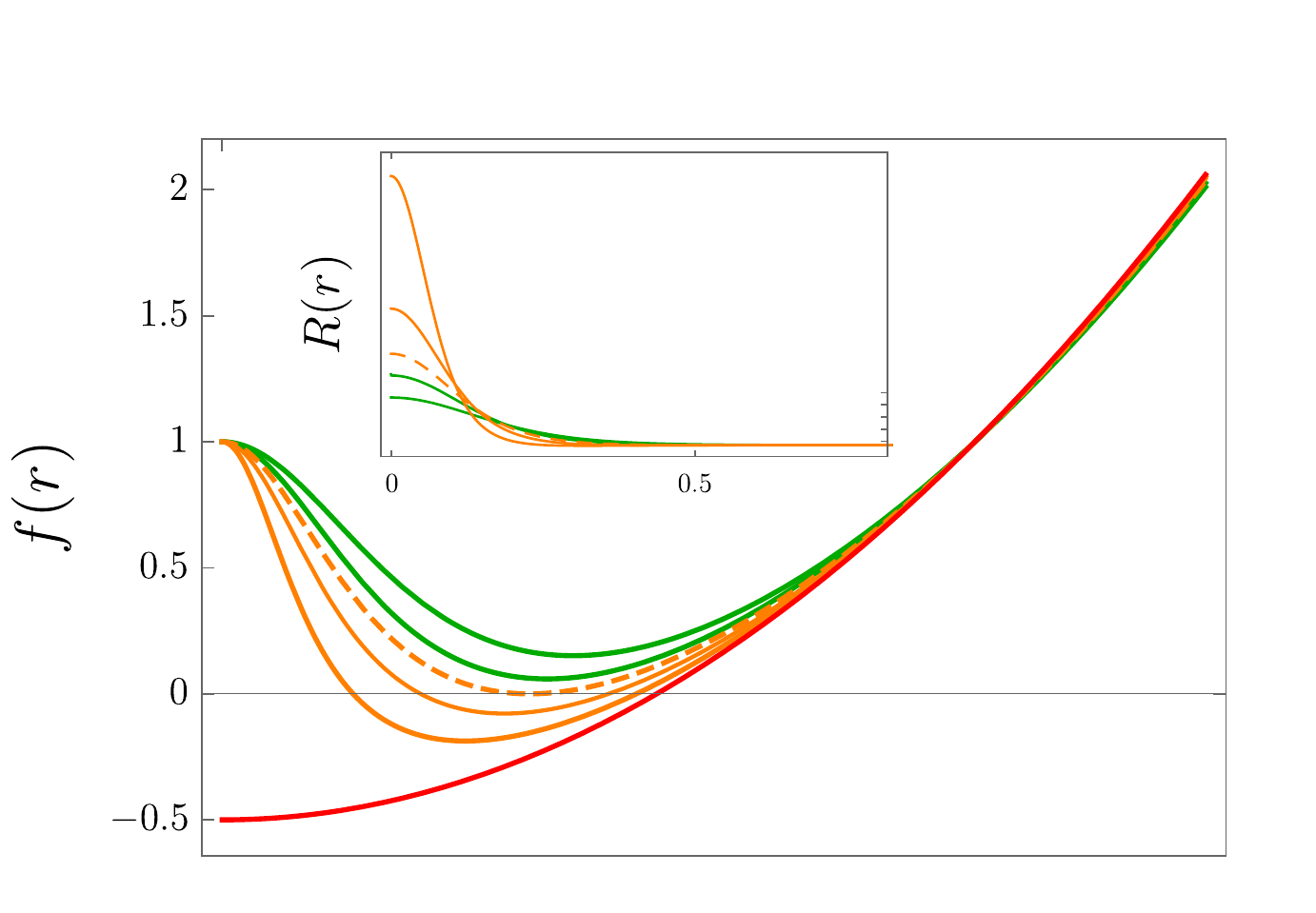}\vspace{-1cm}
\includegraphics[width=1\textwidth]{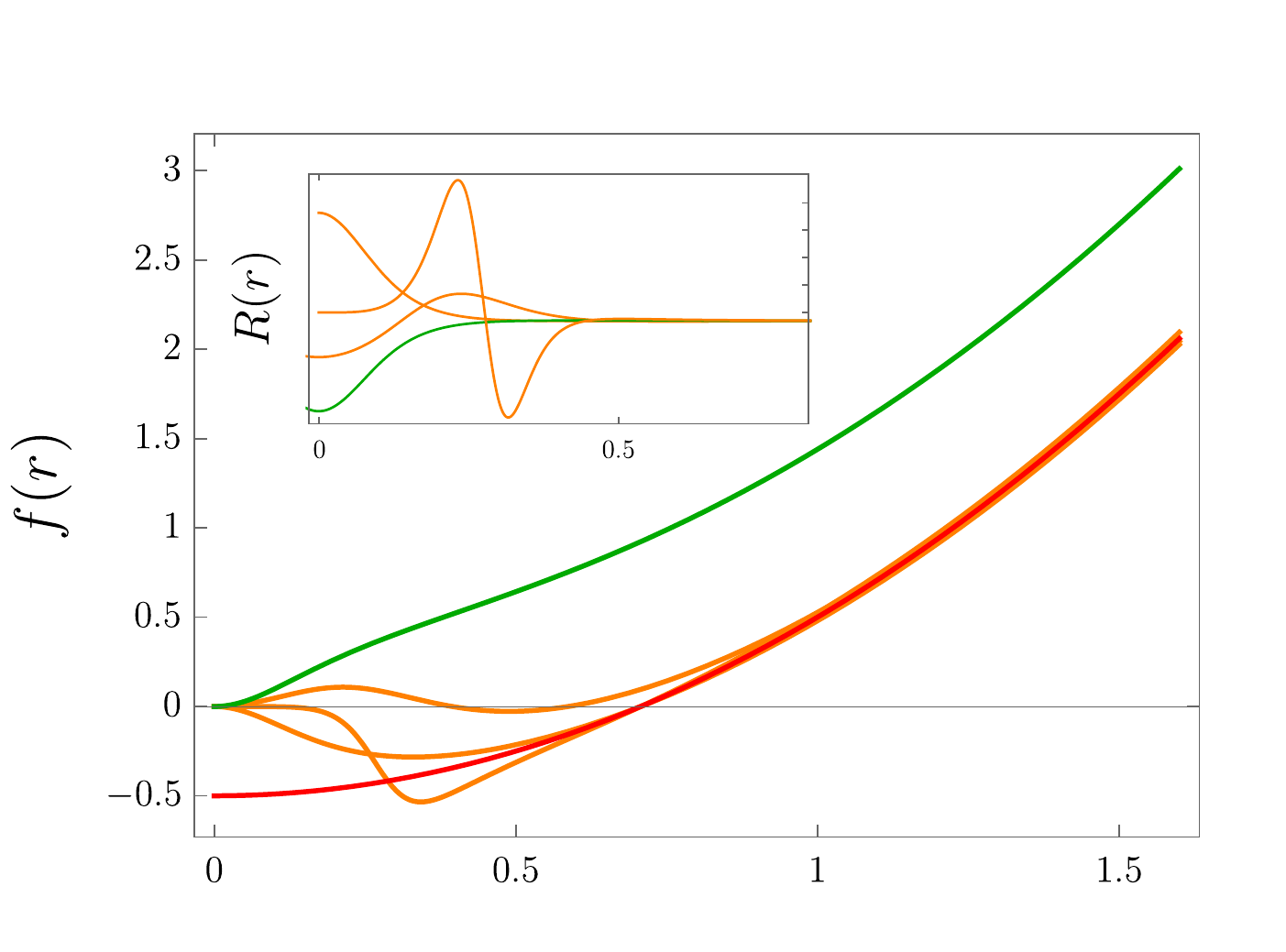}
	\end{minipage}\hfill
\begin{minipage}[c]{0.545\textwidth}
	\caption{\sffamily We plot $f(r)$ and the Ricci scalar (inset) in various cases (we set $L=1$, $\mu=1/2$, $p= 3/8$): (Top) Three black holes with a conical singularity at the origin with one (extremal), two and two horizons respectively, as well as two black holes with  BTZ-like singularities with one horizon  (blue curves from top to bottom: $\beta_0=1/4$, $\alpha_2=512/81\xi$ with $\xi=1;2/3;1/3;-1/3,-2/3$). (Middle) Two black holes with two horizons (thick orange curves), an extremal black hole (dashed) and two regular horizonless solutions (green curves): $\alpha_2=512\xi /81$ and $\beta_0=4\xi/9$ for (lower to upper): $\xi=1/3$; $\xi=2/3$; $\xi=1$; $\xi= 4/3$; $\xi=2$. (Bottom) Three regular black holes with two, one and one horizon respectively and one horizonless solution whose metric functions vanish as even powers at the origin (orange curves: $\alpha_2=54/10$, $\beta_0=8/27$, $\beta_1=1/6$; $\beta_0=8/27$; and $\beta_0=-8/27$, $\beta_1=-1/12$, $\beta_2=1/20$; and green: $\mu=-1/2$, $\beta_0=8/27$). }
	\label{fig:bh2}
		\end{minipage}\hfill
\end{figure}

Our new solutions include non-trivial profiles for which this happens. 
When the spacetime contains at least one horizon, those describe singularity-free black holes. We plot examples of this kind  in the middle plot of figure \ref{fig:bh2}. For instance, if the only active couplings are $\alpha_l$ and $\beta_{l-2}$, the regularity condition (\ref{reg11}) becomes $\beta_{l-2}=\alpha_l p^2/[2(l-1)(2l-3)]$. The simplest example corresponds to the case $\alpha_{1,i\geq 3}=\beta_{j\geq 1}=0$, $p^2=2\beta_0/\alpha_2$. Then, the EMQT action \eqref{eq:EQTG3D} becomes
\begin{equation}\label{sing1}
\frac{\pazocal{Q}}{L^2}=\alpha_2  (\partial \phi)^4-\beta_0 [3 R^{bc}\partial_b\phi\partial_c \phi- (\partial \phi)^2R ]\, ,
\end{equation}
and the solution reads
\begin{align}\label{sing2}
f=\frac{\displaystyle \frac{r^2}{L^2}-\mu +\frac{2 \beta_0^2 L^2}{ \alpha_2 r^2 }}{\displaystyle 1+\frac{2 \beta_0^2 L^2}{ \alpha_2 r^2 }}\, , \quad \phi = \theta \sqrt{\frac{2\beta_0}{\alpha_2}} \, .
\end{align}
For the above metric, the Ricci scalar tends to the constant value $R=3(1+\mu)\alpha_2/( L^2 \beta_0^2)$ at the origin.
As explained at the beginning of this section ---see eq.~\req{rpm}--- this metric can describe up to two horizons  depending on the values of $2p^4 \alpha_2/\mu^2=8\beta_0^2/(\alpha_2 \mu^2)$. 

There are additional ways to achieve singularity-free black holes within the present setup which do not require imposing any constraint at all. The idea is to consider metrics for which $f$ vanishes as some positive power of $r$ near the origin,
\begin{equation}\label{reg22}
f(r) \overset{r\rightarrow 0}{ =} \pazocal{O}(r^{2s})\, , \quad s\geq 1\, ,
\end{equation}
 with the curvature invariants tending to constant values there (being also finite everywhere else). This happens whenever $j_{\rm \ssc max}>i_{\rm \ssc max}-2$ if $i_{\rm \ssc max} \geq 2$; whenever some $\beta_j$ is active and all the $\alpha_i$'s are zero; and whenever $j_{\rm \ssc max} \geq 1$ if $i_{\rm \ssc max}=1$. For those, the point $r=0$ becomes a sort of new asymptotic region. The Ricci scalar behaves there as $R \sim \pazocal{O} (r^{2(j_{\rm \ssc max}-i_{\rm \ssc max}+1)})$ for $i_{\rm \ssc max} \geq 1$ and as $R \sim \pazocal{O} (r^{2m})$ if all the $\alpha_i$'s are turned off. We present examples of this kind in the bottom plot of figure \ref{fig:bh2}. There are of course infinitely many possibilities, but let us mention explicitly the simplest one. This corresponds to setting $\alpha_{i\geq 1}=0$ and $\beta_{j\geq 1}=0$. Then, the EMQT Lagrangian \eqref{eq:EQTG3D} is given now by
\begin{align}\label{sing3}
\frac{\pazocal{Q}}{L^2}=-\beta_0[3 R^{bc}\partial_b\phi\partial_c \phi- (\partial \phi)^2R ]\, ,
\end{align}
and the solution reads
\begin{align}\label{frl}
f(r)=\frac{\displaystyle \frac{r^2}{L^2}-\mu}{\displaystyle 1+\frac{ \beta_0L^2 p^2 }{  r^2 }}\, , \quad \phi= p \theta\, , 
\end{align}
where all the constants: $\mu$, $p$, $\beta_0$ and $L^2$ are free parameters (the only conditions being $\mu>0$, $\beta_0\geq 0$, which ensure that a horizon exists and that there are no poles in the denominator). As expected, curvature invariants are finite everywhere for this solution ---see inset figure in bottom plot of fig.~\ref{fig:bh2}.

 In contrast with most previous attempts at achieving regular black holes, these solutions do not require any sort of: i) complicated functional dependences of the action fields; ii) fine tuning of action parameters or constraint between those and the physical charges; iii) addition of specially selected matter. For the Lagrangian density in eq.~\req{eq:EQTG3D} with $\pazocal{Q}$ given by equation \eqref{sing3}, black holes simply turn out to be singularity-free ---and analogously in the rest of the cases described. 

\subsection{Regular horizonless solutions}
Solutions behaving as equations \eqref{reg11} or \eqref{reg22} with finite curvature invariants everywhere do not necessarily include horizons.\footnote{Examples of three-dimensional regular horizonless solutions for Einstein gravity have been previously constructed \eg in ref.~\cite{Edery:2020kof}. } When they do not, we are left instead with globally regular and horizonless spacetimes. For example, the solutions in  \req{sing2} and in \req{frl} with $\mu \leq 0$ describe configuration of this kind. Examples of regular horizonless solutions are shown in green in the bottom and middle plots of figs.~\ref{fig:bh2}.

\subsection{Adding rotation}\label{rotat}
The static solutions described by the single-function SSS \eqref{eq:EMQTSSSmetric} can be easily generalized into rotating ones by performing a boost in the $t$ and $\theta$ coordinates,
\begin{equation}
t\rightarrow \gamma t-L\omega \theta\, , \quad \theta\rightarrow \gamma\theta -\omega t/L\, .
\end{equation}
The ``trick'' is that this change of variables is only defined locally, so that the global structure of the resulting spacetimes is different from their static counterparts ---see \eg  \cite{Dias:2002ps,CamposDias:2003tv}. Assuming that $\gamma^2-\omega^2=1$ ---so that for $\omega=0$ the metric reduces to the static one---  we get
\begin{align}\label{eq:rotatingmetric}
\diff s^2&=-\frac{r^2f}{\rho^2}\diff t^2+\frac{\diff r^2}{f}+\rho^2\left[\diff \theta+N^{\theta}\diff t\right]^2\, ,\\ \phi&=p\left[ \gamma\theta-\frac{\omega t}{L}\right]\, ,
\end{align}
where
\begin{align}
\rho^2 &\equiv r^2-\omega^2[L^2f-r^2]\, ,\\ N^{\theta} &\equiv \frac{\gamma\omega [L^2f -r^2]}{L \rho^2}\, .
\end{align}
The rotating solution typically has the same horizons as the static one plus an additional one at $r=0$ if $f$ tends to a non-vanishing constant there. 

\section{Discussion}

In this chapter we have studied the quasinormal modes and quasinormal frequencies of the BTZ black hole in arbitrary higher-curvature gravity in full generality. This is possible because in three dimensions it appears as solutions of all higher-order gravities. However, the higher derivative gravities introduce additional degrees of freedom. These are a scalar mode and a massive graviton. 

On the other hand, we presented a new family of solutions in three dimensions to EQT described in sec.~\ref{sec:EQT3D}, which
involving a non-minimally coupled scalar field and describe different kinds of solutions depending on the values of the various couplings and parameters. Among them, we have presented black holes with different numbers of horizons and conical, BTZ-like or curvature singularities appear in some cases. In others, singularities are absent from the geometry and the solutions describe singularity-free black holes or globally regular solutions. This is achieved in two ways, characterized by the behavior of $f$ near the origin ---see eqs.~\req{reg11} and \req{reg22} respectively and orange curves in the middle and bottom plots of fig.~\ref{fig:bh2}. Interestingly, for the second type of solutions, this behavior is encountered without imposing any kind of constraint between the action parameters and/or physical charges.

%

%% file: text/ch5-hentang.tex
\chapter{Shape-dependence in holographic entanglement entropy}\label{chap:sdhee}

Let us now study holographic entanglement entropy dual to Einstein-AdS gravity. As mentioned before, the Kounterterms scheme allows us to isolate the universal contribution to holographic entanglement entropy, canceling the divergences present in this quantity. 

As in this chapter we focus on odd-dimensional CFTs, we first discuss some generalities regarding this particular dimension.

\section{Renormalized holographic entanglement entropy in odd-dimensional CFTs}

In sec.~\ref{sec:holorenS} we reviewed previous results on renormalization of holographic entanglement entropy. A key step in the derivation of the renormalized quantity for even-dimensional manifolds was to add to the Euclidean action the Chern form \eqref{eq:SEEren} and the combination in the orbifold \eqref{eq:finalLM}. However, prior to that, it is convenient for us to express the renormalized Euclidean action in terms of quantities defined in the bulk. To that end we employ the Euler-Chern theorem \eqref{eq:Cherntheo}, which allows to rewrite the renormalized Euclidean action as
\begin{equation}\label{eq:Iren1}
\IEren=\frac{1}{16\pi \GN}\int_{\B}\diff^{d+1}x\sqrt{g}(R-2\Lambda+c_{d}\X_{d+1})-\frac{(-1)^{\frac{d-1}{2}}}{4\GN}\frac{\pi^dL^{d-1}}{\Gamma(d/2)}\chi[\B].
\end{equation}
In ref. \citep{Anastasiou:2018rla}, it was shown that the quantity inside the integral in the above formula can be rewritten in terms of a polynomial of the tensor
\begin{equation}\label{FAds}
\left({F_{\text{AdS}}}\right)\indices{_a_b^c^d}=R\indices{_a_b^c^d}+\frac{1}{L^2}\delta_{ab}^{cd},
\end{equation}
known as ``AdS curvature''. In doing so, the action adopts the form
\begin{equation}\label{eq:Iren2}
I^{\text{ren}}_{\text{E}}=\frac{L^{d-1}}{16\pi \GN}\int_{\B}\diff^{d+1}x\sqrt{g}P_{d+1}(F_{\text{AdS}})-\frac{(-1)^{\frac{d-1}{2}}}{4\GN}\frac{\pi^dL^{d-1}}{\Gamma(d/2)}\chi[\B].
\end{equation}
where the polynomial of the AdS curvature introduced reads
\begin{align}\label{eq:Pol}
P_{d+1}(F_{\text{AdS}})=\frac{1}{2^{\frac{d-1}{2}}(d+1)\Gamma(d)}&\sum_{k=1}^{\frac{d+1}{2}}\frac{(-1)^k[(d+1-2k)]!2^{\frac{d+1-2k}{2}}}{L^{d+1-2k}}\binom{\frac{d+1}{2}}{k}\notag\\ &\times\delta_{b_1\ldots b_{2k}}^{a_1\ldots a_{2k}}\left(F_{\text{AdS}}\right)\indices{_{a_1}_{a_2}^{b_1}^{b_2}}\cdots \left(F_{\text{AdS}}\right)\indices{_{a_{2k-1}}_{a_{2k}}^{b_{2k-1}}^{b_{2k}}}
.
\end{align}
The AdS curvature, of particular convenience in AdS gravity, measures the deviation of the space with respect to global AdS. Notice that the renormalized action consists on the addition of two terms: a topological one, given by the Euler characteristic of the manifold, and another one, characterized by the AdS curvature. This decomposition has been earlier found on the mathematical literature \citep{alexakis2010renormalized} in connection to the concept of renormalized area. It reflects the equivalence between this quantity and renormalized EE, up to a proportionality constant that depends on the dimension of the manifold \citep{Anastasiou:2018rla}. The AdS curvature will be of central importance afterwards for the renormalized EE of deformed entangling surfaces, as information on the deformation is entirely contained in the polynomial.

Once we have the renormalized form of the Euclidean action $I^{\text{ren}}_{\text{E}}$, we evaluate it on the conically singular manifold $\pazocal{M}_{2n}^{(\vartheta)}$ in order to use eq.~\eqref{eq:finalLM}. To do so, we recall the properties of curvature invariants defined on squashed cone manifolds addressed in sec.~\ref{sec:geoentang}. In ref.~\citep{Anastasiou:2018rla}, it was shown that the Einstein-AdS action evaluated on the orbifold consists on the sum of a regular part and a term localized at the conical defect. The explicit form is
\begin{align}
I^{\text{ren}}_{\text{E}}[\B_{\vartheta}]=&\frac{L^{d-1}}{16\pi \GN}\int_{\B}\diff^{d+1}x\sqrt{g}L^{d-1}P_{d+1}(F_{\text{AdS}})-\frac{(-1)^{\frac{d-1}{2}}}{4\GN}\frac{\pi^{d/2}L^{d-1}}{\Gamma(d/2)}\chi[\B]\notag\\
&+T_{\vartheta}\Area^{\text{ren}}[\Gamma_V]+\pazocal{O}\left[(1-\vartheta)^2\right],\label{eq:Iren3}
\end{align}
where, following the notation introduced in ch.~\ref{ch:intro}, $\B\equiv\B_\vartheta\setminus\Gamma_V$ now represents the manifold far away from the conical singularity and $T_\vartheta=\frac{1-\vartheta}{8\GN}$ represents the tension of the codimension-two cosmic brane. The renormalized area of the codimension-two surface reads
\begin{equation}\label{eq:Vol}
\Area^{\text{ren}}\left[\Gamma_V\right]=-\frac{L^{d-1}}{2(d-2)}\int_{\Gamma_V}\diff^{d-1}x\sqrt{{}_{\Gamma_V}g}P_{d-1}({}_{\Gamma_V}F_{\text{AdS}})-\frac{(-1)^{\frac{d-1}{2}}}{4\GN}\frac{\pi^dL^{d-1}}{\Gamma(d/2)}\chi[\Gamma_V],
\end{equation}
is the renormalized area of the codimension-two manifold, where ${}_{\Gamma_V}F_{\text{AdS}}$ is the codimension-two AdS curvature of \eqref{FAds}. It is important to stress that this expression for the renormalized area is generic and not restricted to minimal surfaces. In particular, when $\Gamma_V$ satisfies the minimal condition \eqref{eq:minK}, corresponds to the RT surface. We make that situation explicit by denoting the surface as  $\Gamma_V^{\text{RT}}$.

When introducing the renormalized Euclidean action evaluated in the orbifold \eqref{eq:Iren3} in formula \eqref{eq:finalLM}, the only surviving term corresponds to the renormalized area of the codimension-two surface, \ie
\begin{equation}\label{eq:RTRen}
S^{\text{ren}}_{\EE} \left(V\right )=\frac{\Area^{\text{ren}}\left(\Gamma_V^{\text{RT}}\right)}{4\GN},
\end{equation}
From this expression, we see that the computation of the renormalized entanglement entropy depends on AdS curvature and the Euler characteristic of the codimension-two surface, attending to expression \eqref{eq:Vol}.

This calculation can be equivalently be interpreted as the renormalized area of a tensionless codimension-two brane $\Gamma_V^{\text{RT}}$ embedded in a $d+1$-dimensional AAdS Einstein spacetime, for a  minimal surface $\Gamma_V$ \citep{Dong:2016fnf}.

For a spherical entangling surface $\Sigma=\mathbb{S}^{d-2}$, the polynomial $P_{d-1}({}_{\Gamma_V}F_{\text{AdS}})$ vanishes identically. The contribution to the holographic entanglement entropy is coming uniquely from the topology of the RT surface, which is an hemisphere. Because the Euler characteristic is $\chi\left[\Gamma_V^{\text{RT}}\right]=1$, the finite part of the entanglement entropy of a ball-shaped surface takes the form
\begin{equation}\label{eq:RenEgen}
\suniv(\mathbb{B}^{d-1})=\frac{(-1)^{(d-1)/2}}{4 \GN}\frac{\pi^{d/2}L^{d-1}}{\Gamma(d/2)},
\end{equation}
where we have re-expressed the result in terms of the odd-dimensional $d$ of the CFT. Notice that this result is in agreement with the universal part of the EE \cite{Nishioka:2018khk}. 

As discussed in sec.~\ref{sec:RGflows},  $S^{\text{ren}}_{\EE}(\mathbb{B}^{d-1})=\suniv(\mathbb{B}^{d-1})$ is equivalent to the free energy, $F_0$, of a CFT$_d$ on a spherical background $\mathbb{S}^d$ \eqref{eq:CHMmap}.

 Once the general picture has been discussed, we will illustrate explicitly the duality AdS$_4$/CFT$_3$ in this context by particular examples.

\subsection{Entanglement entropy in AdS$_4$/CFT$_3$ in the global coordinate patch}\label{sec:example}

The use of the Kounterterms in the renormalization of holographic entanglement entropy has been applied for spatial entangling regions embedded on a flat background, in the Poincar\'e-AdS patch \eqref{eq:Poincare}, in the context of the AdS/CFT correspondence \citep{Anastasiou:2017xjr}. In particular, in what follows, we study the entanglement entropy of a polar cap-like entangling region immersed on an Einstein Static Universe background (ESU), \ie $\mathbb{R} \times \mathbb{S}^2$ to account for properties of a CFT$_3$. In this case, the dual bulk geometry is given by global AdS$_4$ spacetime, whose line element is given in eq.\eqref{eq:globalcoord} but particularizing $\diff\Omega_2^2=\diff \theta^2+\sin^2\theta\diff\varphi^2$ as the metric of $\mathbb{S}^2$.

Consider a two-dimensional orthogonal spacetime, homologous and cobordant to $V$, that is spanned along the directions $i=t,r$. The induced metric ${}_{\Gamma_V}g$ of such a surface is given by
\begin{equation}
\diff s^2_{\Gamma_V}=-\left(1+\frac{r^2}{L^2}\right)\diff t^2+\left(r^2+\frac{L^2 {r'}^2}{L^2+r^2}\right)\diff \theta^2+r^2\sin^2\theta\diff\theta^2
\end{equation}
where we have parametrized the geometry with the embedding function $r=r(\theta)$ and $r'=\partial_\theta r(\theta)$. The surface $\Gamma_V$ must satisfy the minimality condition \eqref{eq:minK} in order to be the RT surface.

Solving the second order differential equation that results from eq. \eqref{eq:minK}, we find that the RT surface is characterized by \citep{Hubeny:2007xt,Hubeny:2012wa,Bakas:2015opa} the function
\begin{equation}
r^2(\theta)=\frac{L^2\cos^2\theta_0}{\cos^2\theta\sin^2\theta_0-\sin^2\theta\cos^2\theta_0}.
\end{equation}
For this embedding, the polynomial $P_{2}\left({}_{\Gamma_V^{\text{RT}}}F_{\text{AdS}}\right)$ in eq.~\eqref{eq:Vol} vanishes identically, as it is a constant-curvature subspace. The only nonvanishing part is the topological one, for which the universal part of the entanglement entropy takes the form
\begin{equation}\label{eq:Srenglobalads}
S^{\text{ren}}_{\EE}(\mathbb{B}^2)=-\frac{\pi L^2}{2 \GN}.
\end{equation}
Thus, even though this time the spherical entangling surface is  immersed in the curved background of ESU metric, eq. \eqref{eq:Srenglobalads} matches the one obtained for the flat case \citep{Anastasiou:2017xjr}.

\subsection{Renormalized entanglement entropy of a deformed disk}\label{RenEx}

\begin{figure}
\centering
\includegraphics[width=0.6\textwidth]{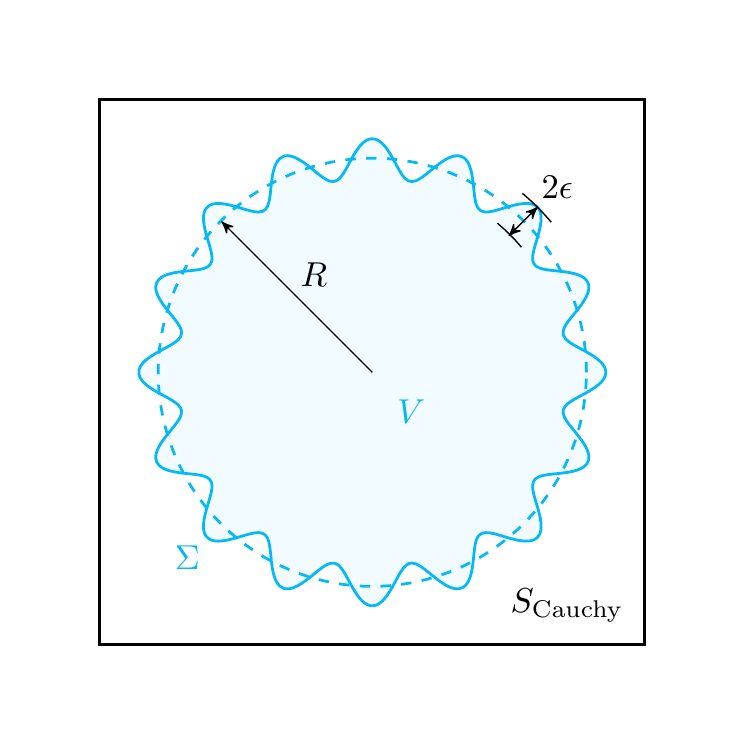}
\caption{\textsf{Small perturbation of the spherical entangling surface of radius $R$ fparametrized by expression \eqref{fiy}.}}\label{fig:mezei}
\end{figure}

We now study the effect of considering small perturbations around the disk-like entangling region. From the results in refs.~\cite{Allais:2014ata,Mezei:2014zla,Faulkner:2015csl} we know that the finite part of entanglement entropy is locally maximized for a disk region in arbitrary CFTs. In particular, let us consider a slightly deformed disk $\mathbb{B}^2_\epsilon$ defined by the polar equation
\begin{equation}\label{fiy}
r(\theta)=R \left[1+ \epsilon \sum_{\ell} \left( \frac{a^{(c)}_{\ell}}{\sqrt{\pi}} \cos (\ell \theta) + \frac{a^{(s)}_{\ell}}{\sqrt{\pi}} \sin (\ell \theta) \right) \right]\, , \quad \epsilon \ll 1\, ,
\end{equation} 
(where $a^{(c)}_{\ell}$, $a^{(s)}_{\ell}$ are some numerical coefficients which determine the shape of the perturbation). This parametrization of the entangling surface is represented in fig~\ref{fig:mezei}. 
The effect of such deformation on $\suniv(\mathbb{B}^2_\epsilon)=-F(\mathbb{B}^2_\epsilon)$ at leading order in $\epsilon$ is given, for general CFTs, by the three-dimensional version of Mezei's formula \cite{Mezei:2014zla,Faulkner:2015csl}
\begin{equation}\label{fmeze0}
F(\mathbb{B}^2_\epsilon)=F_0+ \epsilon^2 \, \frac{\pi^3 C_{\ssc T} }{24} \sum_{\ell} \ell (\ell^2-1)
\left[ (a^{(c)}_{\ell})^2+(a^{(s)}_{\ell})^2 \right] + \pazocal{O}(\epsilon^4) \, , \end{equation}
where $C_{\ssc T}$ is the coefficient which controls the flat-space stress-tensor two-point function charge \eqref{eq:ctdef} and $F_0$ is the finite part of entanglement entropy for a disk without deformation. Notice that, in this expression, everything is determined by the geometry of the entangling region with the exception of $C_{\ssc T}$, which is the only theory-dependent piece.

We shall study the deformations of CFTs dual to Einstein-AdS gravity in two coordinate systems:  polar coordinates (following \citep{Allais:2014ata}) and spherical coordinates (in order to make contact with refs.\citep{Mezei:2014zla} and \citep{Faulkner:2015csl}). Using the Kounterterms, we make contact with the renormalized area of the RT surface \eqref{eq:Vol}, which contains both local (curvature) and global (topological) terms \citep{Anastasiou:2018mfk}. Our analysis below allows us to track the origin of the shape-dependent contributions to the curvature part in eq.\eqref{eq:Vol}.

\subsubsection{Deformed disk in polar coordinates} 
 
\begin{figure}
\centering
\includegraphics[width=0.8\textwidth]{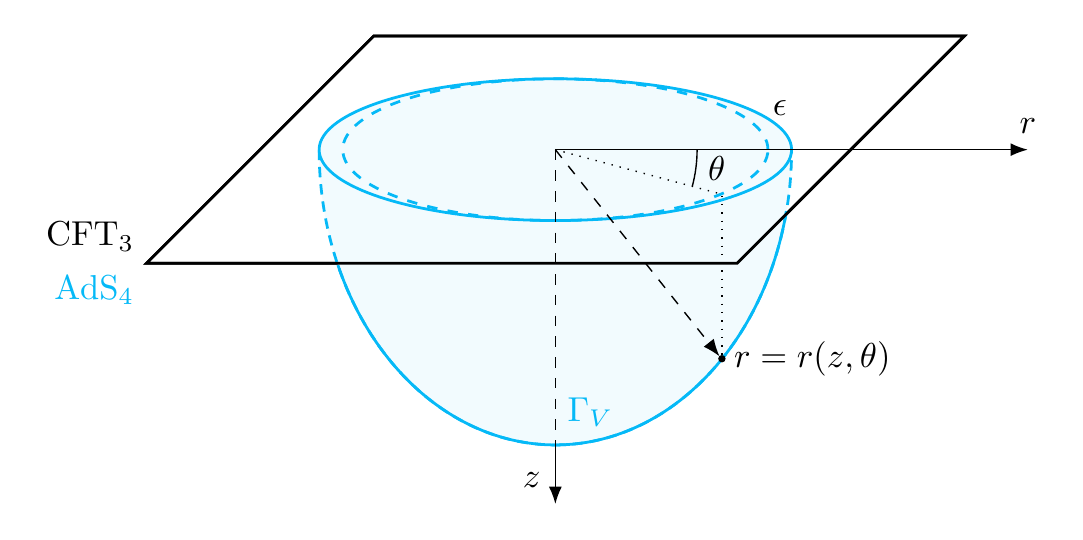}
\caption{\textsf{Time slice of minimal codimension-two surface ${\Gamma_V}$ with an elliptical deformation $\epsilon$ ($\ell=2$).}}\label{fig1}
\end{figure}

Consider the Euclidean version of Poincar\'e-AdS$_4$ spacetime (see eq.~\eqref{eq:Poincare} for the explicit expression), written in polar coordinates.
We define the embedding function of the RT surface $\Gamma_V^{\rt}$ by $r(z,\theta)$, where $r$ and $\theta$ are the radial and the angular coordinate at the boundary, respectively. The deformation breaks the azimuthal symmetry of $\Gamma_V^{\rt}$. Hence, the simplification used in the section \ref{sec:example} is not applicable. In this case, the codimension-two induced metric reads
\begin{equation}\label{indEAdS}
\diff s^2_{\Gamma_V^{\rt}}=\frac{L^2}{z^2}\left[\left(1+{r'}^2\right)\diff z^2+\left(r^2+{{\dot{r}}}^2\right)\diff\theta^2+2r'\dot{r}\diff z\diff \theta\right],
\end{equation}
where $r'=\partial_z r(z,\theta)$ and $\dot{r}=\partial_\theta r(z,\theta)$. It is indeed easy to find the equations of motion of the RT surface following eq. \eqref{eq:minK}. If we consider the binormal directions as $i=t,r$, we find that
\begin{equation}
\K\indices{^{(r)}_z^z}+\K\indices{^{(r)}_\theta^\theta}=0,
\end{equation}
provided that the temporal foliation is constant, what implies into $\K\indices{^{(t)}_z^z}=\K\indices{^{(t)}_\theta^\theta}=0$. This leads to the equations of motion
\begin{equation}\label{eom}
\frac{r\left(1+r'^2\right)}{m z^2}-\partial_z\left(\frac{r^2r'}{m z^2}\right)-\frac{1}{z^2}\partial_\theta\left(\frac{\dot{r}}{m}\right)=0,
\end{equation}
where we have introduced an auxiliary function
\begin{equation}
m=m(z,\theta)=\sqrt{r^2\left(1+{r'}^2\right)+\dot{r}^2}\ .
\end{equation}
In absence of deformations, the embedding function \eqref{eom} is parametrized by a hemisphere of unit radius, $r^2=1-z^2$. The shape can be deformed as linear perturbations around the unitary circle of the form $r(\theta)=\left[1+\epsilon f(\theta)\right]$, where $\epsilon$ is the deformation parameter \citep{Allais:2014ata}. Altogether, we assume that its embedding in AdS$_4$ geometry is given by the ansatz
\begin{equation}\label{ansatz1}
r(z,\theta)=\sqrt{1-z^2}\left[1+\epsilon f(z,\theta)\right],
\end{equation}
for the separation of variables $f(z,\theta)=R(z)\Theta(\theta)$. The corresponding functions satisfy the conditions  $R(0)=1$ and $\Theta(\theta)=\Theta(\theta+2 \pi)$ at the conformal boundary. This is a consequence of the homologous constraint on the RT surface, as it is anchored to the conformal boundary $z=0$. An additional condition comes from the fact that the maximum reach of the embedding does not change when the RT surface is deformed, what leads to $R(1)=0$ \citep{Allais:2014ata,Hubeny:2012ry} (see fig. \ref{fig1}).

Solving eq.~\eqref{eom} for $R(z)$ and $\Theta(\theta)$, we obtain
\begin{equation}\label{solution1}
r(z,\theta)=\sqrt{1-z^2}\left[1+\epsilon\sum_{\ell}\left(\frac{1-z}{1+z}\right)^{\ell/2}\frac{1+\ell z}{1-z^2} \left(a_\ell\cos(\ell\theta)+b_\ell\sin(\ell\theta)\right)\right],
\end{equation}
where $\ell$ is the degree of the harmonic function and labels the deformation with respect to the circle.

Once we have obtained the embedding function \eqref{solution1}, we are able to compute the renormalized entanglement entropy for the perturbed circle by using eq.\eqref{eq:RTRen}. For Einstein gravity in four dimensions this reads
\begin{equation}\label{Sren2}
\suniv(\mathbb{B}_\epsilon^2 )=-\frac{\pi L^2}{2 \GN}\chi[\Gamma_V^\rt]+\frac{L^2}{8 \GN} \int_{\Gamma_V^\rt} \diff^2 x \sqrt{{}_{\Gamma_V^\rt}\gamma}\, {}_{\Gamma_V^\rt} F_{\text{AdS}}.
\end{equation}
As we pointed out before, the information of the deformation is encoded in the second term, characterized by ${}_{\Gamma_V^\rt} F_{\text{AdS}}$. Replacing the embedding function \eqref{solution1} into eq. \eqref{Sren2}, we obtain the expression for the renormalized EE for the perturbation considered, given by
\begin{equation}\label{SEHe}
\suniv \left(\mathbb{B}_\epsilon^2 \right)=-\frac{\pi L^2}{2 \GN}\left[1+\epsilon^2\sum_{\ell}\frac{\ell\left(\ell^2-1\right)}{4}(a^{2}_\ell+b^2_{\ell})+\pazocal{O}(\epsilon^4)\right].
\end{equation}
This result is in agreement with the holographic computation for an arbitrary perturbation of a circle performed in ref.~\cite{Allais:2014ata}.

\subsubsection{Deformed disk in spherical coordinates} 
\begin{figure}
\centering
\includegraphics[width=0.8\textwidth]{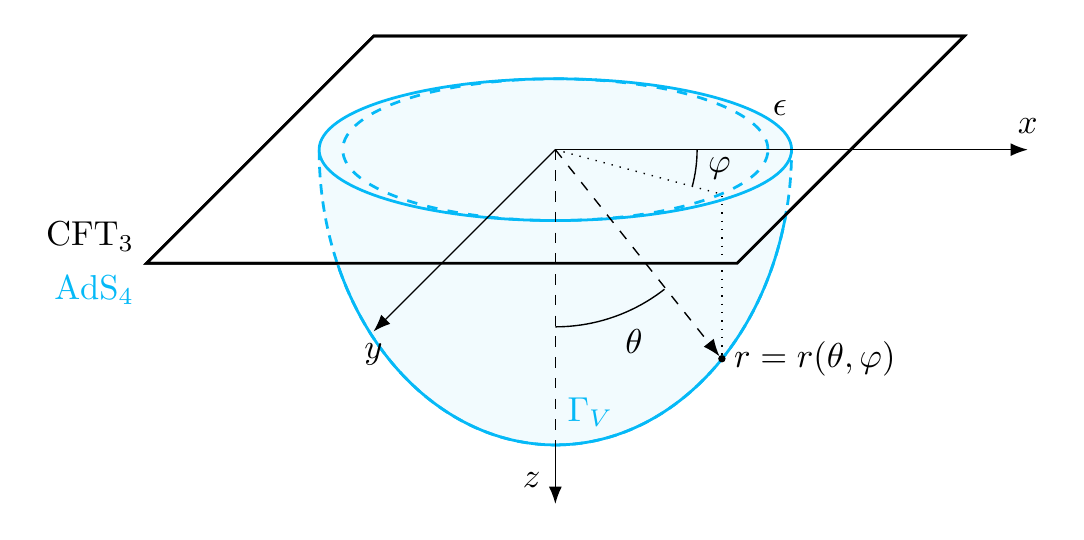}
\caption{\textsf{Time slice of the  extremal codimension-two surface ${\Gamma_V}$ with an elliptical deformation $\epsilon$ ($\ell=2$).}}\label{fig2}
\end{figure}

Consider now the Euclidean version of Poincar\'e-AdS spacetime written in spherical coordinates as
\begin{equation}
\diff s^2=\frac{L^2}{r^2\cos^2\theta}\left(\diff t^2+\diff r^2+r^2\diff\theta^2+r^2\sin^2\theta\, \diff \varphi^2\right).
\end{equation}
Polar and spherical coordinates are mapped into each other by the transformation
\begin{align}\label{sphericalcoords}
r=\sqrt{r^2+z^2},\quad
\theta=\arctan \frac{r}{z}.
\end{align}
In this coordinate system, the embedding function of the minimal surface $\Gamma_V^\rt$ is defined by $r=r(\theta,\varphi)$, such that the induced metric is
\begin{equation}\label{gammaSphe}
\diff s^2_{\Gamma_V^\rt}=\frac{L^2}{r^2\cos^2\theta}\left[\left(1+{r'}^2\right)\diff \varphi^2+\left(1+{\dot{r}}^2\right)\diff\theta^2+2r'\dot{r}\diff\theta\diff\varphi\right],
\end{equation}
where we denoted ${r'}=\partial_\theta r (\theta,\varphi)$ and ${\dot{r}}=\partial_\varphi r (\theta,\varphi)$. The minimality condition \eqref{eq:minK} leads to the equation  for $r(\theta,\varphi)$,
\begin{equation}\label{EOMsphe}
\frac{1}{m r^3\cos^2\theta}\left({r'}^2\sin^2\theta +\dot{r}^2\right)+\partial_\theta\left(\frac{r'\tan^2\theta }{r^2m}\right)+\frac{1}{\cos^2\theta}\partial_\theta\left(\frac{\dot{r}}{r^2 m}\right)=0,
\end{equation}
with the corresponding auxiliary function
\begin{equation}
m=m(\theta,\varphi)=\sin\theta\sqrt{1+\frac{{r'}^2+\dot{r}^2}{r^2}}.
\end{equation}
From eq.~\eqref{EOMsphe}, in the undeformed case, the parametrization of the embedding function of the RT surface is given by the unit hemisphere, $r^2=1$. In a similar fashion as in the previous parametrization, we consider the linear perturbation of the entangling region as \citep{Mezei:2014zla}
\begin{equation}\label{ansatz2}
r(\theta,\varphi)=1+\epsilon f(\theta,\varphi).
\end{equation}
For a choice $f(\theta,\varphi)=\Theta(\theta)\Theta(\theta)$, the boundary conditions correspond to a periodic function $\Phi$ with period $2 \pi$ and $\Theta\rightarrow 1$ at the conformal boundary, \ie $\Theta(\frac{\pi}{2})=1$. Here, the maximal reach of the RT surface implies $\Theta(0)=0$ (see Figure \ref{fig2}). Thus, eq.\eqref{EOMsphe} for the ansatz \eqref{ansatz2} leads to a solution of the form 
\begin{equation}\label{embedding2}
r(\theta,\varphi)=1+\epsilon\sum_\ell\tan^\ell\frac{\theta}{2}(1+\ell\cos\theta)\left[a_\ell\cos(\ell\theta)+b_\ell\sin(\ell\theta)\right]\,.
\end{equation}
For this embedding function, the only nonvanishing AdS curvature component reads
\begin{equation}
\left({}_{\Gamma_V^\rt}F_{\text{AdS}}\right)\indices{^\theta^\varphi_\theta_\varphi}=-\epsilon^2\sum_\ell\frac{\ell^2(\ell^2-1)}{\pi L^2}\left(a_{\ell}+b_\ell\right)\tan\left(\frac{\varphi}{2}\right)^{2\ell}\cot^4\varphi+\pazocal{O}(\epsilon^3).
\end{equation}
Introducing this expression into eq.~\eqref{Sren2}, yields
\begin{equation}
\suniv(\mathbb{B}_\epsilon^2)=-\frac{\pi L^2}{2 \GN}\left[1+\epsilon^2\sum_{\ell}\frac{\ell\left(\ell^2-1\right)}{4}(a^{2}_\ell+b^2_{\ell})+\pazocal{O}(\epsilon^4)\right].
\end{equation}
what matches exactly the result \eqref{SEHe} of the previous subsection, and, in turn, agrees with the formula for $d=3$ in ref.~\citep{Mezei:2014zla}.

A quick analysis of the results above leads to the fact that the $\pazocal{O} \left(\epsilon^2\right)$ contribution is coming only from the curvature part in formula \eqref{Sren2}. Indeed, the information on the deformation of the entangling region is only contained in the polynomial ${}_{\Gamma_V^\rt}P_{2}$. As it shall be discussed below, this behavior can be explained once the equivalence between the renormalized entanglement entropy and the renormalized area of the RT surface \eqref{eq:Vol}, is taken into account.

Continuous perturbations on the hemisphere do not modify its topology, leaving intact the Euler characteristic in eq.~\eqref{eq:Vol}. As a consequence, its shape dependence is encoded only on the local properties of the manifold, which are reflected in the polynomial in the curvature.

The term of the renormalized entanglement entropy that is quadratic in the perturbation carries information on ``entanglement susceptibility'', associated to the change of shape of the entangling region \citep{Nozaki:2013wia,Nozaki:2013vta,Bhattacharya:2014vja,Faulkner:2015csl,Witczak-Krempa:2018mqx}. This quantity contains universal information due to the coefficient $C_{\ssc T}$ of the two-point correlation function of the energy-momentum tensor in a ground state of the  CFT$_3$. Indeed, the subleading term of the formula \eqref{SEHe}, can equivalently be written as
\begin{equation}
\suniv(\mathbb{B}^2_\epsilon)=\frac{\pi^4 C_T}{24}\sum_\ell \ell(\ell^2-1)\left(a_\ell^2+b_\ell^2\right).
\end{equation}
This expression\footnote{In ref.\citep{Mezei:2014zla}, the proportionality constant differs by a factor $\pi$. This corresponds to a different normalization for the spherical harmonics, leading to an overall factor $1/\sqrt{\pi}$ for each one of them.}
 makes manifest the analogy between the entanglement susceptibility and $C_{T}$.
A posteriori, one can say that the AdS curvature of a deformed entangling region is a geometrical probe of $C_{T}$.

From this analysis we see that the leading-order contribution is a shape-independent constant that corresponds to the universal part of the entanglement entropy of a circular entangling surface. Its relation to the sphere free energy \eqref{eq:CHMmap} provides firmer ground to a connection between the topology and the effective number of degrees of freedom of the field theory.

\section{Renormalized area and Willmore energy}\label{sec:Willmore}

\subsection{Minimal and non-minimal surfaces}

The connection between quantum information theoretic measures and geometry can be extended beyond entanglement entropy. Dong in ref.~\cite{Dong:2016fnf} showed that a similar area formula is valid for the calculation of the modular entropy. In this case, the codimension-2 hypersurface $\Gamma_V^{T_\vartheta}$ is not minimal, but its location is determined by the minimization of the Nambu-Goto action of a cosmic brane with tension $T_\vartheta$.

The prescription used for the cancellation of divergences in the holographic entanglement entropy of entangling surfaces is linked to the area renormalization given in the mathematical literature \cite{alexakis2010renormalized}. As shown in ref.~\cite{Anastasiou:2018mfk}, isolating the finite contribution of the modular entropy amounts to the renormalization of the area of ${\Gamma_V^{T_\vartheta}}$
\begin{equation}\label{renmodular}
\tilde{S}_m^\ren =\frac{\Area^\ren\left ({\Gamma_V^{T_\vartheta}}\right )}{4\GN}.
\end{equation}
Interestingly enough, both quantities, entanglement entropy and modular entropy, are described by the same geometrical object, the renormalized area of a codimension-two surface $\Gamma_V$. In four-dimensional Einstein-AdS gravity, the corresponding renormalized area of $\Gamma_V$ is given by eq.~\eqref{eq:Vol} and reads
\begin{equation} \label{renvolume}
\Area^\ren\left(\Gamma_V \right) = -2\pi L^{2}\chi[\Gamma_V] +\frac{L^2}{4}\int_{\Gamma_V}\diff^{2}x\sqrt{{}_{\Gamma_V}g}\, \delta_{kl}^{ij}\left({}_{\Gamma_V}R\indices{_i_j^k^l} +\frac{1}{L^{2}}\delta _{ij}^{kl}\right).
\end{equation} 
This expression matches the renormalized area expression given in ref.~\citep{Fischetti:2016fbh}. Notice that this formula holds whether $\Gamma_V$ is minimal or not.
A physical example of minimal surface is a soap film that spans between two wires. As there is no pressure difference between the sides, the membrane has zero mean curvature. In turn, soap bubbles are non-minimal, due to the difference of pressure at the interface \cite{isenberg1978science, reilly1982mean}. In the latter case, they are constant mean curvature surfaces, and modelled by Helfrich energy \cite{hopf2003differential}.


For extremal surfaces, minimality condition amounts to the vanishing of the trace of the extrinsic curvature  of the surface $\Gamma_V$ \eqref{eq:minK}. For this reason, it is useful to rewrite eq.~\eqref{renvolume} in terms of ${\K^{\left(m\right)}}_{ij} $ using the Gauss-Codazzi relation, for codimension-2 surfaces
\begin{equation}\label{gausscodazzi}
R\indices{_i_j^k^l}={}_{\Gamma_V}R\indices{_i_j^k^l}-2\K\indices{^{(m)}_{[i}^k}\K\indices{^{(m)}_{j]}^l}.
\end{equation}
Taking into account the antisymmetry of Kronecker delta, we find that
\begin{small}
\begin{equation}
\Area^\ren\left (\Gamma_V\right) = -2\pi L^{2}\chi\left (\Gamma_V\right) +\frac{L^{2}}{4}\int_{\Gamma_V}\diff^{2}x\sqrt{{}_{\Gamma_V}}\delta_{kl}^{ij}\left(R\indices{_i_j^k^l}+2\K\indices{^{(m)}_i^k}\K\indices{^{(m)}_j^l}+\frac{1}{L^2}\delta_{ij}^{kl}\right).
\end{equation}
\end{small}
In addition, the Weyl tensor for Einstein-AdS spaces can be written as $ 
W\indices{_a_b^c^d} =R\indices{_a_b^c^d} +\frac{1}{L^{2}}\delta^{ab}_{cd}$,
what allows us to express the renormalized area of $\Gamma_V$ as
\begin{equation}\label{renvol1}
\Area^{\ren}\left (\Gamma_V\right ) = -2\pi L^{2}\chi \left[\Gamma_V\right]+\frac{L^{2}}{4}\int_{\Gamma_V}\diff^{2}x\sqrt{{}_{\Gamma_V}}\delta_{kl}^{ij}\left(W\indices{_i_j^k^l}+2\K\indices{^{(m)}_i^k}\K\indices{^{(m)}_j^l}\right).
\end{equation}
In turn, the extrinsic curvature in eq.~\eqref{renvol1} can be decomposed into its trace ${\K^{(m)}}_{ij}$ and a traceless part ${\K^{(m)}}_{\langle ij\rangle}$ as
\begin{equation}
{\K^{(m)}}_{ij} =\K\indices{^{(m)}_{\langle ij\rangle}}+\frac{1}{2}{}_{\Gamma_V}g_{ij}{\K^{(m)}}.
\end{equation}
We can also replace the trace by the mean curvature $H^{(m)}$, which expresses a linear combination of the eigenvalues of $\K\indices{^{(m)}_i_j}$, that is, $H^{(m)} ={\K^{(m)}}/2$. Armed with these tools, we deduce that the renormalized area in eq.~\eqref{renvolume} can be equivalently written as

\begin{small}
\begin{equation}\label{renvolnonminimal}
\Area^{\ren}\left (\Gamma_V\right ) = -2\pi L^{2}\chi[\Gamma_V] +\frac{L^{2}}{2}\int_{\Gamma_V}\diff^{2}x\sqrt{{}_{\Gamma_V}g}\left(W\indices{_i_j^i^j} +2{H^{(m)}}^{2} -\K\indices{^{(m)}_{\langle ij\rangle}}\K\indices{^{(m)}^{\langle ij\rangle}}\right).
\end{equation}
\end{small}
When a minimal two-dimensional surface $\Gamma_V^\rt$ is considered, the above relation reduces to
\begin{small}
\begin{equation}\label{renvolminimal}
\Area^{\ren}\left (\Gamma_V^\rt\right ) = -2\pi L^{2}\chi [\Gamma_V^\rt]+\frac{L^{2}}{2}\int_{\Gamma_V^\rt}\diff^{2}x\sqrt{{}_{\Gamma_V^\rt}g}\left(W\indices{_i_j^i^j}-\K\indices{^{(m)}_{\langle ij\rangle}}\K\indices{^{(m)}^{\langle ij\rangle}}\right).
\end{equation}
\end{small}
The last two equations are in agreement with the result of Alexakis and Mazzeo in ref.~\cite{alexakis2010renormalized} for the renormalized area of submanifolds. These equivalent expressions allow us to tell between the two prescriptions in eqs.\eqref{eq:RTRen} and \eqref{renmodular}: the RT surface satisfies the minimality condition while the cosmic brane used in modular entropy not. Renormalized area relations in eqs.~\eqref{renvolnonminimal} and \eqref{renvolminimal} can be further simplified when considering entangling regions for a vacuum CFT. Since its gravity dual is global AdS$_{4}$ spacetime, which is conformally flat, the bulk Weyl tensor vanishes identically.

\subsection{Willmore energy of the RT double copy}

The relation between renormalized entanglement entropy and the renormalized area of the RT surface has two key ingredients. On one hand, the topology of the minimal surface, expressed by the Euler characteristic, which captures global properties of $\Gamma_V^\rt$. On the other hand, the local properties of $\Gamma_V^\rt$ are dictated by the AdS curvature term inside the integral in eq.~\eqref{FAds}.

According to the analysis in sec.~\ref{RenEx}, the deformation in the shape of a disk entangling region is encoded only at the curvature part of the renormalized entanglement entropy, leaving the topological contribution unchanged. To analyze the shape-dependence, let us introduce a functional called Willmore energy.

\begin{defi}\
Let us consider a general, smooth, orientable, closed surface, $\Xi$ embedded in $\mathbb{R}^3$, the Willmore energy is defined as \cite{willmore1965note,willmore1996riemannian,Toda2017Willmore}
\begin{equation}\label{eq:Willmore}
\pazocal{W}\left (\Xi\right ) \equiv\int_{\Xi}H^{2}\diff S
\end{equation}
where $H\equiv (k_1+k_2)/2$ is the mean curvature constructed from the principal curvatures  $k_1$ and $k_2$ of ${\Xi}$ and $\diff S$ the surface element.
\end{defi}
This quantity has a lower bound that will become of great relevance in our discussion below. Let us discuss further about it.

\begin{theorem}[Willmore, 1965 \cite{willmore1965note}] 
If $\Xi$ is a simply connected surface, the Willmore energy has a global lower bound, given by
\begin{equation}\label{eq:Willb}
\pazocal{W}\left(\Xi\right)\geq4\pi,
\end{equation}
which is saturated when $\Xi$ is a spherical surface.
\end{theorem}

\noindent
\textit{Proof.}  In order to see why this is the case, one can modify the Willmore functional \eqref{eq:Willmore} subtracting the Gaussian curvature $K=k_1k_2$ as
\begin{equation}\label{WillmoreDef2}
\tilde{\pazocal{W}}\left(\Xi\right)=\int_{\Xi} \left(H^2-K\right)\diff S=\pazocal{W}\left(\Gamma_V\right)-\int_{\Xi} K\diff S .
\end{equation}
The study of both versions of the functional is equivalent. This is because from the Gauss-Bonnet theorem $\int_{\Xi} K\diff S=2\pi \chi[\Xi]$, where $\chi[\Xi]$ is the Euler characteristic of the surface. Hence, for fixed genus, $\tilde{\pazocal{W}}$ is just a constant shift  with respect to $\pazocal{W}$. The interest of the modified functional $\tilde{\pazocal{W}}$ is manifest when written in terms of the principal curvatures
\begin{equation}
\tilde{\pazocal{W}}\left(\Xi\right)=\frac{1}{4}\int_{\Xi} \left(k_1-k_2\right)^2\diff S \, . 
\end{equation}
It immediately follows that $\tilde{\pazocal{W}}\left({\Xi}\right)$ has a sign, $\tilde{\pazocal{W}}\left(\Xi\right)\geq 0$, and also that the bound is exclusively saturated for a totally umbilical surface, $k_1=k_2$, which in the present context is just a fancy name for a sphere \cite{willmore1996riemannian}.
After this discussion, it is straightforward to check that the bound for the original functional \eqref{eq:Willb} is found from \eqref{WillmoreDef2} particularizing the Euler characteristic to the  sphere case, $\chi[\mathbb{S}^2]=2$. $\square$

If the surface is not simply connected, \ie it has a higher genus higher than zero, then the bound is higher. It was recently proven that, for surfaces with genus one ${\Xi}^{g=1}$, the Willmore energy is bounded by $\pazocal{W}\left({\Xi^{g=1}}\right)\geq2\pi^2$ \cite{marques2014willmore}.

These bounds are fundamental for the analysis below, where we establish the connection between Willmore energy and quantum information theoretic measures, through the concept of renormalized area of the entangling surface.

\begin{figure}
\centering
\includegraphics[width=.5\linewidth]{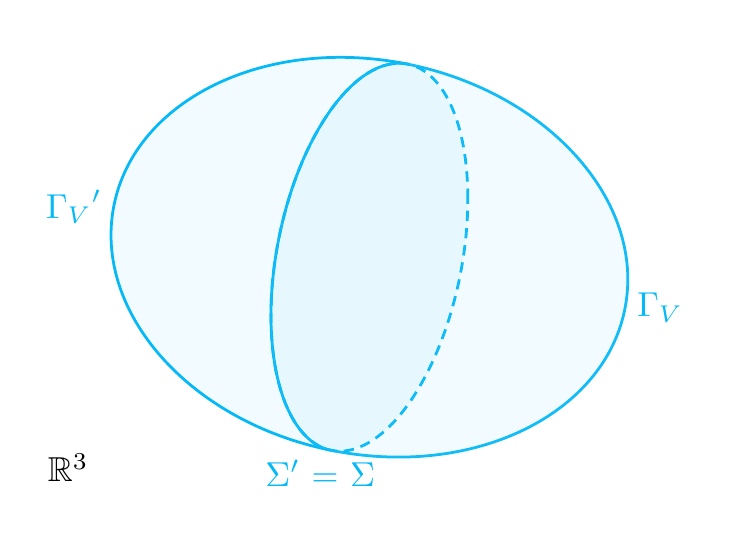}
\caption{\textsf{The doubling of the minimal surface ${\Gamma_V}$ is achieved by gluing a copy ${\Gamma_V}'$ so that they are cobordant $\Sigma=\Sigma'$.}}\label{fig:will}
\end{figure}

We stress that Willmore energy is defined for a closed manifold. Therefore, the first obstacle we face is that the formula \eqref{renvolume} invoves an open two-dimensional surface anchored to the boundary of an AAdS space. In order to overcome this problem, we generalize the field doubling method proposed in ref.~\citep{Fonda:2015nma}, for AAdS manifolds. In this case, we consider the embedding of the codimension-two surface ${\Gamma_V} $ and its reflection with respect to the $z =0$ plane, ${\Gamma_V} ^{ \prime }$. The intersection of ${\Gamma_V} $ and ${\Gamma_V} ^{ \prime }$ is the entangling curve $ \partial V$, at the conformal boundary, such that $\partial{\Gamma_V}=\partial{\Gamma_V}'=\partial V$. Continuity conditions at the interface situated at $z=0$, require the two surfaces to be immersed in a regular spacetime. In this case, its union  produces a closed two-dimensional surface $2 {\Gamma_V}={\Gamma_V} \cup {\Gamma_V}'$, which is embedded into the smooth spacetime ${}_{\Gamma_V}\tilde{g}_{ij}$. A pictorial representation of the method is shown in fig.~\ref{fig:will}.

With this geometrical setup in mind, we examine the rescaling properties of the renormalized area of ${\Gamma_V} $ \eqref{renvolnonminimal} under generic Weyl transformations of the ambient metric $g_{ab} =\e{2\Omega}\tilde{g}_{ab}$. Notice that, when $\Omega=\Omega(z)= -\log(z/L)$, one recovers the AAdS metric. The Euler characteristic is a topological invariant and does not change under metric rescalings. Thus, we focus on the quantities which appear under the integral symbol in eq.~\eqref{renvolnonminimal}.

By definition of the bulk line element in terms of the codimension-two metric $\hat{g}_{ab}$
\begin{equation}
g_{ab} =\left (n\indices{^{\left (m\right )}_{a}}n\indices{^{\left (m\right )}_{b}} +e\indices{^i_a}e\indices{^j_b}\, {}_{\Gamma_V}g_{ij}\right ) ,
\end{equation}
where $n\indices{^{\left (m\right )}_{a}}$ are the corresponding normal vectors and $e\indices{^m_a}$ are the frame vectors, we have that $n\indices{^{(m)}_{a}}=\e{\Omega}\tilde{n}\indices{^{(m)}_{a}}$ and
${}_{\Gamma_V}g_{ij} =\e{2\Omega}{}_{\Gamma_V}\tilde{g}_{ij}$.  Here, the quantities with tilde indicate an embedding with respect to the regular metric $\tilde{g}_{ab}$.

On the other hand, the extrinsic curvature of ${\Gamma_V} $ transforms as
\begin{equation}\label{K}
\K\indices{^{(m)}_i_j}=\e{\Omega}\left (\tilde{\K}\indices{^{(m)}_i_j}+{}_{\Gamma_V}\tilde{g}\,\tilde{n}^{(m)}\cdot\partial\Omega\right ), 
\end{equation}
where we have omitted the indices in the contraction $\tilde{n}\indices{^{\left (m\right )}_i} \partial^{i}\Omega$. This expression allows us to write down the trace of $\K\indices{^{(m)}_i_j}$ as
$\K^{(m)}=\e{ -\theta}\left (\K^{(m)}+2\tilde{n}^{(m)}\cdot\partial\Omega\right ),$
and its traceless part as $\K\indices{^{(m)}_{\langle ij\rangle}}=\e{-\Omega }\tilde{\K}\indices{^{(m)}_{\langle ij\rangle}}$. A Weyl transformation of the area element is given by
$\diff^{2}x\sqrt{{}_{\Gamma_V}g} =\diff^{2}x\sqrt{{}_{\Gamma_V}\tilde{g}}\, \e{2\Omega}$.
Then, it is straightforward to show that the following object is Weyl invariant
\begin{equation}\label{tracelessKweyl}
\int_{{\Gamma_V} }\diff^{2}x\sqrt{{}_{\Gamma_V}g}\K\indices{^{(m)}_{\langle ij\rangle}}\K\indices{^{(m)}^{\langle ij\rangle}}=
\int_{{\Gamma_V} }\diff^{2}x\sqrt{{}_{\Gamma_V}\tilde{g}}\tilde{\K}\indices{^{(m)}_{\langle ij\rangle}}\tilde{\K}\indices{^{(m)}^{\langle ij\rangle}} .
\end{equation}
In turn, the square of the trace of the extrinsic curvature is not Weyl invariant
\begin{small}
\begin{equation}\label{meancurvatureweyl}
\int_{\Gamma_V}\diff^{2}x\sqrt{{}_{\Gamma_V}g}{\K^{\left (m\right )}}^{2} =\int_{\Gamma_V}\diff^{2}x\sqrt{{}_{\Gamma_V}\tilde{g}}\left [{{}\tilde{\K}^{\left (m\right )}}^{2} +4\tilde{\K}^{\left (m\right )} \left(\tilde{n}^{\left (m\right )} \cdot \partial\Omega \right) + 4\left (\tilde{n}^{\left (m\right )} \cdot \partial\Omega \right )^{2}\right ] .
\end{equation}
\end{small}
Altogether, the bulk Weyl tensor satisfies ${}_{\Gamma_V }W\indices{^i_j_k_l} =\tilde{W}\indices{^i_j_k_l}$, what implies the relation
$W\indices{_i_j^k^l}=\e{ -2\Omega }\tilde{W}\indices{_i_j^k^l}$. In doing so, the integral of the double subtrace of the Weyl tensor on the area element is proved to be invariant
\begin{equation}\label{weylsmooth}
\int_{\Gamma_V}\diff^{2}x\sqrt{{}_{\Gamma_V}g}\,W\indices{_i_j^i^j} =\int_{\Gamma_V}\diff^{2}x\sqrt{{}_{\Gamma_V}\tilde{g}}\, \tilde{W}\indices{_i_j^i^j}
\end{equation}
Therefore, the renormalized area \eqref{renvolume}, when expressed in terms of the smooth metric, reads
\begin{align}
\nonumber\label{renarea1}
\Area^{\ren}\left ({\Gamma_V} \right ) =&-2\pi L^{2}\chi[\Gamma_V] +\frac{L^{2}}{2}\int_{\Gamma_V}\diff^{2}x\sqrt{{}_{\Gamma_V}\tilde{g}}\bigg[\tilde{W}\indices{_i_j^i^j}+2{{}\tilde{H}^{\left (m\right )}}^{2}-\K\indices{^{(m)}_{\langle ij\rangle}}\K\indices{^{(m)}^{\langle ij\rangle}} \notag\\
&\left.+2\tilde{K}^{\left (m\right )}\left( \tilde{n}^{\left (m\right )} \cdot \partial\Omega \right)  +2\left(\tilde{n}^{\left (m\right )} \cdot \partial\Omega \right )^{2}\right].
\end{align}
This formula adopts a more compact form by taking Gauss-Codazzi eq.\eqref{gausscodazzi} and the relations between bulk and codimension-2 curvature tensors \citep{Anastasiou:2019ldc}
\begin{IEEEeqnarray}{rl}
\tilde{\K}\indices{^{(m)}_{\langle ij\rangle}}\tilde{\K}\indices{^{(m)}^{\langle ij\rangle}}& =
\K\indices{^{(m)}_i_j}\K\indices{^{(m)}^i^j}-2{{}\tilde{H}^{\left (m\right )}}^{2} , \notag \\
\tilde{\K}\indices{^{(m)}_{\langle ij\rangle}}\tilde{\K}\indices{^{(m)}^{\langle ij\rangle}}&=\tilde{R} +\tilde{R}^{(m)(n)(m)(n)}-2\tilde{R}^{(m)(m)}-{}_{\Gamma_V}\tilde{R} +4{{}\tilde{H}^{\left (m\right )}}^{2} , \notag \\
\tilde{W}\indices{_i_j^i^j}&=\tilde{R}\indices{_i_j^i^j} -2\tilde{S}\indices{_i^i} , \notag \\
\tilde{R}\indices{_i_j^i^j}&=\tilde{R} +\tilde{R}^{(m)(n)(m)(n)}-2\tilde{R}^{(m)(m)}\notag,
\end{IEEEeqnarray}
where $\tilde{S}_{ab}$ is the Schouten tensor\footnote{The definition of the Schouten tensor in arbitrary dimension is given by $S_{ab}=\frac{1}{d-2}\left(R_{ab}-\frac{1}{2(d-1)}g_{ab}R\right)$.} of $g_{ab}$. Combining these expressions, we find that
\begin{equation}
\tilde{\K}\indices{^{(m)}_{\langle ij\rangle}}\tilde{\K}\indices{^{(m)}^{\langle ij\rangle}}=\tilde{W}\indices{_i_j^i^j}+2\tilde{S}\indices{_i^i} -{}_{\Gamma_V}R +2{{}\tilde{H}^{\left (m\right )}}^{2},
\end{equation}
what leads to
\begin{align}
\Area^{\ren}\left ({\Gamma_V} \right ) =-2\pi L^{2}[\Gamma_V] +L^{2}\int_{\Gamma_V}&\diff^{2}x\sqrt{{}_{\Gamma_V}\tilde{g}}\bigg[\frac{1}{2}{}_{\Gamma_V}R +\tilde{\K}^{\left (m\right )} \left(\tilde{n}^{(m)} \cdot \partial\Omega  \right) \notag\\
&+\left(\tilde{n}^{(m)} \cdot \partial\Omega \right )^{2}-\tilde{S}\indices{_i^i}\bigg]. \label{volsigma}
\end{align}
Until now, we have treated the two-dimensional sheet ${\Gamma_V} $ as an open surface anchored to the $z =0$ plane. As discussed above, ${\Gamma_V} $ is also half of the closed surface $2{\Gamma_V} $. For the compact manifold $2{\Gamma_V}$, using the Euler theorem \eqref{eq:Cherntheo} in two dimensions and the relation $\chi[2{\Gamma_V}]=2\chi[\Gamma_V]$ for the Euler characteristic, the previous eq.~\eqref{volsigma} yields
\begin{equation}\label{renvolgeneric}
\Area^{\ren}\left ({2\Gamma_V} \right ) =\frac{^{}L^{2}}{2}\int\limits_{2{\Gamma_V} }\diff^{2}y\sqrt{\tilde{\gamma }}\left [\tilde{K}^{(m)} \left(\tilde{n}^{(m)} \cdot \partial\Omega \right) +\left(\tilde{n}^{(m)} \cdot \partial\Omega \right )^{2} -\tilde{S}\indices{_i^i}\right ] ,
\end{equation}
what is equivalent to the renormalized area formula in eq.~\eqref{renvolume}. As a matter of fact, it is more general as it is valid for both minimal and non-minimal surfaces embedded in an AAdS$_{4}$ spacetime. Any constraint on the shape of ${\Gamma_V} $ can be readily implemented as a relation between the different terms in eq.~\eqref{renvolgeneric}.

For instance, minimality condition \eqref{eq:minK} leads to
\begin{equation}\label{minimality condition}
\K^{(m)} = -2\tilde{n}^{(m)} \cdot \partial\Omega  ,
\end{equation}
when written in terms of the smooth metric $\tilde{G}_{\mu \nu }$.
As a consequence, when a minimal surface ${\Gamma_V} _{\text{min}}$ is considered, the renormalized area reduces to
\begin{equation}
\label{renvolsmoothminimal}
\Area^{\ren}\left (2\Gamma_V^\rt\right ) = -\frac{^{}L^{2}}{2}\int_{2\Gamma_V^\rt}d^{2}x\sqrt{{}_{\Gamma_V^\rt}g}\left({{}\tilde{H}^{(m)}}^{2} +\tilde{S}\indices{_i^i}\right).
\end{equation}
In particular, when $\Gamma_V^\rt$ is a spatial subregion of an AdS$_{4}$ spacetime, which corresponds to the conditions $\tilde{H}^{\left (t\right )} =0$ and $\tilde{S}\indices{_i^i}=0$ ($\tilde{g}_{ab}$ is a locally flat space), the last equation reads
\begin{equation}\label{renvolwillmore}
\Area^{\ren}\left (2\Gamma_V^\rt\right ) = -\frac{L^{2}}{2}\int_{2\Gamma_V^\rt}d^{2}x\sqrt{{}_{\Gamma_V^\rt}g}\tilde{H}^{2}=-\frac{L^2}{2}\W\left (2\Gamma_V^\rt\right ).
\end{equation}
Therefore, it is made explicit the connection between the renormalized area and the Willmore energy. The equivalence of these two geometric concepts lead to interesting inequalities about the AdS curvature of the minimal surface.

Indeed, when the closed surface $2{\Gamma_V} _{\text{min}}$ belongs to the topological class of the sphere ($g=0$), the combination of eqs.~\eqref{renvolume}, \eqref{renvolwillmore} and  $\chi\left[2\Gamma_V\right]=2\chi\left[\Gamma_V\right]$ leads to the inequality
\begin{equation} \label{EEinequality}
\int_{\Gamma_V^\rt}\diff^{2}x\sqrt{{}_{\Gamma_V^\rt}}\, {}_{\Gamma_V^\rt}F \leq 0.
\end{equation}
For a toroidal closed surface ($g=1$), the inequality for $g=1$ gives
\begin{equation} \label{EEinequality2}
\int_{\Gamma_V^\rt}\diff^{2}x\sqrt{{}_{\Gamma_V^\rt}}\, {}_{\Gamma_V^\rt}F \leq -2\pi^2 .
\end{equation}
In this geometry, the bound is saturated by the Clifford torus \cite{Astaneh:2014uba}. Notice that in both cases the integral of the trace of the AdS curvature is non-positive.

From a different starting point, Alexakis and Mazzeo arrived at the same type of inequalities in ref.~\cite{alexakis2010renormalized}.
Note that our derivation of the Willmore energy from renormalized area relies on the existence of an AdS bulk. Thus, the bounds \eqref{EEinequality} 
and \eqref{EEinequality2} cannot be extended to other backgrounds. A generalization of these results to 
generic bulk and boundary geometries can be seen in ref.~\cite{Fischetti:2016fbh}.

\subsection{Bound on holographic entanglement entropy}

The connection between renormalized area and Willmore energy provide us insight on surfaces immersed in a higher-dimensional manifold. In particular, the dependence of EE on the geometry becomes manifest when taking a minimal surface $ {\Gamma_V}_{\text{min}}$.
 
For RT surfaces, the renormalized area is equivalent to the renormalized EE of the subregion $A$ implying that the universal part of holographic entanglement entropy in three dimensions can be written as
\begin{equation} \label{EEwilmore}
\suniv(V) = -\frac{L^{2}}{8G_N}\W\left(2\Gamma_V^\rt\right ).
\end{equation}
One can map the universal part of entanglement entropy of an entangling region for a vacuum state of the CFT$_3$ to the Willmore energy of a closed geometry constructed by gluing two copies of the RT surface.

Inequalities \eqref{eq:Willb} and $\W(\Gamma_V^{g=1})\geq2\pi^2$ set a bound to renormalized holographic entanglement entropy \eqref{EEwilmore} depending on the topology of the entangling region. For a doubled RT surface which correspond to $g=0$ \citep{Fonda:2015nma}, the finite part reads
\begin{equation}\label{renEEbound}
F^{\text{holo}}(V) \leq \frac{\pi L^{2}}{2G_N} ,
\end{equation}
that means the finite part of the entanglement entropy is maximized for a circular surface among all the possible shapes within the same topological class --- or, equivalently, the $F$-term is minimized. The same bound was obtained in ref.\cite{alexakis2010renormalized}.

In a similar fashion, when the closed surface $2{\Gamma_V} _{\rt}$ is of genus $g=1$, the finite term of the finite part of entanglement entropy satisfies $F_{\text{holo}}^{g=1}(V) \leq \frac{\pi ^2L^{2}}{4G_N} $. Therefore, the sphere results as the global maximum of the entanglement entropy between surfaces of genus up to one. Astaneh, Gibbons and Solodukhin arrived at the same conclusion by extending their study to higher dimensional surfaces of genus larger than one in ref.\cite{Astaneh:2014uba}. 

The lower bound of $F$ provides the sphere free energy of the field theory $F_0$ thanks to the CHM map \eqref{eq:CHMmap}. As entanglement entropy is maximized by the disk-entangling region the degrees of freedom of the theory are captured by the sphere free energy. In this analysisl, it is manifest that the origin of the non-local nature of the free energy has a topological , as its depends on the Euler characteristic of the entangling region. 

In sec.~\ref{RenEx} we showed that a measure of the deformation of an entangling surface is given by the trace of the AdS curvature, subjected to the inequalities \eqref{EEinequality} and \eqref{EEinequality2}. This quantity is a holographic geometric probe of entanglement susceptibility in the dual CFT. Interestingly enough, the susceptibility is negative as a consequence of the strong subadditivity property of EE \cite{Nozaki:2013vta,Faulkner:2015csl,Witczak-Krempa:2018mqx}. Hence, strong subadditivity imposes a restriction on the curvature of the RT surface side which reads
\begin{equation} \label{geometricSSA}
\int_{\Gamma_V^{\rt}}\diff^{2}x\sqrt{{}_{\Gamma_V^\rt}g} \left({}_{\Gamma_V^\rt}R + \frac{2}{L^2} \right) \leq 0 .
\end{equation}
An analogous constraint on the spacetime curvature was derived in ref.~\cite{Bhattacharya:2014vja} in the context of covariant EE in AdS$_3$/CFT$_2$.

\section{Discussion}

In this chapter, we have studied the shape dependence of entanglement entropy in three-dimensional CFTs dual to Einstein-AdS gravity. The finite part of the entanglement entropy is expressed as the renormalized area of the RT surface \eqref{eq:RTRen} for CFTs in odd dimensions. It consists on two contributions: a topological part, proportional to the Euler characteristic of $\Gamma_V$ (shape independent); and a curvature term, which encodes the information of the deformation of the entangling region with respect to a constant-curvature condition.
 
Explicit computations on entangling regions with deformations for three-dimensional CFTs are presented, along the line of refs.~\cite{Anastasiou:2017xjr,Anastasiou:2018mfk,Anastasiou:2018rla}. We match the results found in the literature given in refs.~\cite{Mezei:2014zla,Allais:2014ata}. Our analysis shows that the number of degrees of freedom of the field theory is given by the topological part. In turn, the quadratic term in the deformation is coming from the integral of the AdS curvature. This means, that the AdS curvature of the RT surface carries information on the coefficients of the correlation function of the dual CFT$_3$.

We also shown that Willmore energy arises as a special case of renormalized area formula of a two-dimensional surface. Indeed, expression \eqref{renvolume} is general, as there is no distinction between minimal and non-minimal surfaces.
Demanding a minimal surface in a constant time slice of global AdS$_4$ bulk spacetime, makes eq.\eqref{renvolume} equivalent to the Willmore functional. The latter provides a lower bound, saturated by a circular entangling surface. This also shows that renormalized EE of a disk-like entangling region is maximal among all the shapes with the same perimeter. This is in consonance with the observations made in ref.~\cite{Allais:2014ata}, which points out that the universal contribution $s_\text{univ}$ of the entanglement entropy is minimized by a circular entangling surface. At the same time, we know that $s_\text{univ}$ matches the free energy of a CFT$_3$ on a spherical background due to the CHM map \cite{Casini:2011kv}. What we learnt here is that, as prescribed by eq.~\eqref{EEwilmore}, $s_\text{univ}$ can be equivalently seen as the Willmore energy of $\mathbb{S}^2$.

%% file: text/ch6-hentangqcg.tex
\chapter{Holographic entanglement entropy dual to higher-curvature gravity}\label{chap:heeqcg}

We now turn our attention to renormalization of CFTs that are dual to higher-curvature gravity. More concretely, here we explore the cancellation of divergences in quadratic curvature gravity using the Kounterterms prescription. As usual, once properly renormalized, we will be able to read off the universal contributions to holographic entanglement entropy. This can be achieved by considering particular shapes of entangling regions (\ie spheres and cylinders), where the $\C$-function candidates and other holographic quantities like the type B anomaly coefficient $c$ can be directly obtained

\section{Renormalization of QC gravity using Kounterterms}

Following the same principles as described in the previous chapter, we start with the QC gravity action, defined in eq.~\eqref{eq:QCGgen}. As usual, the renormalization of the Euclidean action $I^\text{ren}_{\text{E}}$ is achieved by the addition of the boundary term. However, opposed to the Einstein-AdS case, the coefficient $c_d$ will be different to that in eq.~\eqref{eq:cd} \cite{Giribet:2020aks}.

To obtain the correct coupling one must solve the usual characteristic equation to find the vacuum of the theory \eqref{eq:polych}. In the case of QC gravity, the embedding function is given by eq.~\eqref{eq:pchQC}. From there, the value of $c_d$ is fixed by requiring the action of the pure AdS solution (vacuum) to be finite, thus
finding \cite{Giribet:2020aks}
\begin{equation}\label{eq:cdQCG}
c_{d}=a_dc_d^{\text{E}}
\end{equation}
where the auxiliary function $a_d$ reads\footnote{For simplicity, we dropped the superindices $(2)$, as we will focus in QC gravity throughout the whole chapter.}
\begin{equation}\label{eq:ad} a_d=1-\frac
{2d}{\Ls^2}\left[\left(d+1\right)  \alpha_1
+\alpha_2+\frac{\left(d-2\right)\left(d-1\right)}{d}\alpha_3\right],
\end{equation}
and $c_d^{\text{E}}$ is given in eq.~\eqref{eq:cd} (the superscript E denotes that its validity for Einstein-AdS gravity). In addition, we must take into account that $L$ must be replaced by the effective AdS radius $L_\star$.

In principle, in order to obtain the holographic entanglement entropy functional, we could use the Camps-Dong function \eqref{eq:CDEE} without encountering the splitting problem. However, we can equivalently evaluate the QC Euclidean action directly on the orbifold and find the correct functional emplyoing the relations derived by Fursaev, Patrushev and Solodukhin reviewed in sec.~\ref{sec:geoentang}. By doing so, we find the bare entanglement entropy functional
\begin{align}
 \SEE(V)=&\frac{1}{4G_{N}}\Bigg\{\mathrm{Area}\left(\Gamma_V^\rt\right)+\int_{\Gamma_V^\rt}\diff^{d-1}x\sqrt{{}_{\Gamma_V}g}\bigg[\beta\left(R^{(m)(m)}-\frac{1}{2}\K^{(m)}\K^{(m)}\right)+2\gamma \tilde{R}\nonumber\\
&+2\alpha\left(\tilde{R}
+2R^{(m)(m)}-R^{(m)(n)(m)(n)}-2\K\indices{^{(m)}_{[i}^j}\K\indices{^{(m)}_{j]}^i}\right)
\bigg]\Bigg\}\label{eq:SQCGbare},
\end{align}
Following eq.~\eqref{eq:SEEren}, renormalization is achieved with the addition of the boundary term described in eq.~\eqref{eq:ktS}. Having the renormalized entanglement entropy functional at hand, we evaluate it on certain configurations, i.e. sphere and cylinder, whose universal terms encode significant information for the corresponding CFT.

\section{Holographic entanglement entropy for spheres in vacuum CFT}

As mentioned before, the bulk dual to the vacuum state of a $d$-dimensional CFT is the Euclidean version of pure AdS$_{d+1}$, whose metric in Poincar\'e coordinates is given in eq.~\eqref{eq:Poincare}. For ball-shaped entangling regions of radius $R$ in the CFT, the bulk extremal surface is given by the spherical hemisphere of the same radius \cite{Bhattacharyya:2014yga}, whose embedding is described by
\begin{equation}
\Gamma_V^\rt:\left\{t=\text{const.};r^2+z^2=R^2\right\}\label{ExplicitSphere} \,.
\end{equation}
For the following analysis, it is convenient to foliate pure AdS with warped spherical hemispheres. In order to make the extremal surface explicit, the change of coordinates $r=X\sin U$, $z=X\cos U$ is performed. After this change, metric \eqref{eq:Poincare} reads
\begin{equation}
    \diff s^2=\frac{\Ls^2}{X^2\cos^2 U}\left(\diff t^2+\diff X^2 +X^2\diff U^2 + X^2\sin^2U\diff\Omega_{d-2}^2\right).
\end{equation}
In this metric, the hemispheres are the constant $(t,X)$ codimension-two surfaces and the extremal one is located at $X=R$. Also, the non-zero components of the normal vectors to the surfaces read
\begin{equation}
    n_X^{(X)}=n_t^{(t)}=\frac{\Ls}{X\cos U}.
\end{equation}
Therefore, the non-zero components of the projected Riemann and Ricci tensors along these directions read
\begin{IEEEeqnarray}{rl}
   R^{(X)(X)}&=R^{ab}n\indices{^{(X)}_a}n\indices{^{(X)}_b}=-\frac{d}{\Ls^2}, \quad R^{(t)(t)}=R^{ab}n\indices{^{(t)}_a}n\indices{^{(t)}_b}=-\frac{d}{\Ls^2},\\
    R^{(X)(t)(X)(t)}&=R^{abcd}n\indices{^{(X)}_a}n\indices{^{(t)}_b}n\indices{^{(X)}_c}n\indices{^{(t)}_d}=-\frac{1}{\Ls^2}.
\end{IEEEeqnarray}
Regarding the extrinsic curvatures, since the foliation defines a sphere, they identically vanish, \ie 
\begin{equation}
    \K\indices{^{(X)}_i^j}=\K\indices{^{(t)}_i^j}=0.
\end{equation}
In Poincar\'e coordinates, the induced metric reads
\begin{equation}\label{eq:IndSigma}
    \diff s^2_{\Gamma_V^\rt}={}_{\Gamma_V} g_{ij}\diff x^i \diff x^j=\frac{\Ls^2}{z^2}\left[\frac{R^2\diff z^2}{R^2-z^2}+(R^2-z^2)\diff\Omega^2_{d-2}\right],
\end{equation}
which admits a Fefferman-Graham-like expansion as
\begin{IEEEeqnarray}{rl}\label{eq:IndSigma2}
    \diff s^2_{\Gamma_V^\rt}a&=\frac{\Ls^2}{z^2}\left[1+\frac{z^2}{R^2}+\frac{z^4}{R^4}+\pazocal{O}\left(z^6\right)\right]\diff z^2+\tilde{\gamma}_{\alpha\beta}\diff x^{\alpha}\diff x^{\beta},\\
    \tilde{\gamma}_{\alpha\beta}&=\frac{R^2\Ls^2}{z^2}\left(1-\frac{z^2}{R^2}\right)\Omega_{\alpha\beta},
\end{IEEEeqnarray}
where $\Omega_{\alpha\beta}$ is the metric of the $(d-2)$-dimensional sphere. The induced metric $\tilde{\gamma}_{\alpha\beta}$ is fixed at the regulator $z=\delta$, \ie when $\delta\rightarrow0$.

Under the previous considerations, the quantities present in eq. ~\eqref{eq:SQCGbare} read
\begin{IEEEeqnarray}{rl}
\text{Area}[\Gamma_V]&=\vol\left(\mathbb{S}^{d-2}\right)R\int_\delta^{z_{\max}}\diff z\left(R^2-z^2\right)^{\frac{d-3}{2}}\left(\frac{\Ls}{z}\right)^{d-1} \,, \\
\tilde{R}&=-\frac{(d-1)(d-2)}{\Ls^2} \,, \\
R^{(i)(i)}&=-\frac{2d}{\Ls^2} \,, \\
R^{(i)(j)(i)(j)}&=-\frac{2}{\Ls^2} \,, \\
\K^{(m)}\K^{(m)}=\K\indices{^{(m)}_i^j}\K\indices{^{(m)}_j^i}&=0 \,.
\end{IEEEeqnarray}
The codimenstion-two extrinsic curvatures vanish because the geometry after the foliations describes a sphere.

Therefore, considering all the terms that appear in eq.~\eqref{eq:SQCGbare}, we have that the bare holographic entanglement entropy $\SEE(\mathbb{B}^{d-1})$ is given by
\begin{IEEEeqnarray}{rl}
    \SEE(\mathbb{B}^{d-1})=&\frac{1}{4\GN}\Bigg(\vol\left(\mathbb{S}^{d-2}\right)R\int_\delta^{z_{\max}}\diff z\left(R^2-z^2\right)^{\frac{d-3}{2}}\left(\frac{\Ls}{z}\right)^{d-1}\label{eq:bQCGf} \\
    &+\int_\Sigma\diffmone\deth\left\{-\frac{2\alpha_1}{\Ls^2}\left[(d-1)(d-2)+4d-2\right]-\frac{2d\alpha_2}{\Ls^2}-\alpha_3\frac{(d-1)(d-2)}{\Ls^2}\right\}\Bigg) \nonumber,
\end{IEEEeqnarray}
which can be rearranged, using the definition of $a_d$ in eq.~\eqref{eq:ad}, to
\begin{IEEEeqnarray}{rl}
    \SEE(\mathbb{B}^{d-1})=&\frac{a_d\vol\left(\mathbb{S}^{d-2}\right)}{4G_N}\int_\delta^{z_{\max}}\diff z\ R\left(R^2-z^2\right)^{\frac{d-3}{2}}\left(\frac{\Ls}{z}\right)^{d-1} \nonumber \\
    =& a_d \frac{\text{Area}\left[ \Gamma_V^\rt \right]}{4G_N}.
\end{IEEEeqnarray}
Now, adding the holographic entanglement entropy Kounterterm \eqref{eq:ktS}, we have that
\begin{equation}\label{QCGuniv}
\SEE^{\text{ren}}(\mathbb{B}^{d-1})=\frac{a_d}{4\GN} \left(\text{Area}\left[ \Gamma_V^\rt\right]+ c^{\text{E}}_{d}\left \lfloor{\frac{(d+1)}{2}}\right \rfloor\int_{\Sigma}B_{d-2} \right),
\end{equation}
such that the universal part of the holographic entanglement entropy for ball-shaped entangling regions becomes proportional to the universal part of the area of the minimal surface $\SEE^{\text{ren}}(\mathbb{B}^{d-1})=\frac{a_d}{4\GN} \text{Area}^{\text{ren}}\left[\Gamma_V^\rt\right]$.

Finally, using our results of refs. \cite{Anastasiou:2019ldc,Anastasiou:2018rla}, the renormalized holographic entanglement entropy is given by\footnote{Note the choice of $2R$ as the characteristic scale inside the logarithm of the universal term. This choice allows to absorb the finite term as part of the logarithmically divergent term.}
\begin{equation} \label{areauniv}
\SEE^{\text{ren}}(\mathbb{B}^{d-1})=\frac{a_d}{4\GN}\begin{dcases*}(-1)^{\frac{d-1}{2}}\frac{2^{d-1}\pi^{\frac{d-1}{2}}\Ls^{d-1}}
{\left( d-1 \right)  !} & \text{if $d$ odd},\\
(-1)^{\frac{d}{2}-1}\frac{2\pi^{\frac{d}{2}-1}\Ls^{d-1}}
{\left( \frac{d-2}{2} \right)  !} \log\left(\frac{2R}{\delta}\right)&  \text{if $d$ even,}\end{dcases*}
\end{equation}
where $a_d$ was defined in eq.~\eqref{eq:ad}.
The explicit cancellation of the IR divergences in the area functional and the identification of the universal term are given in Appendix \ref{Appendix A}.

\subsection{$\C$-function candidates in CFTs dual to QC gravity}

As discussed in sec.~\ref{sec:RGflows}, we can compute the $\C$-function candidates for CFTs dual to QC gravity, which are conjectured to decrease along the RG flows. Following the discussion in the case of ball-shaped entangling regions, we can read out the $C$-function candidates for both odd and even dimensional CFTs directly from the expression for $\SEE^{\text{ren}}(\mathbb{B}^{d-1})$ \cite{Myers:2010xs,Myers:2010tj,Nishioka:2018khk}. In particular, we have
\begin{equation}
\SEE^{\text{ren}}(\mathbb{B}^{d-1})=\begin{dcases*}(-1)^{\frac{d-1}{2}}F_0, & \text{$d$ odd},\\
(-1)^{\frac{d}{2}-1}4A \log\left(\frac{2R}{\delta}\right), &  \text{ $d$ even,}\end{dcases*}\label{LogUniv}
\end{equation}
where
\begin{equation}
F_0=a_{d}\frac{2^d\pi^{\frac{d-1}{2}}\Ls^{d-1}}
{8G_{N}\left( d-1 \right)  !} 
\end{equation}
is identified as the sphere free energy according to the CHM map \eqref{eq:CHMmap}, and
\begin{equation}
A=a_{d}\frac{\pi^{\frac{d}{2}-1}\Ls^{d-1}}
{8G_{N}\left( \frac{d}{2}-1 \right)  !}
\end{equation}
is the type A anomaly coefficient. They correspond to quantities that are considered monotonic along RG flows for odd and even-dimensional CFTs respectively \cite{Myers:2010tj,Myers:2010xs}. Therefore, the $\C$-function candidate for CFTs dual to QC gravity is proportional to the one of Einstein-AdS gravity, but multiplied by an overall coefficient that depends on the parameters of the theory $a_{d}$. An identical behavior of the universal terms has been found in other higher-curvature theories of gravity such as Einstein cubic gravity, quasi-topological gravity and Lovelock theories, at least, at perturbative level \cite{Bueno:2020uxs}.

The explanation behind this proportionality becomes clear when eqs.~\eqref{QCGuniv} and \eqref{areauniv} are considered. Note that the computation of $S_{\text{EE}}^{\text{ren}}$ is simplified, for the case of ball-shaped entangling regions, as the resulting functional is proportional to the renormalized area $\text{Area}^{\text{ren}}$ of the minimal surface in the bulk. This is due to the fact that for spheres, the bare entropy functional becomes proportional to the area of $\Gamma_V^\rt$ what is the RT functional. Therefore, the $\C$-function candidates of both QC and Einstein-AdS gravities correspond to the universal terms of the area of $\Gamma_V^\rt$, given by $\text{Area}^{\text{ren}}$, up to a factor that depends on the QC gravity couplings.

The fact that the entropy has to be proportional to the area in the spherical case is universal, and can be inferred directly from the CHM map \cite{Casini:2011kv}. In particular, due to the conformal symmetry of the CFT, the entanglement entropy of the ball-shaped subregion can be mapped to the thermal entropy of the CFT at a certain temperature that depends on the replica index. This entropy can be computed, using AdS/CFT, as the Wald entropy of a hyperbolic black hole of constant curvature, which is trivially proportional to the area of the black hole horizon.

Note also, that in the expression for the log universal term of eq.~\eqref{LogUniv}, one can consider the radius of the sphere $R$ as the characteristic size scale. In which case the logarithmic term can be written as $\log\frac{R}{\delta}+\log 2$. This extra $\log2$ appears in even $d$. Because of the robustness of the term in different dimensions, it is suggestive to consider it as coming from a topological term. Indeed, it can be written as $\log\left(\chi\left[\Sigma \right]\right)$, where $\chi\left[\Sigma \right]$ is the Euler characteristic of the entangling surface in the CFT.

In the following section, we consider  cylinder-shaped entangling regions, from which it is possible to compute the type B anomaly coefficient in four-dimensional CFTs \cite{Bhattacharyya:2014yga,Hung:2011xb}. 

\section{holographic entanglement entropy for a cylinder in vacuum CFT}

In order to characterize the type B anomaly of a CFT, it is useful to consider the log part of the holographic entanglement entropy for a cylindrical entangling region. For instance, in the case of AdS$_{5}$/CFT$_{4}$, this universal term is related to $c$ (the type B anomaly coefficient) according to
\begin{equation}
\suniv(\mathbb{R}\times\mathbb{B}^2)=-\frac{cH}{2l}\log\frac{l}{\delta},\label{SUnivC}
\end{equation}
where $l$ is the radius of the cylinder, $H$ is its length along the axis and $\delta$ is the usual UV cutoff in the CFT \cite{Bhattacharyya:2014yga,Hung:2011xb}. When computing the entanglement entropy holographically, by comparing the obtained result with the previous expression, it is possible to identify the $c$ coefficient in terms of the bulk gravity quantities.

We start by considering the metric of pure AdS$_{5}$ written as
\begin{equation}
    \diff s^2=\frac{\Ls^2}{z^2}\left(\diff t^2+\diff z^2+\diff x_3^2+\diff r^2+r^2\diff\theta^2\right),
\end{equation}
where $\theta$ represents the angular direction of an $\mathbb{S}^{1}$ sphere. For cylindrical entangling regions of radius $l$ in the CFT, with their axis extending infinitely along the $x_3$ direction, the bulk extremal surface, in the near-boundary region, is described by the embedding  
\begin{equation}\label{eq:SigmaCyl}
    \Gamma_V^\rt:\left\{t=\text{const.};r=l\left[1-\frac{z^2}{4l^2}+\pazocal{O}\left(z^4\right)\right]\right\}.
\end{equation}
The normal vectors to the surface read
\begin{equation}
    n\indices{^{(X)}_a}=\frac{\Ls}{\sqrt{4l^2+z^2}}\left(0,1,0,\frac{2l}{z},0\right),\quad n\indices{^{(t)}_a}=\left(\frac{\Ls}{z},0,0,0,0\right).
\end{equation}
In this case, the projected Riemann and Ricci tensors read
\begin{IEEEeqnarray}{rl}
   R^{(t)(t)}&=R^{tt}n\indices{^{(t)}_t}n\indices{^{(t)}_t}=-\frac{4}{\Ls^2},\\
   R^{(X)(X)}&=R^{rr}n\indices{^{(X)}_r}n\indices{^{(X)}_r}+R^{zz}n\indices{^{(X)}_z}n\indices{^{(X)}_z}=-\frac{4}{\Ls^2},\\
   R^{(X)(t)(X)(t)}&= R^{ztzt}n\indices{^{(X)}_z}n\indices{^{(t)}_t}n\indices{^{(X)}_z}n\indices{^{(t)}_t}+R^{rtrt}n\indices{^{(X)}_r}n\indices{^{(t)}_t}n\indices{^{(X)}_r}n\indices{^{(t)}_t}=-\frac{1}{\Ls^2}.
\end{IEEEeqnarray}
The extrinsic curvature along the temporal axis vanishes, \ie $\K\indices{^{(t)}_i^j}=0$ . However, the foliation in the $z$ coordinates gives a non-zero extrinsic curvature whose components read
\begin{IEEEeqnarray}{rl}
\K\indices{^{(X)}_z^z}&=-\frac{z^3}{8 l^3\Ls}+\pazocal{O}\left(z^5\right),\\
  \K\indices{^{(X)}_{x_3}^{x_3}}&=-\frac{z}{2 l\Ls}+\frac{z^3}{16  l^3\Ls}+\pazocal{O}\left(z^5\right),\\
    \K\indices{^{(X)}_\theta^\theta}&=\frac{z}{2 l\Ls}+\frac{3z^3}{16  l^3\Ls}+\pazocal{O}\left(z^5\right).
\end{IEEEeqnarray}
The induced metric in the codimension-two manifold $\Gamma_V$ reads
\begin{equation}
    \diff s^2_{\Gamma_V^\rt}=\frac{\Ls^2}{z^2}\left\{\left[1+\frac{z^2}{4l^2}+\pazocal{O}\left(z^4\right)\right]\diff z^2+\diff x_{3}^2+l^2\left[1-\frac{z^2}{4l^2}+\pazocal{O}\left(z^4\right)\right]^2\diff\theta^2\right\} \,.
\end{equation}
As in the spherical entangling region case, this expression admits a Fefferman-Graham expansion as well. 

Based on these considerations, the geometric quantities appearing in the EE functional in eq.~\eqref{eq:SQCGbare} are given by
\begin{IEEEeqnarray}{rl}
\text{Area}\left[\Gamma_V^\rt\right]&=\int_0^{2\pi}\diff\theta\int_0^{H}\diff x_{3}\int_\delta^{z_\text{max}}\diff z\sqrt{{}_{\Gamma_V}g} \,,\\
 \tilde{R}&=-\frac{6}{\Ls^2}-\frac{z^2}{2l^2\Ls^2}+\pazocal{O}\left(z^4\right) \,,\\
R^{(m)(m)}&=-\frac{8}{\Ls^2} \,,\\
R^{(m)(n)(m)(n)}&=-\frac{2}{\Ls^2} \,,\\
{\K^{(m)}}^2&={\pazocal{O}\left(z^6\right) \,,}\\
\K\indices{^{(m)}_i^j}\K\indices{^{(m)}_i^j}&=\frac{z^2}{2 l^2 \Ls^2}+\pazocal{O}\left(z^4\right) \,.
\end{IEEEeqnarray}
In even-dimensional CFTs, the finite part of the entanglement entropy, $b_0$, is non-universal, and therefore, upon evaluating the integral in the Poincar\'e coordinate of the area functional, the upper limit (at $z_{\text{max}}$) can be neglected. For the lower limit we expand the metric determinant, finding
\begin{equation}
 \sqrt{{}_{\Gamma_V^\rt}g}=\frac{l \Ls^3}{z^3}-\frac{\Ls^3}{8 l z}+\pazocal{O}\left(z\right) \,.
\end{equation}
Now, plugging all these results into the functional, we obtain
\begin{equation}\label{eq:Sbarecyl}
    \SEE(\mathbb{B}^2\times\mathbb{R})=\frac{\pi H\Ls^3}{4l \GN}\left(a_4\frac{l^2}{\delta^2}-\frac{\tilde{a}_4}{4}\log\frac{l}{\delta}\right)+b_0,
\end{equation}
where we define the coefficient
\begin{equation}
\tilde{a}_4=1-\frac{4}{\Ls^2}\left(10\alpha_1+2\alpha_2+\alpha_3\right) \,.\label{b_4}
\end{equation}
This factor differs from $a_4$ defined in eq.~\eqref{eq:ad}. 

Finally, we check that the boundary term cancels the power law term in eq.~\eqref{eq:Sbarecyl}. The induced metric at the boundary $z=\delta$ reads
\begin{equation}
\sqrt{{}_{\Gamma_V^\rt}g}=\frac{l\Ls^2}{\delta^2}-\frac{\Ls^2}{4l}+\pazocal{O}\left(\delta\right) \,,
\end{equation}
yielding
\begin{equation}
S_{\text{kt}}(\mathbb{B}^2\times\mathbb{R})=-\frac{a_4\pi  H l \Ls^3}{4 \GN \delta ^2} \,.
\end{equation}
As we can see, the power law divergence in eq.~\eqref{eq:Sbarecyl} is indeed cancelled by the Kounterterm.

Thus, up to a non-universal finite part, one has that for the cylinder entangling region in $d=4$, the universal part of the HEE is given by
\begin{eqnarray}
    \SEE^{\text{ren}}(\mathbb{B}^2\times\mathbb{R})=-a_{4}\frac{\pi H \Ls^3}{16 l \GN}\log \frac{l}{\delta} \,.
\end{eqnarray}
Finally, comparing this expression with that of eq.(\ref{SUnivC}), we have that
\begin{eqnarray}\label{c_Coeff}
    c=b_{4}\frac{\pi \Ls^3}{8 G_N} \,.
\end{eqnarray}
In the QC gravity case, it is evident from our results that the $A$ and $c$ central charges are different. However, for Einstein-AdS ($\tilde{a}_{4}=a_{4}=1$), they coincide.

\subsection{Extremal surface for the cylinder}

In refs.~\cite{Bhattacharyya:2014yga,Hung:2011xb}, it was shown that the surface \eqref{eq:SigmaCyl} extremizes the holographic functional for Gauss-Bonnet theory. In QC gravity, higher-order terms appear in the entanglement entropy. However, the same embedding function yields the extremal surface at order $\pazocal{O}\left(z^3\right)$. In order to see this, consider an arbitrary surface parametrized with $r=r(z)$. In $d=4$, the holographic entanglement entropy functional for QC gravity reads
\begin{IEEEeqnarray}{cl}\label{eq:Cylfunc}
 S_{\text{EE}}=&\frac{\pi H\Ls}{4\GN}\int_\delta^{z_\text{max}}\diff z\frac{1}{z^3 r
\left(r'^2+1\right)^{5/2}}\Big\{2 \Ls^2 r^2 \left(r'^2+1\right)^3\nonumber\\
&+16\alpha r(r'^2+1)\left[z(z+2rr')r''-(r'^2+1)(r(5+3r'^2)-z r')\right]
\nonumber\\
   &-\beta  \left[\left(r \left(r'^3+r'+z
   r''\right)+z \left(r'^2+1\right)\right)^2+16 r^2 \left(r'^2+1\right)^3\right]\nonumber\\
   &-   4
   \gamma  r \left(r'^2+1\right) \left[2 z \left(2 r r'+z\right) r''+2 \left(3
   r-2 z r'\right) \left(r'^2+1\right)\right]\Big\} \,.
\end{IEEEeqnarray}
This expression is obtained once the terms given by
\begin{IEEEeqnarray}{rl}
\text{Area}\left[\Gamma_V^\rt\right]&=\int_0^{2\pi}\diff\theta\int_0^{H}\diff x_3\int_\delta^{z_\text{max}}\diff z\sqrt{{}_{\Gamma_V^\rt}g}=2\pi H \Ls^3\int_\delta^{z_\text{max}}\diff z \frac{r\sqrt{1+r'^2}}{z^3} \,,\\
\tilde{R}&=-\frac{2 z \left(2 r r'+z\right) r''+2 \left(3 r-2 z r'\right)
   \left(r'^2+1\right)}{\Ls^2 r \left(r'^2+1\right)^2} \,,\\
R^{(m)(m)}&=-\frac{8}{\Ls^2} \,,\\
R^{(m)(n)(m)(n)}&=-\frac{2}{\Ls^2}\,,\\
{\K^{(m)}}^2&=\frac{\left[\left(r r'+z\right) \left(r'^2+1\right)+z r r''\right]^2}{\Ls^2 r^2
   \left(r'^2+1\right)^3} \,,\\
\K\indices{^{(m)}_i^j}\K\indices{^{(m)}_j^i}&=\frac{\left[(z+r r')^2+r^2r'^2\right](1+r'^2)^2+r^2(r'+r'^3-zr'')}{\Ls^2r^2(1+r'^2)^3} \,,
\end{IEEEeqnarray}
are plugged into eq.~\eqref{eq:SQCGbare}. The resulting functional \eqref{eq:Cylfunc} constitutes a Lagrangian $\pazocal{L}=\pazocal{L}(z,r,r',r'')$ that contains second order derivatives of the dynamical function $r(z)$. Because of this, the Euler-Lagrange equation needed to find the extremal surface reads
\begin{equation}
    \frac{\partial \pazocal{L}}{\partial r}-\frac{\diff}{\diff z}\frac{\partial \pazocal{L}}{\partial r'}+\frac{\diff^2}{\diff z^2}\frac{\partial \pazocal{L}}{\partial r''}=0 \,.
    \label{EL_HEE}
\end{equation}
From this expression, an equations of motion containing fourth-derivative terms in the function $r(z)$ is found. However, the ansatz
\begin{equation}
    r=l\left[1-\frac{z^2}{4l^2}+\pazocal{O}\left(z^4\right)\right] \,,
    \label{cylindemb}
\end{equation}
from ref.~\cite{Hung:2011xb} is verified to satisfy the equations of motion up to order $\pazocal{O}\left(z^3\right)$. Because of this, the extremal surface for QC gravity coincides with that for Gauss-Bonnet gravity in this perturbative regime. This is an expected result due to the universality of the second term in the asymptotic expansion of the embedding function, what is linked to the universality of the $\tilde{\gamma}^{(2)}_{\alpha\beta}$ coefficient in the Fefferman-Graham expansion in terms of the Schouten tensor of $\tilde{\gamma}^{(0)}_{\alpha\beta}$, as discussed in ref.~\cite{Schwimmer:2008yh}.

The case for the cylindrical entangling region in $d=4$ is interesting as it isolates the contribution from the type B anomaly in the universal part. The same should be the case for higher-dimensional cylinders, as the coefficient obtained should represent a linear combination of the couplings of different conformal invariants.\footnote{Although little is known about conformal invariants beyond eight dimensions, one may think this computation would provide information on the part of these invariants which is polynomial in the Weyl tensor.} However, the embedding function is not known in the higher-dimensional case as it would require knowledge of the subleading terms in the expansion of eq.~\eqref{cylindemb}, which are not universal. Furthermore, the Kounterterm renormalization procedure has limitations regarding the types of entangling surfaces that it can accommodate, for dual bulk manifolds of dimension greater than 5. In particular, it requires the dimensional continuation of codimension-two conformal invariants at the entangling surface to vanish. For example, to the next-to-leading order, the method only works for surfaces such that
\begin{equation}
W\indices{_i_j^i^j}-\K\indices{^{(m)}_{\langle ij\rangle}}\K\indices{^{(m)}^{\langle ij\rangle}}=0 \,,
\label{Kountershapelimit}
\end{equation}
as shown in ref.~\cite{Anastasiou:2019ldc}. Expression \eqref{Kountershapelimit} is trivially satisfied for spheres, but not for cylinders or arbitrary shapes.\footnote{In bulk dimensions up to 5, the Kounterterm procedure works for arbitrary entangling regions. Furthermore, as proven in \cite{Araya:2021atx}, the procedure correctly renormalizes actions for gravity theories of arbitrary order in the Riemannian curvature, and therefore, it is expected to work for renormalizing holographic entanglement entropy for CFTs dual to said theories as well.}

\section{Discussion}

The results for holographic entanglement entropy for CFTs dual to QC gravity presented here come as the natural blend between the Kounterterm method applied to this gravity theory \cite{Giribet:2018hck,Giribet:2020aks} and a remarkable feature of the boundary term $B_{d}$ when evaluated in spacetimes with a conical defect \cite{Anastasiou:2019ldc}. In this respect, we have recovered the universal part of the holographic entanglement entropy found in the literature  regarding the computation of the $\C$-function candidates \cite{Bueno:2020uxs}. This function captures essential properties of CFTs, which are given by the type A anomaly coefficient in the case of even $d$ and by the generalized $F_0$ quantity for odd $d$ \cite{Myers:2010xs,Nishioka:2018khk,Imbimbo:1999bj}.

The above calculation requires a ball-shaped entangling region in the CFT, where for the case of pure AdS spacetime (dual to the vacuum of the CFT), the embedding of the extremal surface \eqref{ExplicitSphere} for the QC gravity holographic entanglement entropy functional \eqref{eq:SQCGbare} is explicitly given. Then, the Fefferman-Graham-like expansion of all the terms involved in the functional can be obtained. In both, even and odd boundary dimension $d$, it can be seen that the  $\C$-function candidate derived is proportional to the one for Einstein-AdS gravity, but with an overall coupling-dependent factor $a_{d}$, whose form is given in eq.~\eqref{eq:ad}.

We have also obtained the type B anomaly coefficient $c$ in the case of four-dimensional CFTs. In order to perform this computation, we have considered a cylindrical entangling region in the CFT, and the near-boundary expansion (up to cubic order in the Poincar\'e coordinate) of the embedding for the minimal surface. In this situation, we have derived the corresponding Euler-Lagrange equation for the embedding function $r(z)$, by taking variations of the entropy functional of eq.~\eqref{eq:Cylfunc}. We have verified that the same embedding function considered for Einstein-AdS gravity is also a solution of the extremization equation in the QC gravity case, in $d=4$, and up to cubic order. The condition for the minimal surface obtained in this way contains higher-order derivative terms in the dynamical variable (akin to the ``acceleration''). Thus, extra boundary conditions are required beyond setting the border of the surface to coincide with the entangling region. When the surface is extremal (\ie the intersection of the surface with the conformal boundary is orthogonal), the boundary problem is completely fixed. This is the case of the cylinder in $d=4$, due to the fact that the entropy functional becomes proportional to the area up to cubic order. In the result for the $c$ coefficient we also find agreement with the literature \cite{Bueno:2020uxs}. It is evident from the  expression obtained in eq.~\eqref{c_Coeff}, that the type B anomaly coefficient can be written as the one for Einstein-AdS gravity, but multiplied with a factor $\tilde{a}_{4}$, given in eq.~\eqref{b_4}, which incorporates the information on the couplings of QC gravity theory. The fact that $\tilde{a}_{4}$ is different from the $a_{4}$ of eq.~\eqref{eq:ad} allows for different central charges in four-dimensional CFTs dual to QC gravity, unlike the Einstein-AdS case.

All in all, for both even and odd dimensional CFT cases, the Kounterterm procedure allows to isolate the universal part of the HEE of the dual gravity theory. For bulk dimensions lower than 6, the Kounterterm procedure works on entangling regions of arbitrary shape. Also, despite its limitations on the type of entangling regions that can be renormalized in higher dimensions (as discussed in the previous section), the Kounterterms prescription is the only method available so far for renormalizing holographic entanglement entropy in higher-curvature gravity theories.\footnote{The alternative renormalization procedure of \cite{Taylor:2016aoi}, based on Holographic Renormalization \cite{deHaro:2000vlm}, was only applied for Einstein and Einstein-Gauss-Bonnet gravity theories.} 

In isolating the universal part of the holographic entanglement entropy, we have been able to express it as a covariant functional which is given by the standard holographic entanglement entropy functional plus an extrinsic counterterm in codimension-three \eqref{eq:ktS}. In the particular case of spherical entangling regions in pure AdS (vacuum CFTs), the renormalized holographic entanglement entropy functional becomes proportional to the renormalized area \eqref{QCGuniv}, which is logarithmically divergent for even-$d$ and finite for odd-$d$. 

For a cylindrical entangling region in $d>4$, the renormalized entropy functional is no longer proportional to the renormalized codimension-two volume. However, in $d=4$, the expressions coincide --- up to the normalizable order---, albeit with a different proportionality constant than for the spherical case.

We point out that the Kounterterm procedure is a nonperturbative method, in the sense that nowhere it is assumed that the couplings of the quadratic terms are small. In other words, the prescription does not rely on the linearization of the equations of motion such that the theory behaves like Einstein gravity with a modified Newton's constant. One can wonder whether the prescription isolates $\C$-function candidates when computing the holographic entanglement entropy for CFTs dual to higher-than-second-curvature gravity theories. It turns out that the answer is positive, as shown in ref.~\cite{Anastasiou:2021jcv}, when considering the Jacobson-Myers functional for Lovelock gravity.

Moreover, it was recently shown that

%% file: text/ch7-entang.tex
\chapter{Shape-dependence of entanglement entropy in general CFT}\label{ch:gen}

In ch.~\ref{chap:sdhee} we found that the disk-like entangling region maximized the finite part of entanglement entropy (or minimized $F$) for CFTs dual to Einstein-AdS gravity. Moreover, from Mezei's formula we know that this is the case for any CFT, as corrections coming from small perturbations in the entangling surface to $F_0$ are given by a positive-definite geometry-dependent coefficient weighted by the (also positive-definite) stress-tensor $C_{\ssc T}$.

This motivates the question whether this situation is only valid for holographic theories or for general CFTs for arbitrary entangling shapes. In other words, we want to elucidate if the relation
\begin{equation}\label{Fiso}
F(V)/F_0 \geq 1\, , \quad  \text{with} \quad F(V)/F_0=1 \Leftrightarrow  V =\mathbb{B}^{2},
\end{equation}
holds any three-dimensional CFT under consideration. On general grounds, one can also argue that $F$ tends to grow as regions incorporate sufficiently thin sectors, as in that case $F$ includes contributions proportional to the ratio between the corresponding long and short dimensions ---see \eg \cite{Casini:2005zv,Ryu:2006ef}.

An analogous problem has been studied for the universal (logarithmic) contribution to the entanglement entropy in four-dimensions in \cite{Astaneh:2014uba} and \cite{Perlmutter:2015vma}. In that case, however, the essentially local nature of the corresponding term makes it be fully determined in terms of two (theory-independent) local integrals over $\partial A$ weighted by the trace-anomaly charges of each theory \rd{$a$ and $c$. The global minimization  of the entanglement entropy universal term  by the round sphere follows then rather straightforwardly for general CFTs in the case of arbitrary entangling surfaces of genus $g=0$ and $g=1$. On the other hand, as pointed out in \cite{Perlmutter:2015vma}, when $a>c$ it can be shown that the entanglement entropy universal coefficient is actually unbounded from below for sufficiently high genus.\footnote{The argument follows from two facts: i) the EE universal coefficient can be written as a linear combination of the Willmore functional of $\partial A$ plus $(a/c-1)$ times the Euler characteristic of $\partial A$ ---which is proportional to $(1-g)$, where $g$ is the genus of the entangling surface; ii)  for every genus $g$ there exists at least a surface whose Willmore energy lies somewhere between the values $4\pi$ and $8\pi$  \cite{lawson1970complete,kusner1989comparison}. Then, whenever $(a/c-1)>0$, a growing $g$ will make the Euler characteristic piece more and more negative, whereas the Willmore part for certain geometries will remain bounded between the aforementioned values independently of $g$. } }

In this chapter, we argue that conjecture \eqref{Fiso} is valid  for general CFTs in three dimensions. Before providing a proof, in sec.~\ref{elipsq} we study in detail the case of elliptic regions. In particular, we provide analytic approximations of $F$ for such regions for arbitrary values of the eccentricity for general CFTs. These we compare with the exact result obtained numerically for the so-called ``Extensive Mutual Information'' (EMI) model as well as with lattice results for free scalars and fermions finding good agreement for the latter and a small puzzle for the former. The lattice results are obtained using mutual information (MI) as a geometric regulator which requires the evaluation of the MI for two concentric regions and the subtraction of a purely geometric piece weighted by the coefficient controlling the entanglement entropy of a thin strip region. In sec.~\ref{secemi} we evaluate  $F$ for several families of more general entangling regions in the EMI model. This allows to illustrate to what extent $F$ may vary in general as the geometric features of the entangling region non-perturbatively deviate from the most symmetric cases. In sec.~\ref{proof} we prove \req{Fiso} for general theories.

\section{Elliptic entangling regions}\label{elipsq}

Perhaps the simplest (non-perturbative) generalizations of the disk region which come to mind are ellipses. In the holographic context, they have been studied numerically alongside more general regions in ref.~\cite{Fonda:2014cca}. In the same context, some bounds for their corresponding $F$ were obtained in ref.~\cite{Allais:2014ata}. In this section we start by considering the limit in which the elliptic regions are very squashed (eccentricity $e\rightarrow 1$), which leads to a general approximation in terms of the strip coefficient $k^{(3)}$ valid for general theories. Then, using this approximation along with Mezei's formula  for the opposite limit of almost round ellipses, we build a trial function which approximates $F_{(e)}$ for arbitrary values of the eccentricity for a general CFT as long as we know the values of the three coefficients $F_0$, $C_{\ssc T}$, $k^{(3)}$. We present the resulting approximations for the EMI model, holographic Einstein gravity, free scalars and free fermions. For the last two, we perform lattice calculations of  $F_{(e)}$ using mutual information as a regulator for $e=\sqrt{3}/2\simeq 0.866 $, $e=2 \sqrt{2}/3\simeq 0.943$, $e= \sqrt{15}/4\simeq 0.968$ and $e=2\sqrt{6}/5 \simeq 0.98$ which we compare to the approximations. The fermion results agree reasonably well with the trial function, which also approximates the exact EMI model results with a discrepancy lower than a $\sim 5 \%$ for the whole range. On the other hand, the lattice produces results for the scalar which are considerably lower than the ones obtained from the approximation. While this appears to be an artifact of the lattice results (at least to some extent) we introduce an additional trial function which does a better job in approximating the scalar results. Along with the initial one, these two functions confidently bound the possible values of $F_{(e)}$ for a given CFT.  In all cases, $F_{(e)}$ is a monotonically increasing function of the eccentricity.

\subsection{General-CFT approximations for arbitrary eccentricity }
Ellipses of semi-major and semi-minor axes $a$,$b$ and eccentricity $e\equiv \sqrt{1-b^2/a^2}\in [0,1)$ can be parametrized in polar and Cartesian coordinates respectively by
\begin{equation}
r(\theta)=\frac{b}{\sqrt{1-e^2\cos^2\theta}}\, , \quad \text{and } \quad [x(t),y(t)]=[a \cos(t), b \sin(t)]\, ,
\end{equation}
where the ``eccentric anomaly'' $t$ is related to the polar angle $\theta$ by $\tan (\theta) = \tfrac{b}{a} \tan (t)$.
As we vary $e$, we can interpolate between the regime which is very close to the disk (as $e\rightarrow 0$) and the one which approaches the strip-like behavior described eq.~\eqref{strip} (as $e\rightarrow 1$). In the former case, one finds, from Mezei's formula
\begin{equation}\label{eli0}
F_{(e)}  \overset{e\rightarrow 0}{\simeq} F_0 +e^4 \frac{ \pi^4 C_{\ssc T}}{64  }{}+\pazocal{O}(e^8)\, .
\end{equation}
In the opposite regime, we can take advantage of \req{strip}, namely, of the fact that very thin regions are controlled by the strip coefficient $k^{(3)}$ times the ratio of their long and thin dimensions. Using the equation for the width of the squashed ellipse, $2y=2b\sqrt{1-x^2/a^2}$, we can then approximate $F_{(e \rightarrow 1)} $ as \cite{Allais:2014ata}
\begin{equation} \label{elie1}
F_{(e)}  \overset{e\rightarrow 1}{\simeq} k^{(3)} \int_{-a}^{a} \frac{\diff x}{2 b \sqrt{1- x^2/a^2}}=\frac{ k^{(3)}  \pi}{2 \sqrt{1-e^2}} = \frac{ k^{(3)}  \pi}{2\sqrt{2} \sqrt{1-e}} + \pazocal{O}(\sqrt{1-e}) \, .
\end{equation}
This is valid for general CFTs and, at least for some models, it in fact provides a very good approximation to the exact result for not-so-squashed ellipses ---see comments below \req{kemi}.  
Similar formulas valid for general CFTs can be analogously obtained for other closed squashed entangling regions.

Formulas \req{elie1} and \req{eli0} provide us with approximations to $F_{(e)} $ in those two regimes, for general CFTs provided we know the coefficients $k^{(3)}$, $F_0$ and $C_{\ssc T}$. 
Using them, we can try to do better and produce expressions which approximate $F_{(e)} $ for arbitrary values of the eccentricity. 

Our strategy is to build an approximation from a linear combination of test functions with particular relative coefficients. The idea is to fix those coefficients in a way such that eqs.~\req{elie1} and \req{eli0} hold when the full expression is expanded around $e\rightarrow 0$ and $e\rightarrow 1$ respectively.\footnote{Let us mention that a similar strategy was used in ref.~\cite{Bueno:2015qya} (see also \cite{Helmes:2016fcp}) in order to produce high-precision approximations to the entanglement entropy corner functions of general CFTs using the parameters $C_{\ssc T}$ and $k^{(3)}$. In that case, the available (numerical) results for free fields and holography allowed the authors to verify the excellent agreement between the trial functions and the exact curves. } In the former case, this implies setting to zero the $\pazocal{O}(e^1,e^2,e^3,e^5,e^6,e^7)$ coefficients. Naturally, there are many possible candidate functions one may consider, but we find it convenient to choose as building blocks functions of the form $\sim E[e^2]^k/\sqrt{1-e^2}$ with $k$ some positive integer and others $\sim (1-e^2)^{l/2}$ with $l=-1,1,3,\dots$ 
Functions of these types are such that the $\pazocal{O}(e^1,e^3,e^5,e^7)$ coefficients around $e=0$ are automatically vanishing. For a general linear combination of functions,  expanding around $e=0$ and imposing eq.~\req{eli0} fixes four of the coefficients. Expanding around $e=1$ and imposing eq.~\req{elie1} fixes an additional one.  These restrictions considerably constrain the possible behavior of the trial functions ---especially if one requires them to be monotonically increasing functions of the eccentricity--- but still allow for some room of variance for intermediate values of $e$. We construct the following two trial functions,
\begin{align}\label{trialf}
F_{{(e)}}|^{{\rm tri}_{1} }&= \frac{ \alpha_1(e) F_0 +  \alpha_2(e)  C_{\ssc T}+  \alpha_3(e)k^{(3)}}{\alpha_4(e)},\notag\\
F_{{(e)}}|^{{\rm tri}_{2} }&= \frac{\beta_0+ \beta_1(e) F_0 +  \beta_2(e)  C_{\ssc T}+  \beta_3(e)k^{(3)}}{\beta_4(e)}\, ,
\end{align}
where\footnote{$E[x]$ is the complete elliptic integral, $E[x]\equiv \int_0^{\pi/2} \sqrt{1-x \sin^2 \theta} \diff \theta$. }
\begin{align}
\alpha_1(e)\equiv &+ 2 \Big[ \pi^3 \left(-120+\pi [204+\pi (17\pi-114)]\right) \\ \notag & \quad  \quad  - \pi^2 [-688+\pi \left(1088-504\pi +17\pi^3\right)] E[e^2]\\ \notag & \quad  \quad  +6\pi \left(-176+208\pi-84\pi^3+19\pi^4\right) E[e^2]^2 \\   \notag & \quad  \quad +2 \left(288-2\pi^2 [312+17\pi(3\pi-16)]\right)E[e^2]^3 \\  \notag &\quad  \quad +8 \left(-72+\pi [132+\pi (15\pi-86)] \right)E[e^2]^4\Big]\, , \\ 
\alpha_2(e)\equiv & -\pi^4 (\pi-2)^2  \left( \pi -2 E[e^2] \right)^2 \left( E[e^2]-1\right)  \left( \pi (\pi-6)+(20-6\pi) E[e^2]\right) \, , \\ 
\alpha_3(e)\equiv &+\pi^5  \left( \pi -2 E[e^2] \right)^4 \, , \\ 
\alpha_4(e)\equiv &+ \frac{2\pi^4 (\pi- 2)^4 \sqrt{1-e^2}}{E[e^2]} \, ,
\end{align}
and
\begin{align}
\beta_0(e)\equiv & +\frac{2}{5}\Big[ \pi^2 \left( -8 (32-48e^2+e^4)+\pi^3(17-18e^2+e^4) \right. \\  \notag  & \quad  \quad \left. -2\pi^2 (96-80e^2+7e^4)+4\pi (128-112 e^2+9e^4) \right)  \\  \notag &  \quad  \quad+2 \pi(-1024+128\pi+23\pi^3)E[e^2]-4(-1280+64\pi^2+25\pi^3)E[e^2]^2 \\  \notag  & \quad  \quad -8(640-256\pi+15\pi^2)E[e^2]^3\Big] \, , \\
\beta_1(e)\equiv & +2\pi^2 \left(48-184\pi+84\pi^2-9\pi^3 + e^2 (-24+\pi (44-42\pi+9\pi^2))\right)\\  \notag &  +4E[e^2](7\pi^2(32-3\pi^2)+2E[e^2] \left[-64+7\pi^2(5\pi-16)+(64-6\pi^2)E[e^2] \right])\, , \\
\beta_2(e)\equiv & +\pi^6 \left(48-128\pi + 52\pi^2-5\pi^3+(\pi-2)^2 (5\pi-14)e^2\right) \\  \notag &  \quad  \quad +2\pi^4E[e^2] \left[ \pi (128+40\pi-9\pi^3)+2E[e^2](-176+\pi^2(13\pi-20) \right. \\ \notag & \quad  \quad  \left.+2(88+(\pi-32)\pi) E[e^2]) \right] \, , \\
\beta_3(e)\equiv & 2\pi^6 (2e^2-1)-28\pi^5 E[e^2]+72\pi^4 E[e^2]^2-16\pi^3E[e^2]^3  \, , \\
\beta_4(e)\equiv &4\pi^2(\pi^3-14\pi^2+36\pi-8)\sqrt{1-e^2} \, .& 
\end{align}

\begin{figure}[t] \hspace{-0.25cm}
	\includegraphics[scale=0.6]{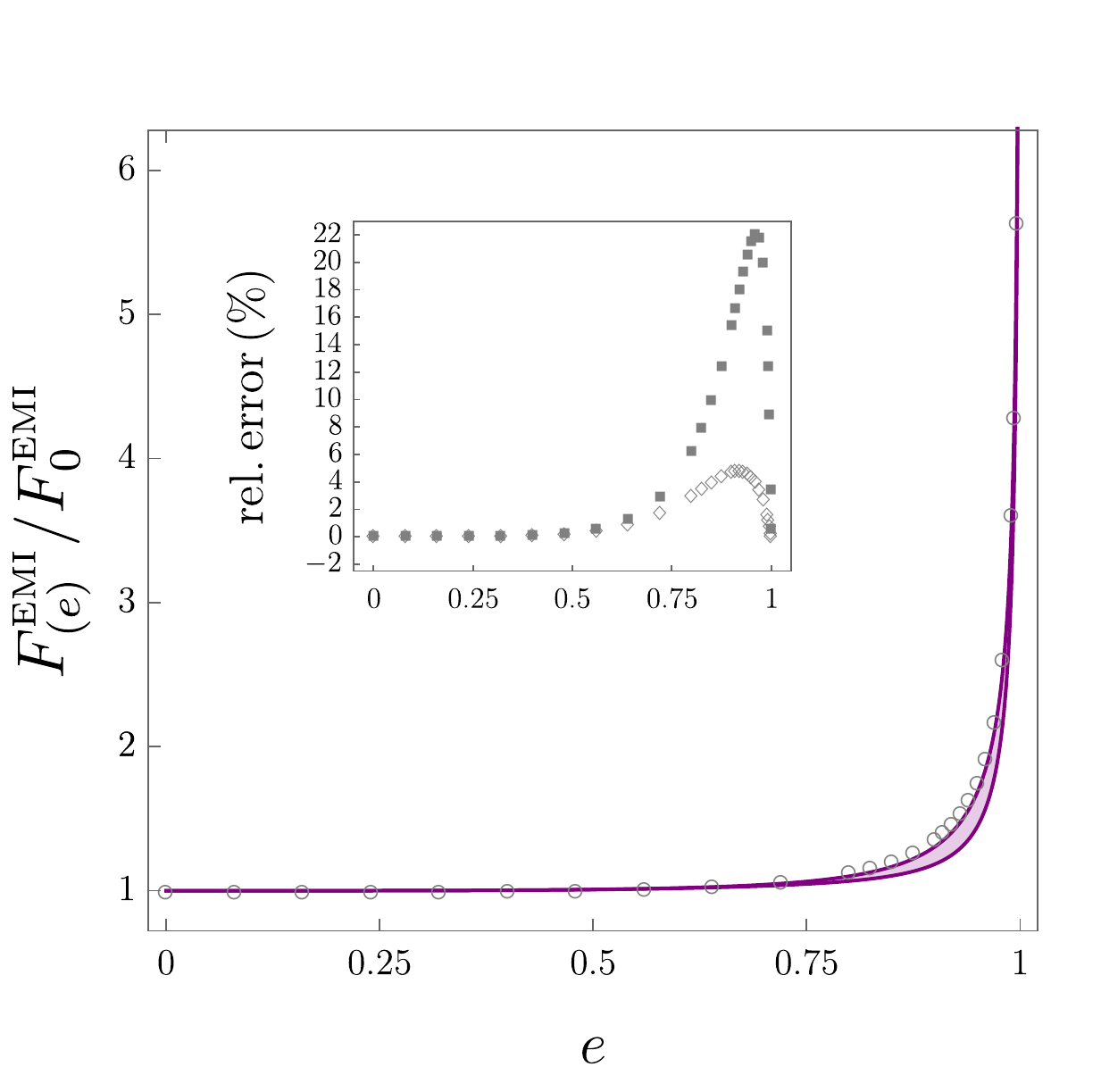}\hspace{-0.2cm}
	\includegraphics[scale=0.6]{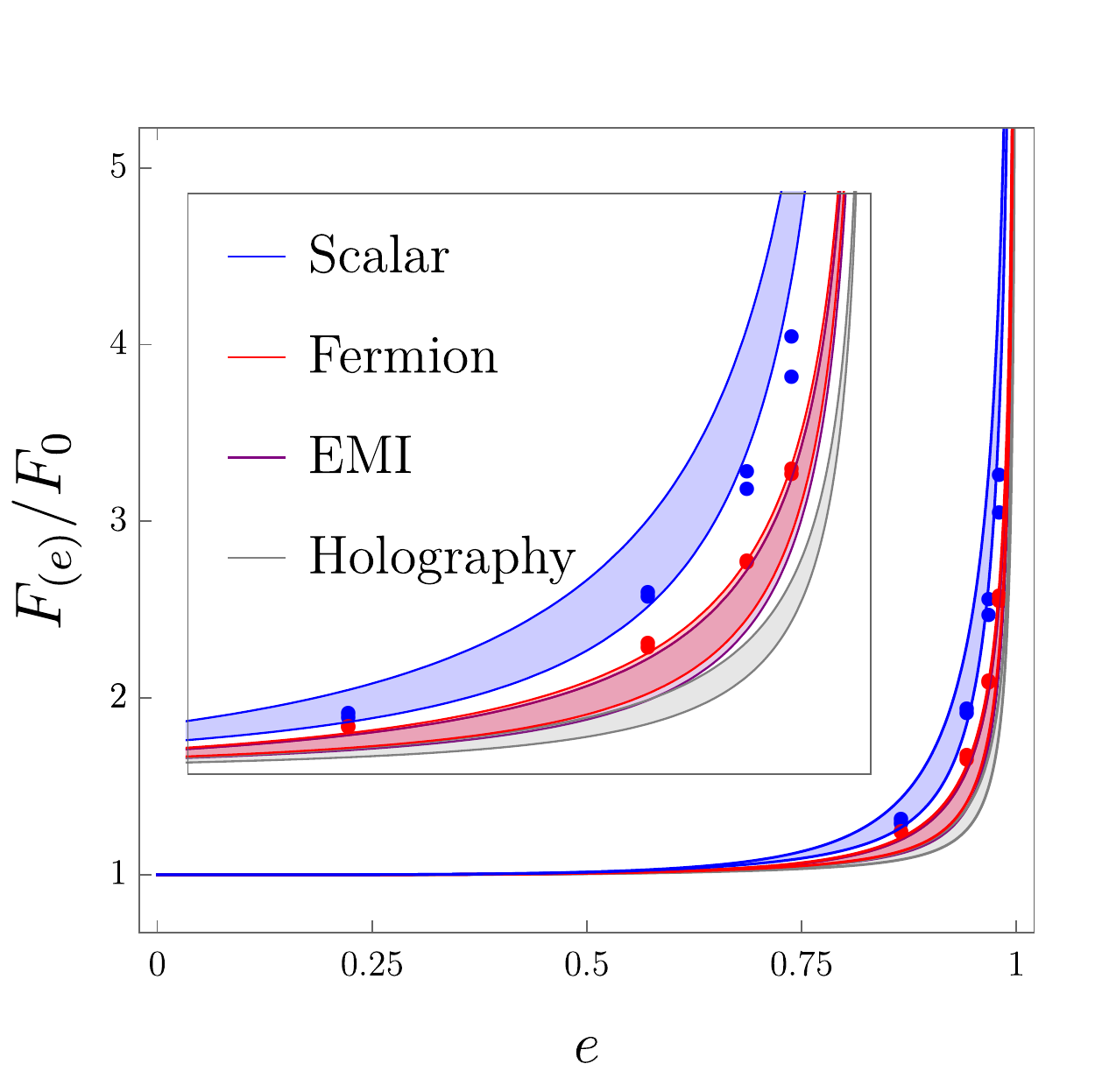}
	\caption{ 	\textsf{We plot the universal contribution to the entanglement entropy corresponding to an elliptic region, normalized by the disk result, $F_0$, as a function of the eccentricity. In the left plot, the data points are (exact) numerical results obtained for the EMI model and the purple curves are the trial functions $F_{(e)}|^{{\rm trial}_{1,2}} $ particularized to this model. The relative error in the approximations is shown in the inset. In the right plot, we present both trial-function curves for: a free scalar, a free fermion, the EMI model and holographic Einstein gravity. The dots correspond to lattice values for the scalar and the fermion. \rd{The inset is a zoom in of the region $e > 0.85$.}}}  
	\label{refiss254}
\end{figure}

Although $F_{(e)}|^{{\rm tri}_{1,2}} $ do not look particularly simple, note that the exact $F_{(e)}$ for a given CFT will in general be an extremely complicated function of $e$ depending on many details of the theory. Our approximations above are completely explicit and depend on the CFT under consideration exclusively through the three constants $k^{(3)}$, $F_0$ and $C_{\ssc T}$. Anticipating results presented in Section \ref{secemi}, we can test the precision of our formulas against numerical calculations for the EMI model. For this, we know $k^{(3)}$, $F_0$ and $C_{\ssc T}$ analytically ---see \req{kemi}, \req{F0emi} and \req{CTEMI} below--- and we can also compute $F_{(e)}$ exactly (up to numerical precision). The first function, $F_{(e)}|^{{\rm trial}_{1}}$ produces a very good approximation to the exact EMI model results. As shown in the inset of the left plot in fig.~\ref{refiss254}, the greatest discrepancy is lower than $\sim 5\%$ and it is much smaller for most values of $e$.
The case of $F_{(e)}|^{{\rm trial}_{2}}$ is rather different. For that one, the discrepancy grows up to a maximum of $\sim 22\%$ for $e\sim 0.96$. In view of this, we expect $F_{(e)}|^{{\rm trial}_{1}}$ to be a better approximation for general CFTs. However, we decided to keep $F_{(e)}|^{{\rm trial}_{2}}$ as it does a much better job in approximating the free scalar results obtained in the lattice, as we explain below. 

As anticipated, there are other CFTs for which we know the values of the three relevant coefficients and for which  we can therefore compute $F_{(e)}|^{{\rm tri}_{1,2}} $. In particular, for a free scalar \cite{Casini:2009sr,Osborn:1993cr,Klebanov:2011gs,Marino:2011nm}, a free fermion \cite{Casini:2009sr,Osborn:1993cr,Klebanov:2011gs,Marino:2011nm} and Einstein gravity holography \cite{Ryu:2006ef,Liu:1998bu} we have
\begin{align}\label{ctf0s}
&k^{(3)}_{\rm ferm}\simeq 0.0722 \, , & & C_{\ssc T}^{\rm ferm}=\frac{3}{16\pi^2} \, , & & F_0^{\rm ferm}=\frac{1}{8}\left[ 2\log 2+\frac{3}{\pi^2}\zeta(3)\right] \, , \\ \label{ctf0s1}
&k^{(3)}_{\rm scal}\simeq 0.0397 \, ,& & C_{\ssc T}^{\rm scal}=\frac{3}{32\pi^2} \, , & &F_0^{\rm scal}=\frac{1}{16} \left[2\log 2 -\frac{3}{\pi^2} \zeta(3)\right] \, , \\
&k^{(3)}_{\rm holo}=\frac{\pi \Gamma(\tfrac{3}{4})^2}{\Gamma(\tfrac{1}{4})^2}\frac{L^2}{\GN} \, , & & C_{\ssc T}^{\rm holo}=\frac{3}{\pi^3}\frac{L^2_\star}{\GN} \, ,& & F_0^{\rm holo}=\frac{\pi L^2}{2\GN} \, . 
\end{align}
Note that the fermion results correspond to a Dirac field, so the values of the coefficients per degree of freedom would require dividing them by $2$.
Using these coefficients, in the right plot of fig.~\ref{refiss254} we present the corresponding $F_{(e)}|^{{\rm tri}_{1,2}}  $ curves, alongside the EMI one. As we can see, the fermion and EMI curves are very close to each other\footnote{This is not surprising. For more on the relation between the free fermion and the EMI model see \cite{Agon:2021zvp}.} and, for each model, both trial functions are also rather similar. The scalar and holography curves are the ones making $F/F_0$ greater and lower, respectively, for general values of $e$. This hierarchy of theories is similar to the one encountered for the entanglement entropy corner function $a(\theta)$ normalized by $C_{\ssc T}$ in ref.~\cite{Bueno:2015rda}. In all cases, $F_{(e)}|^{{\rm tri}_{2}}$ lies notably below $F_{(e)}|^{{\rm tri}_{1}} $ in the whole range. 



 In the present case, we  have also performed lattice calculations for the free scalar and the fermion corresponding to $e=\sqrt{3}/2\simeq 0.866$, $e=2\sqrt{2}/3\simeq 0.943 $, $e \sqrt{15}/4\simeq 0.968$ and $e=2\sqrt{6}/5 \simeq 0.98$. As explained in the following subsection, we obtain two values of $F_{(e)}$ for each eccentricity and model. They are both presented in fig.~\ref{refiss254}. 
 We find
\begin{align} \notag
\bar F_{(\sqrt{3}/2)}^{\rm ferm}|_{\rm \ssc lattice}&\simeq  \{1.25 ,  1.25 \}  \, ,  \quad && \bar F_{(\sqrt{3}/2)}^{\rm scal}|_{\rm \ssc lattice}\simeq  \{1.32,  1.30 \}   \, . \\ \notag
\bar F_{(2\sqrt{2}/3)}^{\rm ferm}|_{\rm \ssc lattice}&\simeq \{1.66,  1.69 \}  \, ,  \quad && \bar F_{(2\sqrt{2}/3)}^{\rm scal}|_{\rm \ssc lattice}\simeq    \{1.95,  1.92 \}  \, , \\ \notag
\bar F_{(\sqrt{15}/4)}^{\rm ferm}|_{\rm \ssc lattice}&\simeq \{2.11,  2.10 \} \, ,  \quad && \bar F_{(\sqrt{15}/4)}^{\rm scal}|_{\rm \ssc lattice}\simeq   \{2.48,  2.57 \}   \, . \\ \label{elips}
\bar F_{(2\sqrt{6}/5)}^{\rm ferm}|_{\rm \ssc lattice}&\simeq  \{2.59,  2.56 \} \, ,  \quad && \bar F_{(2\sqrt{6}/5)}^{\rm scal}|_{\rm \ssc lattice}\simeq  \{3.06,  3.27 \}  \, ,
\end{align}
where we defined $\bar F_{(e)} \equiv  F_{(e)}/F_0$ here to avoid the clutter. Now, for the trial functions we find
 \begin{align} 
\bar F_{(\sqrt{3}/2)}^{\rm ferm}|^{{\rm  tri}_1}&  \simeq  1.21  \, ,  \notag
\quad   \bar F_{(2\sqrt{2}/3)}^{\rm ferm}|^{{\rm  tri}_1}\simeq 1.62  \, , \quad  \bar F_{(\sqrt{15}/4)}^{\rm ferm}|^{{\rm  tri}_1}\simeq 2.10\, , \quad \bar F_{(2\sqrt{6}/5)}^{\rm ferm}|^{{\rm  tri}_1}\simeq 2.60\, , \\ \notag
\bar F_{(\sqrt{3}/2)}^{\rm ferm}|^{{\rm  tri}_2}&  \simeq  1.13  \, ,  
\quad   \bar F_{(2\sqrt{2}/3)}^{\rm ferm}|^{{\rm  tri}_2}\simeq 1.41  \, , \quad  \bar F_{(\sqrt{15}/4)}^{\rm ferm}|^{{\rm  tri}_2}\simeq 1.80\, , \quad \bar F_{(2\sqrt{6}/5)}^{\rm ferm}|^{{\rm  tri}_2}\simeq 2.25\, , \\ \notag
\bar F_{(\sqrt{3}/2)}^{\rm scal}|^{{\rm  tri}_1}&  \simeq 1.43   \, ,  \notag
\quad   \bar F_{(2\sqrt{2}/3)}^{\rm scal}|^{{\rm  tri}_1}\simeq 2.31  \, , \quad  \bar F_{(\sqrt{15}/4)}^{\rm scal}|^{{\rm  tri}_1}\simeq 3.30 \, , \quad \bar F_{(2\sqrt{6}/5)}^{\rm scal}|^{{\rm  tri}_1}\simeq 4.32 \, , \\
\bar F_{(\sqrt{3}/2)}^{\rm scal}|^{{\rm  tri}_2}&  \simeq 1.27  \, ,  
\quad   \bar F_{(2\sqrt{2}/3)}^{\rm scal}|^{{\rm  tri}_2}\simeq 1.87   \, , \quad  \bar F_{(\sqrt{15}/4)}^{\rm scal}|^{{\rm  tri}_2}\simeq 2.67 \, , \quad \bar F_{(2\sqrt{6}/5)}^{\rm scal}|^{{\rm  tri}_2}\simeq 3.57 \, .
\end{align}
We observe that the fermion results never differ from the first trial function prediction more than a $\sim 4\%$ and in most cases the agreement is better. The agreement with the second trial function is much worse, and the discrepancies reach a maximum of $\sim 17\%$. In the case of the scalar, it is the second trial function the one which approximates  better the lattice results, with disagreements lower than $\sim 5\%$ in most cases, except for the last value, which differs by a $\sim 13\%$. On the other hand, the results are clearly off from the ones predicted by the first trial function: in some cases, the discrepancy grows up to $\sim 40\%$.
We expect the results both for the scalar and the fermion to have an uncertainty which is probably not less than $\sim 5\%$. This does not explain however why we cannot fit the results for both theories with a single curve. In fact, we suspect that the lattice is considerably underestimating the actual scalar results. In particular, as we mention below, analogous calculations for the disk region yield $\bar F_0^{\rm ferm}|_{\rm \ssc lattice}\simeq 0.99$ and $\bar F_0^{\rm scal}|_{\rm \ssc lattice}\simeq 0.92$. Namely, while the fermion result approaches the analytical answer very well, the scalar one is an $8\%$ lower. In view of this and of the agreement for the EMI and free fermion results with the first trial function, a reasonable guess is that $F_{{(e)}}|^{{\rm tri}_{1} }$ is actually a very good approximation to the exact curve for general CFTs, including the scalar, whereas  $F_{{(e)}}|^{{\rm tri}_{2} }$ is not so much. We say a bit more about this in the next subsection.


\subsection{Free-field calculations in the lattice}
As anticipated, in this subsection we compute $F_{(e)}/F_0$ for elliptic regions in the lattice for free scalars and fermions. As argued in ref.~\cite{Casini:2015woa}, trying to obtain $F_0$ from a direct calculation of the entanglement entropy in a square lattice does not produce reasonable results\footnote{For satisfactory calculations in radial lattices see ref.~\cite{Liu:2012eea,Klebanov:2012va}.}. Rather, one obtains wildly oscillating answers as the ratio $R/\delta$ varies (here $\delta$ stands for the lattice spacing).  The problem has to do with the fact that we cannot resolve the radius of the disk with a precision better than $\delta$, which means that we cannot distinguish disks with radii $R$ and $R( 1+ a \delta)$, with $a \sim \pazocal{O}(1)$. But such uncertainty will pollute $F$ via the area-law piece in the entanglement entropy \eqref{entro} as $-F \rightarrow -F + 2\pi b_{(1)} a$. As it is clear from this, the issue cannot be resolved by making the disk radius larger in the lattice.

\begin{figure}[t]  \centering
\includegraphics[width=0.4425\textwidth]{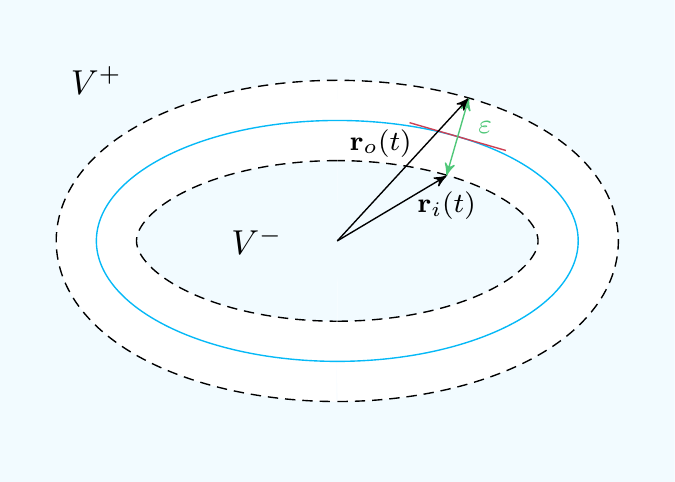}
\includegraphics[width=0.5075\textwidth]{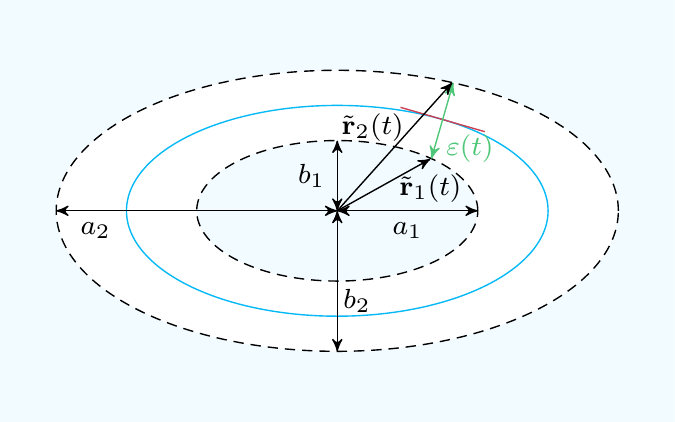}
\caption{ \textsf{(Left) We plot an example of the geometric setup required for the evaluation of the entanglement entropy for an elliptic region $V$ (solid blue curve) using the mutual information of an inner region $V^-$ and the complement of one which contains it, $\overline{V^+}$, as in \req{annulii}. In this case, the inner and outer regions are defined so that $\varepsilon$ is constant for all values of $t$. The points ${\bf r}_i(t)$ and ${\bf r}_o(t)$ are determined by moving a distance $\varepsilon/2$ inwards and outwards respectively from each point of $\Sigma$.  (Right) We show an alternative setup for which $\overline{V^+}$ and $V^-$ are both elliptic regions, which forces $\varepsilon(t)$ to be variable. The points $\tilde {\bf r}_{1,2}(t)$ are determined from the intersection of the normal to $\Sigma$ at a given point with the inner and outer ones.}  }
	\label{refiss34}
\end{figure}

An alternative approach which does produce convergent results consists in using mutual information as a geometric regulator \cite{Casini:2007dk,Casini:2008wt,Casini:2014yca,Casini:2015woa}. Given a region $V$, we consider two concentric ones $V^-$ and $\overline{V^+}$. These are defined by considering a normal to $\Sigma$ at each point and moving a distance $\varepsilon/2$ inwards and outwards along such direction, respectively. $\varepsilon$ can be chosen to be constant for all values of the parameter $s$ defining $\Sigma$ or, alternatively, it can be a function of $s$ if one decides that $V^+$ and $\overline{V^-}$ should be dilated versions of $V$ ---\eg if $V$ is an ellipse, $V^-$ and $\overline{V^+}$ would also be ellipses using this second method.  In both cases, we are left with an inner annulus of width $\varepsilon$ (constant or variable) ---see fig.~\ref{refiss34}.

The idea is then to consider the mutual information
\begin{equation}
I(V^+,V^-)=\see(V^+)+\see(V^-)-\see(V^+ \cup V^-)\, ,
\end{equation}
which, using the purity of the ground state can be rewritten as
\begin{equation}\label{annulii}
I(V^+,V^-)=\see(\overline{V^+})+\see(V^-)-\see( \overline{V^+ \cup V^-})\, ,
\end{equation}
where 
$\overline{V^+ \cup V^-}$ is the annulus region. The three regions appearing in \req{annulii} are finite and are thus defined by a finite number of points in the lattice, so they are suitable for that setup.

 Now the idea is to consider annuli such that $\varepsilon/R\ll 1$ while keeping $\varepsilon/\delta \gg 1$. In that limit, the MI behaves as $I(V^+,V^-) \simeq 2 \see (V)$, where $V$ is the intermediate region. This equality is true up to terms which diverge as $\sim 1/\varepsilon$ and $\sim 1/\delta$ respectively in the corresponding limits. Hence, in order to extract $F$ from the mutual information (\ref{annulii}), one must deal with those terms first. As argued in ref.~\cite{Casini:2015woa}, if we parametrize the boundary  of $V$ by the length parameter $s$, one finds for  $\varepsilon/R\ll 1$,
 \begin{equation}\label{ik3}
 I(V^+,V^-)=k^{(3)} \int_{\Sigma} \frac{\diff s}{\varepsilon (s)}-2 F + \text{subleading in } \varepsilon\, , 
 \end{equation} 
where $k^{(3)}$ is the coefficient characterizing the EE of a thin strip region ---see \req{strip}. In order for this to work, the region $V$ must be chosen precisely half way between $\Sigma^+$ and $\Sigma^-$. As mentioned earlier, $\varepsilon (s)$ corresponds to the separation between $\Sigma^+$ and $\Sigma^-$ at a given point $s$, measured along the normal direction to $\Sigma$ at that point. Therefore, given some region $V$ in the lattice, we can extract its $F$ by computing $I(V^+,V^-)$ as in \req{annulii} and subtracting the first piece in the right hand sinde of \req{ik3}, then dividing by $-2$. For fixed values of $\varepsilon/R$, where $R$ is some characteristic size of $V$, the results will improve their accuracy as $R/\delta \gg 1$.

In the case of disk regions, \req{ik3} simplifies considerably, and one gets
 \begin{equation}\label{ik23}
 F_0=-\frac{1}{2}\left[  I({\rm disk }_{R_2},{\rm disk}_{R_1})-k^{(3)} \frac{2\pi R}{\varepsilon} \right]\, ,
 \end{equation} 
 where $\varepsilon\equiv R_2-R_1$ and $R\equiv (R_1+R_2)/2$. In this case, both methods described above involve a constant $\varepsilon$ and $V^-$, $\overline{V^+}$ are always disks.
  
 For an elliptic region $V$ of eccentricity $e$, on the other hand, we can either choose a constant $\varepsilon$ or, alternatively, force $V^-$ and $\overline{V^+}$ to be ellipses as well. In both cases, 
computing $\varepsilon$ requires obtaining the line which intersects the boundary of the elliptic entangling region normally at the point determined by $t$ ---see fig.~\ref{refiss34}. Assuming the ellipse is centered at the origin of coordinates and parametrized by $[a \cos(t),b \sin(t)]$, this is given by
\begin{equation}
y(x)=\frac{a}{b} \tan(t)  x + \frac{b^2-a^2}{b} \sin(t)\, .
\end{equation}
 A point on the normal can be parametrized as $[a \cos(t),y(a \cos(t))]$ and we want to determine the quantity $\alpha$ that we need to add to both coordinates so that the new point lies a distance $\varepsilon/2$ from $[a \cos(t), b \sin(t)]$. Hence, we need to solve
 \begin{equation}
 [a \cos (t) - (a \cos (t)+ \alpha)]^2 + [b \sin (t) - y(a \cos(t)+ \alpha)]^2 =\varepsilon/2\, ,
 \end{equation}
 for $\alpha$. Solving this equation and plugging back we find that the points ${\bf  r}_{i,o}(t)$ defining the shapes of the inner and outer curves read
 \begin{eqnarray} \label{rio}
{\bf  r}_{i,o}(t)=
 \begin{cases}
\left[a \cos(t)+\frac{ (-1)^i  b \varepsilon}{2\sqrt{b^2+a^2\tan^2(t)}},y\left(a \cos(t)+\frac{ (-1)^i  b \varepsilon}{2\sqrt{b^2+a^2\tan^2(t)}}\right)\right]\quad \text{for} \quad \frac{3\pi}{2}<t\leq\frac{\pi}{2} \, ,\\
\left[a \cos(t)-\frac{ (-1)^i  b \varepsilon}{2\sqrt{b^2+a^2\tan^2(t)}},y\left(a \cos(t)-\frac{  (-1)^i  b \varepsilon}{2\sqrt{b^2+a^2\tan^2(t)}}\right)\right]\quad \text{for} \quad\frac{\pi}{2}< t\leq\frac{3\pi}{2}\, .
\end{cases}
\end{eqnarray}
Here, $(-1)^i=-1,1$ for the inner and outer shapes, respectively. Finally, we can compute the entanglement entropy of the original elliptic region as
 \begin{equation}\label{ik130}
 F_{(e)}=-\frac{1}{2}\left[  I({\rm pseudoellipse }_{i},{\rm pseudoellipse}_{o})-k^{(3)} \frac{4 a E[e^2]}{\varepsilon}\right]\, ,
 \end{equation}  
where $I({\rm pseudoellipse }_{i},{\rm pseudoellipse}_{o})$ is the mutual information of the ``pseudoellipses'' defined by the shapes ${\bf  r}_{i}(t)$ and ${\bf  r}_{o}(t)$ respectively.

 
For the second method, on the other hand, we consider two concentric ellipses of the same eccentricity $e$ parametrized by $[a_{1,2}\cos(t) ,b_{1,2}  \sin(t)]$. Computing $\varepsilon(t)$ requires now obtaining the intersections of the normal line to $\partial A$ with the exterior and interior ellipses. These have equations $y=b_{1,2} \sqrt{1-x^2/a_{1,2}^2}$ where now $a= (a_1+a_2)/2$, $b=(b_1+b_2)/2$. The intersections occur at the points ${\bf \tilde r}_{1,2}(t)$ ---see fig.~\ref{refiss34}--- where
\begin{align} \label{x1}
\tilde x_1&\equiv \frac{a a_1^2(a^2-b^2)\sin(t)\tan(t) + \sqrt{a_1^2 b^2b_1^2\left[b^2b_1^2+a^2a_1^2\tan^2(t)- (a^2-b^2)^2 \sin^2(t)\right]}}{a^2a_1^2\tan^2(t)+b^2b_1^2} \, , \\ \label{x2}
\tilde y_1&\equiv \frac{b^2 b_1^2(b^2-a^2)\sin(t)+a\tan(t)  \sqrt{a_1^2 b^2b_1^2\left[b^2b_1^2+a^2a_1^2\tan^2(t)- (a^2-b^2)^2 \sin^2(t)\right]}}{a^2a_1^2b\tan^2(t)+b^3b_1^2} \, ,
\end{align}
and $(\tilde x_2,\tilde y_2)$ are given by the same expressions replacing $(1\leftrightarrow 2)$  in all labels.
From this, $\varepsilon(t)$ can be obtained as
\begin{equation} \label{epis}
\varepsilon(t)=\sqrt{[\tilde x_1(t)-\tilde x_2(t)]^2+[\tilde y_1(t)-\tilde y_2(t)]^2}\, .
\end{equation}
Putting the pieces together, we find that the EE universal term for an ellipse of eccentricity $e$ with a boundary half way between two concentric ellipses of the same eccentricity can be obtained  as
 \begin{equation}\label{ik13}
 F_{(e)}=-\frac{1}{2}\left[  I({\rm ellipse }_{2},{\rm ellipse}_{1})-k^{(3)} \int_0^{\frac{\pi}{2}} \diff t \frac{2(a_1+a_2)\sqrt{1-e^2 \cos^2 t} }{\varepsilon(t)}\right]\, ,
 \end{equation} 
where $1,2$ are labels referring to the two (inner and outer, respectively) concentric ellipses.

Both eqs.~\req{ik130} and \req{ik13} should produce equivalent approximations to $F_{{(e)}}$ for sufficiently large values of $b/\varepsilon$. However, as we discuss in app.~\ref{EMIMIregu} in the case of the EMI model ---for which we can perform calculations for arbitrary values of $b/\varepsilon$--- the prescriptions are not equally good in their precision for finite values of $b/\varepsilon$ ---see fig.~\ref{ref44}. In particular, we find that the pseudoellipses one produces more stable and accurate answers, specially for values of $e$ close to $1$. In the lattice, we have evaluated $I({\rm pseudoellipse }_{i},{\rm pseudoellipse}_{o})$ as well as  $I({\rm ellipse }_{2},{\rm ellipse}_{1})$ numerically using eq,~\eqref{annulii} and then subtracted the purely geometric pieces proportional to $k^{(3)}$ in each case. Similarly to the EMI case, we observe that the first method produces greater and more stable results, so we only present those here. 

We perform  lattice calculations for free scalars and fermions. For the former, we consider  a set of  fields and conjugate momenta $\phi_i, \pi_j$, $i, j= 1,\dots, N$  labeled by their positions at the square lattice. These satisfy   canonical  commutation relations,  $[\phi_i ,\pi_j]=i\delta_{ij}$ and $[\phi_i,\phi_j]=[\pi_i,\pi_j]=0$. Then, given some  Gaussian state   $\rho$, the entanglement entropy can be computed from the two-point correlators
$
X_{ij}\equiv \tr (\rho \phi_i\phi_j)$ and   $P_{ij}\equiv \tr (\rho \pi_i \pi_j)\, .
$
In particular, one  has \cite{2003JPhA...36L.205P,Casini:2009sr}
\begin{equation}\label{see}
\see(V)=\tr \left[(C_V +1/2) \log (C_V   +1/2)- (C_V -1/2)   \log (C_V-1/2) \right]\, , 
\end{equation}
where $C_V \equiv \sqrt{X_V    P_V}$ and  $(X_V)_{ij}\equiv X_{ij}$, $(P_V)_{ij}=P_{ij}$ (with $i,j\in V$) are the restrictions of the correlators   to the  sites belonging  to region $V$.

The lattice  Hamiltonian reads  in this case
\begin{equation}
H=\frac{1}{2}\sum_{n,m=-\infty}^{\infty}\left[\pi^2_{n,m}   + (\phi_{n+1,m}   -\phi_{n,m})^2  +(\phi_{n,m+1}-\phi_{n,m})^2 \right]\, ,
\end{equation}
where the lattice spacing   is set to one. The   relevant expressions for the correlators $X_{(x_1,y_1),(x_2,y_2)}$ and $P_{(x_1,y_1),(x_2,y_2)}$   corresponding to the vacuum state are given by \cite{Casini:2009sr} 
\begin{align}
X_{(0,0),(i,j)} & = \frac{1}{8\pi^2}\int_{-\pi}^{\pi} \diff  x \int_{-\pi}^{\pi} \diff y \frac{\cos (jy) \cos (ix) }{\sqrt{2(1-\cos x)+2(1-\cos y)}}\, , \\
P_{(0,0),(i,j)} & = \frac{1}{8\pi^2}\int_{-\pi}^{\pi} \diff  x \int_{-\pi}^{\pi} \diff y  \cos(jy)  \cos(ix) \sqrt{2(1-\cos x)+2(1-\cos y)}\, .
\end{align}

In the case of the free fermion, we consider fields  $\psi_i$, $i=1,\dots, N$  defined at the lattice points  and satisfying canonical  anticommutation   relations, $\{\psi_i, \psi_j^{\dagger} \}=\delta_{ij}$. Given a Gaussian state  $\rho$, we define the matrix of correlators  $D_{ij} \equiv \tr \small( \rho \psi_i \psi_j^{\dagger} \small)$. Then, similarly to the scalars case,  the EE for some region $A$ can be obtained from the restriction of $D_{ij}$ to the corresponding  lattice sites as \cite{Casini:2009sr}
\begin{equation}
\see (V)=- \tr   \left[ D_V   \log D_V   + (1-D_V) \log (1-D_V)\right] \, .
\end{equation}
The lattice Hamiltonian is given   by
\begin{equation}
H=-\frac{i}{2}  \sum_{n ,m} \left[  \left(\psi^{\dagger}_{m, n}   \gamma^0     \gamma^1 (\psi_{m+1,n}  -\psi_{m,n})+\psi^{\dagger}_{m,  n}   \gamma^0 \gamma^2   (\psi_{m,n+1}   -\psi_{ m,n}  ) \right)  - h.c.\right] \, ,
\end{equation}
and the correlators in the vacuum state read \cite{Casini:2009sr} 
\begin{equation}
D_{(n,k),(j,l)} =   \frac{1}{2}\delta_{n,j} \delta_{kl}     - \int_{-\pi}^{\pi} \diff x  \int_{-\pi}^{\pi} \diff y   \frac{\sin (x) \gamma^0 \gamma^1+   \sin(y) \gamma^0 \gamma^2}{8\pi^2 \sqrt{   \sin^2 x +   \sin^2 y}} e^{i(x (n-j)+y(k-l))}\, .
\end{equation}

\begin{figure}[t] \hspace{-0.3cm}
	 \includegraphics[scale=0.6]{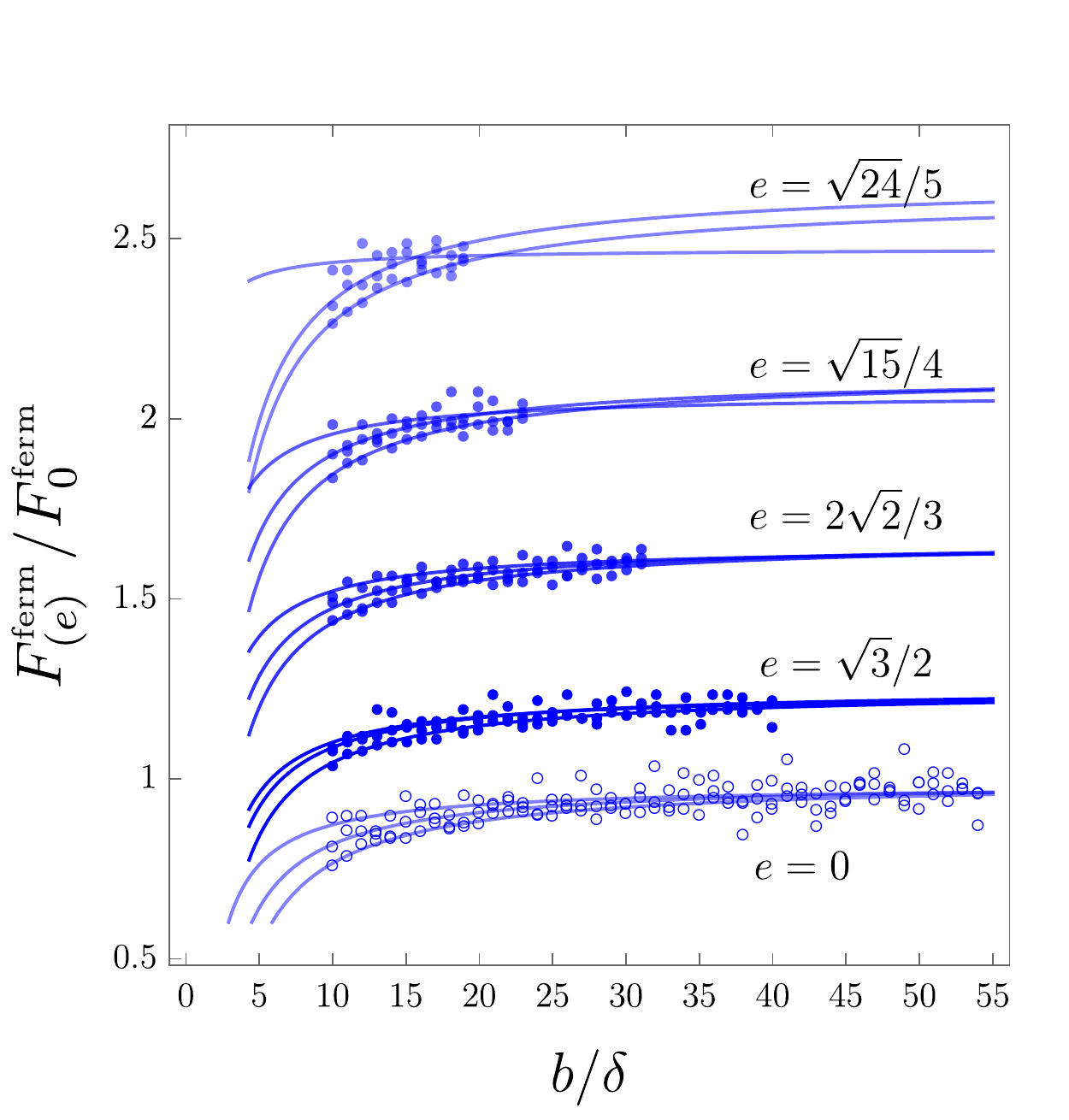}\hspace{-0.2cm}	
	\includegraphics[scale=0.6]{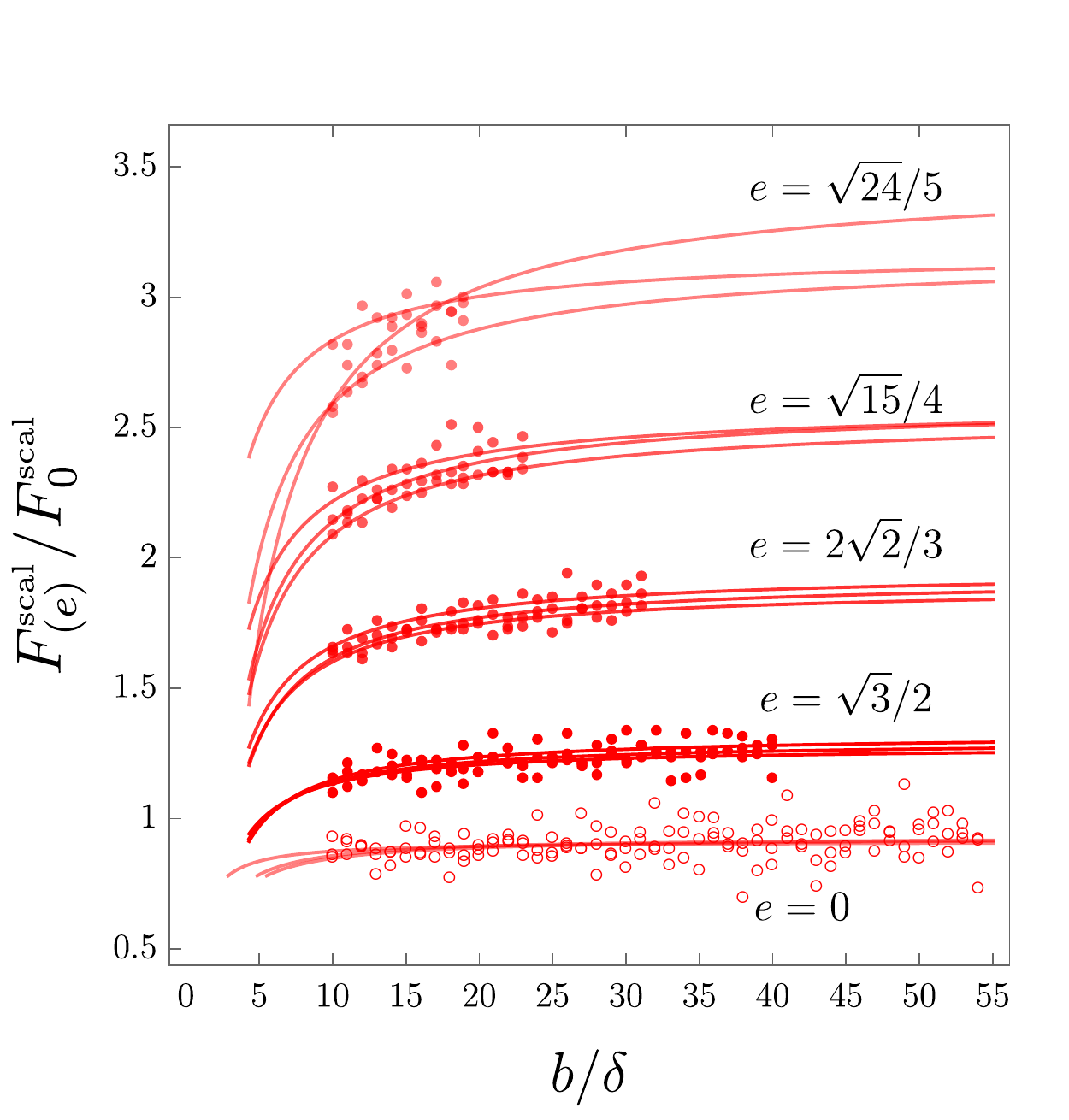}
	\caption{ \textsf{We plot $F_{(e)}$ normalized by the disk result for free fermions and scalars in the lattice for eccentricities $e=0,\sqrt{3}/2\simeq 0.943 ,2\sqrt{2}/3\simeq0.968 ,\sqrt{15}/4\simeq 0.968, 2\sqrt{6}/5\simeq 0.98$, as a function of the ellipses' semiminor axes divided by the lattice spacing. The data points are obtained using the constant-$\varepsilon$ formula \req{ik130} for $\varepsilon=6,8,10$. The solid lines correspond to $\{1/x,1\}$ fits  performed ---for each value of $e$--- to the data points for fixed $\varepsilon$. } 
	 }
	\label{refiss344}
\end{figure}

As a warm up, we perform calculations for disk regions ---see also \cite{Casini:2015woa}, where this was previously done. In that case, the results for $F_0$ are known analytically both for the scalar and the fermion ---see eqs.~\eqref{ctf0s} and \eqref{ctf0s1} above. In the lattice, we consider pairs of disks of different radii and the corresponding annuli and use \req{ik23} to extract our results. The results have a numerical uncertainty associated to the finiteness of both $\varepsilon/R$ and $\delta/\varepsilon$. This comes from the difficulty of achieving $\varepsilon/R \ll 1$ while keeping $\delta/\varepsilon \ll 1$. Naturally, 
 considering sufficiently large values of $\varepsilon$ requires taking disks with large enough radii, which increases the computation time. In fig.~\ref{refiss344} we present various data points of $F_0$ (normalized by the analytic result) for both the scalar and the fermion as a function of $b/\varepsilon$. We take $\varepsilon/\delta=6,8,10$ and values of disk radius up to $R/\delta =58$. As we can see, the results oscillate considerably for both models. In order to extract a number for each of them, we proceed in two ways. First, we take the two values of $F_{(e)}$ corresponding to the largest $R/\delta$ considered for each of the three values of $\varepsilon$, and take the average of those six values. Then, we multiply them by $1$ plus a small correction factor ($\sim 0.05$) which  we obtain in app.~\ref{EMIMIregu} based on the EMI model and which should approximately account for the fact that we are considering not too large values of $R/\delta$ for the given values of $\varepsilon/\delta$.  Secondly, we perform fits  $\{1/x,1 \}$ to the series of data points for each $\varepsilon$ and take the average of the three values.  We find from these two methods, respectively,
\begin{equation}
F_0^{\rm scal}|_{\rm \ssc lattice}\simeq \{0.98,0.92\} F_0^{\rm scal} \, ,  \quad F_0^{\rm ferm}|_{\rm \ssc lattice}\simeq  \{1.02,0.99\} F_0^{\rm ferm}\, .
\end{equation}
The results are in very good agreement with the expectations, although we observe that the scalar one obtained from the fits is still underestimating the actual result.

Moving on to the ellipses, we perform calculations for eccentricities $e=\sqrt{3}/2\simeq 0.866$, $e=2\sqrt{2}/3\simeq 0.943$, $e=\sqrt{15}/4\simeq 0.968$ and $e= 2\sqrt{6}/5\simeq 0.98$. These belong to the region in which the almost-round and very-thin approximations ---eqs.~\eqref{eli0} and \eqref{elie1} respectively--- do not work so well. As mentioned earlier, the pseudoellipses method yields better results than the variable-$\varepsilon$ one, so we use the former to perform our calculations. The results are shown in fig.~\ref{refiss344}. Proceeding similarly to the disks, in order to extract approximations for  large-$b/\delta$ results we make fits to the data for fixed values of $\varepsilon$ and take the average. Alternatively, we also consider for each   $\varepsilon$ the values obtained for the two greatest values of $b/\delta$, take the average of the six and multiply by the correction factor. The results appear  in eq.~\eqref{elips} above. The fermion data points look somewhat more harmonious than the scalar ones and the data series produce more similar predictions. On the other hand, as mentioned in the previous subsection, our candidate trial function $F_{{(e)}}|^{{\rm tri}_{1} }$ ---which yielded a very good agreement with the exact EMI results--- produces results which also agree very well with the fermion ones in the lattice, but not so for the scalar. It is possible that the methods we are using are still insufficient to account for the underestimation of the results associated to the finite-$R/\delta$ limitations in the case of the free scalar. If that is case, the predictions of $F^{\rm scal}_{{(e)}}|^{{\rm tri}_{1} }$ are probably more accurate than our lattice results. Alternatively, it may also be that the idea of accurately reproducing $F_{(e)}$ for any CFT with a single function completely determined by $k^{3}$, $C_{\ssc T}$ and $F_0$ does not work so well and one rather requires a set of functions ---\eg $F^{\rm scal}_{{(e)}}|^{{\rm tri}_{1} }$, $F^{\rm scal}_{{(e)}}|^{{\rm tri}_{2} }$ and the intermediate curves between them, as displayed in fig.~\ref{refiss254}--- to precisely account for all possible theories.

Regarding the fundamental question which is subject of study in the present paper, the evidence  gathered from holography \cite{Fonda:2014cca,Anastasiou:2020smm}, the EMI model, the lattice results for free fields and the two limits ($e \rightarrow 0$ and $e \rightarrow 1$) for general theories makes it clear that $F_{(e)}$ is  a monotonically increasing function of the eccentricity for a given CFT ---and therefore $F_{(e)}\geq F_0$ for all $e$ for general CFTs.  As a final comment for this section, we also point out that the methods utilized here for computing EE using MI should be similarly applicable to other regions beyond ellipses. In that case, the constant-$\varepsilon$ method appears to be the best choice.

\section{More general shapes in the EMI model}\label{secemi}
In order to obtain explicit results for more complicated entangling regions, we consider now the EMI model \cite{Casini:2008wt}. This follows from considering a general formula for the mutual information compatible with the known general axioms satisfied by this measure in a general QFT ---see \eg \cite{Agon:2021zvp}--- plus the additional requirement that it is an extensive function of its arguments \eqref{eq:ExMI}. This model corresponds to a free fermion in $d=2$ \cite{casini2005entanglement} but does not describe the mutual information of any actual theory (or limit of theories) for $d\geq 3$, as recently shown \cite{Agon:2021zvp}. Nonetheless, the fact that it satisfies all known properties expected for a valid mutual information makes it a useful toy model which has shown to produce qualitatively and quantitatively reasonable results in various cases \cite{Casini:2015woa,Bueno:2015rda,Witczak-Krempa:2016jhc,Bueno:2019mex,Estienne:2021hbx}.

In the EMI model,  the entanglement entropy of a region $V$ in a time slice of three-dimensional Minkowski spacetime is given by
\begin{equation} \label{emiee}
\SEMI(V)=\kappa_{(3)}\int_{\Sigma}\diff\mathbf{r}_1\int_{\Sigma}\diff\mathbf{r}_2\, \frac{\mathbf{n}_1\cdot\mathbf{n}_2}{\abs{\mathbf{r}_1-\mathbf{r}_2}^{2}}\, ,
\end{equation}
where the integrals are both along the entangling surface $\Sigma$, $\mathbf{n}$ is a unit normal vector and $\kappa_{(3)}$ is a positive parameter.

There are different ways to regulate the above integrals, which otherwise diverge as $\mathbf{r}_1 \rightarrow \mathbf{r}_2$. For our purposes, it will be convenient to introduce a UV regulator $\delta$ along an auxiliary extra dimension, so that we replace  $\abs{\mathbf{r}_1-\mathbf{r}_2}^{2} \rightarrow \abs{\mathbf{r}_1-\mathbf{r}_2}^{2}+\delta^2$ with $\delta \ll $ all the rest of scales.

Now, in order to obtain a computationally useful formula for the universal constant term, $F^{\rm \ssc EMI}(V)$, note from \req{entro} that, on general grounds,
\begin{equation}\label{delF}
 F(V)=- \left[\delta\frac{\partial \see (V)}{\partial \delta} + \see(V) \right] \, ,
\end{equation}
\ie the combination in the right hand side is equivalent to $F(V)$ up to terms which vanish as the regulator is taken to zero. 

Using eq.~\eqref{delF} we find
\begin{equation}\label{femi}
F^{\rm \ssc EMI}(V)=-\kappa_{(3)}\int_{\Sigma}\diff\mathbf{r}_1\int_{\Sigma}\diff\mathbf{r}_2\,  \frac{| \mathbf{r}_1-\mathbf{r}_2|^2-\delta^2}{(\abs{\mathbf{r}_1-\mathbf{r}_2}^{2}+\delta^2)^2} \mathbf{n}_1\cdot\mathbf{n}_2\, .
\end{equation}
In order to fix $b_1$, we can evaluate the entanglement entropy for a simple region like a strip, or a disk. By doing so, we find that $b_1=\pi$ for the regulator introduced above, and we can write\footnote{One could try to extract a $\delta$-independent expression for $F^{\rm \ssc EMI}(A)$, \eg by rewriting eq.~\eqref{femi} as
\begin{equation}\label{femi0}
F^{\rm \ssc EMI}(V)= -\frac{ \kappa_{(3)}}{2}\int_{\Sigma}\diff\mathbf{r}_1\int_{\Sigma}\diff\mathbf{r}_2\, \frac{\partial^2}{\partial \delta^2} \log \left(|\mathbf{r}_1 - \mathbf{r}_2|^2 + \delta^2\right)  \mathbf{n}_1\cdot\mathbf{n}_2,
\end{equation}
trading the derivatives with the integrals and finally taking the $\delta \rightarrow 0$ limit. Unfortunately, we have not been able to obtain anything too useful from this approach. }
\begin{equation}\label{femi2}
F^{\rm \ssc EMI}(V)=\kappa_{(3)} \left[ \frac{\pi}{\delta}\int_{\Sigma}\diff\mathbf{r}_1 -\int_{\Sigma}\diff\mathbf{r}_1\int_{\Sigma}\diff\mathbf{r}_2\, \frac{\mathbf{n}_1\cdot\mathbf{n}_2}{\abs{\mathbf{r}_1-\mathbf{r}_2}^{2}+\delta^2}\right]\, .
\end{equation}

Now, in order to make progress, let us choose a particular set of coordinates. We parametrize our entangling surfaces as 
\begin{align}
\mathbf{r}_i=&f(\theta_i)  \left(\cos\theta_i,  \sin\theta_i\right)\, , \quad i=1,2\, ,
\end{align}
where $f(\theta_i)$ is a function of the polar angle, $\theta \in [0,2\pi)$.
Naturally, in the case of a disk, $f(\theta_i)=R$, where $R$ is just its radius. 
In general we have
\begin{equation}
\diff \mathbf{r}_i=\diff\theta_i \sqrt{f_i^2+f_i'^2}\quad \text{and}\quad \abs{\mathbf{r}_1-\mathbf{r}_2}^{2}=f_1^2+f_2^2-2f_1f_2\cos\left[\theta_1-\theta_2\right]\, ,
\end{equation}
where we used the notation $f_i \equiv f(\theta_i)$.
As for the normal vectors, one finds
\begin{equation}
\mathbf{n}_i=-\left(\frac{f_i \cos \theta + f_i' \sin\theta}{ \sqrt{f_i^2+f_i'^2}},\frac{f_i \sin \theta- f'_i \cos \theta}{ \sqrt{f_i^2+f_i'^2}}\right)\, .
\end{equation}
Using these we get
\begin{equation}
\diff \mathbf{r}_1 \diff \mathbf{r}_2\,  \mathbf{n}_1 \cdot \mathbf{n}_2= \diff \theta_1 \diff \theta_2 \left[\cos (\theta_1-\theta_2) \left[f_1 f_2+f'_1 f'_2\right]+ \sin (\theta_1-\theta_2) \left[f_1' f_2-f_2' f_1\right] \right]\, .
\end{equation}
Plugging the above expressions back in eqs.~\eqref{femi} or \req{femi2}, we obtain an explicit formula for the universal term.

Let us first consider the simplest possible case, corresponding to a disk region. Then, we have $f_i=R$, and the expression for $F^{\rm \ssc EMI}$ simplifies drastically,
\begin{equation}
F^{\rm \ssc EMI}_0=-\kappa_{(3)} R^2 \int_0^{2\pi} \diff \theta_1  \int_0^{2\pi} \diff \theta_2   \frac{ [2R^2[1-\cos(\theta_1-\theta_2)]-\delta^2] \cos(\theta_1-\theta_2)}{[2R^2[1-\cos(\theta_1-\theta_2)]+\delta^2]^2}+\mathcal{O}(\delta) \, .
\end{equation}
The integrals can be explicitly performed and the result reads
\begin{equation}\label{F0emi}
F^{\rm \ssc EMI}_0=2\pi^2 \kappa_{(3)}\, .
\end{equation}
Alternatively, we can use \req{femi2}. The first integral is trivially $2\pi R$ and, for the second, we can use the symmetry of the disk to fix $\mathbf{r}_2=(R,0)$ $\mathbf{n}_2=(1,0)$. Then, we have
\begin{align}
F^{\rm \ssc EMI}_0&=\kappa_{(3)} \left[\frac{\pi}{\delta } 2\pi R- 2\pi R^2 \int_0^{2\pi} \diff \theta_1  \frac{\cos \theta_1}{ 4 R^2\sin^2(\theta_1/2)+\delta^2} \right]\notag \\
&=\kappa_{(3)} \left[ \frac{2\pi^2 R}{\delta } - \frac{2\pi^2 R}{\delta }+ 2\pi^2\right]= 2\pi^2 \kappa_{(3)},
\end{align}
finding the same answer.

Now, in order to consider more complicated figures, we will numerically integrate eqs.~\eqref{femi} and \eqref{femi2} for various values of $\delta \ll 1$ and obtain a linear fit to the resulting data ---which is indeed approximately linear in $\delta$. The values of $F^{\rm \ssc EMI}$ are in each case obtained as the $\delta=0$ limits of the fits. 

\subsection{Slightly deformed disks}
The next-to-simplest case corresponds to small deformations of the disk region like the ones considered in eq.~\eqref{fiy}. In that case, we can compare our numerical results to Mezei's formula \req{fmeze0}. 

In the case of the EMI model, the stress-tensor two-point function coefficient takes the value \cite{Agon:2021zvp}
\begin{equation}\label{CTEMI}
C^{\rm \ssc EMI}_{\ssc T}=\frac{16 \kappa_{(3)}}{\pi^2}\, .
\end{equation}
This can be obtained by considering an entangling region $V$ with a straight corner which, for general CFTs, contains a universal term of the form \cite{Casini:2006hu,Hirata:2006jx}
\begin{equation}
\suniv(V)= - a(\theta) \log\frac{H}{\delta}\, , 
\end{equation}
where the function of the opening angle $a(\theta)$ satisfies  \cite{Bueno:2015rda}
\begin{equation}
a(\theta)=\frac{\pi^2}{24}C_{\ssc T} (\theta-\pi)^2+ \dots 
\end{equation}
for almost smooth corners. 
In the EMI case, an explicit calculation produces \cite{Casini:2008wt,Swingle:2010jz}
\begin{equation}\label{}
a_{\rm \ssc EMI}(\theta)= 2 \kappa_{(3)} [1 +(\pi-\theta) \cot \theta]\, ,
\end{equation}
from which eq.~\eqref{CTEMI} easily follows. Combining this with eq.~\eqref{fmeze0} we have
\begin{equation}\label{FEMI} 
F^{\rm \ssc EMI}=2\pi^2 \kappa_{(3)}+  \epsilon^2 \, \frac{2 \pi \kappa_{(3)}}{3} \sum_{\ell} \ell (\ell^2-1)
\left[ (a^{(c)}_{\ell})^2+(a^{(s)}_{\ell})^2 \right] +\pazocal{O}(\epsilon^4)\, .
 \end{equation}
In order to test the validity of this formula, we start by inserting eq.~\eqref{fiy} in eq.~\eqref{femi}. Even though we have not succeeded in performing the integrals analytically in full generaly, we manage to do so in a case-by-case basis for individual values of $\ell$. We start by expanding around $\epsilon=0$ inside the integrand and then take the limit $\delta \rightarrow 0$ of the $\pazocal{O}(\epsilon)$, $\pazocal{O}(\epsilon^2)$ and $\pazocal{O}(\epsilon^3)$ terms, which turn out to be finite. The resulting expressions can be integrated analytically, and we find that the  $\pazocal{O}(\epsilon)$ and  $\pazocal{O}(\epsilon^3)$ pieces vanish, as expected on general grounds, and an exact match with eq.~\eqref{FEMI} for the quadratic term.
 
\begin{figure}[t] \centering
\includegraphics[scale=0.8]{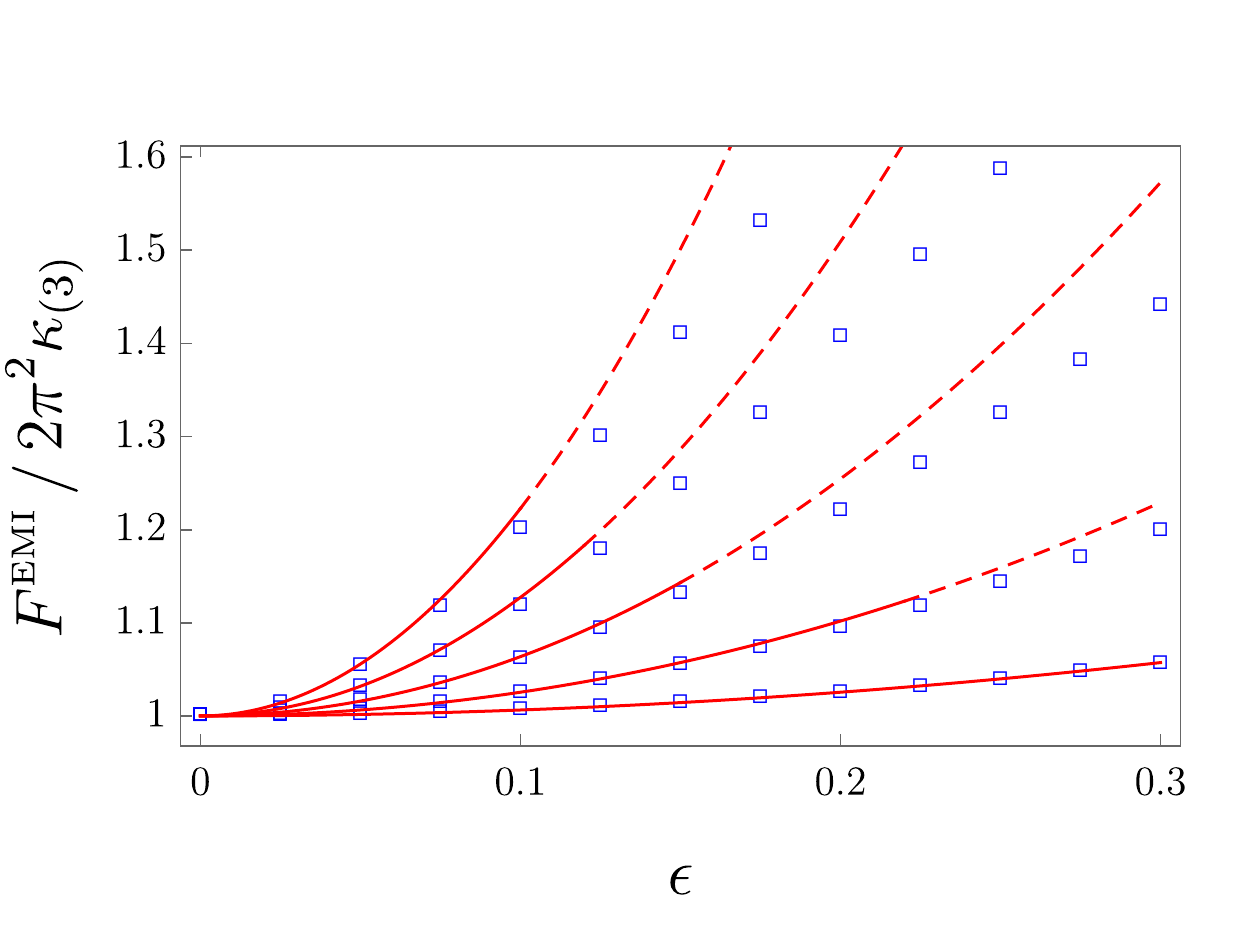}
\caption{ \textsf{We plot the universal piece $F^{\rm \ssc EMI}$ for small deformations of a disk region corresponding to eq.~\eqref{fami1} for $\ell=2,3,4,5,6$ as a function of  the small parameter $\epsilon$. The blue dots are obtained numerically integrating eq.~\eqref{femi} in each case. The red lines correspond to the leading-order parabolic approximations produced by Mezei's formula eq.~\eqref{FEMI}.}}  
	\label{refiss2}
\end{figure}
 
We can also use formula eq.~\eqref{FEMI}  to perform some checks on the numerical methods that we use below for arbitrary (non-perturbative) shapes. In order to extract $F^{\rm \ssc EMI}$ from eq.~\eqref{femi} or eq.~\eqref{femi0} for a given $f(\theta)$, we numerically integrate the corresponding expressions for various small values of $\delta$ and perform a linear fit of the resulting data points. The value of  $F^{\rm \ssc EMI}$ is then obtained as the $y$-intercept of the corresponding line. Doing this for a family of the form 
\begin{equation}\label{famif}
f_{\ell}(\theta)\equiv R \left[ 1 + \frac{\epsilon}{\sqrt{\pi}} \cos (\ell \theta)\right]\, ,
\end{equation}
 for several values of $\ell$ we obtain results (essentially identical for both eq.~\eqref{femi} and eq.~\eqref{femi0}) which are always compatible with Mezei's formula for small enough values of $\epsilon$, as shown in fig.~\ref{refiss2}. There, we observe that as $\ell$ grows, the quadratic approximation becomes worse. This is not surprising: while the coefficient of the quadratic term grows like $\ell^3$, it is natural to expect higher powers of $\ell$  to appear in the quartic and higher-order pieces. Fitting the data points to expansions involving even powers of $\epsilon$, we can extract the quadratic coefficient and compare with eq.~\eqref{FEMI}. The results are
 \begin{align}
&\bar F^{\rm \ssc EMI}|^{\req{femi}}_{f_{2}(\theta)} \simeq 1 +0.637\epsilon^2+\pazocal{O}(\epsilon^4) \, ,  & & \bar F^{\rm \ssc EMI}|^{\rm \ssc Mezei}_{f_{2}(\theta)}=  1 + \frac{2}{\pi} \epsilon^2 \simeq 1 + 0.637 \epsilon^2 \, ,  \\
&\bar F^{\rm \ssc EMI}|^{\req{femi}}_{f_{3}(\theta)} \simeq 1 +2.55 \epsilon^2 +\pazocal{O}(\epsilon^4)\, , & &  \bar F^{\rm \ssc EMI}|^{\rm \ssc Mezei}_{f_{3}(\theta)}=  1 + \frac{8}{\pi} \epsilon^2 \simeq 1 + 2.55\epsilon^2 \, , \\
&\bar F^{\rm \ssc EMI}|^{\req{femi}}_{f_{4}(\theta)} \simeq 1 +6.35  \epsilon^2 +\pazocal{O}(\epsilon^4) \, ,& &  F^{\rm \ssc EMI}|^{\rm \ssc Mezei}_{f_{4}(\theta)}=  1 + \frac{20}{\pi} \epsilon^2 \simeq 1 + 6.37\epsilon^2 \, , \\
&\bar F^{\rm \ssc EMI}|^{\req{femi}}_{f_{5}(\theta)} \simeq 1 + 12.7 \epsilon^2 +\pazocal{O}(\epsilon^4) \, , & & \bar F^{\rm \ssc EMI}|^{\rm \ssc Mezei}_{f_{5}(\theta)}=  1 + \frac{40}{\pi} \epsilon^2 \simeq 1 + 12.7\epsilon^2 \, , \\
&\bar F^{\rm \ssc EMI}|^{\req{femi}}_{f_{6}(\theta)} \simeq 1 + 22.2 \epsilon^2 +\pazocal{O}(\epsilon^4) \, , & & \bar  F^{\rm \ssc EMI}|^{\rm \ssc Mezei}_{f_{6}(\theta)}=  1 + \frac{70}{\pi} \epsilon^2 \simeq 1 + 22.3\epsilon^2 \, , 
 \end{align}
As we can see, the numerics do an excellent job in reproducing the exact coefficients in all cases.  A similar analysis in the case of the other family of basis functions, $\sin (\ell \theta)/\sqrt{\pi}$, displays an analogous match with the exact formula for the quadratic piece. We are therefore confident that we can trust our numerical approach and proceed to study regions with non-perturbative shapes.

\subsection{More general shapes}

Let us move on and consider now more general shapes, which do not correspond to small perturbations of a disk. As anticipated earlier, once we specify the entangling surface via $f(\theta)$, we perform numerical integrations of eqs.~\eqref{femi} and \eqref{femi0} and extract $F^{\rm \ssc EMI}$ from the $\delta \rightarrow 0$ limit. 
It is not obvious at first sight which regions will have a greater $F$. In particular, as mentioned earlier, shapes related by conformal transformations share the same $F$. We do know, nevertheless, that shapes including thin portions will tend to increase $F$ ---see  eq.~\req{strip}--- and that shapes which are similar enough to disks will have small $F$'s. Note though that disks deformed by relatively small bumps can modify $F$ considerably, as shown in fig.~\ref{refiss376}.
\begin{figure}[t] \centering
	\includegraphics[width=1\textwidth]{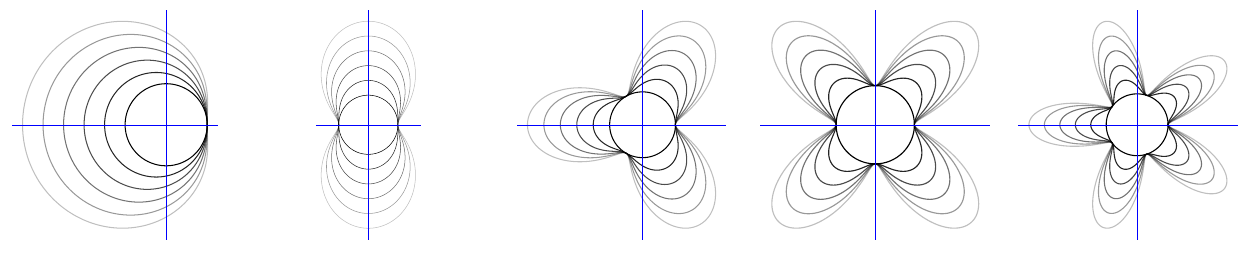}\\ 
	\includegraphics[width=1\textwidth]{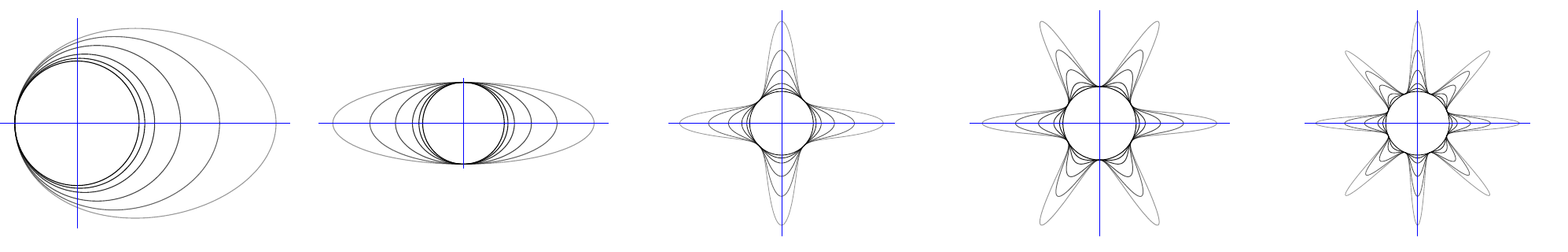}
	\includegraphics[scale=0.64]{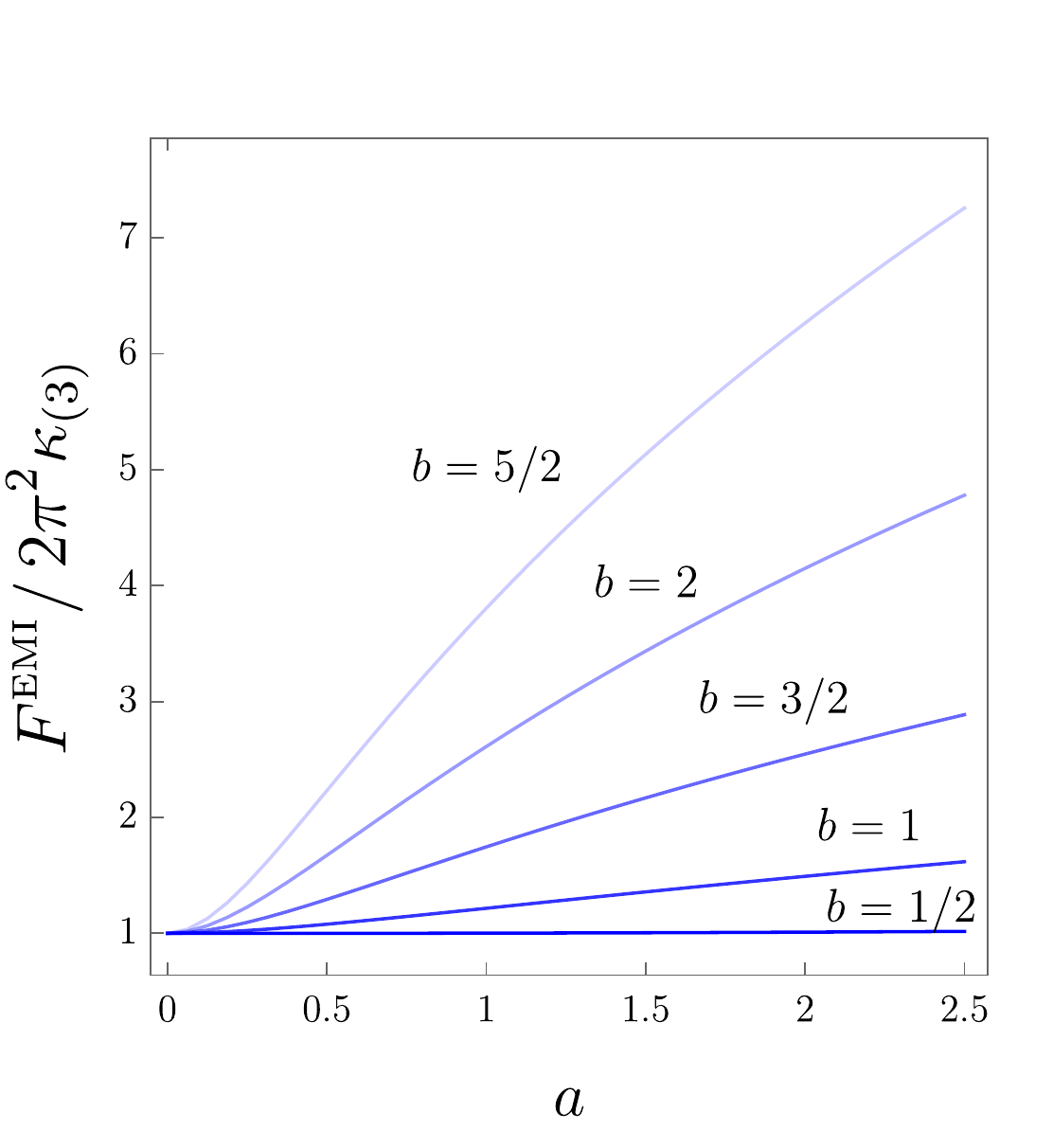}\hspace{-0.2cm}
	\includegraphics[scale=0.64]{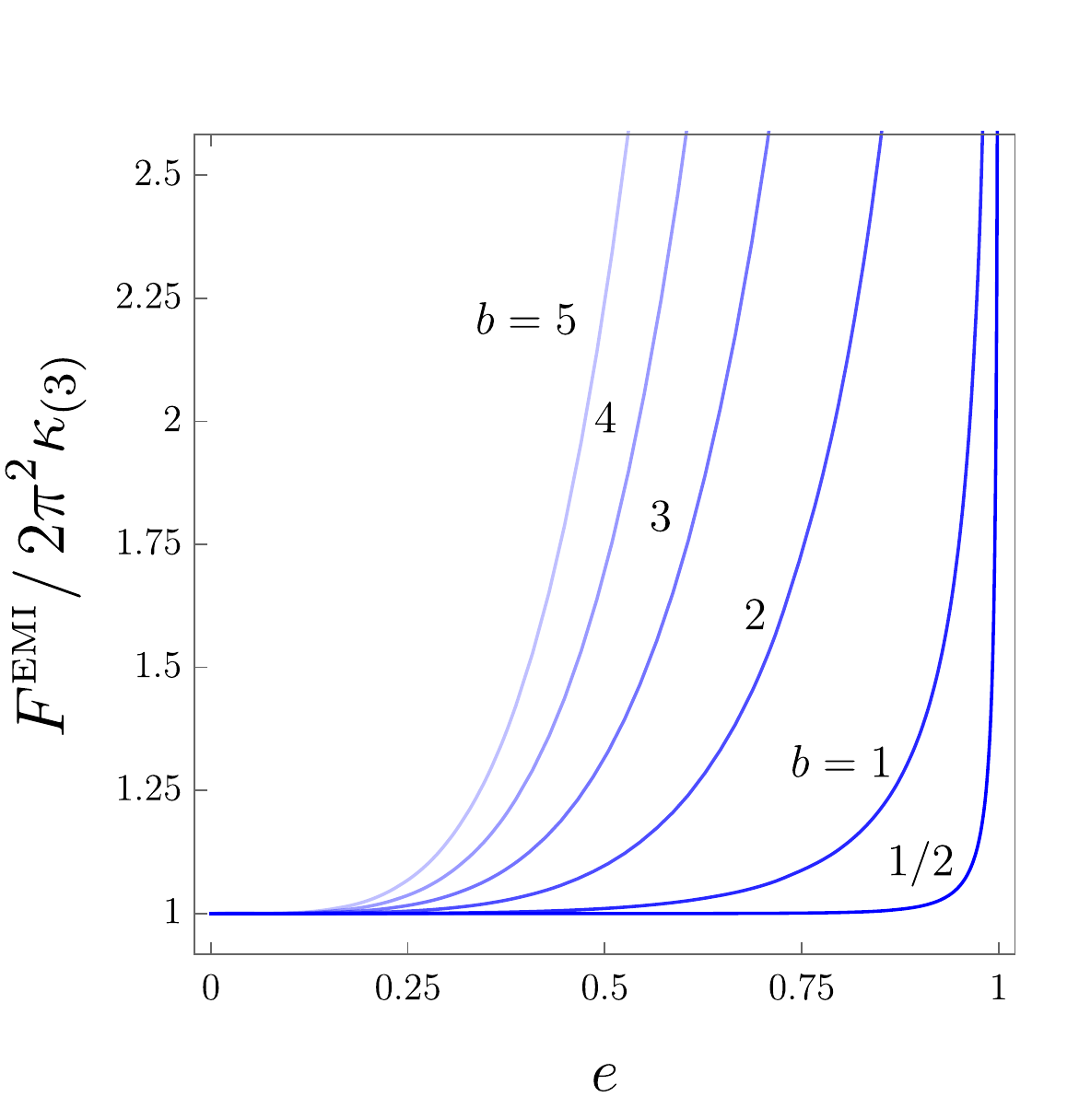}
	\caption{ \textsf{(First row) We plot entangling regions parametrized by the function $f(\theta)=1 + a \sin^2 (b \theta)$ for different values of $a,b$. From left to right we plot five figures corresponding to regions with $b=1/2,1,3/2,2,5/2$ respectively. In each figure, the value $a=0$ corresponds to the disk, and moving outwards we have regions corresponding to $a=1/2,1,3/2,2,5/2$. (Second row) We plot entangling regions corresponding to $f(\theta)=1/\sqrt{1-e^2 \cos^2 (b \theta})$ for $b=1/2,1/2,3,4,5$ (left to right plots). In each plot, the different figures correspond to different values of $ e \in (0,1)$.  (Third row) We plot the entanglement entropy universal coefficient $F^{\rm \ssc EMI}$, normalized by the disk value, as a function of $a$ and $e$, respectively, for various values of $b$ for each family.}}
	\label{refiss}
\end{figure}

As our first family, we consider regions defined by the equation
\begin{equation}\label{fami1}
f(\theta)=1+a \sin^2(b \theta)\, ,
\end{equation}
for various values of $a,b$. We plot the ones corresponding to $a=1/2,1,\dots,5/2$, $b=1/2,1\dots,5/2$ in the upper row of fig.~\ref{refiss}. In the same figure, we show the results obtained for $F^{\ssc \rm EMI}$ for half-integer values of $b$ as a function of $a$. We observe that in all cases, the curves lie above the disk result. The values tend to increase as the figures include more geometric features ---\eg as the number of ``petals'' increases. 

The second family we consider corresponds to functions
\begin{equation}\label{fami2}
f(\theta)=\frac{1}{\sqrt{1-e^2 \cos^2 (b \theta})}\, ,
\end{equation}
which includes ellipses of eccentricity $e$ as particular cases for $b=1$. We plot the results for $F^{\rm \ssc EMI}$  for various values of $b$ as a function of $e$ in the right plot of the lower row of fig.~\ref{refiss}. Once again, we find that all shapes produce results which lie above the disk result and which tend to grow monotonically as the parameters make them become increasingly different from it.

Using the ellipse results and the formulas obtained in sec.~\eqref{elipsq}, we can perform another check of the validity of the numerics beyond the perturbative level. In particular, we know that for sufficiently squashed ellipses, the result should approach eq.~\eqref{elie1} for general CFTs. In the case of the EMI model, the strip coefficient $k^{(3)}$ can be computed analytically, and the result reads 
\begin{equation}\label{kemi}
k^{(3)}_{\rm \ssc EMI}= 2\pi \kappa_{(3)}\, .
\end{equation}
For $e=0.99$ one finds $F^{\rm \ssc EMI}|^{ \req{elie1}}_{e=0.99}\simeq 71.9 \kappa_{(3)} $ whereas our numerics give $F^{\rm \ssc EMI}|_{e=0.99}^{\req{femi}} = 71.3 \kappa_{(3)} $, which is already very close ($\sim 0.8\%$ off). The match improves as $e$ grows. For instance, we have $F^{\rm \ssc EMI}|_{e=0.999}^{\req{elie1}} \simeq 221.6 \kappa_{(3)} $ and $F^{\rm \ssc EMI}|_{e=0.999}^{\req{femi}} = 221.2 \kappa_{(3)} $; $F^{\rm \ssc EMI}|^{\req{elie1}}_{e=0.9999} \simeq 698.2 \kappa_{(3)}$ and  $F^{\rm \ssc EMI}|_{e=0.9999}^{\req{femi}} = 697.9$, which differ by  $\sim 0.2\%$ and  $\sim 0.05\%$ respectively.

\begin{figure}[t!] \centering
	\includegraphics[scale=0.6]{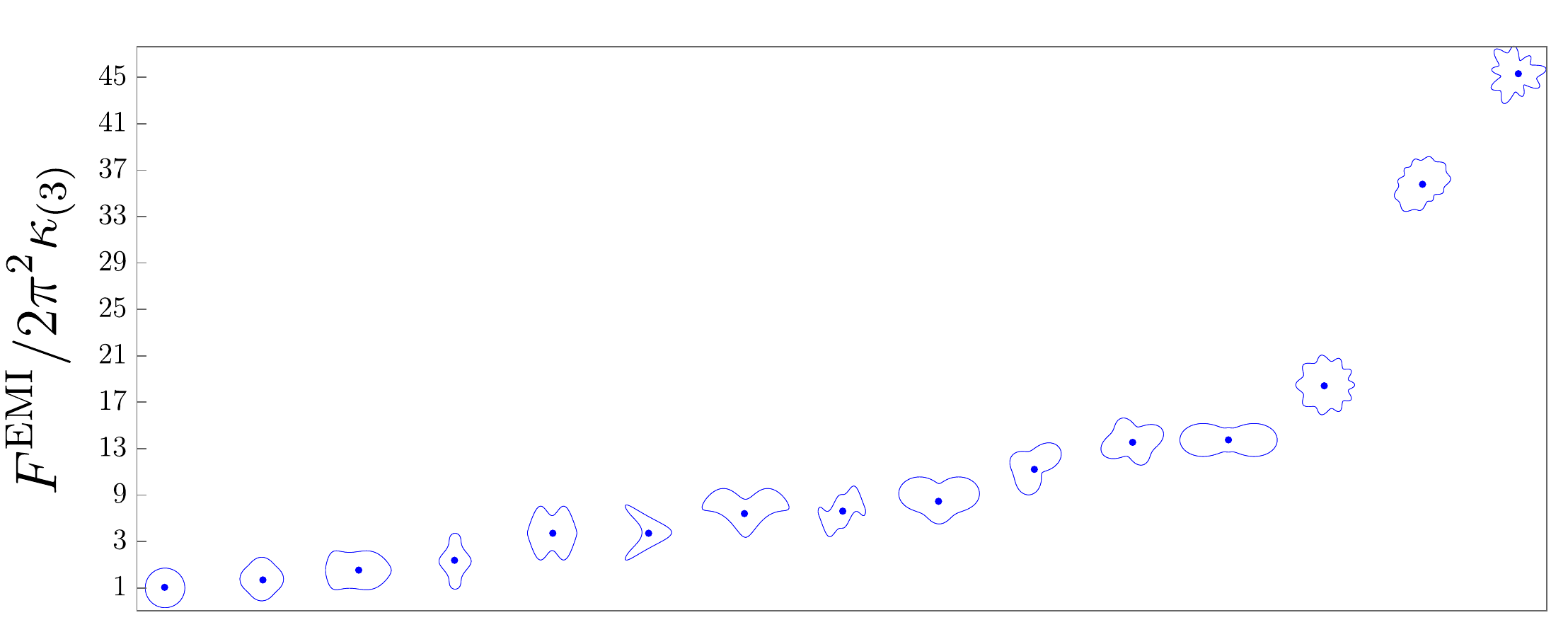}
	\caption{ \textsf{We plot the entanglement entropy universal term normalized by the disk result for the EMI model for various entangling regions. Each data point corresponds to the entangling region displayed on top of it. The shapes are arranged in a way such that $F^{\rm \ssc EMI}$ grows as we move to the right. } }
	\label{refissss2s}
\end{figure}

Naturally, there is a priori no limit in the complexity of the shapes we can probe. We have considered a variety of less symmetric regions and gathered some of the results for $F^{\rm \ssc EMI}$ in fig.~\ref{refissss2s}.
 As we can see, the values can vary considerably from shape to shape in a way which is far from obvious by looking at the geometry of the figures. 
 
 As an approximate guiding rule, it turns out that we can establish an analogy with the quantity 
\begin{equation}
\EuScript{R}(V) \equiv  \frac{ \text{perimeter}(\Sigma)^2}{4\pi \cdot \text{area}(V)}\, .\label{isopk}
\end{equation}
Similarly to the property we wish to test for $F$, this geometric quantity satisfies  the isoperimetric inequality,
\begin{equation}\label{isopp}
\EuScript{R}\geq 1\, , \quad  \text{with} \quad \EuScript{R}=1 \Leftrightarrow  V = \mathbb{B}^2\, .
\end{equation}
Besides, for small deformations of a disk and for very thin strips it behaves, respectively, as
\begin{equation}\label{Rpert}
\EuScript{R}=1+\frac{\epsilon^2}{2\pi}\sum_{\ell\geq 2} \left(\ell^2-1\right)\left[ (a^{(c)}_{\ell})^2+(a^{(s)}_{\ell})^2 \right]+\pazocal{O}(\epsilon^4)\, , \quad \EuScript{R}\simeq  \frac{L}{4\pi r}+ \dots ,
\end{equation}
which are rather similar to the general expressions for $F$ in those regimes. Hence, it is natural to wonder about possible relations between the two quantities. Obviously, while $\EuScript{R}$ is a fixed number for a given $V$, $F$ depends on the theory under consideration, so one can only expect the analogy to be approximate at best. Besides, as opposed to $F$, $\EuScript{R}$ is not conformally invariant. In the case of the EMI model, we find that in most cases we have tested, given two entangling regions $V_1$, $V_2$ such that $\EuScript{R}_{V_1} > \EuScript{R}_{V_2}$, it happens that $F^{\rm \ssc EMI}_{V_1}> F^{\rm \ssc EMI}_{V_2}$ and viceversa. However, this is not a general rule, and we have found counterexamples. For instance, if $V_1$ is an ellipse of eccentricity $e=0.97$ and $V_2$ is a region defined by $f(\theta)=1+15/26\cdot \sin(2\theta)^2$, we have $\EuScript{R}(V)_{V_1}/\EuScript{R}(V)_{V_2}\simeq 1.12514$ whereas $F^{\rm \ssc EMI}_{V_1}/ F^{\rm \ssc EMI}_{V_2} \simeq 0.871263$.

In sum, we observe that the explicit evaluation of $F$ in a concrete model produces results which are always larger than the disk one and which can vary very considerably as we change the entangling region, even without the need of considering shapes with very thin sectors.

\section{General proof using strong subadditivity}\label{proof}
In this section we provide a general proof for conjecture \req{Fiso}. Namely, we establish that $F$ is globally minimized by disk regions for general theories. Before getting there we require some preliminary results regarding general inequalities satisfied by $F$ which follow from the  strong subadditivity of entanglement entropy and the behavior of $F$ under different kinds of geometric deformations for a given entangling region. These are included in the first two subsections. The proof is presented in sec.~\ref{proooof}.  

\subsection{Strong superadditivity of $F$}
 
 The entanglement entropy of two entangling regions  $\gamma_A$ and $\gamma_B$  satisfies the strong subadditivity property\footnote{In this section, we will indistinctly denote entangling regions and their boundaries by $\gamma$. This will avoid some unnecessary clutter in the expressions.} \eqref{eq:SSEEi}
In the case of three-dimensional QFTs, this implies for the universal term
\begin{equation}
F(\gamma_A)+F(\gamma_B)\le F(\gamma_{A}\cap \gamma_{ B})+F(\gamma_{A}\cup \gamma_{B})\,.\label{ine}
\end{equation} 
This follows from eq.~\eqref{entro} because the perimeters cancel in the combination. We call this property of $F$ ``strong supperadditivity'' (SSA), having the opposite sign to strong subadditivity because of the conventional sign of $F$ in eq.~\eqref{entro}. The inequality \eqref{ine} holds even if the intersection and union do not have smooth boundaries. If these boundaries have discontinuous first derivatives, the difference between the two sides of the inequality is in fact infinite. As mentioned earlier, here we will need only be concerned with smooth intersections and unions.    

For pure states, the entropy for a region $\gamma$ is equal to the one of its complement $\bar \gamma$, and this implies 
\begin{equation}
F(\gamma)=F(\bar \gamma)\,. \label{bar}
\end{equation} 
Eqs. (\ref{ine}) and (\ref{bar}) together give rise to a new inequality which reads
\begin{equation}
 F(\gamma_A)+F(\gamma_B)\le F(\gamma_{A} -\gamma_B)+F(\gamma_{B}- \gamma_A)\,,\label{ine2}
\end{equation} 
where the difference $\gamma_A-\gamma_B$ stands for the relative complement $\gamma_A\cap \overline{\gamma_B}$. 

We will consider regions lying in the plane $t=0$. In this case the inequality \eqref{ine} cannot be nontrivially saturated in QFT. This is a consequence of the Reeh-Schlieder property.  The saturation of eq.~\eqref{ine} is called the Markov property and only occurs in special situations such as regions with boundary in the null plane, or in CFTs for regions with boundaries in the null cone \cite{Casini:2017roe,Casini:2017vbe}.

 \subsection{Geometric perturbations of $F$ and $4$-expandable functionals}

In order to show that the property of SSA implies that $F$ is globally minimized for circles in a CFT,  a preliminary step concerns the behavior of the functional $F$ under small perturbations.
 We consider only smooth boundaries for entangling regions and call $\mathfrak{S}$ the space of curves in the plane which are boundaries of compact regions with smooth and finite boundary. Let $s$ be a length parameter along $\gamma$, and call $\eta(s)$ to the outward pointing normal unit vector.  We call a perturbation of $\gamma \in\mathfrak{S}$ to a set of curves $\gamma_\epsilon \in \mathfrak{S}$, $\epsilon\in[0,\epsilon_0]$, given by 
\begin{equation}
\gamma_\epsilon(s)=\gamma(s)+\delta_\epsilon(s)\,\eta(s)\,, 
\end{equation} 
where the smooth function $\delta_\epsilon(s)$ satisfies $\lVert \delta_\epsilon \rVert\le \epsilon$, and where we write the uniform norm $\lVert h \rVert =\max_{s} |h(s)|$.  If the perturbation is given by 
\begin{equation}
\delta_\epsilon(s)=\epsilon\, h(s)\,,
\end{equation} 
for a fixed function $h(s)$, such that the function with all its derivatives go to zero with $\epsilon$ with the same velocity, we should have a power expansion for $F$ of the form
\begin{equation}
F(\gamma_\epsilon)=  F(\gamma)+ \int \diff s\, A_1^\gamma(s)\,\delta_\epsilon(s)+\int \diff s_1\, \diff s_2\, A_2^\gamma(s_1,s_2) \,\delta_\epsilon(s_1)\,\delta_\epsilon(s_2)+{\cal O}(\epsilon^3)\,.   \label{expa0}  
\end{equation} 
The question we want to address is what happens for more general perturbations where the derivatives of $\delta_\epsilon(s)$ do not go to zero with the same velocity, or even  do not go to zero at all.   

 Physically, only ultraviolet entanglement can be sensitive to the derivatives of $\delta_\epsilon$ as the perturbation size goes to zero. Then, it is clear that the non-local part of the functional $F$, which relates the perturbation at a point with the shape of the curve or the perturbation far away, must still satisfy eq.~\eqref{expa0}) with the first two terms going to zero as $\epsilon$ and $\epsilon^2$ respectively. To understand local contributions due to short length entanglement we can forget about the global shape of $\gamma$. These local terms are included for example in the distributional nature of the kernel $A_2^\gamma(s_1,s_2)$ in the vicinity of coincidence points. Locally, scale invariance implies (for dimensional reasons) the form
\begin{equation}
\int \diff s_1\, \diff s_2\, \frac{\delta_\epsilon(s_1)\,\delta_\epsilon(s_2)}{|s_1-s_2|^4} \sim \int \diff s_1\, \diff s_2\, \log(|s_1-s_2|)\,\delta_\epsilon''(s_1)\,\delta_\epsilon''(s_2)\, ,
\end{equation}   
for the most singular possible term for the contribution quadratic in the perturbation. 

Powers of the local curvature have positive dimension and can only soften the contribution.    The definition of the regularization of the distribution on the left hand side of the equation is given by the right hand side.

In momentum space the above term writes for large momentum   
\begin{equation}
\int \diff p \, |\tilde{\delta}_\epsilon(p)|^2\, p^3\,,
\end{equation}  
which is precisely the form of the leading angular momentum term for deformations of disk regions   ---see eq.~\eqref{fmeze0}. The coefficient of this singular term must be independent of the precise form of $\gamma$ and is proportional to $C_{\ssc T}$ \cite{Mezei:2014zla}.  For comparison, the perimeter functional, which is not scale invariant, has the milder leading local term
\begin{equation}
\int \diff p \,  |\tilde{\delta}_\epsilon(p)|^2\, p^2\,.
\end{equation} 
This has dimensions of length since $\tilde{\delta}_\epsilon$ has length-square dimensions.

Analogously, we can analyze the behavior of possible higher non-linear terms in the perturbation. In the scale-invariant case, in momentum space we have a leading behavior for large momentum given by  
\begin{equation}
\int  \diff p\, |\tilde{\delta}_\epsilon(p)|^{n} p^{2 n-1}\,.\label{lab}
\end{equation}    

For the sake of the proof below, we are interested in understanding perturbations of the form $\delta_\epsilon(s)=\lambda^{a}\, h(s/\lambda)$, where we have written $\epsilon=\lambda^a$, $a>0$. For these, the derivatives go as $\delta_\epsilon^{(k)}\sim \lambda^{a-k}$. In momentum space, $\tilde{\delta}_\epsilon(p)\sim \lambda^{a+1}$, $p\sim  \lambda^{-1}$. The leading local term of $n^{\rm th}$ order \eqref{lab} goes as $\lambda^{n (a-1)}$. On the other hand, the non-local pieces scale as $\lambda^{(a+1) n}$.     
For these perturbations, in order to separate the first term ($n=1$) of eq.~\eqref{expa0}), from the second ($n=2$), we need $a>3$.\footnote{For the $n=1$ and $n=2$ terms we have for the non-local and leading local pieces, respectively: $\{ \lambda^4,\lambda^2\}$ and $\{ \lambda^8,\lambda^4\}$ for $a=3$;  $\{ \lambda^5,\lambda^4\}$ and $\{ \lambda^{10},\lambda^6\}$ for $a=4$. Hence, using $a=3$ would mix the non-local $n=1$ contribution with the local $n=2$ one. }  We will use $a=4$ in the proof. For $a=4$, the first term in eq.~\eqref{expa0}) is order $\lambda^5$, while the second (the leading local piece) is order $\lambda^6$. The rest of the contributions start at order $\lambda^7$.  This motivates the following definition. 

\begin{defi}
We call a functional $f:\mathfrak{S}\rightarrow \mathbb{R}$ a-expandable if for any perturbation $\delta_\lambda(s)$ of any $\gamma\in \mathfrak{S}$, such that $\lVert \delta_\lambda^{(k)}\rVert \sim \lambda^{a-k}$ as $\lambda \rightarrow 0$, the following expansion is valid
\begin{equation}
f(\gamma_\lambda)=  f(\gamma)+ \int \diff s\, A_1^\gamma(s)\,\delta_\lambda(s)+\int \diff s_1\, \diff s_2\, A_2^\gamma(s_1,s_2) \,\delta_\lambda(s_1)\,\delta_\lambda(s_2)+\dots \, ,    \label{expa01}  
\end{equation} 
where the second-order term is higher order in $\lambda$ than the linear term, and the rest is higher order than the second term.  
\end{defi}

Naturally, $F$ is an example of a $4$-expandable functional.

 \subsection{$F$ is globally minimized by disk regions} \label{proooof}

With this understanding we are in position to prove the minimality of $F$ for disk regions. In order to highlight the geometric nature of the proof we state it for general functionals. 

\begin{theorem}If an Euclidean invariant\footnote{Namely, invariant under rotations and translations.} 4-expandable functional $F$ on ${\cal S}$ is  strong superadditive and constant for disks, then $F$ is globally minimized for disks.
\end{theorem}

\textit{Proof}:  First we deal with regions with non-trivial topology. Suppose $\gamma$ has more than one connected components $\gamma=\gamma_1\cup \cdots \cup\gamma_n$. From SSA 
\begin{equation}
F(\gamma)\ge \sum_{i=1}^n  F(\gamma_i)\,.  \label{sapa}
\end{equation} 
This shows that multicomponent regions have larger $F$ than certain single-component ones. Then suppose $\gamma$ is a single-component region with holes, $\gamma=\gamma_0-(\gamma_1\cup \dots\cup \gamma_n)$, where $\gamma_0,\gamma_1\,\cdots,\gamma_n$ are single-component simply connected regions. From eq.~\eqref{ine2} it follows
\begin{equation}
F(\gamma) \ge  F(\gamma_0) +\sum_{i=1}^n  F(\gamma_i)\,,
\end{equation} \label{514}
which shows that $F(\gamma)$ is greater or equal than some single-component region without holes. Therefore, in order to prove the theorem, we can from now on restrict ourselves to single-component simply connected regions.

 Consider then a region $\gamma$ and another region $\tilde{\gamma}$ included in it, $\tilde{\gamma}\subseteq \gamma$. Assume $\tilde{\gamma}$ is osculating to $\gamma$, that is, both curves are tangent to each other and at the point of contact ---which we can set to the length parameter $s=0$ for both curves--- their curvatures are equal, $\tilde{\gamma}''(0)=\gamma''(0)$. Since one curve is included inside the other, the third derivatives must also agree $\tilde{\gamma}'''(0)=\gamma'''(0)$, and the deviation between them will be fourth order $|\tilde{\gamma}(s)-\gamma(s)|\sim |s|^4$ near $|s|=0$. 
 
 \begin{figure}[t!] \hspace{-0.35cm}
	\includegraphics[width=1\textwidth]{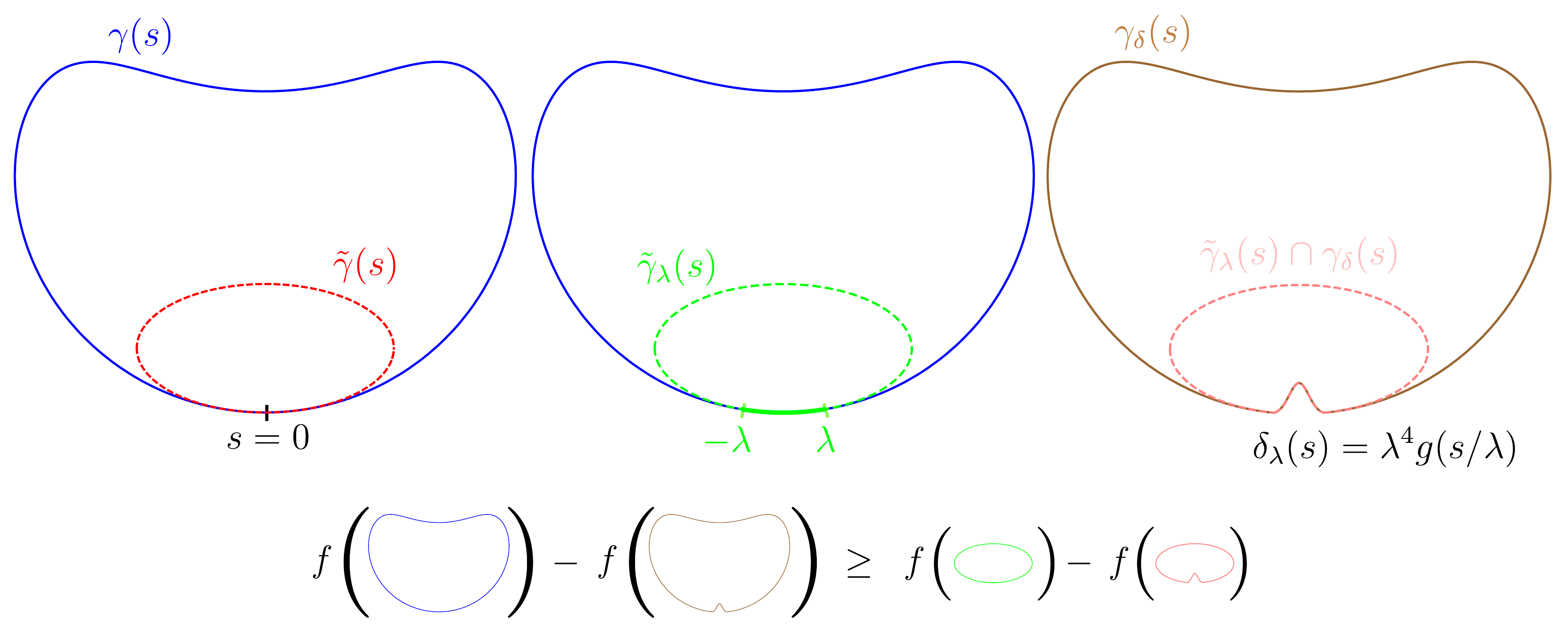}
	\caption{ \textsf{We  show an schematic representation of the geometric setup leading to the inequality \req{dsf}. In the left plot we show two entangling regions, $\tilde \gamma(s) \subseteq \gamma(s)$ such that $\tilde \gamma(s)$ is osculating to $\gamma(s)$ at $s=0$. The second plot includes $\gamma(s)$ and a deformed version of the inner region, $\tilde \gamma_{\lambda}(s)$, which exactly coincides with $\gamma(s)$ for  $s\in (-\lambda,\lambda)$. The third plot shows deformed versions of $\gamma(s)$ and $\tilde \gamma_{\lambda}(s)$ which include a small inwards bump supported within the interval for which both regions coincide. The equation shown below is just \req{fgg} in the particular case considered above.  }}
	\label{refides}
\end{figure}
 
 Around the point $s=0$ we consider a family of deformations $\tilde{\gamma}_\lambda(s)$ of $\tilde{\gamma}$, with deformation function $\tilde{\delta}_\lambda(s)$, such that the deformation has support in $s\in(-2\lambda,2 \lambda)$, and the curve $\tilde{\gamma}_\lambda(s)$ coincides with $\gamma(s)$ for $s\in (-\lambda,\lambda)$. Then, the deformation can be chosen such that $\lVert\tilde{\delta}_\lambda^{(k)}\rVert\sim  \lambda^{4-k}$. With support inside the interval $s\in (-\lambda,\lambda)$ where $\tilde{\gamma}_\lambda$ and $\gamma$ coincide, we can place a deformation of $\gamma$ given by $
\delta_\lambda(s)=\lambda^4\, g(s / \lambda)$, 
where $g$ is a smooth function. We also impose $g< 0$ such the deformation is inwards ---see fig.~\ref{refides} for a representation of the geometric setup just described. We also have $\lVert\delta_\lambda^{(k)}\rVert \sim \lambda^{4-k}$. 
 From SSA we find then\footnote{\rd{In particular, \req{fgg} follows from choosing $\gamma_A\equiv \gamma_{\delta}$ and $\gamma_B=\tilde{\gamma}_{\lambda}$ in  \req{ine}.}}
\begin{equation} \label{fgg}
F(\gamma)-F(\gamma_\delta)\ge F(\tilde{\gamma}_\lambda)-F(\tilde{\gamma}_\lambda\cap \gamma_\delta)\,.
\end{equation}  
Expanding for small $\lambda$, we have for the first-order deviation
\begin{equation} 
A_1^{\gamma}(0)-A_1^{\tilde{\gamma}}(0)\ge 0\,. \label{dsf}
\end{equation} 
That is, to first order, $f$ for the larger region increases more than for the smaller one going outwards at the point of osculation.
 
  For a disk region and an Euclidean invariant $F$, the coefficient $A_1^{\rm disk}(s)$ has to be independent of $s$. Then, taking a first order variation from a disk to a scaled disk, and considering that $F$ is constant on disks, it follows that 
  \begin{equation} 
  A_1^{\rm disk}(s)=0\,.\label{circlea}
\end{equation}   
   This will be useful when combined with eq.~\eqref{dsf}. In order to use this result, we have first to discuss some geometric preliminaries. We will show that for an arbitrary curve $\gamma$ we can always place an osculating circle inside (or outside) it.    
 
\begin{figure}[t!] \centering
	\includegraphics[scale=0.5]{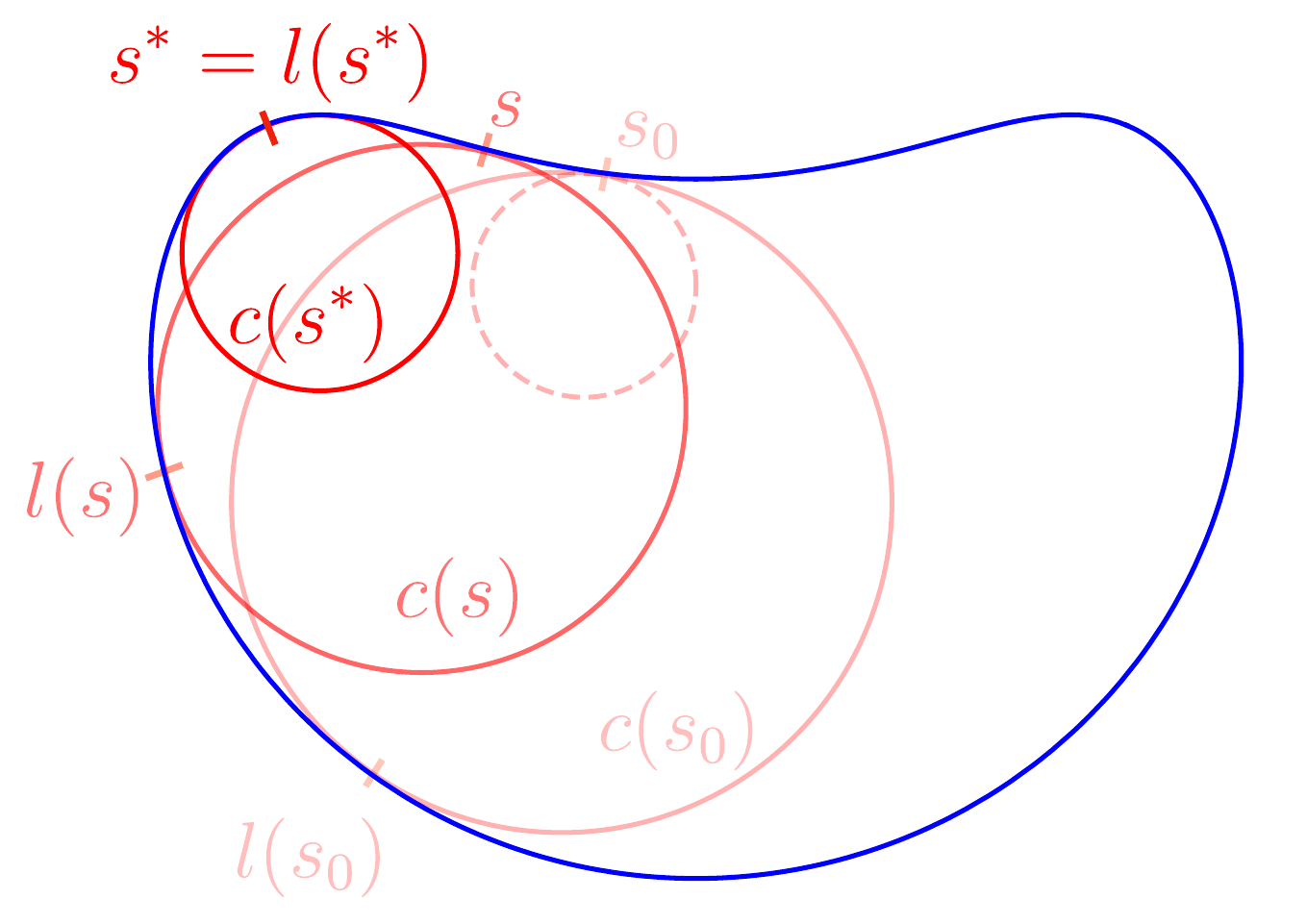}
	\caption{ \textsf{We show the procedure outlined in the text for finding an osculating circle $c(s^*)$ inside a given shape. }}
	\label{refidesss8}
\end{figure}

 To show this, call the length parameter $s\in [0,L]$. Take a point $s_0$ and a small circle internal to $\gamma$ and tangent to $\gamma$ at $s_0$ ---see fig.~\ref{refidesss8}. By increasing the radius of this circle and keeping it tangent to $\gamma$ at $s_0$ we arrive to a unique circle $c(s_0)$ which is still included in $\gamma$ and is tangent to $\gamma$ at $s_0$ and at (at least) another point $l(s_0)$. We can choose the length parameter such that $s_0>0$, $l(s_0)<L$, $s_0<l(s_0)$. We define the function $l(s)$ for another $l(s_0)>s>s_0$ in a similar way, by taking the largest circle $c(s)$ tangent at $s$ and internal to $\gamma$, and where $l(s)$ is the smallest length parameter (greater than $s$) among the points at which the circle $c(s)$ is tangent to $\gamma$. It is clear that moving $s$ from $s_0$ to larger values,  $l(s)$ can only decrease because $\gamma([s,l(s)])$ and a segment of the circle $c(s)$ from the point $s$ to $l(s)$ defines a closed curve that divides the plane in two regions. Any $c(s')$ for $s<s'<l(s)$ is included in this region and will have $l(s')< l(s)$. Then, there is a smallest $s^*>s_0$ such that $s^*=l(s^*)$. This indicates that $c(s^*)$ is an internal osculating circle to $\gamma$. At this point the osculating circle and $\gamma$ have a point of contact of degree four, the curvatures of $\gamma$ and $c$ agree, and the derivative of the curvature of $\gamma$ vanishes. In fact $s^*$ is a local maximum of the curvature. If it where a minimum, $\gamma$ would leave part of the circle outside.

{ We have shown that for any $\gamma$ we can place an osculating circle inside it. It is not difficult to show, using the same ideas, that we can also find an external osculating circle to $\gamma$. A shorter proof follows by using conformal transformations. We can make a conformal inversion $I$ centered at a point inside $\gamma$. As a result $\gamma$ is mapped to a closed curve $\tilde{\gamma}$, and the exterior  of $\gamma$ is mapped to the interior of $\tilde{\gamma}$.  Again there will be a  circle $\tilde{c}$ osculating to $\tilde{\gamma}$ internally. Considering that the osculating condition is conformally invariant, and that circumferences are mapped to circumferences, by inverting back to the original plane we obtain that there is a circumference $c=I(\tilde{c})$ osculating to $\gamma$, and this circumference is completely included in the exterior of $\gamma$. The corresponding circle either lies completely outside $\gamma$, in which case the point of contact is a local minimum of curvature for $\gamma$, with negative curvature, or the circle includes $\gamma$ completely, in which case the contact point is a positive curvature local minimum.}\footnote{{The two cases depend on whether the center of inversion lies inside or outside the osculating circle $\tilde{c}$.}}

{ Hence, we can find a point of local curvature maximum and another of local curvature minimum where we can place internal and external osculating circles. By direct application of the above calculation to the curve $\gamma$ and these osculating circles we get the inequality \eqref{dsf} for the $A_1$ coefficients of both curves. From the eq.~\eqref{circlea} for circles,  we find that for any $\gamma$ there is at least one point of local curvature maximum where $A_1\ge 0$ and one point of local curvature minimum where $A_1\le 0$. If $|A_1|>0$, the sign of $A_1$ is kept constant in a finite interval around the point, whereas for $A_1=0$ it can change sign there.  We can use a perturbation of order $\epsilon$, with all derivatives of the same order, to increase the size of the local curvature minimum or decrease the size of the local curvature maximum, by pushing the curve to the outside or the inside respectively.\footnote{{ This can be done, for example, by replacing a small interval of the curve around the point of osculation by a segment of a circle tangent at two points at each side of it, and smearing up the points of contact to make it smooth. The same can be done if the minimum or maximum of curvature correspond to a segment of a circle rather than a point.}}   
If $|A_1|>0$ we can then use the expansion \eqref{expa01} for this perturbation and check that $F$ always decreases. In the case $A_1=0$ the perturbation can be chosen such that the change in $F$ vanishes to the order of the change in the curvature. 
 For any $\gamma$, we have some curvature maxima $\{r_1^+,\cdots, r_k^+\}$ and minima $\{r_1^-,\cdots,r_k^-\}$, and define $K(\gamma)=\sum_{i=1}^k r_i^+ - \sum_{i=1}^k r_i^-\ge 0$. 
 Then we have shown that for any $\gamma$ there is a $\gamma'$ with $K(\gamma')< K(\gamma)$ and $F(\gamma')\le F(\gamma)$. Hence, the function
 \begin{equation} 
 f(q)=\textrm{min}_{\gamma, K(\gamma)=q}\, F(\gamma) \,,      
 \end{equation} 
has to be non decreasing with $q$. In particular, the absolute minimum of $F$ has to be achieved for $q=0$, corresponding to a circle. }  
 $\square$     

\bigskip

Therefore, the $F$ term in the entanglement entropy of three-dimensional CFTs is globally minimized for disk regions.
Except for the particular continuity property of the functional that we used, which is natural for conformal invariant functionals, we did not need to invoke conformal invariance but just Euclidean invariance, plus the requirement that the functional is constant for disks. 
The proof should go over to the Lorentzian case as well. In that case, however, all deformations of the circle on its light cone have the same $F$.  

{ Another comment is that having an $a$-expandable functional for $a>4$ and not expandability for $a=4$ (\ie more singular functionals) is not enough for the proof, since we need to match the boundaries of a deformed osculating circle with the curve, which requires perturbations with $a=4$. In principle, lower $a$-expandability (softer functionals) ---such as the isoperimetric ratio $\EuScript{R}(V)$ defined in eq. (\ref{isopk}), having $3$-expandability--- will be enough. Note however that $\EuScript{R}(V)$ is not strong superadditive and we cannot use the same proof as for the isoperimetric inequality. If we have $2$-expandability instead, we could modify the proof above taking just tangent circles instead of osculating ones. However, we always have both interior and exterior tangent circles to a given point, and that would imply a vanishing first order coefficient $A_1$ at any point. Then, the only possibility for a SSA functional to be 2-expandable is that it is constant under deformations.}

\section{Conclusions}

We have proven that the disk-like entangling region maximizes entanglement entropy in general CFTs among all possible shapes and topologies.

It is then natural to wonder what happens in three-dimensions with the analogous question for entangling regions with a fixed number of holes. In this case, however, the answer turns out to be more trivial. For instance, for annuli regions defined by pairs of radii from the set $\{R_1,R_2,R_3\}$ with $R_1<R_2<R_3$, one has $F(R_3,R_1)< F(R_2,R_1)$ \cite{Nakaguchi:2014pha}. This implies that $F$ will always decrease as we make the inner radius smaller with respect to the outer one, being $F_0$ the limiting result as $R_1/R_3 \rightarrow 0$. In fact, for regions with non-trivial topology, the general bound \eqref{Fiso} can be improved. Imagine we have a region with $n$ disconnected subregions $\gamma=\cup_j^n \gamma_j $, each one of which has $m_j$ holes, $\gamma_j=\gamma_{j,0}- (\gamma_{j,1} \cup \dots \cup \gamma_{j,m_j})$. Then, using eqs.~\eqref{sapa}, \eqref{514} and \eqref{Fiso} we have 
\begin{equation}
F(\gamma) \geq \sum_j^n F(\gamma_j) \geq  \sum_j^n [F(\gamma_{j,0})+\sum_{i=1}^{m_j} F(\gamma_{j,i})] \geq n F_0 + \sum_j^n m_j F_0=n_B F_0\, .
 \end{equation}
This follows from applying $F \geq F_0$ repeatedly to each single-component simply connected piece. In the last equality we rewrote $n+\sum_j^n m_j$ as $n_B$, which is the total number of boundaries of the region.  In words, for a region with  $n_B$ boundaries, $F$ is bounded below not only by the disk result $F_0$, but by $n_B$ times $F_0$. Saturation of the inequality occurs only for purely topological theories, $F^{\rm topo}(\gamma) = n_B F_0^{\rm topo}$.


It is also natural to wonder about the problem analogous to the one considered here for $d=4+1$ and $d=5+1$ theories. In the latter case, the geometric nature of the universal logarithmic term for smooth entangling regions \cite{Safdi:2012sn,Miao:2015iba} should make it amenable to an analysis similar to the one in ref.~\cite{Astaneh:2014uba,Perlmutter:2015vma} for the $d=3+1$ case. Note however that in that case the sign of the different geometric contributions weighted by the trace-anomaly coefficients is not obvious ---at least at first sight--- so the situation is trickier.  The $d=4+1$ case is more challenging but methods similar to the ones considered here may be useful. Naturally, in all cases the expectation is that entangling regions bounded by round (hyper)spheres globally maximize the EE. This expectation is supported by partial evidence from holographic theories \cite{Astaneh:2014uba}.

%% file: text/ch8-concl.tex
\chapter{Conclusions}

Let us conclude this thesis presenting remarks and general ideas regarding the results we presented.

In the first part, we have provided substancial progress in the understanding of the higher-curvature gravity theories known as GQT gravity. Namely, we have shown that, in dimensions higher than four, at order $n$ there is one and only one QT gravity theory and $n-1$. In addition, we have provided a framework to compute the thermodynamic charges of the black holes in any GQT gravity. This is particularly convenient taking into account that any gravitational effective action involving higher-curvature corrections is equivalent, via metric redefinitions to some GQT gravity. Remarkably, we can study study black hole thermodynamics of arbitrary effective theories in a dramatically simpler scenario provided by the GQT gravity counterpart. 

The situation in three spacetime dimensions differs. In this case,
the fact that the Riemann tensor is proportional to the Ricci tensor constraints enormously the space of possible theories. In turn, this allowed to provide general results, such as the number of possible results, its linearized spectrum, identifying the theories that satisfy an holographic c-theorem and its relation to Born-Infeld gravity, etc. However, among the possible theories we saw that there are no non-trivial GQT gravities. This no-go result can be bypassed by adding matter content to the theory, finding EQT gravities. Interestingly, these theories have as solution a plethora of new analytic black holes that represent continuous generalizations of the BTZ black hole. Among them, one can find metrics displaying curvature, conical, BTZ-like singularities as well as regular black holes.

Motivated by the fact that entanglement entropy generalizes the concept of black hole entropy to situations in which the system does not show a killing horizon, we studied the holographic scenario. In this context,the Kounterterms renormalization scheme isolates the universal contributions, canceling the divergences in entanglement entropy. We showed that in the case of QC gravity, it also allows to compute the type A and type B central charges of the dual CFT theories. Moreover, in the case of Einstein-AdS gravity, we shown that the Kounterterms allow to study the shape-dependence of entanglement entropy in dual odd-dimensional CFTs and separate the contributions into a local (depending on the geometry of the entangling surface) and a non-local (with topological origin). In three dimensions, these parts can be written in terms of the Willmore energy and employing its properties it can be shown that the finite part of entanglement entropy is maximized in the case of a disk-entangling region. A natural question is whether entanglement entropy maximization by the disk-like entangling region is restricted to CFTs dual to Einstein-AdS gravity or general regardless the shape of the entangling region. Supported by previous observations in the literature, we conjectured and proved that entanglement entropy is indeed maximized by the disk region in any CFT.

%% file: text/appendices.tex
%

\chapter{Explicit covariant GQT densities for $n=4,5,6$ in $D=5,6$} \label{densiii}
In this appendix we present explicit GQT covariant densities of each of the $(n-2)$ existing types for $n=4,5,6$ in $D=5$ and $D=6$. 

At quartic order, examples of representatives of the two inequivalent classes of GQT densities in $D=5$ are (we use Roman numbers to label the different types)
\begin{align}
\pazocal{S}^{(D=5,n=4)}_{\rm  I}=&+12 R\indices{^a^b^c^d} R\indices{_a_b^e^f} R\indices{_c^g_e^h} R\indices{_d_g_f_h}+3 R\indices{^a^b^c^d} R\indices{_a^e_c^f} R\indices{_b_g_d_h}
   R\indices{_e^g_f^h}\\
   & -6 R\indices{^a^b^c^d} R\indices{_a^e_c^f} R\indices{_e^g_b^h} R\indices{_f_g_d_h}-9 R\indices{^a^b} R\indices{_c^h_e_a}
   R\indices{_d_h_f_b} R\indices{^c^d^e^f}+R R\indices{_a^c_b^d} R\indices{_e^a_f^b} R\indices{_c^e_d^f},
\nonumber\\
\pazocal{S}^{(D=5,n=4)}_{\rm  II}=&+4 R\indices{^a^b^c^d} R\indices{_a_b^e^f} R\indices{_c_e^g^h} R\indices{_d_f_g_h}+30 R\indices{^a^b^c^d} R\indices{_a_b^e^f} R\indices{_c^g_e^h}R\indices{_d_g_f_h}\\
& -11 R\indices{^a^b^c^d} R\indices{_a^e_c^f} R\indices{_e^g_b^h} R\indices{_f_g_d_h}-16 R\indices{^a^b} R\indices{_c^h_e_a}R\indices{_d_h_f_b} R\indices{^c^d^e^f}-R\indices{^a^b} R\indices{_c_d^h_a} R\indices{_e_f_h_b} R\indices{^c^d^e^f} \nonumber\\
&-3 R\indices{^a^b} R\indices{_a^c_b^d}R\indices{_e_f_h_c} R\indices{^e^f^h_d}+3 R\indices{^a^b} R\indices{^c^d} R\indices{^e_a^f_b} R\indices{_e_c_f_d}
+R\indices{^a^b} R\indices{^c^d} R\indices{^e_a^f_c}R\indices{_e_b_f_d}\nonumber,
\end{align}
which evaluated on the single-function ansatz reduce to linear combinations of $\pazocal{S}{(4,j)}|_f$, as defined in \req{rj}, with
\begin{align}
\tau^{(D=5,n=4)}_{\rm  I}&=4\tau_{(4,1)}+12 \tau_{(4,3)}-6 \tau_{(4,4)} \, , \\  \tau^{(D=5,n=4)}_{\rm  II}&=6 \tau_{ (4,2)}-\tau_{(4,4)}\, ,
\end{align}
respectively. It is straightforward to check that both satisfy conditions \eqref{cond1} and \req{cond2}. In $D=6$, we find
\begin{align}
\pazocal{S}^{(D=6,n=4)}_{\rm  I}=&+15 R\indices{^a^b^c^d} R\indices{_a_b^e^f} R\indices{_c_e^g^h} R\indices{_d_f_g_h}+20 R\indices{^a^b^c^d} R\indices{_a_b^e^f} R\indices{_c^g_e^h}
   R\indices{_d_g_f_h}-4 R\indices{^a^b^c^d} R\indices{_a^e_c^f} R\indices{_b_g_d_h} R\indices{_e^g_f^h}\nonumber\\
   &-36 R\indices{^a^b^c^d} R\indices{_a^e_c^f}
   R\indices{_e^g_b^h} R\indices{_f_g_d_h}+48 R\indices{^a^b} R\indices{_c^h_e_a} R\indices{_d_h_f_b} R\indices{^c^d^e^f}-8 R\indices{^a^b} R\indices{_c_d^h_a}
   R\indices{_e_f_h_b} R\indices{^c^d^e^f}\nonumber\\
   &-8 R R\indices{_a^c_b^d} R\indices{_e^a_f^b} R\indices{_c^e_d^f}+8 R\indices{^a^b} R\indices{^c^d} R\indices{^e_a^f_c}
   R\indices{_e_b_f_d} \, , \\
\pazocal{S}^{(D=6,n=4)}_{\rm  II}=&-5 R\indices{^a^b^c^d} R\indices{_a_b^e^f} R\indices{_c_e^g^h} R\indices{_d_f_g_h}-28 R\indices{^a^b^c^d} R\indices{_a_b^e^f} R\indices{_c^g_e^h}
   R\indices{_d_g_f_h}-20 R\indices{^a^b^c^d} R\indices{_a^e_c^f} R\indices{_b_g_d_h} R\indices{_e^g_f^h}\nonumber\\
   &+52 R\indices{^a^b^c^d} R\indices{_a^e_c^f}
   R\indices{_e^g_b^h} R\indices{_f_g_d_h}-16 R\indices{^a^b} R\indices{_c^h_e_a} R\indices{_d_h_f_b} R\indices{^c^d^e^f}+8 R\indices{^a^b} R\indices{_c_d^h_a}
   R\indices{_e_f_h_b} R\indices{^c^d^e^f}\nonumber\\
   &-8 R\indices{^a^b} R\indices{_a^c_b^d} R\indices{_e_f_h_c} R\indices{^e^f^h_d}+8 R\indices{^a^b} R\indices{^c^d}
   R\indices{^e_a^f_b} R\indices{_e_c_f_d}-8 R\indices{^a^b} R\indices{^c^d} R\indices{^e_a^f_c} R\indices{_e_b_f_d} \, ,
\end{align}
and for those
\begin{align}
\tau^{(D=6,n=4)}_{\rm  I}&=\tau_{(4, 4)} - 2 \tau_{(4, 3)} - 2\tau_{(4, 1)}.\\
  \tau^{(D=6,n=4)}_{\rm  II}&=\tau_{(4, 4)} - 4 \tau_{(4, 3)} - 6\tau_{(4, 2)}.
\end{align}

 At quintic order, examples of the three inequivalent classes read
\begin{align}
\pazocal{S}^{(D=5,n=5)}_{\rm  I}=&+
3235 R^5 - 28409 R^3 R_{a}{}^{b} R_{b}{}^{a} + 46980 R^2 R_{a}{}^{c}R_{b}{}^{a} R_{c}{}^{b} - 93522 R R_{a}{}^{d} R_{b}{}^{a} R_{c}{}^{b} R_{d}{}^{c}\nonumber\\
&+ 11928 R_{a}{}^{b} R_{b}{}^{a} R_{c}{}^{e} R_{d}{}^{c} R_{e}{}^{d} + 98700 R R_{b}{}^{a} R_{d}{}^{b} R_{e}{}^{c} 
R_{ac}{}^{de} + 2870 R^3 R_{ab}{}^{cd} R_{cd}{}^{ab}\nonumber\\
&+ 52080 
R_{a}{}^{b} R_{b}{}^{a} R_{e}{}^{c} R_{f}{}^{d} R_{cd}{}^{ef} - 
151200 R R_{c}{}^{a} R_{d}{}^{b} R_{ab}{}^{ef} R_{ef}{}^{cd}\nonumber\\
& + 137655 
R R_{b}{}^{a} R_{c}{}^{b} R_{ad}{}^{ef} R_{ef}{}^{cd}- 5845 R 
R_{a}{}^{b} R_{b}{}^{a} R_{cd}{}^{ef} R_{ef}{}^{cd} \nonumber\\
&- 23940 
R_{a}{}^{b} R_{b}{}^{a} R_{d}{}^{c} R_{ce}{}^{fg} R_{fg}{}^{de} \, , \\
\pazocal{S}^{(D=5,n=5)}_{\rm  II}=&+10505 R^5 - 98197 R^3 R_{a}{}^{b} R_{b}{}^{a} + 242460 R^2 
R_{a}{}^{c} R_{b}{}^{a} R_{c}{}^{b} - 362526 R R_{a}{}^{d} 
R_{b}{}^{a} R_{c}{}^{b} R_{d}{}^{c}\nonumber\\ 
&+ 77784 R_{a}{}^{b} R_{b}{}^{a} 
R_{c}{}^{e} R_{d}{}^{c} R_{e}{}^{d} + 77700 R R_{b}{}^{a} R_{d}{}^{b} 
R_{e}{}^{c} R_{ac}{}^{de} + 1120 R^3 R_{ab}{}^{cd} R_{cd}{}^{ab}\nonumber\\
&+ 
139440 R_{a}{}^{b} R_{b}{}^{a} R_{e}{}^{c} R_{f}{}^{d} R_{cd}{}^{ef} 
- 173880 R R_{c}{}^{a} R_{d}{}^{b} R_{ab}{}^{ef} R_{ef}{}^{cd} \nonumber\\
& + 
194985 R R_{b}{}^{a} R_{c}{}^{b} R_{ad}{}^{ef} R_{ef}{}^{cd}+ 12355 
R R_{a}{}^{b} R_{b}{}^{a} R_{cd}{}^{ef} R_{ef}{}^{cd} \nonumber\\
&- 104580 
R_{a}{}^{b} R_{b}{}^{a} R_{d}{}^{c} R_{ce}{}^{fg} R_{fg}{}^{de} - 
15120 R R_{b}{}^{a} R_{ad}{}^{bc} R_{ce}{}^{fg} R_{fg}{}^{de}\nonumber\\
&- 3780 
R R_{b}{}^{a} R_{ac}{}^{fg} R_{de}{}^{bc} R_{fg}{}^{de} + 11340 
R_{a}{}^{b} R_{b}{}^{a} R_{cd}{}^{gh} R_{ef}{}^{cd} R_{gh}{}^{ef} \, , \\
\pazocal{S}^{(D=5,n=5)}_{\rm  III}=&
-108751900 R^5 + 1026499979 R^3 R_{a}{}^{b} R_{b}{}^{a} - 2724816480 
R^2 R_{a}{}^{c} R_{b}{}^{a} R_{c}{}^{b}\nonumber\\ &+ 3743976918 R R_{a}{}^{d} 
R_{b}{}^{a} R_{c}{}^{b} R_{d}{}^{c} - 981715812 R_{a}{}^{b} 
R_{b}{}^{a} R_{c}{}^{e} R_{d}{}^{c} R_{e}{}^{d} \nonumber\\ &+ 241948812 R 
R_{b}{}^{a} R_{d}{}^{b} R_{e}{}^{c} R_{ac}{}^{de}+ 11124379 R^3 
R_{ab}{}^{cd} R_{cd}{}^{ab} \nonumber\\ &+ 2523150 R R_{ab}{}^{cd}{}^2 
R_{cd}{}^{ab}{}^2 - 1472417016 R_{a}{}^{b} R_{b}{}^{a} R_{e}{}^{c} 
R_{f}{}^{d} R_{cd}{}^{ef}\nonumber\\
& + 442592640 R R_{c}{}^{a} R_{d}{}^{b} 
R_{ab}{}^{ef} R_{ef}{}^{cd} - 1009017009 R R_{b}{}^{a} R_{c}{}^{b} 
R_{ad}{}^{ef} R_{ef}{}^{cd}\nonumber\\
& - 199666439 R R_{a}{}^{b} R_{b}{}^{a} 
R_{cd}{}^{ef} R_{ef}{}^{cd} + 1327705722 R_{a}{}^{b} R_{b}{}^{a} 
R_{d}{}^{c} R_{ce}{}^{fg} R_{fg}{}^{de}\nonumber\\
&- 7998480 R R_{b}{}^{a} 
R_{ad}{}^{bc} R_{ce}{}^{fg} R_{fg}{}^{de} + 151439400 R R_{b}{}^{a} 
R_{ac}{}^{fg} R_{de}{}^{bc} R_{fg}{}^{de}\nonumber\\
& - 197676360 R_{a}{}^{b} 
R_{b}{}^{a} R_{cd}{}^{gh} R_{ef}{}^{cd} R_{gh}{}^{ef} +35700000 
R_{ab}{}^{cd} R_{cd}{}^{ab} R_{ej}{}^{gh} R_{fh}{}^{ij} R_{gi}{}^{ef} \nonumber\\&
+ 121836960 R_{b}{}^{a} R_{ad}{}^{bc} R_{cf}{}^{de} R_{eg}{}^{hi} 
R_{hi}{}^{fg} - 89250 R_{ab}{}^{cd} R_{cd}{}^{ab} R_{ef}{}^{ij} 
R_{gh}{}^{ef} R_{ij}{}^{gh}\, ,
\end{align}
And for them
\begin{align}
\tau^{(D=5,n=5)}_{\rm  I}&=+2 \tau_{(5,0)}-\tau_{(5,1)}-12\tau_{(5,2)}-10 \tau_{(5,3)}+2 \tau_{(5,4)}+3 \tau_{(5,5)}, \\
\tau^{(D=5,n=5)}_{\rm  II}&=-5 \tau_{(5,0)}+4 \tau_{(5,1)}+18 \tau_{(5,2)}+4 \tau_{(5,3)}-5 \tau_{(5,4)}\, , \\
\tau^{(D=5,n=5)}_{\rm  III}&=+45\tau_{(5,0)}-46 \tau_{(5,1)}+44 (\tau_{(5,3)}-3 \tau _{(5,2)}\, .
\end{align}
For $D=6$, we find
\begin{align}
\pazocal{S}^{(D=6,n=5)}_{\rm  I}=&-123946191482880 R_{a}{}^{b} R_{b}{}^{a} R_{c}{}^{e} R_{d}{}^{c} \
R_{e}{}^{d} + 1472406237369312 R_{a}{}^{d} R_{b}{}^{a} R_{c}{}^{b} \
R_{d}{}^{c} R\nonumber\\ &- 1080277675306560 R_{a}{}^{c} R_{b}{}^{a} R_{c}{}^{b} \
R^2 + 162174148310040 R_{a}{}^{b} R_{b}{}^{a} R^3\nonumber\\
&- 11444059832562 \
R^5 + 1702982503075584 R_{b}{}^{a} R_{d}{}^{b} R_{e}{}^{c} R \
R_{ac}{}^{de}\nonumber\\
&+ 75220642409760 R^3 R_{ab}{}^{cd} R_{cd}{}^{ab} + \
12994390356246 R \left(R_{ab}{}^{cd}{} R_{cd}{}^{ab}{}\right)^2\nonumber\\
&- 941724825600 \
R_{a}{}^{b} R_{b}{}^{a} R_{e}{}^{c} R_{f}{}^{d} R_{cd}{}^{ef} - \
1826681030324352 R_{c}{}^{a} R_{d}{}^{b} R R_{ab}{}^{ef} \
R_{ef}{}^{cd}\nonumber\\
&+ 1161324617394816 R_{b}{}^{a} R_{c}{}^{b} R \
R_{ad}{}^{ef} R_{ef}{}^{cd} - 402058236112056 R_{a}{}^{b} R_{b}{}^{a} \
R R_{cd}{}^{ef} R_{ef}{}^{cd}\nonumber\\
&+ 796036321619712 R_{b}{}^{a} R \
R_{ad}{}^{bc} R_{ce}{}^{fg} R_{fg}{}^{de}\nonumber\\
&- 226245709813248 \
R_{b}{}^{a} R R_{ac}{}^{fg} R_{de}{}^{bc} R_{fg}{}^{de}\nonumber\\
&- \
2713887813611520 R_{ag}{}^{cd} R_{bi}{}^{ef} R_{ce}{}^{ab} \
R_{dj}{}^{gh} R_{fh}{}^{ij}\nonumber\\
&+ 5441837051289600 R_{ag}{}^{cd} \
R_{bi}{}^{ef} R_{ce}{}^{ab} R_{dh}{}^{ij} R_{fj}{}^{gh}\nonumber\\
&- \
8516393811394560 R_{ag}{}^{cd} R_{bh}{}^{ij} R_{ce}{}^{ab} \
R_{di}{}^{ef} R_{fj}{}^{gh}\nonumber\\
&- 9075154990067712 R_{aj}{}^{gh} \
R_{bd}{}^{ij} R_{ce}{}^{ab} R_{fg}{}^{cd} R_{hi}{}^{ef}\, ,\\
\pazocal{S}^{(D=6,n=5)}_{\rm  II}=&-39481565540352000 R_{a}{}^{b} R_{b}{}^{a} R_{c}{}^{e} R_{d}{}^{c} \
R_{e}{}^{d} + 496958473622415360 R_{a}{}^{d} R_{b}{}^{a} R_{c}{}^{b} \
R_{d}{}^{c} R\nonumber\\
&- 366085018636185600 R_{a}{}^{c} R_{b}{}^{a} \
R_{c}{}^{b} R^2 + 56771103624384000 R_{a}{}^{b} R_{b}{}^{a} R^3\nonumber\\
&-4236457006581120 R^5 + 605739537316331520 R_{b}{}^{a} R_{d}{}^{b} \
R_{e}{}^{c} R R_{ac}{}^{de}\nonumber\\
&+ 25066678861324800 R^3 R_{ab}{}^{cd} \
R_{cd}{}^{ab} + 2911274422692480 R \left(R_{ab}{}^{cd}{}
R_{cd}{}^{ab}{}\right)^2 \nonumber\\
&- 9235519903334400 R_{a}{}^{b} R_{b}{}^{a} \
R_{e}{}^{c} R_{f}{}^{d} R_{cd}{}^{ef}\nonumber\\
&-654135376602562560 \
R_{c}{}^{a} R_{d}{}^{b} R R_{ab}{}^{ef} R_{ef}{}^{cd}\nonumber\\
&+ \
384078592166215680 R_{b}{}^{a} R_{c}{}^{b} R R_{ad}{}^{ef} \
R_{ef}{}^{cd}\nonumber\\
&- 128301089938030080 R_{a}{}^{b} R_{b}{}^{a} R \
R_{cd}{}^{ef} R_{ef}{}^{cd}\nonumber\\
&+ 247957574993141760 R_{b}{}^{a} R \
R_{ad}{}^{bc} R_{ce}{}^{fg} R_{fg}{}^{de}\nonumber\\
&- 54410152259543040 \
R_{b}{}^{a} R R_{ac}{}^{fg} R_{de}{}^{bc} R_{fg}{}^{de}\nonumber\\
&- \
915942099386695680 R_{ag}{}^{cd} R_{bi}{}^{ef} R_{ce}{}^{ab} \
R_{dj}{}^{gh} R_{fh}{}^{ij} \nonumber\\
&+ 1855713735622656000 R_{ag}{}^{cd} \
R_{bi}{}^{ef} R_{ce}{}^{ab} R_{dh}{}^{ij} R_{fj}{}^{gh}\nonumber\\
&- \
2983978700100403200 R_{ag}{}^{cd} R_{bh}{}^{ij} R_{ce}{}^{ab} \
R_{di}{}^{ef} R_{fj}{}^{gh}\nonumber\\
&- 3268733794665431040 R_{aj}{}^{gh} \
R_{bd}{}^{ij} R_{ce}{}^{ab} R_{fg}{}^{cd} R_{hi}{}^{ef}\, ,\\
\pazocal{S}^{(D=6,n=5)}_{\rm  III}=&-113245541360640 R_{a}{}^{b} R_{b}{}^{a} R_{c}{}^{e} R_{d}{}^{c} \
R_{e}{}^{d} + 1060631652273264 R_{a}{}^{d} R_{b}{}^{a} R_{c}{}^{b} \
R_{d}{}^{c} R\nonumber\\
&- 903602985933600 R_{a}{}^{c} R_{b}{}^{a} R_{c}{}^{b} \
R^2 + 127080097757820 R_{a}{}^{b} R_{b}{}^{a} R^3 - 8955723921633 R^5\nonumber\\
&+ 1791407446201728 R_{b}{}^{a} R_{d}{}^{b} R_{e}{}^{c} R \
R_{ac}{}^{de} + 65583784852200 R^3 R_{ab}{}^{cd} R_{cd}{}^{ab} \nonumber\\
&+17709732531387 R \left(R_{ab}{}^{cd}{} R_{cd}{}^{ab}{}\right)^2 + 3780034053120 \
R_{a}{}^{b} R_{b}{}^{a} R_{e}{}^{c} R_{f}{}^{d} R_{cd}{}^{ef}\nonumber\\
&-\
2136457519124544 R_{c}{}^{a} R_{d}{}^{b} R R_{ab}{}^{ef} \
R_{ef}{}^{cd} + 1548204355449792 R_{b}{}^{a} R_{c}{}^{b} R
R_{ad}{}^{ef} R_{ef}{}^{cd}\nonumber\\
&- 341027462136492 R_{a}{}^{b} R_{b}{}^{a} \
R R_{cd}{}^{ef} R_{ef}{}^{cd} + 601767492758784 R_{b}{}^{a} R \
R_{ad}{}^{bc} R_{ce}{}^{fg} R_{fg}{}^{de}\nonumber\\
&- 195741719323776 
R_{b}{}^{a} R R_{ac}{}^{fg} R_{de}{}^{bc} R_{fg}{}^{de}\nonumber\\
&-
686045879580672 R_{ag}{}^{cd} R_{bi}{}^{ef} R_{ce}{}^{ab} \
R_{dj}{}^{gh} R_{fh}{}^{ij}\nonumber\\
&- 409211547264000 R_{ag}{}^{cd} \
R_{bi}{}^{ef} R_{ce}{}^{ab} R_{dh}{}^{ij} R_{fj}{}^{gh}\nonumber\\
&-4137732154183680 R_{ag}{}^{cd} R_{bh}{}^{ij} R_{ce}{}^{ab} \
R_{di}{}^{ef} R_{fj}{}^{gh}\nonumber\\
&- 8161945395342336 R_{aj}{}^{gh} \
R_{bd}{}^{ij} R_{ce}{}^{ab} R_{fg}{}^{cd} R_{hi}{}^{ef}\, ,
\end{align}
and for them
\begin{align}
\tau_{(D=6,n=5)}^{\rm  I}&=\tau_{(5,5)}-10\tau_{(5,2)}\, ,\\
\tau_{(D=6,n=5)}^{\rm  II}&=\tau_{(5,3)} - 3 \tau_{(5,2)} + \tau_{(5,1)} \, ,\\
\tau_{(D=6,n=5)}^{\rm  III}&=\tau_{(5, 4)} - \tau_{(5, 3)} - 3 \tau_{(5, 2)}\, .
\end{align}

At order six we have four inequivalent GQT classes. Representatives in $D=5$ are given by
\begin{align}
\pazocal{S}^{(D=5,n=6)}_{\rm  I}=&
-73164000 \left(R_{ab}{} R^{ab}{}\right)^3 - 1714893120 R_{ab} R^{ab} 
R_{c}{}^{e} R_{d}{}^{c} R_{e}{}^{d} R \nonumber\\ &+ 1318812172 R_{ab} R^{ab} 
R_{c}{}^{d} R_{d}{}^{c} R^2+ 271196208 R_{a}{}^{c} R_{b}{}^{a} 
R_{c}{}^{b} R^3 \nonumber\\ &- 317404865 R_{ab} R^{ab} R^4 + 18018062 R^6+ 
300979224 R_{c}{}^{a} R_{d}{}^{b} R^3 R_{ab}{}^{cd} \nonumber\\ &+ 248125440 
R_{b}{}^{a} R_{d}{}^{b} R_{e}{}^{c} R^2 R_{ac}{}^{de} + 170805000 
\left(R_{ef}{} R^{ef}{}\right)^2 R_{abcd} R^{abcd} \nonumber\\ & - 452092811 R_{ef} R^{ef} R^2 
R_{abcd} R^{abcd} + 74766829 R^4 R_{abcd} R^{abcd}  \nonumber\\ & - 139080000 R_{ef} 
R^{ef} \left(R_{abcd}{} R^{abcd}{}\right)^2  + 38179125 R^2 \left(R_{abcd}{} 
R^{abcd}{}\right)^2  \nonumber\\ &+ 35080000 \left(R_{abcd}{} R^{abcd}{}\right)^3  - 2244499440 R_{ab} 
R^{ab} R_{e}{}^{c} R_{f}{}^{d} R R_{cd}{}^{ef} \nonumber\\ & - 445474968 
R_{b}{}^{a} R^3 R_{ac}{}^{de} R_{de}{}^{bc}  - 87720000 
\left(R_{a}{}^{c}{}_{b}{}^{d}{} R_{c}{}^{e}{}_{d}{}^{f}{}
R_{e}{}^{a}{}_{f}{}^{b}{}\right)^2  \nonumber\\ & + 84746910 R^3 R_{ab}{}^{ef} 
R_{cd}{}^{ab} R_{ef}{}^{cd} + 2407239480 R_{ab} R^{ab} R_{d}{}^{c} R 
R_{ce}{}^{fg} R_{fg}{}^{de} \nonumber\\ & - 88583040 R_{b}{}^{a} R^2 R_{ad}{}^{bc} 
R_{ce}{}^{fg} R_{fg}{}^{de} - 410141550 R_{ab} R^{ab} R R_{cd}{}^{gh} 
R_{ef}{}^{cd} R_{gh}{}^{ef}  \nonumber\\ & + 564422400 R_{b}{}^{a} R R_{ad}{}^{bc} 
R_{cf}{}^{de} R_{eg}{}^{hi} R_{hi}{}^{fg}\nonumber\\ & - 61305000 R R_{abcd} 
R^{abcd} R_{ef}{}^{ij} R_{gh}{}^{ef} R_{ij}{}^{gh} \nonumber\\ & + 727920000 R^{de} 
R_{abcd} R^{abc}{}_{e} R_{g}{}^{i}{}_{h}{}^{j} 
R_{i}{}^{k}{}_{j}{}^{l} R_{k}{}^{g}{}_{l}{}^{h}\nonumber\\ & - 578160000 
R_{ab}{}^{cd} R_{cd}{}^{ef} R_{ef}{}^{ab} R_{g}{}^{i}{}_{h}{}^{j} 
R_{i}{}^{k}{}_{j}{}^{l} R_{k}{}^{g}{}_{l}{}^{h}\, , \\
\pazocal{S}^{(D=5,n=6)}_{\rm  II}=&-
137140000 \left(R_{ab}{} R^{ab}{}\right)^3 - 1947491520 R_{ab}
R^{ab} R_{c}{}^{e} R_{d}{}^{c} R_{e}{}^{d} R  \nonumber\\& + 1751816692 R_{ab}
R^{ab} R_{c}{}^{d} R_{d}{}^{c} R^2  + 329051088 R_{a}{}^{c}
R_{b}{}^{a} R_{c}{}^{b} R^3 \nonumber\\& - 432438015 R_{ab} R^{ab} R^4 + 25289682
R^6 + 400229864 R_{c}{}^{a} R_{d}{}^{b} R^3 R_{ab}{}^{cd}\nonumber\\&  + 165181440
R_{b}{}^{a} R_{d}{}^{b} R_{e}{}^{c} R^2 R_{ac}{}^{de} + 272619000
\left(R_{ef}{} R^{ef}{}\right)^2 R_{abcd} R^{abcd} \nonumber\\& - 609591221 
R_{ef} R^{ef} R^2 R_{abcd} R^{abcd}  + 100315219 R^4 R_{abcd} R^{abcd} 
\nonumber\\&- 173400000 R_{ef} R^{ef} \left(R_{abcd}{} R^{abcd}{}\right)^2 + 
46512875 R^2 \left(R_{abcd}{} R^{abcd}{}\right)^2\nonumber\\&  + 35600000 
\left(R_{abcd}{} R^{abcd}{}\right)^3 - 2869300240 R_{ab} R^{ab} 
R_{e}{}^{c} R_{f}{}^{d} R R_{cd}{}^{ef} \nonumber\\& - 484473448 R_{b}{}^{a} R^3 
R_{ac}{}^{de} R_{de}{}^{bc}  - 31320000 
\left(R_{a}{}^{c}{}_{b}{}^{d}{} R_{c}{}^{e}{}_{d}{}^{f}{} 
R_{e}{}^{a}{}_{f}{}^{b}{}\right)^2 \nonumber\\& + 55581410 R^3 R_{ab}{}^{ef} 
R_{cd}{}^{ab} R_{ef}{}^{cd} + 2452251080 R_{ab} R^{ab} R_{d}{}^{c} R 
R_{ce}{}^{fg} R_{fg}{}^{de} \nonumber\\& + 118104960 R_{b}{}^{a} R^2 R_{ad}{}^{bc} 
R_{ce}{}^{fg} R_{fg}{}^{de} - 242892050 R_{ab} R^{ab} R R_{cd}{}^{gh} 
R_{ef}{}^{cd} R_{gh}{}^{ef}\nonumber\\&  + 425990400 R_{b}{}^{a} R R_{ad}{}^{bc} 
R_{cf}{}^{de} R_{eg}{}^{hi} R_{hi}{}^{fg} \nonumber\\&- 77575000 R R_{abcd} 
R^{abcd} R_{ef}{}^{ij} R_{gh}{}^{ef} R_{ij}{}^{gh}\nonumber\\&  + 2129040000 
R^{de} R_{abcd} R^{abc}{}_{e} R_{g}{}^{i}{}_{h}{}^{j} 
R_{i}{}^{k}{}_{j}{}^{l} R_{k}{}^{g}{}_{l}{}^{h} \nonumber\\&- 909840000 
R_{ab}{}^{cd} R_{cd}{}^{ef} R_{ef}{}^{ab} R_{g}{}^{i}{}_{h}{}^{j} 
R_{i}{}^{k}{}_{j}{}^{l} R_{k}{}^{g}{}_{l}{}^{h}\, , \\
\pazocal{S}^{(D=5,n=6)}_{\rm  III}=&-859300000 \left(R_{ab}{} R^{ab}{}\right)^3 - 25179802560 R_{ab} R^{ab} 
R_{c}{}^{e} R_{d}{}^{c} R_{e}{}^{d} R \nonumber\\& + 19703296676 R_{ab} R^{ab} 
R_{c}{}^{d} R_{d}{}^{c} R^2 + 4227840144 R_{a}{}^{c} R_{b}{}^{a} 
R_{c}{}^{b} R^3  \nonumber\\& - 4975158595 R_{ab} R^{ab} R^4 + 291039066 R^6 + 
5123673672 R_{c}{}^{a} R_{d}{}^{b} R^3 R_{ab}{}^{cd} \nonumber\\&+ 1331589120 
R_{b}{}^{a} R_{d}{}^{b} R_{e}{}^{c} R^2 R_{ac}{}^{de} + 2222415000 
\left(R_{ef}{} R^{ef}{}\right)^2 R_{abcd} R^{abcd}\nonumber\\& - 6346768033 R_{ef} R^{ef} 
R^2 R_{abcd} R^{abcd} + 1019618087 R^4 R_{abcd} R^{abcd}   \nonumber\\& - 1819320000 
R_{ef} R^{ef} \left(R_{abcd}{} R^{abcd}{}\right)^2  + 513207375 R^2 \left(R_{abcd}{} 
R^{abcd}{}\right)^2 \nonumber\\& + 450800000 \left(R_{abcd}{} R^{abcd}{}\right)^3 - 33156269520 
R_{ab} R^{ab} R_{e}{}^{c} R_{f}{}^{d} R R_{cd}{}^{ef}\nonumber\\&  - 6081896904 
R_{b}{}^{a} R^3 R_{ac}{}^{de} R_{de}{}^{bc} + 4158600000 
\left(R_{a}{}^{c}{}_{b}{}^{d}{} R_{c}{}^{e}{}_{d}{}^{f}{}
R_{e}{}^{a}{}_{f}{}^{b}{}\right)^2\nonumber\\& + 1060299930 R^3 R_{ab}{}^{ef} 
R_{cd}{}^{ab} R_{ef}{}^{cd}  + 34472700840 R_{ab} R^{ab} R_{d}{}^{c} R 
R_{ce}{}^{fg} R_{fg}{}^{de}\nonumber\\& - 834145920 R_{b}{}^{a} R^2 R_{ad}{}^{bc} 
R_{ce}{}^{fg} R_{fg}{}^{de} - 5503384650 R_{ab} R^{ab} R 
R_{cd}{}^{gh} R_{ef}{}^{cd} R_{gh}{}^{ef} \nonumber\\& + 6734419200 R_{b}{}^{a} R 
R_{ad}{}^{bc} R_{cf}{}^{de} R_{eg}{}^{hi} R_{hi}{}^{fg}  \nonumber\\&- 809475000 R 
R_{abcd} R^{abcd} R_{ef}{}^{ij} R_{gh}{}^{ef} R_{ij}{}^{gh}\nonumber\\& + 
15109200000 R^{de} R_{abcd} R^{abc}{}_{e} R_{g}{}^{i}{}_{h}{}^{j} 
R_{i}{}^{k}{}_{j}{}^{l} R_{k}{}^{g}{}_{l}{}^{h}\nonumber\\&  - 9162000000 
R_{ab}{}^{cd} R_{cd}{}^{ef} R_{ef}{}^{ab} R_{g}{}^{i}{}_{h}{}^{j} 
R_{i}{}^{k}{}_{j}{}^{l} R_{k}{}^{g}{}_{l}{}^{h}\, , \\
\pazocal{S}^{(D=5,n=6)}_{\rm  IV}=&+ 31500000 \left(R_{ab}{} R^{ab}{}\right)^3 - 4028310720 R_{ab} R^{ab} 
R_{c}{}^{e} R_{d}{}^{c} R_{e}{}^{d} R  \nonumber\\& + 2252042612 R_{ab} R^{ab} 
R_{c}{}^{d} R_{d}{}^{c} R^2  + 683314128 R_{a}{}^{c} R_{b}{}^{a} 
R_{c}{}^{b} R^3\nonumber\\& - 555694015 R_{ab} R^{ab} R^4 + 27464642 R^6 + 
877183464 R_{c}{}^{a} R_{d}{}^{b} R^3 R_{ab}{}^{cd} \nonumber\\& - 96706560 
R_{b}{}^{a} R_{d}{}^{b} R_{e}{}^{c} R^2 R_{ac}{}^{de} + 163995000 
\left(R_{ef}{} R^{ef}{}\right)^2 R_{abcd} R^{abcd}\nonumber\\& - 407173621 R_{ef} R^{ef} R^2 
R_{abcd} R^{abcd}  + 62048819 R^4 R_{abcd} R^{abcd} \nonumber\\& - 292440000 R_{ef} 
R^{ef} \left(R_{abcd}{} R^{abcd}{}\right)^2 + 28222875 R^2 \left(R_{abcd}{}
R^{abcd}{}\right)^2\nonumber\\&  + 97600000 \left(R_{abcd}{} R^{abcd}{}\right)^3 - 4629108240 R_{ab} 
R^{ab} R_{e}{}^{c} R_{f}{}^{d} R R_{cd}{}^{ef} \nonumber\\& - 1045548648 
R_{b}{}^{a} R^3 R_{ac}{}^{de} R_{de}{}^{bc} + 2409000000 \left(
R_{a}{}^{c}{}_{b}{}^{d}{} R_{c}{}^{e}{}_{d}{}^{f}{}
R_{e}{}^{a}{}_{f}{}^{b}{}\right)^2 \nonumber\\&  + 280021410 R^3 R_{ab}{}^{ef} 
R_{cd}{}^{ab} R_{ef}{}^{cd}  + 6317083080 R_{ab} R^{ab} R_{d}{}^{c} R 
R_{ce}{}^{fg} R_{fg}{}^{de} \nonumber\\&  - 655655040 R_{b}{}^{a} R^2 R_{ad}{}^{bc} 
R_{ce}{}^{fg} R_{fg}{}^{de}  - 1401052050 R_{ab} R^{ab} R 
R_{cd}{}^{gh} R_{ef}{}^{cd} R_{gh}{}^{ef}\nonumber\\&  + 664070400 R_{b}{}^{a} R 
R_{ad}{}^{bc} R_{cf}{}^{de} R_{eg}{}^{hi} R_{hi}{}^{fg} - 46575000 R 
R_{abcd} R^{abcd} R_{ef}{}^{ij} R_{gh}{}^{ef} R_{ij}{}^{gh} \nonumber\\& + 
1414800000 R^{de} R_{abcd} R^{abc}{}_{e} R_{g}{}^{i}{}_{h}{}^{j} 
R_{i}{}^{k}{}_{j}{}^{l} R_{k}{}^{g}{}_{l}{}^{h}\nonumber\\&  - 1832400000 
R_{ab}{}^{cd} R_{cd}{}^{ef} R_{ef}{}^{ab} R_{g}{}^{i}{}_{h}{}^{j} 
R_{i}{}^{k}{}_{j}{}^{l} R_{k}{}^{g}{}_{l}{}^{h}\, .
\end{align}
And the corresponding $\tau(r)$ are given by
\begin{align}
\tau^{(D=5,n=6)}_{\rm  I}&=+\tau_{(6,0)}+12 \tau_{(6,5)}-8 \tau_{(6,6)}\, , \\
\tau^{(D=5,n=6)}_{\rm  II}&=-5 \tau_{(6,2)}-16 \tau_{(6,5)}+11 \tau_{(6,6)}\, , \\
\tau^{(D=5,n=6)}_{\rm  III}&=-5\tau_{(6,3)}-3 \tau_{(6,5)}+3 \tau_{(6,6)}\, , \\
\tau^{(D=5,n=6)}_{\rm IV}&=+15 \tau_{(6,4)}-2 (6 \tau_{(6,5)}-\tau_{(6,6)}) \, .
\end{align}


For $D=6$ we find
\begin{align}
\pazocal{S}_{ \rm I}^{(D=6,n=6)}=&-14096679060821760 R_{a}{}^{b} R_{b}{}^{c} R_{c}{}^{d} R_{d}{}^{e} \
R_{e}{}^{f} R_{f}{}^{a}\nonumber\\
&+ 14852647970900544 R_{c}{}^{e} R_{d}{}^{c} \
R_{e}{}^{d} R_{i}{}^{j} R_{j}{}^{i} R\nonumber\\
&- 5617985150718012 \
\left(R_{i}{}^{j}{} R_{j}{}^{i}{}\right)^2 R^2 - 1124843605416416 R_{a}{}^{c} \
R_{b}{}^{a} R_{c}{}^{b} R^3\nonumber\\
&+ 1005726172300248 R_{ab} R^{ab} R^4 - \
29156254184830 R^6 \nonumber\\
&- 1438756007591232 R_{c}{}^{a} R_{d}{}^{b} R^3 \
R_{ab}{}^{cd} + 2380028275859520 R^2 R_{b}{}^{a} R_{d}{}^{b} \
R_{e}{}^{c} R_{ac}{}^{de}\nonumber\\
&+ 1254308457170736 R_{ef} R^{ef} R^2 \
R_{abcd} R^{abcd} - 168004022190642 R^4 R_{ab}{}^{cd} R_{cd}{}^{ab}\nonumber\\
&+ \
3230088574927500 R_{i}{}^{j} R_{j}{}^{i} \left(R_{ab}{}^{cd}{} \
R_{cd}{}^{ab}{}\right)^2 - 607399901908371 R^2 \left(R_{ab}{}^{cd}{}
R_{cd}{}^{ab}{}\right)^2\nonumber\\
&+ 721416483693312 R_{e}{}^{c} R_{f}{}^{d} \
R_{i}{}^{j} R_{j}{}^{i} R R_{cd}{}^{ef}+ 1133891404354368 \
R_{b}{}^{a} R^3 R_{ac}{}^{de} R_{de}{}^{bc} \nonumber\\
&- 682346981951712 R^3 \
R_{ab}{}^{ef} R_{cd}{}^{ab} R_{ef}{}^{cd} + 9376966635379200 R^{ab} \
R^{cd} R_{i}{}^{j} R_{j}{}^{i} R_{ecfd} R^{e}{}_{a}{}^{f}{}_{b} \nonumber\\
&- \
8990642116684800 R^{ab} R_{i}{}^{j} R_{j}{}^{i} \
R_{a}{}^{c}{}_{b}{}^{d} R_{efgc} R^{efg}{}_{d} \nonumber\\
&- 6299359808303232 \
R_{d}{}^{c} R_{i}{}^{j} R_{j}{}^{i} R R_{ce}{}^{fg} R_{fg}{}^{de}\nonumber\\
& + \
1901604108792960 R_{b}{}^{a} R^2 R_{ad}{}^{bc} R_{ce}{}^{fg} \
R_{fg}{}^{de} \nonumber\\
&+ 3847116811602240 R_{i}{}^{j} R_{j}{}^{i} R \
R_{cd}{}^{gh} R_{ef}{}^{cd} R_{gh}{}^{ef} \nonumber\\
&- 10134930764312640 \
R_{a}{}^{b} R_{b}{}^{c} R_{c}{}^{d} R_{d}{}^{a} R_{ef}{}^{hi} \
R_{hi}{}^{eg} \nonumber\\
&- 2178824133657600 R_{b}{}^{a} R R_{ad}{}^{bc} \
R_{cf}{}^{de} R_{eg}{}^{hi} R_{hi}{}^{fg} \nonumber\\
&+ 304956123151680 R \
R_{ab}{}^{cd} R_{cd}{}^{ab} R_{ef}{}^{ij} R_{gh}{}^{ef} R_{ij}{}^{gh} \nonumber\\
&
- 1895257162656000 R_{ab}{}^{cd} R_{cd}{}^{ef} R_{ef}{}^{gh} \
R_{gh}{}^{ij} R_{ij}{}^{kl} R_{kl}{}^{ab} \nonumber\\
&+ 1259726446836000 \
R_{ab}{}^{cd} R_{cd}{}^{ef} R_{ef}{}^{gh} R_{gh}{}^{ab} R_{ij}{}^{kl} \
R_{kl}{}^{ij}\, ,\\
\pazocal{S}_{ \rm II}^{(D=6,n=6)}=&-31836692340236160 R_{a}{}^{b} R_{b}{}^{c} R_{c}{}^{d} R_{d}{}^{e} \
R_{e}{}^{f} R_{f}{}^{a}\nonumber\\
&+ 34439506371202464 R_{c}{}^{e} R_{d}{}^{c} \
R_{e}{}^{d} R_{i}{}^{j} R_{j}{}^{i} R - 13557698416858564 \
\left(R_{i}{}^{j}{} R_{j}{}^{i}{}\right)^2 R^2 \nonumber\\
&- 2849781769779440 R_{a}{}^{c} \
R_{b}{}^{a} R_{c}{}^{b} R^3 + 2611991351109630 R_{ab} R^{ab} R^4\nonumber\\
& - \
88399029128845 R^6 - 3677104626840832 R_{c}{}^{a} R_{d}{}^{b} R^3 \
R_{ab}{}^{cd} \nonumber\\
&+ 6132894365769600 R^2 R_{b}{}^{a} R_{d}{}^{b} \
R_{e}{}^{c} R_{ac}{}^{de} + 3157451844617752 R_{ef} R^{ef} R^2 \
R_{abcd} R^{abcd} \nonumber\\
&- 441312471667562 R^4 R_{ab}{}^{cd} R_{cd}{}^{ab} + \
7146998363226150 R_{i}{}^{j} R_{j}{}^{i} \left(R_{ab}{}^{cd}{} \
R_{cd}{}^{ab}{}\right)^2 \nonumber\\
&- 1314167139538110 R^2 \left(R_{ab}{}^{cd}{} \
R_{cd}{}^{ab}{}\right)^2 + 2087392939560192 R_{e}{}^{c} R_{f}{}^{d} \
R_{i}{}^{j} R_{j}{}^{i} R R_{cd}{}^{ef} \nonumber\\
&+ 2787439490093632 \
R_{b}{}^{a} R^3 R_{ac}{}^{de} R_{de}{}^{bc} - 1533575360560320 R^3 \
R_{ab}{}^{ef} R_{cd}{}^{ab} R_{ef}{}^{cd} \nonumber\\
&+ 22373284326307200 R^{ab} \
R^{cd} R_{i}{}^{j} R_{j}{}^{i} R_{ecfd} R^{e}{}_{a}{}^{f}{}_{b} \nonumber\\
&- \
20005506406828800 R^{ab} R_{i}{}^{j} R_{j}{}^{i} \
R_{a}{}^{c}{}_{b}{}^{d} R_{efgc} R^{efg}{}_{d} \nonumber\\
&- 15079987603900032 \
R_{d}{}^{c} R_{i}{}^{j} R_{j}{}^{i} R R_{ce}{}^{fg} R_{fg}{}^{de} \nonumber\\
&+ \
3611210786726400 R_{b}{}^{a} R^2 R_{ad}{}^{bc} R_{ce}{}^{fg} \
R_{fg}{}^{de}\nonumber\\
&+ 8724416327193600 R_{i}{}^{j} R_{j}{}^{i} R \
R_{cd}{}^{gh} R_{ef}{}^{cd} R_{gh}{}^{ef}\nonumber\\
& - 23212861724463840 \
R_{a}{}^{b} R_{b}{}^{c} R_{c}{}^{d} R_{d}{}^{a} R_{ef}{}^{hi} \
R_{hi}{}^{eg}\nonumber\\
&- 3819771274176000 R_{b}{}^{a} R R_{ad}{}^{bc} \
R_{cf}{}^{de} R_{eg}{}^{hi} R_{hi}{}^{fg}\nonumber\\
&+ 616693394116800 R \
R_{ab}{}^{cd} R_{cd}{}^{ab} R_{ef}{}^{ij} R_{gh}{}^{ef} R_{ij}{}^{gh} \
\nonumber\\
&- 4293178347744000 R_{ab}{}^{cd} R_{cd}{}^{ef} R_{ef}{}^{gh} \
R_{gh}{}^{ij} R_{ij}{}^{kl} R_{kl}{}^{ab} \nonumber\\
&+ 2871990255420000 \
R_{ab}{}^{cd} R_{cd}{}^{ef} R_{ef}{}^{gh} R_{gh}{}^{ab} R_{ij}{}^{kl} \
R_{kl}{}^{ij}\, ,\\
\pazocal{S}_{ \rm III}^{(D=6,n=6)}=&-6943970290757760 R_{a}{}^{b} R_{b}{}^{c} R_{c}{}^{d} R_{d}{}^{e} \
R_{e}{}^{f} R_{f}{}^{a}\nonumber\\
&+ 5474732611842144 R_{c}{}^{e} R_{d}{}^{c} \
R_{e}{}^{d} R_{i}{}^{j} R_{j}{}^{i} R - 2281174181020312 \
\left(R_{i}{}^{j}{} R_{j}{}^{i}{}\right)^2 R^2\nonumber\\
&- 355850581360016 R_{a}{}^{c} \
R_{b}{}^{a} R_{c}{}^{b} R^3 + 372382664514798 R_{ab} R^{ab} R^4 - \
8079133402255 R^6 \nonumber\\
&- 724084260087232 R_{c}{}^{a} R_{d}{}^{b} R^3 \
R_{ab}{}^{cd} + 680829460259520 R^2 R_{b}{}^{a} R_{d}{}^{b} \
R_{e}{}^{c} R_{ac}{}^{de} \nonumber\\
&+ 397223834424736 R_{ef} R^{ef} R^2 \
R_{abcd} R^{abcd} - 65621870854892 R^4 R_{ab}{}^{cd} R_{cd}{}^{ab} \nonumber\\
&
+656091001244250 R_{i}{}^{j} R_{j}{}^{i} \left(R_{ab}{}^{cd}{} \
R_{cd}{}^{ab}{}\right)^2 - 121001538886371 R^2 \left(R_{ab}{}^{cd}{} \
R_{cd}{}^{ab}{}\right)^2 \nonumber\\
&+ 2073302769914112 R_{e}{}^{c} R_{f}{}^{d} \
R_{i}{}^{j} R_{j}{}^{i} R R_{cd}{}^{ef} + 671781101071168 R_{b}{}^{a} \
R^3 R_{ac}{}^{de} R_{de}{}^{bc}\nonumber\\
&- 224552737043712 R^3 R_{ab}{}^{ef} \
R_{cd}{}^{ab} R_{ef}{}^{cd}\nonumber\\
&+ 2768431158979200 R^{ab} R^{cd} \
R_{i}{}^{j} R_{j}{}^{i} R_{ecfd} R^{e}{}_{a}{}^{f}{}_{b}\nonumber\\
&- \
2668237319404800 R^{ab} R_{i}{}^{j} R_{j}{}^{i} \
R_{a}{}^{c}{}_{b}{}^{d} R_{efgc} R^{efg}{}_{d}\nonumber\\
&- 3774533321404032 \
R_{d}{}^{c} R_{i}{}^{j} R_{j}{}^{i} R R_{ce}{}^{fg} R_{fg}{}^{de}\nonumber\\
&+ \
523902114552960 R_{b}{}^{a} R^2 R_{ad}{}^{bc} R_{ce}{}^{fg} \
R_{fg}{}^{de}\nonumber\\
&+ 1290427980114240 R_{i}{}^{j} R_{j}{}^{i} R \
R_{cd}{}^{gh} R_{ef}{}^{cd} R_{gh}{}^{ef}\nonumber\\
&- 849014886788640 \
R_{a}{}^{b} R_{b}{}^{c} R_{c}{}^{d} R_{d}{}^{a} R_{ef}{}^{hi} \
R_{hi}{}^{eg}\nonumber\\
&- 965396466777600 R_{b}{}^{a} R R_{ad}{}^{bc} \
R_{cf}{}^{de} R_{eg}{}^{hi} R_{hi}{}^{fg} \nonumber\\
&+ 70932032851680 R \
R_{ab}{}^{cd} R_{cd}{}^{ab} R_{ef}{}^{ij} R_{gh}{}^{ef} R_{ij}{}^{gh} \nonumber\\
&
- 336455941536000 R_{ab}{}^{cd} R_{cd}{}^{ef} R_{ef}{}^{gh} \
R_{gh}{}^{ij} R_{ij}{}^{kl} R_{kl}{}^{ab}\nonumber\\
& + 219456058536000 \
R_{ab}{}^{cd} R_{cd}{}^{ef} R_{ef}{}^{gh} R_{gh}{}^{ab} R_{ij}{}^{kl} \
R_{kl}{}^{ij} \, ,\\
\pazocal{S}_{ \rm IV}^{(D=6,n=6)}=&-14096679060821760 R_{a}{}^{b} R_{b}{}^{c} R_{c}{}^{d} R_{d}{}^{e} \
R_{e}{}^{f} R_{f}{}^{a}\nonumber\\
&+ 14852647970900544 R_{c}{}^{e} R_{d}{}^{c} \
R_{e}{}^{d} R_{i}{}^{j} R_{j}{}^{i} R - 5617985150718012 \
\left(R_{i}{}^{j}{} R_{j}{}^{i}{}\right)^2 R^2\nonumber\\
&- 1124843605416416 R_{a}{}^{c} \
R_{b}{}^{a} R_{c}{}^{b} R^3 + 1005726172300248 R_{ab} R^{ab} R^4 \nonumber\\
&- \
29156254184830 R^6 - 1438756007591232 R_{c}{}^{a} R_{d}{}^{b} R^3 \
R_{ab}{}^{cd}\nonumber\\
&+ 2380028275859520 R^2 R_{b}{}^{a} R_{d}{}^{b} \
R_{e}{}^{c} R_{ac}{}^{de} + 1254308457170736 R_{ef} R^{ef} R^2 \
R_{abcd} R^{abcd}\nonumber\\
&- 168004022190642 R^4 R_{ab}{}^{cd} R_{cd}{}^{ab} + \
3230088574927500 R_{i}{}^{j} R_{j}{}^{i} \left(R_{ab}{}^{cd}{}
R_{cd}{}^{ab}{}\right)^2 \nonumber\\
&- 607399901908371 R^2 \left(R_{ab}{}^{cd}{} \
R_{cd}{}^{ab}{}\right)^2 + 721416483693312 R_{e}{}^{c} R_{f}{}^{d} \
R_{i}{}^{j} R_{j}{}^{i} R R_{cd}{}^{ef}\nonumber\\
&+ 1133891404354368 \
R_{b}{}^{a} R^3 R_{ac}{}^{de} R_{de}{}^{bc}- 682346981951712 R^3 \
R_{ab}{}^{ef} R_{cd}{}^{ab} R_{ef}{}^{cd} \nonumber\\
&+ 9376966635379200 R^{ab} \
R^{cd} R_{i}{}^{j} R_{j}{}^{i} R_{ecfd} R^{e}{}_{a}{}^{f}{}_{b}- \nonumber\\
&\
8990642116684800 R^{ab} R_{i}{}^{j} R_{j}{}^{i} \
R_{a}{}^{c}{}_{b}{}^{d} R_{efgc} R^{efg}{}_{d} \nonumber\\
&- 6299359808303232 \
R_{d}{}^{c} R_{i}{}^{j} R_{j}{}^{i} R R_{ce}{}^{fg} R_{fg}{}^{de} \nonumber\\
&+ \
1901604108792960 R_{b}{}^{a} R^2 R_{ad}{}^{bc} R_{ce}{}^{fg} \
R_{fg}{}^{de}\nonumber\\
& + 3847116811602240 R_{i}{}^{j} R_{j}{}^{i} R \
R_{cd}{}^{gh} R_{ef}{}^{cd} R_{gh}{}^{ef} \nonumber\\
&- 10134930764312640 \
R_{a}{}^{b} R_{b}{}^{c} R_{c}{}^{d} R_{d}{}^{a} R_{ef}{}^{hi} \
R_{hi}{}^{eg} \nonumber\\
&- 2178824133657600 R_{b}{}^{a} R R_{ad}{}^{bc} \
R_{cf}{}^{de} R_{eg}{}^{hi} R_{hi}{}^{fg} \nonumber\\
&+ 304956123151680 R \
R_{ab}{}^{cd} R_{cd}{}^{ab} R_{ef}{}^{ij} R_{gh}{}^{ef} R_{ij}{}^{gh} \
\nonumber\\
&- 1895257162656000 R_{ab}{}^{cd} R_{cd}{}^{ef} R_{ef}{}^{gh} \
R_{gh}{}^{ij} R_{ij}{}^{kl} R_{kl}{}^{ab}\nonumber\\
& + 1259726446836000 \
R_{ab}{}^{cd} R_{cd}{}^{ef} R_{ef}{}^{gh} R_{gh}{}^{ab} R_{ij}{}^{kl} \
R_{kl}{}^{ij}\, ,
\end{align}
and for them
\begin{align}
\tau^{(D=6,n=6)}_{\rm  I}&=\tau_{(6, 6)} - 15 \tau_{(6, 2)} + 4 \tau_{(6, 1)}\, ,\\
\tau^{(D=6,n=6)}_{\rm  II}&=\tau_{(6, 5)} - 10 \tau_{(6, 2)} + 3 \tau_{(6, 1)}\, ,\\
\tau^{(D=6,n=6)}_{\rm  III}&=\tau_{(6, 4)} - 6 \tau_{(6, 2)} + 2 \tau_{(6, 1)}\, ,\\
\tau^{(D=6,n=6)}_{\rm  IV}&=\tau_{(6, 3)} - 3 \tau_{(6, 2)} + \tau_{(6, 1)}\, .
\end{align}

\chapter{Redefining the metric}\label{App:2}

\section*{Changing variables}

Implementing a differential change of variables directly in the action can be problematic if one is not careful enough. In order to see this, let us consider the equations of motion of $\tilde g_{ab}$ ---defined so that $g_{ab}=\tilde g_{ab}+ K_{ab}$--- by computing the variation of the new action $\tilde I[\tilde g_{ab}]=I[g_{ab}]$:\footnote{Note that in the second term we used the chain law for the functional derivative, which is in general  given by
\begin{equation}
\frac{\updelta I}{\delta \phi}\frac{\updelta \phi}{\updelta \psi}=\frac{\updelta I}{\updelta \phi}\frac{\partial \phi}{\partial \psi}-\partial_{a}\left(\frac{\updelta I}{\updelta \phi}\frac{\partial \phi}{\partial_{a}\psi}\right)+\ldots
\end{equation}
}

\begin{equation}\label{eq:badeq}
\frac{\updelta \tilde I}{\updelta \tilde g_{ab}}=\frac{\updelta I}{\updelta g_{ab}}+\frac{\updelta I}{\updelta g_{ef}}\frac{\updelta K_{ef}}{\updelta \tilde g_{ab}} \Bigg|_{g_{ab}=\tilde g_{ab}+K_{ab}}\, .
\end{equation}
Now, it is clear that we can always solve these equations if
\begin{equation}\label{eq:goodeq}
\frac{\updelta I}{\updelta g_{ab}} \Bigg|_{g_{ab}=\tilde g_{ab}+K_{ab}}=0\, .
\end{equation}
In other words, implementing the change of variables directly in the equations of the original theory produces an equation that solves the equations of $\tilde I$. However, the equations of $\tilde I$ contain more solutions. These additional solutions are spurious and appear as a consequence of increasing the number of derivatives in the action, so they should not be considered. 
A possible way to formalize this intuitive argument consists in introducing auxiliary field so that the redefinition of the metric becomes algebraic. Let us consider the following action
\begin{align}
I_{\chi}=\frac{1}{16\pi \GN}\int \diff^Dx\sqrt{-g}\Big[&-2\Lambda+R+f\left(g^{ab}, \chi_{abcd}, \chi_{e_1,abcd},\chi_{e_1e_2,abcd},\ldots\right)\\ \notag&+\frac{\partial f}{\partial \chi_{abcd}}\left(R_{abcd}-\chi_{abcd}\right)+\frac{\partial f}{\partial \chi_{e_1,abcd}}\left(\nabla_{e_1}R_{abcd}-\chi_{e_1,abcd}\right)\\ \notag &+\frac{\partial f}{\partial \chi_{e_1e_2,abcd}}\left(\nabla_{e_1}\nabla_{e_2}R_{abcd}-\chi_{e_1e_2,abcd}\right)+\ldots\Big]\, ,
\end{align}
where we have introduced some auxiliary fields $\chi_{abcd}$, $\chi_{e_{1},abcd}$, \ldots $\chi_{e_{1}\ldots e_{n},abcd}$. Let us convince ourselves that this action is equivalent to eq.~\req{eq:generalhdg}. When we take the variation with respect to $\chi_{e_{1}\ldots e_{i},abcd}$, we get
\begin{equation}
\sum_{j=0}\frac{\partial^2 f}{\partial\chi_{e_1\ldots e_i,a_1b_1c_1d_1} \partial \chi_{f_1\ldots f_j,a_2b_2c_2d_2}}\left(\nabla_{f_1}\ldots\nabla_{f_j}R_{a_2b_2c_2d_2}-\chi_{f_1\ldots f_j,a_2b_2c_2d_2}\right)=0.
\end{equation}
In this way, we get a system of algebraic equations for the variables $\chi_{e_{1}\ldots e_{i},abcd}$ that always has the following solution
\begin{eqnarray}
\label{eq:solchi}
\chi_{abcd}&=&R_{abcd}\, ,\\
\chi_{e_1,abcd}&=&\nabla_{e_1}R_{abcd}\, ,\\
\chi_{e_1 e_2,abcd}&=&\nabla_{e_1}\nabla_{e_2}R_{abcd}\, ,\\
&\ldots&
\end{eqnarray}
This is the unique solution if the matrix of the system is invertible, and this is the expected case if $f$ is general. When we plug this solution back in the action we recover eq.~\req{eq:generalhdg} (with explicit Einstein-Hilbert and cosmological constant terms), so that both formulations are equivalent. 

Now let us perform the following redefinition of the metric in $I_{\chi}$:
\begin{equation}
g_{ab}= \tilde g_{ab}+\alpha K_{ab}\, ,\quad\text{where}\,\,\, K_{ab}=K_{ab}\left(\tilde g^{ef}, \chi_{efcd}, \chi_{a_1,efcd},\ldots\right)\, ,
\end{equation}
this is, $K_{ab}$ is a symmetric tensor formed from contractions of the $\chi$ variables and the metric, but it contains no derivatives of any field. In this way, the change of variables is algebraic and can be directly implemented in the action. We therefore get
\begin{equation}
\tilde{I}_{\chi}\left[\tilde{g}_{ab}, \chi\right]=I_{\chi}\left[\tilde{g}_{ab}+\alpha K_{ab}, \chi\right]\, ,
\end{equation}
where, for simplicity, we are collectively denoting all auxiliary variables by $\chi$.  Now, both actions are equivalent and so are the field equations:
\begin{eqnarray}
\label{eq:eqmetric}
\frac{\updelta \tilde{I}_{\chi}}{\updelta\tilde g_{ab}}&=&\frac{\updelta I_{\chi}}{\updelta g_{ab}}\bigg|_{g_{ab}= \tilde g_{ab}+\alpha K_{ab}}\, ,\\
\label{eq:eqchi}
\frac{\updelta \tilde{I}_{\chi}}{\updelta\chi}&=&\frac{\updelta {I}_{\chi}}{\updelta\chi}+\alpha \frac{\updelta {I}_{\chi}}{\updelta g_{ab}}\frac{\updelta K_{ab}}{\updelta\chi}\bigg|_{g_{ab}= \tilde g_{ab}+\alpha K_{ab}}\, .
\end{eqnarray}
Using the first equation into the second one, we see that the equations for the auxiliary variables become $\updelta I_{\chi}/\updelta\chi=0$, which of course have the same solution as before \req{eq:solchi}. When we take that into account, $K_{ab}$ becomes a tensor constructed from the curvature of the original metric $g_{ab}$, so that we get
\begin{equation}
g_{ab}= \tilde g_{ab}+\alpha K_{ab}\left(\tilde g^{ef}, R_{efcd}, \nabla_{\alpha_1}R_{efcd},\ldots\right)\, .
\end{equation}
Then, according to eq.~\req{eq:eqmetric}, the equation for the metric $\tilde g_{ab}$ is simply obtained from the equation of $g_{ab}$ by substituting the change of variables:
\begin{equation}\label{eq:consistenteq}
\frac{\delta I_{\chi}}{\delta g_{ab}}\bigg|_{g_{ab}= \tilde g_{ab}+K_{ab}}=0\, .
\end{equation}
However, note that this is not the same as substituting \req{eq:solchi} in the action and taking the variation. This would yield instead
\begin{equation}\label{eq:fake}
\frac{\updelta \tilde{I}_{\chi}\left[\tilde{g}_{ab}, \chi(\tilde g_{ab})\right]}{\updelta\tilde g_{ab}}=\frac{\updelta \tilde I_{\chi}}{\updelta \tilde g_{ab}}+\frac{\updelta \tilde{I}_{\chi}}{\updelta\chi}\frac{\updelta\chi}{\updelta\tilde g_{ab}}=\frac{\updelta I_{\chi}}{\updelta g_{ab}}\bigg|_{g_{ab}= \tilde g_{ab}+\alpha K_{ab}}-\alpha \frac{\updelta {I}_{\chi}}{\updelta g_{ef}}\frac{\updelta K_{ef}}{\updelta\chi}\frac{\updelta\chi}{\updelta\tilde g_{ab}}\bigg|_{g_{ab}= \tilde g_{ab}+\alpha K_{ab}}\, .
\end{equation}
This equation is formally different to \req{eq:eqmetric} due to the second term, and it is equivalent to eq.~\eqref{eq:badeq}. The second term appears because the auxiliary variables $\chi(\tilde g_{\mu\nu})$ do not solve the equation $\updelta \tilde{I}_{\chi}/\updelta\chi=0$, but $\updelta {I}_{\chi}/\updelta\chi=0$. However, we must solve $\updelta \tilde{I}_{\chi}/\updelta\chi=0$ in order to get a solution of $\tilde{I}_{\chi}\left[\tilde{g}_{ab}, \chi\right]$, and according to eq.~\eqref{eq:eqchi} this would only happen if $(\updelta {I}_{\chi}/\updelta g_{ab}) (\partial K_{ab}/\partial\chi)=0$, so that the only consistent solutions of eq.~\eqref{eq:fake} are those which satisfy \eqref{eq:consistenteq}. 
This explains why the only solutions of eq.~\eqref{eq:badeq} we should consider are the ones satisfying condition
\eqref{eq:goodeq}. 

\section*{$W^n \nabla W\nabla W$ terms on SSS backgrounds} \label{apricot}
Now, let us show that eq.~\eqref{eq:Inv3} holds. In order to do that, it is convenient to carry out the following change or radial coordinate in the SSS ansatz \eqref{eq:SSS}:
\begin{equation}
\diff \tilde{r}^2=\frac{\diff r^2}{r^2}\, .
\end{equation}
In these coordinates, the SSS metric reads
\begin{equation}
\diff s^2=r(\tilde{r})^2\bigg [ -\tilde{N}(\tilde{r})^2\tilde{f}(\tilde{r})\diff t^2+\diff \tilde{r}^2+\diff \Omega^2_{(D-2)} \bigg],
\end{equation}
where we denoted $\tilde{N}(\tilde{r})=N(r(\tilde{r}))$ and $\tilde{f}(\tilde{r})=f(r(\tilde{r}))$.

We use a tilde to denote tensor components in the new coordinates. Direct computation shows that the components of the Weyl tensor in these new coordinates have formally the same expression as in the original ones, namely,
\begin{equation}
\tensor{ \tilde{W}}{^{ab}_{cd}}=-2\tilde{\chi}(\tilde{r}) \frac{(D-3)}{(D-1)}\tensor{ \tilde{w}}{^{ab}_{cd}}\, ,
\end{equation}
where the tensorial structure $\tensor{ \tilde{w}}{^{ab}_{cd}}$ is given by
\begin{equation}
\tensor{ \tilde{w}}{^{ab}_{cd}}=2\tilde{\tau}^{[a}_{[c} \tilde{\rho}^{b]}_{d]}-\frac{2}{(D-2)} \left(\tilde{\tau}^{[a}_{[c} \tilde{\sigma}^{b]}_{d]}+\tilde{\rho}^{[a}_{[c} \tilde{\sigma}^{b]}_{d]} \right)+\frac{2}{(D-2)(D-3)} \tilde{\sigma}^{[a}_{[c} \tilde{\sigma}^{b]}_{d]}.
\end{equation}
Here, $\tilde{\rho}_a^b$ denotes the projection onto our new radial coordinate $\tilde{r}$ and $\tau_a^b$ and $\sigma_a^b$ are defined in sec.~\ref{sec:numbertheo}. If we define $\tilde{H}_a^b=\tilde{\tau}_a^b+\tilde{\rho}_a^b$, we may express $\tensor{ \tilde{w}}{^{ab}_{cd}}$ as
\begin{equation}\label{mywayw}
\tensor{ \tilde{w}}{^{ab}_{cd}}=\tilde{H}^{[a}_{[c} \tilde{H}^{b]}_{d]}-\frac{2}{(D-2)} \tilde{H}^{[a}_{[c} \tilde{\sigma}^{b]}_{d]}+\frac{2}{(D-2)(D-3)} \tilde{\sigma}^{[a}_{[c} \tilde{\sigma}^{b]}_{d]}\, .
\end{equation}
Consequently, the covariant derivative of the Weyl tensor turns out to be
\begin{equation}
\left. \nabla_e \tensor{\tilde{W}}{^{ab}_{cd}}  \right|_{\rm SSS}=-2\frac{(D-3)}{(D-1)} \left[ \frac{\diff \tilde{\chi}}{\diff \tilde{r}} \delta_e^1 \tensor{ \tilde{w}}{^{ab}_{cd}}+\tilde{\chi}(\tilde{r}) \left. {\nabla}_e \tensor{ \tilde{w}}{^{ab}_{cd}} \right|_{\rm SSS} \right]\, ,
\end{equation}
where we are denoting the components of the covariant derivative of any tensor $T$ in our new coordinates as $\nabla_e \tilde{T}_{ab...}^{cd...}$. Hence we just need to work out $\left. {\nabla}_e \tensor{ \tilde{w}}{^{ab}_{cd}} \right|_{\rm SSS}$. Using eq.~\eqref{mywayw}, we find
\begin{equation}
\begin{split}
\nabla_e \tensor{ \tilde{w}}{^{ab}_{cd}}&=2 \nabla_e \tilde{H}^{[a}_{[c} \tilde{H}^{b]}_{d]}-\frac{2}{(D-2)} \nabla_e \tilde{H}^{[a}_{[c} \tilde{\sigma}^{b]}_{d]}\\&-\frac{2}{(D-2)} \nabla_e \tilde{\sigma}^{[a}_{[c} \tilde{H}^{b]}_{d]}+\frac{4}{(D-2)(D-3)} \nabla_e \tilde{\sigma}^{[a}_{[c} \tilde{\sigma}^{b]}_{d]}\, .
\end{split}
\end{equation}
Since $\nabla_e \tilde{H}_a^b+\nabla_e \tilde{\sigma}_a^b=0$, we just need to compute $\nabla_e \tilde{H}_a^b$. A straightforward calculation produces
\begin{equation}
\nabla_e \tilde{H}_a^b=\frac{1}{(r(\tilde{r}))^3}\frac{\diff r}{\diff \tilde{r}} \tilde{g}_{eg} \tilde{\sigma}_a^f \delta_1^b +\frac{1}{r(\tilde{r})}\frac{\diff r}{\diff \tilde{r}} (D-2) \tilde{\sigma}_e^b \delta_a^1\, .
\end{equation}
Using this, the covariant derivative of the Weyl tensor gives
\begin{align}
\label{cdwtab}
\left. \nabla_e \tensor{ \tilde{W}}{^{ab}_{cd}}  \right|_{\rm SSS}=&-2\frac{(D-3)}{(D-1)}  \frac{\diff \tilde{\chi}}{\diff\tilde{r}} \delta_e^1 \tensor{ \tilde{w}}{^{ab}_{cd}} -2\frac{(D-3)}{(D-1)} \tilde{\chi}(\tilde{r}) \frac{\diff r}{\diff\tilde{r}} \bigg[ \frac{2}{(r(\tilde{r}))^3} \tilde{g}_{ef} \tilde{\sigma}_{[c|}^{f} \delta_1^{[a} \tilde{H}_{| d]}^{b]}\\ \notag &+\frac{2(D-2)}{r(\tilde{r})} \tilde{\sigma}_e^{[a|} \delta_{[c}^1 \tilde{H}_{d]}^{|b]}-\frac{2 }{(D-2)(r(\tilde{r}))^3}\tilde{g}_{ef} \tilde{\sigma}_{[c|}^{f} \delta_1^{[a} \tilde{\sigma}_{| d]}^{b]}\\ \notag &-\frac{2}{r(\tilde{r})}\tilde{\sigma}_e^{[a|} \delta_{[c}^1 \tilde{\sigma}_{d]}^{|b]}  +\frac{2 }{(D-2)(r(\tilde{r}))^3}\tilde{g}_{ef} \tilde{\sigma}_{[c|}^{f} \delta_1^{[a} \tilde{H}_{| d]}^{b]}+\frac{2 }{r(\tilde{r})}\tilde{\sigma}_e^{[a|} \delta_{[c}^1 \tilde{H}_{d]}^{|b]}\\  \notag&-\frac{4}{(D-2)(D-3)(r(\tilde{r}))^3}\tilde{g}_{ef} \tilde{\sigma}_{[c|}^{f} \delta_1^{[a} \tilde{\sigma}_{| d]}^{b]}-\frac{4 }{(D-3)r(\tilde{r})}\tilde{\sigma}_e^{[a|} \delta_{[c}^1 \tilde{\sigma}_{d]}^{|b]} \bigg]\, .
\end{align}
Equipped with expression \eqref{cdwtab}, we may infer the general form of any invariant $\left. \mathcal{R}_2^{\{1,1\}} \right |_{\rm SSS}$ as defined in eq.~\req{eq:Inv2}. Since the $\left. \mathcal{R}_2^{\{1,1\}} \right |_{\rm SSS}$ are scalars, we can obtain them expressed in the original coordinates by performing all calculations in the new ones and then substituting any dependence on $\tilde{r}$ by the initial radial coordinate $r$.

We notice the following facts:
a) any $\left. \mathcal{R}_2^{\{1,1\}} \right |_{\rm SSS}$ will have three types of terms: those carrying  a factor $\tilde{\chi}^n \left (\diff \tilde{\chi}/\diff \tilde{r} \right)^2$,  those involving a factor $\tilde{\chi}^{n+1}(\diff \tilde{\chi}/\diff \tilde{r}) (\diff r/\diff\tilde{r})$ and a third type of terms with the common factor $\tilde{\chi}^{n+2}(\diff r/\diff \tilde{r}  )^2$; b) since $\tilde{r}$ is dimensionless, we infer that the first type of terms is not weighted by any power of $r$, the second type is accompanied by $r^{-1}$ and the third type, by $r^{-2}$. An additional overall factor of $r^{-2}$ is required by dimensional analysis.
Using these observations, it follows that
\begin{equation}
\left. \mathcal{R}_2^{\{1,1\}} \right |_{\rm SSS}=\frac{\tilde{\chi}^n(\tilde{r})}{r(\tilde{r})^2}\bigg[ c_1 \left (\frac{\diff \tilde{\chi}}{\diff \tilde{r}} \right)^2+c_2  \frac{\diff \tilde{\chi}}{\diff\tilde{r}} \frac{\diff r}{\diff\tilde{r}} \frac{\tilde{\chi}(\tilde{r})}{r(\tilde{r})}+c_3  \left ( \frac{\tilde{\chi}(\tilde{r})}{r(\tilde{r})} \right )^2 \left( \frac{\diff r}{\diff \tilde{r}}  \right)^2\bigg]\, ,
\end{equation}
for some constants $c_1$, $c_2$, $c_3$ which will depend on the specific term.
Taking into account that $\diff \tilde{r}/\diff r =1/ (r \sqrt{f(r)})$
we finally find
\begin{equation}
\left. \mathcal{R}_2^{\{1,1\}} \right |_{\rm SSS}=\chi^n f(r)\left ( c_1  (\chi ')^2+c_2\frac{\chi \chi'}{r}+c_3 \frac{\chi^2}{r^2} \right ),
\end{equation}
where $\chi'=\diff \chi/\diff r$.

\chapter{Cancellation of divergences for spherical entangling regions in quadratic curvature gravity}\label{Appendix A}

In this section, we present explicit computations in the Kounterterms scheme and show the cancellation of divergences appearing in the entanglement entropy for CFTs dual to QC gravity when the entangling region is a sphere.

\section*{Three-dimensional case}

For three dimensions, the universal part of the entanglement entropy \eqref{eq:SEEren} reduces to
\begin{IEEEeqnarray}{rl}\label{sd1}
\SEE^{\text{ren}}(\mathbb{B}^2)=\frac{\pi R}{2\GN}\int_{\delta}^{z_{\max}}\frac{\diff z}{z^2}\left(\Ls^2-24\alpha_1-6\alpha_2    \right)+\frac{c_3}{2\GN}\int_{\Sigma}B_1,
\end{IEEEeqnarray}
where the auxiliary function $c_3$ and the boundary term $B_1$, defined in eq. \eqref{eq:Bdminus2odd} in the case of odd-dimensions. They resepectively read
\begin{IEEEeqnarray}{rl}
c_3&=\frac{1}{4}\left(\Ls^2-24\alpha_1-6\alpha_2\right),\\
B_1&=-2\sqrt{{}_{\Sigma}\gamma}\K\,\diff \Omega_1=-\frac{2}{\delta}R\left[1+\pazocal{O}\left(\delta^2\right)\right]\diff \Omega_1.
\end{IEEEeqnarray}
On the other hand, the determinant of the metric $\tilde{\sigma}$ is given by
\begin{equation}\label{eq:dettildeh}
\sqrt{{}_{\Sigma}\gamma}=\frac{R\Ls}{\delta}\sqrt{1-\frac{\delta^2}{R^2}}=\Ls\left[\frac{R}{\delta}-\frac{\delta}{2R}-\frac{\delta^3}{R^3}+\pazocal{O}\left(\delta^4\right)\right].
\end{equation}
Also, $\K$ is the trace of the extrinsic curvature $\K_{\alpha\beta}=-\frac{1}{2\sqrt{{}_{\Sigma}\gamma_{zz}}}\partial_z({}_{\Sigma}\gamma_{zz})$ of the Fefferman-Graham-like expansion \eqref{eq:IndSigma2}, which reads
\begin{equation}
\K_{\alpha\beta}=\Ls\left[\frac{R^2}{\delta^2}-\frac{1}{2}-\frac{\delta^2}{8R^2}+\pazocal{O}\left(\delta^4\right)\right]\Omega_{\alpha\beta}.
\end{equation}
Since the inverse metric reads,
\begin{equation}
{}_{\Sigma}\gamma^{\alpha\beta}=\frac{1}{\Ls^2}\left[\frac{\delta^2}{R^2}+\pazocal{O}\left(\delta^4\right)\right]\Omega^{\alpha\beta},
\end{equation}
then, the expansion of the trace yields
\begin{equation}
    \K=\frac{1}{\Ls}\left[1+\frac{\delta^2}{2R^2}+\pazocal{O}\left(\delta^4\right)\right].
\end{equation}
In consequence, according to eq.~\eqref{eq:ktS} and up to leading order, the Kounterterm in this case reads
\begin{equation}\label{eq:Sd3}
    S_{\text{kt}}(\mathbb{B}^2)=-\frac{\pi R}{2\GN \delta}
    \left(\Ls^2-24\alpha_1-6\alpha_2\right)+\pazocal{O}(\delta).
\end{equation}
Thus, upon performing the integral in eq.~\eqref{sd1} on the extremal surface, along the Poincaré coordinate from $z=\delta$ to $z=R$, one gets
\begin{IEEEeqnarray}{rl}
\SEE^{\text{ren}}(\mathbb{B}^2)=F_0+\left(\Ls^2-24\alpha_1-6\alpha_2\right)\left[\frac{\pi R}{2\GN \delta}-\frac{\pi R}{2\GN \delta}\right]+
    \pazocal{O}(\delta),
\end{IEEEeqnarray}
where $F_0=-\frac{\pi}{2G}\left(\Ls^2-24\alpha_1-6\alpha_2\right)$ is the universal finite part. Thus, it becomes manifest that upon taking the $\delta \rightarrow 0$ limit, $\SEE^{\text{ren}}(\mathbb{B}^2)$ recovers the universal finite part and thus the holographic entanglement entropy is renormalized correctly.

\section*{Four-dimensional case}

In the four-dimensional case, from eq.\eqref{eq:Bdminus2even} it can be seen that the corresponding boundary term is given by
\begin{equation}
    B_{2}=-2\Ls\frac{\sqrt{R^2-\delta^2}}{\delta^2},
\end{equation}
and hence, the Kounterterm expanded around $\delta=0$ reads
\begin{equation}
    S_{\text{kt}}(\mathbb{B}^3)=-\frac{a_4\pi\Ls^{3}}{2\GN}\left(\frac{R^2}{\delta^2}-\frac{1}{2}\right).
\end{equation}
On the other hand, the bare entanglement entropy is given by
\begin{IEEEeqnarray}{rl}
\SEE=&\frac{a_4\pi}{\GN}\int_\delta^{z_{\max}}\diff z\ R\left(R^2-z^2\right)^{\frac{1}{2}}\left(\frac{\Ls}{z}\right)^{3}\\
    =&\frac{a_4\pi\Ls^3}{\GN}\left[\coth^{-1}\left(\frac{R}{\sqrt{R^2-z_\text{max}^2}}\right)-\frac{R}{z_\text{max}^2}\sqrt{R^2-z_\text{max}^2}\right]\nonumber\\
    &+\frac{a_4\pi\Ls^3}{2\GN}\left[\frac{R^2}{\delta^2}-\log\frac{R}{\delta}-\frac{1}{2}\left(1+\log 4\right)\right].\label{eq:S4d}
\end{IEEEeqnarray}
Since the first line of eq.~\eqref{eq:S4d} vanishes in the limit $z_\text{max}\rightarrow R$, then,
after adding the Kounterterm, the renormalized entanglement entropy reads
\begin{equation}\label{eq:RenS4}
    S^{\text{Univ}}_{\text{QCG}}=-\frac{a_4 \pi \Ls^3}{2\GN}\log\frac{2R}{\delta}.
\end{equation}
Notice that the Kounterterm isolates the logarthmic divergence, whose coefficient is universal and related to the type A charge of the CFT. A part of the finite term is cancelled and there is a piece left that is reabsorbed in the logarithmic divergence.

\section*{Bare holographic entanglement entropy in the arbitrary dimensional case}

For arbitrary dimensions, evaluating the quantities present in the entanglement entropy functional for QC gravity \eqref{eq:bQCGf} expression using the metric \eqref{eq:Poincare} yields
\begin{IEEEeqnarray}{rl}
    \SEE(\mathbb{B}^{d-1})=&\frac{1}{4\GN}\Bigg(\vol\left(\mathbb{S}^{d-2}\right)R\int_\delta^{z_{\max}}\diff z\left(R^2-z^2\right)^{\frac{d-3}{2}}\left(\frac{\Ls}{z}\right)^{d-1}\\
    &+\int_\Sigma\diffmone\sqrt{{}_{\Gamma_V}g}\left\{-\frac{2\alpha_1}{\Ls^2}\left[(d-1)(d-2)+4d-2\right]-\frac{2d\alpha_2}{\Ls^2}-\alpha_3\frac{(d-1)(d-2)}{\Ls^2}\right\}\Bigg),\nonumber
\end{IEEEeqnarray}
which can be rearranged, using definition of $a_d$ in eq.~\eqref{eq:ad}, to
\begin{equation}
    \SEE(\mathbb{B}^{d-1})=\frac{a_d\vol\left(\mathbb{S}^{d-2}\right)}{4\GN}\int_\delta^{z_{\max}}\diff z\ R\left(R^2-z^2\right)^{\frac{d-3}{2}}\left(\frac{\Ls}{z}\right)^{d-1}.
\end{equation}
After computing the integral and expanding around $\delta=0$, the holographic entanglement entropy yields
\begin{equation}\label{eq:bareSdodd}
\SEE(\mathbb{B}^{d-1})=\SEE^{\text{ren}}(\mathbb{B}^{d-1})+\frac{a_d\vol\left(\mathbb{S}^{d-2}\right)\Ls^{d-1}}{4\GN}\left[\frac{1}{d-2}\frac{R^{d-2}}{\delta^{d-2}}-\frac{d-3}{2(d-4)}\frac{R^{d-4}}{\delta^{d-4}}+\pazocal{O}\left(\delta^{-(d-6)}\right)\right].
\end{equation}
In the previous expression, the universal term is given by
\begin{equation}
\SEE^{\text{ren}}(\mathbb{B}^{d-1})=\begin{dcases*}(-1)^{\frac{d-1}{2}}\frac{a_d(4\pi)^{\frac{d-1}{2}}\left(\frac{d-1}{2}\right)!\Ls^{d-1}}{4G(d-1)!} & \text{if $d$ odd},\\
(-1)^{\frac{d}{2}-1}\frac{a_d\pi^{\frac{d}{2}-1}\Ls^{d-1}}{2\GN(\frac{d-2}{2})!}\log\frac{2R}{\delta}&  \text{if $d$ even,}\end{dcases*}
\end{equation}
where $z_\text{max}$ is set to the radius $R$ and the volume of the sphere is given by 
\begin{equation}
\vol(\mathbb{S}^{d-2})=\frac{2^d\pi^{\frac{d}{2}-1}(\frac{d}{2})!}{d(d-2)}.
\end{equation}

\section*{Cancellation of divergences for odd dimensions}

Let us now focus on the Kounterterm. In the odd-dimensional case, the boundary form can be read, as before, from eq.\eqref{eq:Bdminus2odd}.

Since we are considering a spherical entangling region, the extrinsic curvature and the Riemann tensor read, respectively,
\begin{equation}\label{eq:sphECRT}
    \K\indices{_\alpha^\beta}=\frac{1}{\Ls}\frac{R}{\sqrt{R^2-\delta^2}}\delta_\alpha^\beta,\quad \R\indices{_\alpha_\beta^\gamma^\delta}=\frac{1}{\Ls^2}\frac{\delta^2}{R^2-\delta^2}\delta_{\alpha\beta}^{\gamma\delta}
\end{equation}
Plugging these values into the boundary form, we find
{\footnotesize \begin{align}
    B_{d-2} & =-(d-1)!\int_0^1\sqrt{{}_{\Sigma}\gamma}\delta^{\alpha_1\cdots \alpha_{d-2}}_{\beta_1\cdots \beta_{d-2}}\delta^{\alpha_1}_{\alpha_1}\frac{1}{\Ls}\frac{R}{\sqrt{R^2-\delta^2}}\left(\frac{1}{2\Ls^2}\frac{\delta^2}{R^2-\delta^2}\delta^{\beta_2\beta_3}_{\alpha_2\alpha_3}-s^2\frac{1}{\Ls^2}\frac{R^2}{R^2-\delta^2}\delta^{\beta_2}_{\alpha_2}\delta^{\beta_3}_{\alpha_3}\right)\times\cdots\notag \\
    & \cdots\times\left(\frac{1}{2\Ls^2}\frac{\delta^2}{R^2-\delta^2}\delta^{\beta_{d-3}\beta_{d-2}}_{\alpha_{d-3}\alpha_{d-2}}-s^2\frac{1}{\Ls^2}\frac{R^2}{R^2-\delta^2}\delta^{\beta_{d-3}}_{\alpha_{d-3}}\delta^{\beta_{d-2}}_{\alpha_{d-2}}\right).
\end{align}}

In this expression, we make use the relations
\begin{IEEEeqnarray}{rl}
&\delta^{\alpha_1\cdots \alpha_{d-3}}_{\beta_1\cdots \beta_{d-3}}\delta^{\beta_2\beta_3}_{\alpha_2\alpha_3}=2\delta^{\alpha_1\cdots \alpha_{d-3}}_{\beta_1\cdots {d-3}}\delta^{\beta_2}_{\alpha_2}\delta^{\beta_3}_{\alpha_3},\nonumber\\
\delta_{\beta_1\cdots \beta_{d-2}}^{\alpha_1\cdots \beta_{d-2}}\delta^{\beta_{d-2}}_{\alpha_{d-2}}&=\delta_{\beta_1\cdots \beta_{d-3}}^{\alpha_1\cdots \alpha_{d-3}},\quad \delta_{\beta_1\cdots \beta_{d-3}}^{\alpha_1\cdots \alpha_{d-3}}\delta^{\beta_1}_{\alpha_1}\cdots\delta^{\beta_{d-3}}_{\alpha_{d-3}}=(d-2)!\ .
\end{IEEEeqnarray}
Besides, writing explicitly the determinant $\sqrt{{}_{\Sigma}\gamma}$ given in eq.~\eqref{eq:dettildeh}, the boundary form reads
\begin{equation}
     B_{d-2}=-(d-1)!\int_0^1\diff s\left[\frac{R^2}{\delta^2}\left(1-\frac{\delta^2}{R^2}\right)\right]^{\frac{d-2}{2}}\frac{R}{\sqrt{R^2-\delta^2}}\left(\frac{\delta^2}{R^2-\delta^2}-s^2\frac{R^2}{R^2-\delta^2}\right)^{\frac{d-3}{2}}.
     \end{equation}
     This expression can be expanded around $\delta=0$ as
     \begin{equation}
         B_{d-2}=-(d-1)!\int_0^1\diff s\left[(-1)^{\frac{d+1}{2}}s^{d-3}\frac{R^{d-2}}{\delta^{d-2}}-(-1)^{\frac{d+1}{2}}s^{d-5}\frac{d-3}{2}\frac{R^{d-4}}{\delta^{d-4}}+\pazocal{O}\left(\delta^{-(d-6)}\right)\right],
     \end{equation}
which after the integration reads
\begin{equation}
    B_{d-2}=-(d-1)!\left[(-1)^{\frac{d+1}{2}}\frac{1}{d-2}\frac{R^{d-2}}{\delta^{d-2}}-(-1)^{\frac{d+1}{2}}\frac{d-3}{2(d-4)}\frac{R^{d-4}}{\delta^{d-4}}+\pazocal{O}\left(\delta^{-(d-6)}\right)\right].
\end{equation}
Therefore the Kounterterm $S_{\text{kt}}$, after using the definition $c_d$ in the odd dimensional case from eq.~\eqref{eq:ad}, becomes
\begin{equation}
   S_{\text{kt}}=-\frac{a_d\vol\left(\mathbb{S}^{d-2}\right)\Ls^{d-1}}{4\GN}\left[\frac{1}{d-2}\frac{R^{d-2}}{\delta^{d-2}}-\frac{d-3}{2(d-4)}\frac{R^{d-4}}{\delta^{d-4}}+\pazocal{O}\left(\delta^{-(d-6)}\right)\right].
\end{equation}
Therefore, adding the previous expression to the bare entanglement entropy found in eq.\eqref{eq:bareSdodd}, the renormalized entanglement entropy reads
\begin{equation}
    \SEE^{\text{ren}}(\mathbb{B}^{d-1})=(-1)^{\frac{d-1}{2}}F_0.
\end{equation}

\section*{Cancellation of divergences in even dimensions}

Proceeding in the same way as in the odd-dimensional case, the extrinsic curvature and the Riemann tensor, given in eq.~\eqref{eq:sphECRT}, are used along with the identities
\begin{IEEEeqnarray}{rl}
\delta_{\beta_1\cdots \beta_{d-3}}^{\alpha_1\cdots \alpha_{d-3}}\delta_{\beta_2\beta_3}^{\alpha_2\alpha_3}=&2\delta_{\alpha_1\cdots \alpha_{d-3}}^{\beta_1\cdots \beta_{d-3}}\delta^{\beta_2}_{\alpha_2}\delta^{\beta_3}_{\alpha_3},\quad
\delta_{\beta_1\cdots \beta_{d-3}}^{\alpha_1\cdots \alpha_{d-3}}\delta^{\beta_{d-3}}_{\alpha_{d-3}}=2\delta_{\beta_1\cdots \beta_{d-4}}^{\alpha_1\cdots \alpha_{d-4}},\nonumber\\
&\delta^{\alpha_1\cdots \alpha_{d-4}}_{\beta_1\cdots \beta_{d-4}}\updelta^{\beta_1}_{\alpha_1}\cdots\updelta^{\beta_{d-4}}_{\alpha_{d-4}}=\frac{(d-2)!}{2}.
\end{IEEEeqnarray}
After implementing all these relations and substituting $\detsigtil$ the boundary form reads
\begin{equation}
    B_{d-2}=-(d-2)^2(d-3)!\Ls\int_0^1\diff s\int_0^s\diff t\ b(t,s,\delta),
\end{equation}
where the  the introduced function $b(t,s)$ reads
\begin{equation}
    b(t,s,\delta)=\frac{R^{d-2}}{\delta^{d-2}}\left(1-\frac{\delta^2}{R^2}\right)^{d-2}\frac{R}{\sqrt{R^2-\delta^2}}\left(\frac{\delta^2}{R^2-\delta^2}-s^2\frac{R^2}{R^2-\delta^2}+t^2\right)^{\frac{d}{2}-2}.
\end{equation}
If the function $b(t,s,\delta)$ is expanded around $\delta=0$, then
{\small\begin{equation}
    b(t,s,\delta)=\left(t^2-s^2\right)^{\frac{d}{2}-2}\frac{R^{d-2}}{\delta^{d-2}}+\frac{1}{2}\left(t^2-s^2\right)^{\frac{d}{2}-3}\left[d-4+s^2-(d-3)t^2\right]\frac{R^{d-4}}{\delta^{d-4}}+\pazocal{O}\left(\delta^{-(d-6)}\right).
\end{equation}}
In this expression, the convergence condition demands that the order $\mathcal{O}\left(\delta^{-(d-2)}\right)$ term appears for $d>2$. Likewise, the order $\pazocal{O}\left(\delta^{-(d-4)}\right)$ term appears when $d>4$ and successively. Now, computing the integrals
\begin{IEEEeqnarray}{rl}
\int_0^1\diff s\int_0^s\diff t\left(t^2-s^2\right)^{\frac{d}{2}-2}&=\frac{(-1)^{\frac{d}{2}}2^{d-2}\left(\frac{d}{2}-1\right)!^2}{(d-2)^2(d-2)!},\\
\int_0^1\diff s\int_0^s\diff t\left(t^2-s^2\right)^{\frac{d}{2}-3}\left[d-4+s^2-(d-3)t^2\right]&=-\frac{(-1)^{\frac{d}{2}}2^{d-2}(d-3)\left(\frac{d}{2}-1\right)!^2}{(d-2)(d-4)(d-2)!},\nonumber\\
\end{IEEEeqnarray}
the boundary term reduces to
\begin{IEEEeqnarray}{rl}
        B_{d-2}=&(-1)^{\frac{d}{2}+1}2^{d-2}\Ls\left[\left(\frac{d}{2}-1\right)!\right]^2\left[\frac{1}{d-2}\frac{R^{d-2}}{\delta^{d-2}}-\frac{d-3}{2(d-4)}\frac{R^{d-4}}{\delta^{d-4}}+\pazocal{O}\left(\delta^{-(d-6)}\right)\right]\nonumber\\
        &+\frac{(d-2)!\Ls}{2}H_{\frac{d}{2}-1}.
\end{IEEEeqnarray}
Notice that in this expression, the last term is finite and it is written in terms of the $\left(\frac{d}{2}-1\right)$-th harmonic number $H_{\frac{d}{2}-1}=\sum_{i=0}^{\frac{d}{2}-1}\frac{1}{i}$. 

Once the boundary term is computed, the Kounterterm $S_{\text{kt}}$ is derived easily from eq.~\eqref{eq:ktS} for the even dimensional case, obtaining
\begin{IEEEeqnarray}{rl}
    \SKt=&-\frac{a_d\vol(\mathbb{S}^{d-2})\Ls^{d-1}}{4\GN}\left[\frac{1}{d-2}\frac{R^{d-2}}{\delta^{d-2}}-\frac{d-3}{2(d-4)}\frac{R^{d-4}}{\delta^{d-4}}+\pazocal{O}\left(\delta^{-(d-6)}\right)\right]\nonumber\\
    &+\frac{(d-2)!}{2}H_{\frac{d}{2}-1}.
\end{IEEEeqnarray}
The structure of power-law divergences is the same as in the odd dimensional case. However, in even-dimensional CFTs, the bare entanglement entropy \eqref{eq:bareSdodd} differs in a finite term and in a log term, whose coefficient is the universal part of the entanglement entropy.  Following the procedure in ref.~\cite{Anastasiou:2019ldc}, the log term is successfully isolated in arbitrary even $d$, reading
\begin{equation}
    \SEE^{\text{ren}}=(-1)^{\frac{d}{2}-1}\frac{a_d\pi^{\frac{d}{2}-1}\Ls^{d-1}}{2\GN(\frac{d}{2}-1)!}\log\frac{2R}{\delta}.
\end{equation}

\chapter{Mutual information regularization of entanglement entropy in the EMI model}\label{EMIMIregu}
In Section \ref{secemi} we computed $F$ numerically for many kinds of entangling regions in the EMI model. In order to do so, we introduced a small regulator $\delta$ along an auxiliary extra dimension, evaluated $F^{\rm \ssc EMI}(A)$ as a function of $\delta$ and then extracted the $\delta \rightarrow 0$ limit. However, as we explained in Section \ref{elipsq} there is an alternative way of regularizing EE that makes use of the MI of concentric regions and which yields converging results in the lattice. In this appendix we explore this alternative method for the EMI model in the case of elliptic entangling regions for the two geometric setups explained in Section \ref{elipsq}, namely: fixing a constant $\varepsilon$, and forcing the concentric regions to be ellipses. This will allow us to test to what extent we may expect both methods to differ in general when the size of the entangling regions cannot be considered to be extremely larger than the separation between the auxiliary regions ---which is always the case in the lattice. We will also verify the convergence of both methods between themselves and with the $\delta$ method used in the main text.

In the three-dimensional EMI model, the mutual information between two regions $V,V_2$ in a fixed time slice is given by
\begin{equation}
I^{\rm \ssc EMI}(V,W)=-2\kappa_{(3)}\int_{\partial V}\diff\mathbf{r}_V\int_{\partial W}\diff\mathbf{r}_W\, \frac{\mathbf{n}_{V}\cdot\mathbf{n}_{W}}{\abs{\mathbf{r}_{V}-\mathbf{r}_{W}}^{2}}\, ,
\end{equation}
where the normal vectors are defined outwards to each region, respectively. This formula is manifestly non-divergent, as $\mathbf{r}_V$ and $\mathbf{r}_W$ correspond now to different regions. For regions $V,W$ parametrized by $[x_{V,W}(t),y_{V,W}(t)]$ we have
\begin{equation}\label{emimi}
I^{\rm \ssc EMI}(V,W)=-2\kappa_{(3)}\int_{\partial V}\diff t_{V}\int_{\partial W}\diff t_{W}\, \frac{ \dot y_V(t_{V}) \dot y_W(t_{W}) + \dot x_V(t_{V}) \dot x_W(t_{W})  }{  [x_V(t_{V})-x_W(t_W)]^2+(y_V(t_V)-y_W(t_W))^2}\, .
\end{equation}

Applied to the setup required for regularizing the entanglement entropy, as in \req{annulii}, we need to evaluate the mutual information $I^{\rm \ssc EMI}(V^+,W^-)$, which involves integrals over $\partial V^+=\partial \overline{V^+}$ and $\partial V^-$. Using the entanglement entropy formula for the EMI model, \req{emiee}, it is easy to see that in $I^{\rm \ssc EMI}(V^+,V^-)$ the contributions from $\see^{\rm \ssc EMI}(\overline{V^+})$ and $\see^{\rm \ssc EMI}(V^-)$ cancel with the terms in $\see^{\rm \ssc EMI}(\overline{V^+ \cup V^-})$ which involve performing both integrals over  $\partial V^+$ or both integrals over $\partial V^-$. The only terms which survive are the ones which involve one integral over $\partial V^+$ and the other one over $\partial V^-$, in agreement with eq.~\eqref{emimi}.

 \begin{figure}[t!] \centering
	\includegraphics[scale=0.7]{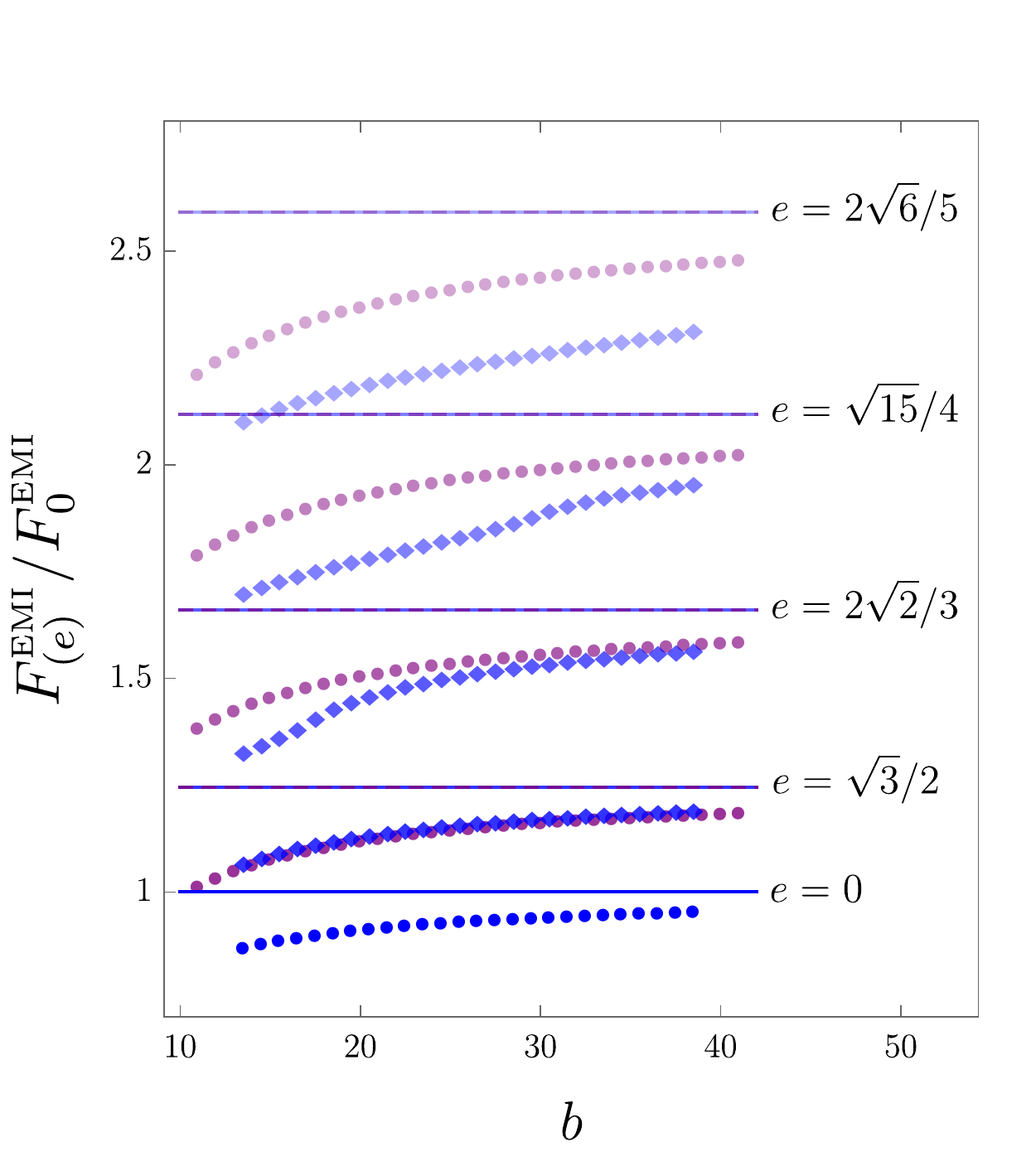}
	\caption{ \textsf{We plot $F_{(e)}^{\rm \ssc EMI}/F_0^{\rm \ssc EMI}$ for different values of the eccentricity $e= 0,\sqrt{3}/2,2\sqrt{2}/3,\sqrt{15}/4,2\sqrt{6}/5$ using the expressions \req{ik130emi} (purple dots) and \req{ik13emi} (blue diamonds). In the case of  \req{ik130emi}, we fix $\varepsilon=5$ and plot the results as a function of the semi-minor axis $b$. In the case of \req{ik13emi} we fix $b_2=b_1+7$ and plot the results as a function of $b=(b_1+b_2)/2$ as well. The straight lines correspond to the exact values obtained using the $\delta$ method, which can be also reproduced using \req{ik130emi} or \req{ik13emi} for sufficiently large values of $b$.}}
	\label{ref44}
\end{figure}

In the case in which we keep $\varepsilon$ constant for all points of the ellipse boundary, the relevant formula is
 \begin{equation}\label{ik130emi}
 F^{\rm \ssc EMI}_{(e)}=-\frac{1}{2}\left[  I^{\rm \ssc EMI}({\rm pseudoellipse }_{i},{\rm pseudoellipse}_{o})-k_{{\rm \ssc EMI}}^{(3)} \frac{4 a E[e^2]}{\varepsilon}\right]\, ,
 \end{equation}  
where the parametrization of the pseudoellipses appears in \req{rio} and $k_{{\rm \ssc EMI}}^{(3)}$ is defined in \req{kemi}. On the other hand, if we force the auxiliary regions to be ellipses, the relevant formula reads
 \begin{equation}\label{ik13emi}
 F^{\rm \ssc EMI}_{(e)}=-\frac{1}{2}\left[  I^{{\rm \ssc EMI}}({\rm ellipse }_{2},{\rm ellipse}_{1})-k_{{\rm \ssc EMI}}^{(3)} \int_0^{\frac{\pi}{2}} \diff t \frac{2(a_1+a_2)\sqrt{1-e^2 \cos^2 t} }{\varepsilon(t)}\right]\, ,
 \end{equation} 
where $a=(a_1+a_2)/2$, $b=(b_1+b_2)/2$ and the formula for $\varepsilon(t)$ appears in  \req{epis}.

As a first check, we have verified that both eqs.~\eqref{ik130emi} and \eqref{ik13emi} produce results essentially identical to the ones obtained using the $\delta$ method for sufficiently large values of $b/\varepsilon$ ---\eg $b/\varepsilon \sim 100$ yields excellent results in all cases. On the other hand, practical limitations force us to consider smaller values of such a quotient when performing calculations in the lattice ($b/\varepsilon \sim 10$). In fig.~\ref{ref44} we have plotted the results achieved using both methods for values of $b$ similar to the ones considered in our lattice calculations in Section \ref{elipsq}. As we can see, in both cases the results tend to underestimate considerably the actual ones ---this underestimation tends to be greater as the eccentricity grows. Importantly, we observe that the pseudoellipses method makes a better job in approximating the actual results than the variable-$\varepsilon$ one. The difference between both methods becomes rather considerable for greater values of the eccentricity. For each eccentricity and each value of $b$, we can observe what is the factor we need to multiply the corresponding EMI result by in order to obtain the exact answer $F_{(e)}^{\rm \ssc EMI}$. We use such factors in sec.~\ref{elipsq} to correct the lattice results obtained for the corresponding values of $b/\delta$.